\documentclass[twoside,12pt]{article}
\usepackage{epsfig}

\newcommand{\be}{\begin{equation}}
\newcommand{\ee}{\end{equation}}
\newcommand{\bea}{\begin{eqnarray}}
\newcommand{\eea}{\end{eqnarray}}

\usepackage{amsmath}
\usepackage{amsfonts}
\usepackage{amssymb}
\usepackage{amsxtra}
\newcommand\spm{\mathrel{\text{\framebox[0.9\width]{$\pm$}}}}
\newcommand\smp{\mathrel{\text{\framebox[0.9\width]{$\mp$}}}}
\newcommand\cpm{\mathrel{\text{\textcircled{\makebox{$\pm$}}}}}
\newcommand\cmp{\mathrel{\text{\textcircled{\makebox{$\mp$}}}}}
\newcommand\sminus{\boxminus}
\usepackage{supertabular}
\begin{document}
\title{ \vspace{1cm}
How does Clifford algebra show the way to the second quantized fermions with
unified spins, charges and families, and with vector and scalar gauge 
fields beyond the {\it standard model} \\
%
}
\author{N.S.\ Manko\v c Bor\v stnik,$^{1}$ H.B.\ Nielsen$^2$\\
\\
$^1$Department of Physics, University of Ljubljana\\
SI-1000 Ljubljana, Slovenia\\
$^2$Niels Bohr Institute, University of Copenhagen\\
Blegdamsvej 17, Copenhagen\O, Denmark}
\maketitle

\begin{abstract}  



 Fifty years ago the {\it standard model} offered an elegant new step
towards understanding elementary fermion and boson fields, making several 
assumptions, suggested by experiments. The assumptions are still waiting 
for an explanation. 
There are many proposals in the literature for the next step. 
The {\it spin-charge-family} theory of one of us (N.S.M.B.) is offering the 
explanation for not only all by the {\it standard model} 
assumed properties of quarks and leptons and antiquarks and antileptons, with the 
families included, of the vectors gauge fields, of the Higgs's scalar and Yukawa 
couplings, but also for the second quantization postulates of Dirac and
for cosmological observations, like there are the appearance of the {\it dark matter},
of {\it matter-antimatter asymmetry}, making several predictions. This theory
proposes a simple starting action in $ d\ge (13+1)$-dimensional space with fermions
interacting with the gravity only (the vielbeins and the two kinds of the spin 
connection fields), what manifests in $d=(3+1)$ as the vector and scalar gauge 
fields, and uses the odd Clifford algebra to describe the internal space of 
fermions, what enables that the creation and annihilation operators for fermions
fulfill the anticommutation relations for the second quantized fields without Dirac's
postulates: Fermions single particle states already anticommute.
We present in this review article a short overview of the {\it spin-charge-family} 
theory, illustrating shortly on the toy model the breaks of the starting symmetries 
in $d=(13+1)$-dimensional space, which are triggered either by scalar fields --- 
the vielbeins with the space index belonging to $d>(3+1)$ --- or by the 
condensate of the two right handed neutrinos, with the family quantum 
number not belonging to the observed families. 
We compare properties and predictions of this theory with the properties and
predictions of $SO(10)$ unifying theories.
.

\end{abstract}



%
\section{Introduction}
\label{introduction}
%


Physicists are gaining knowledge about the laws, that apply to the smallest 
constituents of nature and thus to the whole universe, by observations, 
experiments, we make mathematical models, predict with models new events, 
check predictions by experiments, observations,  all these lead 
to new ideas, new models. 

Repeating this circle again and again we ask ourselves questions like:  
{\bf i.} 
Is the law of 
nature simple and elegant,  manifesting complexity on many fermion and 
boson states in particular at low energies, or ''nature uses mathematics, 
when just appears to be needed''? 
{\bf ii.}  
Is the space-time $(3+1)$-dimensional? Why? Or has it 
many more dimensions? How many? There are namely only two simple and
elegant choices, zero and infinity.
{\bf iii.}
What are elementary constituents and interactions among constituents in our 
universe, in any universe?
{\bf iv.} 
How has the space-time of our universe started?
{\bf v.} And many others.\\

\vspace{0.1cm}

\noindent
Fifty years ago the {\it standard model} offered an elegant new step towards 
understanding elementary fermion and boson fields by postulating: \\
{\bf a.} The existence of massless fermion family members with the  spins 
and charges in the fundamental representation of the groups, {\bf a.i.} the 
quarks as colour triplets and colourless leptons, {\bf a.ii} the left 
handed  members as the weak doublets, the right handed 
weak chargeless members, {\bf a.iii.} the left handed quarks differing from 
the left handed leptons in the hyper charge, {\bf a.iv.} all the right 
handed members differing among themselves in hyper charges, 
{\bf a.v.} antifermions carrying the corresponding anticharges of fermions and 
opposite handedness, {\bf a.vi.} the  families of massless fermions, 
suggested by experiments  and required by the gauge invariance of the boson 
fields (there is no right handed neutrino postulated, 
since it would carry none of the so far observed charges, and correspondingly 
there is also no left handed antineutrino allowed in the {\it standard model}).\\ 
 {\bf b.} The existence of massless vector gauge fields to the observed 
charges of quarks and leptons, carrying charges in the adjoint representations 
of the corresponding charged groups and manifesting the gauge invariance.\\
 {\bf c.}  The existence of the massive weak doublet scalar higgs, {\bf c.i.}
 carrying the weak charge $\pm \frac{1}{2}$  and the hyper charge 
 $\mp \frac{1}{2}$ (as it would be in the fundamental representation of the
 two groups), {\bf c.ii.} gaining at some step of the expanding universe the 
 nonzero vacuum expectation value, {\bf c.iii.} breaking the weak and the 
 hyper charge and correspondingly breaking the mass protection, {\bf c.iv.}
taking care of the properties of fermions and of the weak bosons masses, 
 {\bf c.v.} as well as the existence of the Yukawa couplings. \\
{\bf d.} The presentation of fermions and bosons as second quantized fields.\\
 {\bf e.} The gravitational field in $d=(3+1)$ as independent gauge field.
 (The {\it standard model} is defined  without gravity in
order that it be renormalizable, but yet the {\it standard model} particles 
are ''allowed'' to couple to gravity in the ''minimal'' way.)
\\
 %
 The {\it standard model} assumptions have been experimentally confirmed 
without raising any doubts so far, except for some few and possibly statistical 
fluctuations anomalies~\footnote{%
We think here on the improved {\it standard model}, 
in which neutrinos have non-zero masses, and has no ambition to explain
severe cosmological problems}, but also by offering no explanations for 
the assumptions.  
 The last among the fields postulated by the {\it standard model},  the 
scalar higgs, was detected in June 2012,  the gravitational waves 
 were detected in February 2016.
 
The {\it standard model}  has in the literature several  explanations, mostly 
with many new not explained assumptions. The most  popular seem to be 
the grand unifying theories~\cite{Geor,FritzMin,PatiSal,GeorGlas,Cho,ChoFreu,%
Zee,SalStra,DaeSalStra,Mec,HorPalCraSch,Asaka,ChaSla,Jackiw,Ant,Ramond,Horawa}. 
At least $SO(10)$-unifying theories offer the explanation for the postulates from 
{\bf a.i.} to {\bf a.iv}, partly 
to {\bf b.} by assuming that to all the fermion charges there exist the 
corresponding vector gauge fields --- but does not explain the  assumptions 
{\bf a.v.} up to {\bf a.vi.}, {\bf c.} and {\bf d.}, and does not connect gravity 
with gauge vector and scalar fields.


What questions should we ask ourselves to be able to find a trustworthy next step 
beyond the {\it standard models} of elementary particle physics and cosmology, 
which would offer understanding of not yet understood phenomena? \\
{\bf i.} Where do fermions, quarks and leptons, originate and why do they differ  
from the boson fields in spins, charges and statistics? \\
{\bf ii.} Why are charges of quarks and leptons so different, why have the 
left handed family members so different charges from the right handed ones and
why does the handedness relate charges to anticharges?\\
{\bf iii.} Isn't the first quantization an approximation to the second quantization?
How should one describe the internal degrees of fermions to explain the 
Dirac's postulates of the second quantization?\\
{\bf iv.} Where do families of quarks and leptons originate and how many 
families do exist? \\
{\bf v.} Why do family members -- quarks and leptons --- manifest so different 
masses if they all start as massless?\\
{\bf vi.} How is the origin of the scalar field (the Higgs's scalar) and the 
Yukawa couplings connected with the origin of families and how many scalar 
fields determine properties of the so far (and others possibly be) observed 
fermions and masses of weak bosons? (The Yukawa couplings certainly speak 
for the existence of  several scalar fields with the properties of Higgs's 
scalar.)
Why is  the Higgs's scalar, or are all scalar fields of similar properties 
as the higgs,  if there are several, doublets with respect to the weak and the 
hyper charge? \\
{\bf vii.} Do possibly exist also scalar fields with the  colour charges in the 
fundamental representation (like the Higgs's scalars are doublets with respect 
to the weak charge) and where, if they are, do they manifest?\\
{\bf  viii.} Where do the so far observed  (and others possibly non observed)  
vector gauge fields originate? Do they have anything in common with the scalar
fields and the gravitational fields?\\
{\bf ix.} Where does the {\it dark matter} originate?\\
{\bf x.}  Where does the "ordinary"  matter-antimatter asymmetry originate?\\
{\bf xi.} Where does the  dark energy originate and why is it so small?\\
{\bf xii.} What is the dimension of space? $(3+1)?$, $((d-1)+1)?$, $\infty?$ \\
{\bf xiii.} And many others. 

\vspace{1mm}

\noindent
The assumptions of the {\it standard model}  are still waiting for an explanation.\\

\vspace{1mm}

\noindent
{\it Motivation for using Clifford algebra objects to  describe the internal
space of fermions:}\\

In a long series of works we, mainly one of us N.S.M.B.~(\cite{norma92,%
norma93,IARD2016,n2014matterantimatter,nd2017,n2012scalars,%
JMP2013,normaJMP2015,nh2018} and the references therein), have found 
phenomenological success with the model named 
the {\it spin-charge-family}  theory,  with fermions, the internal space of which 
is described with the Clifford algebra of all linear 
combinations of odd products of $\gamma^a$'s in $d=(13+1)$, 
Sect.~\ref{internalspace},
interacting with only gravity~(\cite{IARD2020} and references 
therein), Sect.~\ref{fermionandgravitySCFT}. The spins of fermions from
 higher dimensions, $d>(3+1)$, manifest 
in $d=(3+1)$ as charges of the {\it standard model}, Sect.~\ref{internalspace},
gravity in higher dimensions manifest as the {\it standard model} gauge vector 
fields as well as the Higgs's scalar and Yukawa 
couplings, Sect.~\ref{fermionandgravitySCFT}~\cite{normaJMP2015,nd2017}.

Let be added that one irreducible representation of $SO(13,1)$ contains, 
if looked from the point of view of $d=(3+1)$,
all the quarks and leptons and antiquarks and antileptons just with the
properties, required by the {\it standard model}, including the relation between 
quarks and leptons and handedness  and antiquarks and antileptons and the 
opposite handedness, Table~\ref{Table so13+1.}. 
All what in the {\it standard model} have to be assumed (extremely efficiently
 "read" from experiments), in the 
{\it spin-charge-family} theory appear as a possibility 
from the starting simple action, Eq.~(\ref{wholeaction}).
This simple starting action offers the explanation for not only the properties 
of quarks and leptons and antiquarks and antileptons, but also for vector 
gauge fields, scalar gauge fields, which represent higgs and explain the 
Yukawa couplings and offers also the scalars, which cause matter/antimatter 
asymmetry, proton decay, the appearance of the dark matter, 
Sect.~\ref{fermionandgravitySCFT}.





It has happened so many times in the history of science that the simpler,  
more elegant, model has shown up as a more ''powerful'' one.
Examples  can be found in nonrelativistic classical physics, after embedded it 
into relativistic physics, in Newton's laws after embedded them in general theory 
of gravity, in symmetries of fermion and gauge fields when embedded 
symmetries into larger groups, etc..\\
%
The working hypotheses of at least one of us (N.S.M.B.) is that the laws of 
nature are simple and correspondingly elegant and that the many body 
systems around the phase transitions look to us complicated at least from 
the point of view of the elementary constituents of fermion and boson fields.

To this working hypotheses belongs also the description of the internal space
of fermions with the Clifford algebra.


There appear in the literature two anticommuting kinds of algebras, the 
Grassmann algebra~\cite{norma93}  and the Clifford algebra~\cite{clifford,%
Lounesto2001,MP2017}. 
We recognized~\cite{norma92,norma93,norma95,norma94,nh02,nh03,
normaBled2020,n2019PIPII,2020PartIPartII} that the Grassmann algebra includes
two  Clifford algebras, anticommuting with each other, each of them expressible 
with  the Grassmann algebra elements and opposite. 

The Grassmann algebra, with elements $\theta^a$, 
and their Hermitian conjugated partners 
$\frac{\partial}{\partial \theta^a}$~\cite{nh2018}, 
can  be used to describe the internal space of fermions with the integer spins 
and charges in the adjoint representations, each of the two Clifford algebras, 
we denote their elements with $\gamma^a$ and 
$\tilde{\gamma}^a$~\cite{norma92,norma93,norma95,norma94,nh02,nh03}, 
can be used to describe the half integer spins and charges in fundamental 
representations. 
The Grassmann algebra  is expressible with the two Clifford algebras and 
opposite, Sects.~\ref{propertiesGrass0},~\ref{propertiesCliff0}.


In both algebras, the Grassmann algebra and the two Clifford algebras, the odd 
products of objects, arranged to be eigenstates of the Cartan subalgebras of the 
corresponding Lorentz algebras, form creation and annihilation operators, which 
fulfill the anticommutation relations postulated by Dirac for the second quantized 
fermions. These creation operators can be used to describe the internal space of 
fermions, Sects~\ref{propertiesGrass0},~\ref{propertiesCliff0}.\\

\vspace{0.2cm}

\noindent
{\it  Steps to the second quantized states with the Clifford or Grassmann algebra:}\\

\noindent
{\bf i.}  The internal space of a fermion, 
Sects.~\ref{internalspace},$\,$ \ref{GrassmannClifford}, is described by either Clifford or 
Grassmann 
algebra of an odd Clifford character (superposition of an odd number of Clifford 
"coordinates" --- operators --- $\gamma^{a}$'s or of an odd number of Clifford
 "coordinates" --- operators --- $\tilde{\gamma}^{a}$'s)
or of an odd Grassmann character (superposition of an odd number of Grassmann 
"coordinates"  --- operators --- $\theta^a$'s).\\
{\bf ii.} The eigenvectors of all the (chosen) Cartan subalgebra members of the 
corresponding Lorentz algebra 
are used to define the "basis vectors" of an odd character. 
(The Cartan subalgebra is in all three cases chosen in the way to be in agreement
 with the ordinary choice.)
The algebraic application of this "basis vectors" on the corresponding vacuum state (either
 Clifford $|\psi_{oc}>$, defined in Eq.~(\ref{vac1}), or Grassmann $|\phi_{og}>$, 
Eq.~(\ref{vactheta}), which  is in the Grassmann case just the identity) generates 
the "basis states", usable for describing the internal space of fermions. The members 
of the "basis vectors" manifest together with their Hermitian conjugated partners 
the properties of creation and annihilation operators,  which anticommute, 
Eq.~(\ref{ijthetaprod}) in the Grassmann case and Eq.~(\ref{almostDirac}) in the 
Clifford case, when
applying on the corresponding vacuum state, due to the algebraic properties of the 
odd products of the algebra elements.\\
{\bf iii.} The plane wave solutions of the corresponding  Weyl equations (either Clifford, 
 Eq.~(\ref{Weyl}) or Grassmann, Eq.~(\ref{Weylgrass})) for free massless fermions are 
the tensor products, $*_{T}$, of the superposition of the members of the "basis vectors" 
and of the momentum basis. The coefficients of the superposition correspondingly 
depend on a 
chosen momentum $\vec{p}$, with $p^0 =|\vec{p}|$, for any of the continuous
values of moments $\vec{p}$.\\
{\bf iv.} The creation operators defined on the tensor products, $*_{T}$, of 
superposition of the finite number of  ''basis vectors'' defining the internal space and 
of the infinite (continuous) momentum space, Eq.~(\ref{Weylp0}) in the Clifford case 
and  Eq.~(\ref{ptheta}) in the Grassmann case, define the infinite basis.  \\ 
{\bf v.} Applied on the vacuum state these creation operators form anticommuting 
single fermion states of an odd Clifford/Grassmann character. \\
{\bf vi.} The second quantized Hilbert space ${\cal H}$ consists of all possible
tensor products, $*_{T_{H}}$,  of any number of single fermion states, starting
with no single single state, 
"Slater determinants" with no single particle state occupied (with no creation operators 
applied on the vacuum state), with one single particle state occupied (with one creation 
operator applied on the vacuum state), with two single particle states occupied  (with 
two creation operator applied on the vacuum state), and so on. \\
%
{\bf vii.} The creation operators together with their Hermitian conjugated partners 
annihilation operators fulfill, due to the oddness of the ''basis vectors'', while the momentum
part commutes, the anticommutation relations, 
postulated by Dirac for second quantized fermion fields, not only when they apply on the
vacuum state, but also when they apply on the Hilbert space ${\cal H}$,  
Eq.~(\ref{H}) in the Clifford case and Eq.~(\ref{ijthetaprodgenHT}) in the Grassmann
case.

\noindent
In the Clifford case this happens only after ''freezing out'' half of the Clifford  space, 
what brings besides the correct anticommutation 
relations also  the ''family'' quantum number to each irreducible representation of 
the Lorentz group of the remaining internal space. 
\noindent
The oddness of the "basis vectors" is transfered  to the creation operators 
forming the single fermion states and further  to the creation operators forming 
the whole Hilbert space of the second quantized fermions in the Clifford case, 
Sect.~\ref{HilbertCliff0}, Eq.~(\ref{ijthetaprodgenH}), and in the Grassmann case, 
Sect.~\ref{actionGrass},  Eq.~(\ref{ijthetaprodgenHT}).\\
%
 %
{\bf viii.} Correspondingly the creation and annihilation operators with the internal 
space described by either odd Clifford or odd Grassmann algebra, since fulfilling 
the anticommutation relations required for the second quantized fermions without 
postulates, explain the Dirac's postulates for the second quantized fermions.\\
%








\noindent
In Sect.~\ref{HNcomparison} we clarify the notation used in the second quantization of 
fermions in literature and in our description of  the internal space of fermions when
using either the Grassmann or the Clifford algebras. The comparison of the
 Dirac and our way of the second quantization of fermions is presented in 
Sect.~\ref{creationannihilationtensor}.\\
\noindent
In Sect.~\ref{internalspace} we present in steps our use of the odd Grassmann 
algebra and the odd Clifford algebras for description of the internal 
space of fermions --- their spins and charges in both algebras, and families in the
Clifford algebra,  in a tensor product with 
the external space of coordinates or momenta,  relating our way of describing the 
internal space of fermons in any $d$, Sect.~\ref{GrassmannClifford}, to the usual 
way   using the group theory, Sect.~\ref{internalspaceordinary}. \\ 
%
Sect.~\ref{propertiesGrass0} presents the Grassmann odd algebra, that is 
the algebra of odd products of coordinates $\theta^a$'s, determining with their 
Hermitian conjugated partners $\frac{\partial}{\partial \theta_a}$'s 
$(\theta^{a \dagger} = \eta^{a a} \frac{\partial}{\partial \theta_{a}}$,  
$\eta^{a b}=diag\{1,-1,-1,\cdots,-1\}$) the internal space of  second 
quantized "fermions" with the integer spin.  Corresponding creation operators
determine in $d=(3+1)$ spins and charges of "fermions" in adjoint representations.\\
In Sect.~\ref{propertiesCliff0} the two anticommuting Clifford odd 
algebras, $\gamma^a$'s and $\tilde{\gamma}^a$'s, are discussed, each of them
are the superposition of $\theta^a$'s and $\frac{\partial}{\partial \theta_a}$'s
( $\gamma^{a}= (\theta^{a} + \frac{\partial}{\partial \theta_{a}})$, $\tilde{\gamma}^{a}= i \,(\theta^{a} -  
\frac{\partial}{\partial \theta_{a}})$~\cite{norma93,nh02,nh03}). They are, up to 
$\eta^{aa}$, Hermitian operators. 
 Each of the two Clifford algebras describes half integer spins, manifesting in $d=(3+1)$ 
spins, charges and families in  the fundamental representations of the corresponding
subgroups.\\
In Sect.~\ref{reduction} we explain the 
reduction of the two Clifford algebras to only one, we make a choice of
$\gamma^a$'s, while $\tilde{\gamma}^a$'s determine the ''family charges'' 
to each irreducible representation of  the generators of the Lorentz algebra in the 
remaining Clifford space  of $\gamma^a$'s and insure that the corresponding 
creation and annihilation operators
have all the properties  of the second quantized fermion fields. 
The reduction of the Clifford two algebras to only one reduces also the 
Grassmann algebra space.\\
%
In Sect.~\ref{poincare} we discuss the tensor products of the internal space of 
fermions and the space of coordinates or momenta in the ordinary second quantization 
procedure, as well as in the case when the internal space is described by the 
Clifford algebra.\\
In Sect.~\ref{creationtensorusual}   the ordinary second quantization procedure 
is presented.            \\
In Sect.~\ref{creationtensorClifford}  the second quantization with the 
Clifford algebra is presented.\\
Sect.~\ref{exampleClifford} demonsrates the solutions of equations of motion
for free massless chargeless fermions in $d=(3+1)$ for the case
 that the internal space is described by the Clifford algebra with $\gamma^a$'s.\\
 Sect.~\ref{creationannihilationtensor}  relates  our way of 
second quantization and the usual, Dirac's, way of second quantization,
 Refs.~\cite{n2019PIPII,2020PartIPartII}.\\
In Sect.~\ref{CPT} the discrete symmetry operators for fermions and  bosons 
(vielbeins and spin connections) in $d=((d-1)+1)$-dimensional space and  
as observed in $d=(3+1)$  as fermion fields, vector gauge fields and 
scalars gauge fields in the {\it spin-charge-family}, Ref.~\cite{nhds}, are presented.
App.~\ref{DSVS} adds more details about discrete symmetries of vector and 
scalar gauge fields.\\
%
\noindent
In Sect.~\ref{actionGrassCliff} the action and the corresponding equations of motion
for free massless  fermions in the {\it standard model}, Sect.~\ref{actionusual} and 
in the {\it spin-charge-family} theory, Sect.~\ref{actionClifford}, is discussed and
solutions expressed with the creation operators in $d=(5+1)$, 
Sect.~\ref{solutions5+1},  in $d=(9+1)$ and $d=(13+1)$ and in 
Sect.~\ref{solutions9+1 13+1}, with families included.
The proposal for the action for free massless "Grassmann fermions" can be found in
 App.~\ref{actionGrass}, where also the solutions are presented, again expressed with
the creation operators. \\
\noindent
In Sect.~\ref{HilbertCliff0} the Hilbert space determined by all possible numbers of 
tensor products of all possible odd Clifford creation operators applying on the 
vacuum state is presented and the anticommutation properties of the 
application of creation and annihilation operators on the Hilbert space is demonstrated, 
Sect.~\ref{HilbertCliffapp}.\\
\noindent
In Sect.~\ref{fermionandgravitySCFT} the simple starting action for massless 
fermions, the internal space of which is described by the odd Clifford algebra, and
which interact with the vielbeins and the two kinds of the spin connection fields only,
and for the gravitational fields is presented and discussed. \\
In Sect.~\ref{fermionactionSCFT} the action for massless fermions as seen in 
$d=(3+1)$ when fermions interact with the dynamical vectors gauge fields, 
Sect.~\ref{vector3+1}
 and the scalar  gauge fields, which for particular scalar space indexes manifest properties 
of Higgs's scalar, explaining the appearance of the scalar higgs and the Yukawa 
couplings of the {\it standard model}, Sect.~\ref{scalar3+1}, while for the rest of the
scalar indexes the scalar triplets and antitriplet fields cause the transition of 
antileptons into quarks and antiquarks into quarks, offering explanation for the 
matter/antimatter asymmetry in the expanding universe and predict the proton decay. \\
\noindent
In Sect.~\ref{fermionsbosonslowE} we compare the assumptions of the {\it standard
model}, Sect.~\ref{SM}, with in the literature proposed unifying theories, 
Sect.~\ref{SO10SCFT}, mostly $SO(10)$-unifying theories, Sect.~\ref{overviewSO(10)},
and the {\it spin-charge-family} theory, Sect.~\ref{overviewSCFT}, discussing besides the
assumptions also the predictions of these theories, 
Sects.~\ref{predictionsSO(10)SCFT}, 
~\ref{predictionSCFT},  searching for the next step beyond 
the {\it standard model}.\\
\noindent
In Sect.~\ref{TDN0} the  toy model is discussed in $d=(5+1)$ of massless odd 
Clifford fermions, interacting with the zweibein  and spin connection fields in $d=(5,6)$, 
of particular properties, which cause the break of the $5+1$-dimensional space into
$3+1$ times an almost $S^2$ sphere as an introduction into the study of the 
{\it spin-charge-family} theory case in $d=(13+1)$, which is under
 consideration~\footnote{
The mass of fermions is in this toy model determined either 
with the dynamics in the $6^{th}$ and the $7^{th}$ dimension or, at low energies, 
by the constant values of the spin connection fields, which break the mass protection.}.\\
In Sect.~\ref{others}  we shortly present other
possibilities for which we have our own understanding of the  topics.
\\
%
%
\noindent
In Sect.~\ref{manybody} we mention possibilities that the realization and  
predictions of the {\i spin-charge-family} theory could influence hadron physics, if at all.\\
\noindent
In Sect.~\ref{conclusions} we shortly overview what one can learn from this
 review article.\\
\noindent 
In App.~\ref{actionGrass} the trial to find the action for the massless integer 
spin "Grassmann fermions" is presented.\\
\noindent
In  App.~\ref{trial} properties of the Clifford even commuting  "basis vectors" are 
discussed, App.~{evenclifford}, and the matrices for $\gamma^a$'s,  
$\tilde{\gamma}^a$'s, $S^{ab}$, $\tilde{S}^{ab}$ are presented and discussed, 
App.~\ref{matrixCliffordDMN}\\. 
\noindent
In App.~\ref{appanomalies} the triangle anomaly cancellation of the {\it standard
model} is explained in the {\i spin-charge-family} theory and in the $SO(10)$-unifying
 theories.

Most of the rest of appendixes offer detail explanations or proofs needed in the 
main text.


%

%

%
%
\section{Second quantization in text books and relation to our way of second quantization}
\label{HNcomparison}
%


This section is to present in a short way how does the main assumption, 
 the decision to describe the internal space of fermions with 
the ''basis vectors'' expressed with the superposition of odd products of 
the anticommuting members of the algebra, as discussed in details in 
Sect.~\ref{internalspace},  either the Clifford one or
the Grassmann one, acting algebraically, $*_{A}$, on the internal 
vacuum state  $|\psi_{o}>$ in each of these algebras, relate to the creation 
and annihilation anticommuting operators of the second quantized  fermion fields, 
discussed in details in Sect.~\ref{poincare},~Sect.~\ref{creationtensorusual},~%
Sect.~\ref{creationtensorClifford}. 


Let us first  tell that the algebraic product $*_A$ is 
 usually not present in other works, and thus has no 
well known physical meaning. It is at first a product by which one can 
multiply two internal "basis vectors", describing internal space of
fields, $\hat{b}^{\dagger}_{i}$ and 
$\hat{b}^{\dagger}_{j}$, with each other,
\begin{eqnarray}
\label{HNA}
\hat{c}^{\dagger}_k&=& \hat{b}^{\dagger}_i *_{A} \hat{b}^{\dagger}_j\,,\nonumber\\
\hat{b}^{\dagger}_i  *_{A} \hat{b}^{\dagger}_j&=& \mp \hat{b}^{\dagger}_j  *_{A} 
\hat{b}^{\dagger}_i \,, 
\end{eqnarray}
the sign $\mp$ depends on whether $\hat{b}^{\dagger}_i$ and $\hat{b}^{\dagger}_j$ 
are products of odd or even number of algebra elements: The sign is $-$ if both 
are (superposition of) odd products of algebra elements, in all other cases
 the sign is $+$.

Let ${\bf R}^{d-1}$ define the external spatial or momentum basis $|\vec{p}>$. 
Then the tensor product $*_{T}$ extends the internal "basis vectors"  
into the creation operator $ \hat{\bf b}^{\dagger}_{\vec{p} ,\,i}$ defined in both spaces, 
Sect.~\ref{poincare}, App.~\ref{continuous}  
\begin{eqnarray}
\label{HNT}
 \hat{\bf b}^{\dagger}_{\vec{p}, \, i} =|\vec{p}>*_{T}\, \hat{b}^{\dagger}_{i}\,,
\end{eqnarray}
where again $\hat{b}^{\dagger}_i $ represent the superposition of products of elements 
of the  anticommuting algebras, in our case either 
$\theta^a$ or $\gamma^a$ or $\tilde{\gamma}^a$, 
used in this paper.

We can make a choice of the orthogonal and normalized basis so that 
$<\hat{\bf b}_{\vec{p}, i} | \hat{\bf b}^{\dagger}_{\vec{p'}, j} > = 
\delta(\vec{p} - \vec{p'}) \,\delta_{ij}$. Let us point out that either 
$\hat{b}^{\dagger}_{i}$ or $\hat{\bf b}^{\dagger}_{\vec{p}, \, i}$ apply algebraically 
on the corresponding vacuum state, $\hat{b}^{\dagger}_i *_{A} |\psi_o>$ 
and $\hat{\bf b}_{\vec{p}, \, i} *_{A}
|\psi_o>$.



To give to  the algebraic product, $*_A$,  and to the tensor product, $*_{T}$, 
defining the space of single particle wave functions depending on the internal space
and external space, the understandable meaning, 
we postulate the connection  between the anticommuting/commuting properties 
of the ''basis  vectors'', expressed with  the odd/even products of the anticommuting 
algebra elements and the corresponding creation operators, creating second 
quantized single fermion/boson states. 

%


%
\begin{eqnarray}
\label{HNb}
\hat{\bf b}^{\dagger}_{\vec{p}, i} *_{A}|\psi_o> &=& |\psi_{\vec{p}, i}>\,,\nonumber\\
\hat{\bf b}^{\dagger}_{\vec{p}, i}  *_{T}\,|\psi_{\vec{p'}, j}> &=& 0\,,\nonumber\\
{\rm if \,} \vec{p} &=& \vec{p'}\, {\rm and} \, i=j\,,\nonumber\\
{\rm in\; all\;other\;cases\;} & &{\rm it\;follows\,}\nonumber\\
\hat{\bf b}^{\dagger}_{\vec{p}, i}  *_{T}\, 
\hat{\bf b}^{\dagger}_{\vec{p'}, j} *_{A}|\psi_o>
&=& \mp \,\hat{\bf b}^{\dagger}_{\vec{p'},j}  *_{T} \, 
\hat{\bf b}^{\dagger}_{\vec{p}, i}  *_{A}|\psi_o>\,,
\end{eqnarray} 
with the sign $\pm $ depending on whether  $\hat{\bf b}^{\dagger}_{\vec{p}, i}$
have both an odd character, the sign is $-$, if this is not the case the sign is $+$.

Not necessarily both, odd and even products of anticommuting algebra elements 
have any physical meaning (any realization in the observed phenomena), if
 any at all. It will be shown that in Grassmann algebra both, odd and even products 
of anticommuting algebra elements exist, behaving as either anticommuting or
commuting  creation and annihilation operators, respectively,  Sect.~\ref{evengrass}, 
while in the Clifford algebra
anticommuting odd products of anticommuting elements of the Clifford algebra
exist with the Hermitian conjugated partners annihilation operators belonging to 
the independent group, Sect.~\ref{evenclifford}~\cite{2020PartIPartII}. 
The Clifford even products of elements commute and have their Hermitian conjugated 
partners within the same group, the paper is in preparation.

To each creation operator $\hat{\bf b}^{\dagger}_{\vec{p}, i}$ its Hermitian 
conjugated partner  represents the annihilation operator $\hat{\bf b}_{\vec{p}, i}$
\begin{eqnarray} 
\label{HNC}
\hat{\bf b}_{\vec{p}, i} &=&
(\hat{\bf b}^{\dagger}_{\vec{p}, i})^{\dagger}\,,
\nonumber\\
{\rm with \; the}&&{\rm property}\nonumber\\
\hat{\bf b}_{\vec{p}, i}\, *_{A}\,|\psi_o> &=&0\,,\nonumber\\
{\rm defining\; the\; } && {\rm vacuum \; state\; as\;}\nonumber\\
|\psi_o> :&=&\sum_i \,{\hat b}_{i}\, *_{A} \,{\hat b}^{\dagger}_{i}
\,|\;I>|0_{\vec{p}}>\,,
\end{eqnarray}
where summation $i$ runs over all different products of annihilation operator 
$\times$ its Hermitian conjugated  creation operator, no matter what 
$\vec{p}$ is chosen, and $|\;I>\,|0_{\vec{p}}>$ represents the identity with the
starting momentum state $|0_{\vec{p}}>$, from where all the momentum 
states $|\vec{p}>$ follow by translation in momentum space, ${\hat b}_{i}$ represents
the Hermitian conjugated annihilation operator to  $({\hat b}^{\dagger}_{i})^{\dagger}$,
Sect.~\ref{poincare}. 


Let the tensor multiplication $*_{T_{H}}$ denotes   the multiplication of any number of
any possible single particle states, and correspondingly of any number of any possible
creation operators, Sect.~\ref{HilbertCliff0}.


What further means that to each single particle wave function we define 
the creation operator $\hat{\bf b}^{\dagger}_{\vec{p}, i} $, applying 
in a tensor product from the left hand side on the second quantized 
Hilbert space --- consisting of all possible products 
of any number of  single particle wave functions --- adding to the Hilbert 
space the single particle wave function created by this particular creation 
operator. In the case of the second quantized fermions, if this particular 
wave function with the quantum numbers ${}_{i}$ and momentum $\vec{p}$  of 
$\hat{\bf b}^{\dagger}_{\vec{p}, i} $ is already 
among the single fermion wave functions of a particular product of the 
fermion wave functions, the action of the creation operator gives zero, 
otherwise the number of the fermion wave functions increases for one,
Sect.~\ref{HilbertCliff0}. 
(In the boson case the number of boson second quantized wave functions 
 increases always  for one.) 

If we apply with the annihilation operator $\hat{\bf b}_{\vec{p}, i}$ 
on the second quantized Hilbert space, then the application gives a 
nonzero contribution only if  the particular products of the single particle 
wave functions do include the wave function with the quantum number 
$i$ and $\vec{p}$, Sect.~\ref{HilbertCliff0}. 

In a Slater determinant formalism the single particle wave functions define
the empty or occupied places of any of infinite numbers of Slater determinants. 
The creation operator $\hat{\bf b}^{\dagger}_{\vec{p}, i} $ applies
on a particular Slater determinant from the left hand side. Jumping over 
occupied states to the place with its $i$ and $\vec{p}$. If this state is 
occupied, the application gives in the fermion case zero. (In the boson case 
the number of particles increase for one.) 
The particular Slater determinant changes sign in the 
fermion case if 
$\hat{\bf b}^{\dagger}_{\vec{p}, i}$ jumps over odd numbers of occupied 
states. (In the boson case the sign of the Slater determinant does not change.) 

When annihilation operator $\hat{\bf b}_{\vec{p}, i} $ applies on 
particular Slater determinant, is jumping over occupied states to its own place,
giving zero, if this place is empty and decreasing the number of occupied states 
if this place is occupied. The Slater determinant changes sign in the fermion 
case, if the number of occupied states before its own place is odd. (In the boson
case the sign does not change.)

Let us stress again that choosing antisymmetry or symmetry of products of 
the algebra elements is possible only in the case of the Grassmann algebra --- 
Grassmann
odd algebra describes second quantized integer spin "fermions", 
Sect.~\ref{propertiesGrass0}, while the commuting Grassmann even algebra can be 
used to describe second quantized integer spin bosons.

In the Clifford case Clifford odd choice, describing the 
half integer fermions, is the only choice, since the Clifford even algebra does not 
offer the description of the second quantized boson states, as explained in 
App.~\ref{trial}. 


To describe the second quantized fermion states we use the "basis vectors", which are 
the superposition of the odd numbers of algebra elements in both algebras, in the two
Clifford algebras and in the Grassmann algebra.

The creation operators and their Hermitian conjugated partners annihilation 
operators therefore in the fermion case anticommute. The single fermion 
states, which are the application of the creation operators on the vacuum 
state $|\psi_o>$, manifest correspondingly as well the oddness. 
The vacuum state, defined as the sum over all different products of 
annihilation $\times$ the corresponding creation operators, have an even
character.

One usually means antisymmetry when talking about Slater-determinants
because otherwise one would not get determinants.

In the present paper~\cite{norma92,norma93,IARD2016,nh02} the choice of 
the symmetrizing versus antisymmetrizing  relates indeed the commutation versus 
anticommutation with respect to the a priori completely different product $*_A$, 
of anticommuting members of the Clifford or Grassmann algebra. The oddness or 
evenness  of these products transfer to quantities to which these algebras
extend.

Let us add up with the recognition: Oddness and eveness of the "basis vectors" to 
describe "fermions" and bosons is meaningful 
only in the case of Grassmann algebra, while the  Clifford algebra offers only the
description of fermions, offering at the same time family quantum numbers for
different irreducible representations of the corresponding Lorentz algebra. This is 
uniqum of the Clifford algebra that it offers the unification of spins, charges and 
families. The approaches with the group theory description of the
internal space of fermions do not offer this kind of unification.


%
\section{Internal and external space of fermions}
\label{internalspace}

Single fermion states are functions of  external coordinates 
and of internal space of fermions. 
The basis for single fermion states are usually chosen to have some symmetries in both spaces,  external and internal. 

If $M^{ab}$  denote infinitesimal generators of the Lorentz algebra in both spaces,
 $M^{ab}= L^{ab} + S^{ab}$, with $L^{ab}= x^a p^b - x^b p^a$, $p^a=
i \frac{\partial}{\partial x_a}$, determining operators in ordinary space, while 
 $S^{ab}$ are equivalent operators in internal space of fermions~\footnote{%
 We here mainly think of the algebra and do not for the moment go into
the problem that if we take these  operators to be represented on
the space of states for a particle then  $x^0$ is not the operator,
it is instead a moment of time, what we do have in mind.}
%
%
\begin{eqnarray}
\label{Lorentzcom}
\{M^{ab}, M^{cd}\}_{-} &=&  
 i \{M^{ad} \eta^{bc} + M^{bc} \eta^{ad} -  M^{ac} \eta^{bd}- \bf M^{bd} \eta^{ac}\}\,,\nonumber\\
\{M^{ab}, p^{c}\}_{-} &=&  -i  \eta^{ac}p^b + i \eta^{cb}p^a\,, \nonumber\\
\{M^{ab}, S^{cd}\}_{-} &=&i \{S^{ad} \eta^{bc} + S^{ad} -  S^{ac} \eta^{bd}- \bf S^{bd} \eta^{ac}\}\,, ,,
\end{eqnarray}
with $L^{ab\,\dagger}=L^{ab}$, provided that $p^{a \dagger} =p^a$, since they  
apply on states which are zero on the boundary~\cite{hermiticity}, while $S^{ab \dagger}
=\eta^{aa}\eta^{bb} S^{ab}$. 
Commutation relation  of Eq.~(\ref{Lorentzcom}) are valid for either $L^{ab}$ or 
$S^{ab}$ in any dimension of space-time, $\{L^{ab}, S^{cd}\}_{-}=0 $. 

We use the metric tensor $\eta^{ab}=diag(1,-1,-1,\dots,-1,-1) $ for 
$a=(0,1,2,3,5,\dots,d)$.

The  usual choice of the Cartan subalgebra of the commuting operators of the 
Lorentz algebra is  
\begin{eqnarray}
M^{03}, M^{12}, M^{56}, \dots, M^{d-1\, d}\,, 
\label{cartanM}
\end{eqnarray}
defining correspondingly the members of the irreducible representations of the 
Lorentz group to be the eigenvectors of all  the members of the Cartan subalgebra
presented in Eqs.~(\ref{cartanM}, \ref{cartangrasscliff}).  

 In the literature the single fermion states usually do not anticommute.
 Dirac postulated anticommuting properties for fermion fields only in the second 
 quantization, by postulating anticomutation rules for creation and annihilation operators, 
 which operating on the vacuum state generate the second quantized states. 
 Although all the elementary fermion constituents, quarks and leptons,
 do carry besides spins also charges and appear in families,  and although
 charges and families are treated as internal degrees of freedom of fermions, yet
  internal space of fermions is not built from spins, charges and families in equivalent
 way, unifying spins, charges and families. 
 
 One finds in the literature~\cite{Geor,FritzMin} several trials to unify all the charges, 
 but not  the spin with the charges. We comment in App.~\ref{appanomalies} that the 
appearance of spins and charges in one irreducible representation of the Lorentz 
group explains the triangle anomalies cancellation in the {\it standard model} better
 than the unification of only charges, since in this first case the handedness and 
charges are already related, this is happening in the {\it spin-charge-family} theory, 
while in the unification of only  charges the relation between spins and charges must
be postulated like in the {\it standard model}.

 We show in what follows that anticommuting properties of spinor states, brought into
the theory by using the anticommuting "basis vectors", bring simplicity, transparency 
and elegance in the theoretical description of fermion states, if spins and  charges 
appear in the same irreducible representation of the group. 
 
Let us point out that in the literature the anticommuting properties of the second 
quantized fermion fields are just postulated. In the ''Grassmann fermions'' and the two 
''Clifford fermions'' the ''basis vectors'' anticommute due to the oddness of the ''basis 
vectors''. The Hermitian conjugated partners of the ''basis vectors'' belong to independent 
representations, while in the usual second quantized fields they are just postulated to
exist.
 
 
 

We shall use in what follows two kinds of anticommuting algebras, the Grassmann
algebra and  the two anticommuting kinds of the Clifford algebras, denoting them by
$\theta^{a}$'s and $p^{\theta a}$'s $= \frac{\partial}{\partial \theta_a}$'s, for the
Grassmann algebra objects, and $\gamma^a$'s  and $\tilde{\gamma}^{a}$'s
for the two kinds of the  Clifford algebra objects. 
The existence of the two kinds of the Clifford algebras is our recognition~\cite{norma92,norma93,norma94,nh02,nh03}. 

We denote the corresponding infinitesimal Lorentz generators as ${\cal {\bf S}}^{ab}$
for the Grassmann algebra, and as $S^{ab}$ and $\tilde{S}^{ab}$ for the two kinds 
of the Clifford algebras.  
%
\begin{eqnarray}
{\cal {\bf S}}^{ab}= i \, (\theta^{a} \frac{\partial}{\partial \theta_{b}} - \theta^{b}
\frac{\partial}{\partial \theta_{a}})\,, \quad 
&& S^{ab} = \frac{i}{4}(\gamma^a \gamma^b - \gamma^b \gamma^a)\,,\quad  
\tilde{S}^{ab} =\frac{i}{4}(\tilde{\gamma}^a \tilde{\gamma}^b - \tilde{\gamma}^b
 \tilde{\gamma}^a)\, , \nonumber\\
{\cal {\bf S}}^{ab} =S^{ab} + \tilde{S}^{ab}\,, \quad&&
\{S^{ab}, \tilde{S}^{ab}\}_{-}=0\,,\nonumber\\
\{{\cal {\bf S}}^{ab}, \, \theta^e\}_{-} = - i \,(\eta^{ae} \,\theta^b - \eta^{be}\,\theta^a)\,, \quad 
&&\{{\cal {\bf S}}^{ab}, \, p^{\theta e}\}_{-} = - i \,(\eta^{ae} \,p^{\theta b} - 
\eta^{be}\,p^{\theta a})\,,
\nonumber\\
\{ S^{ab}, \gamma^c\}_{-}&=& i (\eta^{bc} \gamma^a - \eta^{ac} \gamma^b)\,, \nonumber\\
\{ \tilde{S}^{ab}, \tilde{\gamma}^c \}_{-}&=& i (\eta^{bc} \tilde{\gamma}^a - 
\eta^{ac }\tilde{\gamma}^b)\,,\nonumber\\
\{ S^{ab}, \tilde{\gamma}^c\}_{-}&=&0\,,\quad \{\tilde{S}^{ab}, \gamma^c\}_{-}=0\,,
\label{sabtildesab}
\end{eqnarray}
%
the proof for the relation ${\cal {\bf S}}^{ab} =S^{ab} + \tilde{S}^{ab}$ is
presented in App.~\ref{proofs}, Statement 2..

It then follows for the  Cartan subalgebra of the commuting operators of
the Lorentz algebra for each of the two kinds of the operators of the Clifford algebra, 
$S^{ab}$ and $\tilde{S}^{ab}$ and for the Grassmann algebra ${\cal {\bf S}}^{ab}$
%
\begin{eqnarray}
{\cal {\bf S}}^{03}, {\cal {\bf S}}^{12}, {\cal {\bf S}}^{56}, \cdots, 
{\cal {\bf S}}^{d-1 \;d}\,, \nonumber\\
S^{03}, S^{12}, S^{56}, \cdots, S^{d-1 \;d}\,, \nonumber\\
\tilde{S}^{03}, \tilde{S}^{12}, \tilde{S}^{56}, \cdots,  \tilde{S}^{d-1\; d}\,. 
\label{cartangrasscliff}
\end{eqnarray}


We shall see that the Grassmann algebra  is uniquely expressible with the two Clifford 
algebras  and opposite.

\subsection{Internal space of fermions  in text books}
 \label{internalspaceordinary}

Fermion states can be represented as a superposition of tensor products of basis in 
ordinary (coordinate or momentum) space and in internal space. Fermions carry in 
general as a part of the internal degrees of freedom besides spins also charges. 

The spinor part of the internal space of massless fermions (without charges) can be  
represented by the spinor representation of the Lorentz group $S^{mn}$, $(m,n)
=(0,1,2,3)$. Making the usual choice of the Cartan subalgebra of the Lorentz group,
Eq.~(\ref{cartanM}),
\begin{eqnarray}
\label{cartaninternal}
S^{03}, S^{12}\,,
\end{eqnarray}
one usually  choses  internal part of states to be the eigenstates of 
the spin, $S^{12}$, and the handedness, 
$\Gamma^{(3+1)}=-\frac{i}{3!} \varepsilon_{mnop} S^{mn}S^{op}$, which is
in the case of the Cartan subalgebra of Eq.~(\ref{cartaninternal}) equal to 
$\Gamma^{(3+1)}=i (-2i)^2 S^{03 } S^{12 }$.

There are two possibilities for spin, $S^{12}=\pm \frac{1}{2}$, and two for
$\Gamma^{(3+1)}=\pm 1$, for right and left handedness, respectively. 
Due to the relation $\Gamma^{(3+1)}=i (-2i)^2 S^{03 } S^{12 }$ the spinor 
part can be represented as well as ''eigenstates'' of the Cartan subalgebra members.

These four spinor states, which are the eigenstates of the Cartan subalgebra,
can be  written as presented in Table~(\ref{Table 3+1.}).
 \begin{table}
 \begin{center}
 \begin{minipage}[t]{16.5 cm}
 \caption{Four spinor states in $d=(3+1)$,
 two left handed (chiral) and two right handed (chiral), each with spin 
 $\pm \frac{1}{2}$,  are the eigenstates of the Cartan subalgebra of 
 the Lorentz algebra in the internal space of fermions. }
 \label{Table 3+1.}    
\end{minipage}
 \begin{tabular}{|r|r|r|r|r|}
 \hline
$ h =\Gamma^{(3+1)} 
$&$\psi_{S^{03}, S^{12}}$&
$S^{03}$&$ S^{1 2}$&$\Gamma^{3+1}$ \\
\hline
$R$&$\psi_{+\frac{1}{2}, +\frac{1}{2}}$ &$+\frac{1}{2}$&$+\frac{1}{2}$&$+1$\\ 
$R$&$\psi_{-\frac{1}{2},- \frac{1}{2}}$ &$-\frac{1}{2}$&$-\frac{1}{2}$&$+1$\\ 
$L$&$\psi_{-\frac{1}{2}, +\frac{1}{2}}$ &$-\frac{1}{2}$&$\frac{1}{2}$&$-1$\\ 
$L$&$\psi_{+\frac{1}{2},- \frac{1}{2}}$ &$\frac{1}{2}$&$-\frac{1}{2}$&$-1$\\ 
 \hline 
 \end{tabular}
 \end{center}
 \end{table}
%
 These two representations, the right and  the left one, can be written each
 in terms of Pauli spin matrices $\vec{\sigma}$ and the unit matrix $\sigma^0$.
 \begin{equation}\label{pauli}
   \sigma^1=
 \left(  \begin{array}{cc}
     0 & 1 \\
     1 & 0
    \end{array} \right)\, , \qquad
   \sigma^2=
   \left(  \begin{array}{cc}
     0 & -i\\
     i & 0
    \end{array} \right)\, , \qquad
   \sigma^3=
    \left(  \begin{array}{cc}
     1 & 0\\
     0 & -1
   \end{array} \right)\, ,  \qquad
  \sigma^0=
  \left(  \begin{array}{cc}
     1 & 0\\
     0 & 1
   \end{array} \right)\,.
 \end{equation}
  Let us write down the $4 \times 4$ matrix of handdedness/chirality, 
  $\Gamma^{(3+1)}$, and the corresponding  matrices  of the infinitesimal 
  generators   of the Lorentz transformations in   internal space of fermions, $\vec{S}$ 
  ($\vec{S}=\frac{1}{2}  \varepsilon_{ijk} S^{jk}$) and   $\vec{K}$ ($K^{i}= S^{0i}$), 
  defined in the space of four spinor states, 
\begin{equation}\label{sigma4x4}
 \mathbf{\Gamma^{(3+1)}}_= 
 \left(  \begin{array}{cc}
  \mathbf{1} & 0 \\
   0 &-  \mathbf{1}
  \end{array} \right)\, ,\qquad
  \vec{S}=
  \frac{1}{2} \left(  \begin{array}{cc}
  \vec{\sigma} & 0 \\
   0 &   \vec{\sigma} 
  \end{array} \right)\, ,\qquad
  \vec{K}=
 \frac{i}{2} \left(  \begin{array}{cc}
  \vec{\sigma} & 0 \\
   0 &  - \vec{\sigma} 
  \end{array} \right)\, .
   \end{equation}
Each of the two  ''basis states'' represents two decoupled irreducible representations 
of the Lorentz group. They only can be connected when introducing Dirac matrices 
$(\gamma^0, \gamma^1,\gamma^2,\gamma^3)$ with properties 
\begin{equation}\label{gamma4x4}
 \mathbf{\gamma^{0}}_= 
 \left(  \begin{array}{cc}
  0 & \sigma^0 \\
   \sigma^0& 0
  \end{array} \right)\, ,\qquad
   \mathbf{\gamma^{1}}_= 
 \left(  \begin{array}{cc}
  0 & \sigma^1 \\
  - \sigma^1& 0
  \end{array} \right)\, ,\qquad
   \mathbf{\gamma^{2}}_= 
 \left(  \begin{array}{cc}
  0 & -\sigma^2 \\
   \sigma^2& 0
  \end{array} \right)\, ,\qquad
   \mathbf{\gamma^{3}}_= 
 \left(  \begin{array}{cc}
  0 & \sigma^3 \\
   -\sigma^3& 0
  \end{array} \right)\, .
   \end{equation}

The eigenstates of the Cartan subalgebra, Table~\ref{Table 3+1.}, are in the literature 
not assumed to anticommute, they are rather assumed to be commuting objects.  In Ref.~(\cite{Ramond}, page 17)  P. Ramond mentions 
the possibility to give to spinors the Grassmann  anticommuting character.

We point out in this paper that the internal space of fermions do have anticommuting 
character, either with the integer spins, if represented with the odd Grassmann algebra,
or with the half integer spins, if represented with any of the two kinds of the odd 
Clifford algebras. All the ''basis vectors'' in internal space, either described by
 the Grassmann algebra or in any of the two Clifford algebras, are taken to be 
superposition of products of an odd number of algebra elements and to be at the same
time eigenstates of all the Cartan subalgebra members, to be comparable with the
usual choice of the spin representation.

We show in App.~\ref{matrixCliffordDMN} that using any of the two odd Clifford 
algebras to describe the internal space of fermions,  offers  several irreducible 
representations. However,  the left and the right handed representation in $d=(3+1)$ 
appear in the same irreducible representation (what means that  both
have the odd handedness)  provided that $d>(3+1)$. 
In the case of higher dimensions than $d=(3+1)$ not only the spin and 
handedness but also all the charges contribute to the internal space of fermions, 
determining the anticommuting ''basis vectors''.

In $d=(13 +1)$, for example, in one irreducible representation of the Lorentz group
there are all the quarks and all the leptons and the antiquarks and the antileptons
 with all the properties postulated by the {\it standard model} for quarks and 
 antiquarks and for leptons and antileptons, all carrying the required charges,
 handedness and spins.

 To the oddness of each member of the irreducible representation the spin part
 and the charge part contribute. 
 All  the irreducible representations are equivalent with respect to the Lorentz group.

  
 
\subsection{Internal space of fermions described by Grassmann algebra and two kinds 
       of Clifford algebras}
\label{GrassmannClifford}
We shall pay attention only to even dimensional spaces: $d=2(2n+1)$ and $d=4n$,
$n$ is a non negative integer.

In Grassmann $d$-dimensional space there are $d$ anticommuting operators 
$\theta^{a}$, $\{\theta^{a}, \theta^{b}\}_{+}=0$, $a=(0,1,2,3,5,..,d)$, and 
$d$ anticommuting derivatives with respect to $\theta^{a}$, 
$\frac{\partial}{\partial \theta_{a}}$, $\{\frac{\partial}{\partial \theta_{a}}, 
\frac{\partial}{\partial \theta_{b}}\}_{+} =0$, offering
together $2\cdot2^d$ operators, the half of which are superposition of products of  
$\theta^{a}$ and another half corresponding superposition of 
$\frac{\partial}{\partial \theta_{a}}$.
\begin{eqnarray}
\label{thetaderanti0}
\{\theta^{a}, \theta^{b}\}_{+}=0\,, \, && \,
\{\frac{\partial}{\partial \theta_{a}}, \frac{\partial}{\partial \theta_{b}}\}_{+} =0\,,
\nonumber\\
\{\theta_{a},\frac{\partial}{\partial \theta_{b}}\}_{+} &=&\delta_{ab}\,, (a,b)=(0,1,2,3,5,\cdots,d)\,.
\end{eqnarray}
We define~\cite{nh2018} 
\begin{eqnarray}
(\theta^{a})^{\dagger} &=& \eta^{a a} \frac{\partial}{\partial \theta_{a}}\,,\nonumber\\
{\rm leading  \, to}\nonumber\\
(\frac{\partial}{\partial \theta_{a}})^{\dagger}&=& \eta^{a a} \theta^{a}\,,
\label{thetaderher0}
\end{eqnarray}
the identity is the self adjoint member. We make a choice for the complex properties 
of $\theta^a$, and correspondingly 
of $\frac{\partial}{\partial \theta_{a}}$, as follows
\begin{eqnarray}
\label{complextheta}
\{\theta^a\}^* &=&  (\theta^0, \theta^1, - \theta^2, \theta^3, - \theta^5,
\theta^6,...,- \theta^{d-1}, \theta^d)\,, \nonumber\\
\{\frac{\partial}{\partial \theta_{a}}\}^* &=& (\frac{\partial}{\partial \theta_{0}},
\frac{\partial}{\partial \theta_{1}}, - \frac{\partial}{\partial \theta_{2}},
\frac{\partial}{\partial \theta_{3}}, - \frac{\partial}{\partial \theta_{5}}, 
\frac{\partial}{\partial \theta_{6}},..., - \frac{\partial}{\partial \theta_{d-1}}, 
\frac{\partial}{\partial \theta_{d}})\,. 
\end{eqnarray}
In $d$-dimensional space of anticommuting Grassmann 
coordinates and of their Hermitian conjugated partners derivatives, 
Eqs.~(\ref{thetaderanti0}, \ref{thetaderher0}), 
there exist two kinds of the Clifford coordinates (operators) --- 
$\gamma^{a}$ and $\tilde{\gamma}^{a}$ --- both are expressible in terms of 
$\theta^{a}$ and their conjugate momenta $p^{\theta a}= i \,
\frac{\partial}{\partial \theta_{a}}$ ~\cite{norma93}.
\begin{eqnarray}
\label{clifftheta}
\gamma^{a} &=& (\theta^{a} + \frac{\partial}{\partial \theta_{a}})\,, \quad 
\tilde{\gamma}^{a} =i \,(\theta^{a} - \frac{\partial}{\partial \theta_{a}})\,,\nonumber\\
\theta^{a} &=&\frac{1}{2} \,(\gamma^{a} - i \tilde{\gamma}^{a})\,, \quad 
\frac{\partial}{\partial \theta_{a}}= \frac{1}{2} \,(\gamma^{a} + i \tilde{\gamma}^{a})\,,
\end{eqnarray}
offering together  $2\cdot 2^d$  operators: $2^d$ of those which are products of 
$\gamma^{a}$  and  $2^d$ of those which are products of $\tilde{\gamma}^{a}$.

Taking into account Eqs.~(\ref{thetaderanti0}, \ref{thetaderher0}) ($\{\theta^{a}, \theta^{b}\}_{+}=0$,
$\{\frac{\partial}{\partial \theta_{a}}, \frac{\partial}{\partial \theta_{b}}\}_{+} =0$,
$\{\theta_{a},\frac{\partial}{\partial \theta_{b}}\}_{+} = \delta_{ab}$,
$\theta^{a \dagger} 
=\eta^{aa}\, \frac{\partial}{\partial \theta_{a}} $ and 
$(\frac{\partial}{\partial \theta_{a}})^{\dagger}=\eta^{aa} \theta^{a}$) one finds
\begin{eqnarray}
\label{gammatildeantiher0}
\{\gamma^{a}, \gamma^{b}\}_{+}&=&2 \eta^{a b}= \{\tilde{\gamma}^{a}, 
\tilde{\gamma}^{b}\}_{+}\,, \nonumber\\
\{\gamma^{a}, \tilde{\gamma}^{b}\}_{+}&=&0\,,\quad
 (a,b)=(0,1,2,3,5,\cdots,d)\,, \nonumber\\
(\gamma^{a})^{\dagger} &=& \eta^{aa}\, \gamma^{a}\, , \quad 
(\tilde{\gamma}^{a})^{\dagger} =  \eta^{a a}\, \tilde{\gamma}^{a}\,,\nonumber\\
\gamma^a \gamma^a &=& \eta^{aa}\,, \quad 
\gamma^a (\gamma^a)^{\dagger} =I\,,\quad
 \tilde{\gamma}^a  \tilde{\gamma}^a = \eta^{aa} \,,\quad
 \tilde{\gamma}^a  (\tilde{\gamma}^a)^{\dagger} =I\,,
\end{eqnarray}
where $I$ represents the unit operator.
%

Making a choice for the $\theta^a$ properties as presented in 
Eq.~(\ref{complextheta}), it follows for the Clifford objects
\begin{eqnarray}
\label{complexgamatilde}
\{\gamma^a\}^* &=&  (\gamma^0, \gamma^1, - \gamma^2, \gamma^3, - \gamma^5,
\gamma^6,...,- \gamma^{d-1}, \gamma^d)\,, \nonumber\\
\{\tilde{\gamma}^a\}^* &=&  (- \tilde{\gamma}^0, - \tilde{\gamma}^1, \tilde{\gamma}^2, 
- \tilde{\gamma}^3,  \tilde{\gamma}^5,- \tilde{\gamma}^6,..., \tilde{\gamma}^{d-1}, 
- \tilde{\gamma}^d)\,, 
\end{eqnarray}

The Cartan subalgebra of the Lorentz algebra is presented in 
Eqs.~(\ref{cartangrasscliff}, \ref{cartanM}).

\subsubsection{Grassmann algebra and ''basis vectors'' describing internal space}
\label {propertiesGrass0}
%
 
It appears useful to arrange $2^d$ products of $\theta^{a}$ into  irreducible
representations with 
respect to the Lorentz group with the generators~\cite{norma93}, 
Eq.~(\ref{sabtildesab}), 
\begin{eqnarray}
{\cal {\bf S}}^{a b} &=& i \, (\theta^{a} \frac{\partial}{\partial \theta_{b}} - \theta^{b}
\frac{\partial}{\partial \theta_{a}})\,, \quad  ({\cal {\bf S}}^{a b})^{\dagger} = \eta^{a a}
\eta^{b b} {\cal {\bf S}}^{a b}\, . 
\label{thetasab}
\end{eqnarray} 
%
$2^{d-1}$ members of the representations  have an odd Grassmann character (those which are 
superposition of  odd products of $\theta^{a}$'s).  All the members of any particular odd 
irreducible representation follow from any starting member by the application of 
${\cal {\bf S}}^{a b}$'s.

If we exclude the self adjoint identity there is $(2^{d-1}-1)$ members of an even Grassmann 
character, they are even products of $\theta^{a}$'s. All the members of any particular even 
representation follow from any starting member by the application of ${\cal {\bf S}}^{a b}$'s.

The Hermitian conjugated $2^{d-1}$ odd partners of  odd representations of $\theta^a$'s 
and $(2^{d-1}-1)$  even partners of even representations of $\theta^a$'s are reachable from 
odd and even representations, respectively, by the application of Eq.~(\ref{thetaderher0}).

It  appears useful to make the choice of the Cartan subalgebra of the commuting operators of the Lorentz algebra, 
%
${\cal {\bf S}}^{03}, {\cal {\bf S}}^{12}, {\cal {\bf S}}^{56}, \cdots, 
{\cal {\bf S}}^{d-1 \;d}$\,, Eq.~(\ref{cartangrasscliff}),
%
and choose the members of the irreducible representations of the Lorentz group to be the 
eigenvectors of all  the members of the Cartan subalgebra of Eq.~(\ref{cartangrasscliff}), 
${\cal {\bf S}}^{a b} = i \, (\theta^{a} \frac{\partial}{\partial \theta_{b}} - \theta^{b}
\frac{\partial}{\partial \theta_{a}})$  
\begin{eqnarray}
{\cal {\bf S}}^{ab} \,\frac{1}{\sqrt{2}}\, (\theta^a + \frac{\eta^{aa}}{i k} \theta^b) &=&
k\,\frac{1}{\sqrt{2}} (\theta^a + \frac{\eta^{aa}}{ik} \theta^b) \,, \nonumber\\
{\cal {\bf S}}^{ab} \,\frac{1}{\sqrt{2}}\, (1+ \frac{i}{k}  \theta^a \theta^b) &=&0\,,
 \nonumber\\ &{\rm or} &\;  \nonumber\\
{\cal {\bf S}}^{ab} \,\frac{1}{\sqrt{2}}\,  \frac{i}{k}  \theta^a \theta^b &=&0\,,
\label{eigengrasscartan}
\end{eqnarray}
with $k^2=\eta^{aa}\eta^{bb}$.  The  eigenvector $\frac{1}{\sqrt{2}}\, 
(\theta^0 \mp  \theta^3)$ of   ${\cal {\bf S}}^{03}$ has the eigenvalue $k=\pm i$,
respectively,
the eigenvalues of all the other eigenvectors of the rest of the Cartan subalgebra 
members, Eq.~(\ref{cartangrasscliff}), are  $k=\pm 1$.
In App.~\ref{proofs}, Statement 2a. the proof for Eq.~(\ref{eigengrasscartan}) is presented.

We choose the "basis vectors" to be products of an odd number of nilpotents 
$\frac{1}{\sqrt{2}}\, (\theta^a + \frac{\eta^{aa}}{i k} \theta^b)$ and the 
rest of the even objects $  \frac{i}{k}  \theta^a \theta^b $, with eigenvalues 
$k=\pm i$  and $0$, respectively, so that all the Cartan subalgebra eigenvectors
appear in the "basis vestor".  
 
Let us check how does ${\cal {\bf S}}^{ac}= i (\theta^a \frac{\partial}{\partial \theta_c} - 
\theta^c \frac{\partial}{\partial \theta_a})$ transform the product of two "nilpotents"
$\frac{1}{\sqrt{2}}\, (\theta^a + \frac{\eta^{aa}}{i k} \theta^b)$ and 
$\frac{1}{\sqrt{2}}\, (\theta^c + \frac{\eta^{cc}}{i k'} \theta^d)$.
One finds that ${\cal {\bf S}}^{ac}$ 
$\frac{1}{\sqrt{2}}\, (\theta^a + \frac{\eta^{aa}}{i k} \theta^b)$  
$\frac{1}{\sqrt{2}}\, (\theta^c + \frac{\eta^{cc}}{i k'} \theta^d)$ $ =
 - \frac{\eta^{aa} \eta^{cc}}{2k}$ $(\theta^a \theta^b  + 
 \frac{k}{k'} \theta^c \theta^d)$.
 ${\cal {\bf S}}^{ac}$ transforms the product of two Grassmann odd  
eigenvectors of the Cartan 
subalgebra into the superposition of two Grassmann even eigenvectors. 
 

"Basis vectors" have an odd or an even Grassmann character, if their products contain 
an odd  or an even number of "nilpotents", $\frac{1}{\sqrt{2}}\, (\theta^a + 
\frac{\eta^{aa}}{i k} \theta^b)$, respectively. 
"Basis vectors" are normalized, up to a phase, in accordance with Eq.~(\ref{grassintegral}) of App.~\ref{normgrass}.


The Hermitian conjugated representations of (either an odd or an even) products of $\theta^a$'s 
can be obtained by taking into account  Eq.~(\ref{thetaderher0}) for each "nilpotent"  
\begin{eqnarray}
\frac{1}{\sqrt{2}} (\theta^a + \frac{\eta^{aa}}{i k} \theta^b)^{\dagger} &=&
\eta^{aa}\,\frac{1}{\sqrt{2}} (\frac{\partial}{\partial \theta_{a}} + \frac{\eta^{aa}}{- i k}
 \frac{\partial}{\partial \theta_{b}}) \,, \nonumber\\
  (\frac{i}{k}\, \theta^a \theta^b)^{\dagger} &=& 
\frac{i}{k} \,\frac{\partial}{\partial \theta_{a}}\,\frac{\partial}{\partial \theta_{b}}\,.
\label{grasscartanher}
\end{eqnarray}
%

Making a choice of the identity for the vacuum state, 
\begin{eqnarray}
\label{vactheta}
|\phi_{og}> &=& |\,1>\,, 
\end{eqnarray}
we see that algebraic products --- we shall use  a dot, ${\,}\cdot{\,}$, or without  a dot 
for an algebraic product of eigenstates of the Cartan subalgebra forming ''basis vectors''
and $*_{A}$ for the algebraic product of  ''basis vectors''  --- of different $\theta^a$'s, 
if applied on such a vacuum state, give always nonzero contributions, \\
$$ (\theta^0 \mp \theta^3) \cdot (\theta^1 \pm i \theta^2)
\cdots  (\theta^{d-1} \pm i \theta^d) |\,1>\ne {\rm zero},$$ 
(this is true also, if we substitute any of  nilpotents 
$ \frac{1}{\sqrt{2}} (\theta^a + \frac{\eta^{aa}}{i k} \theta^b)$ or all of them 
with the corresponding  even operators $(\frac{i}{k}\, \theta^a \theta^b)$; in the case
of odd Grassmann irreducible representations at least one nilpotent must remain in the product).
The Hermitian conjugated partners, Eq.~(\ref{grasscartanher}), applied on $|\,1>$, give 
always zero  
$$ (\frac{\partial}{\partial \theta_{0}} \pm
\frac{\partial}{\partial \theta_{3}}) \cdot (\frac{\partial}{\partial \theta_{1}} \mp i 
\frac{\partial}{\partial \theta_{2}})\cdots ( \frac{\partial}{\partial \theta_{d-1}} \mp i 
\frac{\partial}{\partial \theta_{d}}) |\,1> = 0.$$ 

Let us notice the properties of the odd products of $\theta^a$'s and of their Hermitian 
conjugated partners:\\
 {\bf i.}  Superposition of  
 products of different  $\theta^a$'s,  applied on the vacuum state  
$|\,1>$, give nonzero contribution.  To create on the vacuum state the  ''fermion'' states
we make a choice of the ''basis vectors'' of the odd number of $\theta^a$'s, arranging them
to be the eigenvectors of all the Cartan subalgebra elements, Eq.~(\ref{cartangrasscliff}).   \\
{\bf ii.}  The Hermitian conjugated partners of the ``basis vectors'', they are 
products of derivatives $\frac{\partial}{\partial \theta_{a}}$'s,  give, when applied  
on the vacuum state  $|\,1>$, Eq.~(\ref{vactheta}),  zero. Each annihilation operator 
annihilates the corresponding creation operator, due to :
$\frac{\partial}{\partial \theta_{a}}\theta^b = \eta^{ab}$.\\
{\bf iii.} The algebraic product, $*_{A}$, of  a ``basis vector'' by itself gives zero, 
the algebraic anticommutator of any two ''basis vectors'' of an odd Grassmann character
(superposition of an odd products of $\theta^a$'s) gives zero, due to Eq.~(\ref{thetaderanti0}). \\ 
 {\bf iv.}  The algebraic  application of  any annihilation operator on the corresponding Hermitian conjugated ''basis vector'' gives identity, on all the rest of  ''basis vectors'' gives
  zero. Correspondingly the algebraic  anticommutators of the creation operators and their Hermitian conjugated partners, applied on the vacuum state, give identity,  all the rest anticommutators of creation and annihilation operators applied on the vacuum state, 
  give zero.\\
{\bf v.} Correspondingly the ``basis vectors'' and their Hermitian conjugated 
partners, applied on the vacuum state $|\,1>$, Eq.~(\ref{vactheta}), fulfill the properties of  creation  and annihilation operator, respectively, for the second quantized ''fermions''
on the level of one ''fermion'' state. 

\vspace{2mm}

From the eigenvectors of the members of the Cartan subalgebra, 
Eq.~(\ref{cartangrasscliff}), we construct  $ 2^{d-1}$ Grassmann odd "basis vectors" 
and $ 2^{d-1}-1$ (we skip self adjoint identity, which we use to describe the vacuum 
state $|\,1>$)  Grassmann even "basis vectors" as superposition of odd and even 
products of $\theta^a$'s, respectively. Their Hermitian conjugated $ 2^{d-1}$ odd 
and $ 2^{d-1}-1$ even partners are, according to Eqs.~(\ref{thetaderher0}, \ref{grasscartanher}), determined by the corresponding superposition of  odd and even  products of  $\frac{\partial}{\partial \theta_{a}}$'s, respectively~\footnote{  
Relations among operators and their Hermitian conjugated partners in both kinds of 
the Clifford algebra objects are more complicated than in the Grassmann case, where the 
Hermitian conjugated operators follow by taking into account Eq.~(\ref{thetaderher0}).

In the Clifford case 
$\frac{1}{2} (\gamma^a + \frac{\eta^{aa}}{i\,k} \gamma^b)^{\dagger}$  is 
proportional to $\frac{1}{2} (\gamma^a + \frac{\eta^{aa}}{i \,(- k)} \gamma^b) $, 
while $ \frac{1}{\sqrt{2}} (1 +  \frac{i}{k}  \gamma^a \gamma^b)$ are self adjoint. 
This is the case also for representations in the sector of $\tilde{\gamma}^a$'s.}. \\
We present in this Sect.~\ref{propertiesGrass0} the Grassmann anticommuting 
odd "basis vectors". 
In App.~\ref{evengrass} the reader can find illustration of the Grassmann even and 
correspondigly commuting ''basic vectors''.

\vspace{2mm}


%

Let us  in $d=2(2n+1)$-dimensional space-time, $n\ge 0$,  make a choice of the starting 
Grassmann odd "basis vector" $\hat{b}^{\theta 1\dagger}_{1}$, which is the
eigenvector of the Cartan subalgebra of Eqs.~(\ref{cartangrasscliff}, \ref{eigengrasscartan}) 
with the egenvalues
 $(+i, +1, +1, \cdots, +1 )$, respectively, and has the Hermitian conjugated partner 
 equal to $(\hat{b}^{\theta 1\dagger}_{1})^{\dagger}=$ $\hat{b}^{\theta 1}_{1}$,
\begin{eqnarray}
&&\hat{b}^{\theta 1\dagger}_{1}{\bf :}= (\frac{1}{\sqrt{2}})^{\frac{d}{2}} \,
(\theta^0 - \theta^3) (\theta^1 + i \theta^2) (\theta^5 + i\theta^6) \nonumber\\
&& {} \cdots (\theta^{d-1} + i \theta^{d}) \,, \nonumber\\
&& \hat{b}^{\theta 1}_{1}  {\bf :} =
 (\frac{1}{\sqrt{2}})^{\frac{d}{2}}\,
 (\frac{\partial}{\partial \theta^{d-1}} - i \frac{\partial}{\partial \theta^{d}})
{}\cdots (\frac{\partial}{\partial \theta^{0}}
+ \frac{\partial}{\partial \theta^3})\,,\nonumber\\
d &=& 2(2n+1) \,.                                                      
\label{start(2n+1)2theta}
\end{eqnarray}
%

In the case of $d=4n$, $n\ge 0$, the corresponding starting Grassmann odd "basis vector" 
can be chosen as  
\begin{eqnarray}
{\hat b}^{\theta 1 \dagger}_{1}{\bf :} &=&(\frac{1}{\sqrt{2}})^{\frac{d}{2}-1} \,
(\theta^0 - \theta^3) (\theta^1 + i \theta^2) (\theta^5 + i \theta^6)\cdots
                              \nonumber\\
  &&{}\cdots (\theta^{d-3} +
 i \theta^{d-2})  \theta^{d-1} \theta^d\,,\nonumber\\
d &=& 4n \,.              
\label{start4ntheta}
\end{eqnarray}
All the rest of "basis vectors", belonging to the same irreducible representation of the 
Lorentz group, follow by the application of ${\cal {\bf S}}^{ab}$'s.
 
Let us denote the members $m$ of this starting irreducible representation $f$ by 
$\hat{b}^{\theta m \dagger}_{f}$, $m=1, f=1$, 
and their Hermitian conjugated partners by $\hat{b}^{\theta m}_{f}$, with 
$m=1, f=1$.

 "Basis vectors", belonging to different irreducible representations $f'$, will be 
denoted by $\hat{b}^{\theta m' \dagger}_{f'}$, $f'$ and their Hermitian 
conjugated partners by 
$\hat{b}^{\theta m' }_{f'}=(\hat{b}^{\theta m' \dagger}_{f'})^{\dagger} $.

${\cal {\bf S}}^{ac}$'s, which do not belong to the Cartan subalgebra, transform 
step by step the two by two "nilpotents", no matter how many "nilpotents" are 
between the chosen two, up to a constant,  
as follows:\\ 
${\cal {\bf S}}^{ac}
\frac{1}{\sqrt{2}}\, (\theta^a + \frac{\eta^{aa}}{i k} \theta^b) \cdots $ 
$\frac{1}{\sqrt{2}}\, (\theta^c + \frac{\eta^{cc}}{i k'} \theta^d)$
$\propto - \frac{\eta^{aa} \eta^{cc}}{2k} (\theta^a \theta^b  + 
\frac{k}{k'} \theta^c \theta^d) \cdots$,\\
leaving at each step at least one "nilpotent" unchanged, so that the whole irreducible
representation remains odd.\\ 
The superposition of ${\cal {\bf S}}^{bd}$ and $ i {\cal {\bf S}}^{bc}$
transforms $- \frac{\eta^{aa} \eta^{cc}}{2k}$ $(\theta^a \theta^b  +
 \frac{k}{k'} \theta^c \theta^d)$ into $\frac{1}{\sqrt{2}} $
$(\theta^a - \frac{\eta^{aa}}{i k} \theta^b) 
 \frac{1}{\sqrt{2}}\, (\theta^c - \frac{\eta^{cc}}{i k'} \theta^d)$, and not into
$\frac{1}{\sqrt{2}}\, (\theta^a + \frac{\eta^{aa}}{i k} \theta^b)$ 
$\frac{1}{\sqrt{2}}\, (\theta^c - \frac{\eta^{cc}}{i k'} \theta^d)$ or not into
$\frac{1}{\sqrt{2}}\, (\theta^a - \frac{\eta^{aa}}{i k} \theta^b)$ 
$\frac{1}{\sqrt{2}}\, (\theta^c + \frac{\eta^{cc}}{i k'} \theta^d)$.

Therefore we can start another odd representation with the "basis vector" 
$\hat{b}^{\theta m \dagger }_{2}$  as follows
%
\begin{eqnarray}
\hat{b}^{\theta m=1\dagger}_{2}{\bf :} &=& (\frac{1}{\sqrt{2}})^{\frac{d}{2}} \,
(\theta^0 + \theta^3) (\theta^1 + i \theta^2) (\theta^5 + i\theta^6) 
 {} \cdots (\theta^{d-1} + i \theta^{d}) \,, \nonumber\\
(\hat{b}^{\theta m=1 \dagger}_{2})^{\dagger} &=& \hat{b}^{\theta 1}_{2} {\bf :} =
 (\frac{1}{\sqrt{2}})^{\frac{d}{2}}\,
 (\frac{\partial}{\partial \theta^{d-1}} - i \frac{\partial}{\partial \theta^{d}})
{}\cdots (\frac{\partial}{\partial \theta^{0}}
-\frac{\partial}{\partial \theta^3})\,.
\label{start(2n+1)2thetasecond}
\end{eqnarray}
The application of ${\cal {\bf S}}^{ac}$'s determines the whole second irreducible representation 
$\hat{b}^{\theta 2 \dagger }_{j}$.
 
One finds that each of these two irreducible representations has 
${\bf \frac{1}{2}\frac{d!}{\frac{d}{2}! \frac{d}{2}!}}$ members, Ref.~\cite{nh2018}.



%
%
Taking into account Eq.~(\ref{thetaderanti0}), it follows that odd products of $\theta^a$'s 
anticommute and so do the odd products of $\frac{\partial}{\partial \theta_{a}}$'s.
It then follows:\\

{\bf Statement 1. } 
The oddness of the products of  $\theta^a$'s guarantees the 
anticommuting properties of all objects which include odd number of $\theta^a$'s.\\

One further sees that $\frac{\partial}{\partial \theta^{a}} \theta^b = \eta^{ab}$,
while $\frac{\partial}{\partial \theta_{a}}|\,1>=0$, and $\theta^a |\,1>=\theta^a  |\,1>$.
We can therefore conclude
\begin{eqnarray}
\{ \hat{b}^{\theta m}_f, \hat{b}^{\theta m' \dagger}_{f'} \}_{*_{A}+} 
|\,\,1> &=& \delta_{f f'}\; \delta^{m m'}\;|\,\,1>\,,\nonumber\\
\{ \hat{b}^{\theta m}_f, \hat{b}^{\theta m'}_{f'} \}_{*_{A}+}  |\,\,1>
&=& 0\;\cdot\, |\,\,1> \,,\nonumber\\
\{\hat{b}^{\theta m \dagger}_f,\hat{b}^{\theta m' \dagger}_{f'}\}_{*_{A}+} \;|\,\,1>
&=&0\;\cdot\, |\,\,1> \,,\nonumber\\
\hat{b}^{\theta m}_{f} \,*_{A}\,|\,\,1>& =&0\;\cdot\, |\,\,1> \,, \nonumber\\
 |\phi^{m}_{o\, f }> &=& \hat{b}^{\theta m\dagger}_{f} \,|\,1>\,,
\label{ijthetaprod}
\end{eqnarray}
where $\{ \hat{b}^{\theta m}_f, \hat{b}^{\theta m' \dagger}_{f'} \}_{*_{A}+}=$
$ \hat{b}^{\theta m}_f *_{A} \hat{b}^{\theta m' \dagger}_{f'} + 
\hat{b}^{\theta m'}_{f'} *_{A} \hat{b}^{\theta m \dagger}_{f} $.
 
These anticommutation relations of the ''basis vectors'' of the odd Grassmann character, 
manifest on the level of the Grassmann algebra the  anticommutation relations 
required by Dirac~\cite{Dirac} for second quantized fermions.

%
%
The creation operators, determining the ''Grassmann fermions'' carrying the integer 
spins (in adjoint representations, Eq.~(\ref{eigengrasscartan})), are tensor products, 
$*_{T}$, of ''basis vectors'' in internal space and of the (momentum or coordinate) 
''basis'' in ordinary space. Since the oddness of the ''basis vectors''  transfers to 
the creation operators, then the creation operators and their Hermitian conjugated
partners annihilattion operators, fulfill the anticommutation relations postulated 
by Dirac for the second quantized ''Grassmann fermions'' with the integer spin.
 
 The integer spin fermions have not been observed, all the observed  fermions carry the 
 half integer spin. 
 

\vspace{2mm}

$\;\;$ {\it Grassmann integer spin fermions in $d=(5+1)$:}

\vspace{2mm}

 Let us illustrate  anticommuting ''basis vectors'' with spins and charges in adjoint 
 representation in $d=(5+1)$-dimensional space.

\begin{small}
Table~\ref{Table grassdecuplet.} represents two decuplets, which are the 
"egenvectors" of the Cartan subalgbra (${\cal {\bf S}}^{03}$,  
${\cal {\bf S}}^{12}$, ${\cal {\bf S}}^{56}$), Eq.~(\ref{cartangrasscliff}), 
of the Lorentz algebra ${\cal {\bf S}}^{ab}$. 
The two decuplets represent two Grassmann odd  irreducible representations 
of $SO(5,1)$. 

One can read on the same table, from the first to the third and from the 
fourth to the sixth line in each of the two decuplets, two Grassmann even 
triplet representations of $SO(3,1)$, if paying attention on the 
eigenvectors of ${\cal {\bf S}}^{03}$ and  ${\cal {\bf S}}^{12}$ alone,  
while the  eigenvector of ${\cal {\bf S}}^{56}$ has, as a "spectator", the 
eigenvalue  either $+1$ (the first triplet in both decuplets) or $-1$ (the 
second triplet in both decuplets).  Each of the two decuplets contains 
also one "fourplet"  with the "charge" ${\cal {\bf S}}^{56}$ equal to zero 
($(7^{th}, 8^{th}, 9^{th}, 10^{th})$ lines in each of the two decuplets 
(Table II in Ref.~\cite{norma93})).  

Paying attention on the eigenvectors of ${\cal {\bf S}}^{03}$ alone one 
recognizes as well even and odd representations of $SO(1,1)$: 
$\theta^0 \theta^3$ 
and $\theta^0 \pm \theta^3$, respectively.

The Hermitian conjugated "basis vectors" follow  by using  
Eq.~(\ref{grasscartanher}) and is for the first "basis vector" of 
Table~\ref{Table grassdecuplet.} equal to $(-)^2 
(\frac{1}{\sqrt{2}})^3 (\frac{\partial}{\partial \theta_5} -i
\frac{\partial}{\partial \theta_6})\, (\frac{\partial}{\partial \theta_1} - i
\frac{\partial}{\partial \theta_2})\, (\frac{\partial}{\partial \theta_0} + 
\frac{\partial}{\partial \theta_3})$. One correspondingly finds that when $(\frac{1}{\sqrt{2}})^3 
(\frac{\partial}{\partial \theta_5} -i \frac{\partial}{\partial \theta_6})\, (\frac{\partial}{\partial \theta_1} - i
\frac{\partial}{\partial \theta_2})\, (\frac{\partial}{\partial \theta_0} + 
\frac{\partial}{\partial \theta_3}$) applies on $(\frac{1}{\sqrt{2}})^3
 (\theta^{0} - \theta^{3}) (\theta^{1} + i \theta^{2}) (\theta^{5} + i \theta^{6}) $ the result is 
identity. Application of $ (\frac{1}{\sqrt{2}})^3 (\frac{\partial}{\partial \theta_5} -i
\frac{\partial}{\partial \theta_6})\, (\frac{\partial}{\partial \theta_1} - i
\frac{\partial}{\partial \theta_2})\, (\frac{\partial}{\partial \theta_0} + 
\frac{\partial}{\partial \theta_3})$ on all the rest of "basis vectors" of the decuplet $I$ as well as 
on all the "basis vectors" of the decuplet $II$ gives zero. "Basis vectors" are orthonormalized with respect to
 Eq.~(\ref{grassintegral}) of App.~\ref{normgrass}.
 
%
 \begin{table*}
\begin{center}
\begin{minipage}[t]{16.5 cm}
\caption{The two decuplets, the odd eigenvectors of the 
Cartan subalgebra, Eq.~(\ref{cartangrasscliff}),
 (${\cal {\bf S}}^{0 3}, {\cal {\bf S}}^{1 2}$, ${\cal {\bf S}}^{5 6}$) 
 of the Lorentz algebra $SO(5,1)$ in Grassmann 
$(5+1)$-dimensional space, forming two irreducible representations, are presented.
Table is taken from Refs.~\cite{nh2018,2020PartIPartII}.  
 The "basis vectors" within each decuplet are reachable from any member by 
${\cal {\bf S}}^{ab}$'s and are decoupled from another decuplet.
The two operators of handedness,  $\Gamma^{((d-1)+1)}$ for $d=(6,4)$, are invariants 
of the Lorentz algebra, Eq.~(\ref{handedness}) of App.~\ref{handednessGrassCliff}, $\Gamma^{(5+1)}$ for the whole decuplet, 
$\Gamma^{(3+1)}$ for the "triplets" and "fourplets".}
\label{Table grassdecuplet.} 
\end{minipage}
 \begin{tabular}{|c|r|r|r|r|r|r|r|}
 \hline
$I$&$i$ &$\rm{decuplet\; of\; eigenvectors} $&${\cal {\bf S}}^{03}$&${\cal {\bf S}}^{1 2}$&
${\cal {\bf S}}^{5 6}$&$\Gamma^{(5+1)}$&$\Gamma^{(3+1)}$\\
 \hline 
& $1$  & ($\frac{1}{\sqrt{2}})^3 (\theta^{0} - \theta^{3}) (\theta^{1} + i \theta^{2})
 (\theta^{5} + i \theta^{6})$ &$ i$&$ 1$&$1$&$1$&$1$\\
\hline
&$2$  & ($\frac{1}{\sqrt{2}})^2 (\theta^{0} \theta^{3} + i \theta^{1} \theta^{2}) 
 (\theta^{5} + i \theta^{6})$ & $ 0$ & $0 $ &$1$&$1$&$1$\\
\hline
&$3$  & ($\frac{1}{\sqrt{2}})^3 (\theta^{0} +  \theta^{3}) (\theta^{1} - i \theta^{2})
  (\theta^{5} + i \theta^{6})$ &$-i $&$-1$&$1$&$1$&$1$\\
\hline
&$4$  & ($\frac{1}{\sqrt{2}})^3 (\theta^{0} -  \theta^{3}) (\theta^{1} - i \theta^{2})
  (\theta^{5} - i \theta^{6})$ &$ i $&$-1$&$-1$&$1$&$-1$\\
\hline
&$5$  & ($\frac{1}{\sqrt{2}})^2 (\theta^{0} \theta^{3} - i \theta^{1} \theta^{2}) 
 (\theta^{5} - i \theta^{6})$ & $ 0 $& $0 $&$-1$&$1$&$-1$\\
\hline
&$6$  & ($\frac{1}{\sqrt{2}})^3 (\theta^{0} +  \theta^{3}) (\theta^{1} + i \theta^{2})
  (\theta^{5} - i \theta^{6})$ &$-i $&$ 1$&$-1$&$1$&$-1$\\
\hline
& $7$  & ($\frac{1}{\sqrt{2}})^2 (\theta^{0} - \theta^{3}) (\theta^{1} \theta^{2} +
 \theta^{5} \theta^{6})$ &$ i$&$ 0$&$0$&$1$&$0$\\
\hline
& $8$  & ($\frac{1}{\sqrt{2}})^2 (\theta^{0} + \theta^{3}) (\theta^{1} \theta^{2} -
 \theta^{5} \theta^{6})$ &$- i$&$ 0$&$0$&$1$&$0$\\
\hline
& $9$  & ($\frac{1}{\sqrt{2}})^2 (\theta^{0}  \theta^{3} +i \theta^{5} \theta^{6}) 
(\theta^{1}+i \theta^{2}) 
$ &$ 0$&$ 1$&$0$&$1$&$0$\\
\hline
& $10$  & ($\frac{1}{\sqrt{2}})^2 (\theta^{0}  \theta^{3} - i \theta^{5} \theta^{6}) 
(\theta^{1}-i \theta^{2}) 
$ &$ 0$&$- 1$&$0$&$1$&$0$\\
\hline\hline 
$II$&$i$ &$\rm{decuplet\; of\; eigenvectors}$&${\cal {\bf S}}^{03}$&${\cal {\bf S}}^{1 2}$&
${\cal {\bf S}}^{5 6}$&$\Gamma^{(5+1)}$&$\Gamma^{(3+1)}$\\ 
 \hline 
& $1$  & ($\frac{1}{\sqrt{2}})^3 (\theta^{0} + \theta^{3}) (\theta^{1} + i \theta^{2})
 (\theta^{5} + i \theta^{6})$ &$- i$&$ 1$&$1$&$-1$&$-1$\\
\hline
&$2$  & ($\frac{1}{\sqrt{2}})^2 (\theta^{0} \theta^{3} - i \theta^{1} \theta^{2}) 
 (\theta^{5} + i \theta^{6})$ & $ 0$&$ 0 $&$1$&$-1$&$-1$\\
\hline
&$3$  & ($\frac{1}{\sqrt{2}})^3 (\theta^{0} -  \theta^{3}) (\theta^{1} - i \theta^{2})
  (\theta^{5} + i \theta^{6})$ &$ i $&$-1$&$1$&$-1$&$-1$\\
\hline
&$4$  & ($\frac{1}{\sqrt{2}})^3 (\theta^{0} +  \theta^{3}) (\theta^{1} - i \theta^{2})
  (\theta^{5} - i \theta^{6})$ &$- i $&$-1$&$-1$&$-1$&$1$\\
\hline
&$5$  & ($\frac{1}{\sqrt{2}})^2 (\theta^{0} \theta^{3} + i \theta^{1} \theta^{2}) 
 (\theta^{5} - i \theta^{6})$ & $ 0$& $0 $&$-1$&$-1$&$1$\\
\hline
&$6$  &($\frac{1}{\sqrt{2}})^3 (\theta^{0} -  \theta^{3}) (\theta^{1} + i \theta^{2})
  (\theta^{5} - i \theta^{6})$ &$ i $&$ 1$&$-1$&$-1$&$1$\\
\hline
& $7$  & ($\frac{1}{\sqrt{2}})^2 (\theta^{0} + \theta^{3}) (\theta^{1} \theta^{2} +
 \theta^{5} \theta^{6})$ &$- i$&$ 0$&$0$&$-1$&$0$\\
\hline
& $8$  & ($\frac{1}{\sqrt{2}})^2 (\theta^{0} - \theta^{3}) (\theta^{1} \theta^{2} -
 \theta^{5} \theta^{6})$ &$ i$&$ 0$&$0$&$-1$&$0$\\
\hline
& $9$  & ($\frac{1}{\sqrt{2}})^2 (\theta^{0} \theta^{3} - i \theta^{5} \theta^{6}) 
(\theta^{1}+i \theta^{2}) 
$ &$ 0$&$ 1$&$0$&$-1$&$0$\\
\hline
& $10$  & ($\frac{1}{\sqrt{2}})^2 (\theta^{0}  \theta^{3} + i \theta^{5} \theta^{6}) 
(\theta^{1}-i \theta^{2}) 
$ &$ 0$&$- 1$&$0$&$-1$&$0$\\
\hline\hline 
 \end{tabular}
\end{center}
%
 \end{table*} 
Let us notice that $\frac{\partial}{\partial \theta_{a}}$ on a "state" which is just  an identity, $|\,1>$,
gives zero, $\frac{\partial}{\partial \theta_{a}}\,|\,1>=0$, while $\theta^a \,|\,1>$, or 
any superposition of products of different $\theta^a$'s, applied on $|\,1>$, gives the 
"vector" back.

One easily sees that application of 
products of superposition of $\theta^a$'s on $|\,1>$  gives nonzero contribution, while 
application of products of superposition of $\frac{\partial}{\partial \theta^a}$'s on $|\,1>$ gives zero.

The two by ${\cal {\bf S}}^{ab}$ decoupled Grassmann decuplets of 
Table~\ref{Table grassdecuplet.} are the largest two 
irreducible representations of  odd products of $\theta^a$'s. There are $12$  additional Grassmann 
odd "vectors", arranged into irreducible representations of six singlets and six sixplets
\begin{eqnarray} 
&& (\frac{1}{2}\,(\theta^0\mp \theta^3), \frac{1}{2}\,(\theta^1\pm i \theta^2), 
 \frac{1}{2}\,(\theta^5 \pm i \theta^6)\,,\nonumber\\
&&\frac{1}{2}\,(\theta^0 \mp \theta^3)\, \theta^1 \theta^2 \theta^5 \theta^6,
\frac{1}{2}\,(\theta^1\pm i \theta^2)\, \theta^0 \theta^3 \theta^5 \theta^6\,, 
\frac{1}{2}\,(\theta^5 \pm i \theta^6)\,\theta^0 \theta^3 \theta^1 \theta^2)\,.
\label{nondecuplet}
\end{eqnarray}

The algebraic application of products of superposition of 
$\frac{\partial}{\partial \theta^a}$'s on the corresponding Hermitian conjugated partners, which are products of superposition of $\theta^a$'s,
leads to the identity for either even or odd Grassmann character~\footnote{
We shall see that the vacuum states are in the Clifford case, similarly as in the 
Grassmann case, for both kinds of the Clifford algebra objects, $\gamma^a$'s 
and  $\tilde{\gamma}^a$'s, normalized sums of products of the annihilation 
$\times$, $*_{A}$, its Hermitian conjugated creation operators, and correspondingly self adjoint operators, 
but they are not the identity.}.

Besides $32$ Grassmann odd eigenvectors of the Grassmann Cartan subalgebra, 
Eq.~(\ref{cartangrasscliff}), there are $(32-1)$ Grassmann "basis vectors", 
which we  arrange into irreducible representations, which are superposition of 
even products of $\theta^a$'s. The even self adjoint operator identity 
(which is indeed the normalized product of all the annihilation times, $*_{A}$, 
creation operators)  is used to represent the vacuum state. 

%

%


It is not difficult to see that  Grassmann "basis vectors" of an odd Grassmann character 
anticommute among themselves and so do odd products of superposition of 
$\frac{\partial}{\partial \theta^a}$'s, while equivalent even
products commute. 

We shall demonstrate the matrix representations of the operators of the
Lorentz transformations and of the elements of the algebras for the Clifford 
algebra elements $\gamma^a$'s and $\tilde{\gamma}^a$'s when applying 
on ''basis vectors'' of the Clifford algebra of $\gamma^a$'s in 
App.~\ref{matrixCliffordDMN}.


The Grassmann odd algebra (as well as the two odd Clifford algebras) offers, due to the
oddness of the internal space giving oddness as well to the elements of the tensor products
of the internal space and of the momentum space, the 
description of the anticommuting second quantized fermion fields, as postulated by Dirac. 
But  the Grassmann "fermions" carry the integer spins, while the observed fermions ---  
quarks and leptons --- carry half integer spin.
\end{small}
%

%


%
\subsubsection{Clifford algebras and ''basis vectors'' describing internal space}
\label{propertiesCliff0}
%

%
%

We learn in Sect.~\ref{GrassmannClifford}, Eq.~(\ref{clifftheta}), that in 
$d$-dimensional space of anticommuting Grassmann coordinates (and of their 
Hermitian conjugated partners --- derivatives) 
there exist two kinds of the Clifford coordinates (operators) --- 
$\gamma^{a}$ and $\tilde{\gamma}^{a}$ --- both are expressible in terms of 
$\theta^{a}$ and their conjugate momenta $p^{\theta a}= i \,
\frac{\partial}{\partial \theta_{a}}$ ~\cite{norma93}.
\begin{eqnarray}
\label{clifftheta1}
\gamma^{a} &=& (\theta^{a} + \frac{\partial}{\partial \theta_{a}})\,, \quad 
\tilde{\gamma}^{a} =i \,(\theta^{a} - \frac{\partial}{\partial \theta_{a}})\,,\nonumber\\
\theta^{a} &=&\frac{1}{2} \,(\gamma^{a} - i \tilde{\gamma}^{a})\,, \quad 
\frac{\partial}{\partial \theta_{a}}= \frac{1}{2} \,(\gamma^{a} + i \tilde{\gamma}^{a})\,,
\nonumber\\
\end{eqnarray}
offering together  $2\cdot 2^d$  operators: $2^d$ of those which are products of 
$\gamma^{a}$  and  $2^d$ of those which are products of $\tilde{\gamma}^{a}$.

Let us repeat the anticommuting properties of both Clifford algebras, presented 
already  in Eq.~(\ref{gammatildeantiher0}).
\begin{eqnarray}
\label{gammatildeantiher}
\{\gamma^{a}, \gamma^{b}\}_{+}&=&2 \eta^{a b}= \{\tilde{\gamma}^{a}, 
\tilde{\gamma}^{b}\}_{+}\,, \nonumber\\
\{\gamma^{a}, \tilde{\gamma}^{b}\}_{+}&=&0\,,\quad
 (a,b)=(0,1,2,3,5,\cdots,d)\,, \nonumber\\
(\gamma^{a})^{\dagger} &=& \eta^{aa}\, \gamma^{a}\, , \quad 
(\tilde{\gamma}^{a})^{\dagger} =  \eta^{a a}\, \tilde{\gamma}^{a}\,,\nonumber
\end{eqnarray}
with $\eta^{a b}=diag\{1,-1,-1,\cdots,-1\}$, leading to the statement that the
two Clifford algebras form the independent algebras.\\

{\bf  Statement 2.} $\gamma^a$'s  and $\tilde{\gamma}^{a}$'s define two independent Clifford algebras.\\

The proof of this statement can be found in App.~\ref{proofs}, Statement 1. 

\vspace{2mm}

As in the Grassmann case we chose the ''basis vectors'' in each of the two
spaces to beproducts of eigenstates of the Cartan subalgebra members, 
Eq.~(\ref{cartangrasscliff}), of the Lorentz algebras, 
($S^{ab} =\frac{i}{4}(\gamma^a \gamma^b - \gamma^b \gamma^a), 
\tilde{S}^{ab} =\frac{i}{4}(\tilde{\gamma}^a \tilde{\gamma}^b -
 \tilde{\gamma}^b  \tilde{\gamma}^a)$).

First we look for ''eigenstates'' of each of the Cartan subalgebra members, 
Eq.~(\ref{cartangrasscliff}), for each of the two kinds of the Clifford algebras 
separately,   
\begin{eqnarray}
S^{ab} \frac{1}{2} (\gamma^a + \frac{\eta^{aa}}{ik} \gamma^b) &=& \frac{k}{2}  \,
\frac{1}{2} (\gamma^a + \frac{\eta^{aa}}{ik} \gamma^b)\,,\quad
S^{ab} \frac{1}{2} (1 +  \frac{i}{k}  \gamma^a \gamma^b) = \frac{k}{2}  \,
 \frac{1}{2} (1 +  \frac{i}{k}  \gamma^a \gamma^b)\,,\nonumber\\
\tilde{S}^{ab} \frac{1}{2} (\tilde{\gamma}^a + \frac{\eta^{aa}}{ik} \tilde{\gamma}^b) &=& 
\frac{k}{2}  \,\frac{1}{2} (\tilde{\gamma}^a + \frac{\eta^{aa}}{ik} \tilde{\gamma}^b)\,,
\quad
\tilde{S}^{ab} \frac{1}{2} (1 +  \frac{i}{k}  \tilde{\gamma}^a \tilde{\gamma}^b) = 
 \frac{k}{2}  \, \frac{1}{2} (1 +  \frac{i}{k} \tilde{\gamma}^a \tilde{\gamma}^b)\,,
\label{eigencliffcartan}
\end{eqnarray}
 $k^2=\eta^{aa} \eta^{bb}$.
 The proof of Eq.~(\ref{eigencliffcartan}) is presented in App.~(\ref{proofs}), 
 Statement 2a.
The Clifford "basis vectors" --- nilpotents $\frac{1}{2} (\gamma^a + \frac{\eta^{aa}}{ik} \gamma^b), (\frac{1}{2} (\gamma^a + \frac{\eta^{aa}}{ik} \gamma^b))^2=0$ and projectors $ \frac{1}{2} (1 +  \frac{i}{k}  \tilde{\gamma}^a \tilde{\gamma}^b),
( \frac{1}{2} (1 +  \frac{i}{k}  \tilde{\gamma}^a \tilde{\gamma}^b))^2 =
 \frac{1}{2} (1 +  \frac{i}{k}  \tilde{\gamma}^a \tilde{\gamma}^b)$ --- 
of both algebras are normalized,  up to a phase, with respect to 
Eq.~(\ref{grassintegral}) of App.~\ref{normgrass}.
 
Both, nilpotents and projectors, have half integer spins. The "eigenvalues" of the operator $S^{03}$, for example,  
for the "vector" $\frac{1}{2} (\gamma^0 \mp \gamma^3)$ are equal to 
$\pm\, \frac{i}{2}$, respectively, for the "vector" $\frac{1}{2} (1\pm \gamma^0  \gamma^3)$
are $\pm\, \frac{i}{2}$, respectively, while all the rest "vectors" have "eigenvalues" 
$\pm\, \frac{1}{2}$. One finds equivalently for the "eigenvectors" of the operator 
$\tilde{S}^{03}$: for $\frac{1}{2} \,( \tilde{\gamma^0} \mp \tilde{\gamma}^3)$  
the "eigenvalues"
$\pm\, \frac{i}{2}$, respectively, and for the "eigenvectors" $\frac{1}{2} 
(1\pm \tilde{\gamma}^0  \tilde{\gamma}^3)$ the "eigenvalues" $k=\pm\,  \frac{i}{2}$, 
respectively, while all the rest "vectors" have $k=\pm\,  \frac{1}{2}$. 

To make discussions easier let us introduce the notation for the "eigenvectors" of the 
two Cartan subalgebras, Eq.~(\ref{cartangrasscliff}), Ref.~\cite{nh02,norma93}.
\begin{eqnarray}
\stackrel{ab}{(k)}:&=& 
\frac{1}{2}(\gamma^a + \frac{\eta^{aa}}{ik} \gamma^b)\,,\quad 
\stackrel{ab}{(k)}^{\dagger} = \eta^{aa}\stackrel{ab}{(-k)}\,,\quad 
(\stackrel{ab}{(k)})^2 =0\,, \quad \stackrel{ab}{(k)}\stackrel{ab}{(-k)}
=\eta^{aa}\stackrel{ab}{[k]}\nonumber\\
\stackrel{ab}{[k]}:&=&
\frac{1}{2}(1+ \frac{i}{k} \gamma^a \gamma^b)\,,\quad \;\,
\stackrel{ab}{[k]}^{\dagger} = \,\stackrel{ab}{[k]}\,, \quad \quad \quad \quad
(\stackrel{ab}{[k]})^2 = \stackrel{ab}{[k]}\,, 
\quad \stackrel{ab}{[k]}\stackrel{ab}{[-k]}=0\,,
\nonumber\\
\stackrel{ab}{(k)}\stackrel{ab}{[k]}& =& 0\,,\qquad \qquad \qquad 
\stackrel{ab}{[k]}\stackrel{ab}{(k)}=  \stackrel{ab}{(k)}\,, \quad \quad \quad
  \stackrel{ab}{(k)}\stackrel{ab}{[-k]} =  \stackrel{ab}{(k)}\,,
\quad \, \stackrel{ab}{[k]}\stackrel{ab}{(-k)} =0\,,
\nonumber\\
\stackrel{ab}{\tilde{(k)}}:&=& 
\frac{1}{2}(\tilde{\gamma}^a + \frac{\eta^{aa}}{ik} \tilde{\gamma}^b)\,,\quad 
\stackrel{ab}{\tilde{(k)}}^{\dagger} = \eta^{aa}\stackrel{ab}{\tilde{(-k)}}\,,\quad
(\stackrel{ab}{\tilde{(k)}})^2=0\,,\nonumber\\
\stackrel{ab}{\tilde{[k]}}:&=&
\frac{1}{2}(1+ \frac{i}{k} \tilde{\gamma}^a \tilde{\gamma}^b)\,,\quad \;\,
\stackrel{ab}{\tilde{[k]}}^{\dagger} = \,\stackrel{ab}{\tilde{[k]}}\,,
\quad \quad \quad \quad
(\stackrel{ab}{\tilde{[k]}})^2=\stackrel{ab}{\tilde{[k]}}\,,\nonumber\\
\stackrel{ab}{\tilde{(k)}}\stackrel{ab}{\tilde{[k]}}& =& 0\,,\qquad \qquad \qquad 
\stackrel{ab}{\tilde{[k]}}\stackrel{ab}{\tilde{(k)}}=  \stackrel{ab}{\tilde{(k)}}\,, 
\quad \quad \quad
  \stackrel{ab}{\tilde{(k)}}\stackrel{ab}{\tilde{[-k]}} =  \stackrel{ab}{\tilde{(k)}}\,,
\quad \, \stackrel{ab}{\tilde{[k]}}\stackrel{ab}{\tilde{(-k)}} =0\,,
\nonumber\\
\label{graficcliff}
\end{eqnarray}
with $k^2 = \eta^{aa} \eta^{bb}$ for both algebras. 
Let us notice that the ``eigenvectors'' of  the Cartan subalgebras and the eigenvalues 
are equivalent  in both algebras. Both algebras have projectors and nilpotents:
($ (\stackrel{ab}{[k]})^2= \stackrel{ab}{[k]}\,,(\stackrel{ab}{(k)})^{2}=0$),
($ (\stackrel{ab}{\tilde{[k]}})^2= \stackrel{ab}{\tilde{[k]}}\,,
 (\stackrel{ab}{\tilde{(k)}})^{2}=0$).
We pay attention on even dimensional spaces, $d=2(2n+1)$ or $d=4n$, $n\ge0$ only.

The "basis vectors", which are products of $\frac{d}{2}$ either of nilpotents or
of projectors or of both, are ``eigenstate'' of all the members of the Cartan 
subalgebra, Eq.~(\ref{cartangrasscliff}), of the corresponding Lorentz algebra, forming 
$2^{\frac{d}{2}-1}$ irreducible representations with $2^{\frac{d}{2}-1}$ members 
in each of the two  Clifford algebras cases. 

The "basis vectors", which appear  below in  Eq.~(\ref{allcartaneigenvec}), 
are "eigenvectors"  of all the Cartan subalgebra members, Eq.~(\ref{cartangrasscliff}), in
 $d=2(2n+1)$-dimensional space of $\gamma^a$'s.
The first one is the product of  nilpotents only and correspondingly a superposition 
of an odd products of $\gamma^a$'s. The second one belongs to the same 
irreducible representation as the first one, since it follows from the first one by the 
application of $S^{01}$, for example, the third one follows from the second one by
the application of $S^{d-3\,d-1}$.
%
\begin{eqnarray}
\label{allcartaneigenvec}
&&\stackrel{03}{(+i)}\stackrel{12}{(+)}\cdots \stackrel{d-1 \, d}{(+)}\,,\;\;\quad
 \stackrel{03}{[-i]}\stackrel{12}{[-i]} \stackrel{56}{(+)} \cdots 
\stackrel{d-1 \, d}{(+)}\,,  \nonumber\\ 
&&\stackrel{03}{[-i]} \stackrel{12}{[-]} \stackrel{56}{(+)} \cdots 
\stackrel{d-3\,d-2}{[-]}\;\stackrel{d-1\,d}{[-]}\,.
\end{eqnarray}
The number of nilpotents in any of the members remains all the time odd, due to 
the eveness of the generators of the Lorentz transformations.
One finds for their Hermitian conjugated partners, up to a sign,
\begin{eqnarray}
\label{allcartaneigenvecher}
&&\stackrel{03}{(-i)}\stackrel{12}{(-)}\cdots \stackrel{d-1 \, d}{(-)}\,,\;\;\quad
\ \stackrel{03}{[-i]}\stackrel{12}{[-i]} \stackrel{56}{(-)}\cdots 
\stackrel{d-1 \, d}{(-)}\,,\nonumber\\  
&& \stackrel{03}{[-i]} \stackrel{12}{[-]} \stackrel{56}{(-)}\cdots 
\stackrel{d-3\,d-2}{[-]}\;\stackrel{d-1\,d}{[-]}\,.
\nonumber
\end{eqnarray}
The members of the above irreducible representation for the $\tilde{\gamma}^a$'s
follow from Eq.~(\ref{allcartaneigenvecher}) by replacing $\gamma^a$'s with
$\tilde{\gamma}^a$'s.

The "basis vectors" form an orthonormal basis within each of the irreducible 
representations: 
%
\begin{eqnarray}
&& \stackrel{d-1 \, d}{(-)}\cdots \stackrel{12}{(-)} \stackrel{03}{(-i)} *_{A}
\stackrel{03}{(+i)}\stackrel{12}{(+)}\cdots \stackrel{d-1 \, d}{(+)} =1\,.
\label{Cliffnormalization}
\end{eqnarray}
%
This is,  due to Eq.~(\ref{graficcliff}), equal to
$\stackrel{03}{[-i]}\stackrel{12}{[-]}\cdots \stackrel{d-1 \, d}{[-]}$, what we
normalize to one, as we show in Eq.~(\ref{vac1}).  
                                                                                                                                                                                                                                                                                                                                                                                                                                                                                                                                                                                                                                                                                                                                                                                                                                                                                                                                                                

\vspace{2mm}
 

Usually  in the literature, the operators $\gamma^a$'s are represented as matrices. 
We use here $\gamma^a$'s as Clifford objects, which form the ''basis vectors''. 
One can calculate,  as seen in Ref.~\cite{DMN} and in App.~\ref{matrixCliffordDMN}, 
the matrix representations of $\gamma^a$'s when apply on "basis vectors" and 
the generators of the Lorentz transformations $S^{ab}$ when apply on
 "basis vectors" 
 defined in $d$-dimensional space. In App.~\ref{matrixCliffordDMN} the
matrix representations  of $\gamma^a$'s and $S^{ab}$'s in 
$d=(3+1)$-dimensional space-time are presented. 

%
%

In the Grassmann case the $ 2^{d-1}$ odd and $ 2^{d-1}-1$ (we skip the selfadjoint
identity defining the vacuum state, Eq.~(\ref{vactheta})) even Grassmann 
''basis vectors'' (operators), which are superposition of either odd or even products 
of $\theta^a$'s, respectively, are well distinguishable from their $ 2^{d-1}$ odd and 
$ 2^{d-1}-1$ even Hermitian conjugated partners, which are superposition of  
odd and even  products of $\frac{\partial}{\partial \theta_{a}}$'s. 

In the Clifford case the relation 
between "basis vectors" and their Hermitian conjugated partners  (made 
of products of nilpotents ($\stackrel{ab}{(k)}$ or  $\stackrel{ab}{\tilde{(k)}}$) and 
projectors  ($\stackrel{ab}{[k]}$ or  $\stackrel{ab}{\tilde{[k]}}$), Eq.~(\ref{graficcliff}),
 are less transparent (although still easy to be evaluated).
%
%
 This can be noticed in Eq.~(\ref{graficcliff}), since for nilpotents we notice that
$\frac{1}{\sqrt{2}} (\gamma^a + \frac{\eta^{aa}}{i\,k} \gamma^b)^{\dagger}$  
 $=\eta^{aa}\,\frac{1}{\sqrt{2}} (\gamma^a + \frac{\eta^{aa}}{i \,(- k)} 
 \gamma^b) $, while for projectors we find that they are self adjoint objects,
$ (\frac{1}{\sqrt{2}} (1 +  \frac{i}{k}  \gamma^a \gamma^b))^{\dagger}=$
$ \frac{1}{\sqrt{2}} (1 +  \frac{i}{k}  \gamma^a \gamma^b)$. 
This is the case also for representations in the sector of $\tilde{\gamma}^a$'s.

One easily sees that in even dimensional spaces, either in $d=2(2n+1)$ or in $d=4n$, 
the Clifford odd "basis vectors" (they are products of an odd number of nilpotents and 
an even number of projectors) have their Hermitian conjugated partners in another
irreducible representation, since Hermitian conjugation changes an odd number of 
nilpotents (changing at the same time the handedness of the "basis vectors"), while the generators of the Lorentz transformations change two nilpotents at 
the same time (keeping the handedness unchanged).

The Clifford even "basis vectors" have an even number of nilpotents and can have 
an odd or an even number of projectors. Correspondingly an irreducible representation 
of even "basis vectors" have among "basis vectors" the product of projectors only, 
which  is  therefore  selfadjoint.

%
%

Taking into account Eq.~(\ref{graficcliff}) one finds that the product of 
 Hermitian conjugated partner of a ''basis vector''  from Eq.~(\ref{allcartaneigenvec})
 and the corresponding ''basis vector'' is nonzero and normalized to identity, as
 presented in Eq.~(\ref{Cliffnormalization})
($\stackrel{03}{(-i)}\stackrel{12}{(-)}\cdots \stackrel{d-1 \, d}{(-)} *_{A}
\stackrel{03}{(+i)}\stackrel{12}{(+)}\cdots \stackrel{d-1 \, d}{(+)}$). 
This is the case for any irreducible 
representation in the case of the Clifford algebra of $\gamma^a$'s as well in
the Clifford algebra of $\tilde{\gamma}^a$'s. \\

{\bf Statement 3.} 
The product of any Hermitian conjugated ''basis vector'' with the corresponding 
''basic vector'' is the same for all the members of an irreducible representation.\\

The proof is presented in App.~\ref{proofs}, Statement 4.\\

{\bf Statement 4.}
Changing a pair of any two nilpotents $\stackrel{ab}{(k)}\stackrel{cd}{(k')}$ to a pair 
of projectors with the same ($k,k'$) $\stackrel{ab}{[k]}\stackrel{cd}{[k']} $, or of
 any two projectors $\stackrel{ab}{[k']}\stackrel{cd}{[k'']}$ two a pair of two 
nilpotents with the 
same ($k',k''$), or  a pair of any nilpotent (projector) and projector (nilpotent)
$\stackrel{ab}{(k)}\stackrel{cd}{[k']}$ to a pair of the projector (nilpotent) and the 
nilpotent (projektor) with the same ($k,k'$) $\stackrel{ab}{[k]}\stackrel{cd}{(k')}$, 
 the obtained ''basis vector'' belongs to different irreducible representations.,
 since the Lorentz transformations  $S^{ce}$  transform 
 $\stackrel{ab}{(k)}\stackrel{cd}{(k')}$ into $\stackrel{ab}{[-k]}\stackrel{cd}{[-k']}$,
 never to $\stackrel{ab}{[k]}\stackrel{cd}{[k']} $, and equivalently for other  pairs.
 
 This statement is proven in App.~\ref{proofs}, Statement 2b. Due to 
Eq.~(\ref{eigencliffcartan}) the irreducible representations are equivalent with respect
 to the eigenvalues of the Cartan subalgebra members and also with respect to the
 generators of the Lorentz transformations $S^{ab}$.  

Statement is valid also if we  replace $\gamma^a$ with $\tilde{\gamma}^a$.

It can be checked, however, that there are Hermitian conjugated partners of the 
''basis vectors'' of different irreducible representations, which applying on 
$\stackrel{03}{(+i)}\stackrel{12}{(+)}\cdots \stackrel{d-1 \, d}{(+)}$ from
the left hand side, give also nonzero contributions, not the 
identity. Like it is 
$\stackrel{03}{[+i]}\stackrel{12}{[+]}\cdots \stackrel{d-1 \, d}{(-)}*_{A}$
$\stackrel{03}{(+i)}\stackrel{12}{(+)}\cdots \stackrel{d-1 \, d}{(+)}$, giving (due to
Eq.~(\ref{graficcliff}))  
$\stackrel{03}{(+i)}\stackrel{12}{(+)}\cdots \stackrel{d-1 \, d}{[-]}$.
The ''basis vector'' of $\stackrel{03}{[+i]}\stackrel{12}{[+]}\cdots \stackrel{d-1 \, d}{(-)}$
is $\stackrel{03}{[+i]}\stackrel{12}{[+]}\cdots \stackrel{d-1 \, d}{(+)}$, belonging 
to another irreducible representation.


There are several other choices, like
\begin{eqnarray}
\label{notunique0} 
\stackrel{03}{[+i]}\stackrel{12}{(-)}\stackrel{56}{[+]}\cdots \stackrel{d-1 \, d}{(-)} *_{A} 
\stackrel{03}{(+i)}\stackrel{12}{(+)}\cdots \stackrel{d-1 \, d}{(+)}\,,\nonumber\\
\stackrel{03}{[+i]}\stackrel{12}{[+]} \stackrel{56}{[+]}\cdots 
\stackrel{d-1 \, d}{(-)} *_{A} 
 \stackrel{03}{(+i)}\stackrel{12}{(+)}\cdots \stackrel{d-1 \, d}{(+)}\,,
\end{eqnarray}
which also give nonzero contributions duo to Eq.~(\ref{graficcliff}).

Let us recognize:

i. The two Clifford spaces, the one spanned by $\gamma^{a}$'s and the second one
spanned by $\tilde{\gamma}^{a}$'s, are independent vector spaces, each with 
$2^{d}$ "basis vectors" (App.~\ref{proofs}, Statement 1.). 

ii. The Clifford odd "basis vectors" (the superposition of products of 
odd numbers of $\gamma^a$'s or of $\tilde{\gamma}^{a}$'s, respectively)
can be arranged for each kind of the Clifford algebras into two groups, Hermitian 
conjugated to each other, of 
$2^{\frac{d}{2} - 1}$ members of $2^{\frac{d}{2} - 1} $ irreducible 
representations of the corresponding Lorentz group. 

iii. Different irreducible representations are indistinguishable with respect to the 
"eigenvalues" of the corresponding  Cartan subalgebra members, Statement 4. 

iv. The Clifford even part (made of superposition of products of even numbers of  
$\gamma^a$'s or  $\tilde{\gamma}^{a}$'s, respectively) splits as well into 
twice $2^{\frac{d}{2} - 1} \cdot 2^{\frac{d}{2} - 1} $ irreducible representations 
of the Lorentz group. One member of each Clifford  even representation, the one
which is the product of projectors only, is self adjoint. Members of one irreducible 
representation are with respect to the Cartan subalgebra indistinguishable from
all the other irreducible representations for each of the two algebras.

%
v. The  odd $2^{\frac{d}{2}-1}$ members of each of the $ 2^{\frac{d}{2}-1}$
irreducible representations are among themselves orthogonal and so are 
orthogonal $2^{\frac{d}{2}-1} \cdot$ $2^{\frac{d}{2}-1}$
members of their Hermitian conjugated part. For illustration of the 
orthogonality one can look at Table~\ref{cliff basis5+1.},  and recognizes 
that any ''basis vector'' of the first four multiplets ({\it odd I, II, III, IV}),  the
third column, gives zero, if it is multiplied from the left hand side or from the 
right hand side with any other ''basis vector'' from the third column,
according to Eq.~(\ref{graficcliff}). 
The same is true for any ''basis vector'' from the fourth column, Hermitian 
conjugated to the third column. 

Generalization to any even dimension $d$ is straightforward.

vi.  Let us denote the Clifford odd "basis vectors" of the Clifford algebra $\gamma^a$ 
 by $\hat{b}^{m \dagger}_{f}$, where $f$ defines different irreducible 
 representations and $m$ a member in the representation $f$. Then
 their Hermitian conjugate partners is denoted by  $\hat{b}^{m}_{f}$ 
$=(\hat{b}^{m \dagger}_{f})^{\dagger}$, let us start for $d=2(2n+1)$.\\

Let us repeat {\it Statement 2.}:
The algebraic product of $\hat{b}^{m}_{f}{}_{*_A}
\hat{b}^{m \dagger}_{f}$ is the same for all the members of one
irreducible representation. \\


vii. Let us look for the vacuum state $|\psi_{oc}>$ as the sum of the 
products $\hat{b}^{m}_{f}{}_{*_A}
\hat{b}^{m \dagger}_{f}$ over  all the irreducible representations
\begin{eqnarray}
\label{vaccliff}
|\psi_{oc}>= \sum_{f=1}^{2^{\frac{d}{2}-1}}\,\hat{b}^{m}_{f}{}_{*_A}
\hat{b}^{m \dagger}_{f} \,|\,1\,>\,,
\end{eqnarray}
for one of the members $m$, anyone, of the odd irreducible representation $f$,
with $|\,1\,>$, which is the vacuum without any structure
($\hat{b}^{m}_{f}{} |\,1\,>=0$).

If we make a choice of the starting member of the starting irreducible representation
as
\begin{eqnarray}
\label{bmfdagerbmf}
\hat{b}^{m=1 \dagger}_{f=1} \, {\bf :} &=&\stackrel{03}{(+i)}\stackrel{12}{(+)}
\cdots \stackrel{d-1 \, d}{(+)}\,,\nonumber\\
(\hat{b}^{m=1 \dagger}_{f=1})^{\dagger}=\hat{b}^{m=1 }_{f=1} \, {\bf :} &=& \, 
\stackrel{d-1 \, d}{(-)} \cdots \stackrel{12}{(-)}\stackrel{03}{(-i)}\,, 
\end{eqnarray}
%
we recognize for the "basis vectors" of an 
odd Clifford character for each of the two Clifford algebras the properties
\begin{eqnarray}
\label{almostDirac}
\hat{b}^{m}_{f} {}_{*_{A}}|\psi_{oc}>&=& 0\, |\psi_{oc}>\,,\nonumber\\
\hat{b}^{m \dagger}_{f}{}_{*_{A}}|\psi_{oc}>&=&  |\psi^m_{f}>\,,\nonumber\\
\{\hat{b}^{m}_{f}, \hat{b}^{m'}_{f'}\}_{*_{A}+}|\psi_{oc}>&=&
 0\,|\psi_{oc}>\,, \nonumber\\
\{\hat{b}^{m \dagger}_{f}, \hat{b}^{m \dagger}_{f}\}_{*_{A}+}|\psi_{oc}>
&=&|\psi_{oc}>\,,
\end{eqnarray}
where  ${*_{A}}$ represents the algebraic multiplication of 
$\hat{b}^{m \dagger}_{f}$  and $ \hat{b}^{m'}_{f'} $  among themselves and  
with the vacuum state  $|\psi_{oc}>$ of Eq.(\ref{vaccliff}), which takes into account 
Eq.~(\ref{gammatildeantiher0}).

All the products of Clifford algebra elements are up to now the algebraic ones and so 
are also the products in Eq.~(\ref{almostDirac}). Since we use  here 
anticommutation relations, we want to point out with ${}_{*_{A}}$ this algebraic 
character
of the products, to be later distinguished from the tensor product ${}_{*_{T}}$,
when the creation and annihilattion operators  are defined on an extended basis, 
which is the tensor product of the superposition of the ''basis vectors'' of the Clifford 
space and of the momentum basis, applying on the Hilbert space of 
''Slater determinants''. 
The tensor product $*_{T_{H}}$ will be  used  as the product 
mapping a pair of the fermion wave functions into two fermion wave functions 
and further to many fermion wave functions --- that is to the extended algebra 
of many fermion system.

Obviously, $\hat{b}^{m \dagger}_{f}$ and $\hat{b}^{m}_{f}$ have on the 
level of the algebraic products, when applying on the vacuum state $|\psi_{oc}>$,
{\it almost} the properties of creation and annihilation operators of the second 
quantized fermions in the postulates of Dirac,  as it is discussed in the next items. 
We illustrate properties of "basis vectors"  and their Hermitian conjugated 
partners on the example of $d=(5+1)$-dimensional space in 
Sect.~\ref{reduction}.\\
 ${\quad}$ viii.  There is, namely, the  property, which the  second quantized 
fermions should 
fulfill in addition to the relations of Eq.~(\ref{almostDirac}). The anticommutation 
relations of creation and annihilation operators should include:
\begin{eqnarray} 
\label{should}
\{\hat{b}^{m}_{f}, \hat{b}^{m'\dagger}_{f'}\}_{*_{A}+}|\psi_{oc}>&=&
\delta^{m m'} \delta_{f f'} |\psi_{oc}>\,.
\end{eqnarray}
For  any $\hat{b}^{m}_{f}$ and any $\hat{b}^{m'\dagger}_{f'}$ this is not 
the case, as we demonstrated in Eq.~(\ref{notunique0})  
(besides $\hat{b}^{m=1}_{f=1} = \, 
\stackrel{d-1 \, d}{(-)} \cdots  \stackrel{56}{(-)} \stackrel{12}{(-)}\stackrel{03}{(-i)}$, for example, also 
%
$\hat{b}^{m'}_{f'} = \, \stackrel{d-1 \, d}{(-)} \cdots  \stackrel{56}{(-)}
 \stackrel{12}{[+]}\stackrel{03}{[+i]} $ 
%
and several others give, when applied on $\hat{b}^{m=1\dagger}_{f=1}$,
nonzero contributions).
 There are namely $2^{\frac{d}{2}-1}-1$ too many annihilation
operators for each creation operator, which give, applied on the creation operator, 
nonzero contribution. \\
${\quad}$ viii. a. To use the Clifford algebra objects to describe second quantized fermions,
representing
the observed quarks and leptons as well as the antiquarks and antileptons%
~\cite{IARD2016,n2014matterantimatter,nd2017,%
n2012scalars,JMP2013,normaJMP2015,nh2017,nh2018},  {\it the families should exist}. \\
${\quad}$ viii. b. 
The operators should exist, which connect one irreducible 
representation of fermions with all the other irreducible representations.
\\
${\quad}$ viii. c. 
Two independent choices for describing the internal degrees 
of freedom of the observed quarks and leptons are not in agreement with the observed 
properties of fermions. 

We solve these problems, cited in viii. a., viii. b., viii. c., 
by reducing  the degrees 
of freedom offered by  the two kinds of the Clifford algebras, $\gamma^a$'s
and  $\tilde{\gamma}^a$'s, making a choice of one --- $\gamma^a$'s --- to describe 
the internal space of fermions, and using the other one --- 
$\tilde{\gamma}^a$'s --- to describe the "family" quantum number of each irreducible
representation of $S^{ab}$'s in  space  defined by $\gamma^a$'s.

\subsubsection{Reduction of Clifford and Grassmann space and 
appearance of family quantum number}
 \label{reduction}

The creation and annihilation operators of an odd Clifford  algebra of both kinds, of 
either $\gamma^a$'s or  $\tilde{\gamma}^{a}$'s,   would obviously 
obey the anticommutation relations for the second quantized fermions, 
postulated by Dirac, at least on the vacuum state, which is a sum of all the products 
of annihilation times, $*_{A}$, the corresponding creation operators, provided that 
each of the irreducible representations would carry a different quantum number.  

But we know that a particular member $m$ has for all the irreducible representations 
the same quantum numbers, that is the same "eigenvalues" of the 
Cartan subalgebra (for the vector space of either $\gamma^a$'s or  
$\tilde{\gamma}^{a}$'s), Eq.~(\ref{graficcliff}).  

{\it The only possibility to "dress" each irreducible representation of one kind of the
two independent vector spaces with a new, let us say   "family"  quantum number, 
is that we "sacrifice" one of the two vector spaces, let us make a choice of}  
$\tilde{\gamma}^{a}$'s,  {\it and use $\tilde{\gamma}^{a}$'s 
to define the "family" quantum number for each irreducible representation of the 
vector space of} $\gamma^a$'s, while  {\it  keeping the relations of} 
Eq.~(\ref{gammatildeantiher0}) {\it  unchanged:}  $\{\gamma^{a}, 
\gamma^{b}\}_{+}=2 \eta^{a b}= \{\tilde{\gamma}^{a}, 
\tilde{\gamma}^{b}\}_{+}$, $\{\gamma^{a}, \tilde{\gamma}^{b}\}_{+}=0$,
  $ (\gamma^{a})^{\dagger} = \eta^{aa}\, \gamma^{a}$, 
$(\tilde{\gamma}^{a})^{\dagger} =  \eta^{a a}\, \tilde{\gamma}^{a}$, 
$(a,b)=(0,1,2,3,5,\cdots,d)$.\\

We therefore {\it postulate}:\\
  Let  $\tilde{\gamma}^{a}$'s operate on $\gamma^a$'s as follows~%
\cite{nh03,norma93,JMP2013,normaJMP2015,nh2018}
\begin{eqnarray}
\{\tilde{\gamma}^a B &=&(-)^B\, i \, B \gamma^a\}\, |\psi_{oc}>\,,
\label{tildegammareduced}
\end{eqnarray}
with $(-)^B = -1$, if $B$ is (a function of) an odd product of $\gamma^a$'s,
 otherwise $(-)^B = 1$~\cite{nh03}, $|\psi_{oc}>$ is defined in 
Eq.~(\ref{vaccliff}).\\ 

After this postulate the vector  space of $\tilde{\gamma}^{a}$'s is 
 "frozen out". No vector space of $\tilde{\gamma}^{a}$'s 
needs to be taken into account  any longer, in agreement with the observed properties
of fermions. This solves the problems viii.a -  viii. c. of Sect.~\ref{propertiesCliff0}.\\


Taking into account Eq.~(\ref{tildegammareduced})  we can check that
(App.~\ref{proofs}, Statement 3, 3a, 3b):

\vspace{2mm}
\noindent 
{\bf a.} Relations of Eq.~(\ref{gammatildeantiher0}) remain unchanged~\footnote{
Let us show that the relation $ \{\tilde{\gamma}^{a}, 
\tilde{\gamma}^{b}\}_{+}= 2 \eta^{ab}$ remains valid when applied on $B$, if $B$ 
is either an odd or an even product of $\gamma^a$'s:
$ \{\tilde{\gamma}^{a}, \tilde{\gamma}^{b}\}_{+}$ $ \gamma^c=$
$-i \,(\tilde{\gamma}^{a} \gamma^c \gamma^b + \tilde{\gamma}^{b} \gamma^c \gamma^a)=$ $-i\,i\,\gamma^c ( \gamma^b \gamma^a +  \gamma^a \gamma^b)
 = 2 \eta^{ab} \gamma^c$, while $ \{\tilde{\gamma}^{a}, \tilde{\gamma}^{b}\}_{+}$ 
 $ \gamma^c \gamma^d=$ $ i \,(\tilde{\gamma}^{a} \gamma^c \gamma^d  \gamma^b + \tilde{\gamma}^{b} \gamma^c \gamma^d \gamma^a)=$ $i(-i)\,\gamma^c \gamma^d (\gamma^b \gamma^a +  \gamma^a \gamma^b)
 = 2 \eta^{ab} \gamma^c \gamma^d$. The relation is valid for any $\gamma^c$ and 
 $\gamma^d$, even if  $c=d$.}.
 
 \vspace{2mm}
\noindent 
{\bf b.} Relations of Eq.~(\ref{sabtildesab}) 
remain unchanged~\footnote{One easily checks
that $\tilde{\gamma}^{a \dagger} \gamma^c= -i \gamma^c \gamma^{a \dagger}=
-i \eta^{aa}\gamma^c \gamma^a$ $= \eta^{aa} \tilde{\gamma}^a \gamma^c =$ 
$-i \eta^{aa}\gamma^c \gamma^a$.}.

\vspace{2mm}
\noindent 
{\bf c.} The eigenvalues of the operators $S^{ab}$  and $\tilde{S}^{ab}$ 
on nilpotents and projectors of $\gamma^a$'s are after the reduction of Clifford space
equal to
\begin{eqnarray}
\label{signature0}
S^{ab} \,\stackrel{ab}{(k)} = \frac{k}{2}  \,\stackrel{ab}{(k)}\,,\quad && \quad
\tilde{S}^{ab}\,\stackrel{ab}{(k)} = \frac{k}{2}  \,\stackrel{ab}{(k)}\,,\nonumber\\
S^{ab}\,\stackrel{ab}{[k]} =  \frac{k}{2}  \,\stackrel{ab}{[k]}\,,\quad && \quad 
\tilde{S}^{ab} \,\stackrel{ab}{[k]} = - \frac{k}{2}  \,\,\stackrel{ab}{[k]}\,,
\end{eqnarray}
demonstrating that the eigenvalues of $S^{ab}$ on nilpotents and projectors of 
$\gamma^a$'s differ from the eigenvalues of $\tilde{S}^{ab}$, so that 
$\tilde{S}^{ab}$ can be used to denote irreducible representations of $S^{ab}$
 with the ''family'' quantum number, what solves the problems viii. a. and viii. b. 
of Subsubsect.~\ref{propertiesCliff0}.

\vspace{2mm}
\noindent 
{\bf d.} We further recognize 
that $\gamma^a$ transform  $\stackrel{ab}{(k)}$ into  $\stackrel{ab}{[-k]}$, never 
to $\stackrel{ab}{[k]}$, while $\tilde{\gamma}^a$ transform  $\stackrel{ab}{(k)}$ 
into $\stackrel{ab}{[k]}$, never to $\stackrel{ab}{[-k]}$ 
\begin{eqnarray}
&&\gamma^a \stackrel{ab}{(k)}= \eta^{aa}\stackrel{ab}{[-k]},\; \quad
\gamma^b \stackrel{ab}{(k)}= -ik \stackrel{ab}{[-k]}, \; \quad 
\gamma^a \stackrel{ab}{[k]}= \stackrel{ab}{(-k)},\;\quad \;\;
\gamma^b \stackrel{ab}{[k]}= -ik \eta^{aa} \stackrel{ab}{(-k)}\,,\nonumber\\
&&\tilde{\gamma^a} \stackrel{ab}{(k)} = - i\eta^{aa}\stackrel{ab}{[k]},\quad
\tilde{\gamma^b} \stackrel{ab}{(k)} =  - k \stackrel{ab}{[k]}, \;\qquad  \,
\tilde{\gamma^a} \stackrel{ab}{[k]} =  \;\;i\stackrel{ab}{(k)},\; \quad
\tilde{\gamma^b} \stackrel{ab}{[k]} =  -k \eta^{aa} \stackrel{ab}{(k)}\,. 
\label{snmb:gammatildegamma}
\end{eqnarray}

\vspace{2mm}
\noindent 
{\bf e.}
One finds, using  Eq.~(\ref{tildegammareduced}),
\begin{eqnarray}
\stackrel{ab}{\tilde{(k)}} \, \stackrel{ab}{(k)}& =& 0\,, 
\qquad 
\stackrel{ab}{\tilde{(-k)}} \, \stackrel{ab}{(k)} = -i \,\eta^{aa}\,  
\stackrel{ab}{[k]}\,,\quad  
\stackrel{ab}{\tilde{(k)}} \, \stackrel{ab}{[k]} = i\, \stackrel{ab}{(k)}\,,\quad
\stackrel{ab}{\tilde{(k)}}\, \stackrel{ab}{[-k]} = 0\,, \nonumber\\
%
%
\stackrel{ab}{\tilde{[k]}} \, \stackrel{ab}{(k)}& =& \, \stackrel{ab}{(k)}\,, 
\quad 
\stackrel{ab}{\tilde{[-k]}} \, \stackrel{ab}{(k)} = \, 0 \,,  \qquad  \quad 
\quad \;\;\;\,
\stackrel{ab}{\tilde{[k]}} \, \stackrel{ab}{[k]} =  0\,,\qquad \;\;\;
\stackrel{ab}{\tilde{[- k]}} \, \stackrel{ab}{[k]} =  \, \stackrel{ab}{[k]}\,.
\label{graphbinomsfamilies}
\end{eqnarray}

\vspace{2mm}
\noindent 
{\bf f.}
From Eq.~(\ref{snmb:gammatildegamma}) it follows
\begin{eqnarray}
\label{stildestrans}
S^{ac}\stackrel{ab}{(k)}\stackrel{cd}{(k)} &=& -\frac{i}{2} \eta^{aa} \eta^{cc} 
\stackrel{ab}{[-k]}\stackrel{cd}{[-k]}\,,\qquad 
S^{ac}\stackrel{ab}{[k]}\stackrel{cd}{[k]} = \frac{i}{2}  
\stackrel{ab}{(-k)}\stackrel{cd}{(-k)}\,,\,\nonumber\\
\tilde{S}^{ac}\stackrel{ab}{(k)}\stackrel{cd}{(k)} &=& \frac{i}{2} \eta^{aa} \eta^{cc} 
\stackrel{ab}{[k]}\stackrel{cd}{[k]}\,,\qquad \qquad \;\;
\tilde{S}^{ac}\stackrel{ab}{[k]}\stackrel{cd}{[k]} = -\frac{i}{2}  
\stackrel{ab}{(k)}\stackrel{cd}{(k)}\,,\,\nonumber\\
S^{ac}\stackrel{ab}{(k)}\stackrel{cd}{[k]}  &=& -\frac{i}{2} \eta^{aa}  
\stackrel{ab}{[-k]}\stackrel{cd}{(-k)}\,,\,\qquad \quad
S^{ac}\stackrel{ab}{[k]}\stackrel{cd}{(k)} = \frac{i}{2} \eta^{cc}  
\stackrel{ab}{(-k)}\stackrel{cd}{[-k]}\,,\,\nonumber\\
\tilde{S}^{ac}\stackrel{ab}{(k)}\stackrel{cd}{[k]} &=& -\frac{i}{2} \eta^{aa}  
\stackrel{ab}{[k]}\stackrel{cd}{(k)}\,,\qquad \qquad\;\;\,
\tilde{S}^{ac}\stackrel{ab}{[k]}\stackrel{cd}{(k)} = \frac{i}{2} \eta^{cc}  
\stackrel{ab}{(k)}\stackrel{cd}{[k]}\,. 
\end{eqnarray}

\vspace{2mm}
\noindent 
{\bf g.}
Each irreducible representation 
 has now the "family" quantum
number, determined by $\tilde{S}^{ab}$ of the Cartan subalgebra of 
Eq.~(\ref{cartangrasscliff}).
Correspondingly the creation and annihilation operators fulfill algebraically the 
anticommutation relations, postulated by Dirac for the second quantized fermions,
since: {\bf g.a} Different irreducible representations carry different 
"family" quantum numbers and to each "family" quantum member only one 
Hermitian conjugated partner with the same "family" quantum number belongs. 
{\bf g.b} Each summand of the vacuum state,  Eq.~(\ref{vaccliff}), belongs to
 a particular  "family". 
This solves the problem viii. a. -  viii. c. of Sect.~\ref{propertiesCliff0}. 

The anticommutation relations of Dirac, postulated for fermions, are 
fulfilled on the vacuum state,  Eq.~(\ref{vaccliff}), on the algebraic level, without 
postulating them.
They follow by themselves due to  the fact that the creation and 
annihilation operators are superposition of odd products of $\gamma^{a}$'s. \\

{\bf Statement 5. } 
The oddness of superpositions of products of  $\gamma^a$'s, representing the
"basis vectors", guarantees the anticommuting properties of objects which are 
tensor, $*_{T}$, products of these "basis vectors" and the (commuting) basis of 
the ordinary space.\\

We show in Sect.~\ref{actionGrassCliff} that solutions of equations of motion 
must be  the tensor, $*_{T}$, product of the finite number of the 
"basis vectors", $2^{\frac{d}{2}-1}\times 2^{\frac{d}{2}-1}$
 (determining properties of the internal space of fermions), and 
the (continuously)  infinite number of basis vectors in ordinary space. 
The corresponding creation and annihilation operators fulfill the
Dirac's postulates for the second quantized fermions. We shall see
that for positive energy $p^0=|\vec{p}|$ the solutions of equations of motion
for massless fermions of particular handedness and charges  
 the solutions for their antifermions of opposite charges and handedness 
with $-\vec{p}$ correspond, reducing the (continously infinite) number of 
creation operators for the factor of $2$.
 
In Sect.~\ref{HilbertCliff0} we show that the anticommuting property of
creation and annihilation operators  manifest also on 
 the Hilbert space, formed as the tensor products, $*_{T_{H}}$, of all
possible numbers of all possible fermion and antifermion states solving the 
equations of motion, the creation operators of which are  
the tensor products, $*_{T}$, of the 
finite number, $2^{\frac{d}{2}-1}\times 2^{\frac{d}{2}-1}$, of Clifford odd 
''basis vectors'' and of the continuous infinite basis of the ordinary (momentum or 
coordinate) space, 
applying on the vacuum state. 
%

Let us write down the anticommutation relations of Clifford odd "basic vectors",
representing the creation operators in internal space of fermions with half integer
spin represented by $\gamma^a$'s and of the corresponding Hermitian
conjugated partners annihilation operators again. After the reduction of the 
Clifford algebra any irreducible representation carry the family quantum number, 
distinguishing families among themselves.
\begin{eqnarray}
\{ \hat{b}^{m}_{f}, \hat{b}^{m' \dagger}_{f'} \}_{*_{A}+}\, |\psi_{oc}> 
&=& \delta^{m m'} \, \delta_{ff'} \,  |\psi_{oc}>\,,\nonumber\\
\{ \hat{b}^{m}_{f}, \hat{b}^{m'}_{f'} \}_{*_{A}+}  \,  |\psi_{oc}>
&=& 0 \,\cdot\,  |\psi_{oc}>\,,\nonumber\\
\{\hat{b}^{m  \dagger}_{f},\hat{b}^{m' \dagger}_{f'}\}_{*_{A}+} \, |\psi_{oc}>
&=& 0 \, \cdot\, |\psi_{oc}>\,,\nonumber\\
 \hat{b}^{m \dagger}_{f} \,{}_{*_{A}} |\psi_{oc}>&=& |\psi^{m}_{f}>\,, \nonumber\\
 \hat{b}^{m}_{f}   \,{*_{A}}  |\psi_{oc}>&=& 0 \,\cdot\,  |\psi_{oc}>\,,
\label{alphagammatildeprod}
\end{eqnarray}
with ($m,m'$) denoting the "family" members and ($f,f'$) denoting "families",
${*_{A}}$ represents the algebraic multiplication of $ \hat{b}^{\dagger m}_{f} $
and  $ \hat{b}^{m}_{f} $ with the vacuum state $|\psi_{oc}>$ of 
Eq.~(\ref{vaccliff}) and among themselves, taking into account 
Eq.~(\ref{gammatildeantiher0}).

\vspace{2mm}
\noindent 
{\bf h.}
Let us make a choice of  the starting creation operator $\hat{b}^{m=1 \dagger}_{f=1}$ 
of an odd Clifford character and correspondingly of its Hermitian conjugated partner in $d=2(2n+1)$ and $d=4n$, respectively, as we start in Eq.~(\ref{bmfdagerbmf}), 
but now, after reduction of the two kinds of the Clifford algebras into only one, each irreducible representation carries the family quantum number.
%
\begin{eqnarray}
{\hat b}^{1 \dagger}_{1}\, {\bf :} &=& \stackrel{03}{(+i)} \stackrel{12}{(+)} 
\stackrel{56}{(+)}\cdots 
\stackrel{d-3\;d-2}{(+)}\;\;\stackrel{d-1\;d}{(+)}\,,\nonumber\\
({\hat b}^{1 \dagger}_{1})^{\dagger} &=& {\hat b}^{1}_{1}\, {\bf :}= 
\stackrel{d-1\;d}{(-)} \;\;
\stackrel{d-3\;d-2}{(-)}\cdots  \stackrel{56}{(-)}  \stackrel{12}{(-)}  \stackrel{01}{(-i)}\,,
\nonumber\\
d&=&2(2n+1)\,,
 \nonumber\\
{\hat b}^{1 \dagger}_{1} \, {\bf :} &=& \stackrel{03}{(+i)} \stackrel{12}{(+)} 
\stackrel{56}{(+)}\cdots 
\stackrel{d-3\;d-2}{(+)}\;\;\stackrel{d-1\;d}{[+]}\,,\nonumber\\
 ({\hat b}^{1 \dagger}_{1})^{\dagger}&=& {\hat b}^{1}_{1}\, {\bf :} = 
\stackrel{d-1\;d}{[+]} \;\;
\stackrel{d-3\;d-2}{(-)}\cdots  \stackrel{56}{(-)}  \stackrel{12}{(-)}  \stackrel{01}{(-i)}\,,
\nonumber\\
d&=&4n\,.
\label{start(2n+1)2cliffgammatilde4n}
\end{eqnarray}
 All the rest  creation operators in the internal space  described by $\gamma^a$'s
 (the "basis vectors"), belonging to the same Lorentz representation, follow by
the application of the Lorentz generators $S^{ab}$'s. 

\vspace{2mm}
\noindent 
{\bf i.} The representations with different "family" quantum numbers are reachable 
by $\tilde{S}^{ab}$, since, according to Eq.~(\ref{stildestrans}), we recognize 
that $\tilde{S}^{ac}$ transforms two nilpotents 
$ \stackrel{ab}{(k)} \stackrel{cd}{(k)}$ into two projectors 
$ \stackrel{ab}{[k]} \stackrel{cd}{[k]}$, without changing $k$ 
($\tilde{S}^{ac}$ transforms $ \stackrel{ab}{[k]} \stackrel{cd}{[k]}$ into 
$ \stackrel{ab}{(k)} \stackrel{cd}{(k)}$, as well as 
$ \stackrel{ab}{[k]} \stackrel{cd}{(k)}$ into $ \stackrel{ab}{(k)} \stackrel{cd}{[k]}$). 
All the "family" members are reachable from one member of a new family by the 
application of $S^{ab}$'s.

In this way, by starting with the creation operator ${\hat b}^{1 \dagger}_{1}$, 
Eq.~(\ref{start(2n+1)2cliffgammatilde4n}), $2^{\frac{d}{2}-1}$ "families", each with 
$2^{\frac{d}{2}-1}$ "family" members follow. 

Let us find the starting member of the next "family" to the "family" of 
Eq.~(\ref{start(2n+1)2cliffgammatilde4n}) by the application of $\tilde{S}^{01}$
\begin{eqnarray}
{\hat b}^{1 \dagger}_{2}\, {\bf :} &=& \stackrel{03}{[+i]} \stackrel{12}{[+]} 
\stackrel{56}{(+)}\cdots \stackrel{d-3\;d-2}{(+)}\;\;\stackrel{d-1\;d}{(+)}\,,\nonumber\\
{\hat b}^{1}_{2}\,{\bf :} &=& 
\stackrel{d-1\;d}{(-)} \;\;
\stackrel{d-3\;d-2}{(-)}\cdots  \stackrel{56}{(-)}  \stackrel{12}{[+]}  \stackrel{01}{[+i]}\,.
\label{d=2(2n+1)}
\end{eqnarray}


The corresponding annihilation operators, that is the Hermitian conjugated partners of 
$2^{\frac{d}{2}-1}$ "families", each with $2^{\frac{d}{2}-1}$ "family" members, 
following from the starting creation operator ${\hat b}^{1 \dagger}_{1}$ by the application of $S^{ab}$'s --- the family members --- and the application of 
 $\tilde{S}^{ab}$ --- the same family member of another family --- 
can be obtained by Hermitian conjugation.

{\it The creation and annihilation operators of an odd Clifford character, expressed by 
nilpotents and projectors of  $\gamma^a$'s, obey  anticommutation relations of}$\,$ 
Eq.~(\ref{alphagammatildeprod}),
{\it without postulating the second quantized anticommutation relations}. 
%

\vspace{2mm}
\noindent 
{\bf j.} The vacuum state  $|\psi_{oc}>$ for the vector space determined by 
$\gamma^a$'s, 
which is the sum over family quantum numbers of
products of an  annihilation operator with its Hermitian conjugated partner 
creation operator of any family member $m$, 
$\sum_{f=1}^{2^{\frac{d}{2}-1}}\,\hat{b}^{m}_{f}{}_{*_A}
\hat{b}^{m \dagger}_{f}$.
\begin{eqnarray}
|\psi_{oc}>&=& \frac{1}{\sqrt{2^{\frac{d}{2}-1}}} (\stackrel{03}{[-i]} \stackrel{12}{[-]} \stackrel{56}{[-]}\cdots
\stackrel{d-1\;d}{[-]} + \stackrel{03}{[+i]} \stackrel{12}{[+]} \stackrel{56}{[-]} \cdots
               \stackrel{d-1\;d}{[-]} \nonumber\\
  + &&\stackrel{03}{[+i]} \stackrel{12}{[-]} \stackrel{56}{[+]}\cdots
\stackrel{d-1\;d}{[-]} + \cdots) |1>\,, \quad \nonumber\\
&&{\rm for}\; d=2(2n+1)\,,
\nonumber\\
|\psi_{oc}>&=&\frac{1}{\sqrt{2^{\frac{d}{2}-1}}} (\stackrel{03}{[-i]} \stackrel{12}{[-]} \stackrel{35}{[-]}\cdots
               \stackrel{d-3\;d-2}{[-]}\stackrel{d-1\;d}{[+]} \nonumber\\
  + &&\stackrel{03}{[+i]} \stackrel{12}{[+]} \stackrel{56}{[-]}
\cdots \stackrel{d-3\;d-2}{[-]} \;\,\stackrel{d-1\;d}{[+]} + \cdots )|1>\,, 
\quad  \nonumber\\
&&{\rm for}\; d=4n\,, \quad n \,{\rm is \,a\, positive\, integer}\,.
\label{vac1}
\end{eqnarray}
%


The Hermitian conjugated part of the space in the Grassmann case is "freezed out"
together with the "basis vector" space of $\tilde{\gamma}^a$'s.

The even partners of the Clifford odd creation and annihilation operators follow by either the 
application of $\gamma^a$ on the creation operators, leading to  $2^{\frac{d}{2}-1}$
 "families", each with $2^{\frac{d}{2}-1}$ members,  or with the application of 
$\tilde{\gamma}^a$ on the creation  operators,  leading to another group of the 
Clifford even operators, again with the  $2^{\frac{d}{2}-1}$ "families", each with 
$2^{\frac{d}{2}-1}$ members.

It is not difficult to recognize, that each of the Clifford even "families", obtained by  
the application of $\gamma^a$ or by $\tilde{\gamma}^a$ on the creation operators, 
contains one selfadjoint operator, which is the product  of projectors only, contributing 
as a summand to the vacuum state, Eq.~(\ref{vac1}).

%

%






\vspace{3mm}

$\;\;$ {\it  \bf {Clifford half integer spin fermions in $d=(5+1)$:}}

\vspace{3mm}

 We illustrate properties of the Clifford odd, and correspondingly anticommuting,  
creation and their Hermitian conjugated partners annihilation operators, belonging 
to $2^{\frac{6}{2}-1}=4$ "families", each with  $2^{\frac{6}{2}-1}=4$ 
members in $d=(5+1)$-dimensional space. The spin in  the fifth and the sixth 
dimension manifests as the charge in $d=(3+1)$.
%


Half of the Clifford odd "basis vectors" are 
creation operators 
${\hat b}^{m \dagger}_{f}$, appearing in Table~\ref{cliff basis5+1.}  in the
fourth column. Denoted by {\it odd I}, {\it odd II}, {\it odd III} and {\it odd IV}
they represent four "families", 
each family having four members. Their Hermitian conjugated partners 
${\hat b}^{m}_f$ are presented in the fifth column.

All the families have the same eigenvalues  of the Cartan subalgebra  
members ($S^{03}, S^{12}, S^{56}$), the handedness $\Gamma^{(3+1)}
=i (2i)^2 S^{03}S^{12}$, written in the sixth, seventh, eighth and  ninth 
column.

The rest half of the Clifford odd "basis vectors" are their Hermitian conjugated 
partners  ${\hat b}^{m}_f$, presented in the fifth column of
 Table~\ref{cliff basis5+1.}. The last three  columns  of this table determine 
 the family quantum numbers, the eigenvalues of 
 ($\tilde{S}^{03}, \tilde{S}^{12}, \tilde{S}^{56}$), the same for all family
 members.

The normalized vacuum state is the product of ${\hat b}^{m}_f \cdot$ 
${\hat b}^{m \dagger}_{f}$, 
summed over four families.
 \begin{eqnarray}
\label{vac5+1}
 |\psi_{oc}> &=&\frac{1}{\sqrt{2^{\frac{6}{2}-1}}}\,
 (\stackrel{03}{[-i]}\stackrel{12}{[-1]}\stackrel{56}{[-1]}\
+ \stackrel{03}{[+i]}\stackrel{12}{[+1]}\stackrel{56}{[-1]}
                 \nonumber\\
&+&\stackrel{03}{[+i]}\stackrel{12}{[-1]}\stackrel{56}{[+1]}
+ \stackrel{03}{[-i]}\stackrel{12}{[+1]}\stackrel{56}{[+1]}).
\end{eqnarray}
One easily checks, by taking into account Eq.~(\ref{graphbinomsfamilies}), that  
the creation operators ${\hat b}^{m \dagger}_{f}$
and the annihilation operators ${\hat b}^{m}_{f}$ fulfill the 
anticommutation relations of Eq~(\ref{alphagammatildeprod})%
~\footnote{If we would choose for creation operators those from the fifth column,
 then the annihilation operators would be the present creation operators. 
The vacuum state would be  the sum of  products of  the present creation 
operators of the fourth column  times the present annihilation operators of the 
fifth column.}.


All the Clifford even "families" with "family" members of  Table~\ref{cliff basis5+1.} 
can be obtained as algebraic products, $*_{A}$, of the Clifford odd "basis vectors" 
of Table~\ref{cliff basis5+1.}~(\cite{2020PartIPartII}, Table I). 
 

%
\begin{table}
\begin{tiny}
 \begin{center}
\begin{minipage}[t]{16.5 cm}
\caption{The basic creation operators --- 
 $\hat{b}^{m=(ch,s)\dagger}_{f}$ 
($ch$ (charge), the eigenvalue of $S^{56}$, and $s$ (spin), the eigenvalues of 
$S^{03}$ and $S^{12}$, explain the index $m$) --- and
their annihilation partners --- $\hat{b}^{m=(ch,s)}_{f}$ --- are presented for  
$d= (5+1)$-dimensional case. Each basic creation operator is a product of 
projectors and an odd product of nilpotents, and is the "eigenstate" of the 
Cartan subalgebra members, ($S^{03}$, $S^{12}$, $S^{56}$). The 
eigenvalues of  ($\tilde{S}^{03}$, $\tilde{S}^{12}$, $\tilde{S}^{56}$), 
Eq.~(\ref{cartangrasscliff}), presented in the last three columns of the table, 
determine the family quantum numbers.
Operators $\hat{b}^{m=(ch,s) \dagger}_{f}$ 
and $\hat{b}^{m=(ch, s)}_{f}$
fulfill the anticommutaion relations of Eq.~(\ref{almostDirac}).}
\label{cliff basis5+1.}
\end{minipage}
 \begin{tabular}{|r|l r|r|r|r|r|r|r|r|r|r|r|}
 \hline
${\rm family}\, f $&$m $&$=(ch,s)$&$\hat{b}^{ m=(ch,s) \dagger}_f$&
$\hat{b}^{m=(ch,s) }_{f}$&
$S^{03}$&$ S^{1 2}$&$S^{5 6}$&$\Gamma^{3+1}$ &
$\tilde{S}^{03}$&$\tilde{S}^{1 2}$& $\tilde{S}^{5 6}$\\
\hline
$I$&$1$&$(\frac{1}{2},\frac{1}{2})$&$
\stackrel{03}{(+i)}\,\stackrel{12}{(+)}| \stackrel{56}{(+)}$&
$
{\scriptstyle (-)} \stackrel{56}{(-)}|{\scriptstyle (-)} 
\stackrel{12}{(-)} \stackrel{03}{(-i)}$&$\frac{i}{2}$&$\frac{1}{2}$&$\frac{1}{2}$&$1$
&$\frac{i}{2}$&$\frac{1}{2}$&$\frac{1}{2}$\\
$I$ &$2$&$(\frac{1}{2},-\frac{1}{2})$&$
\stackrel{03}{[-i]}\,\stackrel{12}{[-]}|\stackrel{56}{(+)}$&$
 {\scriptstyle (-)} \stackrel{56}{(-)}| \stackrel{12}{[-]} \stackrel{03}{[-i]}$&
$-\frac{i}{2}$&$-\frac{1}{2}$&$\frac{1}{2}$&$1$
&$\frac{i}{2}$&$\frac{1}{2}$&$\frac{1}{2}$\\
$I$ &$3$&$(-\frac{1}{2},\frac{1}{2})$&$
\stackrel{03}{[-i]}\,\stackrel{12}{(+)}|\stackrel{56}{[-]}$&$
 \stackrel{56}{[-]}|{\scriptstyle (-)} \stackrel{12}{(-)} \stackrel{03}{[-i]}$&
$-\frac{i}{2}$&$ \frac{1}{2}$&$-\frac{1}{2}$&$-1$
&$\frac{i}{2}$&$\frac{1}{2}$&$\frac{1}{2}$\\
$I$ &$1$&$(-\frac{1}{2},-\frac{1}{2})$&$
\stackrel{03}{(+i)}\,\stackrel{12}{[-]}|\stackrel{56}{[-]}$&$
\stackrel{56}{[-]}| \stackrel{12}{[-]} \stackrel{03}{(-i)}$&
$\frac{i}{2}$&$- \frac{1}{2}$&$-\frac{1}{2}$&$-1$
&$\frac{i}{2}$&$\frac{1}{2}$&$\frac{1}{2}$\\
\hline 
$II$&$1$&$(\frac{1}{2},\frac{1}{2})$&$
\stackrel{03}{[+i]}\,\stackrel{12}{[+]}| \stackrel{56}{(+)}$&
$
{\scriptstyle (-)} \stackrel{56}{(-)}|\stackrel{12}{[+]} \stackrel{03}{[+i]}$&$\frac{i}{2}$&
$\frac{1}{2}$&$\frac{1}{2}$&$1$&$-\frac{i}{2}$&$-\frac{1}{2}$&$\frac{1}{2}$\\
$II$ &$2$&$(\frac{1}{2},-\frac{1}{2})$&$
\stackrel{03}{(-i)}\,\stackrel{12}{(-)}|\stackrel{56}{(+)}$&$
 {\scriptstyle (-)} \stackrel{56}{(-)}| {\scriptstyle (-)} \stackrel{12}{(+)}
 \stackrel{03}{(+i)}$&
$-\frac{i}{2}$&$-\frac{1}{2}$&$\frac{1}{2}$&$1$
&$-\frac{i}{2}$&$-\frac{1}{2}$&$\frac{1}{2}$\\
$II$ &$3$&$(-\frac{1}{2},\frac{1}{2})$&$
\stackrel{03}{(-i)}\,\stackrel{12}{[+]}|\stackrel{56}{[-]}$&$
 \stackrel{56}{[-]}| \stackrel{12}{[+]} \stackrel{03}{(+i)}$&
$-\frac{i}{2}$&$ \frac{1}{2}$&$-\frac{1}{2}$&$-1$
&$-\frac{i}{2}$&$-\frac{1}{2}$&$\frac{1}{2}$\\
$II$ &$4$&$(-\frac{1}{2},-\frac{1}{2})$&$
\stackrel{03}{[+i]}\, \stackrel{12}{(-)}|\stackrel{56}{[-]}$&$
\stackrel{56}{[-]}|{\scriptstyle (-)}  \stackrel{12}{(+)} \stackrel{03}{[+i]}$&
$\frac{i}{2}$&$- \frac{1}{2}$&$-\frac{1}{2}$&$-1$
&$-\frac{i}{2}$&$-\frac{1}{2}$&$\frac{1}{2}$\\ 
%
%
 \hline
$III$&$1$&$(\frac{1}{2},\frac{1}{2})$&$
\stackrel{03}{[+i]}\,\stackrel{12}{(+)}| \stackrel{56}{[+]}$&
$
 \stackrel{56}{[+]}|{\scriptstyle (-)}\stackrel{12}{(-)} \stackrel{03}{[+i]}$&$\frac{i}{2}$&
$\frac{1}{2}$&$\frac{1}{2}$&$1$&$-\frac{i}{2}$&$\frac{1}{2}$&$-\frac{1}{2}$\\
$III$ &$2$&$(\frac{1}{2},-\frac{1}{2})$&$
\stackrel{03}{(-i)}\,\stackrel{12}{[-]}|\stackrel{56}{[+]}$&$
 \stackrel{56}{[+]}|  \stackrel{12}{[-]} \stackrel{03}{(+i)}$&
$-\frac{i}{2}$&$-\frac{1}{2}$&$\frac{1}{2}$&$1$
&$-\frac{i}{2}$&$\frac{1}{2}$&$-\frac{1}{2}$\\
$III$ &$3$&$(-\frac{1}{2},\frac{1}{2})$&$
\stackrel{03}{(-i)}\,\stackrel{12}{(+)}|\stackrel{56}{(-)}$&$
 {\scriptstyle (-)}\stackrel{56}{(+)}| {\scriptstyle (-)} \stackrel{12}{(-)} \stackrel{03}{(+i)}$&
$-\frac{i}{2}$&$ \frac{1}{2}$&$-\frac{1}{2}$&$-1$
&$-\frac{i}{2}$&$\frac{1}{2}$&$-\frac{1}{2}$\\
$III$ &$4$&$(-\frac{1}{2},-\frac{1}{2})$&$
\stackrel{03}{[+i]} \stackrel{12}{[-]}|\stackrel{56}{(-)}$&$
{\scriptstyle (-)} \stackrel{56}{(+)}|  \stackrel{12}{[-]} \stackrel{03}{[+i]}$&
$\frac{i}{2}$&$- \frac{1}{2}$&$-\frac{1}{2}$&$-1$
&$-\frac{i}{2}$&$\frac{1}{2}$&$-\frac{1}{2}$\\
\hline
$IV$&$1$&$(\frac{1}{2},\frac{1}{2})$&$
\stackrel{03}{(+i)}\,\stackrel{12}{[+]}| \stackrel{56}{[+]}$&
$
 \stackrel{56}{[+]}|\stackrel{12}{[+]} \stackrel{03}{(-i)}$&$\frac{i}{2}$&
$\frac{1}{2}$&$\frac{1}{2}$&$1$&$\frac{i}{2}$&$-\frac{1}{2}$&$-\frac{1}{2}$\\
$IV$ &$2$&$(\frac{1}{2},-\frac{1}{2})$&$
\stackrel{03}{[-i]}\,\stackrel{12}{(-)}|\stackrel{56}{[+]}$&$
 \stackrel{56}{[+]}| {\scriptstyle (-)} \stackrel{12}{(+)}
 \stackrel{03}{[-]}$&
$-\frac{i}{2}$&$-\frac{1}{2}$&$\frac{1}{2}$&$1$
&$\frac{i}{2}$&$-\frac{1}{2}$&$-\frac{1}{2}$\\
$IV$ &$3$&$(-\frac{1}{2},\frac{1}{2})$&$
\stackrel{03}{[-i]}\,\stackrel{12}{[+]}|\stackrel{56}{(-)}$&$
{\scriptstyle (-)} \stackrel{56}{(+)}| \stackrel{12}{[+]} \stackrel{03}{[-i]}$&
$-\frac{i}{2}$&$ \frac{1}{2}$&$-\frac{1}{2}$&$-1$
&$\frac{i}{2}$&$-\frac{1}{2}$&$-\frac{1}{2}$\\
$IV$ &$4$&$(-\frac{1}{2},-\frac{1}{2})$&$
\stackrel{03}{(+i)}\, \stackrel{12}{(-)}|\stackrel{56}{(-)}$&$
{\scriptstyle (-)} \stackrel{56}{(+)}|\stackrel{12}{(+)} \stackrel{03}{(-i)}$&
$\frac{i}{2}$&$- \frac{1}{2}$&$-\frac{1}{2}$&$-1$
&$\frac{i}{2}$&$-\frac{1}{2}$&$-\frac{1}{2}$\\ 
\hline 
 \end{tabular}
 \end{center}
\end{tiny}
\end{table}
Let us notice that Clifford algebra ''basis vectors'' $\hat{b}^{m \dagger}_{f}$
(forming the creation operators and their Hermitian conjugated partners annihilation 
operators in the internal space of fermions)
offer in $d=(3+1)$ the description of only the spin and family degrees of freedom, 
as can be seen in App.~(\ref{matrixCliffordDMN}), while in $d\ge 5$ the 
Clifford ''basis vectors'' enables  the description of additional families and also of 
charges, as seen in Table~\ref{cliff basis5+1.} and in 
Refs.~(\cite{gn2009,IARD2016,n2014matterantimatter,normaJMP2015} and the references
 therein). 

\subsection{Tensor products of ''basis vectors'' of internal space and basis in ordinary 
            space}  
\label{poincare}            

So far we  have treated only the internal space of fermions either in the ordinary 
description of the internal space, Sect.~\ref{internalspaceordinary}, or when 
using the Grassmann algebra, Sect.~\ref{propertiesGrass0}, or when using 
the Clifford algebra, Sect.~\ref{propertiesCliff0}.

The second quantized fermion fields must include beside the "basis vectors", 
describing the internal space of fermions
also the basis in ordinary space. We write the basis in both spaces as tensor 
products, $*_{T}$, of the basis in the internal space of fermions, created by 
$\hat{b}^{s\dagger}_f$,
which are anticommuting objects (operators), and of the basis in ordinary space, 
created by $\hat{b}^{\dagger}_{\vec{p}}$, which are commuting objects, since 
momenta are commuting objects (operators)
 \begin{eqnarray}
\label{wholespacegeneral}
\{{\bf \hat{b}}^{s \dagger}_{f} (\vec{p}) \,&=& 
\hat{b}^{\dagger}_{\vec{p}}\,*_{T}\,\hat{b}^{s \dagger}_{f}\} \,
|\psi_{oc}>\,*_{T}\, |0_{\vec{p}}> \,,                                                                                                 
 \end{eqnarray}
where $\vec{p}$ determines the momentum in ordinary space and $s$ determines 
all the rest of quantum numbers. 
The state written here as $|\psi_{oc} >\,*_{T}\, |0_{\vec{p}} >$ is considered as
the vacuum for a starting  single particle
states from which one obtains the other single particle states by the
operators, like  $\hat{b}_{\vec{p}}$, which pushes the momentum by an
amount $\vec{p}$.

The meaning of the operator $\hat{b}^{\dagger}_{\vec{p}}$, which translates 
(generates) all the momenta from the starting one, is explained and justified in 
App.~\ref{continuous}. Although the operator for translation of the momentum or 
coordinate is well known, we follow to quite a large extent Ref.~\cite{Louisell},  the
operator $\hat{b}^{\dagger}_{\vec{p}}$ can not be found in the literature,
at least not to our knowledge. We 
postulated it in App.~\ref{continuous} 
 to make  easier to present the relationship between the ordinary second 
quantization procedure and our new one defined in Eq.~(\ref{wholespacegeneral}). 

The dimension of space-time can be any 
$d=(d-1)+1$, although usually is taken the observed one $d=(3+1)$. 
Quantum numbers of fermions are denoted by $s$. In the usual second quantization procedure  for free massless fermions the attention is payed only on the spin 
and handedness. 
Since the Euler-Lagrange equations relate  $p_{0}$ and $\vec{p}$, $(p^0)^2=
(\vec{p})^2$, the continuously infinite basis in the momentum space, created by 
$\hat{b}^{\dagger}_{\vec{p}}\,$, depends on only $\vec{p}$, $|0_{\vec{p}}>$
denotes momentum part of the vacuum state on which ${\bf \hat{b}}^{s \dagger}_{f} (\vec{p})$ applies.

In App.~\ref{continuous} we discuss the creation and annihilation 
operators $\hat{b}^{\dagger}_{\vec{p}}$ and $\hat{b}_{\vec{p}}$  in the 
case of continuous spectra in ordinary momentum and coordinate space in details. 
Here we only present the most needed definitions.


Let us repeat the relations  concerning the momentum part of the  single 
fermion states.
\begin{eqnarray}
\label{creatorp}
|\vec{p}>&=& \hat{b}^{\dagger}_{\vec{p}} \,|\,0_{p}\,>\,,\quad 
<\vec{p}\,| = <\,0_{p}\,|\,\hat{b}_{\vec{p}}\,, \nonumber\\
<\vec{p}\,|\,\vec{p}'>&=&\delta(\vec{p}-\vec{p}')=
<\,0_{p}\,|\hat{b}_{\vec{p}}\; \hat{b}^{\dagger}_{\vec{p}'} |\,0_{p}\,>\,, 
\nonumber\\
&&{\rm leading \;to\;} \nonumber\\
\hat{b}_{\vec{p'}}\, \hat{b}^{\dagger}_{\vec{p}} &=&\delta(\vec{p'}-\vec{p})\,,
\end{eqnarray}
since we normalize $<\,0_{p}\, |\,0_{p}\,>=1$ to identity. 
Let us repeat Eq.~(\ref{eigenvalue1}) from App.~\ref{continuous}
 \begin{eqnarray}
 \label{eigenvalue10}
 <\vec{p}\,| \,\vec{x}>&=&<0_{\vec{p}}\,|\,\hat{b}_{\vec{p}}\;
\hat{b}^{\dagger}_{\vec{x}} 
 |0_{\vec{x}}\,>=(<0_{\vec{x}}\,|\,\hat{b}_{\vec{x}}\;
\hat{b}^{\dagger}_{\vec{p}} \,
 |0_{\vec{p}}\,>)^{\dagger}\, \nonumber\\
 \{\hat{b}^{\dagger}_{\vec{p}}\,,  \,
\hat{b}^{\dagger}_{\vec{p}\,'}\}_{-}&=&0\,,\qquad 
\{\hat{b}_{\vec{p}},  \,\hat{b}_{\vec{p}\,'}\}_{-}=0\,,\qquad
\{\hat{b}_{\vec{p}},  \,\hat{b}^{\dagger}_{\vec{p}\,'}\}_{-}=0\,,
\nonumber\\
\{\hat{b}^{\dagger}_{\vec{x}},  \,\hat{b}^{\dagger}_{\vec{x}\,'}\}_{-}&=&0\,,
\qquad 
\{\hat{b}_{\vec{x}},  \,\hat{b}_{\vec{x}\,'}\}_{-}=0\,,\qquad
\{\hat{b}_{\vec{x}},  \,\hat{b}^{\dagger}_{\vec{x}\,'}\}_{-}=0\,,
\nonumber\\
{\rm while}&&\nonumber\\
\{\hat{b}_{\vec{p}},  \,\hat{b}^{\dagger}_{\vec{x}}\}_{-}&=&
 e^{i \vec{p} \cdot \vec{x}} \frac{1}{\sqrt{(2 \pi)^{d-1}}}\,,\qquad,
\{\hat{b}_{\vec{x}},  \,\hat{b}^{\dagger}_{\vec{p}}\}_{-}=
 e^{-i \vec{p} \cdot \vec{x}} \frac{1}{\sqrt{(2 \pi)^{d-1}}}\,,
\end{eqnarray}

If $\hat{b}^{s\dagger}_{f}$ and their Hermitian conjugated partners do fulfill the 
anticommuting properties of Eq.~(\ref{alphagammatildeprod}), then also
${\bf \hat{b}}^{s \dagger}_{f} (\vec{p})$ and 
$({\bf \hat{b}}^{s \dagger}_{f} (\vec{p}))^{\dagger}$, 
Eq.~(\ref{wholespacegeneral}),  fulfill the 
anticommutation relations of Eq.~(\ref{alphagammatildeprod}) due the 
commutativity of operators $\hat{b}^{\dagger}_{\vec{p}} 
= (\hat{b}^{\dagger}_{-\vec{p}})^{\dagger}=\hat{b}_{-\vec{p}}$.

In usual second quantization procedures the internal space does not have the anticommutation properties. Correspondingly the anticommuting creation and 
annihilation operators 
must be postulated~\cite{Bethe,BetheJackiw} separately. 

The odd Clifford algebra 
offers the ''basis vectors'', which together with their Hermitian conjugated partners 
fulfill the  anticommutation relations of Eq.~(\ref{alphagammatildeprod}), explaining 
correspondingly the Dirac's postulates for the second quantized fermions, as we 
shall see in the next subsubsection.

%
\subsubsection{Creation and annihilation operators in internal and ordinary space
 in usual second quantization procedure of Dirac} 
\label{creationtensorusual}

As we see in Sect.~\ref{internalspaceordinary}, in Table~\ref{Table 3+1.}, 
the internal space of fermions is usually presented with states, denoted by their spins 
and handedness (in the case of free massless fermions, in the case of free massive 
fermions the states are superposition of both handedness). These states are treated as 
commuting vectors, and do not have anticommuting character. Correspondingly the 
anticommuting creation and annihilation operators must be postulated.
  
Let us follow  Ref.~\cite{BetheJackiw}. 
In the Dirac case the second quantized field operators in $d=(3+1)$ dimensions,
$ {\bf  \Psi}^{h s\dagger}(\vec{x}, x^0)$, assumed to
fulfill anticommutation relations
\begin{eqnarray}
\label{stateDiracanti}
\{{\bf  \Psi}^{h s\dagger}(\vec{x}, x^0), \,
{\bf \Psi}^{h' s'\dagger}(\vec{x'}, x^0)\}_{+}&=&0=
\{{\bf \Psi}^{h s}(\vec{x}, x^0), \, 
{\bf \Psi}^{h' s'}(\vec{x'}, x^0)\}_{+}\,,\nonumber\\ 
\{{\bf \Psi}^{h s}(\vec{x}, x^0), \,
{\bf \Psi}^{h' s'\dagger}(\vec{x'}, x^0)\}_{+}&=&
\delta_{h h'} \delta^{s s'} \delta(\vec{x}-\vec{x'})\,,
\end{eqnarray}
are postulated as follows
\begin{eqnarray}
\label{stateDirac}
{\bf \Psi}^{h s}(\vec{x}, x^0)& =&\sum_{m, \vec{p}_k}
 \hat{{\bf a}}^{h s\dagger}_m (\vec{p}_k, x^0)\, 
v^{h s}_m (\vec{p}_k, \vec{x},x^0) \,, \quad
{\bf \Psi}^{h s\dagger}(\vec{x}, x^0) = \sum_{m, \vec{p}_k} 
\hat{{\bf a}}^{h s}_m (\vec{p}_k, x^0)\, 
v^{h s\dagger}_m (\vec{p}_k, \vec{x},x^0)\,.
\end {eqnarray}
Here $v^{h s\dagger}_m (\vec{p}_k, \vec{x},x^0)=(v^{h s}_m (\vec{p}_k, \vec{x},x^0))^*$, with ${}^*$ meaning the complex conjugated values,
$v^{h s}_m (\vec{p}_k,\vec{x},x^0)= u^{h s}_m (\vec{p_k})$ 
$ e^{-i(p^0 x^0- \varepsilon \vec{p}_k \cdot \vec{x})}$ are the two  left handed  
($\Gamma^{(3+1)}=-1=h$) and  the two right handed ($\Gamma^{(3+1)}= 1=h$) 
two-component column matrices, $m=(1,2)$, 
representing twice two solutions $s$ of the Weyl equation for free massless 
fermions of particular momentum 
$|\vec{p}_k|= |p_{k}^{0}|$~(Ref.~\cite{BetheJackiw},  Eqs.~(20-49) - (20-51)), 
the factor $\varepsilon =\pm1$ depends on the product of  handedness and spin.
 (Massive fermions are represented by four-vectors, the superposition of two 
two-vectors of both handedness.)

Vectors $v^{h s}_m (\vec{p}_k,\vec{x},x^0)$ are in the case of discretized  
momenta for a fermion in a box orthogonalized   as follows
\begin{eqnarray}
\int d\vec{x}  \sum_{m} v^{h s \dagger}_m (\vec{p}_k,\vec{x},x^0)\, 
v^{h' s' }_m (\vec{p}_{k'},\vec{x},x^0)&=& \delta_{k k'} \,\delta^{s s'} \,\delta^{h h'}\,.
\label{vvdagger}
\end{eqnarray}

Taking into account the anticommutation relations of Eq.~(\ref{stateDiracanti}), 
the orthogonality of $v^{h s\dagger}_m (\vec{p}_{k}, \vec{x},x^0)$, Eq.~(\ref{vvdagger}), the orthogonality of continuous spectra, Eq~(\ref{creatorp}), 
and Eq.~(\ref{stateDirac}),  the anticommutation relations~(Ref.~\cite{BetheJackiw}, Eqs.~(20-49) - (20-51)) follow for by Dirac 
postulated creation operators $\hat{{\bf a}}^{h \dagger}_m (\vec{p}_k)$ 
and their Hermitian conjugated partners annihilation 
operators $\hat{{\bf a}}^{h}_m (\vec{p}_k)$ 
 \begin{eqnarray}
 \label{comDirac}
\{\hat{{\bf a}}^{h\dagger}_m (\vec{p}_k, x^0), \,\hat{{\bf a}}^{h'\dagger}_n 
(\vec{p}_l, x^0)\}_{*_{T}+}&=&
0= \{\hat{{\bf a}}^{h}_m (\vec{p}_k, x^0), \,\hat{{\bf a}}^{h'}_n (\vec{p}_l, x^0)\}_{*_{T}+}\,,\nonumber\\ 
\{\hat{{\bf a}}^{h}_m (\vec{p}_k, x^0), 
\,\hat{{\bf a}}^{h'\dagger}_n (\vec{p}_l, x^0)\}_{*_{T}+} &=&
\delta_{mn} \,\delta^{h h'} \,\delta_{\vec{p}_k \vec{p}_l}\,.
\end{eqnarray}

In Eq.~(\ref{wholespacegeneral}) presented creation operators  
${\bf \hat{b}}^{s \dagger}_{f} (\vec{p})$ are the tensor products of the basis in 
internal space of fermions, $\hat{b}^{s\dagger}_{f}$, and the basis in momentum
space $\hat{b}^{\dagger}_{\vec{p}}\,$. 
The creation operators $\hat{{\bf a}}^{h s\dagger}_m (\vec{p}_k, x^0)$ 
themselves already have the properties of 
${\bf \hat{b}}^{s \dagger}_{f} (\vec{p})$. They carry all the needed quantum 
numbers of $u^{h s}_m (\vec{p_k})$, while  $v^{h s}_m (\vec{p}_k,\vec{x},x^0)$
''support'' $\hat{{\bf a}}^{h s\dagger}_m (\vec{p}_k, x^0)$  with  
$ e^{-i(p^0 x^0- \varepsilon \vec{p}_k \cdot \vec{x})}$ in addition.

We discuss the continuous momentum basis and the continuous  coordinate basis in 
App.~\ref{continuous}.

We make in Sect.~\ref{creationannihilationtensor} the comparison 
between the above second quantization procedure and our way of second 
quantization, which is presented in the next Sect.~\ref{creationtensorClifford}.
Let us add that the usual second quantized procedure does not take care of charges 
and families of fermions, although both can be included by assuming additional 
quantum numbers and relations among them.



%
%
%

%
\subsubsection{Creation and annihilation operators in internal and ordinary space 
with ''basis vectors'' described by Clifford algebra}
\label{creationtensorClifford}

Our Clifford algebra ''basis vectors'' $\hat{b}^{m \dagger}_{f}$, the 
superposition of products of odd number of $\gamma^a$'s, fulfilling the 
anticommutation relations of Eq.~(\ref{alphagammatildeprod}), have in 
$d$-dimensional spaces $2^{\frac{d}{2}-1}$ families, each with 
$2^{\frac{d}{2}-1}$ members. We present in Sect.~\ref{matrixCliffordDMN} 
the matrix representations of $\gamma^a$'s, $S^{ab}$'s and 
$\tilde{S}^{ab}$'s if space-time is $d=(3+1)$.

Since we observe  only $d=(3+1)$ space-time, it means that
the $(d- 4)$-dimensional space must be directly (dynamically) unobservable, 
although we do observe the charges of quarks and leptons and  of antiquarks 
and antileptons as superposition of spins in $d-4$ of  $d\ge (13+1)$ and at 
least some (so far at least three) of families. 
Quarks and leptons and antiquarks and antileptons, as well as all the 
compositions of elementary fields (like mesons and baryons) experience 
at observable energies the momentum space with
$p^a=(p^0, p^1, p^2, p^3, 0, 0, \cdots, 0)$.
%

To describe quarks and leptons and antiquarks and antileptons we need 
to take into account besides the internal space of fermions also the momentum  
or coordinate space with continuously infinite number of basis vectors, as 
suggested in Eq.~(\ref{wholespacegeneral}), where
${\bf \hat{b}}^{s \dagger}_{f} (\vec{p})$ is the tensor product of the basis in 
internal space of fermions, $\hat{b}^{s\dagger}_f$, and the basis in momentum
space $\hat{b}^{\dagger}_{\vec{p}}\,$, as presented in Eq.~(\ref{creatorp}).\\  

{\bf Statement 6.} 
To derive the anticommutation relations for the Clifford 
fermions, which would replace (being comparable with) the anticommutation 
relations of 
the second quantized fermions, we need to define the tensor product of the 
Clifford odd ''basis vectors''  and the momentum basis first for the single
 fermion states. \\
 
 We pay 
attention here to free massless fields. Since the Euler-Lagrange equations 
for the second quantized free massless field, Sect.~\ref{actionGrassCliff},
relate the energy $p_0$ and the momentum $\vec{p}$, we let the
creation (and correspondingly annihilation) operators to depend only on
$\vec{p}$, with $|p^0|= |\vec{p}|$. Let us therefore define 
\begin{eqnarray}
\{{\hat{\bf b}}^{s  \dagger}_{f} (\vec{p})|_{p^0=|\vec{p}|}\, 
& \stackrel{\mathrm{def}}{=}& 
\sum_{m} c^{m s }{}_{f}\; (\vec{p},|p^0|=|\vec{p}|) \, \, 
\hat{b}^{\dagger}_{\vec{p}}\,*_{T}\,
\hat{b}^{ m \dagger}_{f}\} |\psi_{oc}> |0_{\vec{p}}>\,,
\label{Weylp0}
\end{eqnarray}
with the coefficients $ c^{ ms}{}_{f}\; (\vec{p},|p^0|=|\vec{p}|) $ 
chosen so, that ${\hat{\bf b}}^{ s \dagger}_{f} (\vec{p})|_{p^0=|\vec{p}|}
\cdot e^{-i p_a x^a} $ solve the Weyl equations for free massless fermions,
presented in Eq.~(\ref{Weyl}) in Sect.~\ref{actionGrassCliff} for a chosen 
momentum $\vec{p}$. The creation operator in momentum space is defined 
in Eq.~(\ref{creatorp}).
The tensor product of Eq.~(\ref{wholespacegeneral}) is in  Eq.~(\ref{Weylp0}) 
rewrittwn in a more explicit way. 
In Eq.~(\ref{usefulcontinuous}) the relation $<\vec{p}|\psi_{\vec{p}}>=f(\vec{p})$
concerns only the momentum space, while in Eq.~(\ref{Weylp0}) both basis are related, 
the internal ''basis vectors'' $\hat{b}^{ m \dagger}_{f} $ are multiplied by the continuous
momentum basis, $\,*_{T}\hat{b}_{\vec{p}}\,|0_{\vec{p}}>$, so that the 
superposition of both spaces 
is made, $\sum_{m} c^{m s }{}_{f}\; (\vec{p},|p^0|=|\vec{p}|) \, 
\hat{b}_{\vec{p}}\,*_{T}\,\hat{b}^{ m \dagger}_{f}\} \,|\psi_{oc>} |0_{\vec{p}}>$. 
In such superposition the  function $f(\vec{p})$ of Eq.~(\ref{usefulcontinuous}) in
 App.~\ref{continuous} carries here the quantum numbers of the internal space and is 
replaced by $\sum_{m} c^{m s }{}_{f}\; (\vec{p},|p^0|=|\vec{p}|) $.  


The new {\it basis vectors} are the tensor product, ${}*_{T}$, of the ''basis 
vectors'' in the internal space of fermions, described by the Clifford algebra 
of odd products of $\gamma^a$'s, Eq.~(\ref{alphagammatildeprod}), with 
creation operators $\hat{b}^{ m \dagger}_{f}$ ($m$ denoting the family 
member quantum number, $f$ the family quantum number and $s$ 
different orthonormalized  solutions of the equations of motion) 
and $\hat{b}^{\dagger}_{\vec{p}}$ the ''basis'' in the ordinary space of momenta, 
while $|\psi_{oc}>$ and $|0_{\vec{p}}>$ form the ''vacuum state'', the first 
defined in Eq.~(\ref{vaccliff}), and the second ''vacuum state'' defining the 
starting momentum from which one obtains  other single particle states with 
the same internal part by the operators $\hat{b}^{\dagger}_{\vec{p}}$.
Then it follows for  single fermion states of particular 
$\vec{p}$ and $|p^0|=|\vec{p}|$ and any family quantum numbers $f$ 
and for any solution $s$
\begin{eqnarray}
|\bf {\psi^{s }}_{f} (\vec{p}, p^0)>|_{p^0=|\vec{p}|}  
&=& \int dp^0 \delta(p^0 -|\vec{p}|) \,\hat{\bf b}^{s  \dagger}_{f} (\vec{p}) \, 
 \, {}*_{T}|\psi_{oc}>{}*_{T}|0_{\vec{p}}>\nonumber\\
&=& {\hat{\bf b}}^{ s \dagger}_{f} (\vec{p})
 \,{}*_{T}\,|\psi_{oc}>{}*_{T}|0_{\vec{p}}>\,,
 \label{Weylp1}
 \end{eqnarray}
 while the corresponding state in the coordinate representation is
 \begin{eqnarray}
|\bf{\psi^{s }}_{f} (\vec{x}, x^0)>&=&  \int_{- \infty}^{+ \infty} \,
\frac{d^{d-1}p}{(\sqrt{2 \pi})^{d-1}} \, \hat{\bf b}^{s  \dagger}_{f} (\vec{p})\, 
e^{-i (p^0 x^0- \varepsilon \vec{p}\cdot \vec{x})}|_{p^0=|\vec{p}|} \,{}*_{T}\,|\psi_{oc}>{}*_{T}|0_{\vec{p}}>\,
 \nonumber\\
 &=& {\hat{\bf b}}^{ s \dagger}_{f} (\vec{x})
 \,{}*_{T}\,|\psi_{oc}>{}*_{T}|0_{\vec{x}}>\,,
\label{Weylp2}
\end{eqnarray}
where we take into account Eqs.~(\ref{creatorp}, \ref{eigenvalue10}).
Since the ``basis vectors'' in internal space of fermions are orthogonal and normatized according to 
Eq.~(\ref{almostDirac}), 
\begin{eqnarray}
\label{ccorthogonal0}
 \{\hat{b}^{ m}_{f}\,{}_{*_{A}}\,,\,
 \hat{b}^{ m' \dagger}_{f'}\,{}_{*_{A}}\}_{+}|\psi_{oc}>&=&
 \hat{b}^{ m}_{f}\;{}_{*_{A}}\, 
 \hat{b}^{ m' \dagger}_{f'}\,{}_{*_{A}}|\psi_{oc}> = \delta^{m m'} \delta_{f f'}\,
 |\psi_{oc}>\,,
\end{eqnarray}
it follows by taking into account Eq.~(\ref{usefulcontinuous}) of 
App.~\ref{continuous}~\footnote{In Eq.~(\ref{usefulcontinuous}) one finds
the useful relation $<\vec{p}|f^{*}(\hat{\vec{p}})\, f(\hat{\vec{p}'})|\vec{p}'>=
 f^{*}(\hat{\vec{p}})\, f(\hat{\vec{p}'}) \delta(\vec{p}-\vec{p}')$.} 
 and App.~\ref{proofs}, Statement 8. (there the below relation is derived),
  the  expression 
%
\begin{eqnarray}
\label{ccorthogonal01}
&& <\bf {\psi^{s' }}_{f'} (\vec{p'})| \bf {\psi^{s }}_{f} (\vec{p})>=
<0_{\vec{p}}|<\psi_{oc}|\hat{\bf b}^{s'}_{f'} (\vec{p}'){}*_{T}
\hat{\bf b}^{s  \dagger}_{f} (\vec{p}) \,|\psi_{oc}>|0_{\vec{p}}>
\nonumber\\
&&= <0_{\vec{p}}|<\psi_{oc}|\sum_{m,m'} c^{s'm' *}{}_{f'} (\vec{p}') \,
\hat{b}_{\vec{p}'} \, \hat{b}^{m'}_{f'}\, c^{s m}{}_{f} (\vec{p}) \,
\hat{b}^{m \dagger }_{f}\,\hat{b}^{\dagger}_{\vec{p}}
\,|\psi_{oc}>|0_{\vec{p}}>\nonumber\\
&&= \delta^{s s'} \delta_{f f'}\,\delta(\vec{p}'-\vec{p})\,, 
 \end{eqnarray}
 %
due to
  $<0_{\vec{p}}|\hat{b}_{\vec{p}'} \hat{b}^{\dagger}_{\vec{p}} |0_{\vec{p}}>=
  \delta(\vec{p}'-\vec{p})$,  App.~\ref{continuous},  while
%
%
\begin{eqnarray}
\label{corthogonal1}
&& \sum_{m} c^{ms *}{}_{f}  (\vec{p},|p^0|=|\vec{p}| )\;\; 
c^{m s' }{}_{f'} (\vec{p},|p^0|=|\vec{p}| ) =
 \delta^{s s'} \delta_{f f'}\, \,.
 \end{eqnarray}
 We shall from now on leave out the algebraic product, $*_{A}$,  among the
 ''basis vectors'' in internal space of fermions and the tensor product, $*_{T}$, 
 when creating the  single fermion states out of the ''basis vectors'' and the
 momentum (or coordinate) basis. 
 We  shall use the tensor product, $*{}_{T}$, again for generating the Hilbert
 space out of single fermion states.
 




%
%

 It remains to evaluate  the scalar product $<{\bf \psi^{s'}}_{f'} (\vec{x'}, x^0)\,
  |\,{\bf \psi^{s}}_{f} (\vec{x}, x^0)>$, taking into account that the scalar 
  product is evaluated at a time  $x^0$ 
 %
 \begin{eqnarray}
 \label{xxprodscalar}
&&<{\bf \psi^{s'}}_{f'} ( \vec{x'}, x^0) |\,{\bf \psi^{s}}_{f} (\vec{x}, x^0)> 
=<0_{\vec{x}}|\, <\psi_{oc}| \hat{\bf b}^{s'}_{f `}(\vec{x'}, x^0) \,
 |\hat{\bf b}^{s \dagger}_{f }(\vec{x}, x^0)\,|\psi_{oc}>|0_{\vec{x}}>=
\nonumber\\
&& \int \frac{dp^0}{\sqrt{2\pi}}\, \int \frac{dp'^0}{\sqrt{2\pi}} 
\delta(p^0 -p'^0) \,\int_{- \infty}^{+ \infty} \,
\frac{d^{d-1}p'}{(\sqrt{2 \pi})^{d-1}} \, 
\,\int_{- \infty}^{+ \infty} \,
\frac{d^{d-1}p}{(\sqrt{2 \pi})^{d-1}} \delta(p^0 -|\vec{p}|)\,
\delta(p'^0 -|\vec{p'}|)\,\nonumber\\ 
&&<0_{\vec{x}}|<\psi_{oc}| ( \hat{\bf {b}}^{s'}_{f'} (\vec{p'}, p'^0) \,
  \hat{\bf b}^{s \dagger}_{f }(\vec{p}, p^0) )\,{}_{*_{A}}\, |\psi_{oc}>|0_{\vec{p}}>
 e^{i p'_a x'^a}\, e^{-i p_a x^a} =\nonumber\\
 &&\int \frac{dp^0}{\sqrt{2\pi}}\,
\int_{- \infty}^{+ \infty} \frac{d^{d-1}p'}{(\sqrt{2 \pi})^{d-1}} \, 
\delta(p^0 -|\vec{p'}|)
\int_{- \infty}^{+ \infty} \frac{d^{d-1}p}{(\sqrt{2 \pi})^{d-1}} \, 
\delta(p^0 -|\vec{p}|)\,\nonumber\\
&&<0_{\vec{x}}|<\psi_{oc}| (\hat{\bf {b}}^{s' }_{f'} (\vec{p}, p^0) \,
  \hat{\bf b}^{s \dagger}_{f} (\vec{p'}, p^0))\,{}_{*_{A}} \, |\psi_{oc}>|0_{\vec{x}}>
 e^{i (p^0x^0-\vec{p}\cdot \vec{x})}\,  
 e^{-i (p^0x'^0-\vec{p'}\cdot \vec{x})} =\nonumber\\
 && \delta^{ss'} \,\delta_{ff'}\delta(\vec{x}' - \vec{x})\,,
 \end{eqnarray}
with $<0_{\vec{x}}|<\psi_{oc}|\psi_{oc}>|0_{\vec{x}}>=1$.

The scalar product 
$<{\bf \psi^{s'}}_{f'} (\vec{x'}, x^0)\,  |\,{\bf \psi^{s}}_{f} (\vec{x}, x^0)>$ has 
obviously the desired properties of the second quantized states.

Let us now evaluate  the commutation relation for 
$ \{  \hat{\bf b}^{s' }_{f `}(\vec{p'})\,,\, 
\hat{\bf b}^{s \dagger}_{f }(\vec{p}) \}_{+} |\psi_{oc}> |0_{\vec{p}}>$,
by taking into account that $\hat{\bf b}^{s \dagger}_{f }(\vec{p})=
\sum_{m} c^{ms}{}_{f }\,\hat{b}^{m\dagger}_{f} \hat{b}^{\dagger}_{\vec{p}}$, 
while $\hat{b}^{m}_{f}|\psi_{oc}>=0$, Eq.~(\ref{alphagammatildeprod}),
$\hat{b}_{\vec{p}'}\hat{b}^{\dagger}_{\vec{p}}=\delta(\vec{p}' - \vec{p})$,
Eq.~(\ref{creatorp}), and $\sum_{m} c^{ms *}{}_{f}  (\vec{p'} ) 
c^{m s' }{}_{f'} (\vec{p}) \delta(\vec{p}' - \vec{p})=
 \delta^{s s'} \delta_{f f'}\delta(\vec{p}' - \vec{p})$, Eq.~(\ref{corthogonal1}).
One obtains
\begin{eqnarray}
\label{creationp0}
&&\{  \hat{\bf b}^{s' }_{f `}(\vec{p'})\, ,\, 
\hat{\bf b}^{s \dagger}_{f }(\vec{p}) \}_{+} |\psi_{oc}> |0_{\vec{p}}>=
\{ \sum_{m'} c^{m' s' *}{}_{f `} (\vec{p}') \,\hat{b}^{m'}_{f'} \,\hat{b}^{}_{\vec{p}}\,,
\,\sum_{m} c^{ms}{}_{f }(\vec{p})\,\hat{b}^{m\dagger}_{f} \,
\hat{b}^{\dagger}_{\vec{p}}\,\}_{+}|\psi_{oc}> |0_{\vec{p}}> =\nonumber\\
&&\sum_{m,m'} \hat{\bf b}^{m' }_{f `}\,
\hat{b}^{m\dagger}_{f} \,\,\hat{b}_{\vec{p}'}\, \hat{b}^{\dagger}_{\vec{p}}\,\,
c^{m' s' *}{}_{f `}(\vec{p}') \, c^{ms}{}_{f }(\vec{p})\,|\psi_{oc}> |0_{\vec{p}}>=
\delta^{s s'} \delta_{f f'}\,\delta(\vec{p}' - \vec{p}) \,.
\end{eqnarray}
Let us now write down all the anticommutation relations for 
$\hat{\bf b}^{s }_{f}(\vec{p})$ and $\hat{\bf b}^{s \dagger}_{f }(\vec{p})$
by taking into account Eq.~(\ref{alphagammatildeprod})
\begin{eqnarray}
\{  \hat{\bf b}^{s' }_{f `}(\vec{p'})\,,\, 
\hat{\bf b}^{s \dagger}_{f }(\vec{p}) \}_{+} \,|\psi_{oc}> |0_{\vec{p}}>&=&
\delta^{s s'} \delta_{f f'}\,\delta(\vec{p}' - \vec{p})\, |\psi_{oc}> |0_{\vec{p}}>
\,,\nonumber\\
\{  \hat{\bf b}^{s' }_{f `}(\vec{p'})\,,\, 
\hat{\bf b}^{s}_{f }(\vec{p}) \}_{+} \,|\psi_{oc}> |0_{\vec{p}}>&=&0\,
 |\psi_{oc}> |0_{\vec{p}}>
\,,\nonumber\\
\{  \hat{\bf b}^{s' \dagger}_{f '}(\vec{p'})\,,\, 
\hat{\bf b}^{s \dagger}_{f }(\vec{p}) \}_{+}\, |\psi_{oc}> |0_{\vec{p}}>&=&0
\,|\psi_{oc}> |0_{\vec{p}}>
\,,\nonumber\\
 \hat{\bf b}^{s \dagger}_{f }(\vec{p}) \,|\psi_{oc}> |0_{\vec{p}}>&=&
|\psi^{s}_{f}(\vec{p})>\,\nonumber\\
 \hat{\bf b}^{s}_{f }(\vec{p}) \, |\psi_{oc}> |0_{\vec{p}}>&=&0
 \,|\psi_{oc}> |0_{\vec{p}}>\nonumber\\
 |p^0| &=&|\vec{p}|\,.
\label{Weylpp'comrel}
\end{eqnarray}
The creation operators $  \hat{\bf b}^{s \dagger}_{f }(\vec{p}, p^0) )$  and 
their Hermitian conjugated partners annihilation operators  
$\hat{\bf b}^{s}_{f }(\vec{p}, p^0) )$, creating and annihilating the single fermion state, 
respectively, fulfill when applying on the vacuum state,  
$|\psi_{oc}>|0_{\vec{p}}>$, the anticommutation relations for the second quantized fermions, Eq.~(\ref{comDirac}).

 Let us use Eqs.~(\ref{Weylp0}, \ref{Weylp2})  to  write  the creation operators 
 in the coordinate representation in a more explicit form: 
 \begin{eqnarray}
|\bf{\psi^{s }}_{f} (\vec{x}, x^0)>&=& 
\sum_{m} \,\hat{b}^{ m \dagger}_{f} \, |\psi_{oc}> \int_{- \infty}^{+ \infty} \,
\frac{d^{d-1}p}{(\sqrt{2 \pi})^{d-1}} \, c^{m s }{}_{f}\; 
(\vec{p}) \;  \hat{b}^{\dagger}_{\vec{p}}\;
e^{-i (p^0 x^0- \varepsilon \vec{p}\cdot \vec{x})} 
|0_{\vec{p}}>\,\nonumber\\
&=& \sum_{m} \,\hat{b}^{ m \dagger}_{f} \, |\psi_{oc}>
c^{m s }{}_{f}\; (\vec{x}) \;  \hat{b}^{\dagger}_{\vec{x}}\;|0_{\vec{x}}>
=\hat{\bf b}^{s \dagger}_{f }(\vec{x}) \, |\psi_{oc}> |0_{\vec{x}}>
\,.
\label{Weylx3}
\end{eqnarray}
As expected, the ''basis vectors'' in the internal space remain the same. there are 
the momentum parts $\hat{b}^{\dagger}_{\vec{p}}$, which Fourier transform,
$\int_{- \infty}^{+ \infty} \,\frac{d^{d-1}p}{(\sqrt{2 \pi})^{d-1}} \, c^{m s }{}_{f}\; 
(\vec{p}) \;  \hat{b}^{\dagger}_{\vec{p}}\;
e^{-i (p^0 x^0- \varepsilon \vec{p}\cdot \vec{x})} $, 
into the corresponding coordinate operators,  $\hat{b}^{\dagger}_{\vec{x}}$, and 
so do transform the coefficients which now depend on the coordinates.

We can write down the commutation relations in the coordinate representation
in the equivqlent way, just by  replacing in Eq.~(\ref{Weylpp'comrel}) 
$\hat{\bf b}^{s \dagger}_{f }(\vec{p}) )$ with 
$\hat{\bf b}^{s \dagger}_{f }(\vec{x}, x^0) )$, $\delta(\vec{p}' - \vec{p})$ 
with $\delta(\vec{x}' - \vec{x})$, 
$|0_{\vec{p}}>$ with $|0_{\vec{}x}>$ and drop $ |p^0| =|\vec{p}|$. 
\begin{eqnarray}
\{  \hat{\bf b}^{s' }_{f `}(\vec{x'})\,,\, 
\hat{\bf b}^{s \dagger}_{f }(\vec{x}) \}_{+} \,|\psi_{oc}> |0_{\vec{x}}>&=&
\delta^{s s'} \delta_{f f'}\,\delta(\vec{x}' - \vec{x})\, |\psi_{oc}> |0_{\vec{x}}>
\,,\nonumber\\
\{  \hat{\bf b}^{s' }_{f `}(\vec{x'})\,,\, 
\hat{\bf b}^{s}_{f }(\vec{x}) \}_{+} \,|\psi_{oc}> |0_{\vec{x}}>&=&0\,
 |\psi_{oc}> |0_{\vec{p}}>
\,,\nonumber\\
\{  \hat{\bf b}^{s' \dagger}_{f '}(\vec{x'})\,,\, 
\hat{\bf b}^{s \dagger}_{f }(\vec{x}) \}_{+}\, |\psi_{oc}> |0_{\vec{x}}>&=&0
\,|\psi_{oc}> |0_{\vec{x}}>
\,,\nonumber\\
 \hat{\bf b}^{s \dagger}_{f }(\vec{x}) \,|\psi_{oc}> |0_{\vec{x}}>&=&
|\psi^{s}_{f}(\vec{x})>\,\nonumber\\
 \hat{\bf b}^{s}_{f }(\vec{x}) \, |\psi_{oc}> |0_{\vec{x}}>&=&0
 \,|\psi_{oc}> |0_{\vec{x}}>
 \,.
\label{Weylxx'comrel}
\end{eqnarray}
%
$\hat{\bf b}^{s \dagger}_{f}(\vec{p})$  creates on the vacuum state 
$|\psi_{oc}>|0_{\vec{p}}>$ in Eqs.~(\ref{creationp0}, \ref{Weylpp'comrel}) the single fermion state. 
We can multiply, using  this time the tensor product $*_{T}$  for multiplying 
single particle states, an arbitrary number of such single particle states, what means 
that we multiply an arbitrary number of creation operators 
$\hat{\bf b}^{s \dagger}_{f}(\vec{p})*_{T}$
$\hat{\bf b}^{s'  \dagger}_{f'}(\vec{p'})*_{T}\cdots$
$*_{T}\hat{\bf b}^{s'' \dagger}_{f''}(\vec{p''})$, 
applying on $|\psi_{oc}>\,|0_{\vec{p}}>$, which gives nonzero contributions, 
provided that they distinguish among themselves
in at least one of the properties $(s,f,\vec{p})$, in the internal space quantum 
numbers $(s,f)$ or in momentum part $\vec{p}$, due to the orthonormal property of the
continuous spectrum.

 All new creation operators $\hat{\bf b}^{s \dagger}_{f}(\vec{p})$ 
 are  the superposition of tensor 
products of the ''basis vectors'', described by the odd products of the Clifford 
$\gamma^a$'s, and correspondingly anticommuting, and of the commuting 
''basis'' in momentum (or coordinate in the case of 
$\hat{\bf b}^{s \dagger}_{f}(\vec{x})$ space, and correspondingly fulfilling the 
anticommutation relations of the second quantized fermions, due to 
anticommuting properties of the ''basis vectors'' of the internal space and the 
evenness (commuting properties) of the ''basis'' in coordinate space. 

We can replace the ''basis vectors'' described by $\gamma^a$'s  with the ''basis 
vectors'' described by $\theta^a$'s, thus replacing $\hat{b}^{m \dagger}_{f} $ 
with $\hat{b}^{\theta m \dagger}_{f} $.
But after the reduction of the space of the two odd Clifford algebras, $\gamma^a$'s 
and $\tilde{\gamma}^a$'s  to only  the odd Clifford $\gamma^a$'s,  with  
$\tilde{S}^{ab}$'s used to find all the irreducible representations  and to equip 
them by the family quantum numbers, also the 
Grassmann algebra of $\theta^a$'s and $\frac{\partial}{\partial \theta_a}$'s
reduces to the Clifford odd algebra of $\gamma^a$'s, as we learn in 
Sect.~\ref{reduction}.








%
\subsubsection{Example for solutions of equation of motion for free massless 
fields using Clifford algebra}   
 \label{exampleClifford}   

Let us discuss the simplest case, the case which is the closest to the case discussed 
in Sect.~\ref{internalspaceordinary}.  This is the case of free massless  Clifford 
fermions in $d=(3+1)$,
where $\hat{\bf b}^{s \dagger}_{f}$, applying on the vacuum state $|\psi_{oc}>
|0_{\vec{p}}>$, determine the $s^{th}$ solution of the 
Eq.~(\ref{Weyl}) of the $f^{th}$ family. Fermions carry in this case no charge, 
since there are in our Clifford algebra representation of the internal space of fermions superposition of spins in $d>(3+1)$ which manifest in $d=(3+1)$ the charges. 
There are  $2^{\frac{4}{2}-1}=2$ Clifford odd ''basis vectors'' appearing in 
$2^{\frac{4}{2}-1}=2$ families: 
$\hat{b}^{m=1 \dagger}_{f=1} = \stackrel{03}{[+i]}\,\stackrel{12}{(+)}$,
$\hat{b}^{m=2 \dagger}_{f=1} = \stackrel{03}{(-i)}\,\stackrel{12}{[-]}$,
$\hat{b}^{m=1 \dagger}_{f=2} = \stackrel{03}{(+i)}\,\stackrel{12}{[+]}$,
$\hat{b}^{m=2 \dagger}_{f=2} = \stackrel{03}{[-i]}\,\stackrel{12}{(-)}$,
represented in Table~\ref{3+1} together with their Hermitian conjugated 
partners. (These  two by two  ''basis vectors'' can be ''recognized'' in 
Table~\ref{cliff basis5+1.},  representing the case of $d=(5+1)$, in the 
thirteen, fourteen, seventeen and eighteen line if we ''drop the
charges''  represented with $ \stackrel{56}{(+)}$ and pay attention on only
the odd Clifford parts in $d=(3+1)$.) 
 \begin{table}
 \begin{center}
\begin{minipage}[t]{16.5 cm}
 \caption{The ''basic vectors'' --- the creation operators  
 $\hat{b}^{m \dagger}_{f}$, which on the normalized vacuum state 
 $|\psi_{oc}> =\frac{1}{\sqrt{2}} (\stackrel{03}{[+i]}\stackrel{12}{[-1]}
+ \stackrel{03}{[-i]}\stackrel{12}{[+1]}) $  defines the internal part of states, 
 and  their annihilation partners $\hat{b}^{3}_{f}$ --- are presented for  
$d= (3+1)$-dimensional case. The basic creation operators are the odd products 
of nilpotents and projectors, which are the "eigenstates" of the Cartan subalgebra 
members, ($S^{03}$, $S^{12}$). The eigenvalues of   
($\tilde{S}^{03}$, $\tilde{S}^{12}$, $\tilde{\Gamma}^{3+1}$), Eqs.~(\ref{cartangrasscliff}, \ref{signature0}), 
presented in the last three columns of the table, determine the family quantum 
numbers of the ''basis vectors'', any two of the three values, since
 $\tilde{\Gamma}^{3+1} =-4i \tilde{S}^{03} \,\tilde{S}^{12\,}$.
Operators $\hat{b}^{m \dagger}_{f}$ and $\hat{b}^{m}_{f}$, appearing in
two families, fulfill the anticommutaion relations of Eq.~(\ref{almostDirac}).  
They do not carry any charges.}
\label{3+1}
\end{minipage}
   \begin{tabular}{|r|r|r|r|r|r|r|r|r|r|r|}
 \hline
${\rm family}\, f $&$m $&$
\hat{b}^{ m \dagger}_f$&$
\hat{b}^{m}_{f}$&$
S^{03}$&$ S^{1 2}$&$\Gamma^{3+1}$ &$
\tilde{S}^{03}$&$\tilde{S}^{1 2}$& $\tilde{\Gamma}^{3+1}$\\
$1$ &$1$&$ 
\stackrel{03}{[+i]}\,\stackrel{12}{(+)}$&$
{\scriptstyle (-)}\stackrel{12}{(-)} \stackrel{03}{[+i]}$&$
\frac{i}{2}$&$\frac{1}{2}$&$1$&$
-\frac{i}{2}$&$\frac{1}{2}$&$-1$\\
$1$ &$2$&$
\stackrel{03}{(-i)}\,\stackrel{12}{[-]}$&$
 \stackrel{12}{[-]} \stackrel{03}{(+i)}$&$
 -\frac{i}{2}$&$-\frac{1}{2}$&$1$&$
 -\frac{i}{2}$&$\frac{1}{2}$&$-1$\\
\hline
$2$&$1$&$ 
\stackrel{03}{(+i)}\,\stackrel{12}{[+]}$&$
\stackrel{12}{[+]} \stackrel{03}{(-i)}$&$
\frac{i}{2}$&$\frac{1}{2}$&$1$&$
\frac{i}{2}$&$-\frac{1}{2}$&$-1$\\
$2$ &$2$&$ 
\stackrel{03}{[-i]}\,\stackrel{12}{(-)}$&$
 {\scriptstyle (-)} \stackrel{12}{(+)} \stackrel{03}{[-]}$&$
 -\frac{i}{2}$&$-\frac{1}{2}$&$1$
&$\frac{i}{2}$&$-\frac{1}{2}$&$-1$\\
\hline 
 \end{tabular}
 \end{center}
 \end{table}

There are ${\hat{\bf b}}^{s  \dagger}_{f} (\vec{p})|_{p^0=|\vec{p}|}
e^{-i p_a x^a} $ $= 
\sum_{m} c^{m s }{}_{f}\; (\vec{p}) \, \hat{b}^{m \dagger}_{f}
e^{-i p_a x^a} $, with the appropriately chosen coefficients 
$ c^{ ms}{}_{f}\, (\vec{p})$ and with $p^0=|\vec{p}| $, which solve
the Weyl equations for free massless fermions,
Eq.~(\ref{Weyl}) in Sect.~\ref{actionGrassCliff}, for a chosen 
momentum $\vec{p}$. 

Let us write down two solutions $s=(1,2)$ for each of the two families,  
distinguishing from each other in the family quantum numbers 
($\tilde{S}^{03},\,\tilde{S}^{12}$) 
\begin{small}
\begin{eqnarray}
\label{weylgen0}
p^0&=&|p^0|\,, \quad f {\rm \,is\, determined\, by}\, \tilde{S}^{03}
=-\frac{i}{2}, \tilde{S}^{12}=\frac{1}{2}\,,\nonumber\\
\hat{\bf b}^{s=1  \dagger}_{f=1}\,(\vec{p})  &=&
\beta\, \left( \stackrel{03}{[+i]}\,\stackrel{12}{(+)} + 
\frac{p^1 +i p^2}{ |p^0| + |p^3|} \stackrel{03}{(-i)}\,\stackrel{12}{[-]}
\right)\,
e^{-i(|p^0| x^0 - \vec{p}\cdot\vec{x})}\,,\nonumber\\
\hat{\bf b}^{s=2 \dagger}_{f=1}\,(\vec{p})&=& \beta^*\, 
\left(\stackrel{03}{(-i)}\,\stackrel{12}{[-]} - \frac{p^1 -i p^2}{ |p^0| + |p^3|}\,
 \stackrel{03}{[+i]}\,\stackrel{12}{(+)}\right)\,
e^{-i(|p^0| x^0 + \vec{p}\cdot\vec{x})}\,,\nonumber\\
%
p^0&=&|p^0|\,,\quad f {\rm \,is\, determined\, by\,} \tilde{S}^{03}
=  \frac{i}{2}, \tilde{S}^{12}= -\frac{1}{2}\,,\nonumber\\
\hat{\bf b}^{s=1 \dagger}_{f=2}\,(\vec{p})  &=&  \beta \, 
\left( \stackrel{03}{(+i)}\,\stackrel{12}{[+]}  + \frac{p^1 +i p^2}{ |p^0| + |p^3|}
\stackrel{03}{[-i]}\,\stackrel{12}{(-)}\right)\,
e^{-i(|p^0| x^0 - \vec{p}\cdot\vec{x})}\,,\nonumber\\
\hat{\bf b}^{s=2 \dagger}_{f=2}\,(\vec{p})&=& 
 \beta^*\,\left(\stackrel{03}{[-i]}
\,\stackrel{12}{(-)} - \frac{p^1 -i p^2}{ |p^0| + |p^3|} 
\stackrel{03}{(+i)}\,\stackrel{12}{[+]} \right)\,
e^{-i(|p^0| x^0 + \vec{p}\cdot\vec{x})}\,,
\end{eqnarray}
\end{small}
with $\beta^* \beta= \frac{|p^0| + |p^3|}{2|p^0|} $ taking 
care that the corresponding states are normalized. 
We emphasize again, that using the odd Clifford algebra (after the reduction 
of the two Clifford algebras to only one, Sect.~\ref{reduction}) for the 
description of the internal space of fermions, there appear families already in 
$d=(3+1)$, with the family quantum numbers. Both families have with respect 
to the Lorentz algebra, $S^{ab}$, identical quantum numbers.

We shall latter discuss the appearing of charges in more details, as well as 
the appearance of scalar fields, which are responsible for masses of fermions.


\subsubsection{Comparison between creation and annihilation operators  of 
Dirac  and  of those offered by the odd Clifford algebra } 
         %
\label{creationannihilationtensor}  

Let us compare the usual, the Dirac way of the second quantization and our way 
using the odd Clifford algebra objects.

In Table~\ref{Table 3+1.} the basis states for  single fermions in 
$d=(3+1)$-dimensional  space are presented, two left handed and two
right handed basis states. We treat free massless fermions, looking for solutions 
of the Weyl equation, Eq.~\ref{Weyl}, then we need either the left or the right 
handed  basis states. 

Let us make a choice of the right handed solutions. Then we  have the 
the solution with the superposion of the first two states in Table~\ref{Table 3+1.} 
with spin $S^{12}=\pm\frac{1}{2}$. 

We can write down the creation operator which would, applying on a vacuum 
state, generate such a single fermion state. Eq.~(\ref{stateDirac}) suggests
that the corresponding creation operator of particular momentum $\vec{p}$
can be written as $\hat{\bf a}^{h s \dagger} (\vec{p}) \stackrel{def}{=}\sum_{m}
 \hat{\bf a}^{h s \dagger}_m (\vec{p})\, u^{h s}_m (\vec{p})$, which is our 
 definition.
 
 Table~\ref{3+1} of Sect.~\ref{exampleClifford} represents the ''basis 
 vectors'', that is the creation operators  in $d=(3+1)$  for two families, offered
 by the Clifford algebra. Both are right handed and both give the solution of 
 the Weyl equations, Eq.~\ref{Weyl}. (If we start from the right handed ''basis 
 vectors'', we  can not generate the left handed ones either by the application
 of $S^{ab}$ or $\tilde{S}^{ab}$ on the ''basis vectors''. But ''basis  vectors'' 
 in $d=(5+1)$, for example, do have with respect to $d=(3+1)$ both kinds of 
 ''basis  vectors'', those with $\Gamma^{3+1}=1$ as well as those with 
 $\Gamma^{3+1}=-1$, as we can see in Table~\ref{cliff basis5+1.} in the 
 third and fourth lines and in the seventh and eighth lines.)
  We choose one of these two families, they both have the same properties 
with respect to the solutions  of  the Weyl equations, Eq.~\ref{Weyl}, and with 
respect to $S^{ab}$. They only differ in the family quantum numbers, 
$\tilde{S^{03}}$ and $\tilde{S^{12}}$. 
The creation operators, which on the vacuum state generate the particular 
solution $s$ of the Weyl equation for a chosen momentum $\vec{p}$ in the 
odd Clifford algebra case, are  $\hat{\bf b}^{s \dagger}_{f} (\vec{p}) 
=\sum_{m} c^{ms}{}_{f} (\vec{p})\,\hat{b}^{m \dagger}_f \,
\hat{b}^{\dagger}_{\vec{p}}$.

 Let us write down both kinds of creation operators, the Dirac one and ours,
 both for the right handed case, leaving out  therefore the index  describing 
 handedness $h$ in the Dirac case and  $f$, describing family, in our case
%
\begin{eqnarray}
\label{creationDN}
\hat{\bf a}^{s \dagger} (\vec{p})& \stackrel{def}{=}&\sum_{m}
 \hat{\bf a}^{ \dagger}_m (\vec{p})\, u^{s}_m (\vec{p})\, ,\quad
 \hat{\bf b}^{s \dagger} (\vec{p}) 
=\sum_{m} c^{ms}{} (\vec{p})\,\hat{b}^{m \dagger}\,
\hat{b}^{\dagger}_{\vec{p}}
\,,
\end {eqnarray}
where  we introduced a new creation operator, related to the one proposed 
by Dirac. Comparing both sides of Eq.~(\ref{creationDN}) one recognizes that 
the creation operator of Dirac  $\hat{\bf a}^{\dagger}_m (\vec{p})$ 
(where $v^{s}_{m}(\vec{p}, \vec{x})= u^{s}_{m}(\vec{p})  \,
 e^{i \vec{p} \cdot\,\vec{x}}$ solve the Dirac equation for two choices of $s$), 
 multiplied by  $u^{s}_{m}(\vec{p})$   and summed over $m$ --- $\hat{\bf a}^{s \dagger} (\vec{p}) =\sum_m u^{s}_{m}(\vec{p}) \hat{a}^{\dagger}_m (\vec{p})$  ---
has to be related to 
$\hat{\bf b}^{s \dagger}(\vec{p})$ $ =\sum_m c^{ms} (\vec{p})\,
\hat{b}^{m \dagger}\,\hat{b}^{\dagger}_{\vec{p}}$, for any of the two families  
($f=(\tilde{S}^{03}=-\frac{i}{2}, \tilde{S}^{12}=\frac{i}{2})$ and
$f=(\tilde{S}^{03}=\frac{i}{2}, \tilde{S}^{12}=-\frac{i}{2})$), 
%
\begin{eqnarray}
\label{relationDN}
\hat{\bf a}^{s \dagger} (\vec{p})=
\sum_m  \hat{\bf a}^{ \dagger}_m (\vec{p})\, u^{s}_m (\vec{p})\ \,\quad 
{\rm to\; be\; related\; to}\quad  \hat{\bf b}^{s \dagger}(\vec{p})=
\sum_m c^{ms} (\vec{p}) \,\hat{b}^{m \dagger}\,\hat{b}^{\dagger}_{\vec{p}}\,.
 \end{eqnarray}
 Both creation operators,  $\hat{\bf a}^{s \dagger} (\vec{p})$ and 
 $ \hat{\bf b}^{s \dagger}(\vec{p})$, fulfill the anticommutation relations of 
 Eq.~(\ref{Weylpp'comrel}). $\hat{\bf a}^{s \dagger} (\vec{p})$ fulfil also
 the anticommutation relations of Eq.~(\ref{comDirac}), due to the orthogonality 
  relations of solutions of the Weyl equations, Eq.~(\ref{vvdagger}). An example 
  of the coefficients $c^{ms} (\vec{p}) $ can be read in 
  Eq.~(\ref{weylgen0}) in Sect.~\ref{exampleClifford}.

Dirac equipped the creation operators (and correspondingly also the annihilation 
operators) with the quantum numbers $(s, m)$ and with $\vec{p}$. He postulated 
 for such creation and annihilation operators anticommutation  relations. 
 Our creation and annihilation operators, $ \hat{\bf b}^{s \dagger}(\vec{p})$ and
 $ \hat{\bf b}^{s}(\vec{p})$, have anticommuting properties due to the
 anticommutativity of $ \hat{\bf b}^{m \dagger}$ and $ \hat{\bf b}^{m}$,
 which are Clifford odd objects.

We conclude:  The by Dirac postulated creation operators, 
$ {\hat {\bf a}}^{h s \dagger}_{m} (\vec{p})$, and their annihilation partners, 
$ {\hat {\bf a}}^{h s}_{m} (\vec{p})$, Eqs.~(\ref{stateDiracanti}, \ref{comDirac}), 
related 
in Eq.~(\ref{relationDN}) to the Clifford odd creation and annihilation operators, 
manifest that  the  odd Clifford algebra offers the explanation for  the second 
quantization postulates of Dirac.

\subsubsection{Discrete symmetries  in $d=((d-1)+1)$  with the observed 
properties in $d=(3+1)$}
\label{CPT}  
%


When treating representations of particle and antiparticle states the discrete 
symmetry operators in the second quantized picture are needed. We follow 
here Refs.~\cite{nhds,TDN}.

We have treated so far free massless fermions with the internal space described by 
Grassmann and Clifford algebras. After the reduction of the Clifford algebras to only 
 the algebra generated by $\gamma^a$'s, Sect.~\ref{reduction}, we pay 
attention to only the "Clifford fermions", since besides the Clifford algebras space also 
the Grassmann algebra space is reduced and has no more desired properties 
for the description of the internal space of the integer "fermion states".  
In Refs.~\cite{prd2018,n2019PIPII,2020PartIPartII} the discrete symmetry operators 
for the Grassmann integer spin "fermions" and the Clifford half integer spin fermions
are presented. The fermion anticommuting "nature" of creation and annihilation 
operators originates 
in both cases, the Grassmann case and the Clifford case, in the anticommuting 
properties of the "basis vectors" describing the internal spaces,  
Eq.~(\ref{ijthetaprod}) in the Grassmann case and Eq.~(\ref{alphagammatildeprod},
\ref{Weylpp'comrel}) in the Clifford case.
(Fermions in both spaces are in superposition of eigenstates of the 
Cartan subalgebra  operators of ${\cal {\bf S}}^{ab}$ in the Grassmann case,  
in the Clifford case they are in superposition
 of the Cartan subalgebra operators of $S^{ab}$ as well as of $\tilde{S}^{ab}$.) 

In theories of the Kaluza-Klein kind, as also the {\it spin-charge-family} theory 
is, there are spins or total angular moments in higher dimensions $d>(3+1)$ which 
manifest as charges in the observable $d=(3+1)$ space. 
The charge conjugation requirement, if following the prescription in $d=(3+1)$,
is in contradiction with what we observe for the anti-particles. In Ref.~\cite{nhds}
we redefined the discrete symmetries so that we stay within the subgroups of
the starting group of symmetries, while we require that the angular moments
in higher dimensions manifest as charges in $d=(3+1)$. In the low energy 
regime the angular momenta in higher dimensions manifest only  spins.
As in the whole paper also in this part we pay attention on spaces with 
even $d$ only. 

A theory, which would in the low energy regime explain all the observed 
phenomena, must have the concept of the discrete symmetries $(C,P,T)$ well 
understood, manifesting the Lorentz invariance, causality and $CPT$ symmetry
in $d=(3+1)$~%
\footnote{The concept of what the symmetries C , P and T separately mean is 
in effective theories somewhat a matter of definition partly arranged so as to
make them conserved if possible.}.

We start with the  definition of the discrete symmetries  as they follow from the
prescription in  $d=(3+1)$, defining the ${\cal C}_{{\cal H}}= 
\prod_{\gamma^a \in \Im} \gamma^a \,\, K$, which transforms any single 
particle state $\psi^{s \dagger}_{f} (\vec{p})$ created by 
$\hat{\bf b}^{s \dagger}_{f}$, Eq.~(\ref{creationtensorClifford}), on the vacuum 
state $|\psi_{oc}> |0_{\vec{p}}>$, Eq.~(\ref{Weylp1}), solving the Weyl equation,
Eq.~(\ref{Weyl}), for a free massless spinor with a positive energy  into the 
charge conjugated one with the negative energy. We define the new operator 
$\mathbb{C}_{{\cal H}}$ as the operator which creates the antiparticle state with 
the positive energy and all the properties of $\psi^{s \dagger}_{f} (\vec{p})$.
We do the same for the other two discrete operators, the time reversal 
${\cal T}_{{\cal H}}$ and parity  ${\cal P}^{(d-1)}_{{\cal H}}$ operators,
again following the $d=(3+1)$ prescription 
\begin{eqnarray}
\label{CPTemptH}
\mathbb{C}_{{\cal H}} :& =& 
 (-)^{\frac{d}{2}+1} \prod_{\Im \gamma^a}\gamma^a \,
\Gamma^{(d)} K  \, \prod_{\gamma^a \in \Im} \gamma^a \,K\,\nonumber\\
{\cal T}_{{\cal H}} :&=& \gamma^0 \prod_{\gamma^a \in \Re} 
\gamma^a \,\, K\, I_{x^0}\,,\nonumber\\
{\cal P}^{(d-1)}_{{\cal H}}: &=& \gamma^0\,I_{\vec{x}}\,,
\nonumber\\
I_{x} x^a &=&- x^a\,, \quad I_{x^0} x^a = (-x^0,\vec{x})\,, \quad I_{\vec{x}}
 \vec{x} = -\vec{x}\,, \nonumber\\
I_{\vec{x}_{3}} x^a &=& (x^0, -x^1,-x^2,-x^3,x^5, x^6,\dots, x^d)\,.
\end{eqnarray}
The product $\prod \, \gamma^a$ is meant in the ascending order in $\gamma^a$.

These discrete symmetry operators do not lead to the desired properties of the 
observed anti-particle states.
 
Ref.~\cite{nhds} proposes new operators by  transforming the above  defined 
discrete symmetries so that, while remaining within the same groups of symmetries, 
the redefined discrete symmetries manifest the experimentally acceptable properties
 in $d=(3+1)$, which is of the essential importance  for all the Kaluza-Klein  
theories~\cite{KaluzaKlein,zelenaknjiga,Witten,mil} in which fermions carry only
 the spin and family quantum numbers~\cite{norma95,normaJMP2015} 
\begin{eqnarray}
\label{CNsq}
{\cal C}_{{\cal N}}  &= & {\cal C}_{{\cal H}}
 \, {\cal P}^{(d-1)}_{{\cal H}} \,e^{i \pi J_{1\,2}}\,
e^{i \pi J_{3\,5}}\, e^{i \pi J_{7\,9}}\,e^{i \pi J_{11\,13}},\dots, 
e^{i \pi J_{(d-3)(d-1)}}\,,\nonumber\\
{\cal T}_{{\cal N}}  &= & {\cal T}_{{\cal H}} \,
 {\cal P}^{(d-1)}_{{\cal H}} \,e^{i \pi J_{1\,2}}\,
e^{i \pi J_{3\,6}}\, e^{i \pi J_{8\,10}}\,e^{i \pi J_{12\,14}},\dots, 
e^{i \pi J_{(d-2)d}}\,,\nonumber\\
{\cal P}^{(d-1)}_{{\cal N}}  &= & {\cal P}^{(d-1)}_{{\cal H}} \,e^{i \pi J_{5\,6}}\,
e^{i \pi J_{7\,8}}\, e^{i \pi J_{9\,10}}\,e^{i \pi J_{11\,12}}\,
e^{i\pi J_{13\,14}},\dots, e^{i \pi J_{(d-1)d}}\,,
\nonumber\\
\mathbb{C}_{{ \cal N}} &=& 
\mathbb{C}_{{ \cal H}}\, {\cal P}^{(d-1)}_{{\cal H}} \,e^{i \pi J_{1\,2}}\,
e^{i \pi J_{3\,5}}\, e^{i \pi J_{7\,9}}\,e^{i \pi J_{11\,13}},\dots, 
e^{i \pi J_{(d-3)(d-1)}}\,,
\end{eqnarray}
%
%
%
The rotations ($e^{i \pi J_{1\,2}}\,e^{i \pi J_{3\,5}}\, e^{i \pi J_{7\,9}}\,\dots, 
e^{i \pi J_{(d-3)(d-1)}})$ 
together  with  (multiplied by) ${\cal P}^{(d-1)}_{{\cal H}}$, which are included in 
 $\mathbb{C}_{{\cal N}}$ and ${\cal C}_{{\cal N}} $,
keep $p^{i}$  for $i=(1,2,3)$ unchanged,  
while they transform a state so that  all the eigenvalues of the Cartan subalgebra  
except $S^{03}$ and $J^{12}$ (or at the low energy regime $S^{12}$) change 
sign~\footnote{Since in our extra 
dimension picture $J_{3\,5}$ is no longer a symmetry (for the metric taken as a 
background field) in coordinate 
space, the operation $e^{i \pi J_{3\,5}}$ looks suspicious as being
not a symmetry, but it is. Indeed, the operation $e^{i \pi J_{3\,5}}$ is in the 
coordinate part  composed  
just of a mirror reflection around the $x^3=0$ plane in usual space and reflection 
in the extra dimension space around the surface $x^5=0$. }. 
Correspondingly a second quantized state transforms into  the anti-particle state 
with the same four momentum as the 
starting state but with the opposite values of the total angular momentum 
(or at the low energy regime rather the spin) determined by the Cartan 
subalgebra eigenvalues, except 
for $S^{03}$ and $J^{12}$ (or at the low energy regime $S^{12}$).
The parity operator $ {\cal P}^{(d-1)}_{{\cal N}} $ changes $p^i$ into 
$-p^i$ only for $i=(1,2,3)$, while the time reversal operator corrects all the 
properties of the new 
$\mathbb{C}_{{\cal N}}$) and ${\cal P}^{(d-1)}_{{\cal N}}$ so that
\begin{eqnarray}
\label{CPTNsq}
{\cal C}_{{\cal N}} {\cal P}^{(d-1)}_{{\cal N}} {\cal T}_{{\cal N}} &=&
{\cal C}_{{\cal H}} {\cal P}^{(d-1)}_{{\cal H}} {\cal T}_{{\cal H}} \rightarrow 
\Gamma^{(d)}\, I_{x}\,.
\end{eqnarray}
The above for the Clifford case proposed discrete symmetry operators~\cite{nhds}, 
commute among themselves as also the old ones do. If manifesting dynamics only 
in $d=(3+1)$ space they can be written also, 
 up to a phase, as 
\begin{eqnarray}
\label{CPTNlowE}
{\cal C}_{{\cal N}}  &= & \prod_{\Im \gamma^m, m=0}^{3} \gamma^m\,\, \,
\Gamma^{(3+1)} \,
K \,I_{x^6,x^8,\dots,x^{d}}  \,,\nonumber\\
{\cal T}_{{\cal N}}  &= & \prod_{\Re \gamma^m, m=1}^{3} 
\gamma^m \,\,\,\Gamma^{(3+1)}\,K \,
I_{x^0}\,I_{x^5,x^7,\dots,x^{d-1}}\,,\nonumber\\
{\cal P}^{(d-1)}_{{\cal N}}  &= & \gamma^0\,\Gamma^{(3+1)}\, 
\Gamma^{(d)}\, I_{\vec{x}_{3}}
\,,\nonumber\\
\mathbb{C}_{{ \cal N}} &=& 
\prod_{\Re \gamma^a, a=0}^{d} \gamma^a \,\,\,K
\, \prod_{\Im \gamma^m, m=0}^{3} \gamma^m \,\,\,\Gamma^{(3+1)} \,
K \,I_{x^6,x^8,\dots,x^{d}}= 
\prod_{\Re \gamma^s, s=5}^{d} \gamma^s \, \,I_{x^6,x^8,\dots,x^{d}}\,,
\nonumber\\
\mathbb{C}_{{ \cal N}}{\cal P}^{(d-1)}_{{\cal N}}  &= &\gamma^0
\prod_{\Im \gamma^a, a=5}^{d} \gamma^a \,
I_{\vec{x}_{3}} \,I_{x^6,x^8,\dots,x^{d}}
\end{eqnarray}
Operators $I$ operates as follows in $d=2n$: 
$I_{\vec{x}_{3}} x^a = (x^0, -x^1,-x^2,-x^3,x^5, x^6,\dots, x^d)$,\\ 
$I_{x^5,x^7,\dots,x^{d-1}}$ 
$(x^0,x^1,x^2,x^3,x^5,x^6,x^7,x^8,
\dots,x^{d-1},x^d)$ $=(x^0,x^1,x^2,x^3,-x^5,x^6,-x^7,\dots,-x^{d-1},x^d)$,\\
 $I_{x^6,x^8,\dots,x^d}$ 
$(x^0,x^1,x^2,x^3,x^5,x^6,x^7,x^8,\dots,x^{d-1},x^d)$
$=(x^0,x^1,x^2,x^3,x^5,-x^6,x^7,-x^8,\dots,x^{d-1},-x^d)$. 

\vspace{3mm}

Let us take a simple case that $p^a= (p^0,0,0,p^3,0,\dots,d)$ and 
$d=(13 +1)$. The solution of the Weyl equation, Eq.~(\ref{Weyl}), for free massless 
right handed $u$-quark with the spin $\frac{1}{2}$ and the colour charge 
$C^1= (\frac{1}{2}, \frac{1}{2\sqrt{3}})$, presented in Table~\ref{Table so13+1.}
 in the first line with the creation operator $\hat{u}_{R}^{c1\dagger}$,
applying on the vacuum state $|\psi_{oc}> |0_{\vec{p}}>\,$, is: 
$\hat{u}_{R}^{c1\dagger}$ $= \beta \stackrel{03}{(+i)}\,\stackrel{12}{[+]}|
\stackrel{56}{[+]}\,\stackrel{78}{(+)}
||\stackrel{9 \;10}{(+)}\;\;\stackrel{11\;12}{[-]}\;\;\stackrel{13\;14}{[-]} $.

The application of the Clifford odd operator as it is $\mathbb{C}_{{ \cal N}}$ 
would not belong to the same irreducible representation.
Clifford even operator $\mathbb{C}_{{ \cal N}}{\cal P}^{(d-1)}_{{\cal N}}$
transforms $\hat{u}_{R}^{c1\dagger}$  as follows
$$\mathbb{C}_{{ \cal N}}{\cal P}^{(d-1)}_{{\cal N}} \,
\hat{u}_{R}^{c1\dagger} \, \cdot e^{-i p^0 +i p^3 x^3} = 
\beta \stackrel{03}{[-i]}\,\stackrel{12}{[+]}|
\stackrel{56}{(-i)}\,\stackrel{78}{[-]}
||\stackrel{9 \;10}{[-]}\;\;\stackrel{11\;12}{(+)}\;\;\stackrel{13\;14}{(+)} 
\cdot e^{-i p^0 -i p^3 x^3},$$
which is  the positive energy solution for the left handed $u$-antiquark, the 
creation operator of which is presented in the $35^{th}$ line of 
Table~\ref{Table so13+1.}.
%




The operators $\mathbb{C}_{{\cal H}}, {\cal P}^{(d-1)}_{{\cal H}} $
and ${\cal T}_{{\cal H}}$, 
Eq.~(\ref{CPTemptH}), 
indexed by ${\cal H}$, are good symmetries only when also 
boson fields, in the Kaluza-Klein theories the gravitational fields,   
in higher than $(3+1)$ dimensions are correspondingly transformed and not  
considered as {\em background} fields.
However, the operators $\mathbb{C}_{{\cal N}}, {\cal P}^{(d-1)}_{{\cal N}} $ and 
${\cal T}_{{\cal N}}$ with index ${\cal N}$ will be good symmetries even if we take 
it that there is a {\em background} field depending only on the extra dimension 
coordinates - independent of whether the extra dimension space is compactified or 
not - so that they are not transformed~\footnote{ 
One can namely
easily see that the transformations of the coordinates of the extra dimensions  
are {\em cancelled} between the $\pi$-rotations and the actions of e.g. 
$P_{\cal H}$ on the extra dimensional coordinates. Thus it can be easily seen 
that even if we consider a background gravitational field for the 
extra dimensions - but the $(3+1)$ dimensional space is either flat or  their 
gravitational field is considered dynamical 
so as to be also transformed - these operators with index ${\cal N}$, 
$\mathbb{C}_{{\cal N}}, {\cal P}^{(d-1)}_{{\cal N}} $ and 
$ {\cal T}_{{\cal N}}$, are good symmetries with respect to the space-time
 transformations.They  are indeed good symmetries 
according to their action on the Weyl field. 
The crucial point really is that the ${\cal N}$-indexed operators 
$\mathbb{C}_{{\cal N}}, {\cal P}^{(d-1)}_{{\cal N}} $ 
and $ \mathbb{T}_{{\cal N}}$ with their associated $x$-transformations do 
{\em not transform the extra $(d-4)$coordinates} so that background fields
 depending on these extra dimension coordinates do not matter.}.

 

 
 
 
The properties of the gravitational field, described by vielbeins and spin connections,
under the discrete symmetry operators $\mathbb{C}_{{\cal N}}$, 
${\cal P}^{(d-1)}_{{\cal N}} $ and $ \mathbb{T}_{{\cal N}}$,  
Eq.~(\ref{CPTNlowE}), observed in $d=(3+1)$ as vector and scalar
gauge fields, are discussed in App.~\ref{DSVS}.

\section{From action to equations of motion}
\label{actionGrassCliff} 

 Let us start with the real action functional of local fields $\Psi$ (their Hermitian 
 conjugated fields $\Psi^{\dagger}$) and their 
 first derivative with respect to coordinates  $\frac{\partial \Psi}{\partial x_{a}}$ 
 (and $\frac{\partial \Psi^{\dagger}}{\partial x_{a}}$) in 
$d=(d-1)+1$-dimensional space-time, which is invariant under Poincare 
transformations, 
\begin{eqnarray}
 {\cal A}= \int d^{d}x \,{\cal L}(\Psi, \frac{\partial \Psi}{\partial x_{a}})\,,
 \label{actiongen}
\end{eqnarray}
the integration is limited to the space of interest. 
(In the case that we pay attention to free massless fermions, as we have 
done so far, the integration goes over the whole $d$-dimensional space.)
The arbitrary change of $\Psi $ to $ \Psi+\delta \Psi$ causes the corresponding
change of the action $\cal{A}$
\begin{eqnarray}
\delta {\cal A} &=&\frac{1}{2}  \int d^d x {\bf [} 
\{ \frac{\stackrel{\rightarrow}{\partial {\cal L}}}{\partial \Psi^{\dagger}} -  
 \hat{p}_{a} \,\frac{\stackrel{\rightarrow}{\partial {\cal L}}}{\partial \hat{p}_a \Psi^{\dagger}}\} \delta \Psi^{\dagger}+
 %
\{\frac{\stackrel{\leftarrow}{\partial {\cal L}}}{\partial \Psi }-  
 \hat{p}_{a} \,\frac{\stackrel{\leftarrow}{\partial {\cal L}}}{\partial \hat{p}_a \Psi} \} \delta \Psi {\bf ]}\,. 
 \label{LRA}
\end{eqnarray}
We assume that the two surface terms are equal to zero since the fields are zero 
on the boundary:  $\int d^{d} x \hat{p_a} {\bf [}
\frac{\stackrel{\rightarrow}{\partial {\cal L}}}{\partial  \hat{p}_a \Psi^{\dagger}} \delta \Psi^{\dagger} {\bf ]}=0$
and 
$\int d^{d} x \hat{p_a} {\bf [}
\frac{\stackrel{\leftarrow}{\partial {\cal L}}}{\partial  \hat{p}_a 
\Psi} \delta \Psi {\bf ]}=0$.

We pay in the above Eq.~(\ref{LRA})  attention on the possibility that the fields 
can be fermion fields, and according to our recognition that the fermion fields,
described by either
the odd Grassmann or the odd Clifford algebra, anticommute, the left derivative
is not the same as the right one.

Requiring that the action is stationary with respect to an arbitrary change of
fields, with the boundary terms equal to zero, the Euler-Lagrange equations follow
\begin{eqnarray}
\label{EL}
\frac{\stackrel{\rightarrow}{\partial {\cal L}}}{\partial \Psi^{\dagger}} -  
 \hat{p}_{a} \,\frac{\stackrel{\rightarrow}{\partial {\cal L}}}{\partial \hat{p}_a \Psi^{\dagger}} &=&0\,, \quad 
 \frac{\stackrel{\leftarrow}{\partial {\cal L}}}{\partial \Psi }-  
 \hat{p}_{a} \,\frac{\stackrel{\leftarrow}{\partial {\cal L}}}{\partial \hat{p}_a \Psi}
 =0\,.
 \end{eqnarray}
 %
%
Let us write down the momentum canonically conjugate to the fields, 
$\Pi_{\Psi}=\frac{\delta {\cal A}}{\delta \frac{\partial \Psi}{\partial x_0}}=$  $\frac{\partial {\cal L}}{\partial \frac{\partial \Psi}{\partial x_0}}= i  
\frac{\stackrel{\rightarrow}{\partial {\cal L}}}{\partial \hat{p}_0 \Psi} $, 
	$\Pi_{\Psi^{\dagger}}=\frac{\delta {\cal A}}{\delta \frac{\partial \Psi^{\dagger}}{\partial x_0}}=$  
	$\frac{\partial {\cal L}}{\partial \frac{\partial \Psi^{\dagger}}{\partial x_0}}=i  
\frac{\stackrel{\leftarrow}{\partial {\cal L}}}{\partial \hat{p}_0 \Psi^{\dagger}}$, 
and the Hamilton density 
${\cal H}=\frac{1}{2}\{\Pi_{\Psi} \hat{p}_0 \Psi 
 + \hat{p}_0 \Psi ^{\dagger}\,\Pi_{\Psi^{\dagger}}\} 
- {\cal L}$.
\begin{eqnarray}
\label{EH}
\Pi_{\Psi}&=&
i \frac{\stackrel{\rightarrow}{\partial {\cal L}}}{\partial \hat{p}_0 \Psi}\,, \quad 
\Pi_{\Psi^{\dagger}}=
i \frac{\stackrel{\leftarrow}{\partial {\cal L}}}{\partial \hat{p}_0 \Psi^{\dagger}}\,, \nonumber\\
{\cal H}&=&-i \frac{1}{2}\{\Pi_{\Psi}\, \hat{p}_0 \Psi + \hat{p}_0 \Psi ^{\dagger} \,\Pi_{\Psi^{\dagger}} \} - {\cal L}\,, 
 \end{eqnarray}
in the last line we took again into account that the fields can have an
odd Clifford  (or an odd Grassmann) character.

%
\subsection{Action for free massless fermion fields in standard model and beyond}
\label{actionusual}

 In this paper we treat so far only free massless fermions. The Lagrange action
 is for  general cases, therefore in $d=(d-1)+1$ spaces, equal to
\begin{eqnarray}
{\cal A}\,  &=& \int \; d^d x \; \frac{1}{2}\, (\Psi^{\dagger}\gamma^0 \, \gamma^a \hat{p}_{a} \Psi) +  h.c. =\int \; d^d x \; \frac{1}{2}\, (\Psi^{\dagger}\gamma^0 \, \gamma^a \hat{p}_{a} \Psi - \hat{p}_{a} \Psi^{\dagger} \gamma^0 \, \gamma^a 
 \Psi)\,, 
\label{actionWeylgen}
\end{eqnarray}
up to a surface term, which is assume to contribute nothing. 

The action is invariant under the Lorentz transformations and translations and it is 
real~\footnote{$\gamma^0$ appears in the 
action to pay attention that the action is Lorentz invariant:
%
$S^{ab \dagger} \,\gamma^0 = \gamma^0\, S^{ab}\,,
S^{\dagger}\gamma^0 = \gamma^0 S^{-1}$,
$S=e^{- \frac{i}{2} \omega_{ab} (S^{ab} + L^{ab})}$.
}.

The solutions for free massless fermions are in usual theories, in the {\it standard
model} and in all the theories, which introduce charges and families with the additional
groups, or even the unifying groups for charges, like $SO(10)$ and $SU(5)$,  
 while the space is $d=(3+1)$-dimensional, the superposition of the states, presented 
in Table~\ref{Table 3+1.}, for each handedness separately. The left and the right 
handed solutions can not be obtained by the Lorentz transformations in the 
internal space of fermions.  Families need to be introduced by a separate, additional, 
group.

The right handed solutions of Table~\ref{Table 3+1.} are for the two families, 
existing in $d=(3+1)$ in the case that the internal space is described by the 
odd Clifford algebra (they are reachable from each other by the application of 
$\tilde{S}^{ab}$'s), presented in Eq.~(\ref{weylgen0}). They are superposition
of the ''basis vectors'', presented in Table~\ref{3+1}.

We present in the next Sect.~\ref{actionClifford} the solutions of the equations 
of motion for $d=(d-1)+1$, $d=(5+1)$ and $d=(13+1)$ in the case that only
$p^a= (p^0, p^1,p^2,p^3,0,0,0,0,\dots,0)$ is different from zero --- this is 
expected in the low energy, the observable, regime.

 

%
\subsection{Action  for free massless Clifford fermions 
and solutions of Weyl equations}
\label{actionClifford} 

%

We take the simplest action, the one from Eq.~(\ref{actionWeylgen}),
\begin{eqnarray}
{\cal A}\,  &=& \int \; d^d x \; \frac{1}{2}\, (\psi^{\dagger}\gamma^0 \, \gamma^a \hat{p}_{a} \psi - \hat{p}_{a} \psi^{\dagger} \gamma^0 \, \gamma^a 
 \psi) \,, 
\label{actionWeyl}
\end{eqnarray}
$\hat{p}_{a} = i\, \frac{\partial}{\partial x^a}$, leading to the equations of motion 
\begin{eqnarray}
\label{Weyl}
\gamma^a \hat{p}_{a}  |\psi>&= & 0\,, 
\end{eqnarray}
and fulfilling the Klein-Gordon equation
\begin{eqnarray}
\label{LtoKG}
\gamma^a p_{a} \gamma^b p_b |\psi>&= &   
p^a p_a |\psi>=0\,,\nonumber\\
\end{eqnarray}
for each of the   ${\hat{\bf b}}^{s  \dagger}_{f} (\vec{p})|_{p^0=|\vec{p}|}
e^{-i p_a x^a}|\psi_{0c}>$, with  
${\hat{\bf b}}^{s  \dagger}_{f} (\vec{p})|_{p^0=|\vec{p}|}$ from
Eqs.~(\ref{Weylp0},\ref{Weylp1}), fulfilling the anticommutation relations of 
Eqs.~(\ref{Weylpp'comrel}) on the vacuum state $|\psi_{oc}> |0_{\vec{p}}>$, 
Eq.~(\ref{vac1}).
\begin{eqnarray}
\{ {\hat{\bf b}}^{s  \dagger}_{f} (\vec{p})|_{p^0=|\vec{p}|}\, 
& =& 
\sum_{m} c^{m s }{}_{f}\; (\vec{p},|p^0|=|\vec{p}|) \, \, 
\hat{b}^{\dagger}_{\vec{p}}\,
\hat{b}^{ m \dagger}_{f}\} |\psi_{oc}> |0_{\vec{p}}>\,,\nonumber
\end{eqnarray}

The Euler-Lagrange equations, Eq.~(\ref{EL}), 
$\frac{\stackrel{\rightarrow}{\partial {\cal L}}}{\partial \Psi^{\dagger}} -  
 \hat{p}_{a} \,\frac{\stackrel{\rightarrow}{\partial {\cal L}}}{\partial \hat{p}_a \Psi^{\dagger}} =0$ and 
$ \frac{\stackrel{\leftarrow}{\partial {\cal L}}}{\partial \Psi }-  
 \hat{p}_{a} \,\frac{\stackrel{\leftarrow}{\partial {\cal L}}}{\partial \hat{p}_a \Psi}
 =0$, lead for the Lagrange density,  
\begin{eqnarray}
{\cal L}_{C}\,  &=&  \frac{1}{2} \{\psi^{\dagger} \, \gamma^{0}\,\gamma^a\, 
\hat{p}_a \psi -  \hat{p}_a \psi^{\dagger}\, \gamma^{0}\,\gamma^a \,\psi \, \}\,,
\label{LDWeyl0}
\end{eqnarray}
to
\begin{eqnarray}
\gamma^0 \gamma^a \,\hat{p}_a\,\psi &=& 0\,,\quad 
 - \hat{p}_a \,\psi^{\dagger} \gamma^0 \,\gamma^a=0\,.
\label{LDWeyl2}
\end{eqnarray}
%

Taking into account Eq.~(\ref{EH}) one finds  the momenta and the Hamilton density
\begin{eqnarray}
\label{EHC}
\Pi_{\psi}&=&
i \frac{\stackrel{\rightarrow}{\partial {\cal L}}}{\partial \hat{p}_0 \psi}=
\frac{i}{2} \psi^{\dagger}\,, \quad 
\Pi_{\psi^{\dagger}}=
i \frac{\stackrel{\leftarrow}{\partial {\cal L}}}{\partial \hat{p}_0 \psi^{\dagger}}
=- \frac{i}{2} \psi\,, \nonumber\\
{\cal H}&=&
 \psi^{\dagger}\,\gamma^0 \gamma^i \hat{p}_i\, \psi\,,\qquad  \quad 
 \;\, i=1,2,3,4,\dots,d\,.
 \end{eqnarray}
The solutions ${\hat{\bf b}}^{s  \dagger}_{f} (\vec{p})|_{p^0=|\vec{p}|}
e^{-i p_a x^a} $, 
with  
${\hat{\bf b}}^{s  \dagger}_{f} (\vec{p})|_{p^0=|\vec{p}|} =
\sum_{m} c^{m s }{}_{f}\; (\vec{p},|p^0|=|\vec{p}|) \, \, 
\hat{b}^{\dagger}_{\vec{p}}\:
\hat{b}^{ m \dagger}_{f}$, applying on the vacuum states 
$|\psi_{oc}> |0_{\vec{p}}>$  for $d=((d-2)+1)$, for $d=(5+1), (9+1)$ and $(13+1)$,
will be discussed in Sect.~\ref{examplesClifford}.\\

{\bf Statement 7.} 
All the states, belonging to different values of the Cartan subalgebra ---  they differ 
at least in one value of either the set of $S^{ab}$ or the set of $\tilde{S}^{ab}$, Eq.~(\ref{cartangrasscliff})  --- are 
orthogonal according to the scalar product,  defined as the integral over the Grassmann coordinates, 
Eq.~(\ref{grassintegral}), for a chosen vacuum state $|1\,>$. \\

{\bf Statement 8.} 
Spaces with $d=(d-1)+1$,  $d=2(2n+1)$, $n\ge 1$, have  the special property:
Each family contains fermions and antifermions --- with the properties in $d=(3+1)$ as observed
for quarks and antiquarks and leptons and antileptons --- what is not the case in $d=4n$.\\

This in not difficult to understand:
Each antifermion carries the same spin, determined by the member $S^{12}$ of the Cartan subalgebra, as does the fermion, the opposite handedness in $d=(3+1)$  as the fermion
and the opposite charges, determined by the spins in $d> 3+1$ as the fermion. To achieve 
this in the Clifford description of the internal space of fermions
$\frac{d}{2}$ must be odd in order that $S^{ab}$ change an even number of eigenvalues 
of Cartan subalgebra  members. In Table~\ref{cliff basis5+1.} the ''basis states'' for 
$d=(5+1)$-dimensional space, with $2^{\frac{d}{2}-1}=4$  family members in 
$2^{\frac{d}{2}-1}=4$ families are presented.  One can see that each family includes  
charged fermions and their antifermions of the opposite charge and opposite handedness in $d=(3+1)$.
We shall comment this case and two more cases in the next subsection.



%
\subsection{Examples for solutions of equation of motion for free massless fields 
using Clifford odd algebra 
}
\label{examplesClifford}

In any $d$-dimensional space each family contains $2^{\frac{d}{2}-1}$ members. 
We are in particular interested in spaces with $d=2(2n+1)$, since any family  contains
fermions and antifermions with the property of the observed quarks and leptons and 
antiquarks and antileptons. The fermions and antifermions carry: {\bf i.} the same spin
(the same eigenvalue of the Cartan subalgebra member $S^{12}$), {\bf ii.} the opposite
handedness in $d=(3+1)$, {\bf iii.} the opposite charges (which in our case origin in
spins in $d> (3+1)$). In ordinary theories the charges of fermions and antifermions
must be related by ''hand'', while in the {\it spin-charge-family} theory, using the odd 
Clifford algebra,  the relation between handedness in $d=(3+1)$ and the charges
appear by itself. We discuss in the previous subsection the case of $d=(5+1)$. Here 
we discuss the three cases: The case with $d=(13+1)$, the case with $d=(9+1)$ and the simplest one, the case with $d=(5+1)$.
 
The first case, $d=(13+1)$,  is of the particular interest since one family contains 
all the creation operators for all the quarks and leptons and antiquarks and antileptons, 
with the weak, colour and hypercharges required by the {\it standard model} for massless quarks and leptons before the electroweak transition. There appear after the break of the 
starting symmetry, caused by the condensate of neutrino and antineutrino as explained in
Table.~\ref{Table con.}, two (at low energies almost decoupled) groups of four families, the 
fourth of the lower four families belonging to the observed three, the upper four families
contributing to the dark matter in the universe\footnote{In Ref.~\cite{NH2005,NHD,ND012,familiesNDproc}  $(d=5+1)$-dimensional space is studied 
as a toy model to manifest that the break of symmetry from the higher dimensional space 
to the $(3+1)$-dimensional space {\it can} lead to massless fermions. Fermions were 
described in Clifford space. Here we briefly follow these references, and Refs.~\cite{nhds,TDN},  adding  new observations. }.

The second case, $d=(9+1)$ is only for demonstration, as the case which is the
subgroup of $SO(13+1)$. 

The simplest one, $d=(5+1)$,  is also only for illustration.

We analyse properties of creation operators (of the states when creation operators 
apply on
the vacuum state $|\psi_{oc}>\,|0_{\vec{p}}>$) from the point of view of 
$d=(3+1)$-dimensional space, with  the momentum in ordinary space 
$p^a= (p^0, p^1, p^2, p^3, 0, 0, \cdots, 0)$, 
so that the charges "seen" in $d=(3+1)$ are determined by the Cartan subalgebra of the Lorentz algebra in the internal space --- $S^{st}, (s,t)= (5,6,7,\cdots, d)$. 

We discuss one family in details (let be reminded that the generators $S^{ab}$ connect 
all the members  belonging to one family, while $\tilde{S}^{ab}$ transform a particular 
member of one family into the same member of another family), and comment also on 
the appearance of families (all the families are reachable by $\tilde{S}^{ab}$) and 
present them briefly.

The discrete symmetry operators are  in the Clifford case presented in Eq.~(\ref{CPTNlowE}).


We start with examples in  $d=(5+1)$-dimensional space, with charges determined by $S^{st}, (s,t)= (5,6)$.  




%
\subsubsection{ Clifford fermions and antifermions in $d=(5+1)$}
\label{solutions5+1}

We look for solutions of the Weyl equation, Eq.~(\ref{Weyl}), by taking into account 
four basis creation operators of the first family, $f=1$, in Table~\ref{cliff basis5+1.}. 
 Assuming that  moments in the fifth and the sixth dimensions are zero, 
$p^a=(p^0,p^1,p^2,p^3,0,0)$,  the following four plane wave solutions for 
$p^0 = |\vec{p}|$, can be found, two with the positive charge $\frac{1}{2}$ 
and with spin $S^{12}$ either equal to $\frac{1}{2}$ or to  $-\frac{1}{2}$, and two  
with the negative charge $-\frac{1}{2}$ and again with  $S^{12}$ either  
$\frac{1}{2}$ or  $-\frac{1}{2}$. Since we assume that the nonzero 
components of the momenta are only those in $d=(3+1)$, and we treat only 
free fermions, the dynamics (kinematic) seems similar as the one from 
Eq.~(\ref{weylgen0}), but in that case the space has $d=(3+1)$ and the Clifford 
algebra  offers two families
and no charge. In $d=(5+1)$ case the spin $S^{56}$ determines the  charge 
in $d=(3+1)$. We treat now only one family  out of four  families from 
Table~\ref{cliff basis5+1.}, and each family has four members. Half of them
represent fermions and half antifermions, each fermion has the same $S^{12}$
and the opposite handedness than its antifermion, as expected.
\begin{small}
%
\begin{eqnarray}
\label{weylgen05+1}
&& {\rm Clifford\; odd \;creation \; operators\;} {\rm in \;  d=(5+1)}
\, \nonumber\\
p^0&=&|p^0|\,, \;\; S^{56}= \frac{1}{2}\,, \;\;\Gamma^{(3+1)}=1\,,\nonumber\\
\Bigg( {\hat{\bf b}}^{s=1 \dagger}_{f=1} (\vec{p}) \, 
&=&
\beta\, \left( \stackrel{03}{(+i)}\,\stackrel{12}{(+)}| \stackrel{56}{(+)} + 
\frac{p^1 +i p^2}{ p^0 + p^3} \stackrel{03}{[-i]}\,\stackrel{12}{[-]}|\stackrel{56}{(+)}
\right) \Bigg) \cdot \,\nonumber\\
&&e^{-i (p^0 x^0 - \vec{p}\cdot\vec{x})}\,, \nonumber\\
\Bigg( {\hat{\bf b}}^{s=2  \dagger}_{f=1} (\vec{p}) \, 
&=& \beta^*\, \left(\stackrel{03}{[-i]}\,
\stackrel{12}{[-]}|\stackrel{56}{(+)} - \frac{p^1 -i p^2}{ p^0 + p^3}\,
 \stackrel{03}{(+i)}\,\stackrel{12}{(+)}|\stackrel{56}{(+)}\right) \Bigg) \cdot\,\nonumber\\
&&e^{-i(p^0 x^0 + \vec{p}\cdot\vec{x})}\,,\nonumber\\
&& {\rm Clifford\; odd \;creation \; operators\;} {\rm in \;  d=(5+1)}
\, \nonumber\\
p^0&=&|p^0|\,, \;\;S^{56}=- \frac{1}{2}\,,\;\;\Gamma^{(3+1)}=-1\,,\nonumber\\
\Bigg(  {\hat{\bf b}}^{s=3 \dagger}_{f=1} (\vec{p}) \,
&=& - \beta \, 
\left( \stackrel{03}{[-i]}\,\stackrel{12}{(+)}| \stackrel{56}{[-]} +
 \frac{p^1 +i p^2}{ p^0 + p^3}
\stackrel{03}{(+i)}\,\stackrel{12}{[-]}|\stackrel{56}{[-]}\right) \Bigg) \cdot
\,\nonumber\\
&&e^{-i(p^0 x^0 + \vec{p}\cdot\vec{x})}\,,\nonumber\\
\Bigg(  {\hat{\bf b}}^{s=4  \dagger}_{f=1} (\vec{p}) \,
&=& - \beta^*\,\left(\stackrel{03}{(+i)}
\,\stackrel{12}{[-]}| \stackrel{56}{[-]} - \frac{p^1 -i p^2}{ p^0 + p^3} 
\stackrel{03}{[-i]}\,\stackrel{12}{(+)}|\stackrel{56}{[-]}\right) \Bigg) 
\cdot \,\nonumber\\
&&e^{-i(p^0 x^0 - \vec{p}\cdot\vec{x})}\,,
\end{eqnarray}
\end{small}
Index ${}^{s=(1,2,3,4)}$ counts different solutions of the Weyl equations, 
index ${}_{f=1}$ denotes the family quantum number, all solutions belong to 
the same family, while $\beta^* \beta= \frac{p^0 + p^3}{2p^0} $ takes 
care that the corresponding states are normalized. 

All four superposition of ${\hat{\bf b}}^{s  \dagger}_{f} (\vec{p})|_{p^0=|\vec{p}|}\, 
= \sum_{m} c^{m s }{}_{f}\; (\vec{p},p^0=|\vec{p}|) \, \, 
\hat{b}^{\dagger}_{\vec{p}}\,
\hat{b}^{ m \dagger}_{f} $
 $ e^{-i(p^0 x^0 -\varepsilon \vec{p}\cdot \vec{x})}$, Eq.~(\ref{Weylp0}),
with $m=(1,2)$ for the 
first two states, and with $m=(3,4)$ for the second two states, 
Table~\ref{cliff basis5+1.}, $s=(1,2,3,4)$, are orthogonal and 
correspondingly normalized. 

 The vacuum state $|\psi_{oc}>$, Eq.~(\ref{vac1}), is the sum
of selfajoint operators ($\stackrel{03}{[-i]}\,\stackrel{12}{[-]}| \stackrel{56}{[-]}$, 
$\stackrel{03}{[+i]}\,\stackrel{12}{[+]}| \stackrel{56}{[-]}$, 
$\stackrel{03}{[+i]}\,\stackrel{12}{[-]}| \stackrel{56}{[+]}$,
and  $\stackrel{03}{[-i]}\,\stackrel{12}{[+]}| \stackrel{56}{[+]}$), 
needed that the first, second, third and fourth family creation operators, respectively, 
applying on the vacuum state $|\psi_{oc}>|0_{\vec{p}}>$, give nonzero value, 
while their Hermitian conjugated partners 
${\hat{\bf b}}^{s  \dagger}_{f} (\vec{p})|_{p^0=|\vec{p}|}$, applying on 
$|\psi_{oc}>|0_{\vec{p}}>$, give  zero.

The discrete symmetry operator $\mathbb{C}_{{ \cal N}}{\cal P}^{(d-1)}_{{\cal N}}$,
Eq.~(\ref{CPTNlowE}), which is in this particular case equal to $\gamma^0 \gamma^5
I_{\vec{x}_3}$, transforms $ {\hat{\bf b}}^{s=1 \dagger}_{f=1} (\vec{p})$
into $ {\hat{\bf b}}^{s=3 \dagger}_{f=1} (\vec{p})$ and
$ {\hat{\bf b}}^{s=2 \dagger}_{f=1} (\vec{p})$ into 
$ {\hat{\bf b}}^{s=1 \dagger}_{f=1} (\vec{p})$.

\subsubsection{ Clifford fermions and antifermions in $d=(9+1)$ and $d=(13+1)$}
\label{solutions9+1 13+1}

We look for in both cases, in the case of $d=(9+1)$ and  in the case of $d=(13+1)$, 
the solutions of the Weyl equation, Eq.~(\ref{Weyl}), under the assumption that  
moments in all higher dimensions, $d\ge 5$,  are in the low energy regime equal to zero, 
$p^a=(p^0,p^1,p^2,p^3,0,0,\dots,0 )$.  We  make a 
choice in both cases of only one family out of eight families 
presented in Table~\ref{Table III.}.

Let us first look at the properties of one family (one irreducible representation)  of 
$SO(13,1)$, with $2^{\frac{d}{2}-1}= 64$ members, presented in 
Table~\ref{Table so13+1.}.  This family contains also $SO(9,1)$ family with 
$2^{\frac{d}{2}-1}= 16 $ members,  which can be recognized  in
Table~\ref{Table so13+1.}.

In Table~\ref{Table so13+1.} the ''basis vectors'' --- the creation operators in 
internal space of fermions with the quantum numbers of the observed quarks and 
leptons and antiquarks and antileptons, 
$\hat{b}^{m\dagger}_{f}$, $m=(1,2,\dots, 2^{\frac{d}{2}-1}=64$, 
--- are presented. The 
whole table follows from the starting $\hat{b}^{m=1\dagger}_{f=1}$ 
$=\hat{u}^{c1\dagger}_{R}$, representing the right handed $u$-quark with $S^{12}=
\frac{1}{2}$, the colour charge $(\tau^{33}=\frac{1}{2}, \tau^{38}=\frac{1}{2\sqrt{3}}$), 
the weak charge $\tau^{13}=0$, the hypercharge $Y=\frac{2}{3}$, the 
electromagnetic charge $Y=\frac{2}{3}$ and the ''fermion charge'' $\tau^4=\frac{1}{6}$. 

The reader can calculate all the quantum numbers  of Table~\ref{Table so13+1.} and
Table~\ref{Table III.}, if taking into account
the generators of the two $SU(2)$  ($\subset SO(3,1)$ $\subset SO(7,1) \subset SO(13,1)$) groups, describing  spins of fermions and  the corresponding family quantum numbers
\begin{eqnarray}
\label{so1+3}
&&\vec{N}_{\pm}(= \vec{N}_{(L,R)}): = \,\frac{1}{2} (S^{23}\pm i S^{01},
S^{31}\pm i S^{02}, S^{12}\pm i S^{03} )\,,\quad
\vec{\tilde{N}}_{\pm}(=\vec{\tilde{N}}_{(L,R)}): =
 \,\frac{1}{2} (\tilde{S}^{23}\pm i \tilde{S}^{01}\,,
\end{eqnarray}
the generators of the two $SU(2)$ ($SU(2)$ $\subset SO(4)$ $\subset SO(7,1) 
\subset SO(13,1)$) groups, describing  the weak charge, $\vec{\tau}^{1}$, and
the second kind of the weak charge, $\vec{\tau}^{2}$,  of fermions and 
the corresponding family quantum numbers
%
 \begin{eqnarray}
 \label{so42}
 \vec{\tau}^{1}:&=&\frac{1}{2} (S^{58}-  S^{67}, \,S^{57} + S^{68}, \,S^{56}-  S^{78} )\,,
 \quad
 \vec{\tau}^{2}:= \frac{1}{2} (S^{58}+  S^{67}, \,S^{57} - S^{68}, \,S^{56}+  S^{78} )\,,
 \nonumber\\
 \vec{\tilde{\tau}}^{1}:&=&\frac{1}{2} (\tilde{S}^{58}-  \tilde{S}^{67}, \,\tilde{S}^{57} + 
 \tilde{S}^{68}, \,\tilde{S}^{56}-  \tilde{S}^{78} )\,, \quad 
 \vec{\tilde{\tau}}^{2}:=\frac{1}{2} (\tilde{S}^{58}+  \tilde{S}^{67}, \,\tilde{S}^{57} - 
 \tilde{S}^{68}, \,\tilde{S}^{56}+  \tilde{S}^{78} ),\,\,\;\;
 \end{eqnarray}
and the generators of $SU(3)$ and $U(1)$ subgroups of $SO(6)$ $\subset SO(13,1)$, describing  the colour charge and the ''fermion'' charge  of fermions as well as the corresponding 
family quantum number $\tilde{\tau}^4$
%
 \begin{eqnarray}
 \label{so64}
 \vec{\tau}^{3}: = &&\frac{1}{2} \,\{  S^{9\;12} - S^{10\;11} \,,
  S^{9\;11} + S^{10\;12} ,\, S^{9\;10} - S^{11\;12}\, ,  
  S^{9\;14} -  S^{10\;13} ,\,  \nonumber\\
  && S^{9\;13} + S^{10\;14} \,,  S^{11\;14} -  S^{12\;13}\,, 
  S^{11\;13} +  S^{12\;14} ,\,  \frac{1}{\sqrt{3}} ( S^{9\;10} + S^{11\;12} - 
 2 S^{13\;14})\}\,,\nonumber\\
 \tau^{4}: = &&-\frac{1}{3}(S^{9\;10} + S^{11\;12} + S^{13\;14})\,,\;\;\nonumber\\
 \tilde{\tau}^{4}: = &&-\frac{1}{3}(\tilde{S}^{9\;10} + \tilde{S}^{11\;12} + \tilde{S}^{13\;14})\,.
 \end{eqnarray}
The (chosen) Cartan subalgebra operators, determining the commuting operators in the 
above equations,
is presented in Eq.~(\ref{cartangrasscliff}). 

The  hypercharge $Y$ and the electromagnetic charge $Q$ and the corresponding family
 quantum numbers then follows as
 \begin{eqnarray}
 \label{YQY'Q'andtilde}
 Y:= \tau^{4} + \tau^{23}\,,\;\; Q: =  \tau^{13} + Y\,,\;\; 
 Y':= -\tau^{4}\tan^2\vartheta_2 + \tau^{23}\,, 
 \;\; Q':= -Y \tan^2\vartheta_1 + \tau^{13} \,,&&,\nonumber\\
  \tilde{Y}:= \tilde{\tau}^{4} + \tilde{\tau}^{23}\,,\,\;\tilde{Q}:= 
  \tilde{Y} + \tilde{\tau}^{13}\,,\;\;
   \tilde{Y'}:= -\tilde{\tau}^{4} 
  \tan^2 \vartheta_2 + \tilde{\tau}^{23}\,,\;
  \;\; \tilde{Q'}= -\tilde{Y} \tan^2 \vartheta_1 
  + \tilde{\tau}^{13}\,. &&\,.
  \end{eqnarray}
 %
 \begin{small}
Below are some of the above expressions written in terms of  nilpotents and projectors
 \begin{eqnarray}
\label{plusminus}
 N^{\pm}_{+}         &=& N^{1}_{+} \pm i \,N^{2}_{+} = 
 - \stackrel{03}{(\mp i)} \stackrel{12}{(\pm )}\,, \quad N^{\pm}_{-}= 
 N^{1}_{-} \pm  i\,N^{2}_{-} = 
   \stackrel{03}{(\pm i)} \stackrel{12}{(\pm )}\,,
\nonumber\\
 \tilde{N}^{\pm}_{+} &=& - \stackrel{03}{\tilde{(\mp i)}} 
 \stackrel{12}{\tilde{(\pm )}}\,, \quad 
 \tilde{N}^{\pm}_{-}= 
   \stackrel{03} {\tilde{(\pm i)}} \stackrel{12} {\tilde{(\pm )}}\,,\nonumber\\ 
 \tau^{1\pm}         &=& (\mp)\, \stackrel{56}{(\pm )} \stackrel{78}{(\mp )} \,, \quad   
 \tau^{2\mp}=            (\mp)\, \stackrel{56}{(\mp )} \stackrel{78}{(\mp )} \,,\nonumber\\ 
 \tilde{\tau}^{1\pm} &=& (\mp)\, \stackrel{56}{\tilde{(\pm )}} 
 \stackrel{78}{\tilde{(\mp )}}\,,\quad   
 \tilde{\tau}^{2\mp}= (\mp)\, \stackrel{56}{\tilde{(\mp )}} \stackrel{78}{\tilde{(\mp )}}\,.
 \end{eqnarray}
\end{small}

The corresponding  annihilation operators in internal space of fermions $\hat{b}^{m}_{f}$
are the Hermitian conjugated partners of the creation operators,  
$\hat{b}^{m \dagger}_{f}$, of Table~\ref{Table so13+1.} and Table~\ref{Table III.}.

One sees that  quarks of all three colours as well as the colourless leptons have the same 
$SO(7,1)$ part and the same is true for the antiquarks and antilepons.  The operator 
$S^{09}$, for 
example, transforms $ \hat{u}_{R}^{c1\dagger}$ into $ \hat{\bar{e}}^{ \dagger}_{L}$,
presented on $57$ line of Table~\ref{Table so13+1.},
while the discrete symmetry operator $\mathbb{C}_{{ \cal N}} {\cal P}^{(d-1)}_{{\cal N}}  =  \gamma^0\,\gamma^{5} \,
\cdots  \gamma^{d-1 }\,  I_{\vec{x}_{3}} \,I_{x^6,x^8,\dots,x^{d}} $, presented in 
Eq.~(\ref{CPTNlowE}) of Sect.~\ref{CPT}, transforms $ \hat{u}_{R}^{c1\dagger}$ into 
$\bar{u}_{L}^{\bar{c1} \dagger}$, which is the left handed anti $u$-quark of the same spin 
$S^{12}$ and all the charges of opposite values,  appearing in this table on the line $35$.

Table~\ref{Table III.} presents ''basis vectors'' ($\hat{b}^{m \dagger}_{f}$, Eq.~(\ref{alphagammatildeprod})) for eight families of the right handed $u$-quark 
of  the colour  $(\frac{1}{2}, \frac{1}{2\sqrt{3}})$ and the right handed colourless
$\nu$ --- the SO(7,1) content of the $SO(13,1)$ group are in both cases identical, they 
distinguish only in the $SU(3)$ and $U(1)$ subgroups of $SO(6)$. All the members 
of any of these eight families of Table~\ref{Table III.} follows from either the $u$-quark 
or the $\nu$-lepton  by the application of $S^{ab}$. Each family carries the family quantum
numbers, determined by the Cartan subalgebra of $\tilde{S}^{ab}$ in Eq.~(\ref{cartangrasscliff}) and presented in Table~\ref{Table III.}. When we treat the
$d=(9+1)$ case, the families can be assumed, determined in this case with the
Cartan subalgebra members of $\tilde{\tau}^{13}, \tilde{\tau}^{23}, \tilde{N}^{3}_L,
\tilde{N}^{3}_R$ and $\tilde{S}^{9 \,10}$. This is not the  case, which would be 
realized in nature, at least it is not yet observed.

The Weyl equations for free massless fermions, Eq.~(\ref{Weyl}), offer the same 
solutions for all the  families. Only when fermions  interact with the scalar fields, 
which ''see''  the family quantum numbers, the solutions depend on the family
quantum numbers.

 \begin{table}
 \begin{center}
   \begin{tiny}
\begin{minipage}[t]{16.5 cm}
\caption{Eight families of the ''basis vectors'', the creation operators  
$\hat{b}^{m \dagger}_{f}$,  
of $\hat{u}^{c1 \dagger}_{R}$ --- the right 
handed $u$-quark with spin $\frac{1}{2}$ and the colour charge $(\tau^{33}=1/2$, 
$\tau^{38}=1/(2\sqrt{3}))$, appearing in the first line of Table~\ref{Table so13+1.} --- 
and of  the colourless right handed neutrino $\hat{\nu}^{\dagger}_{R}$ 
of spin $\frac{1}{2}$ , appearing 
in the $25^{th}$ line of Table~\ref{Table so13+1.} ---   
are presented in the  left and in the right part of this table, respectively. Table is taken 
from~\cite{normaJMP2015}. 
Families belong to two groups of four families, one ($I$) is a doublet with respect to 
($\vec{\tilde{N}}_{L}$ and  $\vec{\tilde{\tau}}^{(1)}$) and  a singlet with respect 
to ($\vec{\tilde{N}}_{R}$ and  $\vec{\tilde{\tau}}^{(2)}$), Eqs.~(\ref{so1+3}, \ref{so42}), 
the other ($II$) is a singlet with respect to ($\vec{\tilde{N}}_{L}$ and  
$\vec{\tilde{\tau}}^{(1)}$) and  a doublet with respect to 
($\vec{\tilde{N}}_{R}$ and  $\vec{\tilde{\tau}}^{(2)}$), Eqs.~(\ref{so1+3}, \ref{so42}).
All the families follow from the starting one by the application of the operators 
($\tilde{N}^{\pm}_{R,L}$, $\tilde{\tau}^{(2,1)\pm}$), Eq.~(\ref{plusminus}).  
The generators ($N^{\pm}_{R,L} $, $\tau^{(2,1)\pm}$), Eq.~(\ref{plusminus}),
transform $\hat{u}^{\dagger}_{1R}$ to all the members of one family of the same colour charge. 
The same generators transform equivalently the right handed   neutrino 
$\hat{\nu}^{\dagger}_{1R}$  to all the colourless members of the same family.
}
\label{Table III.}
\end{minipage}
 \begin{tabular}{|r|c|c|c|c|c c c c c|}
 \hline
 &&&&&$\tilde{\tau}^{13}$&$\tilde{\tau}^{23}$&$\tilde{N}_{L}^{3}$&$\tilde{N}_{R}^{3}$&
 $\tilde{\tau}^{4}$\\
 \hline
 $I$&$\hat{u}^{c1 \dagger}_{R\,1}$&
   $ \stackrel{03}{(+i)}\,\stackrel{12}{[+]}|\stackrel{56}{[+]}\,\stackrel{78}{(+)} ||
   \stackrel{9 \;10}{(+)}\;\;\stackrel{11\;12}{[-]}\;\;\stackrel{13\;14}{[-]}$ & 
    $\hat{\nu}^{\dagger}_{R\,1}$&
   $ \stackrel{03}{(+i)}\,\stackrel{12}{[+]}|\stackrel{56}{[+]}\,\stackrel{78}{(+)} ||
   \stackrel{9 \;10}{(+)}\;\;\stackrel{11\;12}{(+)}\;\;\stackrel{13\;14}{(+)}$ 
  &$-\frac{1}{2}$&$0$&$-\frac{1}{2}$&$0$&$-\frac{1}{2}$ 
 \\
  $I$&$\hat{u}^{c1 \dagger}_{R\,2}$&
   $ \stackrel{03}{[+i]}\,\stackrel{12}{(+)}|\stackrel{56}{[+]}\,\stackrel{78}{(+)} ||
   \stackrel{9 \;10}{(+)}\;\;\stackrel{11\;12}{[-]}\;\;\stackrel{13\;14}{[-]}$ & 
   $\hat{\nu}^{\dagger}_{R\,2}$&
   $ \stackrel{03}{[+i]}\,\stackrel{12}{(+)}|\stackrel{56}{[+]}\,\stackrel{78}{(+)} ||
   \stackrel{9 \;10}{(+)}\;\;\stackrel{11\;12}{(+)}\;\;\stackrel{13\;14}{(+)}$ 
  &$-\frac{1}{2}$&$0$&$\frac{1}{2}$&$0$&$-\frac{1}{2}$
 \\
  $I$&$\hat{u}^{c1 \dagger}_{R\,3}$&
   $ \stackrel{03}{(+i)}\,\stackrel{12}{[+]}|\stackrel{56}{(+)}\,\stackrel{78}{[+]} ||
   \stackrel{9 \;10}{(+)}\;\;\stackrel{11\;12}{[-]}\;\;\stackrel{13\;14}{[-]}$ & 
    $\hat{\nu}^{\dagger}_{R\,3}$&
   $ \stackrel{03}{(+i)}\,\stackrel{12}{[+]}|\stackrel{56}{(+)}\,\stackrel{78}{[+]} ||
   \stackrel{9 \;10}{(+)}\;\;\stackrel{11\;12}{(+)}\;\;\stackrel{13\;14}{(+)}$ 
  &$\frac{1}{2}$&$0$&$-\frac{1}{2}$&$0$&$-\frac{1}{2}$
 \\
 $I$&$\hat{u}^{c1 \dagger}_{R\,4}$&
  $ \stackrel{03}{[+i]}\,\stackrel{12}{(+)}|\stackrel{56}{(+)}\,\stackrel{78}{[+]} ||
  \stackrel{9 \;10}{(+)}\;\;\stackrel{11\;12}{[-]}\;\;\stackrel{13\;14}{[-]}$ & 
   $\hat{\nu}^{\dagger}_{R\,4}$&
  $ \stackrel{03}{[+i]}\,\stackrel{12}{(+)}|\stackrel{56}{(+)}\,\stackrel{78}{[+]} ||
  \stackrel{9 \;10}{(+)}\;\;\stackrel{11\;12}{(+)}\;\;\stackrel{13\;14}{(+)}$ 
  &$\frac{1}{2}$&$0$&$\frac{1}{2}$&$0$&$-\frac{1}{2}$
  \\
  \hline
  $II$& $\hat{u}^{c1 \dagger}_{R\,5}$&
        $ \stackrel{03}{[+i]}\,\stackrel{12}{[+]}|\stackrel{56}{[+]}\,\stackrel{78}{[+]}||
        \stackrel{9 \;10}{(+)}\;\;\stackrel{11\;12}{[-]}\;\;\stackrel{13\;14}{[-]}$ & 
         $\hat{\nu}^{\dagger}_{R\,5}$&
        $ \stackrel{03}{[+i]}\,\stackrel{12}{[+]}|\stackrel{56}{[+]}\,\stackrel{78}{[+]}|| 
        \stackrel{9 \;10}{(+)}\;\;\stackrel{11\;12}{(+)}\;\;\stackrel{13\;14}{(+)}$ 
        &$0$&$-\frac{1}{2}$&$0$&$-\frac{1}{2}$&$-\frac{1}{2}$
 \\ 
  $II$& $\hat{u}^{c1 \dagger}_{R\,6}$&
      $ \stackrel{03}{(+i)}\,\stackrel{12}{(+)}|\stackrel{56}{[+]}\,\stackrel{78}{[+]}||
      \stackrel{9 \;10}{(+)}\;\;\stackrel{11\;12}{[-]}\;\;\stackrel{13\;14}{[-]}$ & 
    $\hat{\nu}^{\dagger}_{R\,6}$&
      $ \stackrel{03}{(+i)}\,\stackrel{12}{(+)}|\stackrel{56}{[+]}\,\stackrel{78}{[+]}|| 
      \stackrel{9 \;10}{(+)}\;\;\stackrel{11\;12}{(+)}\;\;\stackrel{13\;14}{(+)}$ 
      &$0$&$-\frac{1}{2}$&$0$&$\frac{1}{2}$&$-\frac{1}{2}$
 \\ 
 $II$&$\hat{u}^{c1 \dagger}_{R\,7}$&
 $ \stackrel{03}{[+i]}\,\stackrel{12}{[+]}|\stackrel{56}{(+)}\,\stackrel{78}{(+)}||
 \stackrel{9 \;10}{(+)}\;\;\stackrel{11\;12}{[-]}\;\;\stackrel{13\;14}{[-]}$ & 
      $ \hat{\nu}^{\dagger}_{R\,7}$&
      $ \stackrel{03}{[+i]}\,\stackrel{12}{[+]}|\stackrel{56}{(+)}\,\stackrel{78}{(+)}|| 
      \stackrel{9 \;10}{(+)}\;\;\stackrel{11\;12}{(+)}\;\;\stackrel{13\;14}{(+)}$ 
    &$0$&$\frac{1}{2}$&$0$&$-\frac{1}{2}$&$-\frac{1}{2}$
  \\
   $II$& $\hat{u}^{c1 \dagger}_{R\,8}$&
    $ \stackrel{03}{(+i)}\,\stackrel{12}{(+)}|\stackrel{56}{(+)}\,\stackrel{78}{(+)}||
    \stackrel{9 \;10}{(+)}\;\;\stackrel{11\;12}{[-]}\;\;\stackrel{13\;14}{[-]}$ & 
    $\hat{\nu}^{\dagger}_{R\,8}$&
    $ \stackrel{03}{(+i)}\,\stackrel{12}{(+)}|\stackrel{56}{(+)}\,\stackrel{78}{(+)}|| 
    \stackrel{9 \;10}{(+)}\;\;\stackrel{11\;12}{(+)}\;\;\stackrel{13\;14}{(+)}$ 
    &$0$&$\frac{1}{2}$&$0$&$\frac{1}{2}$&$-\frac{1}{3}$
 \\ 
 \hline 
 \end{tabular}
 \end{tiny}
 \end{center}
 \end{table}

It is the assumption that the eight families from Table~\ref{Table III.} remain massless  
after the break of symmetry from 
$SO(13,1)$ to $SO(7,1) \times SO(6)$,  
made after e  proved for the toy model~\cite{NHD,ND012} that the break of symmetry 
can leave  some  families of fermions massless, while the rest become massive. 
But we have not yet proven the masslessness  of the $2^{\frac{7+1}{2}-1}$ 
families after the break from $SO(13,1)$ to $SO(7,1) \times SO(6)$. 
The assumed break from the starting symmetry $SO(13,1)$ to 
$SO(7,1) \times SU(3)\times U(1)$is supposed to  be caused by the appearance 
of the condensate 
of two right handed neutrinos with the family quantum numbers of the upper four 
families, that is of the four  families, which do not contain the three so far observed 
families, at the energy of  $\ge 10^{16}$ GeV.
This condensate is presented in Table~\ref{Table con.}. 
 \begin{table}
 \begin{center}
\begin{minipage}[t]{16.5 cm}
\caption{The condensate of the two right handed neutrinos $\nu_{R}$,  with the quantum numbers
of the $VIII^{th}$ 
family, coupled to spin zero and belonging to a triplet with respect to the generators 
$\tau^{2i}$, is presented, together with its two partners. 
The condensate carries $\vec{\tau}^{1}=0$, $\tau^{23}=1$, 
$\tau^{4}=-1$ and $Q=0=Y$. The triplet carries $\tilde{\tau}^4=-1$, $\tilde{\tau}^{23}=1$
 and $\tilde{N}_{R}^3 = 1$, $\tilde{N}_{L}^3 = 0$,  
$\tilde{Y}=0 $, $\tilde{Q}=0$. 
The family quantum numbers of quarks and leptons are presented in
Table~\ref{Table III.}. }
\label{Table con.}
\end{minipage}
 \begin{tabular}{|c|c c c c c c c |c c c c c c c|}
 \hline
 state & $S^{03}$& $ S^{12}$ & $\tau^{13}$& $\tau^{23}$ &$\tau^{4}$& $Y$&$Q$&
$\tilde{\tau}^{13}$&
 $\tilde{\tau}^{23}$&$\tilde{\tau}^4$&$\tilde{Y} $& $\tilde{Q}$&$\tilde{N}_{L}^{3}$& 
$\tilde{N}_{R}^{3}$
 \\
 \hline
 ${\bf (|\nu_{1 R}^{VIII}>_{1}\,|\nu_{2 R}^{VIII}>_{2})}$
 & $0$& $0$& $0$& $1$& $-1$ & $0$& $0$& $0$ &$1$& $-1$& $0$& $0$& $0$& $1$\\ 
 \hline
 $ (|\nu_{1 R}^{VIII}>_{1}|e_{2 R}^{VIII}>_{2})$
 & $0$& $0$& $0$& $0$& $-1$ & $-1$& $-1$ & $0$ &$1$& $-1$& $0$& $0$& $0$& $1$\\ 
 $ (|e_{1 R}^{VIII}>_{1}|e_{2 R}^{VIII}>_{2})$
 & $0$& $0$& $0$& $-1$& $-1$ & $-2$& $-2$ & $0$ &$1$& $-1$& $0$& $0$& $0$& $1$\\ 
 \hline 
 \end{tabular}
 \end{center}
 \end{table}
%

 
In the case that the kinematics of quarks and leptons are determined by moments in 
$d=(3+1)$, the Weyl equation,  Eq.~(\ref{Weyl}), connects  the creation operators 
with spin up and down, that is the first two lines in Table~\ref{Table so13+1.}, when
we treat the right handed weak chargeless $u$-quark of the colour charge $(\frac{1}{2}, 
\frac{1}{2\sqrt{3}})$, with $\tau^4 =\frac{1}{6}$, $Y=\frac{2}{3}$ and $Q=\frac{2}{3}$. When we treat their  antiparticles,  then the $35^{th}$ and the
$36^{th}$ lines of Table~\ref{Table so13+1.} contribute to the solutions of the Weyl 
equation. They are the left handed weak chargeless $u$-antiquark of the colour charge 
$(-\frac{1}{2}, -\frac{1}{2\sqrt{3}})$, the ''fermion'' charge $\tau^4 =-\frac{1}{6}$, 
the hyper charge $Y=-\frac{2}{3}$ and the electromagnetic charge  $Q=-\frac{2}{3}$.
The solutions then look as follows 
%
\begin{small}
%
\begin{eqnarray}
\label{weylgen013+1}
&& {\rm Clifford\; odd \;creation \; operators\;} {\rm in \;  d=(13+1)}
\, \nonumber\\
p^0&=&|p^0|\,, \;\; c_{1}= (\frac{1}{2}, \frac{1}{2\sqrt{3}})\,, \;\;\Gamma^{(3+1)}=1\,,\nonumber\\
\Bigg( {\hat{\bf u}}^{ c1 s=1 \dagger}_{R f=1} (\vec{p}) \, 
&=&
\beta\, \left( \stackrel{03}{(+i)}\,\stackrel{12}{[+]}  + 
\frac{p^1 +i p^2}{ p^0 + p^3} \stackrel{03}{[-i]}\,\stackrel{12}{(-)}
\right)  \cdot \stackrel{56}{[+]}\,\stackrel{78}{(+)}
||\stackrel{9 \;10}{(+)}\;\;\stackrel{11\;12}{[-]}\;\;\stackrel{13\;14}{[-]}\Bigg)\cdot \,\nonumber\\
&& e^{-i(p^0 x^0 - \vec{p}\cdot\vec{x})}\,, \nonumber\\
\Bigg(  {\hat{\bf u}}^{ c1 s=2 \dagger}_{R f=1} (\vec{p})\, 
&=& \beta^*\, \left(\stackrel{03}{[-i]}\,
\stackrel{12}{(-)} - \frac{p^1 -i p^2}{ p^0 + p^3}\,
 \stackrel{03}{(+i)}\,\stackrel{12}{[+]} \right)  \cdot \stackrel{56}{[+]}\,\stackrel{78}{(+)}
||\stackrel{9 \;10}{(+)}\;\;\stackrel{11\;12}{[-]}\;\;\stackrel{13\;14}{[-]}\Bigg) \cdot\,\nonumber\\
&&e^{-i(p^0 x^0 + \vec{p}\cdot\vec{x})}\,,\nonumber\\
&& {\rm Clifford\; odd \;creation \; operators\;} {\rm in \;  d=(13+1)}
\, \nonumber\\
p^0&=&|p^0|\,, \;\; \bar{c}_{1}= (-\frac{1}{2}, -\frac{1}{2\sqrt{3}})\,,\;\;
\Gamma^{(3+1)}=-1\,,\nonumber\\
\Bigg( {\hat {\bar{\bf u}}}_{L f=1}^{\bar{c1} s=35 \dagger} (\vec{p}) \,
&=& - \beta \, \left( \stackrel{03}{[-i]}\,\stackrel{12}{[+]} +
 \frac{p^1 +i p^2}{ p^0 + p^3} \,\stackrel{03}{(+i)}\,\stackrel{12}{(-)}\right) 
 \cdot \stackrel{56}{(-)}\,\stackrel{78}{[-]}
||\stackrel{9 \;10}{[-]}\;\;\stackrel{11\;12}{(+)}\;\;\stackrel{13\;14}{(+)}\Bigg) \cdot
\,\nonumber\\
&&e^{-i(p^0 x^0 + \vec{p}\cdot\vec{x})}\,,\nonumber\\
\Bigg( {\hat{\bar{\bf u}}}_{L f=1}^{\bar{c1} s=36 \dagger} (\vec{p}) \,
&=& - \beta^{*} \,  \left(\stackrel{03}{(+i)}\, \stackrel{12}{(-)} - 
\frac{p^1 -i p^2}{ p^0 + p^3}\, \stackrel{03}{[-i]}\, \stackrel{12}{[+]} \right) 
\cdot \stackrel{56}{(-)}\,\stackrel{78}{[-]}
||\stackrel{9 \;10}{[-]}\;\;\stackrel{11\;12}{(+)}\;\;\stackrel{13\;14}{(+)}\Bigg) \cdot
\,\nonumber\\
&&e^{-i(p^0 x^0 - \vec{p}\cdot\vec{x})}\,.
\end{eqnarray}
\end{small}
Let us recognize that the discrete symmetry operator 
$\mathbb{C}_{{ \cal N}}{\cal P}^{(d-1)}_{{\cal N}}$, Eq.~(\ref{CPTNlowE}), which 
is in this case equal to $\gamma^0 \gamma^5 \gamma^7 \gamma^9 \gamma^{11}
 \gamma^{13} I_{\vec{x}_3}$, transforms $ {\hat{\bf u}}^{ c1 s=1 \dagger}_{R f=1}
 (\vec{p})$ into ${\hat {\bar{\bf u}}}_{L f=1}^{\bar{c1} s=35 \dagger} (\vec{p})$ and
$  {\hat{\bf u}}^{ c1 s=2 \dagger}_{R f=1} (\vec{p})$ into 
$ {\hat{\bar{\bf u}}}_{L f=1}^{\bar{c1} s=36 \dagger} (\vec{p})$.

We can  look for the solutions for leptons in the same way. The left handed neutrino,
for example,
with momentum $p^a=(p^0, \vec{p}, 0, 0, \dots,0)$, with $\vec{p}=(p^1, p^2, p^3)$,
would be the superposition of the two creation operators appearing in $31^{st}$ and
$32^{nd}$ lines of the Tabele~\ref{Table so13+1.},  with the same coefficients as 
presented in Eq.~(\ref{weylgen013+1}), but the handedness and  the weak charge 
would be  changed and its  colour  chargeless would be represented by 
$\stackrel{9 \;10}{(+)}\;\;\stackrel{11\;12}{(+)}\;\;\stackrel{13\;14}{(+)}$.
The righ handed antineutrino would be superposition of the lines $61$ and $62$.

To treat  the $d=(9+1)$-dimensional case the same table, Table~\ref{Table so13+1.},
can be used. The first eight lines and the lines from $33-40$ belong to one irreducible representation, sharing  the quantum numbers with respect to $SO(7,1)$ with 
quarks and leptons, and antiquarks and antileptons. This  $d=(9+1)$ case distinguishes 
from the $d=(13 +1)$ case  in the wealth of the colour charges: There is only one 
possibility for the ''colour'' charge. 
The solutions of the Weyl equation under the assumption that the momentum 
$p^a=(p^0, \vec{p}, 0, 0, \dots,0)$, with $\vec{p}=(p^1, p^2, p^3)$, are changed
with respect to Eq.~(\ref{weylgen013+1}) only in the colour part.

Table~\ref{Table so13+1.} represents in the {\it spin-charge-family} theory the  
''basis states'' (the creation operators) for internal space of the observed {\it quarks 
and leptons and antiquarks and antileptons} for the first of the eight families of
 Table~\ref{Table III.}. 
Hermitian conjugation of the  creation operators of Table~\ref{Table so13+1.} 
generates the corresponding annihilation operators, fulfilling together with the 
creation operators, if applying on the vacuum state $|\psi_{oc}>$, Eq.~(\ref{vac1}), 
the anticommutation relations of Eq.~(\ref{alphagammatildeprod}).

Creation and annihilation operators  for quarks and leptons
and antiquarks and antileptons are the tensor products of the finite ''basis vectors'' in internal space and the continuously infinite basis in ordinary space, as presented in 
Eq.~(\ref{Weylp0}). Since the ''basis vectors'' in internal space transfer their oddness
to creation and annihilation operators, the creation and annihilation operators for quarks
and leptons and antiquarks and antileptons obey on the vacuum state
$|\psi_{oc}> |0_{\vec{p}}>$ the anticommutation relations of 
Eq.~(\ref{Weylpp'comrel}), as discussed in Sect.~\ref{creationtensorClifford}. 

All the creation operators and their Hermitean conjugated partners, presented 
in Eqs.~(\ref{weylgen05+1}, \ref{weylgen013+1}), fulfill the 
anticommutation relations for the second quantized fermion fields of 
Eq.~(\ref{Weylpp'comrel}).
Tables~\ref{Table so13+1.} and~\ref{Table III.} represent $64$ ''basic states'' 
in the internal space of fermions, appearing in eight families. If applying on the 
vacuum state $|\psi_{oc}>$,
these creation operators describe the internal space of quarks and lepons and 
antiquarks and antileptons, which carry one of the
two handedness,  one of the two $S^{12}$ spins, one of the two weak 
charges $\tau^{13}$, one of the two $SU(2)_{II}$ charges $\tau^{23}$, 
either one of the three colour charges ($\tau^{33}$, $\tau^{38}$)  or one 
of the three colour anticharges or they carry the 
colourless or anticolourless quantum numbers of leptons. Quarks carry the 
''fermion'' quantum number $\tau^4=\frac{1}{6}$,  antiquarks carry the 
''fermion'' quantum number $-\frac{1}{6}$,
leptons carry  the ''fermion'' quantum number $-\frac{1}{2}$,  antileptons 
carry  the ''fermion'' quantum number $\frac{1}{2}$. 
The reader can clearly 
see in Table~\ref{Table so13+1.} the strong correlation among the ''fermion'' 
quantum number, charges and handedness. Each of the members carries in addition 
the family quantum numbers, the same for all the members of one irreducible 
representation of $S^{ab}$. The eight possibilities, presented in 
Table~\ref{Table III.}, are 
clustered into two groups. Each of the two groups of four families manifests
 its own $SU(2)\times SU(2)$ structure as seen in the last fifth columns of
Table~\ref{Table III.}. The first four families are doublets with respect to
$\tilde{\vec{\tau}}^1$  and $\tilde{\vec{N}}_{L}$ and singlets with respect
$\tilde{\vec{\tau}}^2$  and $\tilde{\vec{N}}_{R}$, the last four families 
are doublets with respect to $\tilde{\vec{\tau}}^2$  and $\tilde{\vec{N}}_{R}$
and singlets with respect to $\tilde{\vec{\tau}}^1$  and $\tilde{\vec{N}}_{L}$.
They all have the same $\tilde{\tau}^{4}$.

One can take subgroups $SO(3,1)$, $SO(4)$, $SU(3)$  and $U(1)$ of $SO(13,1)$ 
and add also the family groups. But the correlation among spin, handedness, 
charges and families for fermions and antifermions must in such a case be 
postulated and also the  
symmetry of families must be chosen. In this case oddness or evenness of 
creation operators must be postulated as well, in the same way as Dirac did, 
Eqs.~(\ref{stateDiracanti}, \ref{stateDirac}).

All the creation operators, creating quarks and leptons and antiquarks and antileptons,  
have an odd Clifford character. Together with their Hermitian conjugated partners
they take care of the anticommutation properties of creation operators,
determining the states in the internal and momentum (or coordinate) space, as
presented in Eqs.~(\ref{weylgen05+1}, \ref{weylgen013+1}).

Let us at the end point out again that the description of the internal space of 
fermions in $d=2(2n+1)$-dimensional space   
(the one describing spin and handedness in $d=(3+1)$ and the one representing
in $d=(3+1)$  the fermion charges,  originating in $d\ge 5$) with odd Clifford 
algebra objects makes handedness related to charges, takes care of the families,
as well as of the anticommuting properties of fermions. 
In the {\it standard model}, as well as   in the unified theories, the relations among
spins, handedness, charges and families of fermions and
antifermions must be postulated, as well as the anticommutation relations
among creation and annihilation operators, the last one in the way Dirac 
did~\footnote{In App.~\ref{evenclifford} the properties of even Clifford algebra
objects are  discussed, as well as the relation in this case between creation 
operators and their
Hermitian conjugated partners.}.


\tablecaption[16.5cm]{\label{Table so13+1.}%
\small{
The left handed ($\Gamma^{(13,1)} = -1$),
multiplet of  creation operators of fermions --- the members of the first family
in Table~\ref{Table III.} (each family represents one fundamental representation of the 
$SO(13,1)$ group), manifesting the subgroup $SO(7,1)$
 of the colour charged quarks and antiquarks and the colourless
leptons and antileptons --- is presented in the massless basis as odd products of nilpotents and the rest of projectors (together are $\frac{d}{2}=7$ nilpotents and projectors). 
The multiplet contains the left handed  ($\Gamma^{(3+1)}=-1$
weak ($SU(2)_{I}$) charged ($\tau^{13}=\pm \frac{1}{2}$, 
($\vec{\tau}^{1}= \frac{1}{2} (S^{58}- S^{67}, S^{57}+ S^{68}, S^{56}- S^{78})$)
and $SU(2)_{II}$ chargeless ($\tau^{23}=0$, 
$\vec{\tau}^{2}= \frac{1}{2} (S^{58}+ S^{67}, S^{57}- S^{68}, S^{56}+ S^{78})$)
quarks and leptons and the right handed  ($\Gamma^{(3+1)}=1$), 
 weak  ($SU(2)_{I}$) chargeless and $SU(2)_{II}$
charged ($\tau^{23}=\pm \frac{1}{2}$) quarks and leptons, both with the spin 
$ S^{12}$  up and down ($\pm \frac{1}{2}$, respectively). 
The creation operators of quarks distinguish from those of leptons only in the 
$SU(3) \times U(1)$ part: Quarks are triplets of three colours  ( $ (\tau^{33}, \tau^{38})$ 
$ = [(\frac{1}{2},\frac{1}{2\sqrt{3}}),
(-\frac{1}{2},\frac{1}{2\sqrt{3}}), (0,-\frac{1}{\sqrt{3}}) $], 
($\vec{\tau}^{3}= \frac{1}{2}(S^{9\,12}- S^{10\,11},S^{9\,11}+ S^{10\,12},S^{9\,10}-
S^{11\,12},$ $S^{9\,14}- S^{10\,13},S^{9\,13}+ S^{10\,14},S^{11\,14}- S^{12\,13},$
$S^{11\,13}+ S^{12\,14},\frac{1}{\sqrt{3}}(S^{9\,10}+ S^{11\,12} - 2S^{13\,14})$),
carrying  the "fermion charge" ($\tau^{4}=\frac{1}{6}$, 
$=-\frac{1}{3}(S^{9\,10}+ S^{11\,12}+ S^{13\,14})$. 
The colourless leptons carry the "fermion charge" ($\tau^{4}=-\frac{1}{2}$).
To the same multiplet of creation operators the left handed weak 
($SU(2)_{I}$)  chargeless and $SU(2)_{II}$ charged antiquarks and antileptons and the right handed weak ($SU(2)_{I}$) charged and $SU(2)_{II}$ chargeless antiquarks and 
antileptons belong.
Antiquarks distinguish from antileptons again only in the $SU(3) \times U(1)$ part: Anti-quarks are antitriplets, 
 carrying  the "fermion charge" ($\tau^{4}=-\frac{1}{6}$).
The anti-colourless antileptons carry the "fermion" charge ($\tau^{4}=\frac{1}{2}$).
 $Y=(\tau^{23} + \tau^{4})$ is the hyper charge, the electromagnetic charge
is $Q=(\tau^{13} + Y$).
The creation operators of opposite charges (antifermion creation operators) are 
reachable  from the fermion ones besides by $S^{ab}$  also by the application of the discrete symmetry operator
${\bf \mathbb{C}}_{{\cal N}}$ ${\cal P}_{{\cal N}}$, presented in Refs.~\cite{nhds,TDN}.
%
The reader can find this  Weyl representation also in
Refs.~\cite{n2014matterantimatter,pikan2003,pikan2006,normaJMP2015} and in the references therein. }
}
\tablehead{\hline
i&$$&$^a\hat{b}^{\dagger}_i $&$\Gamma^{(3+1)}$&$ S^{12}$&
$\tau^{13}$&$\tau^{23}$&$\tau^{33}$&$\tau^{38}$&$\tau^{4}$&$Y$&$Q$\\
\hline
&& ${\rm (Anti)octet},\,\Gamma^{(7+1)} = (-1)\,1\,, \,\Gamma^{(6)} = (1)\,-1$&&&&&&&&& \\
&& ${\rm of \;(anti) quarks \;and \;(anti)leptons}$&&&&&&&&&\\
\hline\hline}
\tabletail{\hline \multicolumn{12}{r}{\emph{Continued on next page}}\\}
\tablelasttail{\hline}
\begin{center}
\tiny{
\begin{supertabular}{|r|c||c||c|c||c|c||c|c|c||r|r|}
1&$ \hat{u}_{R}^{c1\dagger}$&$ \stackrel{03}{(+i)}\,\stackrel{12}{[+]}|
\stackrel{56}{[+]}\,\stackrel{78}{(+)}
||\stackrel{9 \;10}{(+)}\;\;\stackrel{11\;12}{[-]}\;\;\stackrel{13\;14}{[-]} $ &1&$\frac{1}{2}$&0&
$\frac{1}{2}$&$\frac{1}{2}$&$\frac{1}{2\,\sqrt{3}}$&$\frac{1}{6}$&$\frac{2}{3}$&$\frac{2}{3}$\\
\hline
2&$\hat{u}_{R}^{c1 \dagger}$&$\stackrel{03}{[-i]}\,\stackrel{12}{(-)}|\stackrel{56}{[+]}\,\stackrel{78}{(+)}
||\stackrel{9 \;10}{(+)}\;\;\stackrel{11\;12}{[-]}\;\;\stackrel{13\;14}{[-]}$&1&$-\frac{1}{2}$&0&
$\frac{1}{2}$&$\frac{1}{2}$&$\frac{1}{2\,\sqrt{3}}$&$\frac{1}{6}$&$\frac{2}{3}$&$\frac{2}{3}$\\
\hline
3&$\hat{d}_{R}^{c1 \dagger}$&$\stackrel{03}{(+i)}\,\stackrel{12}{[+]}|\stackrel{56}{(-)}\,\stackrel{78}{[-]}
||\stackrel{9 \;10}{(+)}\;\;\stackrel{11\;12}{[-]}\;\;\stackrel{13\;14}{[-]}$&1&$\frac{1}{2}$&0&
$-\frac{1}{2}$&$\frac{1}{2}$&$\frac{1}{2\,\sqrt{3}}$&$\frac{1}{6}$&$-\frac{1}{3}$&$-\frac{1}{3}$\\
\hline
4&$\hat{d}_{R}^{c1 \dagger} $&$\stackrel{03}{[-i]}\,\stackrel{12}{(-)}|
\stackrel{56}{(-)}\,\stackrel{78}{[-]}
||\stackrel{9 \;10}{(+)}\;\;\stackrel{11\;12}{[-]}\;\;\stackrel{13\;14}{[-]} $&1&$-\frac{1}{2}$&0&
$-\frac{1}{2}$&$\frac{1}{2}$&$\frac{1}{2\,\sqrt{3}}$&$\frac{1}{6}$&$-\frac{1}{3}$&$-\frac{1}{3}$\\
\hline
5&$\hat{d}_{L}^{c1\dagger}$&$\stackrel{03}{[-i]}\,\stackrel{12}{[+]}|\stackrel{56}{(-)}\,\stackrel{78}{(+)}
||\stackrel{9 \;10}{(+)}\;\;\stackrel{11\;12}{[-]}\;\;\stackrel{13\;14}{[-]}$&-1&$\frac{1}{2}$&
$-\frac{1}{2}$&0&$\frac{1}{2}$&$\frac{1}{2\,\sqrt{3}}$&$\frac{1}{6}$&$\frac{1}{6}$&$-\frac{1}{3}$\\
\hline
6&$\hat{d}_{L}^{c1 \dagger} $&$  \stackrel{03}{(+i)}\,\stackrel{12}{(-)}|\stackrel{56}{(-)}\,\stackrel{78}{(+)}
||\stackrel{9 \;10}{(+)}\;\;\stackrel{11\;12}{[-]}\;\;\stackrel{13\;14}{[-]} $&-1&$-\frac{1}{2}$&
$-\frac{1}{2}$&0&$\frac{1}{2}$&$\frac{1}{2\,\sqrt{3}}$&$\frac{1}{6}$&$\frac{1}{6}$&$-\frac{1}{3}$\\
\hline
7&$ \hat{u}_{L}^{c1\dagger}$&$  \stackrel{03}{[-i]}\,\stackrel{12}{[+]}|\stackrel{56}{[+]}\,\stackrel{78}{[-]}
||\stackrel{9 \;10}{(+)}\;\;\stackrel{11\;12}{[-]}\;\;\stackrel{13\;14}{[-]}$ &-1&$\frac{1}{2}$&
$\frac{1}{2}$&0 &$\frac{1}{2}$&$\frac{1}{2\,\sqrt{3}}$&$\frac{1}{6}$&$\frac{1}{6}$&$\frac{2}{3}$\\
\hline
8&$\hat{u}_{L}^{c1 \dagger}$&$\stackrel{03}{(+i)}\,\stackrel{12}{(-)}|\stackrel{56}{[+]}\,\stackrel{78}{[-]}
||\stackrel{9 \;10}{(+)}\;\;\stackrel{11\;12}{[-]}\;\;\stackrel{13\;14}{[-]}$&-1&$-\frac{1}{2}$&
$\frac{1}{2}$&0&$\frac{1}{2}$&$\frac{1}{2\,\sqrt{3}}$&$\frac{1}{6}$&$\frac{1}{6}$&$\frac{2}{3}$\\
\hline\hline
\shrinkheight{0.25\textheight}
9&$\hat{u}_{R}^{c2 \dagger}$&$ \stackrel{03}{(+i)}\,\stackrel{12}{[+]}|
\stackrel{56}{[+]}\,\stackrel{78}{(+)}
||\stackrel{9 \;10}{[-]}\;\;\stackrel{11\;12}{(+)}\;\;\stackrel{13\;14}{[-]} $ &1&$\frac{1}{2}$&0&
$\frac{1}{2}$&$-\frac{1}{2}$&$\frac{1}{2\,\sqrt{3}}$&$\frac{1}{6}$&$\frac{2}{3}$&$\frac{2}{3}$\\
\hline
10&$\hat{u}_{R}^{c2 \dagger}$&$\stackrel{03}{[-i]}\,\stackrel{12}{(-)}|\stackrel{56}{[+]}\,\stackrel{78}{(+)}
||\stackrel{9 \;10}{[-]}\;\;\stackrel{11\;12}{(+)}\;\;\stackrel{13\;14}{[-]}$&1&$-\frac{1}{2}$&0&
$\frac{1}{2}$&$-\frac{1}{2}$&$\frac{1}{2\,\sqrt{3}}$&$\frac{1}{6}$&$\frac{2}{3}$&$\frac{2}{3}$\\
\hline
11&$\hat{d}_{R}^{c2 \dagger}$&$\stackrel{03}{(+i)}\,\stackrel{12}{[+]}|\stackrel{56}{(-)}\,\stackrel{78}{[-]}
||\stackrel{9 \;10}{[-]}\;\;\stackrel{11\;12}{(+)}\;\;\stackrel{13\;14}{[-]}$
&1&$\frac{1}{2}$&0&
$-\frac{1}{2}$&$ - \frac{1}{2}$&$\frac{1}{2\,\sqrt{3}}$&$\frac{1}{6}$&$-\frac{1}{3}$&$-\frac{1}{3}$\\
\hline
12&$ \hat{d}_{R}^{c2 \dagger} $&$\stackrel{03}{[-i]}\,\stackrel{12}{(-)}|
\stackrel{56}{(-)}\,\stackrel{78}{[-]}
||\stackrel{9 \;10}{[-]}\;\;\stackrel{11\;12}{(+)}\;\;\stackrel{13\;14}{[-]} $
&1&$-\frac{1}{2}$&0&
$-\frac{1}{2}$&$-\frac{1}{2}$&$\frac{1}{2\,\sqrt{3}}$&$\frac{1}{6}$&$-\frac{1}{3}$&$-\frac{1}{3}$\\
\hline
13&$\hat{d}_{L}^{c2 \dagger}$&$\stackrel{03}{[-i]}\,\stackrel{12}{[+]}|\stackrel{56}{(-)}\,\stackrel{78}{(+)}
||\stackrel{9 \;10}{[-]}\;\;\stackrel{11\;12}{(+)}\;\;\stackrel{13\;14}{[-]}$
&-1&$\frac{1}{2}$&
$-\frac{1}{2}$&0&$-\frac{1}{2}$&$\frac{1}{2\,\sqrt{3}}$&$\frac{1}{6}$&$\frac{1}{6}$&$-\frac{1}{3}$\\
\hline
14&$\hat{d}_{L}^{c2 \dagger} $&$  \stackrel{03}{(+i)}\,\stackrel{12}{(-)}|\stackrel{56}{(-)}\,\stackrel{78}{(+)}
||\stackrel{9 \;10}{[-]}\;\;\stackrel{11\;12}{(+)}\;\;\stackrel{13\;14}{[-]} $&-1&$-\frac{1}{2}$&
$-\frac{1}{2}$&0&$-\frac{1}{2}$&$\frac{1}{2\,\sqrt{3}}$&$\frac{1}{6}$&$\frac{1}{6}$&$-\frac{1}{3}$\\
\hline
15&$ \hat{u}_{L}^{c2 \dagger}$&$  \stackrel{03}{[-i]}\,\stackrel{12}{[+]}|\stackrel{56}{[+]}\,\stackrel{78}{[-]}
||\stackrel{9 \;10}{[-]}\;\;\stackrel{11\;12}{(+)}\;\;\stackrel{13\;14}{[-]}$ &-1&$\frac{1}{2}$&
$\frac{1}{2}$&0 &$-\frac{1}{2}$&$\frac{1}{2\,\sqrt{3}}$&$\frac{1}{6}$&$\frac{1}{6}$&$\frac{2}{3}$\\
\hline
16&$\hat{u}_{L}^{c2 \dagger}$&$\stackrel{03}{(+i)}\,\stackrel{12}{(-)}|\stackrel{56}{[+]}\,\stackrel{78}{[-]}
||\stackrel{9 \;10}{[-]}\;\;\stackrel{11\;12}{(+)}\;\;\stackrel{13\;14}{[-]}$&-1&$-\frac{1}{2}$&
$\frac{1}{2}$&0&$-\frac{1}{2}$&$\frac{1}{2\,\sqrt{3}}$&$\frac{1}{6}$&$\frac{1}{6}$&$\frac{2}{3}$\\
\hline\hline
17&$ \hat{u}_{R}^{c3 \dagger}$&$ \stackrel{03}{(+i)}\,\stackrel{12}{[+]}|
\stackrel{56}{[+]}\,\stackrel{78}{(+)}
||\stackrel{9 \;10}{[-]}\;\;\stackrel{11\;12}{[-]}\;\;\stackrel{13\;14}{(+)} $ &1&$\frac{1}{2}$&0&
$\frac{1}{2}$&$0$&$-\frac{1}{\sqrt{3}}$&$\frac{1}{6}$&$\frac{2}{3}$&$\frac{2}{3}$\\
\hline
18&$\hat{u}_{R}^{c3 \dagger}$&$\stackrel{03}{[-i]}\,\stackrel{12}{(-)}|\stackrel{56}{[+]}\,\stackrel{78}{(+)}
||\stackrel{9 \;10}{[-]}\;\;\stackrel{11\;12}{[-]}\;\;\stackrel{13\;14}{(+)}$&1&$-\frac{1}{2}$&0&
$\frac{1}{2}$&$0$&$-\frac{1}{\sqrt{3}}$&$\frac{1}{6}$&$\frac{2}{3}$&$\frac{2}{3}$\\
\hline
19&$\hat{d}_{R}^{c3 \dagger}$&$\stackrel{03}{(+i)}\,\stackrel{12}{[+]}|\stackrel{56}{(-)}\,\stackrel{78}{[-]}
||\stackrel{9 \;10}{[-]}\;\;\stackrel{11\;12}{[-]}\;\;\stackrel{13\;14}{(+)}$&1&$\frac{1}{2}$&0&
$-\frac{1}{2}$&$0$&$-\frac{1}{\sqrt{3}}$&$\frac{1}{6}$&$-\frac{1}{3}$&$-\frac{1}{3}$\\
\hline
20&$\hat{d}_{R}^{c3 \dagger} $&$\stackrel{03}{[-i]}\,\stackrel{12}{(-)}|
\stackrel{56}{(-)}\,\stackrel{78}{[-]}
||\stackrel{9 \;10}{[-]}\;\;\stackrel{11\;12}{[-]}\;\;\stackrel{13\;14}{(+)} $&1&$-\frac{1}{2}$&0&
$-\frac{1}{2}$&$0$&$-\frac{1}{\sqrt{3}}$&$\frac{1}{6}$&$-\frac{1}{3}$&$-\frac{1}{3}$\\
\hline
21&$\hat{d}_{L}^{c3 \dagger}$&$\stackrel{03}{[-i]}\,\stackrel{12}{[+]}|\stackrel{56}{(-)}\,\stackrel{78}{(+)}
||\stackrel{9 \;10}{[-]}\;\;\stackrel{11\;12}{[-]}\;\;\stackrel{13\;14}{(+)}$&-1&$\frac{1}{2}$&
$-\frac{1}{2}$&0&$0$&$-\frac{1}{\sqrt{3}}$&$\frac{1}{6}$&$\frac{1}{6}$&$-\frac{1}{3}$\\
\hline
22&$\hat{d}_{L}^{c3 \dagger} $&$  \stackrel{03}{(+i)}\,\stackrel{12}{(-)}|\stackrel{56}{(-)}\,\stackrel{78}{(+)}
||\stackrel{9 \;10}{[-]}\;\;\stackrel{11\;12}{[-]}\;\;\stackrel{13\;14}{(+)} $&-1&$-\frac{1}{2}$&
$-\frac{1}{2}$&0&$0$&$-\frac{1}{\sqrt{3}}$&$\frac{1}{6}$&$\frac{1}{6}$&$-\frac{1}{3}$\\
\hline
23&$ \hat{u}_{L}^{c3 \dagger}$&$  \stackrel{03}{[-i]}\,\stackrel{12}{[+]}|\stackrel{56}{[+]}\,\stackrel{78}{[-]}
||\stackrel{9 \;10}{[-]}\;\;\stackrel{11\;12}{[-]}\;\;\stackrel{13\;14}{(+)}$ &-1&$\frac{1}{2}$&
$\frac{1}{2}$&0 &$0$&$-\frac{1}{\sqrt{3}}$&$\frac{1}{6}$&$\frac{1}{6}$&$\frac{2}{3}$\\
\hline
24&$\hat{u}_{L}^{c3 \dagger}$&$\stackrel{03}{(+i)}\,\stackrel{12}{(-)}|\stackrel{56}{[+]}\,\stackrel{78}{[-]}
||\stackrel{9 \;10}{[-]}\;\;\stackrel{11\;12}{[-]}\;\;\stackrel{13\;14}{(+)}$&-1&$-\frac{1}{2}$&
$\frac{1}{2}$&0&$0$&$-\frac{1}{\sqrt{3}}$&$\frac{1}{6}$&$\frac{1}{6}$&$\frac{2}{3}$\\
\hline\hline
25&$ \hat{\nu}^{ \dagger}_{R}$&$ \stackrel{03}{(+i)}\,\stackrel{12}{[+]}|
\stackrel{56}{[+]}\,\stackrel{78}{(+)}
||\stackrel{9 \;10}{(+)}\;\;\stackrel{11\;12}{(+)}\;\;\stackrel{13\;14}{(+)} $ &1&$\frac{1}{2}$&0&
$\frac{1}{2}$&$0$&$0$&$-\frac{1}{2}$&$0$&$0$\\
\hline
26&$\hat{\nu}^{ \dagger}_{R}$&$\stackrel{03}{[-i]}\,\stackrel{12}{(-)}|\stackrel{56}{[+]}\,\stackrel{78}{(+)}
||\stackrel{9 \;10}{(+)}\;\;\stackrel{11\;12}{(+)}\;\;\stackrel{13\;14}{(+)}$&1&$-\frac{1}{2}$&0&
$\frac{1}{2}$ &$0$&$0$&$-\frac{1}{2}$&$0$&$0$\\
\hline
27&$\hat{e}^{ \dagger}_{R}$&$\stackrel{03}{(+i)}\,\stackrel{12}{[+]}|\stackrel{56}{(-)}\,\stackrel{78}{[-]}
||\stackrel{9 \;10}{(+)}\;\;\stackrel{11\;12}{(+)}\;\;\stackrel{13\;14}{(+)}$&1&$\frac{1}{2}$&0&
$-\frac{1}{2}$&$0$&$0$&$-\frac{1}{2}$&$-1$&$-1$\\
\hline
28&$\hat{e}^{ \dagger}_{R} $&$\stackrel{03}{[-i]}\,\stackrel{12}{(-)}|
\stackrel{56}{(-)}\,\stackrel{78}{[-]}
||\stackrel{9 \;10}{(+)}\;\;\stackrel{11\;12}{(+)}\;\;\stackrel{13\;14}{(+)} $&1&$-\frac{1}{2}$&0&
$-\frac{1}{2}$&$0$&$0$&$-\frac{1}{2}$&$-1$&$-1$\\
\hline
29&$\hat{e}^{ \dagger}_{L}$&$\stackrel{03}{[-i]}\,\stackrel{12}{[+]}|\stackrel{56}{(-)}\,\stackrel{78}{(+)}
||\stackrel{9 \;10}{(+)}\;\;\stackrel{11\;12}{(+)}\;\;\stackrel{13\;14}{(+)}$&-1&$\frac{1}{2}$&
$-\frac{1}{2}$&0&$0$&$0$&$-\frac{1}{2}$&$-\frac{1}{2}$&$-1$\\
\hline
30&$\hat{e}^{ \dagger}_{L} $&$  \stackrel{03}{(+i)}\,\stackrel{12}{(-)}|\stackrel{56}{(-)}\,\stackrel{78}{(+)}
||\stackrel{9 \;10}{(+)}\;\;\stackrel{11\;12}{(+)}\;\;\stackrel{13\;14}{(+)} $&-1&$-\frac{1}{2}$&
$-\frac{1}{2}$&0&$0$&$0$&$-\frac{1}{2}$&$-\frac{1}{2}$&$-1$\\
\hline
31&$ \hat{\nu}^{ \dagger}_{L}$&$  \stackrel{03}{[-i]}\,\stackrel{12}{[+]}|\stackrel{56}{[+]}\,\stackrel{78}{[-]}
||\stackrel{9 \;10}{(+)}\;\;\stackrel{11\;12}{(+)}\;\;\stackrel{13\;14}{(+)}$ &-1&$\frac{1}{2}$&
$\frac{1}{2}$&0 &$0$&$0$&$-\frac{1}{2}$&$-\frac{1}{2}$&$0$\\
\hline
32&$\hat{\nu}^{ \dagger}_{L}$&$\stackrel{03}{(+i)}\,\stackrel{12}{(-)}|\stackrel{56}{[+]}\,\stackrel{78}{[-]}
||\stackrel{9 \;10}{(+)}\;\;\stackrel{11\;12}{(+)}\;\;\stackrel{13\;14}{(+)}$&-1&$-\frac{1}{2}$&
$\frac{1}{2}$&0&$0$&$0$&$-\frac{1}{2}$&$-\frac{1}{2}$&$0$\\
\hline\hline
33&$\hat{\bar{d}}_{L}^{\bar{c1} \dagger}$&$ \stackrel{03}{[-i]}\,\stackrel{12}{[+]}|
\stackrel{56}{[+]}\,\stackrel{78}{(+)}
||\stackrel{9 \;10}{[-]}\;\;\stackrel{11\;12}{(+)}\;\;\stackrel{13\;14}{(+)} $ &-1&$\frac{1}{2}$&0&
$\frac{1}{2}$&$-\frac{1}{2}$&$-\frac{1}{2\,\sqrt{3}}$&$-\frac{1}{6}$&$\frac{1}{3}$&$\frac{1}{3}$\\
\hline
34&$\hat{\bar{d}}_{L}^{\bar{c1} \dagger}$&$\stackrel{03}{(+i)}\,\stackrel{12}{(-)}|\stackrel{56}{[+]}\,\stackrel{78}{(+)}
||\stackrel{9 \;10}{[-]}\;\;\stackrel{11\;12}{(+)}\;\;\stackrel{13\;14}{(+)}$&-1&$-\frac{1}{2}$&0&
$\frac{1}{2}$&$-\frac{1}{2}$&$-\frac{1}{2\,\sqrt{3}}$&$-\frac{1}{6}$&$\frac{1}{3}$&$\frac{1}{3}$\\
\hline
35&$\bar{u}_{L}^{\bar{c1} \dagger}$&$  \stackrel{03}{[-i]}\,\stackrel{12}{[+]}|\stackrel{56}{(-)}\,\stackrel{78}{[-]}
||\stackrel{9 \;10}{[-]}\;\;\stackrel{11\;12}{(+)}\;\;\stackrel{13\;14}{(+)}$&-1&$\frac{1}{2}$&0&
$-\frac{1}{2}$&$-\frac{1}{2}$&$-\frac{1}{2\,\sqrt{3}}$&$-\frac{1}{6}$&$-\frac{2}{3}$&$-\frac{2}{3}$\\
\hline
36&$ \bar{u}_{L}^{\bar{c1} \dagger} $&$  \stackrel{03}{(+i)}\,\stackrel{12}{(-)}|
\stackrel{56}{(-)}\,\stackrel{78}{[-]}
||\stackrel{9 \;10}{[-]}\;\;\stackrel{11\;12}{(+)}\;\;\stackrel{13\;14}{(+)} $&-1&$-\frac{1}{2}$&0&
$-\frac{1}{2}$&$-\frac{1}{2}$&$-\frac{1}{2\,\sqrt{3}}$&$-\frac{1}{6}$&$-\frac{2}{3}$&$-\frac{2}{3}$\\
\hline
37&$\hat{\bar{d}}_{R}^{\bar{c1} \dagger}$&$\stackrel{03}{(+i)}\,\stackrel{12}{[+]}|\stackrel{56}{[+]}\,\stackrel{78}{[-]}
||\stackrel{9 \;10}{[-]}\;\;\stackrel{11\;12}{(+)}\;\;\stackrel{13\;14}{(+)}$&1&$\frac{1}{2}$&
$\frac{1}{2}$&0&$-\frac{1}{2}$&$-\frac{1}{2\,\sqrt{3}}$&$-\frac{1}{6}$&$-\frac{1}{6}$&$\frac{1}{3}$\\
\hline
38&$\hat{\bar{d}}_{R}^{\bar{c1} \dagger} $&$  \stackrel{03}{[-i]}\,\stackrel{12}{(-)}|\stackrel{56}{[+]}\,\stackrel{78}{[-]}
||\stackrel{9 \;10}{[-]}\;\;\stackrel{11\;12}{(+)}\;\;\stackrel{13\;14}{(+)} $&1&$-\frac{1}{2}$&
$\frac{1}{2}$&0&$-\frac{1}{2}$&$-\frac{1}{2\,\sqrt{3}}$&$-\frac{1}{6}$&$-\frac{1}{6}$&$\frac{1}{3}$\\
\hline
39&$\hat{\bar{u}}_{R}^{\bar{c1} \dagger}$&$\stackrel{03}{(+i)}\,\stackrel{12}{[+]}|\stackrel{56}{(-)}\,\stackrel{78}{(+)}
||\stackrel{9 \;10}{[-]}\;\;\stackrel{11\;12}{(+)}\;\;\stackrel{13\;14}{(+)}$ &1&$\frac{1}{2}$&
$-\frac{1}{2}$&0 &$-\frac{1}{2}$&$-\frac{1}{2\,\sqrt{3}}$&$-\frac{1}{6}$&$-\frac{1}{6}$&$-\frac{2}{3}$\\
\hline
40&$\hat{\bar{u}}_{R}^{\bar{c1} \dagger}$&$\stackrel{03}{[-i]}\,\stackrel{12}{(-)}|\stackrel{56}{(-)}\,\stackrel{78}{(+)}
||\stackrel{9 \;10}{[-]}\;\;\stackrel{11\;12}{(+)}\;\;\stackrel{13\;14}{(+)}$
&1&$-\frac{1}{2}$&
$-\frac{1}{2}$&0&$-\frac{1}{2}$&$-\frac{1}{2\,\sqrt{3}}$&$-\frac{1}{6}$&$-\frac{1}{6}$&$-\frac{2}{3}$\\
\hline\hline
41&$ \hat{\bar{d}}_{L}^{\bar{c2} \dagger}$&$ \stackrel{03}{[-i]}\,\stackrel{12}{[+]}|
\stackrel{56}{[+]}\,\stackrel{78}{(+)}
||\stackrel{9 \;10}{(+)}\;\;\stackrel{11\;12}{[-]}\;\;\stackrel{13\;14}{(+)} $
&-1&$\frac{1}{2}$&0&
$\frac{1}{2}$&$\frac{1}{2}$&$-\frac{1}{2\,\sqrt{3}}$&$-\frac{1}{6}$&$\frac{1}{3}$&$\frac{1}{3}$\\
\hline
42&$\hat{\bar{d}}_{L}^{\bar{c2} \dagger}$&$\stackrel{03}{(+i)}\,\stackrel{12}{(-)}|\stackrel{56}{[+]}\,\stackrel{78}{(+)}
||\stackrel{9 \;10}{(+)}\;\;\stackrel{11\;12}{[-]}\;\;\stackrel{13\;14}{(+)}$
&-1&$-\frac{1}{2}$&0&
$\frac{1}{2}$&$\frac{1}{2}$&$-\frac{1}{2\,\sqrt{3}}$&$-\frac{1}{6}$&$\frac{1}{3}$&$\frac{1}{3}$\\
\hline
43&$\hat{\bar{u}}_{L}^{\bar{c2} \dagger}$&$  \stackrel{03}{[-i]}\,\stackrel{12}{[+]}|\stackrel{56}{(-)}\,\stackrel{78}{[-]}
||\stackrel{9 \;10}{(+)}\;\;\stackrel{11\;12}{[-]}\;\;\stackrel{13\;14}{(+)}$
&-1&$\frac{1}{2}$&0&
$-\frac{1}{2}$&$\frac{1}{2}$&$-\frac{1}{2\,\sqrt{3}}$&$-\frac{1}{6}$&$-\frac{2}{3}$&$-\frac{2}{3}$\\
\hline
44&$ \hat{\bar{u}}_{L}^{\bar{c2} \dagger} $&$  \stackrel{03}{(+i)}\,\stackrel{12}{(-)}|
\stackrel{56}{(-)}\,\stackrel{78}{[-]}
||\stackrel{9 \;10}{(+)}\;\;\stackrel{11\;12}{[-]}\;\;\stackrel{13\;14}{(+)} $
&-1&$-\frac{1}{2}$&0&
$-\frac{1}{2}$&$\frac{1}{2}$&$-\frac{1}{2\,\sqrt{3}}$&$-\frac{1}{6}$&$-\frac{2}{3}$&$-\frac{2}{3}$\\
\hline
45&$\hat{\bar{d}}_{R}^{\bar{c2} \dagger}$&$\stackrel{03}{(+i)}\,\stackrel{12}{[+]}|\stackrel{56}{[+]}\,\stackrel{78}{[-]}
||\stackrel{9 \;10}{(+)}\;\;\stackrel{11\;12}{[-]}\;\;\stackrel{13\;14}{(+)}$
&1&$\frac{1}{2}$&
$\frac{1}{2}$&0&$\frac{1}{2}$&$-\frac{1}{2\,\sqrt{3}}$&$-\frac{1}{6}$&$-\frac{1}{6}$&$\frac{1}{3}$\\
\hline
46&$\hat{\bar{d}}_{R}^{\bar{c2} \dagger} $&$  \stackrel{03}{[-i]}\,\stackrel{12}{(-)}|\stackrel{56}{[+]}\,\stackrel{78}{[-]}
||\stackrel{9 \;10}{(+)}\;\;\stackrel{11\;12}{[-]}\;\;\stackrel{13\;14}{(+)} $
&1&$-\frac{1}{2}$&
$\frac{1}{2}$&0&$\frac{1}{2}$&$-\frac{1}{2\,\sqrt{3}}$&$-\frac{1}{6}$&$-\frac{1}{6}$&$\frac{1}{3}$\\
\hline
47&$ \hat{\bar{u}}_{R}^{\bar{c2} \dagger}$&$\stackrel{03}{(+i)}\,\stackrel{12}{[+]}|\stackrel{56}{(-)}\,\stackrel{78}{(+)}
||\stackrel{9 \;10}{(+)}\;\;\stackrel{11\;12}{[-]}\;\;\stackrel{13\;14}{(+)}$
 &1&$\frac{1}{2}$&
$-\frac{1}{2}$&0 &$\frac{1}{2}$&$-\frac{1}{2\,\sqrt{3}}$&$-\frac{1}{6}$&$-\frac{1}{6}$&$-\frac{2}{3}$\\
\hline
48&$\hat{\bar{u}}_{R}^{\bar{c2} \dagger}$&$\stackrel{03}{[-i]}\,\stackrel{12}{(-)}|\stackrel{56}{(-)}\,\stackrel{78}{(+)}
||\stackrel{9 \;10}{(+)}\;\;\stackrel{11\;12}{[-]}\;\;\stackrel{13\;14}{(+)}$
&1&$-\frac{1}{2}$&
$-\frac{1}{2}$&0&$\frac{1}{2}$&$-\frac{1}{2\,\sqrt{3}}$&$-\frac{1}{6}$&$-\frac{1}{6}$&$-\frac{2}{3}$\\
\hline\hline
49&$ \hat{\bar{d}}_{L}^{\bar{c3} \dagger}$&$ \stackrel{03}{[-i]}\,\stackrel{12}{[+]}|
\stackrel{56}{[+]}\,\stackrel{78}{(+)}
||\stackrel{9 \;10}{(+)}\;\;\stackrel{11\;12}{(+)}\;\;\stackrel{13\;14}{[-]} $ &-1&$\frac{1}{2}$&0&
$\frac{1}{2}$&$0$&$\frac{1}{\sqrt{3}}$&$-\frac{1}{6}$&$\frac{1}{3}$&$\frac{1}{3}$\\
\hline
50&$\hat{\bar{d}}_{L}^{\bar{c3} \dagger}$&$\stackrel{03}{(+i)}\,\stackrel{12}{(-)}|\stackrel{56}{[+]}\,\stackrel{78}{(+)}
||\stackrel{9 \;10}{(+)}\;\;\stackrel{11\;12}{(+)}\;\;\stackrel{13\;14}{[-]} $&-1&$-\frac{1}{2}$&0&
$\frac{1}{2}$&$0$&$\frac{1}{\sqrt{3}}$&$-\frac{1}{6}$&$\frac{1}{3}$&$\frac{1}{3}$\\
\hline
51&$\hat{\bar{u}}_{L}^{\bar{c3} \dagger}$&$  \stackrel{03}{[-i]}\,\stackrel{12}{[+]}|\stackrel{56}{(-)}\,\stackrel{78}{[-]}
||\stackrel{9 \;10}{(+)}\;\;\stackrel{11\;12}{(+)}\;\;\stackrel{13\;14}{[-]} $&-1&$\frac{1}{2}$&0&
$-\frac{1}{2}$&$0$&$\frac{1}{\sqrt{3}}$&$-\frac{1}{6}$&$-\frac{2}{3}$&$-\frac{2}{3}$\\
\hline
52&$ \hat{\bar{u}}_{L}^{\bar{c3} \dagger} $&$  \stackrel{03}{(+i)}\,\stackrel{12}{(-)}|
\stackrel{56}{(-)}\,\stackrel{78}{[-]}
||\stackrel{9 \;10}{(+)}\;\;\stackrel{11\;12}{(+)}\;\;\stackrel{13\;14}{[-]}  $&-1&$-\frac{1}{2}$&0&
$-\frac{1}{2}$&$0$&$\frac{1}{\sqrt{3}}$&$-\frac{1}{6}$&$-\frac{2}{3}$&$-\frac{2}{3}$\\
\hline
53&$\hat{\bar{d}}_{R}^{\bar{c3} \dagger}$&$\stackrel{03}{(+i)}\,\stackrel{12}{[+]}|\stackrel{56}{[+]}\,\stackrel{78}{[-]}
||\stackrel{9 \;10}{(+)}\;\;\stackrel{11\;12}{(+)}\;\;\stackrel{13\;14}{[-]} $&1&$\frac{1}{2}$&
$\frac{1}{2}$&0&$0$&$\frac{1}{\sqrt{3}}$&$-\frac{1}{6}$&$-\frac{1}{6}$&$\frac{1}{3}$\\
\hline
54&$\hat{\bar{d}}_{R}^{\bar{c3} \dagger} $&$  \stackrel{03}{[-i]}\,\stackrel{12}{(-)}|\stackrel{56}{[+]}\,\stackrel{78}{[-]}
||\stackrel{9 \;10}{(+)}\;\;\stackrel{11\;12}{(+)}\;\;\stackrel{13\;14}{[-]} $&1&$-\frac{1}{2}$&
$\frac{1}{2}$&0&$0$&$\frac{1}{\sqrt{3}}$&$-\frac{1}{6}$&$-\frac{1}{6}$&$\frac{1}{3}$\\
\hline
55&$ \hat{\bar{u}}_{R}^{\bar{c3} \dagger}$&$\stackrel{03}{(+i)}\,\stackrel{12}{[+]}|\stackrel{56}{(-)}\,\stackrel{78}{(+)}
||\stackrel{9 \;10}{(+)}\;\;\stackrel{11\;12}{(+)}\;\;\stackrel{13\;14}{[-]} $ &1&$\frac{1}{2}$&
$-\frac{1}{2}$&0 &$0$&$\frac{1}{\sqrt{3}}$&$-\frac{1}{6}$&$-\frac{1}{6}$&$-\frac{2}{3}$\\
\hline
56&$\hat{\bar{u}}_{R}^{\bar{c3} \dagger}$&$\stackrel{03}{[-i]}\,\stackrel{12}{(-)}|\stackrel{56}{(-)}\,\stackrel{78}{(+)}
||\stackrel{9 \;10}{(+)}\;\;\stackrel{11\;12}{(+)}\;\;\stackrel{13\;14}{[-]} $&1&$-\frac{1}{2}$&
$-\frac{1}{2}$&0&$0$&$\frac{1}{\sqrt{3}}$&$-\frac{1}{6}$&$-\frac{1}{6}$&$-\frac{2}{3}$\\
\hline\hline
57&$ \hat{\bar{e}}^{ \dagger}_{L}$&$ \stackrel{03}{[-i]}\,\stackrel{12}{[+]}|
\stackrel{56}{[+]}\,\stackrel{78}{(+)}
||\stackrel{9 \;10}{[-]}\;\;\stackrel{11\;12}{[-]}\;\;\stackrel{13\;14}{[-]} $ &-1&$\frac{1}{2}$&0&
$\frac{1}{2}$&$0$&$0$&$\frac{1}{2}$&$1$&$1$\\
\hline
58&$\hat{\bar{e}}^{ \dagger}_{L}$&$\stackrel{03}{(+i)}\,\stackrel{12}{(-)}|\stackrel{56}{[+]}\,\stackrel{78}{(+)}
||\stackrel{9 \;10}{[-]}\;\;\stackrel{11\;12}{[-]}\;\;\stackrel{13\;14}{[-]}$&-1&$-\frac{1}{2}$&0&
$\frac{1}{2}$ &$0$&$0$&$\frac{1}{2}$&$1$&$1$\\
\hline
59&$\hat{ \bar{\nu}}^{ \dagger}_{L}$&$  \stackrel{03}{[-i]}\,\stackrel{12}{[+]}|\stackrel{56}{(-)}\,\stackrel{78}{[-]}
||\stackrel{9 \;10}{[-]}\;\;\stackrel{11\;12}{[-]}\;\;\stackrel{13\;14}{[-]}$&-1&$\frac{1}{2}$&0&
$-\frac{1}{2}$&$0$&$0$&$\frac{1}{2}$&$0$&$0$\\
\hline
60&$\hat{ \bar{\nu}}^{ \dagger}_{L} $&$  \stackrel{03}{(+i)}\,\stackrel{12}{(-)}|
\stackrel{56}{(-)}\,\stackrel{78}{[-]}
||\stackrel{9 \;10}{[-]}\;\;\stackrel{11\;12}{[-]}\;\;\stackrel{13\;14}{[-]} $&-1&$-\frac{1}{2}$&0&
$-\frac{1}{2}$&$0$&$0$&$\frac{1}{2}$&$0$&$0$\\
\hline
61&$\hat{ \bar{\nu}}^{ \dagger}_{R}$&$\stackrel{03}{(+i)}\,\stackrel{12}{[+]}|\stackrel{56}{(-)}\,\stackrel{78}{(+)}
||\stackrel{9 \;10}{[-]}\;\;\stackrel{11\;12}{[-]}\;\;\stackrel{13\;14}{[-]}$&1&$\frac{1}{2}$&
$-\frac{1}{2}$&0&$0$&$0$&$\frac{1}{2}$&$\frac{1}{2}$&$0$\\
\hline
62&$\hat{\bar{\nu}}^{ \dagger}_{R} $&$  \stackrel{03}{[-i]}\,\stackrel{12}{(-)}|\stackrel{56}{(-)}\,\stackrel{78}{(+)}
||\stackrel{9 \;10}{[-]}\;\;\stackrel{11\;12}{[-]}\;\;\stackrel{13\;14}{[-]} $&1&$-\frac{1}{2}$&
$-\frac{1}{2}$&0&$0$&$0$&$\frac{1}{2}$&$\frac{1}{2}$&$0$\\
\hline
63&$ \hat{\bar{e}}^{ \dagger}_{R}$&$\stackrel{03}{(+i)}\,\stackrel{12}{[+]}|\stackrel{56}{[+]}\,\stackrel{78}{[-]}
||\stackrel{9 \;10}{[-]}\;\;\stackrel{11\;12}{[-]}\;\;\stackrel{13\;14}{[-]}$ &1&$\frac{1}{2}$&
$\frac{1}{2}$&0 &$0$&$0$&$\frac{1}{2}$&$\frac{1}{2}$&$1$\\
\hline
64&$\hat{\bar{e}}^{ \dagger}_{R}$&$\stackrel{03}{[-i]}\,\stackrel{12}{(-)}|\stackrel{56}{[+]}\,\stackrel{78}{[-]}
||\stackrel{9 \;10}{[-]}\;\;\stackrel{11\;12}{[-]}\;\;\stackrel{13\;14}{[-]}$&1&$-\frac{1}{2}$&
$\frac{1}{2}$&0&$0$&$0$&$\frac{1}{2}$&$\frac{1}{2}$&$1$\\
\hline
\end{supertabular}
}
\end{center}
%




%
\section{Hilbert space of Clifford fermions}
\label{HilbertCliff0}

The Clifford odd creation operators $\hat{\bf b}_{f}^{s  \dagger} (\vec{p})$, 
 with $p^0=|\vec{p}|$, are defined in Eq.~(\ref{Weylp0}) on the
tensor products, $*_{T}$, of the $(2^{\frac{d}{2}-1})^2 $ ''basis vectors''
 $\hat{b}^{m \dagger}_{f}$ (describing the
internal space of fermion fields) and of the (continuously) infinite basis 
in the momentum space, $\hat{b}^{\dagger}_{\vec{p}}$, applying on the 
vacuum state $|\psi_{oc}>|0_{\vec{p}}>$. The solutions of the Weyl equation, 
Eq.~(\ref{Weyl}), are plane waves  of particular momentum $\vec{p}$ and
 with the energy related to the momentum,  $p^0=|\vec{p}|$.

 The creation operator 
$\hat{\bf b}_{f}^{s  \dagger} (\vec{p})$ defines, when applying 
on the vacuum state $|\psi_{oc}>|0_{\vec{p}}>$, the $s^{th}$ of the 
$2^{\frac{d}{2}-1}$ plane
wave solutions of a particular momentum $\vec{p }$ belonging to the $f^{th}$ of
the $2^{\frac{d}{2}-1}$ ''families''. They  fulfill together with the Hermitian
conjugated partners annihilation operators  
$\hat{\bf b}_{f}^{s } (\vec{p})$ 
 the anticommutation relations of Eq.~(\ref{Weylpp'comrel}). 
 
 The Hilbert space of the  second quantized fermions consists of any number 
of tensor products, $*_{T_{H}}$, of all possible
$\hat{\bf b}_{f}^{s  \dagger} (\vec{p})$, with the finite number of different 
$(s,f)$ --- $(2^{\frac{d}{2}-1})^2 $ --- and continuous many momentum 
$\vec{p}$.

The tensor products, $*_{T_{H}}$, of single fermion states offer any number 
of possibilities, starting with the empty case (empty ''Slater determinant'', 
with none of possible single fermion states occupied) up to 
\begin{eqnarray}
N_{{\cal H}_{\vec{p}}}& =&  2^{2^{d-2}} \,
\label{NHp}
\end{eqnarray}
possibilities for a chosen momentum $\vec{p}$,
since any of the fermion states of particular $\vec{p}$ can be or is not
among the chosen states  
(creating $ 2^{2^{d-2}}$ "Slater 
determinants" with none single fermion state occupied up to all the $2^{d-2}$ 
single fermion states of particular $\vec{p}$ occupied), up to infinite many 
possibilities when any  momentum $\vec{p}$ can be empty or occupied
\begin{eqnarray}
N_{\cal H}& =& \prod_{\vec{p}}^{\infty} 2^{2^{d-2}}\,.
\label{NH}
\end{eqnarray}
%
{\bf Statement 9.} The Hilbert space of Clifford fermions is generated by the 
tensor product multiplication, $*_{T_{H}}$, of 
any number of the Clifford odd fermion states of all possible internal quantum 
numbers and all possible momenta (that is of any number of 
$ \hat {\bf b}^{s \, \dagger}_{f} (\vec{p})$ of any
 $(s,f, \vec{p})$).

The Hilbert space of a particular momentum $\vec{p}$, ${\cal H}_{\vec{p}}$, 
contains the finite number of ''Slater determinants'', $2^{2^{d-2}} $. 

The total Hilbert space of  anticommuting Clifford fermions is the product  
$\otimes_{N}$ of the Hilbert spaces of particular $\vec{p}$
\begin{eqnarray}
{\cal H}& =& \prod_{\vec{p}}^{\infty}\otimes_{N} {\cal H}_{\vec{p}}\,.
\label{H}
\end{eqnarray}
The total Hilbert space ${\cal H}$ is correspondingly infinite. 

Before starting to comment the application of the creation operators  
$\hat{\bf b}_{f}^{s \, \dagger} (\vec{p})$ and their Hermitian conjugated 
partners annihilation operators $\hat{\bf b}_{f}^{s } (\vec{p})$ on the Hilbert space
${\cal H}$ (determined by all possible ''Slater determinants'' of all possible 
occupied and empty fermion states of all possible $(s,f,\vec{p})$, what means 
the tensor products $*_{T_{H}}$ of all possible single fermion states of all 
possible $(s,f,\vec{p})$, with the identity --- the empty "Slater deteminant" 
included) 
let us discuss properties of creation and annihilation operators, the 
anticommutation relations of which are presented in Eq.~(\ref{Weylpp'comrel}).
%

The creation operators $\hat {\bf b}^{s \, \dagger}_{f} (\vec{p})$ and 
the annihilation operators $\hat {\bf b}^{s' }_{f'} (\vec{p'})$, having an odd 
Clifford character, anticommute, manifesting the properties as follows
%
\begin{eqnarray}
\label{tensorproperties}
\hat {\bf b}^{ s \, \dagger }_{f} (\vec{p})*_{T_H} 
\hat {\bf b}^{ s' \, \dagger }_{f'} (\vec{p}{\,}')&=& - 
\hat {\bf b}^{ s' \, \dagger }_{f'} (\vec{p}{\,}')*_{T_H} 
\hat {\bf b}^{ s \, \dagger }_{f} (\vec{p})\,, \nonumber\\
\hat {\bf b}^{ s }_{f} (\vec{p})*_{T_{H}} 
\hat {\bf b}^{ s' }_{f'} (\vec{p}{\,}')&=& - 
\hat {\bf b}^{ s' }_{f'}  (\vec{p}{\,}')*_{T_{H}} 
\hat {\bf b}^{ s }_{f} (\vec{p})\,, \nonumber\\
\hat {\bf b}^{ s }_{f} (\vec{p})*_{T_{H}} 
\hat {\bf b}^{ s' \, \dagger }_{f'} (\vec{p}{\,}')&=& - 
\hat {\bf b}^{ s' \, \dagger }_{f'} (\vec{p}{\,}')*_{T_{H}} 
\hat {\bf b}^{ s }_{f} (\vec{p})\,, \nonumber\\ 
 {\rm if \;\, at \;\, least \,\; one \,\; of \,} (s,f, \vec{p}) && {\rm is \; different\,\;from\;} 
(s',f', \vec{p}{\,}') \,,\nonumber\\
\hat {\bf b}^{ s \, \dagger }_{f} (\vec{p})*_{T_{H}} 
\hat {\bf b}^{ s \, \dagger }_{f} (\vec{p}) &=& 0\,,\nonumber\\
\hat {\bf b}^{ s  }_{f}  (\vec{p})*_{T_{H}} 
\hat {\bf b}^{ s  }_{f} (\vec{p}) &=& 0\,,\nonumber\\ 
\hat {\bf b}^{ s }_{f}  (\vec{p})*_{T_{H}}
\hat {\bf b}^{ s \, \dagger }_{f} (\vec{p}) &=& 1\, ({\rm identity})
\,,\nonumber\\
\hat {\bf b}^{ s \, }_{f} (\vec{p}) |\psi_{oc}> |0_{\vec{p}}>&=& 0\,.
\end{eqnarray}
The above relations following from the anticommutation relations of 
Eq.~(\ref{Weylpp'comrel}),  determine the rules of the application of  creation and 
annihilation operators on  ''Slater determinants'': \\
{\bf i.} The creation operator 
$\hat {\bf b}^{ s \, \dagger }_{f} (\vec{p})$
jumps over the creation operators determining the occupied state of another 
kind (that is over the occupied state distinguishing from the jumping creation operator
 one in any of the 
internal quantum numbers ($s,f$) or in $\vec{p}$) up to the last step when it comes 
to its own empty state with the quantum numbers ($f,s$) and $\vec{p}$,  
occupying this empty state, or, if this state is already occupied, gives zero.  Whenever
$\hat {\bf b}^{ s \, \dagger}_{f} (\vec{p})$ jumps over an occupied state,  
changes the sign of the ''Slater determinant''.\\
{\bf ii.} The annihilation operator changes the sign whenever jumping over the 
occupied state carrying different internal quantum numbers ($s,f$) or different 
$\vec{p}$, 
unless it comes to the occupied state with its own internal quantum numbers  
($s,f$) and its own $\vec{p}$, emptying this state, or, if this state is empty, 
gives zero.

 Let us point out again that the Clifford odd  creation operators,  
$\hat {\bf b}^{ s \, \dagger}_{f} (\vec{p})$, and annihilation operators, 
$\hat {\bf b}^{ s'}_{f'} (\vec{p'})$, fulfill the anticommutation 
relations of Eq.~(\ref{Weylpp'comrel}) for any  $\vec{p}$ and any  $(s,f)$
 due to the anticommuting character (the Clifford oddness)  of the ''basis vectors'', 
$\hat{b}^{m \dagger}_{f}$ and their Hermitian conjugated partners 
$\hat{b}_{f}^{m}$, Eqs.~(\ref{alphagammatildeprod}, \ref{start(2n+1)2cliffgammatilde4n}, \ref{d=2(2n+1)}), 
what means that  the anticommuting character  of creation and annihilation operators 
is not postulated, it origins in the Clifford oddness describing the internal space of 
fermions.  


The total Hilbert space ${\cal H}$ has infinite  number of degrees of freedom 
(of ''Slater determinants'') due to the infinite number of Hilbert spaces 
${\cal H}_{\vec{p}}$ of particular $\vec{p}$, ${\cal H} $ 
$= \prod_{\vec{p}}^{\infty}\otimes_{N} {\cal H}_{\vec{p}}$, while the Hilbert 
space ${\cal H}_{\vec{p}}$ of particular momentum ${\vec{p}}$ has the finite 
dimension $2^{2^{d-2}}$.

Let us write down the number operator, counting the number of fermions with 
particular choice of quantum numbers $(s,f)$ and particular $\vec{p}$,
$\hat{N}^{s f}_{\vec{p}}$,
%
%
 \begin{eqnarray}
\label{NOSD}
\hat{N}^{s f}_{\vec{p}}&=& \hat{\bf b}_{f}^{s \, \dagger} (\vec{p})\,
 \hat{\bf b}_{f}^{s } (\vec{p})\,.
\end{eqnarray}
$\hat{N}^{s f}_{\vec{p}}$ is obviously the Clifford even operator,
and when jumping over the occupied states of with  $(s',f')$ and $\vec{p'}$ not
equal to  $(s,f)$ and $\vec{p}$ at least in one of the three properties, no sign is
changed.

It follows that the application of the number operator $\hat{N}^{s f}_{\vec{p}}$ 
on the vacuum state $ |\psi_{oc}> |0_{\vec{p}}>$ and on all the  "Slater 
determinants" with the particular fermion state, defined by the creation operator,
$\hat{\bf b}_{f}^{s \, \dagger} (\vec{p})$, unoccupied (denoted in 
Eq.~(\ref{NOSD1})  as $ {\bf 0^{s f}}_{\vec{p}}$), gives zero contribution, 
while the application of the number operator $\hat{N}^{s f}_{\vec{p}}$ on 
all the rest "Slater determinants", with this particular state occupied 
 (denoted in 
Eq.~(\ref{NOSD1})  as $ {\bf 1^{s f}}_{\vec{p}}$), gives identity
\begin{eqnarray}
\label{NOSD1}
\hat{N}^{s f}_{\vec{p}}\, |\psi_{oc}> |0_{\vec{p}}>&=& 0\cdot |\psi_{oc}>\,
|0_{\vec{p}}>\,,\quad\;
\quad \hat{N}^{s f}_{ \vec{p}}\, {\bf 0^{s f}}_{\vec{p}}=0\,,\nonumber\\
\hat{N}^{s f}_{ \vec{p}} \,{\bf 1^{s f}}_{\vec{p}}&=&
1\,\cdot {\bf 1^{s f}}_{\vec{p}} \,, \quad
\hat{N}^{s f}_{ \vec{p}} \, \hat{N}^{s f}_{ \vec{p}} \, 
{\bf 1^{s f}}_{\vec{p}}\,=
1\cdot {\bf 1^{s f}}_{\vec{p}} \,.
\end{eqnarray}

One can  check the above relations on the example of $d=(5+1)$, with the
"basis vectors" for $f=(1,2,3,4)$ presented in 
Table~\ref{cliff basis5+1.} and with the solution for Weyl equation, Eq.~(\ref{Weyl}),
presented in Eq.~(\ref{weylgen05+1}). 

In Sect.~\ref{HilbertCliffapp} 
 the properties of Hilbert spaces are discussed in more details and also illustrated.

%


%
\subsection{Application of $\hat{\bf b}_{f}^{s \, \dagger}
 (\vec{p})$ and $\hat{\bf b}_{f}^{s} (\vec{p})$ on Hilbert 
 space of Clifford fermions}
\label{HilbertCliffapp}

Let us write down first the Hilbert space of second quantized fermions 
${\cal H}_{\vec{p}}$, of particular $\vec{p}$. The generalization
 to the total Hilbert space, ${\cal H}$, is presented in Eq.~(\ref{H}).
Let us use the simplified  notation by denoting for   $f=1$  empty states 
as ${\bf 0_{r p}}$, and occupied states as ${\bf 1_{r p}}$, with 
$r=(1,\dots,  2^{\frac{d}{2}-1})$, 
for $f=2$ we count $r= 2^{\frac{d}{2}-1} +1,\cdots, 
2\cdot 2^{\frac{d}{2}-1}$, anding up with $r= 2^{d-2}$.
Correspondingly we can represent  ${\cal H}_{\vec{p}}$ as follows
\begin{eqnarray}
\label{SD}
|{\bf 0_{1 p}}, {\bf 0_{2 p}}, {\bf 0_{3 p}}, \dots, {\bf 0_{2^{d-2} p}}>|_{1}&&,\nonumber\\
|{\bf 1_{1 p}}, {\bf 0_{2 p}}, {\bf 0_{3 p}}, \dots, {\bf 0_{2^{d-2} p}}>|_{2}&&,\nonumber\\
|{\bf 0_{1 p}}, {\bf 1_{2 p}}, {\bf 0_{3 p}}, \dots, {\bf 0_{2^{d-2} p}}>|_{3}&&,\nonumber\\
|{\bf 0_{1 p}}, {\bf 0_{2 p}}, {\bf 1_{3 p}}, \dots, {\bf 0_{2^{d-2} p}}>|_{4}&&,\nonumber\\
\vdots &&\nonumber\\
|{\bf 1_{1 p}}, {\bf 1_{2 p}}, {\bf 0_{3 p}}, \dots, {\bf 0_{2^{d-2} p}}>|_{2^{d-2} +2}&&,\nonumber\\%
\vdots &&\nonumber\\|{\bf 1_{1 p}}, {\bf 1_{2 p}}, {\bf 1_{3 p}}, \dots, {\bf 1_{2^{d-2} p}}>|_{2^{2^{d-2}} }&&\,,
\end{eqnarray}
with a part with none of states occupied ($N_{r p} =0$ for all $r=1,\dots, 2^{d-2}$),
with a part with only one of states occupied ($N_{r p}=1$ for a particular 
$r=(1,\dots, 2^{d-2})$, while  $N_{r' p}=0$ for all the others  $r' \ne r$),
with a part with only two of states occupied ($N_{r p}=1$ and $N_{r' p}=1$, where
$r$ and $r'$ run from $(1,\dots, 2^{d-2}$), and so on. The last part has all the states
occupied.

It is not difficult to see that the creation and annihilation operators, when applied on this
Hilbert space ${\cal H}_{\vec{p}}$, fulfill the anticommutation relations for the second
quantized Clifford fermions.
\begin{eqnarray}
\{ \hat{\bf b}_{f}^{s } (\vec{p})\,, \hat{\bf b}_{f'}^{s' \, \dagger} (\vec{p})\}_{+}
*_{T_{H}} {\cal H}_{\vec{p}} &=& \delta^{s s'}\; \delta^{f f'} 
{\cal H}_{\vec{p}}\,,\nonumber\\
\{ \hat{\bf b}_{f}^{s } (\vec{p})\,, \hat{\bf b}_{f'}^{s' } (\vec{p})\}_{+}
*_{T_{H}} {\cal H}_{\vec{p}} &=& 0\;\cdot {\cal H}_{\vec{p}} \,,\nonumber\\
\{ \hat{\bf b}_{f}^{s } (\vec{p})\,, \hat{\bf b}_{f'}^{s' } (\vec{p})\}_{+}
*_{T_{H}} {\cal H}_{\vec{p}} &=& 0\;\cdot {\cal H}_{\vec{p}}\,.
\label{CliffcomrelHp}
\end{eqnarray}
The proof for the above relations easily follows if one takes into account that 
whenever the creation or annihilation operator jumps over an odd products of 
occupied states the sign of the ''Slater determinant'' changes due to the 
oddness of each of the occupied states. The contribution of the application of 
$\hat{\bf b}_{f}^{s\, \dagger} (\vec{p}) \,*_{T}\,$
$\hat{\bf b}_{f'}^{s' } (\vec{p})\;*_{T}\,$, $(s,f) \ne (s',f')$,  on 
$ {\cal H}_{\vec{p}}$ has the opposite sign than the contribution of  
$\hat{\bf b}_{f'}^{s'} (\vec{p})$
$\,*_{T}\,\hat{\bf b}_{f}^{s \, \dagger} (\vec{p})\; *_{T}\,$ on
${\cal H}_{\vec{p}}$, due to exchanged places of both operators. The sum of both
contributions gives  therefore zero.
If the creation and annihilation operators are Hermitian conjugated to each other, 
the result follows
\[(\hat{\bf b}_{f}^{s} (\vec{p})\,*_{T}\,
\hat{\bf b}_{f}^{s\,f\dagger} (\vec{p}) +
\hat{\bf b}_{f}^{s \, \dagger} (\vec{p})\, *_{T}\,
\hat{\bf b}_{f}^{s } (\vec{p}) \,)\,*_{T}\, {\cal H}_{\vec{p}}= 
{\cal H}_{\vec{p}}\,,\] manifesting that this application on ${\cal H}_{\vec{p}}$
 gives the whole ${\cal H}_{\vec{p}}$ back. Each of the two summands operates 
 on their own half of ${\cal H}_{\vec{p}}$. Jumping together over an even number 
 of occupied states,  $\hat{\bf b}_{f}^{s} (\vec{p})$ and 
 $\hat{\bf b}_{f}^{s f\,\dagger} (\vec{p})$ do 
 not change the sign of the particular "Slater determinant''.  
(Let us add that  $\hat{\bf b}_{f}^{s }  (\vec{p})$  reduces for the 
particular $s$ and $f$ the Hilbert space ${\cal H}_{\vec{p}}$ for the factor  
$\frac{1}{2}$, and so does 
$\hat{\bf b}^{s \dagger}_{f} (\vec{p})$. 
The sum of both, applied on ${\cal H}_{\vec{p}}$, reproduces the whole 
${\cal H}_{\vec{p}}$.) 

 \vspace{3mm} 

The generalization of the anticommutation relations to all possible $\vec{p}$ is due 
to the orthogonality of momentum basis, Eq.~(\ref{creatorp}), straightforward.

Let us repeat that the number of ''Slater determinants'' in the Hilbert space of particular 
momentum $\vec{p}$, ${\cal H}_{\vec{p}}$, in $d$-dimensional space is finite  and equal to
$N_{{\cal H}_{\vec{p}}} =  2^{2^{d-2}}$\,.
%






%
%

The total Hilbert space of  anticommuting fermions is the continuously infinite 
product of the Hilbert spaces of particular $\vec{p}$, Eq.~(\ref{H}), 
%
$ {\cal H}= \prod_{\vec{p}}^{\infty}\otimes_{N} {\cal H}_{\vec{p}}\,$,
%
with the empty ''Slater determinant'' counted.

Due to the Clifford odd character of creation and annihilation operators, 
Eq.~(\ref{Weylpp'comrel}), and the orthogonality of the solutions of 
different momenta $\vec{p}\,$, App.~\ref{continuous},  it follows that 
$\hat{\bf b}_{f}^{s \, \dagger} (\vec{p})$ 
$*_{T_{H}}\,\hat{\bf b}_{f}^{s \, \dagger} (\vec{p}{\,}') \,*_{T_{H}}\,{\cal H} \ne 0 $, 
$\vec{p}\ne \vec{p}{\,}'$,  while $\{\hat{\bf b}_{f}^{s \, \dagger} (\vec{p})
\,*_{T_{H}}\, $  $\hat{\bf b}_{f}^{s \, \dagger} (\vec{p}{\,}') +$ 
$\hat{\bf b}_{f}^{s \, \dagger} (\vec{p}{\,}')\, *_{T_{H}}\, $ 
$\hat{\bf b}_{f}^{s \, \dagger} (\vec{p})\,\}\,*_{T_{H}}\, {\cal H} =0$, 
$\vec{p}\ne \vec{p}{\,}'$. This can be proven if taking into account 
Eq.~(\ref{tensorproperties}).  
The number of ``Slater determinants'' in the Hilbert space ${\cal H}$ in 
$d$-dimensional space is infinite 
%
$N_{\cal H} = \prod_{\vec{p}}^{\infty} 2^{2^{d-2}}\,$.
%

Since the creation operators $\hat{\bf b}_{f}^{s \, \dagger} (\vec{p})$ and 
the annihilation operators $\hat{\bf b}_{f'}^{s'} (\vec{p}{\,}')$ fulfill for particular 
$\vec{p}$ the anticommutation relations on ${\cal H}_{\vec{p}}$, 
Eq.~(\ref{CliffcomrelHp}),  and since the momentum states, the 
plane wave solutions, are 
orthogonal, and correspondingly the creation and annihilation operators defined
on the tensor products of the internal basis and the momentum basis, representing
 fermions, anticommute, Eq.~(\ref{Weylpp'comrel}) (the Clifford odd objects 
$\hat{\bf b}_{f}^{s \, \dagger} (\vec{p})$ demonstrate their 
oddness also with respect to 
$\hat{\bf b}_{f}^{s \, \dagger} (\vec{p}{\,}')$), 
the anticommutation relations follow also for the application of 
$\hat{\bf b}_{f}^{s \, \dagger} (\vec{p})$ and 
$\hat{\bf b}_{f}^{s } (\vec{p})$
on ${\cal H}$
\begin{eqnarray}
\{\hat{\bf b}_{f}^{s} (\vec{p}) \,, 
\hat{\bf b}_{f'}^{s' \, \dagger} (\vec{p}{\,}') \}_{*_{T_{H}}+} \,{\cal H} &=&
 \delta^{s s'}\; \delta_{f f'}\; \delta (\vec{p} -\vec{p}{\,}')\;{\cal H}\,,\nonumber\\
\{\hat{\bf b}}_{f}^{s \, \dagger} (\vec{p}),{\hat{\bf b}_{f'}^{s' \, \dagger} 
(\vec{p}{\,}')\}_{*_{T_{H}}+}\; {\cal H}&=& 0\;\cdot {\cal H} \,,\nonumber\\
\{\hat{\bf b}_{f}^{s \, \dagger} (\vec{p}),
\hat{\bf b}_{f'}^{s' \, \dagger} (\vec{p}{\,}') \}_{*_{T_{H}}+}\; {\cal H}&=&
0\;\cdot{\cal H}\,.
\label{ijthetaprodgenH} 
\end{eqnarray}
%
%


%
%
%

\vspace{3mm}


\begin{small}

Let us illustrate the properties of 
${\cal H}$ and the application of the creation operators on ${\cal H}$ with a simple 
case of $d=(1+1)$ dimensional space in a toy model with two discrete 
momenta ($p^1_1, p^1_2$). Generalization to many momenta is straightforward.

The internal space of fermions contains only one creation operator, one 
"basis vector" $\hat{b}^{1\dagger}_{1}$ $=\stackrel{01}{(+i)}$, one family 
member $m=1$  of the only family $f=1$. 
Correspondingly the creation operators  multiplied by the plane waves
(what solve the equations of motion),
$\hat{\bf b}^{1  \,\dagger}_{1} (\vec{p^1_i})|_{p^0=|p^1_{i}|}\,
e^{-i(p^0 x^0 -p^1_{i} x^1)} \, {\bf :}\,$  
$=\stackrel{01}{(+i)} \,e^{-i(p^0 x^0 -p^1_{i} x^1)}|_{p^0_i  =|p^1_i| }$  
 differ  only in momentum space. 
Their Hermitian conjugated annihilation operators are 
$\hat{\bf b}^{1}_{1} (\vec{p^1 _i})_{p^0=|p^1_{i}|}$, while the vacuum 
state is  $|\psi_{oc}>\,|0_{p^1_i}>$  $=\stackrel{01}{(-i)}\cdot
\stackrel{01}{(+i)}=\stackrel{01}{[-i]} $. 

The whole Hilbert space for this toy model has correspondingly four members,
four "Slater determinants", numerated by
$|\quad>_{i}, i=(1,2,3,4)$
\begin{eqnarray}
(|{\bf 0_{p_1}} {\bf 0_{p_2}}>|_1\,, \, |{\bf 1_{p_1}} {\bf 0_{p_2}}>|_2\,, \,
|{\bf 0_{p_1}} {\bf 1_{p_2}}>|_3\,, \, |{\bf 1_{p_1}} {\bf 1_{p_2}}>|_4)\,,\nonumber
\end{eqnarray}
${\bf 0_{p^1_i}}$ represents an empty state and ${\bf 1_{p^1_i}}$ the 
corresponding occupied state.
Let us evaluate the application of $\{\hat{\bf b}^{1 }_{1} (\vec{p^1_1})\,,$ 
$\hat{\bf b}^{1 \,\dagger}_{1} (\vec{p^1_2})\}_{*_{T_H}+}$ on the
Hilbert space ${\cal H}$.
It follows
\begin{eqnarray}
&&\{\hat{\bf b}^{1 }_{1} (\vec{p}^1_1)\,, 
       \hat{\bf b}^{1  \,\dagger}_{1} (\vec{p}^1_2)\}_{*_{T_{H}}+} {\cal H}=
\nonumber\\
&&\hat{\bf b}^{1 }_{1} (\vec{p}^1_1)\,*_{T_{H}}\, 
(|{\bf 0_{p_1}} {\bf 1_{p_2}}>|_{1\to 3} \,,\,
-|{\bf 1_{p_1}} {\bf 1_{p_2}}>|_{2\to 4}) + \nonumber\\
&&\hat{\bf b}^{1  \,\dagger}_{1} (\vec{p}^1_2)\, *_{T_{H}}\, 
( |{\bf 0_{p_1}} {\bf 0_{p_2}}>_{2\to 1}\,, \, 
 +  |{\bf 0_{p_1}} {\bf 1_{p_2}}>_{4\to3})=\nonumber\\
&& (-|{\bf 0_{p_1}} {\bf 1_{p_2}>}_{2\to 4 \to 3}+
|{\bf 0_{p_1}} {\bf 1_{p_2}}>_{2\to 1\to 3})\,=0\,. \nonumber
\label{l{ijCliffprodgenHT}}
\end{eqnarray}
\end{small}
%


%
\section{Simple action for  interacting fields in $d=(13+1)$  in {\it spin-charge-family} 
theory}
 \label{fermionandgravitySCFT}

In Sect.~\ref{actionGrassCliff} we discussed actions for free  massless fermions
in $d$-dimensional spaces, Eq.~(\ref{actionWeyl}), using odd Clifford algebra,
represented by $\gamma^a$'s, to describe internal space of fermions. We found 
that $d$-dimensional space, $d > (3+1)$, offers the description of spins, families  
and charges of fermions. Of particular interest is $d=(13+1)$-dimensional space,
offering spins, families and charges postulated by the {\it standard model}
before the electroweak break. In the internal space quarks and leptons and 
antiquarks and antileptons appear together in the same irreducible representation 
of the generators of the Lorentz group $S^{ab}$,
with spins and handedness of quarks and leptons related to their charges, and spins 
and handedness of antiquarks and antileptons related to charges
as required by the {\it standard model}, 
each irreducible representation carrying the family quantum number. 

In order that fermions manifest the observed charges of quarks and leptons in 
$d=(3+1)$ the symmetry must be broken 
from $SO(13 +1)$ first to $SO(7,1)$ $\times SU(3)\times U(1)$. This break 
is (assumed to be) caused by the condensate of the two right handed neutrinos, 
 presented in Table~\ref{Table con.}. 
%
%
The description of the internal space of fermions with the Clifford odd algebra of 
one of the two kinds ---  $\gamma^a$'s determine internal space while 
the second kind of the Clifford algebra objects, $\tilde{\gamma}^a$'s, 
determines the family 
quantum number of fermions --- offers that the corresponding creation operators 
and their Hermitian conjugated partners annihilation operators fulfill the 
anticommutation relations on the vacuum state and on the whole Hilbert space 
of the second quantized fermions without postulates. There are
the anticommuting "basis vectors" of the odd Clifford algebra which take care 
of the anticommuting properties of the second quantized fermions.

Let us now generalize the action for free massless fermions to massless 
interacting fermions under the requirement that the theory remains simple and 
therefore elegant.

The {\it spin-charge-family} theory of one of us, N.S.M.B.,  
(\cite{norma92,norma93,IARD2016,n2014matterantimatter,nd2017,n2012scalars,
JMP2013,normaJMP2015,nh2017}, and the references therein) offers a simple 
(and accordingly elegant) starting action for fermions, coupled in 
$d= (13 + 1)$-dimensional space to only gravitational field through the vielbeins 
$f^{\alpha}{}_a$, the gauge fields of momenta, and the two kinds of the spin 
connection fields, $\omega_{ab \alpha}$ and $\tilde{\omega}_{ab \alpha}$, the 
gauge fields of the two kinds of the generators of the Lorentz transformations 
of the two Clifford algebras,  $S^{ab}$ and $\tilde{S}^{ab}$, respectively,
Eq.~(\ref{sabtildesab})~\footnote{%
Let us remind the reader that after the 
reduction of the Clifford space to the part generated only by $\gamma^{a}$'s, 
Sect.~\ref{reduction}, the generators $\tilde{S}^{ab }$'s determine the 
family properties of fermions, and $\tilde{S}^{ab}$'s transform a family member
of particular family into the same family member of the rest of families.}. 

Let the action for interacting second quantized massless fermions and the 
corresponding gauge fields be in $d=(13+1)$-dimensional space as proposed by the 
{\it spin-charge-family} theory
\begin{eqnarray}
{\cal A}\,  &=& \int \; d^dx \; E\;\frac{1}{2}\, (\bar{\psi} \, \gamma^a p_{0a} \psi) 
+ h.c. +
\nonumber\\  
               & & \int \; d^dx \; E\; (\alpha \,R + \tilde{\alpha} \, \tilde{R})\,,
\nonumber\\
               p_{0a } &=& f^{\alpha}{}_a p_{0\alpha} + \frac{1}{2E}\, \{ p_{\alpha},
E f^{\alpha}{}_a\}_- \,,\nonumber\\
          p_{0\alpha} &=&  p_{\alpha}  - \frac{1}{2}  S^{ab} \omega_{ab \alpha} - 
                    \frac{1}{2}  \tilde{S}^{ab}   \tilde{\omega}_{ab \alpha} \,,
                    \nonumber\\                    
R &=&  \frac{1}{2} \, \{ f^{\alpha [ a} f^{\beta b ]} \;(\omega_{a b \alpha, \beta} 
- \omega_{c a \alpha}\,\omega^{c}{}_{b \beta}) \} + h.c. \,, \nonumber \\
\tilde{R}  &=&  \frac{1}{2} \, \{ f^{\alpha [ a} f^{\beta b ]} 
\;(\tilde{\omega}_{a b \alpha,\beta} - \tilde{\omega}_{c a \alpha} \,
\tilde{\omega}^{c}{}_{b \beta})\} + h.c.\,.               
\label{wholeaction}
\end{eqnarray}
Here~\footnote{$f^{\alpha}{}_{a}$ are inverted vielbeins to 
$e^{a}{}_{\alpha}$ with the properties $e^a{}_{\alpha} f^{\alpha}{\!}_b = 
\delta^a{\!}_b,\; e^a{\!}_{\alpha} f^{\beta}{\!}_a = \delta^{\beta}_{\alpha} $, 
$ E = \det(e^a{\!}_{\alpha}) $.
Latin indices  
$a,b,..,m,n,..,s,t,..$ denote a tangent space (a flat index),
while Greek indices $\alpha, \beta,..,\mu, \nu,.. \sigma,\tau, ..$ denote an Einstein 
index (a curved index). Letters  from the beginning of both the alphabets
indicate a general index ($a,b,c,..$   and $\alpha, \beta, \gamma,.. $ ), 
from the middle of both the alphabets   
the observed dimensions $0,1,2,3$ ($m,n,..$ and $\mu,\nu,..$), indexes from 
the bottom of the alphabets
indicate the compactified dimensions ($s,t,..$ and $\sigma,\tau,..$). 
We assume the signature $\eta^{ab} =
diag\{1,-1,-1,\cdots,-1\}$.} 
$f^{\alpha [a} f^{\beta b]}= f^{\alpha a} f^{\beta b} - f^{\alpha b} f^{\beta a}$.

It is shown in Ref.~\cite{nd2017,IARD2020} that the spin connection gauge fields 
manifest in $d=(3+1)$ as the ordinary gravity, the known vector gauge fields and 
the scalar gauge fields, offering the (simple) explanation for the origin of higgs  
assumed by the {\it standard model}, explaining as well the Yukawa 
couplings. 

The theory predicts new vector and scalar gauge fields, 
Sect.~\ref{vectorscalar3+1}, what offers explanation for the 
{\it dark matter}~\cite{gn2009,IARD2016}, Sect.~\ref{predictionSCFT} and 
for the {\it matter-antimatter asymmetry}~\cite{n2014matterantimatter} in the 
universe, Sect.~\ref{scalar3+1}.


The appearance of the scalar condensate (so far just assumed, not yet proven
 that it appears spontaneously)  of the two right 
handed neutrinos with the family quantum numbers of the group of four families, 
which does not include the observed three families (Table~\ref{Table III.}),
Sect.~\ref{vectorscalar3+1}, brings 
masses of the scale $\propto 10^{16}$ GeV or higher to all the vector and scalar 
gauge fields,  which interact with the condensate~\cite{n2014matterantimatter},
Sect.~\ref{actionGrassCliff}. 

Since the left handed spinors couple differently (with respect to $M^{(7+1)}$) to 
scalar fields than the right handed ones, the break can leave massless and mass 
protected $2^{((7+1)/2-1)}(= 8)$ families~\cite{NHD}, Sect.\ref{TDN0},
 Eq.~(\ref{weylTDN}). 
The rest of families get heavy masses~\footnote{%
A toy model~\cite{NHD,ND012,nh2008} was studied in 
$d=(5+1)$-dimensional space with the action presented in Eq.~(\ref{wholeaction}),
Sect.\ref{TDN0}, Eq.~(\ref{weylTDN}). The break from $d=(5+1)$ 
to $d=(3+1) \times$ an almost $S^{2}$ was studied. For a particular choice of 
vielbeins and for a class 
of spin connection fields the manifold $M^{(5+1)}$ breaks into $M^{(3+1)}$ 
times an almost $S^2$, while $2^{((3+1)/2-1)}$ families remain massless and 
mass protected. Equivalent assumption, although not yet proved how does it 
really work, is made also for the $d=(13+1)$ case. This study is in progress quite 
some time.}. 

 The manifold $M^{(7+1)}$  breaks further by 
the scalar fields, presented in Sect.~\ref{scalar3+1}, 
 to $M^{(3+1)} \times$  $SU(2)\times SU(2)$
at the electroweak break. This happens since the 
scalar fields with the space index $(7,8)$,  Subsubsect.~\ref{scalar3+1}, they
 are the part of  a simple 
starting action~Eq.(\ref{wholeaction}), gain the constant values, independent of the 
coordinates in $d=(3+1)$ (or as interpreted usually gain the nonzero vacuum 
expectation values).
These scalar fields carry with respect to the space index the weak charge 
$\pm \frac{1}{2}$ and the hyper charge 
$\mp \frac{1}{2}$~\cite{n2014matterantimatter,IARD2016}, 
Sect.~\ref{scalar3+1}, just as required by the {\it standard model}, 
manifesting with respect to $\tilde{S}^{ab}$ and $S^{ab}$ additional quantum 
numbers.

Let us point out that all the fermion fields (with the families of fermions 
and the neutrinos
forming the condensate included), the vector and the scalar gauge fields, offering
explanation for  by the {\it standard model}  postulated ones, origin in the 
simple starting action. 

The starting action, Eq.~(\ref{wholeaction}), has only a few parameters.
It is assumed that the  coupling of fermions to $\omega^{ab}{}_{c}$'s 
can differ  from the coupling of fermions to $\tilde{\omega}^{ab}{}_{c}$'s, 
The reduction of the Clifford space, Sect.~\ref{reduction}, causes
this difference. The additional breaks of symmetries influence the coupling 
constants in addition. This is under consideration for quite a long time 
and  has not yet been finished. 

All the observed properties of fermions, of vector gauge fields 
and scalar gauge fields are the part of the simple starting action, while the 
breaks of symmetries dictate the properties of fermions and boson fields
after these breaks. 

In next subsections the properties of fermions, vector gauge fields and 
scalar gauge fields will be discussed and  achievements and predictions shortly
presented.   




%
\subsection{Properties of massless interacting fermions as manifesting 
in $d=(3+1)$ before electroweak break  }
\label{fermionactionSCFT}

Let the  fermion part of the action, Eq.~(\ref {wholeaction}), be rewritten in the 
way that the fermion action manifests in $d=(3+1)$, that is in the low energy regime
before the electroweak break,
by the {\it standard model} postulated properties of: 
$\;\;$ {\bf i.} Fermions, their spins, handedness, charges and family quantum
numbers, Eqs.~(\ref{so1+3}, \ref{so42}, \ref{so64}), determined by the 
Cartan subalgebra of $S^{ab}$ and $\tilde{S}^{ab}$, while the internal space 
of fermions is described by the Clifford "basis vectors". $\;\;$ {\bf ii.} Couplings
of fermions to the vector gauge fields, which are the superposition of 
gauge fields $\omega^{st}{}_m $, Sect.~\ref{vector3+1}, 
with the space index $m=(0,1,2,3)$ and with 
charges determined by the Cartan subalgebra of $S^{ab}$ and $\tilde{S}^{ab}$ 
($S^{ab} \omega^{cd}{}_{e}= i (\omega^{ad}{}_{e} \eta^{b c} - 
\omega^{bd}{}_{e} \eta^{ac}$) and equivalently for the other two indexes of
$\omega^{cd}{}_{e}$ gauge fields,  manifesting the symmetry of space 
$(d-4)$), and  to the  scalar gauge fields~\cite{IARD2016,%
normaBled2020,JMP2013,normaJMP2015,pikan2003,pikan2006,norma92,%
norma93,gmdn2008,gn2009,gn2013,IARD2020} with the space index $s\ge5$
and the charges determined  by the Cartan subalgebra of $S^{ab}$ 
and $\tilde{S}^{ab}$ (as explained in the case of the vector gauge fields), and 
which are superposition of either $\omega^{st}{}_s $ or 
$\tilde{\omega}^{abt}{}_s $, Sect.~\ref{scalar3+1}
\begin{eqnarray}
\label{faction}
{\mathcal L}_f &=&  \bar{\psi}\gamma^{m} (p_{m}- \sum_{A,i}\; g^{Ai}\tau^{Ai} 
A^{Ai}_{m}) \psi + \nonumber\\
               & &  \{ \sum_{s=7,8}\;  \bar{\psi} \gamma^{s} p_{0s} \; \psi \} +
 \nonumber\\ 
& & \{ \sum_{t=5,6,9,\dots, 14}\;  \bar{\psi} \gamma^{t} p_{0t} \; \psi \}
\,, 
\end{eqnarray}
where $p_{0s} =  p_{s}  - \frac{1}{2}  S^{s' s"} \omega_{s' s" s} - 
                    \frac{1}{2}  \tilde{S}^{ab}   \tilde{\omega}_{ab s}$, 
$p_{0t}   =    p_{t}  - \frac{1}{2}  S^{t' t"} \omega_{t' t" t} - 
                    \frac{1}{2}  \tilde{S}^{ab}   \tilde{\omega}_{ab t}$,                    
with $ m \in (0,1,2,3)$, $s \in (7,8),\, (s',s") \in (5,6,7,8)$, $(a,b)$ (appearing in
 $\tilde{S}^{ab}$) run within  either $ (0,1,2,3)$ or $ (5,6,7,8)$, $t$ runs 
$ \in (5,\dots,14)$, 
$(t',t")$ run either $ \in  (5,6,7,8)$ or $\in (9,10,\dots,14)$. 
The spinor function $\psi$ represents all family members of all the 
$2^{\frac{7+1}{2}-1}=8$ 
families.


\vspace{3mm}

$\;\;$ The first line of Eq.~(\ref{faction}) determines in $d=(3+1)$ the kinematics and 
dynamics of fermion fields, coupled to the vector gauge 
fields~\cite{nd2017,normaJMP2015,IARD2016}. 
The vector gauge fields are the superposition of the spin connection fields 
$\omega_{stm}$, $m=(0,1,2,3)$, $(s,t)=(5,6,\cdots,13,14)$, and are the
gauge fields of $S^{st}$, Sect.~\ref{vector3+1}.


The operators $\tau^{Ai}$ ($\tau^{Ai} = \sum_{a,b} c^{Ai}{ }_{ab}\, 
S^{ab}$, $S^{ab}$ are the generators of the Lorentz transformations in the 
Clifford space of $\gamma^a$'s) are presented in Eqs.~(\ref{so42}, \ref{so64}) 
of Sect.~\ref{examplesClifford}. They represent the colour charge, 
$\vec{\tau}^3$, the weak charge, $\vec{\tau}^1$, and the hyper charge,
$Y=\tau^4 + \tau^{23}$, $\tau^4$ is the fermion charge, originating in 
$SO(6)\subset SO(13,1)$, $\tau^{23}$ belongs together with $\vec{\tau}^1$ of 
$SU(2)_{weak}$ 
to $SO(4)$ ($\subset SO(13+1)$). 

{\it One fermion irreducible representation of the Lorentz group contains}, as seen in 
Table~\ref{Table so13+1.}, {\it quarks and leptons and antiquarks and antileptons}, 
belonging to the first family in Table~\ref{Table III.}. One notices that the 
$SO(7,1)$ subgroup content of the $SO(13,1)$ group is the same for the quarks
and leptons and the same for the antiquarks and antileptons. Quarks distinguish 
from leptons, and antiquarks from antileptons, only in the $SO(6)\subset SO(13,1)$
part, that is in the colour $(\tau^{33},\tau^{38})$ part and in the "fermion" quantum
number $\tau^4$. The quarks distinguish  from antiquarks, and leptons from 
antileptons, in the handedness, in the $SU(2)_{I}\,\rm (weak), \,SU(2)_{II}$,  colour 
and in the $\tau^4$ part, explaining 
the relation between handedness and charges of fermions and antifermions, postulated 
in the {\it standard model}, App.~\ref{appanomalies}%
~\footnote{ 
Ref.~\cite{nh2017} points out that the connection between handedness and charges 
for fermions and antifermions, both appearing in the same irreducible representation, 
explains the triangle anomalies in the {\it standard model} with no need to connect 
''by hand'' the handedness and charges of fermions and antifermions.}.

All the vector gauge fields, which interact with the condensate, presented 
in Table~\ref{Table con.}, become massive, Sect.~\ref{vector3+1}.
The {\it vector gauge fields not interacting  with the condensate --- the weak, 
colour, hyper charge and electromagnetic vector gauge fields --- remain 
massless}, in agreement with by the {\it standard model} assumed gauge fields 
before the electroweak break of the mass protection~%
\footnote{The superposition of the scalar gauge 
fields $\tilde{\omega}^{st}{}_{7}$ and $\tilde{\omega}^{st}{}_{8}$, which at
the electroweak break gain constant values in $d=(3+1)$, bring masses 
to all the vector gauge fields, which couple to these scalar fields.}. 

%
%
 
After the electroweak break, caused by the scalar fields, 
the only conserved charges are the colour and the electromagnetic 
charge $ Q =  \tau^{13} + Y$   ($ Y= \tau^{4} + \tau^{23}$).

\vspace{3mm}

 $\;\;$ The second line of Eq.~(\ref{faction}) is the mass term, responsible 
in $d=(3+1)$  for the masses of fermions and of the weak gauge field 
(originating in spin connection fields $\omega^{s t}{}_{m}$). 
The interaction of fermions with the scalar fields with the space index 
$s=(7,8)$ (to these scalar fields particular superposition of the spin connection 
fields $\omega^{a b}{}_{s}$ and all the superposition of
$\tilde{\omega}^{a b}{}_{s}$  with the space index $s=(7,8)$ and
$(a,b)=(0,1,2,3)$ or $(a,b)=(5,6,7,8)$),  which gain the constant  
values in $d=(3+1)$, makes fermions and antifermions massive. 
%
{\it The scalar fields, presented in the second line of Eq.~(\ref{faction}), are
in the {\t standard model} interpreted as the higgs and  the Yukawa couplings}, 
Sect.~\ref{scalar3+1}, predicting in the {\it spin-charge-family} theory
that there must exist several scalar fields~\footnote{The requirement of the 
{\it standard model} that there exist the Yukawa couplings, speaks by itself
that there must exist several scalar fields explaining the Yukawa couplings.}. 




These  scalar gauge fields split into two groups of scalar fields, 
one group of two triplets and three singlets manifesting the symmetry ---
$\widetilde{SU}(2)_{(\widetilde{SO}(3,1), L)}$ 
$\times \widetilde{SU}(2)_{(\widetilde{SO}(4), L)}$ $\times U(1)$  --- and the 
other group of another two triplets and the same three singlets manifesting the 
symmetry --- $\widetilde{SU}(2)_{(\widetilde{SO}(3,1), R)}$ 
$\times \widetilde{SU}(2)_{(\widetilde{SO}(4), R)}$ $\times U(1)$. 

The three  $U(1)$ singlet scalar gauge fields are superposition of  
$\omega_{s' t' s}$, $s=(7,8)$, $(s',t')=(5,6,7,8,9,\cdots,14)$, with the  sum of 
$S^{s' t'}$ arranged into 
superposition of $\tau^{13}$, $\tau^{23}$ and $\tau^4$. The three triplets
interact with both groups of quarks and leptons and antiquarks and 
antileptons~\cite{mdn2006,gmdn2007,gmdn2008,gn2009,gn2014,NA2018,%
NH2017newdata}.

Families of fermions from Table~\ref{Table III.}, interacting with these scalar 
fields, split  as well into two groups of four families,  each of these two groups are 
coupled to one of the two groups of scalar triplets while all eight families 
couple to the same three singlets. The scalar gauge fields, manifesting 
$\widetilde{SU}(2)_{L,R} \times \widetilde{SU}(2)_{L,R}$, are the superposition 
of the gauge fields 
$\tilde{\omega}_{ab s}$, $s=(7,8), (a,b) = $ either $(0,1,2,3)$ or $(5,6,7,8)$, 
manifesting as twice two triplets. 


%



\vspace{3mm}

 $\;\;$  The third line of Eq.~(\ref{faction}) represents the scalar fields, which
 cause transitions from antileptons and antiquarks into quarks and leptons and back, 
offering the explanation for the matter/antimatter asymmetry  in the expanding 
universe at non equilibrium conditions~\cite{n2014matterantimatter} and for the
proton decay. These scalar fields are  colour triplets with respect to the space 
index equal to $(9,10,11,12,13,14)$, while they carry the quantum numbers with 
respect to the superposition of ${\cal S}^{ab}$  in adjoint representations, as can 
be seen in  Table~\ref{Table bosons.} and in Fig. ~\ref{proton is born1.} of 
Sect.~\ref{scalar3+1}.

%


\subsection{Vector and scalar gauge fields before electroweak break}
\label{vectorscalar3+1}

This subsection partly follows Ref.~(\cite{IARD2020} and references therein).

The second line of the starting action, Eq.~(\ref{wholeaction}),  represents the 
action for gauge fields  in $d=(13+1)$-dimensional space, vector and scalar ones, 
written explicitly in the fifth and the sixth line, and here repeated as 
${\cal A}_{gf }$, with the index $ {}_{gf}$ denoting gauge fields
\begin{eqnarray}
 {\cal A}_{gf } &= & \int \; d^dx \; E\; (\alpha \,R + \tilde{\alpha} \, \tilde{R})\,,
\nonumber\\
R &=&  \frac{1}{2} \, \{ f^{\alpha [ a} f^{\beta b ]} \;(\omega_{a b \alpha, \beta} 
- \omega_{c a \alpha}\,\omega^{c}{}_{b \beta}) \} + h.c. \,, \nonumber\\ 
\tilde{R} & =&  \frac{1}{2} \, \{ f^{\alpha [ a} f^{\beta b ]} 
\;(\tilde{\omega}_{a b \alpha,\beta} - \tilde{\omega}_{c a \alpha} \,
\tilde{\omega}^{c}{}_{b \beta})\} + h.c.\,,
\label{wholevectorscalar}
\end{eqnarray}
the notation $ f^{\alpha [ a} f^{\beta b ]}$ means that the two indexes $a$ and 
$b$ must be exchanged and the exchanged value taken with the negative sign. 
Whenever in Eq.~(\ref{wholevectorscalar}) two indexes are equal the summation 
over these two is meant.

In  the {\it spin-charge-family} theory, as in all  the Kaluza-Klein theories, the vector
 gauge fields  and the scalar gauge fields --- the gauge fields of the charges 
originating in higher $(d-4)$-dimensional spaces --- are represented through the 
vielbeins $f^{\sigma}{}_{m},\, m=(0,1,2,3)$ and $f^{\sigma}{}_{s},\,s\ge5$, 
respectively. We proved in Ref.~\cite{nd2017} that the vector and the scalar 
gauge fields manifest in $d=(3+1)$, after the break of the starting symmetry, as 
the superposition of spin connection fields, when the space $(d-4)$ manifest the
assumed symmetry.

$f^{\beta}{}_{a}$  and $e^{a}{}_{\alpha}$ are vielbeins and inverted  vielbeins 
respectively, 
%
$e^{a}{}_{\alpha}f^{\beta}{}_{a} =\delta^{\beta}_{\alpha}$, 
$e^{a}{}_{\alpha}f^{\alpha}{}_{b}= \delta^{a}_{b}$,
%
$E =det(e^{a}{}_{\alpha})$.

Varying the action of Eq.~(\ref{wholevectorscalar})  with respect to the spin 
connection fields, the expression for the spin connection fields $\omega_{ab}{}^e$ 
follows
\begin{eqnarray}
\label{omegaabe}
\omega_{ab}{}^{e} &=& 
 \frac{1}{2E} \{   e^{e}{}_{\alpha}\,\partial_\beta(Ef^{\alpha}{}_{[a} f^\beta{}_{b]} )
      - e_{a\alpha}\,\partial_\beta(Ef^{\alpha}{}_{[b}f^{\beta e]})
{} - e_{b\alpha} \partial_\beta (Ef^{\alpha [e} f^\beta{}_{a]})\}
                     \nonumber\\
                  &+& \frac{1}{4}   \{\bar{\Psi} (\gamma^e \,S_{ab} - 
 \gamma_{[a}  S_{b]}{}^{e} )\Psi \}  \nonumber\\
                  &-& \frac{1}{d-2}  
   \{ \delta^e_{a} [
\frac{1}{E}\,e^d{}_{\alpha} \partial_{\beta}
             (Ef^{\alpha}{}_{[d}f^{\beta}{}_{b]})
                        + \bar{\Psi} \gamma_d  S^{d}{}_{b} \,\Psi ] 
{}  - \delta^{e}_{b} [\frac{1}{E} e^{d}{}_{\alpha} \partial_{\beta}
             (Ef^{\alpha}{}_{[d}f^{\beta}{}_{a]} )
            + \bar{\Psi} \gamma_{d}  S^{d}{}_{a}\, \Psi ]\}\,. 
                        \end{eqnarray}
If replacing $S^{ab}$ in Eq.~(\ref{omegaabe}) with $\tilde{S}^{ab}$, the  expression
for the spin connection fields  $\tilde{\omega}_{ab}{}^{e}$ follows.

In Ref.~\cite{nd2017} it is proven~%
\footnote{We presented in 
Ref.~\cite{nd2017} the proof, that the vielbeins $f^\sigma{}_m$ (Einstein index 
$\sigma \ge 5$, $m=0,1,2,3$)  lead in $d=(3+1)$ to the vector gauge fields, 
which are the superposition of the spin connection fields $\omega_{st m}$: 
 $f^\sigma{}_m=  \sum_{A}\,  \vec{A}^{A}_m$ 
$\vec{\tau}^{A \sigma}{}_{\tau}\, x^{\tau}$, 
with $A^{Ai}_{m}=\sum_{s,t} c^{Ai}{}_{st}\, \omega^{st}{}_{m}$, when 
the metric in $(d-4)$, $g_{\sigma \tau}$, is invariant under the coordinate 
transformations $x^{\sigma'} = x^{\sigma} + \sum_{A,i,s,t} \varepsilon^{A i}\,
(x^{m})\,c^{A i}{}_{st}$  $E^{\sigma s t} (x^{\tau})$ and 
$\sum_{s,t} c^{A i}{}_{st} \, E^{\sigma s t} = \tau^{A i \sigma}$, while  
$\tau^{A i \sigma}$  solves the Killing equation: 
$D_{\sigma}\, \tau^{A i}_{\tau} + D_{\tau} \tau^{A i}_{\sigma} =0\,$
($D_{\sigma}\,  \tau^{A i}_{\tau} = \partial_{\sigma}\, \tau^{A i}_{\tau} - 
\Gamma^{\tau'}_{\tau \sigma} \tau^{Ai}_{\tau'})$. And similarly also for 
the scalar gauge fields.} that in spaces with the desired symmetry 
the vielbein can be expressed with the gauge fields,
%
$f^{\sigma}{}_{m}= \sum_{A}\,\vec{\tau}^{A\sigma}\, \vec{A}^{A}_{m}$, 
 and $\tau^{Ai \sigma} =  \sum_{st}\, c^{Ai}{}_{st}\,  (e_{s  \tau}\,
 f^{\sigma}{}_{t} - e_{t  \tau}\,
f^{\sigma}{}_{s}) x^{\tau}$,
%
and
\begin{eqnarray}
\label{taua} 
 A^{Ai}_{m}&=& \sum_{st} \,c^{Ai}{}_{st} \, \omega^{st}{}_{m}\,, 
\nonumber\\
\tau^{Ai} &=& \sum_{st}\, c^{Ai}{}_{st}\,S^{st} \,
\,,\nonumber\\
\{\tau^{Ai}, \tau^{Bj}\}_- &=& i \delta^{AB} f^{Aijk} \tau^{Ak}\,, 
\end{eqnarray}
$m=(0,1,2,3)$ denotes the vector gauge fields. If $m$ is replaced by $s$, 
$s=(5,6,\dots,14)$ the equivalent expression follows for the scalar gauge fields.

If fermions are not present then spin connections of both kinds are uniquely 
determined by vielbeins, as can be noticed from Eq.~(\ref{omegaabe}).
If fermions are present, carrying both --- family members and family --- 
quantum numbers, then vielbeins and both kinds of spin connections 
are influenced by the presence of fermions. 

The vector gauge fields $ A^{Ai}_{m}$ of  $\tau^{Ai}$  represent in the 
{\it spin-charge-family} theory all the observed gauge fields, as well as the 
additional non observed vector gauge fields, which interacting with the 
condensate gain heavy masses.

The scalar (gauge) fields, carrying the space index $s=(5,6,\dots,d)$,  offer 
in the {\it spin-charge-family} for $s=(7,8)$ the explanation for the origin of 
the Higgs's scalar and the Yukawa couplings of the {\it standard model}, 
while scalars with the space index $s=(9,10,\dots, 14)$ offer  the 
explanation for the proton decay, as well as for the matter/antimatter 
asymmetry in the universe. 

The explicit expressions for $c^{Ai}{ }_{ab}$, and correspondingly for $\tau^{Ai}$, 
and $ A^{Ai}_{a}$, are written in Sects.~\ref{vector3+1} and~\ref{scalar3+1}. 



%
\subsubsection{Vector gauge fields in $d=(3+1)$} 
%
\label{vector3+1}
 %


In the {\it spin-charge-family} theory the simple starting action (in which fermions, 
the internal space  of which is described by the odd Clifford algebra, interact with 
gravity only --- vielbeins and two kinds of spin connection fields),
 Eq.~(\ref{wholeaction})%
~\footnote{In $SO(13+1)$ 
there are seven members of Cartan subalgebra, Eq.~(\ref{cartangrasscliff}), which 
determine in Table~\ref{Table so13+1.}  the spin, handedness, weak charges of
two kinds, colour charge and "fermion" charge $\tau^{4}$ of quarks and leptons
before the electroweak break. 
As discussed in Sect.~\ref{vectorscalar3+1} the superposition of spin connection
fields $\omega^{st}{}_{m}$, $m=(0,1,2,3)$, manifest in $d=(3+1)$ as the vector 
gauge fields of $\tau^{Ai}$, that is $\vec{A}^{3}_{m}$, $\vec{A}^{1}_{m}$, 
$A^{Q}_{m}$ and $A^{Y}_{m}$, with the coefficients $c^{Ai}{}_{st}$ 
determined in Eq.~(\ref{taua}) (and as ordinary gravity for $(a,b)$ of 
$\omega^{ab}{}_{m}$ belonging to index $m=(0,1,2,3)$).},
manifests in $d=(3+1)$ before the electroweak break besides as massless 
gravity, colour, weak $SU(2)_{I}$ and hyper charge vector gauge fields, also 
as massive second
$SU(2)_{II}$ weak  and $U(1)_{\tau^4}$, "fermion" vector gauge fields.
The gauge fields 
$\vec{A}^{2}_{m}$ and $A^{4}_{m}$ obtain masses in interaction with the 
condensate of two right handed neutrinos, Table~\ref{Table con.}. All the vector 
and scalar gauge fields of  $S^{ab}$ and of $\tilde{S}^{ab}$, 
which interact with the condensate,  
become massive.

The $U(1)_{\tau^4}$ vector gauge  field is the vector gauge field of 
$\tau^{4}(= -\frac{1}{3} (S^{9\,10} + S^{11\,12} +S^{13\,14}))$, with  
$\tau^4$ denoting the "fermion" charge. 
The hyper charge vector gauge field of the {\it standard model} is the superposition 
of the third component of the second $SU(2)_{II}$ vector gauge fields and the
$U(1)_{\tau^4}$ vector gauge field ($A^{Y}_m=\cos\theta_2 A^{\tau^4}_{m} 
+ \sin \theta_2 A^{23}_m $, $\theta_2 $ is the angle of the break of the 
$SU(2)_{II}\times U(1)_{\tau^4}$ symmetry to $U(1)_{Y}$ at the scale 
$\ge 10^{16}$ GeV~(\cite{normaJMP2015} and references therein), caused 
by the condensate, Table~\ref{Table con.}).
Also the two components of the second $SU(2)_{II}$  vector gauge fields and the 
superposition  $A^{Y'}_{m}= - \sin \theta_2 A^{4}_{m} 
+ \cos \theta_2 A^{23}_m $, which is the gauge field of  
$Y' (=- \tan^2 \theta_{2} \tau^{4} + \tau^{23})$  gain as well masses
due to the interaction with the condensate. 

 If there are no fermions present (like it is the condensate) the spin connection 
fields are expressible with only vielbeins and opposite~\cite{nd2017}. 

 At the electroweak break 
all the scalar gauge fields, $\vec{\tilde{A}}^{Ai}_{s}, s=(7,8)$, together 
with $A^{Q}_{s}$ and $A^{Y}_{s}$, gain constant values (nonzero vacuum 
expectation values), making massive (or contributing to the masses of)
all the rest of $\vec{\tilde{A}}^{Ai}_{m}$, which interact with these
scalar fields, 
leaving massless only gravity, $\vec{A}^{3}_{m}$ and $A^{Q}_{m}$.




All the vector gauge fields
are expressible with the spin connection fields  $\omega_{stm}$ 
as
\begin{eqnarray}
 A^{Ai}_{m} &=& \sum_{s,t} \;c^{Ai}{ }_{st} \; \omega^{st}{}_{m}\,, 
 \label{vector}
 \end{eqnarray}
with $\sum_{A,i} \tau^{Ai} \,A^{Ai}_{m} = \sum^*_{a,b} S^{ab} 
\omega^{ab}{}_{m}$,${}^*$ means that summation runs over $(a,b)$  
respecting the symmetry
$SO(7,1)\times SU(3)\times U(1)$.


 Let us present expressions for the two $SU(2)$ vector gauge fields, $SU(2)_{I}$
 and $SU(2)_{II}$, before the break of symmetries
 \begin{eqnarray}
 \label{su2IandII}
\vec{\cal{A}}^{1}_{m} &=& \vec{A}^{1}_{m} = (\omega_{58m} - \omega_{67m}, 
\omega_{57m} + \omega_{68m}, \omega_{56m} - \omega_{78m})\, ,\nonumber\\
\vec{\cal{A}}^{2}_{m} &=&\vec{A}^{2}_{m} = (\omega_{58m} + \omega_{67m}, 
\omega_{57m} - \omega_{68m}, \omega_{56m} + \omega_{78m})\, .
\end{eqnarray}
It is demonstrated in Ref.~\cite{nd2017}  for the case when $SO(7,1)$ breaks into 
$SO(3,1)\times SU(2)\times SU(2)$ that $\sum_{A,i} \tau^{Ai}\,  A^{Ai}_{m}= 
\sum_{s,t} S^{s t}\,\omega_{s t m}$ and that the effective action in flat $(3+1)$ 
space (with no gravitational field) for the vector gauge fields is  
$\int \,d^{4} x \, \{ - \frac{1}{4} \, F^{A i}{}_{m n} $ $F^{A i m n}\,\} $, where  
$F^{Ai}{}_{mn} = \partial_{m} A ^{Ai}_{n}- \partial_{n} A ^{Ai}_{m} - 
 i f^{Aijk} \, A^{Aj}_{m} $ $A^{Ak}_{n}$, and $f^{Aijk}$ are the 
structure constants of the corresponding gauge groups. 

 The reader can similarly construct all the other vector gauge fields from the
 coefficients for the corresponding charges, or find the expressions in 
Refs.~\cite{IARD2020,JMP2013,n2014matterantimatter,normaJMP2015} or 
references therein. 

The generalization of the break of $SO(13,1)$ into 
$SO(3,1)\times SU(2)\times SU(2) \times SU(3) \times U(1)$, used in the 
{\it spin-charge-family} theory, goes equivalently. 
In a general case 
one has $\sum_{A,i} \tau^{Ai}\, 
 A^{Ai}_{m}= \sum^{*}_{s,t} S^{s t}\,\omega_{s t m}$, where ${}^{*}$  
means that the summation concerns only those $(s,t)$, which appear in 
$\tau^{Ai}= \sum_{s,t} c^{Ai}{}_{st} \,S^{st}$. These vector gauge fields 
$A^{Ai}_{m}$, expressible with the spin connection fields, 
$A^{Ai}_{m}= \sum_{s,t} c^{Ai st} \,\omega_{s t m}$, offer an elegant 
explanation for the appearance of the vector gauge fields in the observed 
$(3+1)$ space.

The effective action for all the massless vector gauge fields
before the electroweak break is in the general case of all the vector gauge fields,
which do not interact with the condensate and remain therefore massless, 
equal to  $\int \,d^{4} x \, \{ - \frac{1}{4} \,
 F^{A i}{}_{m n} $ $F^{A i m n}\,\} $, with  the structure constants  $f^{Aijk}$
concerning the colour $SU(3)$, weak SU(2) and hyper charge $U(1)$ groups.  

All these relations are valid as long as spinors and vector gauge fields  are weak 
fields in comparison with the fields which force $(d-4)$ space to be (almost) curled, 
Sect.~\ref{TDN0}. When 
all these fields, with the scalar gauge fields included,  start to be comparable with 
the fields (spinors or scalars), which determine the symmetry of $(d-4)$ space, 
the symmetry of the whole space changes. 

The electroweak break, caused by the constant (non zero vacuum expectation) 
values of the scalar gauge fields, carrying the space index $s=(7,8)$, makes 
the weak and the hyper charge gauge fields massive. The only vector gauge 
fields which remain massless are, besides the  gravity, the 
electromagnetic and the colour vector gauge fields --- the observed three massless
gauge fields. 


 %
\subsubsection{Scalar gauge fields in $d=(3+1)$}
\label{scalar3+1}

Scalar fields, taking care of the masses of quarks and leptons in the 
{\it spin-charge-family} theory, have  the space index $s=(7,8)$ and carry 
with respect to this space index the weak charge
$\tau^{13} =\pm \frac{1}{2}$ and the hyper charge $Y=\mp \frac{1}{2}$,
Table~\ref{Table doublets.}, Eq.~(\ref{checktau13Y}). 
With respect to the index determined by 
$\tau^{Ai}= \sum_{ab}  c^{Ai}{}_{ab} S^{ab}$ and 
$\tilde{\tau}^{Ai}= \sum_{ab}  c^{Ai}{}_{ab} \tilde{S}^{ab}$, that is with
respect to $S^{ab}$ and $\tilde{S}^{ab}$, they carry charges and family 
charges in adjoint representations, Eq.~(\ref{bosonspin0}).

There are in the starting action of the {\it spin-charge-family} theory,
 Eq.~(\ref{wholeaction}), scalar fields, which transform antileptons and 
antiquarks into quarks and leptons and back. They carry space index 
$s=(9,10,\dots,14)$, They are with respect to the space index colour 
triplets and antitriplets, while they carry charges  $\tau^{Ai}$ and
$\tilde{\tau}^{Ai}$ in adjoint representations. 

Following partly Refs.~\cite{normaJMP2015,IARD2020} we shall review both 
kinds of scalar fields.

Let us demonstrate how do the  infinitesimal generators ${\cal S}^{ab}$ 
apply on the spin connections fields $\omega_{bd e}$ ($= f^{\alpha}{}_{e}\, $ 
$\omega_{bd \alpha}$) and $\tilde{\omega}_{\tilde{b} \tilde{d} e}$ 
($= f^{\alpha}{}_{e}\,$ $\tilde{\omega}_{\tilde{b} \tilde{d} \alpha}$), on 
either the space index $e$ or any of the
indices $(b,d,\tilde{b},\tilde{d})$ 
\begin{eqnarray}
\label{bosonspin0}
{\cal S}^{ab} \, A^{d\dots e \dots g} &=& i \,(\eta^{ae} \,A^{d\dots b \dots g} - 
\eta^{be}\,A^{d\dots a \dots g} )\,,
\end{eqnarray}
(Section~IV. and Appendix~B in Ref.~\cite{normaJMP2015}).





The scalar fields  low energy action~\cite{nd2017} is 
proportional to $\int E\, d^{d-4} x\,$ $ R$, where 
%
\begin{eqnarray*} 
R&=& \{ \Gamma^{\sigma}{}_{ \tau [\tau', \sigma]}
+  \Gamma^{\sigma}{}_{\tau'' [\sigma}\,  \Gamma^{\tau''}{}_{\tau \tau']}\,\}\,
g^{\tau \tau'}\nonumber\\
 &=&\frac{1}{2} \, \{ f^{\sigma [ s} f^{\tau t ]} \;(\omega^{s t}{}_{ \tau, \sigma} 
+ \omega_{st' \sigma}\,\omega^{t'}{}_{t  \tau}) \} + h.c.\,,
\end{eqnarray*}
and $\Gamma^{\sigma}{}_{\tau \sigma'} =\frac{1}{2}\,  
g^{\sigma \tau'}\, (g_{\sigma' \tau'},_{\tau} + g_{\tau \tau'},_{\sigma'} - 
g_{\tau \sigma'},_{\tau'})$.
Similar relation follows also for the superposition of the spin connection fields. 

If $\omega_{st' \sigma}$ depend on $x^{m}$ ($x^{m}$ are coordinates in 
($3+1$) space), the scalar fields are the dynamical fields in $(3+1)$, explaining, 
for example, after the break of the starting symmetry, the appearance of the 
Higgs's scalars and the Yukawa couplings~%
\cite{IARD2016,n2014matterantimatter,normaJMP2015,JMP2013,n2012scalars}, 
as well as the proton decay and the appearance of the matter-antimatter 
asymmetry in our universe.

%

\vspace{4mm}

{\it {\bf a. Scalar gauge fields determining scalar higgs and Yukawa couplings}}

\vspace{4mm}

To be in agreement with the experiments (and with the {\it standard model} 
assumptions) the {\it spin-charge-family} theory chooses the space index for 
scalars, gaining constant values and causing correspondingly the electroweak
 phase transition, equal to $s=(7,8)$ (the choice of $(s=5,6)$ would also work).
 All the family quantum numbers of eight families, Table~\ref{Table III.}, that 
is all the superposition of $\tilde{\omega}_{\tilde{a}\tilde{b}s}$ are allowed, 
while with respect to $\omega_{abs}$ only the  superposition representing 
the scalar gauge fields $A^{Q}_{s}$,  $A^{Y}_{s}$ and $A^{4}_s$, 
$s=(7,8)$ (or any three superposition of these three scalar fields) may 
contribute. 

It is convenient to use the common notation  $ A^{Ai}_{s}$ for all the scalar 
gauge fields with $s=(7,8)$, independently of whether  they originate in 
$\omega_{abs}$ --- in this case $ Ai$ $=(Q$,$Y,\tau^4$) --- or in 
$\tilde{\omega}_{\tilde{a}\tilde{b}s}$. All these gauge fields  contribute to 
the masses of quarks and leptons and antiquarks and antileptons after 
gaining constant values (nonzero vacuum expectation values).
\begin{eqnarray}
\label{commonAi}
 A^{Ai}_{s} &{\rm represents}& (\,A^{Q}_{s}\,,A^{Y}_{s}\,, A^{4}_{s}\,, 
 \vec{\tilde{A}}^{\tilde{1}}_{s}\,, 
 \vec{\tilde{A}}^{\tilde{N}_{\tilde{L}}}_{s}\,, \vec{\tilde{A}}^{\tilde{2}}_{s}\,, 
 \vec{\tilde{A}}^{\tilde{N}_{\tilde{R}}}_{s}\,)\,,\nonumber\\
\tau^{Ai} &{\rm represents}& (Q,\,Y,\,\tau^4, \,\vec{\tilde{\tau}}^{1},\,
 \vec{\tilde{N}}_{L},\,\vec{\tilde{\tau}}^{2},\,\vec{\tilde{N}}_{R})\,.
\end{eqnarray}
Here $\tau^{Ai}$ represent all the operators, which apply on fermions.
These scalars with the space index $s=(7,8)$, they are  scalar gauge fields 
of the generators $\tau^{Ai}$ and 
$\tilde{\tau}^{Ai}$, 
are expressible in terms of the spin connection fields (Ref.~\cite{normaJMP2015},
 Eqs.~(10, 22, A8, A9)).
 

Let us demonstrate~\cite{normaJMP2015} that all the scalar fields with the 
space index $(7,8)$ carry with respect to this space index the weak and the 
hyper charge ($\mp \frac{1}{2}$, $\pm \frac{1}{2}$), respectively. 
This means that all these scalars have properties as required
for the higgs in the {\it standard model}. 

To compare the properties of the scalar fields with those of the Higgs's scalar
of the {\it standard model} we make the scalar fields the eigenstates of 
$\tau^{13}= \frac{1}{2}({\cal S}^{56} - {\cal S}^{78})$.

For this purpose we need to apply the operators $\tau^{13}$ 
$(= \frac{1}{2}({\cal S}^{56} - {\cal S}^{78})$,   
$Y$ $(= \tau^{4} + \tau^{23})$ and $Q$ $( =  \tau^{13} + Y)$, 
Eqs.~(\ref{so42}, \ref{so64}, \ref{YQY'Q'andtilde}) on the scalar fields with
the space index $s=(7,8)$, taking into account 
Eq.~(\ref{bosonspin0}).

 Let us  rewrite the second line of Eq.~(\ref{faction}), ignoring 
the momentum $p_{s}$, $s=(5,6,\dots,d)$, since it is 
expected that solutions with nonzero momenta in higher dimensions do not 
contribute to the masses of fermion fields at low energies 
in $d=(3+1)$. We pay correspondingly no attention to the momentum
 $p_{s}\,, s\in(5,\dots,8)$, when having in 
mind the lowest energy solutions, manifesting at low energies.
\begin{eqnarray}
\label{eigentau1tau2}
 & &\sum_{s=(7,8), A,i}\, \bar{\psi} \,\gamma^s\, ( - \tau^{Ai} \,A^{Ai}_{s}\,)
\,\psi = \nonumber\\
 & &-\sum_{A,i} \bar{\psi}\,\{\,\stackrel{78}{(+)}\, \tau^{Ai} \,(A^{Ai}_{7} - i   
 \,A^{Ai}_{8})\, + \stackrel{78}{(-)}(\tau^{Ai} \,(A^{Ai}_{7} + i \,A^{Ai}_{8})\,\}
\,\psi\,,  \nonumber\\
 & &\stackrel{78}{(\pm)} = \frac{1}{2}\, (\gamma^{7} \pm \,i \, \gamma^{8}\,)
\,,\quad  A^{Ai}_{\scriptscriptstyle{\stackrel{78}{(\pm)}}}: = (A^{Ai}_7 \,\mp i\, 
A^{Ai}_8)\,,
\end{eqnarray}
with the summation over $A$ and $i$ performed, since $A^{Ai}_s$ represent the 
scalar fields ($A^{Q}_{s}$, $A^{Y}_{s}$, $A^{4}_{s}$) determined by 
$\omega_{s',s'',s}\;$, as well as  ($\tilde{A}^{\tilde{4}}_{s}$, 
$\vec{\tilde{A}}^{\tilde{1}}_{s}$, $\vec{\tilde{A}}^{\tilde{2}}_{s}$, 
$\vec{\tilde{A}}^{\tilde{N}_{R}}_{s}$ and $\vec{\tilde{A}}^{\tilde{N}_{L}}_{s}$),
determined by $\tilde{\omega}_{a,b,s}\,, s=(7,8)$.

The application of the operators $ \tau^{13}$, $Y$ ($Y= \frac{1}{2} 
({\cal S}^{56} +{\cal S}^{78})  -\frac{1}{3} ({\cal S}^{9\,10}
 +  {\cal S}^{11\,12} +{\cal S}^{13\,14})$) and $Q$
on the scalar fields ($A^{Ai}_{7}\mp i\,A^{Ai}_{8})$ with respect to the space 
index $s=(7,8)$, by taking into account Eq.~(\ref{bosonspin0})  to  make the 
application of the generators ${\cal S}^{ab}$ on the space indexes, gives  
\begin{eqnarray}
\label{checktau13Y}
\tau^{13}\,(A^{Ai}_7 \,\mp i\, A^{Ai}_8)&=& \pm \,\frac{1}{2}\,(A^{Ai}_7 \,
\mp i\, A^{Ai}_8)\,,\nonumber\\
Y\,(A^{Ai}_7 \,\mp i\, A^{Ai}_8)&=& \mp \,\frac{1}{2}\,(A^{Ai}_7 \,
\mp i\, A^{Ai}_8)\,,\nonumber\\
Q\,(A^{Ai}_7 \,\mp i\, A^{Ai}_8)&=& 0\,.
\end{eqnarray}
Since  $ \tau^{4}$, $Y$, $\tau^{13}$ and 
$\tau^{1 +},  \tau^{1 -}$ give zero if 
applied on  ($A^{Q}_{s}$, $A^{Y}_{s}$ and $A^{4}_{s}$) with  respect to the 
quantum numbers ($Q, Y, \tau^4$), 
 and since $Y,Q,\tau^4$ and $\tau^{13}$ commute with 
the family quantum numbers, one sees that the scalar fields $A^{Ai}_{s}$ 
( =($A^{Q}_{s}$, $A^{Y}_{s}$, $A^{Y'}_{s}$, 
$\tilde{A}^{\tilde{4}}_{s}$, $\tilde{A}^{\tilde{Q}}_{s}$, 
$\vec{\tilde{A}}^{\tilde{1}}_{s}$, 
$\vec{\tilde{A}}^{\tilde{2}}_{s}$, $\vec{\tilde{A}}^{\tilde{N}_{R}}_{s}$,  
$\vec{\tilde{A}}^{\tilde{N}_{L}}_{s}$)), $s=(7,8)$, rewritten as  %
$A^{Ai}_{\scriptscriptstyle{\stackrel{78}{(\pm)}}} $ 
$= (A^{Ai}_7 \,\mp i\, A^{Ai}_8)\,$,
%
are eigenstates of $\tau^{13}$ and $Y$, having the quantum numbers of the 
{\it standard model} Higgs's  scalar.

These superposition of $A^{Ai}_{\scriptscriptstyle{\stackrel{78}{(\pm)}}}$ are 
presented in 
Table~\ref{Table doublets.} as two doublets with respect to the weak charge 
${\cal \tau}^{13}$,  with the eigenvalue of ${\cal \tau}^{23}$  (the second 
$SU(2)_{II}$ charge) 
equal to either $-\frac{1}{2}$ or $+\frac{1}{2}$, respectively. 
\begin{table}
\begin{center}
\begin{minipage}[t]{16.5 cm}
\caption{The two scalar weak doublets, one with $ {\cal \tau}^{23}=- \frac{1}{2}$  
and the other with $ {\cal \tau}^{23}=+ \frac{1}{2}$, both with the "fermion"
quantum number ${\cal \tau}^{4}$ $=0$, are presented. 
In this table all the scalar fields carry besides the quantum numbers determined by 
the space index also  the quantum numbers $A$ and $i$ from 
Eq.~(\ref{commonAi}). The table is taken from Ref.~\cite{normaJMP2015}.}
\label{Table doublets.}
\end{minipage}  
{ \begin{tabular}{|c|c| c c c c r|}
 \hline
 name & superposition & ${\cal \tau}^{13}$& $ {\cal \tau}^{23}$ & spin& ${\cal \tau}^{4}$& $ Q$\\
 \hline
 $A^{Ai}_{\scriptscriptstyle{\stackrel{78}{(-)}}}$ & $A^{Ai}_{7}+iA^{Ai}_{8}$& $+
\frac{1}{2}$& 
 $-\frac{1}{2}$& 0&0& 0\\
 $A^{Ai}_{\scriptscriptstyle{\stackrel{56}{(-)}}}$ & $A^{Ai}_{5}+iA^{Ai}_{6}$& 
$-\frac{1}{2}$& 
 $-\frac{1}{2}$& 0&0& -1\\
 \hline 
$A^{Ai}_{\scriptscriptstyle{\stackrel{78}{(+)}}}$ & $A^{Ai}_{7}-iA^{Ai}_{8}$& 
$-\frac{1}{2}$& 
$+\frac{1}{2}$& 0&0& 0\\
$A^{Ai}_{\scriptscriptstyle{\stackrel{56}{(+)}}}$ & $A^{Ai}_{5}-iA^{Ai}_{6}$& $+
\frac{1}{2}$& 
$+\frac{1}{2}$& 0& 0&+1\\ 
\hline
\end{tabular}
}
 \end{center}
 \end{table}
The operators ${\cal \tau}^{1\spm} = {\cal \tau}^{11}\pm i {\cal \tau}^{12}\, $,
%
$\tau^{1\spm} = \frac{1}{2} [({\cal S}^{58}- {\cal S}^{67})\,\spm \,i\,
({\cal S}^{57}+ {\cal S}^{68})]\,$,
%
transform one member of a doublet from Table~\ref{Table doublets.} into another
member of the same doublet, keeping
$ \tau^{23}$  ($= \frac{1}{2}\,({\cal S}^{56}+ {\cal S}^{78})$) 
unchanged, clarifying the above statement.



It is not difficult to show that the scalar fields 
$A^{Ai}_{\scriptscriptstyle{\stackrel{78}{(\pm)}}}$  
are {\it triplets} as the gauge fields of the  family quantum numbers 
($\vec{\tilde{N}}_{R}, \,$ $\vec{\tilde{N}}_{L},\,$ $ \vec{\tilde{\tau}}^{2},\,$ 
$\vec{\tilde{\tau}}^{1}$;
 Eqs.~(\ref{so1+3}, \ref{so42}, \ref{bosonspin0})) or singlets as the gauge fields of 
$Q=\tau^{13}+Y, \,Q'= -\tan^{2}\vartheta_{1} Y$ $ + \tau^{13} $ and
 $Y' = -\tan^2 \vartheta_{2} \tau^{4} + \tau^{23}$. 

We show this in App.~\ref{scalarsandvectorsapp},  Eq.~(\ref{checktildeNL3Q}), 
concluding 
\begin{eqnarray}
\label{checktildeNL3Q}
\tilde{N}_{L}^{3}\,
\tilde{A}^{\tilde{N}_{L}\spm}_{\scriptscriptstyle{\stackrel{78}{(\pm)}}} 
&=& \spm \tilde{A}^{\tilde{N}_{L}\spm}_{\scriptscriptstyle{\stackrel{78}{(\pm)}}}\,,
\quad \tilde{N}_{L}^{3}\,
\tilde{A}^{\tilde{N}_{L}3}_{\scriptscriptstyle{\stackrel{78}{(\pm)}}}=0
\,,\nonumber\\
Q \,A^{Q}_{\scriptscriptstyle{\stackrel{78}{(\pm)}}} &=&0\,,\nonumber
\end{eqnarray}
%
%
taking into account $ Q={\cal S}^{56} + {\cal \tau}^{4}= {\cal S}^{56} -
\frac{1}{3}({\cal S}^{9\,10}+
{\cal S}^{11\,12} + {\cal S}^{13\,14})$, and with 
${\cal \tau}^{4}$ defined in Eq.~(\ref{so64}), if replacing 
$S^{ab}$ by ${\cal S}^{ab}$ from Eq.~(\ref{bosonspin0}). 
Similarly one finds properties with respect to the $Ai$ quantum 
numbers for all the scalar fields $A^{Ai}_{\scriptscriptstyle{\stackrel{78}{(\pm)}}}$. 


%
%
%


We demonstrated that these scalar fields, three singlets --- $A^{Q}_s, 
A^{Y}_s$ and $A^{4}_s $, the scalar gauge fields of 
$(Q,\,Y,\,\tau^4)$ --- and two groups of two triplets ---
$(\vec{\tilde{A}}^{\tilde{1}}_{s}\,, 
 \vec{\tilde{A}}^{\tilde{N}_{\tilde{L}}}_{s}\,)$ and 
$(\vec{\tilde{A}}^{\tilde{2}}_{s}\,, 
\vec{\tilde{A}}^{\tilde{N}_{\tilde{R}}}_{s}\,)$, the scalar gauge fields of
$(\vec{\tilde{\tau}}^{1},\,
 \vec{\tilde{N}}_{L})$ and $(\vec{\tilde{\tau}}^{2},\,\vec{\tilde{N}}_{R}))$,
 respectively --- all with the 
space index $s=(7,8)$ do  behave as Higgs's scalars. In   
Ref.~\cite{nd2017} it is proven that either vielbeins or spin connection fields 
with the scalar index with respect to $d=(3+1)$ do manifest in $d=(3+1)$  
as several scalar fields. 

{\it The necessity to postulate the Yukawa couplings into the {\it standard model}
is the strong signal that there must be several scalar fields, manifesting as scalar 
higgs and Yukawa couplings}. Although the severe warnings against more than 
one scalar higgs can be found in the literature and also in our 
Sect.~\ref{FCNCproton}, the appearance of several scalar fields in the 
{\it spin-charge-family} theory, origining  in the simple starting action, 
Eq.~(\ref{wholeaction}), and offering the explanations for all the assumptions 
of the {\it standard model}, supports (together with the estimations made) the 
optimism that the theory does explain how and why the flavour changing neutral 
currents and the proton decay have not been observed.

Table~\ref{Table III.} represents two groups of four families. Let us check how 
do the scalar fields $\vec{\tilde{\tau}}^{1}\,\cdot 
\vec{\tilde{A}}^{\tilde{1}}_{\scriptscriptstyle{\stackrel{78}{(\pm)}}}$,
$ \vec{\tilde{N}}_{L}\, \cdot 
\vec{\tilde{A}}^{\tilde{N}_{\tilde{L}}}_{\scriptscriptstyle{\stackrel{78}{(\pm)}}}$,
 $\vec{\tilde{\tau}}^{2}\,\cdot 
\vec{\tilde{A}}^{\tilde{2}}_{\scriptscriptstyle{\stackrel{78}{(\pm)}}}$,
$ \vec{\tilde{N}}_{R}\,\cdot 
\vec{\tilde{A}}^{\tilde{N}_{\tilde{R}}}_{\scriptscriptstyle{\stackrel{78}{(\pm)}}}$ 
influence the two groups of four families, presented in this table.

Taking into account Eqs.~(\ref{plusminus}, \ref{graphbinomsfamilies}) one finds
that $ \tilde{N}^{\pm}_{+}$ and $ \tilde{\tau}^{1 \pm}$ transform the first 
four families among themselves, leaving the second group of four families 
untouched, while $ \tilde{N}^{\pm}_{-}$ and $ \tilde{\tau}^{2 \pm}$  do not 
influence  the first four families and transform the second four families among 
themselves. The operators 
($Q \cdot A^{Q}_{\scriptscriptstyle{\stackrel{78}{(\pm)}}},
Y \cdot A^{Q]Y}_{\scriptscriptstyle{\stackrel{78}{(\pm)}}},
\tau^4 \cdot A^{4}_{\scriptscriptstyle{\stackrel{78}{(\pm)}}}$) are diagonal,
their application on the family members depend on family members quantum 
numbers and on properties of scalar fields. 
All the scalar fields with $s=(7,8)$ "dress" the right handed quarks and leptons
with the hyper charge and the weak charge so that they manifest  charges of 
the left handed partners. 

How strong is the  influence of scalar fields on the masses of quarks and leptons,
depends on the coupling constants  and the masses of the scalar fields.
But  we see  that in both groups of four families, the  mass matrices $4 \times 4$ 
have the symmetry $SU(2)\times SU(2)\times U(1)$ of the form~\footnote{%
The symmetry $SU(2)\times SU(2)\times U(1)$  of the mass matrices, 
Eq.~(\ref{M0}), is expected to remain in all loop corrections~\cite{NA2018}.}
\begin{small}
 \begin{equation}
 \label{M0}
 {\cal M}^{\alpha} = \begin{pmatrix} 
 - a_1 - a & e     &   d & b\\ 
 e*     & - a_2 - a &   b & d\\ 
 d*     & b*     & a_2 - a & e\\
 b*     &  d*    & e*   & a_1 - a
 \end{pmatrix}^{\alpha}\,,
 \end{equation}
 \end{small}
with $\alpha$ representing family members --- quarks and leptons~%
\cite{mdn2006,gmdn2007,gmdn2008,gn2013,gn2014,NH2017newdata,bereziani}. 
In App.~\ref{M0SCFT} the symmetry of mass matrices of Eq.~(\ref{M0})
is discussed. 

The {\it spin-charge-family} theory treats quarks and leptons in equivalent 
way. The difference among family members occur due to the scalar fields
($Q \cdot A^{Q}_{\scriptscriptstyle{\stackrel{78}{(\pm)}}},
Y \cdot A^{Q]Y}_{\scriptscriptstyle{\stackrel{78}{(\pm)}}},
\tau^4 \cdot A^{4}_{\scriptscriptstyle{\stackrel{78}{(\pm)}}}$)~
\cite{gn2014,NH2017newdata}.
%




Since we measure the coupling constants of the vector gauge fields and since 
the break of symmetries influences both coupling constants, the ones of the
vector gauge fields and the one of the scalar gauge fields, we 
should  be able to demonstrate  how do the coupling constants among scalar 
fields at low energies manifest in the estimated, not yet derived, Lagrange 
density after the electroweak break
\begin{eqnarray}
\label{interactingphi}
{\cal L}_{sg} &=& E\,\sum_{A,i} \,\{(p_{m} \,A^{Ai}_{s})^{\dagger} \,(p^{m} \,
A^{Ai}_{s}) - (\lambda^{Ai} - (m'_{Ai})^2)) \,A^{Ai \dagger}_{s} A^{Ai}_{s} 
\nonumber\\ 
&+& \sum_{B,j}\, \Lambda^{Ai Bj}\,A^{Ai \dagger}_{s} A^{Ai}_{s} \;
A^{Bj \dagger}_{s} A^{Bj}_{s}\}\,.
\end{eqnarray}
This work is under consideration for already few years.
%




Let us add that at the electroweak break, causing by constant values of the 
scalar fields with the space index $s=(7,8)$, the mass matrices of the two groups
 of four families manifest either $\widetilde{SU}(2)_{\widetilde{SO}(3,1) L}\times 
\widetilde{SU}(2)_{\widetilde{SU}(4) L}\times U(1)$ symmetry, this is the case 
for the lower four families of the eight families, presented in Table~\ref{Table III.}, 
or $\widetilde{SU}(2)_{\widetilde{SO}(3,1) R}\times 
\widetilde{SU}(2)_{\widetilde{SU}(4) R}\times U(1)$ symmetry, this is the case
for the higher four families, presented in Table~\ref{Table III.}. The two 
$SU(2)$ triplet fields are for each of the two groups different, although manifesting
the same symmetries.  
The same three $U(1)$ singlet fields  contribute to the masses of both groups.

The mass matrix of a family member --- of quarks and leptons --- are $4\times 4$ 
matrices. The observed three families of quarks and leptons form the $3\times 3$ 
submatrices of the $4\times 4$ matrices. The symmetry of the mass matrices,
manifesting in all orders~\cite{NA2018}, limits the number of free parameters.

The mass matrices of the upper four families have the same symmetry as the mass
matrices of the lower four families, but the scalar fields determining the masses 
of the upper four families have different properties (masses and coupling 
constants) than those of the lower four, giving to 
quarks and leptons of 
the upper four families much higher masses in comparison with the lower four
families of quarks and leptons, what offers the explanation 
for the appearance of the {\it dark matter}, studied at Refs.~\cite{gn2009,nm2015}.

Let us conclude:
Twice four families of Table~\ref{Table III.}, with the two groups of two 
triplets applying each on one of the two groups of four families, and one group 
of three singlets applying on all eight families, {\bf i.} offer the explanation for  the 
appearance of the Higgs's scalar and Yukawa couplings of the observed three
families, predicting the fourth family to the observed three families and 
several scalar fields, {\bf ii.}  predict that the stable of the additional four 
families with much higher masses that the lower four families contributes to the 
{\it dark matter}.

\vspace{4mm}

{\bf b.} {\it {\bf Scalar gauge fields causing transitions from antileptons and antiquarks 
 into quarks and leptons}}~\cite{n2014matterantimatter}

\vspace{4mm}


This part follows to a great deal the similar part from Ref.~\cite{IARD2016}. 

Besides the scalar fields with the space index $s=(7,8)$, which manifest
in $d=(3+1)$ as scalar gauge fields with the weak and hyper charge 
$\pm \frac{1}{2}$ and $\mp \frac{1}{2}$, respectively, and which gaining at low 
energies constant values cause masses of families of
quarks and leptons and of the weak gauge field, there are in the starting
action, Eqs.~(\ref{wholeaction}, \ref{faction}), additional scalar gauge
fields with the space index $t=(9,10,11,12,13,14)$. They are with respect
to the space index $t$ either triplets or antitriplets causing transitions from 
antileptons into quarks and from antiquarks into quarks and back. 
These scalar fields are in Eq.~(\ref{faction}) presented in the third line. 


Like in the case of scalar gauge fields with the properties of Higgs's scalars included
 in the second line of Eq.~(\ref{faction}) we can also here rearrange the third
line of ${\cal L}_{f}$ in  Eq.~(\ref{faction})  so that $\omega_{s t t'}$ and 
$\tilde{\omega}_{s t t'}$  are eigenvectors of the Cartan subalgebra 
operators, Eq.~(\ref{cartanM}), with respect to the space index $t'$,  where
the application of ${\cal S}^{t t'}$ on $\omega_{s t t'}$ and
 $\tilde{\omega}_{s t t'}$
is determined by Eq.~(\ref{bosonspin0}) (${\cal S}^{ab} \, A^{d\dots e \dots g} =
 i \,(\eta^{ae} \,A^{d\dots b \dots g} - \eta^{be}\,A^{d\dots a \dots g} )$).
One finds that (${\cal S}^{910} (\omega_{s t 9}\mp i \omega_{s t 10}=\pm 
 (\omega_{s t 9}\mp i \omega_{s t 10}$. Taking this into account 
we can rewrite the third line of the fermion action Eq.~(\ref{faction})   
${\mathcal L}_{f'}=  
\psi^{\dagger} \,\gamma^0 \,\gamma^{t}\,\{ \sum_{t=(9,10,\dots 14)}\,  
 \Large{[}p_{t} - (\,\frac{1}{2}\, S^{s' s"} \,\omega_{s' s" t}+ \,
 \frac{1}{2} \, S^{t' t''} \,\omega_{t' t" t} 
 +\frac{1}{2} \,\tilde{S}^{ab}\, \tilde{\omega}_{ab t}\,)  
\Large{]} \} \, \; \psi,$
after taking into account that in the low energy regime $p_{t}, t\ge5$,  
can be neglected, as follows
\begin{eqnarray}
{\mathcal L}_{f" }&=&  \psi^{\dagger} \,\gamma^0 (-)\,\{  
\sum_{+,-}\,\sum_{(t\,t')} \,\stackrel{t t'}{(\cpm)}\,{\bf \cdot}\nonumber\\
&&[\,
\tau^{2+}\,A^{2+}_{\scriptscriptstyle{\stackrel{t t'}{(\cpm)}}} 
+ \tau^{2-}\,A^{2-}_{\scriptscriptstyle{\stackrel{t t'}{(\cpm)}}} + \tau^{23}\,
A^{23}_{\scriptscriptstyle{\stackrel{t t'}{(\cpm)}}} 
+ 
\tau^{1+}\,A^{1+}_{\scriptscriptstyle{\stackrel{t t'}{(\cpm)}}} + 
\tau^{1-}\,A^{1-}_{\scriptscriptstyle{\stackrel{t t'}{(\cpm)}}} + \tau^{13}\,
A^{13}_{\scriptscriptstyle{\stackrel{t t'}{(\cpm)}}}  \nonumber\\
&+& 
\tilde{\tau}^{2+}\,\tilde{A}^{2+}_{\scriptscriptstyle{\stackrel{t t'}{(\cpm)}}} + 
\tilde{\tau}^{2-}\,\tilde{A}^{2-}_{\scriptscriptstyle{\stackrel{t t'}{(\cpm)}}} + 
\tilde{\tau}^{23}\,\tilde{A}^{23}_{\scriptscriptstyle{\stackrel{t t'}{(\cpm)}}} 
+  
\tilde{\tau}^{1+}\,\tilde{A}^{1+}_{\scriptscriptstyle{\stackrel{t t'}{(\cpm)}}} + 
\tilde{\tau}^{1-}\,\tilde{A}^{1-}_{\scriptscriptstyle{\stackrel{t t'}{(\cpm)}}} + 
\tilde{\tau}^{13}\,\tilde{A}^{13}_{\scriptscriptstyle{\stackrel{t t'}{(\cpm)}}} \nonumber\\
&+&
\tilde{N}^{+}_{R}\,\tilde{A}^{N_{R}+}_{\scriptscriptstyle{\stackrel{t t'}{(\cpm)}}} + 
\tilde{N}^{-}_{R}\,\tilde{A}^{N_{R}-}_{\scriptscriptstyle{\stackrel{t t'}{(\cpm)}}} + 
\tilde{N}^{3}_{R}\,\tilde{A}^{N_{R}3}_{\scriptscriptstyle{\stackrel{t t'}{(\cpm)}}}
+ 
\tilde{N}^{+}_{L}\,\tilde{A}^{N_{L}+}_{\scriptscriptstyle{\stackrel{t t'}{(\cpm)}}} +
\tilde{N}^{-}_{L}\,\tilde{A}^{N_{L}-}_{\scriptscriptstyle{\stackrel{t t'}{(\cpm)}}} + 
\tilde{N}^{3}_{L}\,\tilde{A}^{N_{L}3}_{\scriptscriptstyle{\stackrel{t t'}{(\cpm)}}} \nonumber\\
&+& 
\sum_{i} \tau^{3i}\,A^{3i}_{\scriptscriptstyle{\stackrel{t t'}{(\cpm)}}}\, + 
\tau^{4}\, A^{4}_{\scriptscriptstyle{\stackrel{t t'}{(\cpm)}}} 
+ 
\sum_{i} \tilde{\tau}^{3i}\,\tilde{A}^{3i}_{\scriptscriptstyle{\stackrel{t t'}{(\cpm)}}}\, + 
\tilde{\tau}^{4}\, \tilde{A}^{4}_{\scriptscriptstyle{\stackrel{t t'}{(\cpm)}}} \,] 
%
%
\,\}\, \psi\,,
%
%
\label{factionMaM1}
\end{eqnarray}
where $(t,t')$ run in pairs over $[(9,10),(11,12),(13,14)]$ and the summation 
must go also over $+$ and $-$ of ${}_{\scriptscriptstyle{\stackrel{t t'}{(\cpm)}}}$,
so that we take into account that $\gamma^9 \omega_{s t 9} + 
\gamma^{10} \omega_{s t 10}= \frac{1}{2}\{ (\gamma^9+i \gamma^{10})
(\omega_{s t 9} - i \omega_{s t 10} ) + (\gamma^9-i \gamma^{10})
(\omega_{s t 9} + i \omega_{s t 10} )\}$ and similarly for all the other terms. 

 In Eq.~(\ref{factionMaM1})  the relations below are used
\begin{eqnarray}
\sum_{t,s',s''}\,\gamma^{t}\,\frac{1}{2}\, S^{s' s"} \,\omega_{s' s" t} &=&
\sum_{+,-}\,\sum_{(t\,t')} \,\,\stackrel{t t'}{(\cpm)}\,\frac{1}{2}\, S^{s' s"} \, 
\omega_{\scriptscriptstyle{s" s" \stackrel{t t'}{(\cpm)}}}\,,\nonumber\\
\omega_{\scriptscriptstyle{s" s" \stackrel{t t'}{(\cpm)}}}: &=& 
\omega_{\scriptscriptstyle{s" s" \stackrel{t t'}{(\pm)}}} = 
(\omega_{s' s" t}\,\mp \,i \,\omega_{s' s" t'})\,, \nonumber\\ 
\stackrel{t t'}{(\cpm)}: &=&  = \frac{1}{2}\, 
(\gamma^{t} \pm \gamma^{t'})\,, \nonumber\\
\sum_{+,-}\,\sum_{(t\,t')} \stackrel{t t'}{(\cpm)}\,\frac{1}{2}\, S^{s' s"} \,  
\omega_{\scriptscriptstyle{s" s" \,\stackrel{t t'}{(\cpm)}}}&=&
\stackrel{t t'}{(\cpm)}\,\{\, \tau^{2+}\,A^{2+}_{\scriptscriptstyle{\stackrel{t t'}{(\cpm)}}} 
+ \tau^{2-}\,A^{2-}_{\scriptscriptstyle{\stackrel{t t'}{(\cpm)}}} + \tau^{23}\,
A^{23}_{\scriptscriptstyle{\stackrel{t t'}{(\cpm)}}}\nonumber\\
&+& \tau^{1+}\,A^{1+}_{\scriptscriptstyle{\stackrel{t t'}{(\cpm)}}} + 
\tau^{1-}\,A^{1-}_{\scriptscriptstyle{\stackrel{t t'}{(\cpm)}}} + \tau^{13}\,
A^{13}_{\scriptscriptstyle{\stackrel{t t'}{(\cpm)}}}\, \} \,,\nonumber\\
A^{2\spm}_{\scriptscriptstyle{\stackrel{t t'}{(\cpm)}}} &=& 
(\omega_{\scriptscriptstyle{58 \stackrel{t t'}{(\cpm)}}}+ 
\omega_{\scriptscriptstyle{67 \stackrel{t t'}{(\cpm)}}})\,\smp \, i 
(\omega_{\scriptscriptstyle{57 \stackrel{t t'}{(\cpm)}}}- 
\omega_{\scriptscriptstyle{68 \stackrel{t t'}{(\cpm)}}})\,,
\quad A^{23}_{\scriptscriptstyle{\stackrel{t t'}{(\cpm)}}}= 
(\omega_{\scriptscriptstyle{56 \stackrel{t t'}{(\cpm)}}}+ 
\omega_{\scriptscriptstyle{78 \stackrel{t t'}{(\cpm)}}})\,, 
\nonumber\\
A^{1\spm}_{\scriptscriptstyle{\stackrel{t t'}{(\cpm)}}} &=& 
(\omega_{\scriptscriptstyle{58 \stackrel{t t'}{(\cpm)}}}- 
\omega_{\scriptscriptstyle{67 \stackrel{t t'}{(\cpm)}}})\smp \, i 
(\omega_{\scriptscriptstyle{57 \stackrel{t t'}{(\cpm)}}}+ 
\omega_{\scriptscriptstyle{68 \stackrel{t t'}{(\cpm)}}})\,,\quad 
A^{13}_{\scriptscriptstyle{\stackrel{t t'}{(\cpm)}}}= 
(\omega_{\scriptscriptstyle{56 \stackrel{t t'}{(\cpm)}}}- 
\omega_{\scriptscriptstyle{78 \stackrel{t t'}{(\cpm)}}})\,,\nonumber\\
(t\,t') &\in& ((9\,10), (11\,12),(13\,14))\,. 
\label{factionMaMpart10}
\end{eqnarray}
The rest of expressions in Eq.~(\ref{factionMaMpart10})  are obtained in a similar way.
They are presented in Eq.~(\ref{factionMaMpart20}) of 
App.~\ref{scalar3+1matterantimatterapp}.


The scalar fields  with the space index $s=(9,10,\cdots,14)$, presented in 
Table~\ref{Table bosons.}, carry one of the triplet colour charges and the "fermion" 
charge equal to twice the quark "fermion" charge, or the antitriplet colour charges and 
the "antifermion" charge. 
They carry in addition the quantum numbers of the adjoint representations originating 
in $S^{ab}$ or in 
$\tilde{S}^{ab}$~\footnote{Although carrying the colour charge in  one of the 
triplet or antitriplet states, these fields can not be interpreted as superpartners of 
the quarks since they do  not have quantum numbers as required by, let say, 
the $N=1$ supersymmetry. The hyper charges  and the electromagnetic charges 
are namely not those required by the 
supersymmetric partners to the family members.}.

 \begin{table}
 \begin{tiny}
 \begin{center}
\begin{minipage}[t]{16.5 cm}
\caption{Quantum numbers of the scalar gauge fields carrying the space index 
 $t =(9,10,\cdots,14)$,  appearing in  Eq.~(\ref{faction}), are presented. The space 
 degrees  of freedom contribute one of the triplets values  to the colour charge of all 
 these scalar  fields. These scalars are  with respect to the two $SU(2)$ charges, ($\tau^{13}$  and $\vec{\tau}^2$), and the two $\widetilde{SU}(2)$  charges, 
 ($\vec{\tilde{\tau}}^1$  and $\vec{\tilde{\tau}}^2$),  triplets  (that is in the adjoint representations of the  corresponding groups), and they all carry twice the 
 "fermion" number ($\tau^{4}$) of the quarks.  The quantum numbers of
 the two vector gauge fields,  the colour and the  $U(1)_{II}$ ones,
 are added. }
\label{Table bosons.}%
\end{minipage}
 \begin{tabular}{|c|c|c|c|c|c|c|c|c|c|c|c|c|c|}
 \hline
 ${\rm field}$&prop. & $\tau^4$&$\tau^{13}$&$\tau^{23}$&($\tau^{33},\tau^{38}$)&$Y$&$Q$&$\tilde{\tau}^4$ 
 &$\tilde{\tau}^{13}$&$\tilde{\tau}^{23}$&$\tilde{N}_{L}^{3}$ &$\tilde{N}_{R}^{3}$  \\
 \hline
 $A^{1\spm}_{\scriptscriptstyle{\stackrel{9\,10}{(\cpm)}}}$& scalar&  $\cmp \frac{1}{3}$&$\spm 1$&$0$ 
 & ($\cpm\frac{1}{2},$ $\cpm \frac{1}{2\sqrt{3}}$)& $\cmp \frac{1}{3}$&$\cmp \frac{1}{3}+ \smp 1$&$0$
 &$0$&$0$&$0$&$0$\\ 
 $A^{13}_{\scriptscriptstyle{\stackrel{9\,10}{(\cpm)}}}$   & scalar&  $\cmp \frac{1}{3}$&$0$&$0$ 
 & ($\cpm\frac{1}{2},$ $\cpm \frac{1}{2\sqrt{3}}$)& $\cmp \frac{1}{3}$&$\cmp \frac{1}{3}$&$0$&$0$&$0$&$0$&$0$\\ 
 $A^{1\spm}_{\scriptscriptstyle{\stackrel{11\,12}{(\cpm)}}}$& scalar&  $\cmp \frac{1}{3}$&$\smp 1$&$0$ 
  & ($\cmp\frac{1}{2},$ $\cpm \frac{1}{2\sqrt{3}}$)& $\cmp \frac{1}{3}$&$\cmp \frac{1}{3}+ \smp 1$&$0$
  &$0$&$0$&$0$&$0$\\ 
  $A^{13}_{\scriptscriptstyle{\stackrel{11\,12}{(\cpm)}}}$   & scalar&  $\cmp \frac{1}{3}$&$0$&$0$ 
 & ($\cmp\frac{1}{2},$ $\cpm \frac{1}{2\sqrt{3}}$)& $\cmp \frac{1}{3}$&$\cmp \frac{1}{3}$&$0$&$0$&$0$&$0$&$0$\\ 
 $A^{1\spm}_{\scriptscriptstyle{\stackrel{13\,14}{(\cpm)}}}$& scalar&  $\cmp \frac{1}{3}$&$\smp 1$&$0$ 
   & ($0,$ $\cmp \frac{1}{\sqrt{3}}$)& $\cmp \frac{1}{3}$&$\cmp \frac{1}{3}+ \smp 1$&$0$
   &$0$&$0$&$0$&$0$\\ 
   $A^{13}_{\scriptscriptstyle{\stackrel{13\,14}{(\cpm)}}}$   & scalar&  $\cmp \frac{1}{3}$&$0$&$0$ 
 & ($0,$ $\cmp \frac{1}{\sqrt{3}}$)& $\cmp \frac{1}{3}$&$\cmp \frac{1}{3}$&$0$&$0$&$0$&$0$&$0$\\ 
 \hline
 $A^{2\spm}_{\scriptscriptstyle{\stackrel{9\,10}{(\cpm)}}}$& scalar&  $\cmp \frac{1}{3}$&$0$&$\spm 1$ 
  & ($\cpm\frac{1}{2},$ $\cpm \frac{1}{2\sqrt{3}}$)& $\cmp \frac{1}{3}+ \smp 1$&$\cmp \frac{1}{3}+ \smp 1$&$0$
  &$0$&$0$&$0$&$0$\\  
  $A^{23}_{\scriptscriptstyle{\stackrel{9\,10}{(\cpm)}}}$   & scalar&  $\cmp \frac{1}{3}$&$0$&$0$ 
 & ($\cpm\frac{1}{2},$ $\cpm \frac{1}{2\sqrt{3}}$)& $\cmp \frac{1}{3}$&$\cmp \frac{1}{3}$&$0$&$0$&$0$&$0$&$0$\\ 
 $\cdots$&&&&&&&&&&&&\\
 \hline
 $\tilde{A}^{1 \spm}_{\scriptscriptstyle{\stackrel{9 10}{(\cpm)}}}$& scalar& $\cmp \frac{1}{3}$&$0$&$0$ 
 & ($\cpm\frac{1}{2},$ $\cpm \frac{1}{2\sqrt{3}}$)& $\cmp \frac{1}{3}$&$\cmp \frac{1}{3}$&$0$
 &$\spm 1$&$0$&$0$&$0$\\ 
 $\tilde{A}^{13}_{\scriptscriptstyle{\stackrel{9 10}{(\cpm)}}}$& scalar& $\cmp \frac{1}{3}$&$0$&$0$ 
  & ($\cpm\frac{1}{2},$ $\cpm \frac{1}{2\sqrt{3}}$)& $\cmp \frac{1}{3}$&$\cmp \frac{1}{3}$&$0$
 &$0$&$0$&$0$&$0$\\ 
  $\cdots$&&&&&&&&&&&&\\
 \hline
 $\tilde{A}^{2 \spm}_{\scriptscriptstyle{\stackrel{9 10}{(\cpm)}}}$& scalar& $\cmp \frac{1}{3}$&$0$&$0$ 
  & ($\cpm\frac{1}{2},$ $\cpm \frac{1}{2\sqrt{3}}$)& $\cmp \frac{1}{3}$&$\cmp \frac{1}{3}$&$0$
  &$0$&$\spm 1$&$0$&$0$\\ 
  $\tilde{A}^{23}_{\scriptscriptstyle{\stackrel{9 10}{(\cpm)}}}$& scalar& $\cmp \frac{1}{3}$&$0$&$0$ 
   & ($\cpm\frac{1}{2},$ $\cpm \frac{1}{2\sqrt{3}}$)& $\cmp \frac{1}{3}$&$\cmp \frac{1}{3}$&$0$
  &$0$&$0$&$0$&$0$\\ 
  $\cdots$&&&&&&&&&&&&\\
 \hline
 $\tilde{A}^{N_{L} \spm}_{\scriptscriptstyle{\stackrel{9 10}{(\cpm)}}}$& scalar& $\cmp \frac{1}{3}$&$0$&$0$ 
   & ($\cpm\frac{1}{2},$ $\cpm \frac{1}{2\sqrt{3}}$)& $\cmp \frac{1}{3}$&$\cmp \frac{1}{3}$&$0$
   &$0$&$0$&$\spm 1$&$0$\\ 
   $\tilde{A}^{N_{L} 3}_{\scriptscriptstyle{\stackrel{9 10}{(\cpm)}}}$& scalar& $\cmp \frac{1}{3}$&$0$&$0$ 
    & ($\cpm\frac{1}{2},$ $\cpm \frac{1}{2\sqrt{3}}$)& $\cmp \frac{1}{3}$&$\cmp \frac{1}{3}$&$0$
   &$0$&$0$&$0$&$0$\\ 
   $\cdots$&&&&&&&&&&&&\\
 \hline
 $\tilde{A}^{N_{R} \spm}_{\scriptscriptstyle{\stackrel{9 10}{(\cpm)}}}$& scalar& $\cmp \frac{1}{3}$&$0$&$0$ 
    & ($\cpm\frac{1}{2},$ $\cpm \frac{1}{2\sqrt{3}}$)& $\cmp \frac{1}{3}$&$\cmp \frac{1}{3}$&$0$
    &$0$&$0$&$0$&$\spm 1$\\ 
    $\tilde{A}^{N_{R} 3}_{\scriptscriptstyle{\stackrel{9 10}{(\cpm)}}}$& scalar& $\cmp \frac{1}{3}$&$0$&$0$ 
     & ($\cpm\frac{1}{2},$ $\cpm \frac{1}{2\sqrt{3}}$)& $\cmp \frac{1}{3}$&$\cmp \frac{1}{3}$&$0$
    &$0$&$0$&$0$&$0$\\ 
    $\cdots$&&&&&&&&&&&&\\
 \hline
 $A^{3 i}_{\scriptscriptstyle{\stackrel{9\,10}{(\cpm)}}}$& scalar&  $\cmp \frac{1}{3}$&$0$&$0$ 
  & ($\spm 1 + \cpm\frac{1}{2},$ $\cpm \frac{1}{2\sqrt{3}}$)& $\cmp \frac{1}{3}$&$\cmp \frac{1}{3}$&$0$
 &$0$&$0$&$0$&$0$\\ 
    $\cdots$&&&&&&&&&&&&\\
  \hline
 $A^{4}_{\scriptscriptstyle{\stackrel{9\,10}{(\cpm)}}}$& scalar&  $\cmp \frac{1}{3}$&$0$&$0$ 
  & ($ \cpm\frac{1}{2},$ $\cpm \frac{1}{2\sqrt{3}}$)& $\cmp \frac{1}{3}$&$\cmp \frac{1}{3}$&$0$
 &$0$&$0$&$0$&$0$\\ 
    $\cdots$&&&&&&&&&&&&\\   
\hline 
$\vec{A}^{3}_{m}$& vector&  $0$&$0$&$0$ 
  & octet& $0$&$0$&$0$
 &$0$&$0$&$0$&$0$\\ 
\hline
 $A^{4}_{m}$& vector&  $0$&$0$&$0$ 
  & $0$& $0$&$0$&$0$
 &$0$&$0$&$0$&$0$\\ 
\hline   
 \end{tabular}
 \end{center}
 \end{tiny}
 %
  \end{table}

\begin{figure}
\includegraphics{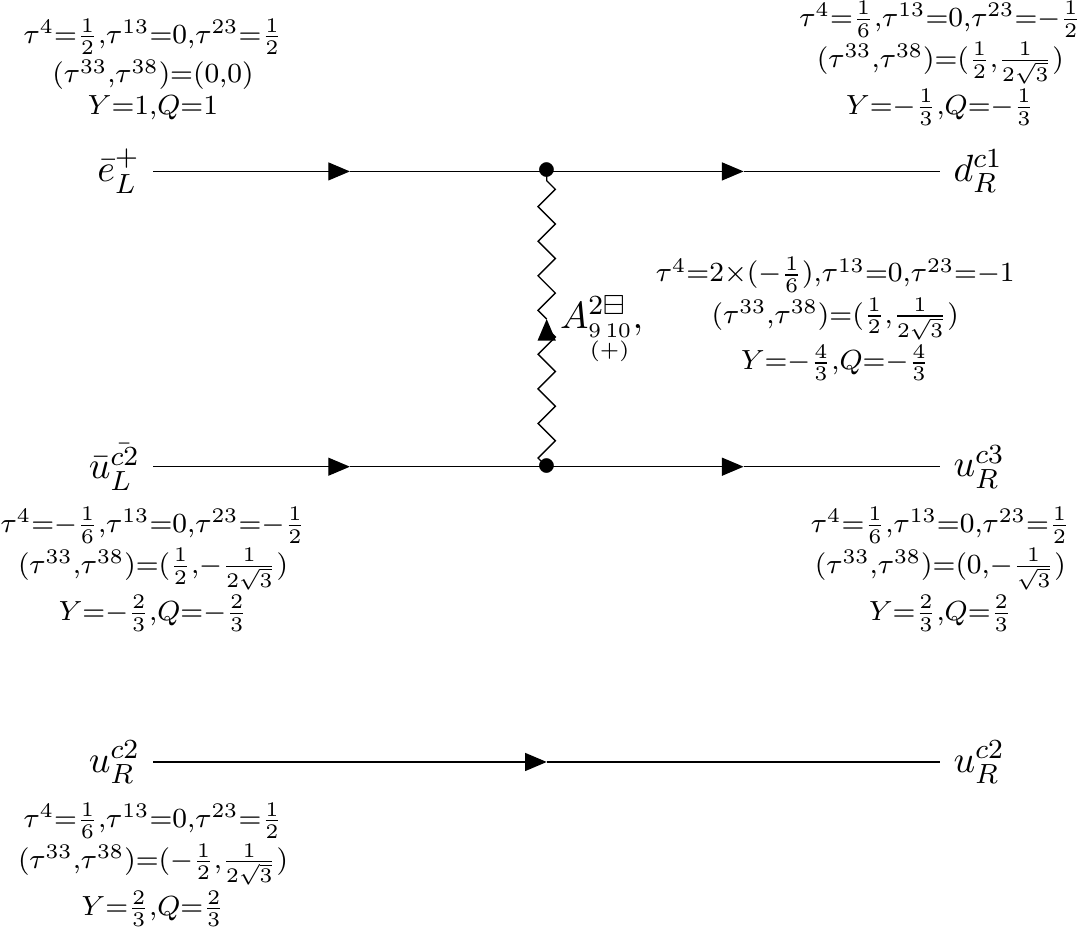}
\centering
\caption{\label{proton is born1.} The birth of a "right handed proton" out of an 
positron%
~$\bar{e}^{\,\,+}_{L}$, 
antiquark $\bar{u}_L^{\bar{c2}}$ and quark (spectator) $u_{R}^{c2}$.  
The family quantum number can be any out of the first four families, presented
in Table~\ref{Table III.}.} 
\end{figure}

If the antiquark $ \bar{u}_{L}^{\bar{c2}}$, from the line $43$ presented in 
Table~\ref{Table so13+1.},  with the "fermion" charge $\tau^{4}=-\frac{1}{6}$, 
the weak charge $\tau^{13} =0$, the second $SU(2)_{II}$ charge 
$\tau^{23} =-\frac{1}{2}$, the colour charge 
$(\tau^{33},\tau^{38})=(\frac{1}{2},-\frac{1}{2\sqrt{3}})$, 
the hyper charge $Y(=\tau^{4}+\tau^{23}=$) $-\frac{2}{3}$ and the 
electromagnetic charge $Q (\,=Y + \tau^{13}=$) $ -\frac{2}{3}$ submits the 
$A^{2 \sminus}_{\scriptscriptstyle{\stackrel{9\,10}{(\oplus)}}} $ scalar 
field, it transforms into $u_{R}^{c3}$ from the line $17$ of 
Table~\ref{Table so13+1.}, carrying the quantum numbers 
$\tau^{4}=\frac{1}{6}$, $\tau^{13} =0$, 
$\tau^{23} =\frac{1}{2}$, $(\tau^{33},\tau^{38})=(0,-\frac{1}{\sqrt{3}})$, 
$Y=\frac{2}{3}$ and $Q=\frac{2}{3} $. 
These two quarks, $d_{R}^{c1} $ and $u_{R}^{c3}$ can bind  together 
with $u_{R}^{c2}$ from the $9^{th}$ line of the same table (at low enough energy, 
after the electroweak transition, and if they belong to a superposition with the left 
handed partners to the first family) -into the colour chargeless baryon - a proton.
This transition is presented in Fig.~\ref{proton is born1.}.

The opposite transition at low energies would make the proton decay.

\vspace{4mm}



Let us conclude this section with the recognition that these scalar triplet and
antitriplet  gauge fields,  since transforming quarks into leptons and
back, have the properties appearing in the literature as leptoquarks. 

%
\section{Fermions and bosons at observable energies}
\label{fermionsbosonslowE}

We confront in  this section 
the {\it standard model} with its rather simple, elegant --- and waiting to be explained ---
assumptions about the properties of fermions coupled to the corresponding vector 
gauge fields and to the Higgs's scalar field with in the literature proposed unifying theories, 
mainly those built on $SO(10)$~\cite{FritzMin}, which unifies the charge groups of quarks 
and leptons separately and  of antiquarks and antileptons separately, unifying 
correspondingly also the vector gauge fields of the charge groups of quarks 
and leptons, and the {\it spin-charge-family} theory of one of us (N.S.M.B.). 
 The unifying theories built on $SU(4)\times SU(2) \times SU(2)$ group, Ref.~\cite{PatiSal}, 
and on $SU(5)$,  Ref.~\cite{GeorGlas}, which are subgroups of the group $SO(10)$, are 
counted under the $SO(10)$-unifying theories. 

All the unifying theories wait to be confirmed by experiments.

Although the $SO(10)$-unifying theories made an important step in understanding 
the origin of charges of quarks and leptons and antiquarks and antileptons, as well
as the origin of  vector gauge fields, they  do not offer the explanation  either 
for the correspondence between the handedness and the charges of quarks and
leptons and the handedness and the charges of antiquarks and antileptons, or 
for the  appearance of families of quarks and leptons and do not unify all the 
vector and scalar fields with gravity. Correspondingly
the $SO(10)$-unifying theories do not offer explanation for the appearance 
of the Higgs's scalar and the Yukawa couplings, except by assuming also the 
existence of scalar fields with the quantum numbers of the vector gauge fields. 

\noindent
One can, namely, just repeat the requirement that not only the vector gauge fields but 
also the scalar gauge fields should exist with the same charges in 
adjoint representations and with the hyper and weak charge of the Higgs's scalar.

We compare the offer of  the $SO(10)$-unifying theories with the offer of
 the {\it spin-charge-family} theory, presented and discussed in this article (and the
articles cited in this article),  to explain the assumptions of the 
{\it standard model}.

We do not discuss supersymmetric models and only  mention shortly  the string theories. 

All the assumptions of the {\it standard model}, cleverly "read" from the 
experimental data, are still waiting to be explained. 
Although there are no evidences  either in experiments or in 
observations, which would help to  show the clear step towards a  simple 
and elegant theory,  the {\it standard model} itself requires  the next step ---  
a simple and elegant theory, which would explain all its assumptions and
would manifest at low 
energy in (the observable) $d=(3+1)$ dimensions what observations and 
experiments suggest.  
%

We namely learn from the scientific experiences  so far that each next step
to simpler and more elegant explanation of the observed phenomena has 
helped to understand better the low of nature.

We hope that the {\it spin-charge-family} theory, which we presented shortly
in this review article, is this new  appropriate
step beyond the {\it standard model}.

\subsection{Standard model of  fermion fields, vector gauge fields,   
and scalar fields} 
\label{SM}

More than 50 years ago the "electroweak and colour" {\it standard model} 
offered an elegant new step in understanding the origin of fermion and boson
fields by postulating properties of massless fermions and 
 vector gauge fields and of massive scalar field before the electroweak break:\\
$\;\;$ {\bf i.} The existence  of several families 
of massless quarks and leptons of the left or right handedness, with the 
charges in the fundamental representations of the weak $SU(2)$ charge, 
colour $SU(3)$ charge, and hyper $U(1)$ charge groups, and of 
massless antiquarks and antileptons, of the same spin $S^{12}$  and of the 
opposite handedness and the opposite charges as quarks and leptons, like it 
is presented in Table~\ref{Table SMF.} for one family, are postulated. Since the  
$\nu_{R}$ lepton and $\bar{\nu}_{L}$ antilepton carry in the {\it standard 
model} no charge at all, as seen in 
Table~\ref{Table SMF.}, the {\it standard model} does not include these
 two members among regular members of a family. 
The hyper charge $Y $ is chosen so that the electromagnetic charges of 
quarks and leptons are in accordance with the measured onces.

Since each quark or lepton carries  either spin $S^{12}= \frac{1}{2}$  or 
$S^{12}=- \frac{1}{2}$, there are $30$ quarks and leptons in each 
family and the same number of antiquarks and antileptons, if one counts
states with the same $p^0=|\vec{p}|$~\footnote{For a chosen 
momentum $\vec{p}$ fermions with spin $S^{12}=- \frac{1}{2}$  carry
opposite momentum $- \vec{p}$, as one can see in Eqs.~(\ref{weylgen0},
 \ref{weylgen05+1}, \ref{weylgen013+1}). In Table~\ref{Table so13+1.} 
the "basis vectors" of one irreducible representation of the Lorentz generator
$S^{ab}$ in $d=13+1$, that is of one family, is presented. Each family
includes  all "basis vectors", those with spin $S^{12}= \frac{1}{2}$ and those
with spin $S^{12}= - \frac{1}{2}$. The requirement that the superposition of
the tensor product, $*_{T}$, of the "basis vectors" and the basis in momentum
or coordinate space solve the equation of motion reduces the number of states
for the factor of $2$.}. 
Counting states with a chosen $\vec{p}$
only, as usually is done, there are correspondingly $15$ quarks and leptons in 
each family and the same number of antiquarks and antileptons. 
Counting $\nu_{R}$ and $\bar{\nu}_{L}$ as regular members of a family, 
there would be for $p^0=|\vec{p}|$ $32$ members of quarks and 
leptons and $32$ members of antiquarks and antileptons in each family. 
Requirement that  $\vec{p}$ and $-\vec{p}$  are treated separately,
the number of quarks and leptons and antiquarks and antileptons  in one
 family is then $32$. 
The relation between handedness and charges of quarks and leptons
and between handedness and charges of antiquarks and antileptons 
is in the {\it standard model} postulated.

The sum of any of these gauge charges over the family members 
quarks and leptons  
is equal to zero, and the same is true also for the gauge charges of 
the antiquarks and antileptons. The family gauge charges are just assumed 
to exist, they are not explained.

The reader can notice the two additional gauge charges in
Table~\ref{Table SMF.}, named ${}^{\tau^{23}}$ and 
${}^{\tau^{4}}$, respectively, separated from the rest of data 
by vertical lines. These two gauge charges do not appear 
in the {\it standard model} at all. Both appear, however, in the 
$SO(10)$-unifying theories and in the {\it spin-charge-family} theory --- 
$\tau^{23}$ as the third component of the second $SU(2)_{II}$
charge, and $\tau^{4}$ as the "fermion" gauge charge.  We shall 
comment them when discussing these two suggested steps beyond
the {\it standard model}.
In  these two kinds of theories the hyper charge $Y$ is equal to $Y=\tau^{4} +
\tau^{23}$, and the electromagnetic charge $Q=Y+\tau^{13}=
\tau^{4} + \tau^{23} + \tau^{13}$. 



The gauge invariance requirement leads to masslessness of quarks and 
leptons and  all vector gauge fields before the electroweak transition. 
It is obviously an 
elegant requirement, since zero and infinity do not need to be justified, as 
both are natural possibility. The massless elementary fermions and gauge 
fields indicate that we are looking at a low energy manifestation of the 
elegant and simple theory, due to which at some scale all the bosons and 
fermions are massless fields, with the Higgs's field included.

The fact that the left and the right handed quarks and leptons  
differ in the gauge charges takes care of their mass protection up to
the point when Higgs's scalar gains a constant value, "dressing" 
the right handed quarks and leptons with the appropriate weak and 
hyper  charge ($\tau^{13}=\frac{1}{2}, Y=-\frac{1}{2}$ for $u_{R}$ 
and $\nu_{R}$, and $\tau^{13}=-\frac{1}{2}, Y=\frac{1}{2}$ for 
$d_{R}$ and $e_{R}$), while the left handed antiquarks and antileptons 
are dressed  with the appropriate weak and hyper charge
($\tau^{13}=-\frac{1}{2}, Y=\frac{1}{2}$ for $\bar{u}_{L}$ and $\bar{\nu}_{L}$, 
and $\tau^{13}=\frac{1}{2}, Y=-\frac{1}{2}$ for 
$\bar{d}_{L}$ and $\bar{e}_{L}$) as well. 
\\

%
\begin{table}
\begin{center}
\begin{minipage}[t]{16.5 cm}  
\caption{Members of each,  $i=(1,2,3)$, of the so far observed families, before the 
electroweak break, are presented.  Each family contains: {\bf i.} The left 
handed weak charged quarks and leptons and the right handed weak 
chargeless quarks and leptons ($\tau^{13}=0$), quarks carrying one of 
the triplet colour charges 
 $(1/2,1/(2\sqrt{3}))$, $(-1/2,1/(2\sqrt{3}))$, $(0,-1/(\sqrt{3})) $,  
leptons are colourless, with the hyper charge $Y$  (and with $Y$ also the 
electromagnetic charge $Q=(\tau^{13}+Y)$) depending on handedness 
of fermions and on  whether fermions carry the colour charge (quarks)
or they are colourless (leptons).
{\bf ii.} The corresponding antiquarks and antileptons carry the same spin 
$S^{12}$ and opposite  handedness and  charges as quarks and leptons.  
The right handed $\nu^{i}_{R}$ and the left handed $\bar{\nu}^{i}_{L}$, 
which would carry all the charges equal to zero, are not accepted as the 
regular members of the antifamily. They are therefore presented in 
Table~\ref{Table SMF.} as $\;{}^{\nu^{i}_{R}}$ and 
${}^{\bar{\nu}^{i}_{L}}$.
%
The hyper charges $Y$ are determined so that the electromagnetic 
charges agree with the observable values. The two columns within the
vertical lines, ${}^{\tau^{23}}$ and ${}^{\tau^{4}}$ have no meaning 
in the {\it standard model}. They are presented here
in purpose for later comparison with the $SO(10)$ theories and with the 
{\it spin-charge-family} theory.}
\label{Table SMF.}
\end{minipage}
\begin{tiny}
\begin{tabular}{|r c r|r| r|r| r r |}
\hline
&{\bf  handedness} &{\bf  weak charge} &  &{\bf colour charge}&& 
{\bf hyper charge}& {\bf  elm. charge }\\
%
{\bf name}    &${\bf -4i S^{03} S^{12}}$&${\bf  \tau^{13}}$  
&${}^{\tau^{23}}$ &$({\bf \tau^{33}, \tau^{38})}$&
${}^{\tau^{4}}$&${\bf  Y}$&    ${\bf Q}=\tau^{13} +Y$\\
\hline
$ {\bf  u^{i}_{L}}$&${\bf -1}$&${\bf \frac{1}{2}}$&$ {}^{0} $ 
&{\bf colour triplet} &${}^{\frac{1}{6}}$&${\bf \frac{1}{6}}$&${\bf \frac{2}{3}}$\\
${\bf  d^{i}_{L}}$&${\bf  -1}$&${\bf -\frac{1}{2}}$& $ {}^{0} $&
{\bf colour triplet}&
${}^{\frac{1}{6}}$&${\bf \frac{1}{6}}$&${\bf -\frac{1}{3}}$\\
\hline
${\bf \nu^{i}_{L}}$&${\bf -1}$&${\bf \frac{1}{2}}$&$ {}^{0} $&{\bf colourless}&
${}^{-\frac{1}{2}}$&${\bf  -\frac{1}{2}}$&${\bf 0}$  \\
${\bf e^{i}_{L}}$&${\bf -1}$ &$ {\bf -\frac{1}{2}}$&$ {}^{0} $ &{\bf colourless}&
${}^{-\frac{1}{2}}$&${\bf -\frac{1}{2}}$&${\bf -1}$ \\
\hline
${\bf  u^{i}_{R}}$&${\bf 1}$&${\bf 0}$ &${}^{ \frac{1}{2}}$&{\bf colour triplet}&
${}^{\frac{1}{6}}$&${\bf  \frac{2}{3}}$&${\bf \frac{ 2}{3}}$\\
${\bf d^{i}_{R}}$&${\bf 1}$& ${\bf 0}$ &${}^{-\frac{1}{2}}$&{\bf colour triplet}&
${}^{\frac{1}{6}}$&${\bf -\frac{1}{3}}$&${\bf -\frac{1}{3}}$\\
\hline
${}^{\nu^{i}_{R}}$&${}^{ 1}$& ${}^{0}$ &${}^{ \frac{1}{2}}$&
${}^{colourless}$&${}^{-\frac{1}{2}}$ & $ {}^{0} $&${}^{0}$           \\
$ {\bf e^{i}_{R}}$&${\bf 1}$& ${\bf 0}$ &${}^{-\frac{1}{2}}$&{\bf colourless}&
${}^{-\frac{1}{2}}$& ${\bf -1} $&${\bf -1}$ \\
\hline\hline
${\bf \bar{ u}^{i}_{R}}$&${\bf 1}$&${\bf - \frac{1}{2}}$&${}^{0}$&
{\bf colour antitriplet}&${}^{-\frac{1}{6}}$&${\bf - \frac{1}{6}}$&
${\bf -\frac{2}{3}}$\\
${\bf  \bar{d}^{i}_{R}}$&${\bf 1}$&${\bf \frac{1}{2}}$&${}^{0}$&
{\bf colour antitriplet}&${}^{-\frac{1}{6}}$& ${\bf -\frac{1}{6}}$&$
{\bf \frac{1}{3}}$\\
\hline
${\bf \bar{\nu}^{i}_{R}}$&${\bf 1}$&${\bf - \frac{1}{2}}$&${}^{ 0}$&
{\bf colourless}&${}^{\frac{1}{2}}$&${\bf \frac{1}{2}}$&${\bf 0}$  \\
${\bf \bar{e}^{i}_{R}}$&${\bf 1}$ &${\bf \frac{1}{2}}$&${}^{0}$&
{\bf colourless}&${}^{\frac{1}{2}}$&${\bf \frac{1}{2}}$&${\bf 1}$ \\
\hline
$ {\bf \bar{u}^{i}_{L}}$&$ {\bf -1}$&${\bf 0}$ &${}^{-\frac{1}{2}}$&
{\bf colour antitriplet}&
${}^{-\frac{1}{6}}$&${\bf  -\frac{2}{3}}$&${\bf -\frac{ 2}{3}}$\\
${\bf \bar{d}^{i}_{L}}$&${\bf - 1}$&${\bf 0}$ &${}^{ \frac{1}{2}}$& 
{\bf colour antitriplet}&
${}^{-\frac{1}{6}}$&${\bf \frac{1}{3}}$&${\bf \frac{1}{3}}$\\
\hline
${}^{\bar{\nu}^{i}_{L}}$&${}^{-1}$& ${}^{0}$ &${}^{-\frac{1}{2}}$&
${}^{colourless}$&${}^{\frac{1}{2}}$& $ {}^{0} $& ${}^{0}$           \\
${\bf \bar{e}^{i}_{L}}$&${\bf -1}$& ${\bf 0}$&${}^{ \frac{1}{2}}$&
{\bf colourless}&
${}^{ \frac{1}{2}}$&$ {\bf 1} $&${\bf 1}$ \\
\hline\hline
\end{tabular}
\end{tiny}
\end{center}
%
\end{table}
$\;\;$ {\bf ii.} The existence  of massless vector fields before the 
electroweak break --- the gauge fields of the observed charges of the family
members quarks and leptons and antiquarks and antileptons --- carrying 
charges in the adjoint representations of the charge groups are postulated: {\bf a.}
 The hyper photon carrying no charge, yet interacting through the 
hyper charge with quarks and leptons. {\bf b.} The  weak charge triplet 
boson, carrying the triplet electromagnetic charge (through relation  $Q= Y 
+ \tau^{13}$, with $Y=0$). {\bf c.} The colour octet  gauge field. 
These vector gauge fields are presented in Table~\ref{Table SMV.}.
For future discussions there appear in the same table two additional 
vector gauge fields, suggested by the $SO(10)$-unifying theories as well 
by the {\it spin-charge-family} theory: The additional $SU(2)$ triplet 
vector gauge field  and the $U(1)$ singlet vector gauge fields. These
vector gauge fields are discussed in Sect.~\ref{vector3+1} from the
point of view of the {\t spin-charge-family} theory.\\ 
%
\begin{table}
\begin{center}
\begin{minipage}[t]{16.5 cm}  
\caption{Massless vector gauge fields in $d=(3+1)$ before 
the electroweak break are presented. They are the gauge 
fields of the hyper, weak  and colour charges, carrying all the charges  in 
the adjoint representations. None of them carries the second weak charge
and "fermion" charge, presented in the fourth and the last column, 
respectively, added because of the two additional vector fields, 
presented below the first three lines and separated 
from the above three by the horizontal line. Both vector gauge fields 
appear in $SO(10)$-unifying theories and the {\it spin-charge-family} theory. 
One of these two vector gauge fields,  the triplet gauge field of the
(second $SU(2)_{II}$) $\vec{\tau}^{2}$ charge, is with respect to the 
{\it standard model} the new field. The second one, the singlet vector 
gauge field of the "fermion" charge $\tau^{4}$, explains 
the vector gauge field of the hyper charge $Y$  of the 
{\it standard model}, since it appears  as the superposition of the $\tau^4$
vector gauge field and the gauge vector field of $\tau^{23}$, the third 
component of 
the new $\vec{\tau}^{2}$ triplet gauge field. We present these two gauge 
fields here for later comments on the {\it standard model} assumptions 
from the point of view the 
$SO(10)$-unifying theories and the {\it spin-charge-family} theory, 
Sect.~\ref{vector3+1}.}
\label{Table SMV.}
\end{minipage}
\begin{tabular}{|r|c|r|r|r|r|c|c|}
\hline
name & hand- &  weak &${}^{weak}$& hyper   
&colour & elm &${}^{"fermion"}$ \\ 
&edness &charge&${}^{ charge_{II}}$&charge&charge&charge&${}^{charge}$\\
\hline
 hyper photon&$ 0$              &$ 0$   &${}^{0}$&  $0$ & colourless&$0$&
${}^{0}$\\
&&&&&&&\\
 weak bosons &$ 0$              & triplet &${}^{0}$&  $0$& colourless&
  triplet&${}^{0}$  \\
&&&&&&&\\
 gluons    & $0$              &$0$    &  ${}^{0}$&$0$   &  colour octet&$0$&${}^{0}$ \\
\hline
&&&&&&&\\
weak bosons$_{I}$ &$ 0$ &$0$ & triplet &  $0$ & colourless& 
 triplet &$0$\\
&&&&&&&\\
hyper photon$_{I}$ &$ 0$ &$0$ & $0$    &  $0$& colourless &
 $0$   &$0$\\
&&&&&&&\\
\hline
\end{tabular}
  \end{center}
%
\end{table}
$\;\;$ {\bf iii.} The  existence of the scalar gauge field, with the weak and 
hyper charges $\pm \frac{1}{2}$ and  $\mp \frac{1}{2}$, respectively, 
coupled to different families of quarks and leptons according to Yukawa 
couplings is postulated.
The Higgs's scalar field, after gaining constant values at the electroweak break, 
gives masses  to quarks and leptons and weak charge gauge fields, due to the
interactions of the weak gauge fields and quarks and leptons and antiquarks 
and antileptons with the weak and hyper charge of the  scalar 
field with the nonzero constant value~%
\footnote{The interaction of 
quarks and leptons with the constant scalar field resembles  the 
interaction of electrons with the ions of the crystal during tunneling of the
electrons  through it, what causes the change of the mass of electrons.}. 
The scalar higgs is presented in Table~\ref{Table SMS.}.  There are only 
 components of two doublets presented in the Table~\ref{Table SMS.},
which gain constant values at the electroweak break, presented 
in the table in bold phase.
In the same table the two triplets and three singlets, discussed in
Sect.~\ref{scalar3+1}, are presented for future discussions
on the next step beyond the {\it standard model}, offered
by the  {\it spin-charge-family} theory. There are three singlets 
and two triplets, which explain the origin of the Higgs's scalar and the
Yukawa couplings, Eq.~(\ref{commonAi}).\\ 
\begin{table}
\begin{center}
\begin{minipage}[t]{16.5 cm}   
\caption{The Higgs's scalar is a massive scalar field in 
$d=(3+1)$, with the charges like there would be in the fundamental 
representation of the weak and hyper charge groups. 
Only  ${\bf < Higgs_{u}>}$ and
 ${\bf < Higgs_{d}>}$ gain (nonzero vacuum expectation) constant values, 
the two other components do  not obtain the constant values. The explanation
for the appearance of the scalar gauge fields is offered by the {\it spin-charge-%
family} theory in Sect.~\ref{scalar3+1}. The two triplets,
$ \vec{\tilde{A}}^{\tilde{1}}_{\scriptscriptstyle{\stackrel{78}{(\pm)}}}$,
$ \vec{\tilde{A}}^{\tilde{N}_{\tilde{L}}}_{\scriptscriptstyle{\stackrel{78}{(\pm)}}}
$, and the three singlets, 
$A^{Q}_{\scriptscriptstyle{\stackrel{78}{(\pm)}}},
A^{Y}_{\scriptscriptstyle{\stackrel{78}{(\pm)}}}, 
A^{4}_{\scriptscriptstyle{\stackrel{78}{(\pm)}}}$, 
Eq.~(\ref{eigentau1tau2}), presented below the double line, replace and explain the 
higgses and Yukawa couplings of the 
{\it standard model}.}
\label{Table SMS.}
\end{minipage}
\begin{tiny}
\begin{tabular}{|r|c|r|r|r|c|}
\hline
{\bf name} &{\bf handedness} &{\bf weak charge} &{\bf hyper charge}& 
{\bf colour charge}&{\bf  elm charge }\\
\hline
&&&&&\\
${}^{Higgs_{u}}$&${}^{0}$&${}^{ \frac{1}{2}}$& 
${}^{\frac{1}{2}}$&${}^{colourless}$
 &${}^{1}$\\
                                        &&&&&\\
${\bf <Higgs}$${\bf_{d}>}$&$ {\bf 0}$&${\bf - \frac{1}{2}}$&
${\bf \frac{1}{2}}$&{\bf colourless}&${\bf 0}$\\                                        
\hline
%
&&&&&\\
 ${\bf < Higgs_{u}>}$&${\bf 0}$&${\bf \frac{1}{2}}$&${\bf  -\frac{1}{2}}$&
{\bf colourless }&${\bf 0}$\\
                                        &&&&&\\
 ${}^{ Higgs_{d}}$ &${}^{0}$&${}^{- \frac{1}{2}}$&
                                        ${}^{ -\frac{1}{2}}$& 
                                        ${}^{colourless}$&${}^{-1}$\\                                        
\hline\hline
&&&&&\\
$A^{Q}_{\scriptscriptstyle{\stackrel{78}{(\pm)}}},
A^{Y}_{\scriptscriptstyle{\stackrel{78}{(\pm)}}}, 
A^{4}_{\scriptscriptstyle{\stackrel{78}{(\pm)}}}$&$0$&
$\mp\frac{1}{2}$&$\pm\frac{1}{2}$&colourless&$0$\\
$ \vec{\tilde{A}}^{\tilde{1}}_{\scriptscriptstyle{\stackrel{78}{(\pm)}}}$&$0$&
$\mp\frac{1}{2}$&$\pm\frac{1}{2}$&colourless&$0$\\
&&&&&\\ 
$ \vec{\tilde{A}}^{\tilde{N}_{\tilde{L}}}_{\scriptscriptstyle{\stackrel{78}{(\pm)}}}
$&$0$&$\mp\frac{1}{2}$&$\pm\frac{1}{2}$&colourless&$0$\\
&&&&&\\
\hline
\end{tabular}
\end{tiny}
\end{center}
%
\end{table}
%


%
%
The Lagrange densities for fermions ($\psi $ includes all the quarks and leptons 
and antiquarks and antileptons of all the observed families) 
and for the vector gauge fields $A^{Ai}_{m}$ of the {\it standard model} 
before the electroweak break is
\begin{eqnarray}
\label{actionSMFV}
{\mathcal L}_f &=&  \bar{\psi}\gamma^{m} (p_{m}- \sum_{A,i}\; 
g^{Ai}\tau^{Ai} 
A^{Ai}_{m}) \psi \,, \nonumber\\
\{\tau^{Ai}, \, \tau^{Bj} \}_{-}&=& \delta^{Ab}\, i \,f^{Aijk} \tau^{k}\,,\nonumber\\
{\mathcal L}_{vg} &=& -\sum_{Ai}\, \frac{\varepsilon^A}{4} \, F^{A i}{}_{m n} \,
F^{A i m n}\,,\nonumber\\
  F^{A i}{}_{m n} &= & -i (p_{m} A^{Ai}_n - p_{n} A^{Ai}_m) -f^{Aijk}
A^{Aj}_m A^{Ak}_n\,,\nonumber\\
\end{eqnarray}
%
 with  the structure constants  $f^{Aijk}$
concerning the colour $SU(3), A=3$, weak $SU(2),A=1$ and hyper charge $U(1), 
A=2$ groups.  

The Lagrange density of the  Higgs's scalar field is before the electroweak phase 
transition equal to
\begin{eqnarray}
\label{SMShB}
{\cal L}_{sH} &=& \{(p_{m} \,\phi)^{\dagger} \,(p^{m} \,
\phi) - ( m_{\phi}^2 \,\phi^{ \dagger} \phi 
+  \lambda\,(\phi^{\dagger} \phi )^2\,)\,.
\end{eqnarray}
After  the electroweak phase transition $p_{m} \phi$ become $p_{0m} \phi$ 
$= (p_m +g^1 \vec{\tau}^{1} \, \vec{A}_m^{1} +g^Y YA_m^Y)\,\phi$,
since the constant higgs field, ${\bf <Higgs}$${\bf_{(d,u)}>}$, causes 
coupling to the weak vector gauge field.
The postulated Yukawa couplings together with the higgs $\phi$,  gaining 
the constant value, require coupling of quarks and leptons to the higgs as well
\begin{eqnarray}
\label{YH}
{\cal L}_{YH}&=& - \sum_{f f' i}\, Y^{f f'}_{i}\,\psi^{i \dagger}_{Lf}\gamma^0 \,
<Higgs_{i}>\, \psi^{i}_{Rf'}\,,
\end{eqnarray}
with $Y^{f f'}_{i}$ which is for each family member $i=(u,d,\nu, e)$ 
$3\times 3$ matrix, $(f,f')$ represent one of the three families of the family 
members, and $<Higgs_{i}>=<Higgs_{u}>$  for $i=(u, \nu) $
and  $<Higgs_{i}>$ $=<Higgs_{d}>$  for $i=(d, e) $. At the electroweak
break  $( m_{\phi}^2 \,\phi^{ \dagger} \phi  +
 \lambda\,(\phi^{\dagger} \phi )^2\,)$ (is assumed to) changes to 
$(- m_{\phi}^2 \,\phi^{ \dagger} \phi 
+ \lambda\,(\phi^{\dagger} \phi )^2\,)$, with the higgs obviously gaining 
the imaginary mass.








%
\subsection{Discussions on  $SO(10)$ unifying theories and 
{\it spin-charge-family} theory}
\label{SO10SCFT}



In both theories, the $SO(10)$-unifying theories and  the {\it spin-charge-family} 
theory, the right handed neutrino is the regular member of  quarks and leptons 
and the left handed antineutrino is the regular member of  antiquarks and 
antileptons, with the difference, that in the {\it spin-charge-family} theory one
irreducible representation of the Lorentz group $SO(13+1)$ contains both quarks
and leptons and antiquarks and antileptons with the handedness and charges
uniquely related. (The Lorentz group $SO(13+1)$ contains as the subgroups
$SO(3+1)$ and $SO(10)$.) This is not the case for the $SO(10)$-unifying 
theories, in which the relation between handedness and charges must be 
assumed, like in the {\it standard model}.
 
Right handed neutrino and left handed antineutrino namely carry the 
$\tau^{23}$ charge of the additional $SU(2)_{II}$ charge 
group, $\nu_{R}$ has $\tau^{23}=\frac{1}{2}$ and $\bar{\nu}_{L}$ has 
$\tau^{23}=-\frac{1}{2}$, as presented in Table~\ref{Table SMF.} (the same 
can be seen also in Table~\ref{Table so13+1.}, representing  in the 
{\it spin-charge-family} theory the "basis vectors" of all the family 
members of one particular family out of two times two groups of four 
families, presented in Table~\ref{Table III.}).

In both theories fermions carry  also the "fermion" gauge charge $\tau^{4}$,
quarks  have $\tau^4 =\frac{1}{6}$, antiquarks $\tau^4 =-\frac{1}{6}$,
leptons have $\tau^4 =-\frac{1}{2}$, antileptons have $\tau^4 =\frac{1}{2}$.
The hyper charge is in both cases equal to $Y=\tau^4 + \tau^{23}$.

Both theories predict the existence of the corresponding triplet, 
$\vec{A}_{m}^{2}$, vector gauge field and  the singlet, $A_{m}^{4}$,  
vector gauge fields, the superposition of which is observable at low energies 
as the hyper charge $Y= (\tau^{23} + \tau^4)$ vector gauge field,
$A_{m}^{Y}$, assumed by the {\it standard model}.



The main differences between these two searches --- $SO(10)$-unifying
theories~\cite{FritzMin} (with $SU(5)$~\cite{GeorGlas} and $SU(4) \times SU(2)$
$\times SU(2)$~\cite{PatiSal} as the subgroups of the $SO(10)$ group included) 
and the {\it spin-charge-family} theory~\cite{norma93,norma95,norma2001,IARD2016,%
prd2018,n2019PIPII} --- for steps beyond the {\it standard model},
 which could explain the {\it standard model} assumptions, are: \\
{\bf i.} The $SO(10)$-unifying theories use, as all the literature does, the Dirac's 
second quantized theory for the description of massless fermions in 
$d=(3+1)$-dimensional space-time, as discussed in 
Sects.~\ref{internalspaceordinary} and \ref{poincare}, and 
Sects.~\ref{creationtensorusual},%
~\ref{creationannihilationtensor} of this paper. 

The {\it spin-charge-family} theory
describes the internal space of massless fermions with the Clifford algebra 
in $d=(13+1)$-dimensional space-time, Sects.~\ref{internalspace}, 
\ref{GrassmannClifford}, of this paper, 
what not only offers the explanation for the second quantization postulates of 
Dirac, Sect.~\ref{creationannihilationtensor}, but offers at the same 
time the explanation for the appearance of families of fermions, as we explain 
step by step in Sect.~\ref{internalspace},  and in particular in%
~\ref{reduction},~\ref{creationtensorClifford} and~\ref{creationannihilationtensor}. \\
{\bf ii.} The $SO(10)$-unifying theories unify all the charges by analyzing the
subgroups of the  $SO(10)$ group from the point of view of the 
{\it standard model} groups,  assuming the existence of vector gauge fields 
to the corresponding charge groups, unifying consequently also
all the vector gauge fields, but they  do not explain either the 
appearance of families of fermions or the appearance of the Higgs's 
scalar and Yukawa couplings, or they suggest the common origin of all the
forces with gravity included. 

The {\it spin-charge-family} theory assumes the 
simple starting action, Eq.~(\ref{wholeaction}), in $d=(13 +1)$-dimensional
space, with fermions interacting with gravity only --- with the vielbeins and the 
two kinds of the spin connection fields~%
\footnote{The spin connection fields are expressible uniquely with vielbeins if 
there are no fermion condensates present, Eq.~(\ref{omegaabe})), 
Sect.~\ref{fermionandgravitySCFT}.},  
unifying gravity and vector and scalar gauge fields. Fermions with the spins defined 
in  $SO(13 +1)$ manifest in $d=(3+1)$ 
the ordinary spin and handedness, charges and families unified, 
Sect.~\ref{internalspace}.
The starting action, Eq.~(\ref{wholeaction}), manifests in $d=(3+1)$ all the 
vector gauge fields, Sect.~\ref{vector3+1}, assumed  by the 
{\it standard model}, as well as the scalar gauge fields, 
Sect.~\ref{scalar3+1}, with the properties of the Higgs's scalar, what 
explains  the appearance of the Higgs's scalar and the Yukawa couplings
of the {\it standard model}, 
predicting the number of families and the symmetry of mass matrices, discussed
in Sect.~\ref{scalar3+1},  Eq.~(\ref{M0}).\\
{\bf iii.} In both theories, the $SO(10)$-unifying theories and the 
{\it spin-charge-family} theory, the breaks of the starting symmetry lead at low 
energies to  the {\it standard model} content of the charges and 
correspondingly also of the vector gauge fields of charges~%
\footnote{The break of the starting symmetry should appear spontaneously 
during the expansion of the universe. To evaluate this from the simple starting
action assumed in the {\it spin-charge-family} theory, without knowing the 
boundary conditions, is impossible, but yet we can do a lot, since we know
the low energy manifestation of quarks and leptons, antiquarks and
antileptons, of  the vector gauge fields and of the scalar gauge fields.}. 

The $SO(10)$ group can break into $SU(3)\times U(1)\times SU(2) \times SU(2)$
either over $SO(6)\times SO(4)$ or over $SU(4)\times SO(4)$ or over 
$SU(5)\times U(1)$.

In order to keep the handedness and charges in 
relations as presented in Table~\ref{Table so13+1.}  and what the 
{\it standard model} assumes, the $SO(13 +1)$ group must  break first to 
$SO(7,1)\times SU(3) \times U(1)$ 
(in both sectors, in $SO(13 +1)$ sector determining the spins, handedness and 
charges of family members of different irreducible representations (families), 
and in $\widetilde{SO}(13 +1)$ sector, which  connects particular 
member of a particular family with the same family member of all the families,
equipping different irreducible representations of $SO(13 +1)$ with the family 
quantum numbers),  and then further to  
$SO(3,1) \times  SU(2) \times SU(2) \times SU(3)\times U(1)$~%
\footnote{If we would make the approximation 
that all the $(14-4)$ extra dimensions are curved with almost spheres, 
Sect.~\ref{TDN0}, of the same radius, or even with different radii, 
we could have a kind of $SO(10)$ theory, but in the way of the 
{\it spin-charge-family} theory, that is  with 
the gravity as the only gauge field manifesting in $d=(3+1)$ as the ordinary gravity,
the vector and the scalar gauge fieds and with fermions manifesting families, the
massess  of which are determined by the scalar gauge fields, 
Sect~\ref{overviewSCFT}, paragraph {\bf v.}.}. 
Sect.~\ref{TDN0} explains in the case of a toy model how the breaks
of symmetries can spontaneously  appear in the {\it spin-charge-family} theory.\\
%
{\bf iv.} The way of breaking symmetries determines in both cases 
the coupling constants of fermions to the corresponding vector gauge fields,
manifesting at the low energy regime. In the {\it spin-charge-family} theory 
the way of breaking of symmetries determines also the coupling constants 
of fermions to the scalar gauge fields.%
\\
{\bf v.} The "miraculously" cancellation of the triangle anomalies of the
{\it standard model} both, the $SO(10)$-unifying theories and the 
{\it spin-charge-family} theory, explain, App.~\ref{appanomalies}, the first  
one after relating charges and handedness by assumption, in the second one
 the handedness and charges are related due to the unification of
spins and charges.




\subsubsection{Short overview of  $SO(10)$ {\it unifying 
theories}~\cite{FritzMin,PatiSal,GeorGlas}}.
\label{overviewSO(10)}

The $SO(10)$-unifying theories offer the explanation for some of the postulates 
of the {\it standard model}. Namely, assuming the $SO(10)$  unifying charge 
group and the existence of the corresponding vector gauge fields with the charges
in the adjoint representations, the theories do treat all the charges in an unique 
way and correspondingly also all the vector gauge fields in an unique way, 
postulating the existence of the charges and vector gauge fields only
once.
The theories could repeat this game also for the scalar gauge fields, but the
scalar index of the {\it spin-charge-family} theory, which naturally appears 
from the simple starting action, Sect.~\ref{vectorscalar3+1}, since the gravity 
in $d=(13+1)$ manifests in $d=(3+1)$ as ordinary gravity and all the
vector and scalar gauge fields appearing in the starting action,
 Eq.~(\ref{wholeaction}), has to be in the $SO(10)$-unifying theories  
postulated like in the {\it standard model}.


{\bf i.} In  the Dirac's second quantization procedure there are in $d=(3+1)$
massless fermions with the half integer spins described by the fundamental 
representation of the Lorentz group $SO(3,1)$, with the spins, $S^{12}$ and 
$S^{03}$, which are the members of the (chosen)  Cartan subalgebra 
of the $SO(3,1)$ algebra, defining as well the handedness, $-4i \,S^{03} S^{12}$, 
as presented in Sect.~\ref{creationannihilationtensor} of this paper.

In the $SO(10)$ unifying theories the second quantized fermions carry spins and 
handedness as in the {\it standard model},  charges are in the 
fundamental representation of the subgroups of the group $SO(10)$, carrying in 
addition to the charges of the {\it standard model}, presented in 
Table~\ref{Table SMF.} with the eigenvalues of the operators 
$\tau^{13}$ (the weak charge), $(\tau^{33}, \tau^{38})$ (the colour charge)
and the hyper charge $Y$, also the charges $\vec{\tau}^{2}$  and $\tau^{4}$, 
presented as well in the same table~\ref{Table SMF.}~%
\footnote{The group $SO(10)$ has
$5$ commuting operators, 5 members of the Cartan subalgebra, presented in 
Table~\ref{Table SMF.}: $\tau^{13}$ (the weak $SU(2)$ group of the
$\vec{\tau}^{1}$ charge), $\tau^{23}$ (the second $SU(2)$ group of the
 $\vec{\tau}^{2}$ charge, the weak $SU(2)$ group and the second $SU(2)$ 
group are the subgroups of the $SO(4)$ group), 
$(\tau^{33}, \tau^{38})$  (the colour SU(3) group of $\vec{\tau}^{3}$ 
charge) and $\tau^{4}$ (the $U(1)$ group, called in  this paper the "fermion" 
charge, the $U(1)$  group and $SU(3)$ group are the subgroups of the 
$SO(6) $ group~\cite{FritzMin} (their expressions with the infinitesimal generators of 
the  Lorentz group are written in Eqs.~(\ref{so42}, \ref{so64}),  their relations
to the hyper charge and the electromagnetic charge are written in 
Eq.~(\ref{YQY'Q'andtilde}) of this paper), or the subgroup of 
$SU(4)$~\cite{PatiSal}.  
The hyper charge is in both cases equal to  $Y=\tau^{23}+ \tau^4$.}.\\ 
%
%
%
{\bf ii.} To all the charges of fermions the corresponding massless vector
gauge fields correspond, carrying charges in adjoint representations: the weak 
$SU(2)$ is the triplet, the second $SU(2)_{II}$ is the triplet, the colour $SU(3)$ 
is the octet and the new one $U(1)_{I}$, which is the singlet, representing
the {\it standard model} hyper charge photon singlet as the superposition 
of the $A^{23}_{m}$ vector gauge field (of  $\tau^{23}$ charge) and this 
second $U(1)$  singlet vector
gauge field. In Table~\ref{Table SMV.} the triplet  vector gauge field of the
second $SU(2)_{II}$ charge is presented as {\it weak bosons$_{I}$} and
the singlet $U(1)_{I}$ as the {\it hyper photon$_{I}$}.\\
{\bf iii.} The $SO(10)$ unifying theories can postulate the  
appearance of families, similarly as the {\it standard model} does, that is 
by introducing additional, this time the family, group and correspondingly 
do not explain the appearance of the Higgs's scalar fields and Yukawa couplings 
of the {\it standard model}, Table~\ref{Table SMS.}. \\
%
{\bf iv.} The relations among the coupling constants and correspondingly 
the unification scale depend on the way of breaking the starting $SO(10)$ to the 
{\it standard model} $SU(3)\times U(1) \times SU(2)$ symmetry groups. 
These theories must postulate the scalar fields to break these symmetries
as well.\\
%
{\bf v.} In usual $SO(10)$-unifying theories one gets in the first approximation 
the same requirement for the gauge couplings being unified in the same way as in
SU(5) (which is a subgroup of the $ SO(10)$) after with the renormalization 
group extrapolated coupling constants to the unification scale~\footnote{
Let us comment that either in the $SO(10)$-unifying theories or 
in the {\it spin-charge-family} theory there is no need that all three 
coupling constants should meet at the same point, due to several breaks 
of symmetries~\cite{NH2002}.}.

%

\subsubsection{ Short overview of the {\it spin-charge-family} 
theory
~\cite{norma93,norma95,norma2001,pikan2003,pikan2006,IARD2016,%
prd2018,n2019PIPII}}
\label{overviewSCFT}

The {\it spin-charge-family} theory of one of us (N.S.M.B.)%
~\cite{norma93,norma95,norma2001,pikan2003,pikan2006,IARD2016,%
prd2018,n2019PIPII} assumes in $d=(13+1)$-dimensional space simple
action, Eq.~(\ref{wholeaction}), for massless fermions  and for massless 
vielbeins and two kinds of the spin connection fields, with which fermions
 interact. Description of the internal degrees of fermions by 
the odd Clifford algebra objects offers the unique explanation of spins,
charges and families, Sect.~\ref{internalspace}, explaining the second 
quantization postulates of Dirac, Sect.~\ref{poincare}.


%


{\bf i.} {\it The odd Clifford algebra describes the whole internal space of 
fermions, Sect.~\ref{internalspace}.}\\
The {\it spin-charge-family} theory assumes massless fermions
in $d=(13 +1)$-dimensional space, describing the internal space of fermions 
by the odd Clifford algebra, which offers  besides the description of spins
and with respect to $d=(3+1)$ spins and charges also the appearance of 
families of fermions, Sects.~\ref{internalspace}, \ref{reduction}, predicting
the number of families and the symmetry of their mass matrices, 
Sect.~\ref{scalar3+1}.\\
{\bf i.a.} {\it The "basis vectors"} of the odd Clifford algebra presentations of 
the Lorentz group {\it anticommute}, 
Sects.~\ref{GrassmannClifford}, \ref{HilbertCliffapp}.
Anticommutativity of the "basis vectors" dictates the anticommuting properties 
to the creation 
operators and their Hermitian conjugated partners annihilation operators,
Sects.~\ref{creationtensorClifford}, \ref{actionClifford}, 
what explains the Dirac's second quantization postulates, 
Sect.~\ref{creationannihilationtensor}, Eq.~(\ref{relationDN}).\\ 
{\bf i.b.} {\it Each irreducible representation of the Lorentz group}, analyzed 
with respect to the {\it standard model} groups, Table~\ref{Table so13+1.},
  includes quarks and leptons and antiquarks and antileptons related to 
handedness as required by the {\it standard model}. (The $SO(10)$ unifying
theories must relate charges and handedness "by hand", following the {\it standard 
model}.)\\
{\bf i.c.} {\it The spins in} $d>(3+1)$, if analyzed with respect to the 
{\it standard model} groups, Eqs.~(\ref{so1+3}, \ref{so42}, \ref{so64}),
 manifest
at low energies charges and spins and handedness of quarks and leptons 
and antiquarks and antileptons assumed by the {\it standard model}. 

Since  $SO(10)$ group is together with the $SO(3,1)$ the subgroup of 
$SO(13,1)$, in both groups,  $SO(10)$ and  $SO(13,1)$, the $SU(2)_{II}$ 
charge group appears together with the weak $SU(2)_{I}$ charge group
as the subgroups of the $SO(4)$ group. The $U(1)$ charge group 
appears  together with the $SU(3)$ colour charge group  as subgroups 
of $SO(6)$ group (while in Ref.~\cite{PatiSal} the charge group $U(1)$ 
belongs together with the $SU(3)$ colour charge group  to the $SU(4)$ charge 
group). The infinitesimal generators of the groups $SU(2)_{II}$ and $U(1)$
are denoted in Table~\ref{Table SMF.} and in Table~\ref{Table so13+1.} as 
$\tau^{23}$ and $\tau^4$ charges, respectively%
\footnote{The group $SO(13+1)$ has $\frac{d}{2}=7$ members 
of the Cartan subalgebra, unifying spins and charges and relating handedness
of quarks and leptons and antiquarks and antileptons to charges.}. \\

{\bf ii.} {\it Fermions interact with the gravity only, manifesting in $d=(3+1)$
all the observed fermion fields coupled to the observed vector and scalar 
gauge fields) as well as gravity}, Ref.~\cite{nd2017}. \\
Gravity is represented by the vielbeins (the 
gauge fields of momenta) and the two kinds of the spin connection fields 
(the gauge fields of $S^{ab}$ and $\tilde{S}^{ab}$), discussed in 
Sect.~\ref{fermionandgravitySCFT}. In Eq.~(\ref{wholeaction}) the 
simple starting action in $d=(13 +1)$ is presented, manifesting in $d=(3+1)$
all the observed fermion fields coupled to the observed vector and scalar 
gauge fields) as well as gravity. \\
{\bf ii.a.} In Sect.~\ref{fermionactionSCFT}, 
in Eq.~(\ref{faction}) 
{\it the fermion Lagrange density is presented, manifesting the couplings to the
vector gauge fields and the scalar gauge fields}. As seen in Table~%
\ref{Table III.} there remain after the (assumed spontaneous) break of the
starting  symmetry $SO(13,1)$ to $SO(7,1)\times SU(1)\times U(1)$ two groups 
of four massless families of quarks and leptons.\\
{\bf ii.b.}
{\it Both groups of four families manifest $SU(2)\times SU(2)\times U(1)$ 
symmetry} as presented in Eq.~(\ref{M0}) of Sect.~\ref{scalar3+1}. 
The three $U(1)$ singlets, $A^{Q}_{\scriptscriptstyle{\stackrel{78}{(\pm)}}},
A^{Y}_{\scriptscriptstyle{\stackrel{78}{(\pm)}}},
A^{4}_{\scriptscriptstyle{\stackrel{78}{(\pm)}}}$, contributing to both
groups of four families, are responsible for the  difference in masses of different
 family members, of ($u$, $d$,  $\nu$, $e$). \\

{\bf iii.} {\it The vielbeins and the two kinds of the spin onection fields manifest
in $d=(3+1)$ all the vector gauge fields and also the scalar gauge fields. }\\
In Ref.~\cite{nd2017} it is shown how 
do the vielbeins and spin connections (if there is no condensate present
any of three determines the rest two, as seen in Eq.~(\ref{omegaabe}))
 manifest either as vector gauge 
fields, Sect.~\ref{vectorscalar3+1}, Sect.~\ref{vector3+1},
offering the explanation for the origin of the {\it standard model} vector 
gauge fields, or the scalar gauge fields, Sect.~\ref{scalar3+1}, 
offering the explanation for the Higgs's scalar and Yukawa couplings, and 
additional scalar gauge fields~%
\footnote{The way of breaking symmetries predicts the number of massless 
families~\cite{familiesNDproc} before the electroweak break, and the 
symmetries of mass matrices, as well as the existence of the additional 
scalar fields, Eq.~(\ref{faction}).}.\\
{\bf iii.a.}
In Table~\ref{Table SMV.} {\it the triplet  vector gauge field of the second 
$SU(2)_{II}$ charge} (denoted by $\vec{\tau}^{2}$) 
is presented 
as the {\it weak bosons$_{I}$}
and {\it the singlet $U(1)_{I}$ vector gauge field of the "fermion" charge 
$\tau^{4}$} is presented as the {\it hyper photon$_{I}$}.

 The $SU(2)_{II}$ 
 vector gauge field appears in addition to the {\t standard model} gauge fields, 
while the {\it standard model} hyper charge vector gauge field is the 
superposition of $\tau^{23}$ component of the $SU(2)_{II}$ 
 vector gauge field and the $U(1)_{I}$ singlet vector gauge field~%
\footnote{The condensate, Table~\ref{Table con.}, of the two right 
handed neutrinos with the
family quantum numbers of the four families which do not belong to 
the first four families in Table~\ref{Table III.}, couple to the triplet  
vector gauge field of the second $\vec{\tau}^{2}$ charge  and
to the singlet  vector gauge field of the "fermion" charge 
$\tau^{4}$, making one of the superposition of these two vector gauge fields
massive, while the other remains massless as the hyper charge vector
gauge field. The weak charge vector gauge field, the colour charge vector 
gauge field and this hyper charge vector gauge fields remain massless, since
they do not couple to the condensate. The condensate has the corresponding
gauge charges equal to zero.}. \\
{\bf iii.b.} {\it The scalar gauge fields with the space index  $s=(7,8)$
determine after the electroweak break masses of two times four families.}

 The scalar gauge fields, which are the superposition of $\tilde{\omega}_{abs}, 
s=(7,8),$  with either $(a,b)=(0,1,2,3)$ or $(a,b)=(5,6,7,8)$,  or the
superposition of $\omega_{t,t',s}$, carrying the charges $(Q,Y,\tau^4)$ with 
respect to $(a, b)$, Sect.~\ref{scalar3+1}, form two groups of two 
triplets (they are superposition of $\tilde{\omega}_{ab s}$) and three singlets 
(they are superposition of $\omega_{t,t',s}$), Eq.~(\ref{commonAi}). 

All the triplets and the singlets with the space index ($s=(7,8)$) have the weak 
and the hyper charge equal
to either ${\cal \tau}^{13}=\frac{1}{2}$ and $ Y=- \frac{1}{2}$ or
 ${\cal \tau}^{13}=-\frac{1}{2}$ and $ Y= \frac{1}{2}$, 
Table~\ref{Table doublets.}. 

One group of two triplet scalar fields and the three singlet scalar fields are 
presented in Table~\ref{Table SMS.} in the lowest part of
the table. The three singlets are denoted by 
$A^{Q}_{\scriptscriptstyle{\stackrel{78}{(\pm)}}},
A^{Y}_{\scriptscriptstyle{\stackrel{78}{(\pm)}}}, 
A^{4}_{\scriptscriptstyle{\stackrel{78}{(\pm)}}}$ and the two triplets are
denoted by $ \vec{\tilde{A}}^{\tilde{1}}_{\scriptscriptstyle{\stackrel{78}{(\pm)}}}$
 and  
$ \vec{\tilde{A}}^{\tilde{N}_{\tilde{L}}}_{\scriptscriptstyle{\stackrel{78}{(\pm)}}}$.
They all origin in the simple starting action, Eq.~(\ref{wholeaction}), and manifest in 
$d=(3+1)$ as superposition of the spin connection fields of both kinds, 
$\omega_{t t' s}, s= (7,8)$ and $\tilde{\omega}_{ab s}, s= (7,8)$, 
Eq.~(\ref{commonAi}).\\
{\bf iii.c.}
{\it The scalar gauge fields with the space index $s=(7,8)$ are expected to gain
a constant values}.

The scalar gauge fields, the two  groups of triplets and the three singlets, 
are expected to gain at low energies the imaginary masses (on the present stage 
of studying properties of the scalar gauge fields this break of symmetry is 
assumed as in the {\it standard model}, although the 
{\it spin-charge-family} does propose the origin of the scalar gauge fields and 
 their starting couplings and consequently their properties) as presented in 
Eq.~(\ref{interactingphi}). This makes 
these scalar fields manifesting   constant values, breaking correspondingly mass 
protection of quarks and leptons, since the constant values of the
vacuum carry the weak and the hyper charge of $\pm \frac{1}{2}, 
\mp \frac{1}{2}$, respectively, Table~\ref{Table doublets.}.  This work is in 
progress for a few years already. 


The three singlets, $A^{Q}_{\scriptscriptstyle{\stackrel{78}{(\pm)}}},
A^{Y}_{\scriptscriptstyle{\stackrel{78}{(\pm)}}},  
A^{4}_{\scriptscriptstyle{\stackrel{78}{(\pm)}}}$ and the two 
triplets, $ \vec{\tilde{A}}^{\tilde{1}}_{\scriptscriptstyle{\stackrel{78}{(\pm)}}}$  
$ \vec{\tilde{A}}^{\tilde{N}_{\tilde{L}}}_{\scriptscriptstyle{\stackrel{78}{(\pm)}}}$, 
coupling to the four of the eight families of fermions presented in 
Table~\ref{Table III.}, determine the mass matrices of the observed three
 families of quarks and leptons, predicting the fourth family to the observed three.

The same three singlets, $A^{Q}_{\scriptscriptstyle{\stackrel{78}{(\pm)}}},
A^{Y}_{\scriptscriptstyle{\stackrel{78}{(\pm)}}},  
A^{4}_{\scriptscriptstyle{\stackrel{78}{(\pm)}}}$ and the second kind of 
two triplets, $ \vec{\tilde{A}}^{\tilde{2}}_{\scriptscriptstyle{\stackrel{78}{(\pm)}}}$  
$ \vec{\tilde{A}}^{\tilde{N}_{\tilde{R}}}_{\scriptscriptstyle{\stackrel{78}{(\pm)}}}$, 
couple to the rest four families of Table~\ref{Table III.}, determining the mass 
matrices of another group of quarks and leptons. 
The stable family of this group form the stable baryons, which offer the 
explanation for the existence of the heavy {\it dark matter} candidates.\\

{\bf iii.d.} {\it There are additional scalar fields
with the space index $s=(9,10,11,12,13,14)$, which are 
 triplets or antitriplets with respect to the space index $s$}.\\
There are additional scalar fields in the {\it spin-charge-family} theory, those 
with the space index $s=(9,10,11,12,13,14)$, Eq.~(\ref{faction}), manifesting 
as a triplets or antitriplets with respect to the space index $s$. They cause 
transitions of antileptons into quarks and back and leptons into antiquarks and 
back, what might be responsible in the expanding universe for the 
matter/antimatter asymmetry and also for 
the proton decay, Sect.~\ref{scalar3+1}.\\

\vspace{3mm}



%

%
{\bf iv.} {\it The break of the starting symmetry $SO(13+1)$ into $SO(3,1)$ 
$\times SU(2)\times SU(3)\times U(1)$ is needed.}\\

\vspace{1mm}

This problem is still not solved although we have the spontaneous breaks,
first from $SO(13,1)$ (and $\widetilde{SO}(13+1)$) to $SO(7,1) \times 
SU(3)\times U(1)$ ($\widetilde{SO}(7+1) \times \widetilde{SU}(3)\times 
\widetilde{U}(1)$) and then further to  $SO(3,1) \times SU(2) \times SU(3)
\times U(1)$ (and $\widetilde{SO}(3,1) \times \widetilde{SU}(2) \times$ 
$\widetilde{SU}(3) \times \widetilde{U}(1)$) all the time present.
%

In Refs.~\cite{nh2008,NHD,NHD2004,NH2006,NH2007Majorana,ND012,%
familiesNDproc} we study the toy model of $d=(5+1)$, in which the spin 
connection fields with the space index $s=(5,6)$ force the infinite surface  
in the fifth and the sixth dimension to form an almost $S^2$ sphere, 
keeping rotational symmetry of the surface around one point, while fermions 
in $d=(3+1)$ of particular handedness keep their masslessness. These 
happens for all the families of fermions of particular 
handedness~\cite{familiesNDproc}. 

We have not yet successfully repeated the $d=(5+1)$ toy model of the 
spontaneous "compactification" in the case of $SO(13,1)$ 
(we put compactification into quotation marks, since an almost $S^{n}$ 
sphere, as also $S^{2}$ sphere in the toy model of $d=(5+1)$ has 
the singular points in all infinities). 
%


\vspace{2mm}

\begin{small}
What we have to study is first to almost "compactify" the space of
 $s=(9,10,11,12,13,14)$ so that the space has the symmetry allowing the 
infinitesimal  transformations of the kind around the center of the symmetry~%
\cite{nd2017,nh2008,NHD,ND012,familiesNDproc}
\begin{eqnarray}
\label{deltaxsigmagen}
x'^{\mu} &=& x^{\mu}\,, \quad 
 x'^{\sigma} = 
 x^{\sigma} - i  \, \sum_{A,i, s,t} \varepsilon^{Ai} (x^{\mu})\, 
c_{Ai}{}^{st }M_{st} \,
x^{\sigma}\,,\nonumber
\end{eqnarray}
with
\begin{eqnarray}
\label{taua1} 
\tau^{Ai} &=& \sum_{s,t}\, c^{Ai}{}_{st}\,M^{st} \,,\quad 
\{\tau^{Ai}, \tau^{Bj}\}_{-} =  i f^{Aijk} \tau^{Ak}\, \delta^{A B}\,,\nonumber\\
\vec{\tau}^{A} &=& \vec{\tau}^{A \sigma}\, p_{\sigma} =
 \vec{\tau}^{A\sigma}{}_{\tau}\,  x^{\tau}\,p_{\sigma}\,, 
 \tau^{Ai \sigma} = \sum_{s,t}\, -i c^{Ai}{}_{st}\,M^{st\sigma}\nonumber\\
 &=& \sum_{s,t}\, c^{Ai}{}_{st}\,  (e_{s  \tau}\, f^{\sigma}{}_{t} - e_{t  \tau}\,
 f^{\sigma}{}_{s}) x^{\tau}= E^{\sigma}_{Ai}\,,
\end{eqnarray}
which lead to massless spinors/fermions of one particular charge and of one 
handedness only for each of all the $2^{\frac{7+1}{2}-1}$ families in the 
remaining space $d=(7+1)$.
The generator of the Lorentz transformations $M_{st}$ is equal to 
$M_{st} =  E^{\sigma}_{st}\,  p_{\sigma}$,  while the Killing vectors
$ E^{\sigma}_{st}$  fulfill the Killing equations, Ref.~\cite{nd2017}, Eq.~(8).

We need to find the solutions for the equations of motion for massless fermions
in $d\ge 9$ with vielbeins and the two kinds of the spin connection fields, which 
curve the infinite six dimensional space  into an almost $S^6$, with the symmetry
of $SU(3)\times U(1)$, in a way that they allow in $d=(7+1)$ only one massless
normalizable spinor state for each $SU(3)$ coloured or colouress state of particular
$U(1)$ "fermion" charges ($\tau^{4}=\frac{1}{6}$ for colour triplet fermion 
states and $-\frac{1}{2}$ for colourless fermion states), as well as only one 
massless normalizable 
spinor state for each $SU(3)$ anticoloured or anticolourless state of particular  
$U(1)$ "antifermion" charges ($\tau^{4}=- \frac{1}{6}$ for colour antitriplet 
states and $\frac{1}{2}$ for colourless states), Table~\ref{Table so13+1.}.

These we did in Refs.~\cite{nh2008,NHD,NHD2004,NH2006,NH2007Majorana,%
ND012,familiesNDproc} for the toy model, in which the several choices of 
vielbeins and spin connection fields take care of breaking the symmetry 
$SO(5+1)$ to $SO(3+1)$, leaving massless states of one handedness for 
each charge and all the families.
\end{small}
\vspace{2mm}

%
%
%

Then we must further break  the  symmetry $SO(7+1)$, 
and at the same time the  symmetry  $\widetilde{SO}(7+1)$, 
in equivalent way as we did the first break, to $SO(3,1)$ 
$\times SU(2)_{I}\times SU(2)_{II}$,
 keeping correspondingly the relations among the spin, 
handedness and charges of quarks and leptons, and the spin, handedness and 
anticharges of antiquarks and antileptons, as presented in 
Table~\ref{Table so13+1.} and as assumed in Table~\ref{Table SMF.}.\\

{\bf iv.a.} {\it The symmetry $SO(13,1)$ (and $\widetilde{SO}(13+1)$) 
first breaks into $SO(7,1) \times SU(3)\times U(1)$}.\\

The reader can 
notice in Table~\ref{Table so13+1.} that the $SO(7,1)$ content of 
$SO(13,1)$ is identical for quarks and leptons and identical for antiquarks and 
antileptons. Quarks distinguish from leptons and antiquarks from antileptons 
only in the $SO(6)$ content of $SO(13,1)$, that is in the colour charge and 
in the "fermion" charge. These two charges are described in the Clifford 
algebra presentation (suggested by the authors in Refs.~\cite{nh02,nh03}) 
for leptons as a factor $\stackrel{9\, 10}{(+)}$$\stackrel{11\, 12}{(+)}$
$\stackrel{13\,14}{(+)}$ (Eq.~(\ref{so64}) dictates for leptons $\tau^{33}=0$,
$\tau^{38}=0$ and $\tau^{4}=- \frac{1}{2}$), and for antileptons as a factor
$\stackrel{9\, 10}{[-]}$ $\stackrel{11\, 12}{[-]}$$\stackrel{13\, 14}{[-]}$
(Eq.~(\ref{so64}) dictates for $\tau^{33}=0$, $\tau^{38}=0$ and 
$\tau^{4}= \frac{1}{2}$). Quarks have three colour charge possibilities 
with $\tau^{4}= \frac{1}{6}$ and antiquarks three colour anticharge 
possibilities with $\tau^{4}=- \frac{1}{6}$.
The relations of handedness to charges and handedness to anticharges follow
from the Lorentz rotation of the kinds ($S^{0 9}, S^{3 9},
S^{0\, 11}, S^{0\, 13}$), which rotate fermions to antifermions within
the same Lorentz irreducible representation. 
Quarks are related to antiquarks and leptons to antileptos also by the discrete
 symmetry operators $\mathbb{C}_{{ \cal N}}{\cal P}^{(d-1)}_{{\cal N}}$, 
presented in Subsubsect,~\ref{CPT}, Eq.~(\ref{CPTNlowE}). \\
%
{\bf iv.b.}
 In the next break of symmetry the {\it $SO(7,1)$ symmetry must break to 
$SO(3,1) \times$ $SU(2)\times SU(2)$.}\\

 In this break the infinite space of  
$d=(5,6,7,8)$ must be curved into an almost $S^4$ sphere with the 
symmetry $SU(2)\times SU(2)$. 
Since the left handed spinors have different $SU(2)$ weak and $SU(2)_{II}$ 
charge (the left handed quarks and leptons have $\tau^{13}=\pm \frac{1}{2}$,
$\tau^{23}=0$) than the right handed spinors (the right handed quarks and 
leptons have $\tau^{13}=0$, $\tau^{23}=\pm \frac{1}{2}$), spinors are
mass protected.\\
%

\vspace{2mm}

{\bf v.} {\it The condensate  makes massive the $SU(2)_{II}$ 
vector  gauge fields and $U(1)_{\tau^4}$ vector gauge fields, as well as all the 
scalar gauge fields}.\\

The appearance of the condensate, Table~\ref{Table con.}, of two right handed 
neutrinos with family quantum numbers of the lower four families of eight 
families presented in  Table~\ref{Table III.},   makes massive the $SU(2)_{II}$ 
vector  gauge fields and $U(1)_{\tau^4}$ vector gauge fields, as well as all the 
scalar gauge fields, leaving massless only the $SU(3)$ colour, $U(1)_{Y}$ and 
$SU(2)$ weak vector gauge fields, Sect.~\ref{vectorscalar3+1}, as well as the
gravity in $d=(3+1)$. 

Fermions --- quarks and leptons and antiquarks and antileptons --- remain 
massless and mass protected, with
the spin, handedness, $SU(3)$ triplet or singlet charges, weak $SU(2)$ charge,
hyper charge and family charge as presented in Tables~\ref{Table SMF.},~%
\ref{Table so13+1.}, "waiting for" spontaneous break of mass protection at the
electroweak break.\\
{\bf v.a.} {\it Spontaneous break of mass protection needs to be studied}.\\
Although the simple starting action of the {\it spin-charge-family} theory, 
Eqs.~(\ref{wholeaction}, \ref{faction}),  offers three singlet and twice two triplet
scalar gauge fields, Sect.~\ref{scalar3+1}, with the space index 
$s=(7,8)$  to break the mass protection of fermions causing the electroweak 
break (carrying with respect to the space index the weak
and the hyper charge as required for Higgs's scalar in the {\it standard model}),
yet we should show how does the electroweak break spontaneously occur when
making the masses of the scalar gauge fields  
 imaginary (in the {\it standard model} this is just assumed). We have not yet 
finish this project. 





%
\subsection{ Predictions of $SO(10)$ unifying theories and {\it spin-charge-family}
theory}
\label{predictionsSO(10)SCFT}

There are many attempts in the literature to explain the {\it standard model}
assumptions, using the gauge groups, discussed in Ref.~\cite{PierreRamond} 
for ether the gauge charges or families or both.

There are many attempts in the literature to reconstruct mass matrices of 
quarks and leptons 
out of the observed masses and mixing matrices in order to learn more about 
properties of the 
fermion families~\cite{FRI,FRI1,FRI2,FRI3,FRI4,FRI5,FRI6,FRI7,FRI8,FRI9,
FRI10,FRI11,AstriAndOthers,AstriAndOthers1,Astri}. The
most popular is the $n\times n$ mass matrix, %
most often with $n=3$,
close to the democratic one, predicting that $(n-1)$ families must be very 
light in comparison with the $n^{\rm th}$ one. 
Most of attempts treat neutrinos differently than the other family members, 
introducing the 
Majorana part and the "sea-saw" mechanism. Most often are the number of 
families taken to be equal to the number of the so far observed families,
 while symmetries of mass matrices are 
chosen in several different ways~\cite{AstriAndOthers,AstriAndOthers1,Astri,lugri}. 
Also possibilities with four families are discussed~\cite{four, four1,four2}.

The existence of the {\it dark matter} is in the literature described mostly with
the invention of new particles~\cite{DM1} and with the primordial black 
holes~\cite{BH1,BH2,BH3}.%


The {\it spin-charge-family} theory~\cite{mdn2006,gmdn2007,gmdn2008,%
gn2009,gn2013,gn2014} predicts two groups of four  families of 
quarks and leptons, each of these two groups of four families manifesting 
the same $\widetilde{SU}(2)\times \widetilde{SU}(2)$ $\times U(1)$ symmetry of 
 mass matrices, Eq.~(\ref{commonAi}), the same for all family members, 
quarks and leptons. Twice two triplets of $\widetilde{SU}(2)\times$ $\widetilde{SU}(2)$
distinguish among the upper and lower four families, the three singlets determining
$U(1)$ distinguish among family members, Sect.~\ref{scalar3+1}.
To the lower of these 
two groups of four families the observed three families belong. The stable of the 
upper four families offer the explanation for the existence of the {\it dark matter}.

The {\it spin-charge-family} theory  predicts
the existence of additional scalar gauge fields with the properties of the Higgs's
scalar explaining the existence of the Yukawa couplings~%
\footnote{ The need for the appearance of the Yukawa couplings in the {\it 
standard model} by itself predicts additional scalar fields.}, 
%
the existence of the scalar fields which are with respect to the scalar index
the colour triplets, offering the explanation for the matter-antimatter 
asymmetry of the ordinary matter, to which the first group of four families 
(mainly the lightest one) contributes, as well as the decay of protons. 

The more work is put into the {\it spin-charge-family} theory, the more 
predictions is the theory offering.


%
%
%



%
\subsubsection{Predictions of {\it spin-charge-family} theory}
\label{predictionSCFT}

We have learned in the previous sections that the simple starting action, 
Eq.~(\ref{wholeaction}), in which fermions interact with gravity only --- the 
vielbeins and the two kinds of the spin connection fields --- offers the 
explanation for all the assumptions of the {\it standard model} presented 
in Sect.~\ref{SM}, and in a short way in Tables~\ref{Table SMF.},~%
\ref{Table SMV.},~\ref{Table SMS.} and unifies all the so far known forces,  
with gravity included, predicting new vector gauge fields, 
new scalar gauge fields and new families of fermions, offering also the 
explanation for the second quantized postulates of Dirac.\\

\vspace{2mm}

{\bf i.} The existence of the lower group of four families predicts the 
fourth family to the observed three, which should be seen in next 
experiments. The masses of quarks of these four families are 
determined by several scalar fields, all with the properties of the scalar 
higgs, what should also be observed.\\

The symmetry~\cite{NA2018,gn2014}, Eq.~(\ref{M0}), and the values 
of mass matrices 
of the lower four families are determined with two triplet scalar fields,  
$ \vec{\tilde{A}}^{\tilde{1}}_{\stackrel{78}{(\pm)}}$
and $\vec{\tilde{A}}^{\tilde{N}_{\tilde{L}}}_{\stackrel{78}{(\pm)}}$, 
and  three singlet scalar fields, $A^{Q}_{\stackrel{78}{(\pm)}}$, 
$A^{Y}_{\stackrel{78}{(\pm)}}$, $ A^{4}_{\stackrel{78}{(\pm)}}$,
Eq.~(\ref{commonAi}), explaining  the Higgs's scalar and  Yukawa 
couplings of the {\it standard model}, 
Refs.~\cite{gn2014,normaJMP2015,IARD2016,NH2017newdata,normaBled2020} 
and references therein.

Any accurate $3\times 3$ submatrix of the $4 \times 4$ unitary matrix 
determines the $4 \times 4$ matrix uniquely. Since neither the quark and 
(in particular) nor the lepton $3\times 3$ mixing matrix are measured 
accurately enough to be able to determine three complex phases of the 
$4 \times 4$ mixing matrix, we assume (what also simplifies the 
numerical procedure)~\cite{mdn2006,gmdn2007,gmdn2008,%
gn2009,gn2013,gn2014} that the mass matrices are symmetric and real 
and correspondingly the mixing matrices are orthogonal. We fitted the 
$6$ free parameters of each family member mass matrix, Eq.~(\ref{M0}),  
to twice three measured masses ($6$) of each pair of either quarks or 
leptons and to the $6$ (from the experimental data extracted) parameters 
of the corresponding $4 \times 4$ mixing matrix.

We present in this paper the results for quarks only, taken from  
Refs.~\cite{gn2013,gn2014}.  
The accuracy of the experimental data for leptons are not yet large 
enough that would allow a meaningful prediction~%
\footnote{The numerical procedure, explained in the papers~\cite{gn2013,gn2014}, 
to fit free parameters of the mass matrices to the experimental data within 
the experimental inaccuracy of masses of the so far observed quarks and 
first of all within the
inaccuracy of the mixing matrix elements, is very tough.}. 
It turns out that the experimental~\cite{datanew}
 inaccuracies are for the mixing matrices too large to tell 
trustworthy mass intervals for the quarks masses of the fourth family 
members~\footnote{We have not 
yet succeeded to repeat the calculations presented in Refs.~\cite{gn2014} with 
the newest data from Ref.~\cite{PDG2020}. Let us say that the accuracy of the
mixing matrix even for quarks remains  in Ref.~\cite{PDG2020} far 
from needed to predict the masses of the fourth two quarks. For the chosen 
masses of the four family quarks the mixing  matrix elements are expected to  
slightly change in the direction proposed by Eq.~(\ref{vudoldexp}).}. 
Taking into account the calculations of Ref.~\cite{datanew},
fitting the experimental data (and the meson decays evaluations in the 
literature,  as well as our own evaluations) the authors of the 
paper~\cite{gn2014} very roughly  estimate that the fourth family quarks 
masses might be pretty above $1$ TeV. 

Since the matrix elements of the $3 \times 3$ submatrix of the $4 \times 4$ 
mixing matrix depend weakly on the fourth family masses, the 
calculated mixing matrix
offers the prediction to what values will more accurate measurements move
the present experimental data and also the fourth family mixing matrix 
elements in dependence of the fourth family masses, Eq.~(\ref{vudoldexp}): 
%
 %
 %
  %
%
$V_{u d}$ will stay the same or will very slightly decrease;  
$V_{u b}$ and $V_{c s}$, will still lower;
$V_{t d}$ will lower, and $V_{t b}$  will lower; 
$V_{u s}$ will slightly increase; 
$V_{c d}$ will (after decreasing) slightly rise; 
$V_{c b}$ will still increase and $V_{t s}$ will (after decreasing) increase. 
 
In Eq.~(\ref{vudoldexp}) the matrix elements of the $4\times 4$ 
mixing matrix for quarks obtained when the $4\times 4$ mass matrices respect 
the symmetry of Eq.~(\ref{M0}) while the parameters of the mass 
matrices are fitted to 
the  ($exp$) experimental data~\cite{datanew}, Ref.~\cite{gn2014}, 
are presented for two choices of the fourth 
family quark masses:  $m_{u_4}= m_{d_4}=700$ GeV  ($scf_{1}$) and 
$m_{u_4}= m_{d_4}=$ $1\,200$ GeV ($scf_{2}$). 
In parentheses, $(\;)$ and $[\;\,]$, the  changes of the matrix elements are 
presented, which are due to the changes of the top mass
within the experimental inaccuracies: with the $m_{t} =$ $(172 + 3\times 0.76)$ 
GeV and $m_{t} =$ $(172 - 3\times 0.76)$, respectively  (if there are one, 
two or more numbers in parentheses the last one or more numbers 
are different, if there is no parentheses no numbers are different) 
[arxiv:1412.5866].\\


\vspace{0.1cm}

\begin{small}
\begin{equation}
\label{vudoldexp}
      |V_{(ud)}|= \begin{pmatrix}
    %
     exp  &    0.97425 \pm 0.00022    &  0.2253 \pm 0.0008 
&  0.00413 \pm 0.00049&   \\
     \hline
     scf_1  &    0.97423(4)            &  0.22539(7)          
&  0.00299  &     0.00776(1)\\  
      scf_2  &    0.97423[5]            &  0.22538[42]        &  0.00299  
&  0.00793[466]\\ 
     \hline 
     exp  &  0.225   \pm 0.008      &  0.986  \pm 0.016     
&  0.0411  \pm 0.0013&   \\
    \hline
    scf_1  &  0.22534(3)       &  0.97335              &  0.04245(6) 
&   0.00349(60) \\  
    scf_2  &  0.22531[5]       &  0.97336[5]          &  0.04248     
&   0.00002[216] \\ 
    \hline
    exp  &  0.0084  \pm 0.0006     &  0.0400 \pm 0.0027    
&  1.021   \pm 0.032&     \\
   \hline
    scf_1  &  0.00667(6)            &  0.04203(4)         &  0.99909   
&     0.00038\\  
     scf_2  &  0.00667                &  0.04206[5]         &  0.99909   
&     0.00024[21] \\
   \hline
    scf_1   & 0.00677(60) & 0.00517(26)    & 0.00020    & 0.99996\\
   scf_2   & 0.00773      & 0.00178   & 0.00022  & 0.99997[9]
     \end{pmatrix}\,.
     \end{equation}
\end{small}
Let us conclude that according to Ref.~\cite{gn2014} the masses of the  fourth 
family lie  much above the known three.
%
The larger are  masses of the fourth family the larger are $V_{u_1 d_4}$
in comparison with $V_{u_1 d_3}$ and the more is valid that 
$V_{u_2 d_4} <V_{u_1 d_4}$, $V_{u_3 d_4}<V_{u_1 d_4}$. 
The flavour changing neutral currents are correspondingly weaker.

Let be noticed that the prediction of  Ref.~\cite{BBB}, 
$V_{u_1 d_4}> V_{u_1 d_3}$, $V_{u_2 d_4}<V_{u_1 d_4}$,  
$V_{u_3 d_4}<V_{u_1 d_4}$, agrees with the
prediction of Refs.~\cite{gn2013,gn2014}.

In Ref.~\cite{NH2017newdata} the authors discuss the question why the existence
of the fourth family is not (at least yet) in contradiction with the 
experimental data.\\

\vspace{2mm}

{\bf ii.} 
The theory predicts the existence of besides the additional  
scalar fields also the additional vector gauge fields, 
Sect.~\ref{vectorscalar3+1}, Subsubsects.~\ref{vector3+1},
 \ref{scalar3+1}, Eqs.~(\ref{wholevectorscalar}, \ref{su2IandII}).

\vspace{2mm}

\begin{small}
Let us comment here the report of LHCb~\cite{LHCb} from March 
$25^{th}$ $2021$ on measurements of lepton decays of mesons $B$, 
showing that the  decay of $B$-mesons to 
$\mu^+ \mu^-$ is for $\approx 15\%$ weaker than to $e^+ +e^-$, while the 
{\it standard model} predicts that they should be  very close or to be equal 
(with the higher corrections  taken into account included). 
We can (so far) only say, that since the {\it spin-charge-family}
theory  predicts two kinds of $Z^0_{m}$ bosons [due to the two 
 vector gauge fields, $A^{Q'}_{m} = \cos \vartheta_{1} \,A^{13}_{m} - 
\sin \vartheta_{1} \,A^{Y}_{m}$,  $A^{Q'}_{m}$ represents $Z^{0}_{m}$ field, 
and $A^{Y'}_{m}=\cos \vartheta_{2} \,A^{23}_{m} - 
\sin \vartheta_{2} \,A^{4}_{m}$, $A^{Y'}_{m}$ represents the second 
$Z'^{0}_{m}$ field,
where $\vartheta_i, i=(1,2)$, are the two angles of breaking symmetries,
%
 Eq.(\ref{Aomegas}) (where the scalar index $s\in (7,8)$ must be replaced 
by the vector index $m\in(0,1,2,3)$, and $A^{4}_{m} 
= -(\omega_{9\,10\,m} + \omega_{11\,12\,m} + \omega_{13\,14\,m})$,
 $A^{13}_{m}=(\omega_{56 m}- \omega_{78 m})$, 
$A^{23}_{m}=(\omega_{56 s}+\omega_{78 m})$,
$A^{Q}_{m} = \sin \vartheta_{1} \,A^{13}_{m} + 
\cos \vartheta_{1} \,A^{Y}_{m}$,
$A^{Y}_{m}  =(g^{56}\,\omega_{56 m} +A^{4}_{m}$)], and several scalar 
gauge field, [two triplets and three singlets, Eq.~(\ref{commonAi}), all with
the properties of the Higgs's scalar] there is no reason that the decay of
the meson $B$  to two leptons belonging to two different families are the same. 
We have not estimated yet these two decays and can not predict the ratio of
the measured decays, but due to our experience it seems very unlikely that the 
difference would be so large.
\end{small}\\

\vspace{2mm}



{\bf iii.} The theory predicts the existence of the upper four families of 
quarks and leptons and antiquarks and antileptons, with the same family 
members charges, Tables~\ref{Table so13+1.},~\ref{Table III.}, as 
the lower four families, interacting correspondingly with the same vector 
gauge fields.  At low energies the upper four families are decoupled from 
the lower four families. 

The masses of the upper four families are determined by the two 
triplets ($ \vec{\tilde{A}}^{\tilde{2}}_{\stackrel{78}{(\pm)}}, 
   \vec{\tilde{A}}^{\tilde{N}_{\tilde{R}}}_{\stackrel{78}{(\pm)}}$) and 
three singlets ($A^{Q}_{\stackrel{78}{(\pm)}}, 
A^{Q'}_{\stackrel{78}{(\pm)}}, A^{Y'}_{\stackrel{78}{(\pm)}}$), the 
same singlets contribute also to masses of the lower four families,
Sect.~\ref{scalar3+1}.

 The stable of the upper four families offers the explanation 
for the appearance of the {\it dark matter} in our universe. 

Since 
the masses of the upper four families are much higher than the masses 
of the lower four families, the "nuclear" force among the baryons and mesons
of these quarks  and antiquarks differ a lot from the nuclear force of the 
baryons and fermions of the lower four families.

A rough estimation of properties of  baryons  of the stable fifth family 
members, of their behaviour  during the evolution of the universe and when 
scattering on the ordinary matter, as well as a study of possible limitations 
on the family properties due to the cosmological and direct experimental 
evidences are done in Ref.~\cite{gn2009}. 

In Ref.~\cite{nm2015} the weak and "nuclear"   scattering  of such very 
heavy baryons by ordinary nucleons is studied, showing that the 
cross-section for such scattering is very small and therefore consistent 
with the observation of experiments so far, provided that the quark mass of this
 baryon is about 100 TeV or above.

In Ref.~\cite{gn2009} a simple hydrogen-like model is used to evaluate 
properties of baryons of these heavy quarks, with one gluon  exchange
determining the force among the constituents of the fifth family baryons~%
\footnote{The weak force and the electromagnetic force start to be at small 
distances due to heavy masses of quarks of the same order of 
magnitude as the colour force.}.

\begin{figure}[h]
\begin{center}
\includegraphics[width=15cm,angle=0]{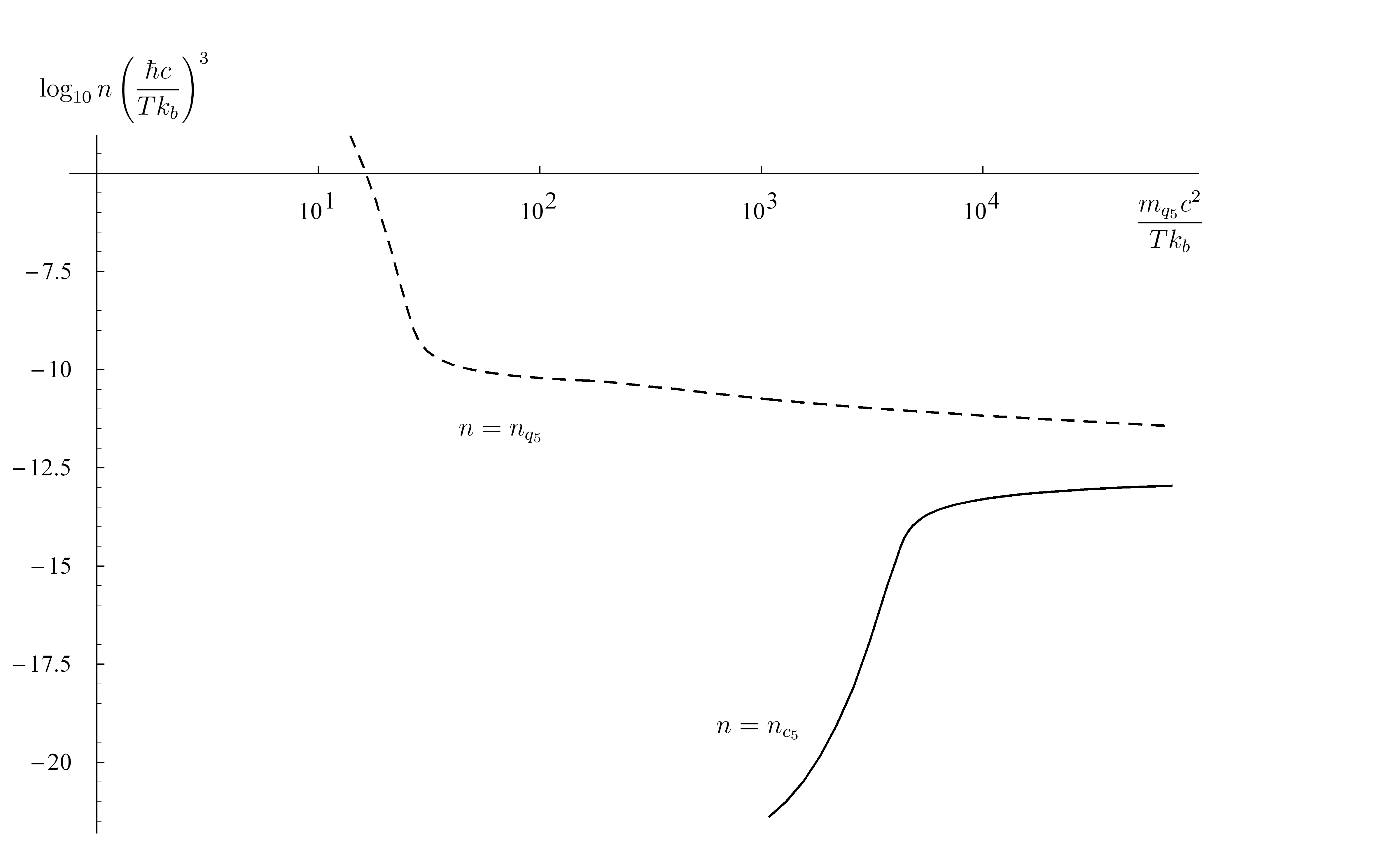}
\caption{The dependence of the two number densities, $n_{q_5}$ of the fifth family
 quarks and $n_{c_5}$ of the fifth family clusters of quarks, as functions of 
$\frac{m_{q_5} \, c^2}{ k_b \, T}$ is presented 
for the special values $m_{q_5} = 71 \,{\rm TeV}$. The estimated 
scattering cross sections, entering into Boltzmann equation, are presented in
Ref.~\cite{gn2009}, Eqs.~(2,3,4.5),
In the treated energy (temperature $ k_b T$) interval 
the one gluon exchange gives the  main 
contribution to the scattering cross sections 
entering into the Boltzmann equations for 
$n_{q_5}$ and $n_{c_5}$. 
}
\end{center}
\label{DiagramI.}
\end{figure}

The authors of Ref.~\cite{gn2009} study the freeze out procedure of 
the fifth family quarks and anti-quarks and the formation of  baryons and 
anti-baryons up to the 
temperature  $ k_b T= 1$ GeV,  when the colour phase transition 
starts which depletes almost all the fifth family quarks and anti-quarks,
 while the colourless
fifth family neutrons with very small scattering cross section decouples 
long before (at $ k_b T= 100$ GeV), Fig.~\ref{DiagramI.}.
The cosmological evolution 
suggests for the mass limits the range $10$ TeV $< m_{q_5}  < 
{\rm a \, few} \cdot 10^2$ TeV 
and for the  scattering cross sections 
$ 10^{-8}\, {\rm fm}^2\, < \sigma_{c_5}\, <   10^{-6} \,{\rm fm}^2  $. 
The measured density of  the  dark matter 
does not put much limitation on the properties of heavy enough clusters %
\footnote{   
In the case that the weak interaction determines the  cross section of  the neutron 
$n_5$, the interval for the  fifth family quarks would be 
$10\; {\rm TeV} < m_{q_5} \, c^2< 10^5$ TeV. }.

\vspace{3mm}


\vspace{3mm}

The DAMA/LIBRA experiments~\cite{RitaB} limit (provided that they measure 
the heavy fifth family clusters) the quark mass in the interval:
 $ 200 \,{\rm TeV} < m_{q_{5}} < 10^5\, {\rm TeV}$, Ref.~\cite{gn2009}.

Baryons of the   fifth family are heavy, forming small 
enough  clusters with small enough scattering amplitude among themselves 
and with the ordinary matter to be the candidate for the dark matter.  

Masses of the stable fifth family of quarks and leptons are much above 
the fourth family members.

Although the upper four families carry the weak (of two kinds) and the 
colour charge, these group of four families are completely decoupled from 
the lower four families up to the $<10^{16}$  GeV, unless the breaks 
of symmetries recover.\\

\vspace{2mm}

{\bf iv.} The spread of masses of quarks and leptons of two groups of 
four families from $10^{-11}$GeV
($\nu$ of the first family) to $10^{15}$GeV ($u$ and $d$ of the fourth family of the
upper four families)  offers the explanation for the "hierarchy problem".
We have not yet in the {\it spin-charge-family} theory  
study loop corrections  to the  squared masses of scalar fields,  
of twice two triplets and three singlets which determine mass matrices
of the two groups of four families, what is needed
to analyze the ''hierarchy'' problem.

But we have various scales in our theory because we have
(as the only choice) to accept different ``compactification''
scales for different dimensions~\footnote{%
The fact that we succeeded to fit the parameters of mass 
matrices to experimental data can mean, 
and in this way we speak about the offer of the theory to solve the 
''hierarchy problem'', that ''nature does solve this problem'' and the spread
of fitted masses approves our hope that we have only to find how ''nature
does solve the problem''.}, what could help to solve 
this ''hierarchy'' problem.









%
%
%

%
%
%
%

%
\subsection{Massless and massive odd Clifford fermions in toy model with
$d=(5+1)$}
\label{TDN0}
%


To illustrate the difference between our description of the internal space of fermions
and the usual one, we demonstrate in this section on the toy model in 
$d=(5+1)$-dimensional space, when starting from  massless fermions interacting with
the gravity only and then breaking symmetries from $d=(5+1)$ to $d=(3+1)$ 
times an almost $S^2$ sphere, how to end up with the massless and  charged  
fermions and antifermions, while a further break of symmetry, caused by scalar 
fields gaining constant values, leads to massive chargeless fermions manifesting
as  Majorana fermions, Ref.~\cite{TDN}.

There are zweibeins and spin connection fields of particular properties in the
$5^{th}$ and $6^{th}$  dimensions which make the ${\cal M}^{5+1}$ manifold
to  break into ${\cal M}^{3+1}$ $\times$ an almost $S^2$ 
sphere~\cite{NHD,ND012}.  The mass  of  spinors (fermions)  
originates either in the dynamics or in the vacuum expectation 
values of zweibein and spin connection fields in $5^{th}$ and $6^{th}$  dimensions.


We learn in Sects.~\ref{internalspace} - \ref{fermionandgravitySCFT}
that creation operators and their Hermitian conjugated annihilation operators 
anticommute, due to the odd character of the "basis vectors" describing the internal
space of fermions. Since the character of the vacuum state is even all the single 
fermion states have an odd character. The equations of motion, either for free or 
for interacting Clifford fermions, have an even character. The solutions of 
equations of motion are correspondingly superposition of creation operators 
belonging to the members of the same irreducible representation, the same 
family, or to different families, each of them having an odd character. The operators 
$S^{ab}$ and $\tilde{S}^{ab}$, connecting members of the same family  
or different families, respectively, have an even Clifford character and do not 
change the oddness of creation operators and correspondingly of fermion 
states. The operators $\gamma^a$'s and $\tilde{\gamma^a}$'s, both having
an odd Clifford character, change the character of states from 
odd to even. 

When describing massive states in usual cases, the ones of Dirac,   
$\gamma^a$'s are used to transform the left handed fermion states in 
$d=(3+1)$ into the right handed ones to generate basis for the massive 
states. Charges are in all these cases added separately, by the choice of 
appropriate charge groups. Vectors representing states are just numbers 
which commute.

In the {\it spin-charge-family} theory, describing the internal space of 
fermions with the odd Clifford  algebra creation operators, operating on the 
vacuum state, the multiplication of the odd 
creation operators, Eqs.~(\ref{weylgen05+1}, \ref{mlessF}), by odd 
$\gamma^a$'s leads to even Clifford algebra creation operators, which 
commute and do not describe fermions, as presented in App.~\ref{evenclifford}. 

But if one looks at the solutions of  equations of motion for chargeless massive 
states in the {\it spin-charge-family} theory case, Eq.~(\ref{Weylmsolsimple}), 
whille neglecting the charge part and correspondingly the oddness
of states, what the usual way of presenting  fermions does, then one easier 
understands the difference of the {\it spin-charge-family} theory  way and the 
Dirac's way of the second quantization, presented in Sect.~\ref{creationtensorusual}.


We present in App.~\ref{trial}, following Ref.~\cite{DMN}, the matrix 
representations of the operators $\gamma^a$'s,  $\tilde{\gamma^a}$'s, 
$S^{ab}$ and $\tilde{S}^{ab}$, discussing their properties in our case when
using odd Clifford algebra to describe the internal space of fermions appearing 
in families, and in the usual case when using Dirac matrices, 
Sect.~\ref{matrixCliffordDMN}.



%
\subsubsection{Massless and massive solutions of equations of motion in 
$d=(5+1)$}
\label{TDN1}

Let us again use for our discussions the simple toy model with odd Clifford algebra 
"basis vectors" describing the internal space
of fermions in $d=(5+1)$-dimensional space. In Table~\ref{cliff basis5+1.} the 
"basis vectors" determining the creation operators and those determining the
annihilation operators are presented. 


In Table~\ref{Table Clifffourplet.} besides
the Clifford odd basis vectors also the Clifford even "basis vectors" are presented.
One can notice that the application of the operators $\gamma^a$'s or
$\tilde{\gamma}^a$'s transforms the Clifford odd anticommuting 'basis vectors" 
into the Clifford even commuting "basis vectors".  

How can the use of the odd Clifford algebra operators $\gamma^a$'s 
in the usual first quantized theory be explained from the point of view of 
the {\it spin-charge-family} theory, 
if $\gamma^a$'s transform creation operators, members of an odd irreducible 
representation, into members of an even irreducible representation, which neither 
have always the Hermitian conjugated partners which would applying on the 
vacuum state give zero (the product of projectors is a self adjoint operator) 
nor has the  anticommuting properties?


We start with the massless solutions of the Weyl equation in $d=(5+1)$ with the 
"basis vectors" presented in Table~\ref{cliff basis5+1.}, and with the spin (or
 the total angular momentum) in extra dimensions, $d>(3+1)$, determining 
the charge in $d=(3+1)$. We then let  the ${\cal M}^{5+1}$ manifold to 
break into ${\cal M}^{3+1}$ $\times$ an almost $S^2$ 
sphere~\cite{NHD,ND012} due to the zweibein and spin connection fields in the
$5^{th}$ and $6^{th}$  dimensions. The mass  of the spinor (fermion) 
originates either in the dynamics in the higher dimensions or in the vacuum 
expectation values of the scalar fields, which are the gauge fields with the 
scalar index with respect to $d=(3+1)$, and therefore well defined.

With this illustration we want to make evident the fact that without knowing the 
action which leads to massless and massive solutions in $d=(3+1)$, the 
comparison of the Dirac solutions of the equation of motion which already 
assumes the nonzero mass and includes charges by assuming additional 
charge groups and the solution of the {\it spin-charge-family} theory does 
not seem meaningful.

Let us here repeat the Weyl equation of the action~(Eq.(\ref{wholeaction})), the 
internal space of solution is described by the odd Clifford algebra objects 
in the case of our toy model of $d=(5+1) $ as follows
 \begin{eqnarray}
 \label{weylTDN}
 &&(\gamma^m p_{m} + \stackrel{56}{(+)}p_{0+} + \stackrel{56}{(-)} p_{0-}) \psi=0\,,\nonumber\\
 &&p_{0\pm}= p_{0}^5 \mp i\,p_{0}^6\,,\nonumber\\
 &&p_{0s}= f^{\sigma}_{s}(p_{\sigma} -\frac{1}{2} S^{ab} \omega_{ab \sigma}) + 
 \frac{1}{2E} \{p_{\sigma},f^{\sigma}_{s} E \}_{-} \,,\nonumber\\ 
 &&\stackrel{56}{(\pm)} = \frac{1}{2}\,(\gamma^5 \pm i \gamma^6)\,.
 \end{eqnarray}
 %


There are $2^{\frac{d}{2}-1}$ ($4$ in our case of $d=6 $) odd "basic vectors" 
$\hat{b}^{m \dagger}_{f}$, $m=(1,2,3,4)$ appearing in $2^{\frac{d}{2}-1}=4$ 
families, $f=(I,II,III,IV)$, presented in Table~\ref{cliff basis5+1.}.
$S^{ab}= \frac{i}{4}(\gamma^a \gamma^b -\gamma^b \gamma^a)$ connect 
family members within each family, $\tilde{S}^{ab}= \frac{i}{4}(\tilde{\gamma^a} 
\tilde{\gamma}^b - \tilde{\gamma}^b \tilde{\gamma}^a)$ transform one family 
member of a particular family into the same family member of another family. 
We look for solutions of the equation of motion by superposition of the creation 
operators applying on the vacuum  state, $\hat{b}^{s \dagger}_{f}
 |\psi_{oc}> |0_{\vec{p}}> = \psi^s_f(\vec{p})$, while the 
application of the annihilation operators on the vacuum state gives zero,
 $\hat{b}^{s}_{f} (\vec{p})|\psi_{oc}>|0_{\vec{p}}>=0$.

We treat here only one family, since we only want to clear up the relation between 
our way of describing the internal space of fermions and the usual (Dirac's) way.

Let us make a choice of the first family, $f=1$, presented in 
Table~\ref{cliff basis5+1.} as $\hat{b}^{m \dagger}_{f}$, with $m=(1,2,3,4)$, 
with $(S^{56} =\pm \frac{1}{2}, S^{12}=\pm frac{1}{2})$, 
and let us skip the family quantum number $f$ in this section. 
The Hermitian conjugated partners of the first family member are presented as 
$\hat{b}^{m}_{f}=$$(\hat{b}^{m \dagger}_{f})^{\dagger}$.  
In Table~\ref{cliff basis5+1.} we read that two of the "basis vectors", the first two, 
are right handed with the charge $\frac{1}{2}$ and the second two are left handed
with the charge $-\frac{1}{2}$
\begin{eqnarray}
\hat{b}^{1 \dagger} &=&  \stackrel{03}{(+i)} \stackrel{12}{(+)} 
\stackrel{56}{(+)}\,,\quad \hat{b}^{2 \dagger} =\stackrel{03}{[-i]} 
\stackrel{12}{[-]} \stackrel{56}{(+)} \,,\quad
\hat{b}^{3 \dagger}=  \stackrel{03}{[-i]} \stackrel{12}{(+)} \stackrel{56}{[-]}\,, \quad
\hat{b}^{4 \dagger} =\stackrel{03}{(+i)} \stackrel{12}{[-]}\stackrel{56}{[-]}\,. 
\label{weylrep}
\end{eqnarray}
The vacuum state $|\psi_{oc}>$ is in this case, when we do not pay attention 
on other families, equal to
$|\psi_{oc}>=\stackrel{03}{[-i]} \stackrel{12}{[-]}\stackrel{56}{[-]}$,
Eq.~(\ref{vac5+1}),
 so that the four "basis vectors" of Eq.~(\ref{weylrep}) are normalized, 
$\hat{b}^{i} \hat{b}^{j \dagger}= \delta^{i}_{j}$, giving a nonzero contribution
when applying on  $|\psi_{oc}>$.

All the basic states are eigenstates of the Cartan subalgebra (of the Lorentz 
transformation Lie algebra), for which we take: $S^{03}, S^{12}, S^{56}$, with the
 eigenvalues, which can be read from Eq.~(\ref{weylrep}) as $\frac{1}{2}$ 
times the numbers $\pm i$ or $\pm 1$ in the parentheses of nilpotents 
$\stackrel{ab}{(k)} $ and projectors $\stackrel{ab}{[k]} $: 
$S^{ab} \stackrel{ab}{(k)} =$ $\frac{k}{2} \stackrel{ab}{(k)} $, 
$S^{ab} \stackrel{ab}{[k]} =$ $\frac{k}{2} \stackrel{ab}{[k]}$. 

One notices that two of the states are right handed
 ($\psi_{1}= \hat{b}^{1 \dagger} |\psi_{oc}> $ and 
$\psi_{2} = \hat{b}^{2 \dagger} |\psi_{oc}> $) 
and two left handed ($\psi_{3}= \hat{b}^{3 \dagger} |\psi_{oc}> $) and 
$\psi_{4} = \hat{b}^{4 \dagger} |\psi_{oc}>$) with respect to $d=(3+1)$ 
(while all four carry the same, left, handedness with respect to $d=(5+1)$). 
The operator of handedness is defined in Eq.~(\ref{handedness}).

With the following choice of the zweibein fields we achieve  that the infinite surface 
$d=(5,6)$ curls into an almost $S^2$ (with one hole~\cite{NHD})
\begin{eqnarray}
e^{s}{}_{\sigma} &=& f^{-1}
\begin{pmatrix}1  & 0 \\
 0 & 1 
 \end{pmatrix},
f^{\sigma}{}_{s} = f
\begin{pmatrix}1 & 0 \\
0 & 1 \\
\end{pmatrix}\,,
\label{fzwei}
f = 1+ (\frac{\rho}{2 \rho_0})^2\,,\nonumber\\ 
E &=& \det(e^s{\!}_{\sigma})=f^{-2}\,, e^s{\!}_{\sigma}\,f^{\sigma}{\!}_{t}=
 \delta^{s}_{t}\,,
\nonumber\\ 
x^{(5)} &=& \rho \,\cos \phi,\quad  x^{(6)} = \rho \,\sin \phi\,, \nonumber
\end{eqnarray}
while $d=(3+1)$ space remains flat ($f^{\mu}{\,}_{m}= \delta^{\mu}_{m}$). 
We choose the spin connection fields on this $S^2$ as 
\begin{eqnarray}
  f^{\sigma}{}_{s'}\, \omega_{st \sigma} &=& i F\, f \, \varepsilon_{st}\; 
  \frac{e_{s' \sigma} x^{\sigma}}{(\rho_0)^2}\, , \quad 
 0 <2F \le 1\, 
  ,\quad s=5,6,\,\,\; \sigma=(5),(6)\,, 
\label{omegas56}
\end{eqnarray}
in order to guarantee that there manifests in $d=(3+1)$ only one massless and 
correspondingly mass 
protected state~\cite{NHD,TDN}, while the rest of states are all massive. There 
is the whole interval 
for the constant $F$ ($0 <2F \le 1$), which fulfills the condition of only one 
massless state of 
the right handedness in $d=(3+1)$, which is square integrable. 

When requiring that the solutions of Eq.~(\ref{weylTDN}) have the angular 
moments in $d=(5,6)$ manifesting the charges in $d=(3+1)$ 
($M^{56}= x^5 p^{6}- x^6 p^{5} + S^{56}=$ 
$- i\frac{\partial}{\partial \phi} + S^{56}$), we write the wave functions  
$\psi^{(6)}_{n+1/2}$ 
for the choice of the coordinate system $p^a= (p^0,0,0,p^3,p^5,p^6)$ as follows 
\begin{eqnarray}
\psi^{(6) }_{n+1/2}= ({\cal A}_{n}\,\stackrel{03}{(+i)} \stackrel{12}{(+)} 
\stackrel{56}{(+)}  
+ {\cal B}_{n+1}\, e^{i \phi}\, \stackrel{03}{[-i]}\stackrel{12}{(+)}\,
\stackrel{56}{[-]})\,\cdot e^{in \phi}
e^{-i(p^0 x^0- p^3 x^3)} |\psi_{oc}> |0_{\vec{p}}>\,, 
\label{mabpsi}
\end{eqnarray}
index ${}^{(6)}$ in $\psi^{(6) }_{n+1/2}$ is to point out that we treat the dynamics
in $d=(5,6)$.
Besides one massless ($\psi^{(6) }_{1/2}$) solution with $n=0$ there is the whole 
series of massive solutions
 manifesting in $d=(3+1)$ the (Kaluza-Klein) charge ${n+1/2}$: $M^{56}\, 
\psi^{(6) }_{n+1/2}=(n+1/2)\,\psi^{(6) }_{n+1/2} $, and solving  Eq.~(\ref{weylTDN}),
 provided that  
${\cal A}_{n}$ and ${\cal B}_{n+1}$ are the solutions of the equations
\begin{eqnarray}
&&-if \,\{ \,(\frac{\partial}{\partial \rho} + \frac{n+1}{\rho})  -   
  \frac{1}{2\, f} \, \frac{\partial f}{\partial \rho}\, (1+ 2F)\}  {\cal B}_{n+1} +
 m {\cal A}_n = 0\,,  
\nonumber\\
&&-if \,\{ \,(\frac{\partial}{\partial \rho} - \quad \frac{n}{\rho}) -   
  \frac{1}{2\, f} \, \frac{\partial f}{\partial \rho}\, (1- 2F)\}  {\cal A}_{n} + 
m {\cal B}_{n+1} = 0\,.
\label{equationm56}
\end{eqnarray}
The  massless positive energy solution with spin $\frac{1}{2}$,
 left handedness (Eq.~(\ref{handedness}) in 
$d=(5+1)$, the charge in $d=(3+1)$ equal to $\frac{1}{2}$ and right handed
 with respect to 
$\Gamma^{(3+1)}$ is equal to
\begin{eqnarray}
\psi^{(6)}_{\frac{1}{2}} ={\cal N}_0  \; f^{-F+1/2} 
\stackrel{03}{(+i)}\stackrel{12}{(+)}\stackrel{56}{(+)}\,e^{-i(p^0 x^0-p^3x^3)}
|\psi_{oc}> |0_{\vec{p}}>\,. 
\label{Massless}
\end{eqnarray} 
For the special choice of $F=\frac{1}{2}$ (from the interval in Eq.~(\ref{omegas56}) 
allowing only right handed square integrable massless states) the solution 
of Eq.~(\ref{Massless})) simplifies 
to
\begin{equation}
\label{mlessF}
\psi^{(6)}_{\frac{1}{2}} ={\cal N}_0  \;  
\stackrel{03}{(+i)}\stackrel{12}{(+)}\stackrel{56}{(+)}\,e^{-i(p^0 x^0-p^3x^3)}\,.
\end{equation}

Massive solutions are in this special case~\cite{NHD,ND012} expressible in terms of 
the associate 
Legendre function $P^{l}_n(x)$, $x= \frac{1-u^2}{1+u^2}$, 
$u= \frac{\rho}{2 \rho_0}$, where $\rho_0$ 
is the radius of (an almost) $S^2$, as follows
\begin{eqnarray}
\label{massF}
{\cal A}^{l(l+1)}_n &=& P^{l}_n\,,\quad 
{\cal B}^{l(l+1)}_{n+1} = \frac{-i}{\rho_0 m} \,\sqrt{1-x^2}\, 
\left(\frac{d}{dx} + \frac{n}{1-x^2} \right)\, {\cal A}^{l(l+1)}_n \,,
\end{eqnarray}
with the masses~\footnote{In the case that $d=(5,6)$ is a compact $S^2$ sphere 
these massive 
solutions would make infinite spectrum with quantum numbers $(l,n)$, 
$l$ defining in $d=(3+1)$ 
the mass and $n+ \frac{1}{2}$ the Kaluza-Klein charge. In the case of an 
almost $S^2$ the spectrum 
start to stop when the energy approaches the strengths of the source which 
causes the vielbein 
leading to an almost $S^2$.} determined by $(\rho_0\, m)^2 =l(l+1)$ and 
$l=1,2,3,\dots$, $0 \le n < l$. 

\vspace{3mm}

{\it {\bf Massive chargeless solutions of the Weyl equation}}
%

\vspace{3mm}

Let us now assume that the scalar fields, the gauge fields of $S^{56}$, that is  
$f^{\sigma}_{s}$ $ \omega_{56 \sigma}$, with $s=(5,6)$ and $\sigma=((5),(6))$, 
gain  constant values (non zero vacuum expectation values). These two scalar fields
 are the analogy to the complex higgs scalar of the {\em standard 
model}: The Higgs's scalar carries in the {\it standard model}  the weak and the 
hyper charge, while our scalar fields carry  only the "hyper" charge $S^{56}$.  
The charge, which is the spin in $d=(5,6)$,  
is after the scalar fields gain nonzero vacuum expectation values no longer the 
conserved quantity and mass protection is correspondingly removed. 

In this case we replace in the Weyl equation~(\ref{weylTDN}) the quantities  
$p_{0\pm}$ with their  
constant values, the vacuum expectation values, $<p_{0\pm}>$, so that the 
equations of motion follow as
 \begin{eqnarray}
 \label{Weylmvac}
 &&<p_{0\pm}> = m_{\pm} \,,\quad 
 (\gamma^m p_{m}  + \stackrel{56}{(+)} \,m_{+}  + \stackrel{56}{(-)}\, m_{-} )\,
 \psi^{(6)}=0
 \,.
 \end{eqnarray}
To simplify  further discussions, the coordinate system in $d=(3+1)$  with  
$\vec{p}=0$ is chosen. 
Then  Eq.~(\ref{Weylmvac}) reads 
 \begin{eqnarray}
 \label{Weylmvacsimple}
  &&\{p_{0} + \gamma^0 (\stackrel{56}{(+)} \,m_{+}  + \stackrel{56}{(-)}\, m_{-} )\}
 \psi^{(6)}=0
  \,.
 \end{eqnarray}
 %
The two positive fermion massive, $m$, solutions  
 with the spin in $d=(3+1)$ equal to $\pm \frac{1}{2}$, respectively, 
both with non conserved  charges $S^{56}$,
Eq.~(\ref{Weylmvacsimple}), are  
 \begin{eqnarray}
 \label{Weylmsolsimple}
\psi^{(6)}_{\;\frac{1}{2},m}&=& (\stackrel{03}{(+i)} \stackrel{12}{(+)} 
\stackrel{56}{(+)} + 
\frac{m\;}{m_{+}} \,\stackrel{03}{[-i]} \stackrel{12}{(+)} \stackrel{56}{[-]})\,
 e^{-imx^0} \,|\psi_{oc}> |0_{\vec{p}}> \,, 
\nonumber\\
\psi^{(6)}_{-\frac{1}{2},m}&=& (\stackrel{03}{[-i]} \stackrel{12}{[-]} 
\stackrel{56}{(+)} 
+ \frac{m\;}{m_{+}} \stackrel{03}{(+i)} \stackrel{12}{[-]} \stackrel{56}{[-]})\, 
e^{-imx^0}\, |\psi_{oc}> |0_{\vec{p}}> \,,
 \stackrel{03}{[-i]} \stackrel{12}{[-]} \stackrel{56}{(+)})\, e^{-imx^0}\nonumber\\
 m^2&=& m_{+}\ m_{-}\,,\quad  m_{+} = - m_{-}\,, \quad 
(p_{0})^2 = m^2\,.
\end{eqnarray}
Since the  scalar gauge fields $f^{\sigma}{}_s \,\omega_{t t' \sigma}$ gain  
constant values, $\mathbb{C}_{{\cal N}}\cdot {\cal P}^{(d-1)}_{{\cal N}}$ from
Sect.~\ref{CPT} is 
no longer the symmetry of the equations of motion and correspondingly does
not transform fermions into antifermions.  It is not difficult to find
the new operator which replaces 
$\mathbb{C}_{{\cal N}}\cdot {\cal P}^{(d-1)}_{{\cal N}}$ after the 
break~\cite{TDN}.
It is $(-i)\,\Gamma^{(6)}\Gamma^{(3+1)}\,\mathbb{C}_{{\cal N}}\cdot 
{\cal P}^{(d-1)}_{{\cal N}}$.
It is easy to see that in our $d=(5+1)$ 
case $(-i)\,\Gamma^{(6)}\Gamma^{(3+1)}\,\mathbb{C}_{{\cal N}}\cdot 
{\cal P}^{(d-1)}_{{\cal N}}=\gamma^5 \gamma^6 \gamma^0 
\gamma^5 I_{\vec{x}_3} I_{x^6}$, after taking into account that 
$m_{-}=- m_{+}$. 
We can check that 
 \begin{eqnarray}
 \label{CPNeffnew}
  &&(-i)\,\Gamma^{(6)}\Gamma^{(3+1)}\,\mathbb{C}_{{\cal N}}\cdot 
{\cal P}^{(d-1)}_{{\cal N}}\;
  (\gamma^0\,\gamma^m p_{m}  + \gamma^0\,(\stackrel{56}{(+)} \,m_{+}  + 
\stackrel{56}{(-)}\, m_{-} ))\; 
 ((-i)\,\Gamma^{(6)} \Gamma^{(3+1)}\,\mathbb{C}_{{\cal N}}\cdot
 {\cal P}^{(d-1)}_{{\cal N}})^{-1} 
 \nonumber\\
 && =(\gamma^0\,\gamma^m p_{m}  + \gamma^0\,((-) \stackrel{56}{(-)} \,m_{+}  
+  (-)\stackrel{56}{(+)}\, m_{-}))= 
 (\gamma^0\,\gamma^m p_{m}  + \gamma^0\,(\stackrel{56}{(-)} \,m_{-}  + 
 (\stackrel{56}{(+)}\, m_{+}))\,, \nonumber\\
 &&{\rm since} \quad m_{+}= - m_{-}\,.
 \end{eqnarray}
Let us now find the antifermion states to the two positive energy states in
 Eq.~(\ref{Weylmsolsimple})
\begin{eqnarray}
 \label{Weylmsolsimpleanti}
 -i\,\Gamma^{(6)} \Gamma^{(3+1)} \mathbb{C}_{{\cal N}}\cdot 
{\cal P}^{(d-1)}_{{\cal N}} \;
\psi^{(6)}_{\;\frac{1}{2},m}&=& ( (-i)\stackrel{03}{[-i]} \stackrel{12}{(+)}
 \stackrel{56}{[-]} + 
\frac{m\;}{m_{+}}\,i \,\stackrel{03}{(+i)} \stackrel{12}{(+)} \stackrel{56}{(+)})\, 
e^{-imx^0} \,, \nonumber\\
-i\,\Gamma^{(6)} \Gamma^{(3+1)} \mathbb{C}_{{\cal N}}\cdot 
{\cal P}^{(d-1)}_{{\cal N}} 
\psi^{(6)}_{-\frac{1}{2},m}&=& ((-i)\stackrel{03}{(+i)} \stackrel{12}{[-]}
 \stackrel{56}{[-]} + 
i \,\frac{m\;}{m_{+}} \stackrel{03}{[-i]} \stackrel{12}{[-]} \stackrel{56}{(+)})\, 
e^{-imx^0} \,,
\nonumber\\
 m^2&=& m_{+}\, m_{-}\,,\quad m_{+} = -m_{-}\,, \quad (p_{0})^2 = m^2\,,
\end{eqnarray}
which means that $-i\,\Gamma^{(6)} \Gamma^{(3+1)}$ 
$ \mathbb{C}_{{\cal N}}\cdot $ ${\cal P}^{(d-1)}_{{\cal N}}\, 
\psi^{(6)}_{\pm \,\frac{1}{2},m}$ $=i \frac{m\;}{m_{+}}\,$ 
$\psi^{(6)}_{\pm \,\frac{1}{2},m}$. 

\vspace{3mm}

The two positive solutions of the effective Weyl equations, Eq.~(\ref{Weylmvacsimple}),  
representing particles carrying no 
charge,  are indistinguishable from the two positive energy solutions for the 
corresponding two antiparticles. They are indeed the Majorana particles.

In the case that there is no conserved charge due to the break of 
the mass protection (caused by the scalar fields $\omega_{56 s}$, which gain 
constant values) the Majorana fermions are  described by the sum of the fermion 
and the corresponding antifermion state, Eq.~(\ref{Weylmsolsimpleanti}). In the 
 simple case that $\vec{p}=0$ the Majorana fermion states 
$\psi^{(6)}_{\pm \frac{1}{2},\tiny{M}} $  can be written as
\begin{eqnarray}
\label{Majoranastates}
\psi^{(6)}_{\pm \frac{1}{2},\tiny{M}} &=&\frac{1}{\sqrt{2}} \,
(\psi^{(6)}_{\pm \frac{1}{2},m}\,
+(\pm)\, (i)\,\Gamma^{(6)} \Gamma^{(3+1)} \mathbb{C}_{{\cal N}}\cdot 
{\cal P}^{(d-1)}_{{\cal N}} \;
\psi^{(6)}_{\pm \frac{1}{2},m}(\vec{p}=0)) \,,\nonumber\\
(\pm)\,, \;\, {\rm if}\; m_{+}= (\mp)\, i\,m\,,  
\end{eqnarray}
and have  the mass equal to $m$~\cite{TDN}.


Me have started in our particular case, the toy model in $d=(5+1)$, with the chain
of massive fermions and the massless  charged fermions of particular handedness 
and the corresponding antifermions, and ended with no massless states and the 
massive Majorana particles (in addition to the chain of massive particles).

%
%


\subsection{Trials to make next step beyond {\it standard model} in 
 theories}
\label{others}
	 

Both authors have participated in holding through $23$ years 
a little workshop at Bled entitled "What comes beyond the standard models",
 at which many works were seeking to build extensions of both standard models, 
the electroweak and colour {\it standard model} and the cosmological model, 
which could explain the phenomena, which the electroweak and colour 
{\it standard model } can not,
 or  even find the next step beyond both standard models.

Most attempts try to guess how to extend the {\it standard model} 
assumtions/details
so that the predictions of their extensions will show up, when the accelerators
reach yet higher energies, or when some particular experiments could measure
their prediction.










There are attempts, like the one called "multiple point principle" of H.B.N. and his 
collaborators~\cite{MPP,MPP1,MPPfit,MPP2,MPP3}~\footnote{The collaborators  are D. Bennett and C.D. Froggatt 
after a long previous development involving also N. Brene, I. Picek, and L. Laperashvili.
}, which try to understand the {\it standard model} 
parameters read from experiments without extending the {\it standard model}, 
just by trying to understand "why has nature made a choice" of the observed 
properties of quarks and leptons.



Some theories are searching for steps beyond the 
{\it standard model}, 
 like there are $SO(10)$-unifying theories, the string theories, the Kaluza-Klein like
theories, as well as the  {\it spin-charge-family} theory, which is also a kind of 
Kaluza-Klein theory, with the ambition  (starting  from one parameter) to explain 
all the parameters and assumptions of the {\it standard model} and the 
cosmological models.
(Although the  "phase transitions" --- breaks of symmetries spontaneously ---
are so hard job, which even if possible, requires the knowledge of boundary 
conditions of expanding universe at a very moment of the break of a symmetry.) 

Having an appropriate theory one can do a lot with the help of observations and 
experiments.

In this Sect.~\ref{others} we intend to present only (some of) possibilities,
some of the ones for which we have our own understanding of the  topics.


\subsubsection{To understand assumptions by string theories}
\label{string}
Those string theories, which
at all have a chance to be useful as models beyond the {\it standard model},
must have special fermionic modes - fields of fermions
running along the string - in addition to the position and momentum
degrees of freedom for the relativistic string. Otherwise there would 
be no fermionic states of the string, and no fermion to correspond to the
fermions in the {\it standard model}.

But 
what in particular makes the (super) string theories being
able to claim that there are (extremely) few assumptions once the string theory
is chosen is that there are several quantum anomalies, 
often the symmetries, that are seemingly put in by hand by assumption, and
do not come out after the quantum corrections are evaluated.

An example is that the rotational symmetry of the string theory
does usually not come out from the quantum calculation, except in the
case of  a very special dimensionality of space time~\footnote{%
If one  uses the 
infinite momentum frame formalism in which a special coordinate axis
is selected and treated in a special way, one has rotational
invariant theory described in a non-rotational invariant formalism.
}.

The rotational symmetry becomes a true breaking symmetry, unless the 
dimension of  the string without the fermion mode is $d=(25 +1)$ and
with the fermion mode is $d=(9+1)$~\footnote{%
Such a trouble that a symmetry that 
should be there, because we put it in, gets spoiled by the quantum 
calculation, is called a quantum anomaly~\cite{Alvarez,AlvarezWitten}.}. 
The choice of the dimensionality 
makes the theory more predictive.  It still remains to prevent that a particle with
a negative squared mass, that is called a tachyon, would appear. The
superstring with the fermionic modes offers this possibility --- in a quite
 tricky way.
 
 It then turned out that there are  only $5$ 
satisfactory string  theories 
  with no anomaly problems left and the
  usual physical principles such as with no tachyons and the Lorents invariance.
%



The extra dimensions must be curled up 
to end up with $d=(3+1)$ 
observable so far. 
This kind of 
compactification 
appears also in the {\it spin-charge-family} theory. 

The quantization of gravity might be the main
strong point of the string theory; and then the somewhat
related point that it in higher than four dimensions
can have a renormalizable or meaningfull theory with
quantum corrections. 

But  with respect to  predicting phenomena at low energy physics
can the string theory hardly be competitive with the 
{\it spin-charge-family} theory, which is so
close to being free theory that one essentially avoids quantum 
corrections~\footnote{%
The work of one of us, H.B.N. and M. Ninomiya might help making sensible theories in higher
dimensions~\cite{stringHM}.}.


In getting out from string theory a detailed way of breaking the
original string groups $SO(32)$ or $E_8\times E_8$ down to the group
to be observed at low energy involves typically so many
choices, that it becomes hard to seriously claim that the string theory
lead to the standard model group. There are also possibilities
to construct models inside string theory giving the standard model group
not by using the gauge groups in the string theory but 
brane bunches that can be made up easily to give $U(N)$ groups.
But the more possibilities the more difficult to claim any clean and
unique prediction.

In the {\it spin-charge-family} theory  there is  some freedom in
the way how to curl up (essentially compactify) non observed dimensions
one after the other. But keeping to the assumption of letting the fermions in
practice sit at the origo in the extra dimensions and basically making
all but a small region near this origo become unimportant, the possible
choices are reduced to how big - in number of dimensions - are the
successive ``compactified'' dimensions, and that sounds giving less
possibilities than in the string theory where one does not restrict the
compactification in the just mentioned way, and has a larger number of
even massless particles at the outset, because of the fundamental
gauge symmetry.

The Kaluza Klein way of obtaining gauge fields is in principle also
awailable in string theory, but not so popular becuase one often has in
mind truly compactified extra dimensions in which case we have a theorem
by Witten making it impossible to achieve chiral fermions in the low
energy theory in this way~\cite{WittenNOGO}.


%

%

\subsubsection{Flavour changing neutral currents and proton decay in theories
with several scalar fields (higgss)}
\label{FCNCproton}

{\bf i. } {\it Comments on the flavour changing neutral currents}

\vspace{3mm}

In theories with several scalar fields (higgses), as it is the {\it spin-charge-family}
theory, one expects difficulties in assuring that transitions from a family member
of a particular family to the same family member of another family would 
agree with the experiments, since such transitions, known as flavour 
changing neutral currents, are not 
observed~\footnote{%
 Due to A.H.G., the coauthor of N.S.M.B. in 
Ref.~\cite{NA2018},  any theory looking for a way beyond the 
{\it standard model} can have serious difficulties in suppressing the 
flavour changing neutral currents, since these processes require proper 
coupling constants and heavy masses of the exchange mediators.
%
%
%
Even if in all orders of corrections the symmetry of mass
 matrices  remains $\widetilde{SU}(2)\times \widetilde{SU}(2) \times U(1)$ 
and the theory reproduces the right masses and mixing matrices, this  by 
itself does not guarantee the suppression of the flavour changing neutral currents.}.
These transitions 
are in the {\it standard model} prevented with one higgs only and with
Yukawa couplings. In the {\it spin-charge-family} theory with the group of four 
families and two triplet scalar fields, coupled to family quantum 
numbers (the flavours), and three singlet scalar fields, coupled to family members 
quantum numbers, the flavour conserving neutral currents 
can be explained as follows:\\ 
When non-zero constants (the vacuum expectation
values) of this higgses  give masses to,  let say, $u_i$ quarks, $i$ counts families,
of $u$-quarks, the mass matrix gets proportional to the Yukawa coupling matrix. 
After diagonalizing the mass matrix of $u_i$ quarks each eigenvector of 
the mass matrix represents one family of particular quantum number (flavour).  
In this new basis 
 the Yukawa coupling matrix becomes  diagonal  in the family quantum 
numbers (flavours), and thus the coupling to the single higgs can not break 
the family quantum number (flavour)  conservation. 
 So the higgs exchange, the one which diagonalizes the mass matrix of
$u$-quarks, since there is only one higgs, cannot break the flavour conservation. 
But because the
definition of the flavours were made for each member or say charge value
of the quarks or leptons separately, one can easily risk that the
heavy gauge bosons, which connect corresponding quarks in a quite different
basis from the one defining the flavours, can violate flavour conservation.
But the heavy gauge bosons of course give so called "charged'' currents
and they will not be conserved, only the "neutral currents'' get
conserved.

If there are more than one higgs acting by being exchanged in the
experimentally accessible region of energies, then at least with a random
coupling assumption two of the higgses will have couplings that
cannot be made diagonal and thus flavour conserving with the same
basis of flavour eigenstates.

The supersymmetry 
works so that the $u_i$-quarks, $i$ denotes $(u,c,t,..)$ quarks, get masses from 
different higgses than the $d_i$-quarks, $i$ denotes $(d,s,b,..)$ quarks. 

The models with no supersymmetry and with two higgses~\cite{Botella,2HDM,2HDM1,2HDM2}  
typically introduce some discrete, like $Z_2$, symmetries to prevent  
couplings of some of fermions to more than one higgs. 

\vspace{3mm}

\noindent
{\bf ii.} {\it Comments on the proton decay}

\vspace{3mm}

The present lower bound for the life time of the proton decay of the kind
$p \rightarrow e^++\pi^0$  is $10^{34}$ years.  In the {\it spin-charge-family},
and in the $SO(10)$-unifying theories without any supersymmetry, 
this life time  is roughly estiated to be up to $10^{35}$ years.

Let be added that in the $SO(10)$-unifying theories the breaking of the 
$SO(10)$ group to subgroups is achieved by some Higgs's scalars, which
contribute to  the proton decay. In the {\it spin-charge-family} theory 
the space is (almost) compactified as discussed in Sect.~\ref{TDN0},
and the investigations how this compactification influences the proton decay 
is still in process. 

One should as well carefully study in all the theories, whether baryon number
conservation, which in the standard model is an accidental or non-trivially 
derived symmetry~\cite{Weinberg},  and other symmetries~\cite{HC}
coming out of the assumed gauge symmetries, work. 


%
\subsubsection{Small representation requirement.} 
\label{small}
%


  There are examples, like it is the "explanation without a model
  behind'', or at least where the model behind is less beautiful than
the assumption made to get the gauge group taken alone, but
  still explaining the gauge group of the {\it standard model}. We may
mention
  the work of one of us (H.B.N.) and the coauthor D. Bennett~\cite{smallrep4,%
smallrep1,smallrep2,smallrep3}.
  In this work the authors,  Bennett and one of us
  set up
  a method of defining, what can be called the "size of a representation
  of a Lie group'', which is essentially the volume of such a
  representation relative to the in the same way defined volume of the
  adjoint representation. The latter concept has to be extended
  a bit to cope with groups that are not semisimple.

  Remarkably the in this way defined smallest faithful representation
  turns out to occur just for the gauge groups of the {\it standard
model}.
  So we can claim that this looking for the smallest true  group
representation
  "explains'' the standard model gauge groups.
(We here emphasize the word group, because it is important
 that one thinks about the Lie group rather than only
the Lie algebra as it is a priori the only thing that matters for a gauge
theory~\footnote{%
O` Rafaitaigh~\cite{ORafaitaigh} proposes a way to use
some knowledge about the representations to select a Lie  group
among the usually several Lie groups corresponding to the same
Lie algebra. But this O'Rafaitaigh extension really means that our
prediction gives a little more information than if it would be only
the Lie algebra.}).

The detailed definition of the quantity which points out the
{\it standard model} group is up to a bit of
choice for the Lie groups, which are not simple, but
for simple groups it is simply the ratio of the
quadratic Casimir operator expectation value for the
representation in question to that of the adjoint representation
\begin{eqnarray}
\label{def1}
   \frac{Cas_r}{Cas_{Adj}}&&,
\end{eqnarray}
a quantity that is ``balanced'' w.r.t. it dimensional dependence in the
sence that  for the smallest true representations it goes to a
constant
for very big dimensionalities of the Lie group.

The winner of our extremization of this quantity among the non-abelian
simple groups is the  $SU(2)$ group, one of the ingredients in the
standard model.

The other property, which our election cares for, than the requirement of
having preferably gauge group
components and representations with small values of $
\frac{Cas_r}{Cas_{Adj}}$, 
is there has been divided out of the center of the say covering group an
as large discrete subgroup as possible. This  is  equivalent to
that the representations of the various representations allowed for
the model having the winning gauge  have favourably most strong
restrictions of the type like triality for the $SU(3)$ of the
representation is connected with the allowed values for the weak hyper charge.
In fact one has in the standard model
\begin{eqnarray}
\label{def2}
   y/2 + \frac{1}{3}``triality'' + ``weak isospin''&=& 0 \hbox{(mod 1)}.
   \end{eqnarray}
Our winner group is strongly favoured by this type of relation being
roughly so complicated as possible.

So even if one should begin to allow oneself to look for other quantities
for
the one we constructed, then just that the standard model is doing very
well
w.r.t. to low  $\frac{Cas_r}{Cas_{Adj}}$ strong rules between the
integerness of the weak hypercharge and the representation of the
non-abelian
groups, brings the choice of the winner close to the standard model.

But why should one then have such prinicple of small
representations?

``The truth behind history of the work of one of us and''
and it is told by me to continue:
`` Don Bennett
leading to suggesting the principle of smallest representaions
is that we worked on a rather complicated idea of a model
beyond the standard model based on the dream of ``Random Dynamics'',
which really means we assume the fundamental physics to be enourmously
complicated with random couplings or parameters, but that somehow by
taking a limit of say looking at only low energy expriments you can
nevertheless extract some consequencies. We piled up more and more
helping assumtions among which the ``Multiple Point Principle''
and then found that the quantity that would favour the chance for a
given Lie group $G$ to be one appearing at low energy in the ``Coulomb
phase''(i.e. with essentially/approximately massless gauge particles)
rather than being confined or Higgesed was this ratio
$\frac{Cas_r}{Cas_{Adj}}$.

That is to say, that thinking back on the way we strictly speaking arrived
to the idea, we are suggested to look for a theory behind the
postulate of ``small representation(s)'' of the type of such
small representations favouring the group to appear as a gauge group
in a fundamental theory put up randomly. I.e. we would say like this:

We would formally completely opposite to the model in the present article
assume that the fundamental theory is extremely complicated and has lots
of parameters with random values.
(It sounds opposite but in a way the point should be that assuming a
random model like this is about simpler than even a very simple
model, and thus in a way of thinking the simplest you can propose).

Let us give the argument how a random theory might favour the
``small representation'' at least in some in words way:

In random theory there are only (gauge) symmetries by accident, and the
main importance is that the Lagrangian or Hamiltonian - which is
now  a random function - varies very little when the one transforms the
fields for some particle with the gauge transformation. But under such a
variation the field moves around on the manifold or metric space
the volume of which is what we call the size or volume of the
representation. Obviously one would say: the smaller this volume the
better the chance that the variation is small, and so the better the
chance for an accidental gauge invariance popping up. So the gauge
symmetries most likely to pop up by accident are the ones with the
smallest volumes for the representations. The natural measure relative to
which to normalize this volume is the corresponding volume of the
adjoint representations, which is in a way the representation on the Lie
algebra itself, and thus only involving the group itself.

It would be very natural to say that varying the gauge group a unit little
step should be normalized by varying it a unit step in the measure of the
adjoint representation. Then for the groups with ``small representations''
in our sense the corresponding step in the representation, of the fields, is small.

In any case it looks that the standard model group - in O'Raifeartaighs
sense - has characteristic that the fields in the representations
vary in our sense exceptionally little under variation of the gauge.
So it would be the easiest one to get by accident.''

\section{Influence of second quantization  in quantum physics of many body 
systems} %

%
\label{manybody}

Most of the recognitions and predictions of  the  {\it spin-charge-family} theory  
might and hopefully will influence the high energy experiments and interpretation of 
the high energy experiments, if   the  {\it spin-charge-family} theory offers the 
right next step beyond the {\it standard model}. \\   
Could the recognitions and predictions presented in this paper influence 
the experiments and the theoretical evaluations
of bound and decaying states in hadron and nuclear many body systems?
Some of them might.\\
{\bf a.} The fact that single particle states, describing quarks and leptons, 
anticommute, 
explains the necessity to use Slater determinants when describing and studying 
the bound states of quarks and quarks and antiquarks in hadron, nuclear and
atomic physics.\\
{\bf b.} Several scalar fields determining the masses of the
observed families of quarks and leptons and the weak bosons, some of them 
 with smaller masses than the Higss's scalar, might influence 
dynamics in bound and decaying states of hadrons (mesons and nucleons)
and of nuclei~\footnote{%
There are warnings in Sect.~\ref{FCNCproton}  that
several higgses, in particular those of smaller masses, predict the flavour changing 
neutral currents (the transitions of a particular family member from one family to 
another) what is not in agreement with the observations. One of the authors, 
N.S.M.B., remains optimistic after making rough estimations.}.

It might still be that physicists studying properties of the bound and decaying 
states of the first group of (so far observed) families, bound in hadrons, will
recognize that there are problems, which they could solve by taking into account the 
predictions of the {\it spin-charge-family} theory that there exist scalar fields
with the properties of the Higgs's scalar, having lighter masses than it is 
the mass of the Higgs's scalar. \\
{\bf c.} The existence of the second group of four families, Sect.~\ref{scalar3+1},
predicted to contribute to the observed {\it dark matter} in the universe, 
manifesting the "new nuclear force" among the corresponding "hadrons" 
of the upper four family members, might be a challenge for those who have 
a great experience in hadron and nuclear physics of the ordinary matter. 
Since the mass of nucleons (mostly of the fifth family) bound in nuclei have
very high masses  to this "nuclear force" the largest contribution comes from  
the one gluon exchange (also the weak and electromagnetic  interactions start 
to contribute almost as much as the colour 
interaction~\cite{gn2009,nm2015}~\footnote{%
In Ref.~\cite{gn2009,nm2015} the evaluation of the bound states of
quarks of the fifth family into nucleons is made with the simple Bohr's model.
The two triplets, the gauge fields of $\tilde{SU}(2)$ and $\tilde{SU}(2) $ 
together with the three singlets, the gauge fields of three $U(1)$, bring 
masses to the upper four families of quarks and leptons.
Due to the interaction of the upper four families with the condensate, 
Table~\ref{Table con.}, appearing at very high energies, Sect.~\ref{scalar3+1},
the masses of these quarks and leptons are expected to be much 
higher than the masses of the members (quarks and leptons) of the 
lower four families.}.\\

The existence of the fourth member of the observed three families 
with masses larger than $1$ TeV might hardly influence the low energy
regions in which properties of hadrons and nucleons are measured.
Also the existence of additional scalar fields which are with respect to the 
space index colour triplets and antitriplets, causing proton decay and might be
responsible for the matter/antimatter asymmetry in the expanding universe could 
not be appropriate candidates to attract physicists with experiences in
 hadron and nuclear physics.\\




 %

\section{Conclusions}
\label{conclusions}

The {\it standard model} certainly made, when postulated, a large step towards 
better understanding  the law of nature. Now a new large step is needed to explain 
all the assumptions on which the {\it standard model} is built. {\it The theory is 
needed, which would explain all the {\it standard model} assumptions in an unique 
unified way.} Not only since the explanation will help to understand the law of nature 
but also since further experiments can hardly be correctly interpreted without 
making the next large enough step. 

In the literature most of suggestions how to explain the assumptions of the 
{\it standard model} rely on embedding  the {\it standard model} gauge groups 
into an unifying group, if possible, like there are $SO(10)$ or larger exceptional
groups. When this does not seem possible or meaningful, as it is with the
family group, then the multiplication of the unifying groups with the family group 
is often suggested. Correspondingly the vector gauge fields of the charged 
subgroups of the unifying group can be assumed following the {\it standard model} 
suggestions how to take into account more charges and in addition the 
observed number of families. And when needed the additional scalar fields 
are assumed. 

Is the group theory approach, on which $50$ years ago the {\it standard model} 
was so innovative and elegantly built, the only way or at least the best way to make
the next step towards understanding the law of nature on the level of elementary 
fermion and boson fields? 

Although the group theory approach~\cite{PierreRamond} might answer several 
open questions which the {\it standard model} leaves unanswered, the authors 
of this review article are convinced, at least (S.N.M.B.), that the law of nature 
is on the level of elementary fields simple and elegant, offering in an elegant 
way all the answers to the questions which the {\it standard model} leaves open. 


In this review article the authors explain step by step, illustrating steps with several 
examples, the building blocks of the {\it spin-charge-family} theory, and also its 
achievements so far and predictions, comparing this theory with the $SO(10)$-%
unifying theories:\\
{\bf A} The simple starting action, Eq.~(\ref{wholeaction}), is in $d=((d-1)+1)$-%
dimensional space, $d\ge (13+1)$, assumed for massless fermions, the
internal space of which is described by the odd Clifford algebras, and for gravity 
as massless boson fields, the only bosons appearing in the theory: The vielbeins 
(the gauge fields of momenta), the spin connection fields (the gauge fields of the 
generators of the Clifford algebra operators $S^{ab}$ and $\tilde{S}^{ab}$)~%
\footnote{Let us repeat that if there are no condensate of fermions present
the spin connections are expressible by vielbeins.}.\\ 
{\bf A.a} There are two kinds of the Clifford algebra objects determining properties 
of fermions in the theory. We use one kind, $\gamma^a$'s, to describe the internal 
space of fermions, and the second kind, $\tilde{\gamma}^a$'s, to generate 
families, equipping each family with the "family charge".\\ 
{\bf A.a.i} The odd Clifford algebra of $\gamma^a$'s describes the internal space
of fermions. The corresponding creation operators and their Hermitian conjugated 
partners annihilation operators fulfill the Dirac's postulates of the second quantized 
fermions, offering explanation for the Dirac's postulates, 
Sect.~\ref{creationannihilationtensor}. Single fermion states are 
correspondingly anticommuting objects, manifesting that the first quantization 
already deals with anticommuting objects. 
This offers the explanation for the necessity of using Slater determinants in the
first quantized many body wave functions.\\
{\bf A.a.ii} In $d=(13+1)$ the spin and charges and families (families are unified 
with the spins and charges by means of our two Clifford algebras), 
Sect.~\ref{internalspace}, explain the appearance of quarks and leptons 
and antiquarks and antileptons, related to handedness as assumed by the
 {\it standard model}, appearing in families. \\
The Clifford operators $S^{ab}$ transform any member of a particular family to
all the other members of the same family, including fermions and antifermions,
Table~\ref{Table so13+1.}.\\
The Clifford operators $\tilde{S}^{ab}$ transform a family member of one 
family into the same family member of another family of fermions, 
Table~\ref{Table III.}.\\ 
In $d$-dimensional  space there are $2^{\frac{d}{2}-1}$ families with 
$2^{\frac{d}{2}-1}$ family members each~\footnote{
When we require that quarks and leptons and antiquarks and antileptons fulfill
for particular energy $p^0 =|\vec{p}|$ the equations of motion, 
Eq.~(\ref{Weyl}), then to a fermion with momentum $\vec{p}$ and particular 
spin the anti-fermion with momentum $-\vec{p}$ and the same spin belongs,
Sect.~\ref{CPT}, reducing the number of "physical states", 
Sects.~\ref{solutions5+1},~\ref{solutions9+1 13+1}.}. Spins, charges and 
families are correspondingly unified.\\
{\bf A.b} Vielbeins and spin connections manifest in $d=(3+1)$, carrying the 
charges of subgroups of the unifying $SO(13+1)$ group, as vector and scalar 
gauge fields, Sect.~\ref{vectorscalar3+1}. The ordinary gravity is the tensor 
gauge field of the "spinor charge" in $d=(3+1)$.\\
{\bf A.b.i} Vector gauge fields, Sect.~\ref{vector3+1}, the gauge fields of the 
charge groups, the gauge subgroups of $SO(13+1)$, carry in $d=(3+1)$ the 
space index $m=0,1,2,3$, Sect.~\ref{vector3+1}.\\
{\bf A.b.ii} Scalar gauge fields, Sect.~\ref{scalar3+1}, with the space index 
$(7,8)$ explain the appearance of the Higgs's scalar and Yukawa couplings in 
the {\it standard model}. Scalar gauge fields with the space index 
$s=(9,10,11,12,13,14)$, transforming antileptons into quarks and antiquarks 
into quarks, and back, offer the explanation for the proton decay and the 
{\it matter-antimatter asymmetry} in the expanding universe. All the scalar
gauge fields have all the charges with respect to $S^{ab}$ and $\tilde{S}^{ab}$ 
in the adjoint representations, as also the vector gauge fields do, while with 
respect to the scalar index they manifest as the weak and hyper charge 
doublets and colour triplets and antitriplets. There are no additional
scalar fields in the simple action in $d=(13+1)$.\\
{\bf A.b.iii} The $SO(3,1)$ "charge subgroup" of $SO(13,1)$ group determines
properties of  vielbeins and spin connections  in $d=(3+1)$, $m=(0,1,2,3)$,
representing the ordinary gravity. \\
{\bf A.b.iv} The {\it spin-charge-family} theory is a free, 
renormalizable theory, except for gravity.\\
{\bf B} The symmetry of the starting manifold $M^{13 +1}$ must break first to
$M^{7 +1} \times SU(3) \times U(1)$ to keep the starting relation among 
handedness and charges, what reduces the number of families to 
$2^{\frac{7+1}{2}-1}$ --- this is done by the condensate of two right handed 
neutrinos with the family quantum number not belonging to the family quantum 
numbers of the observed families of quarks and leptons and antiquarks and 
antileptons, Table~\ref{Table con.} --- and then further to $M^{3 +1}$ 
$\times SU(2) \times SU(3) \times U(1)$.
The further break to $M^{3 +1}$ 
$\times SU(3) \times U(1)$ is caused by the scalar fields with space index $(7,8)$,
Sect.~\ref{scalar3+1}.\\

We discuss this break in the case of the toy model, in which the vielbeins and spin 
connections of particular properties cause the break from $M^{5 +1}$ to
$M^{3 +1} \times $ an almost $S^{2}$ sphere which ensures massless fermions 
also after the break with charges from $d=(5,6)$, Sect.~\ref{TDN0}. The 
masslessness of fermions is broken when scalar fields gain constant values.
The realistic case is under consideration, and it is not yet finished. \\
{\bf C} The {\it spin-charge-family} theory offers several predictions, 
Sect.~\ref{predictionSCFT}:\\
{\bf C.a} The two groups of four families are predicted, with mass matrices 
manifesting the $\widetilde{SU}(2)\times \widetilde{SU}(2) \times U(1)$ symmetry, 
what reduces the number of free parameters of mass matrices. \\
{\bf C.a.i} To the three observed families of quarks and leptons the fourth family
with masses above $1$TeV is predicted. \\
{\bf C.a.ii} The lowest of the upper four families is (almost, up to $10^{14}$GeV 
or  higher) stable, offering explanation for the existence of the 
{\it dark matter} with masses larger then a few 10 TeV.
Due to high masses of quarks and leptons of the upper group of four families, 
"hadrons"  of (mostly) the fifth family quarks experience
"nuclear force", which strongly differs from the so far observed nuclear force.
\\
{\bf C.b} The {\it spin-charge-family} theory predicts the existence of several scalar
 fields, determining Higgs's scalar and Yukawa couplings of the {\it standard model} 
as the superposition of two triplets and three singlets, some of them with  
smaller masses than the Higgs's scalar, as well as the  new vector triplet gauge 
field --- the second $SU(2)$ triplet  with the mass close to $10^{14} $GeV or 
higher, since it couples to the condensate of two right handed neutrinos appearing 
at this scale, Table~\ref{Table con.}.\\

\vspace{3mm}

 The heavy $SU(2)$ triplet is predicted also by the $SO(10)$-unifying
theories, if assuming the appropriate scalar fields, which break the symmetry of 
$SO(10)$. \\

\vspace{3mm}



The authors hope that the reader will recognize that the next step beyond the 
{\it standard model} might not necessarily use the group theory 
aproaches~\cite{PierreRamond}, on which already  the {\it standard model}
was built, assuming larger groups and correspondingly 
unifying charges in $d=(3+1)$, as well as their vector gauge fields.
The {\it spin-charge-family} theory offers, namely, much more, 
while  starting with the simple and correspondingly elegant action in $d\ge (13+1)$: 
$d=(13 +1)$ is the smallest dimension, which offers the next step beyond the  
{\it standard model} in the Kaluza-Klein like theories, manifesting in $d=(3+1)$ all the 
properties of fermions of vector and scalar gauge fields observed so far. The
{\it spin-charge-family} theory
unifies spins, charges and families, as well as all the elementary interactions with the
gravity included, explaining the second quantization postulates and treating already
the first quantized fermions with anticommuting states, manifesting that 
there are only the second quantized fields, which in a simplified way can be 
presented  as the first quantized fermions arranged in Slater determinants.\\
In Ref.~\cite{NH2000} authors of this article  prove, assuming 
that equations of motion are Hermitian and that the solutions 
form the irreducible representations of the Lorentz group, that only the 
metrics with the signature corresponding to $q$ time and $d-q$ space 
dimensions with $q$ odd can exist.  Correspondingly in $d=(3+1)$ the only 
possibility is q=1. \\
%

There are open problems in the {\it spin-charge-family} theory, some of 
them shared with the Kaluza-Klein like theories~\cite{KaluzaKlein,Witten,Duff,App,%
SapTin,Wetterich,zelenaknjiga,mil,nh2017}, waiting to be solved, like:\\
{\bf i.} The compactification of higher dimensions in the presence of families. 
Although the compactification problem is solved for the toy model in $d=(5+1)$ 
without the presence of families, Sect.~\ref{TDN0}, and also with the presence of 
families~\cite{familiesNDproc}, the breaks of the starting symmetry in steps, leading
to the observed properties of quarks and leptons and antiquarks and antileptons, to
the observed properties of the scalar fields, need  to be done also for 
$d=(13+1)$-dimensional case.
\\
{\bf ii.} 
It must be demonstrated how do scalar triplets and singlet fields, 
Sect.~\ref{scalar3+1}, with the properties of the Higgs's scalar, after gaining
 masses through interaction with the condensate, 
spontaneously gain imaginary masses and correspondingly the constant values, 
what breaks the weak and hyper charge symmetry.\\
{\bf iii.} Although at low energies there is no difficulties with the behaviour of the 
vector gauge fields, the gravitational gauge field needs to be quantized.\\
{\bf iv.} And other problems, common to all the theories, for some of  which we 
almost see the way how to solve them and other problems which we do not yet 
know how to look for the solution.

The more effort and work is put into  the  {\it spin-charge-family} theory 
the more answers to the open questions the theory offers. Because of this
 at least one of us, N.S.M.B., remains optimistic. 
The working hypotheses of the authors of this paper (in particular of N.S.M.B.) 
is, since the higher dimensions used in the {\it spin-charge-family} theory 
offer in an elegant (simple) way explanations for the so many observed 
phenomena in elementary particle physiscs and cosmology, that higher 
dimensions should not be excluded by the renormalization and anomaly 
arguments. 

At least the low energy behavior of the spin connections and vielbeins as vector 
and scalar gauge fields manifest as the known and more or less well defined 
theory.

The gravity can not be neglected just because we cannot treat it at the very
high energy scales. But at  low energies we can rather neglect the
higher order terms and treat the {\it spin-charge -family} theory as 
renormalizable in $d=(3+1)$ dimensions only, showing the next 
step beyond the {\it standard model}.

\appendix

\section{Action for free massless "Grassmann fermions" with integer spin~%
\cite{IARD2020,nh2018,n2019PIPII}}
\label{actionGrass}
%



In this appendix the action for the integer spin "Grassmann fermions" is presented, 
taken from Refs.~\cite{IARD2020,nh2018,n2019PIPII}, which offer the anticommuting
"basis vectors", Eq.~(\ref{ijthetaprod}), if they are products of an odd number of 
"eigenvectors" of the Cartan subalgebra, Eq.~(\ref{cartangrasscliff}). 
 
"Basis vectors", which are  products of an even  number of 
"eigenvectors" of the Cartan subalgebra, commute. They are also presented in this
appendix.

After the reduction of the Clifford space also the Grassman space is reduced,
Sect.~\ref{reduction}, and the integer spin "Grassman fermions" might have
no "physical" meaning after this reduction, that is, since we do observe half 
integer fermions --- quarks and leptons and antiquarks and atileptons of several 
families --- the internal degrees of which is describable with one kind of the 
Clifford algebra objects, $\gamma^a$'s, there might be after this  reduction
no possibilities for the integer spin "Grassmann fermions."  Yet we present here
a possible  action for  the integer spin "Grassmann fermions".

\subsection{Action for free massless anticommuting ''Grassmann fermions" with integer 
spin}
\label{actionGrassodd}
In  the Grassmann case  the ''basis vectors'' of an odd Grassmann character, 
chosen to be the eigenvectors of the Cartan subalgebra of the Lorentz algebra in 
Grassmann space, Eq.~(\ref{cartangrasscliff}), manifest the anticommutation 
relations of Eq.~(\ref{ijthetaprod}) on the algebraic level.  
  
To compare the properties of creation and annihilation operators for ''integer spin 
fermions'', for which the internal degrees of freedom are described by the odd 
Grassmann algebra,   
with the creation and annihilation operators postulated by Dirac for the second 
quantized 
fermions depending on the spins of the internal space of fermions in $d=2(2n+1)$ or 
$4n$ (n is non negative integer) dimensional space and on the
momentum space, we need to define the tensor product $*_{T}$  of the odd 
''Grassmann basis states'', described by the superposition of odd products of 
$\theta^a$'s (with the finite degrees of freedom) and of the momentum
 (or coordinate) space (with the infinite degrees of freedom), taking as the new
 basis the tensor product of both
spaces.

{\bf Statement 1:} 
For deriving the anticommutation relations for the ''Grassmann fermions'',
to be compared to anticommutation relations of the second quantized fermions, 
we define the tensor product of the Grassmann odd ''basis vectors''  and the 
momentum space
\begin{equation}
\label{thetaptensor}
{\rm {\bf basis}}_{(p^a, \theta^a)} = |p^a>\, *_{T} \,|\theta^a>\,.
\end{equation}

We need even more, we need to find the Lorentz invariant action for, let say, 
free massless "Grassmann fermions" to define such a ''basis'', that would manifest
 the relation $p^0=|\vec{p}|$. We follow here the suggestion of 
one of us (N.S.M.B.) from Ref.~\cite{nh2018}.
%
\begin{eqnarray}
{\cal A}_{G}\,  &=&  \int \; d^dx \;d^d\theta\; \omega \, \{\phi^{\dagger} \,
 \gamma^{0}_{G}
 \,\frac{1}{2}\,
\theta^a p_{a} \phi \}+ h.c.\,,\nonumber\\
\omega &=&\prod^{d}_{k=0}(\frac{\partial}{\;\,\partial \theta_k} + 
\theta^{k})\,,
\label{actionWeylGrass}
\end{eqnarray}
with $ \gamma^{a}_{G}=(1-2\theta^a \frac{\partial}{\partial \theta_a})$, 
$(\gamma^{a}_{G})^{\dagger}= \gamma^{a}_{G}$, for each 
$a=(0,1,2,3,5,\cdots,d)$.
We use the integral over $\theta^a$ coordinates with the weight function 
$\omega$ from Eq.~(\ref{grassintegral}, \ref{grassnorm}).  
Requiring the Lorentz invariance we add after $\phi^{\dagger}$ the operator 
$\gamma^0_{G}$,
which takes care of the Lorentz  invariance. Namely
\begin{eqnarray}
\label{Linvariancegrass}
{\cal {\bf S}}^{ab \dagger}\, (1-2\theta^0 \frac{\partial}{\partial \theta^0}) &= & 
 (1-2\theta^0 \frac{\partial}{\partial \theta^0})\,{\cal {\bf S}}^{ab}\,,\nonumber\\
{\cal {\bf S}}^{\dagger} \, (1-2\theta^0 \frac{\partial}{\partial \theta^0})&=& 
(1-2\theta^0 \frac{\partial}{\partial \theta^0})\, {\cal {\bf S}}^{-1}\,,\nonumber\\
{\cal {\bf S}} &=& e^{-\frac{i}{2} \omega_{ab} (L^{ab} + {\cal {\bf S}}^{ab})}\,,
\end{eqnarray}
while $\theta^a,  \frac{\partial}{\partial \theta_a}$ and $p^a$ transform as 
Lorentz vectors.

The Lagrange density is up to the surface term equal to~\footnote{%
Taking into account the relations  $\gamma^a= (\theta^{a} + 
\frac{\partial}{\partial \theta_a})$, $\tilde{\gamma}^a=i \,(\theta^{a} - 
\frac{\partial}{\partial \theta_a}$), from where one obtains $ \gamma^{0}_{G}
= - i \eta^{aa} \gamma^a \tilde{\gamma}^{a},$ the Lagrange density can be
 rewritten as
${\cal L}_{G}\,  = -i \frac{1}{2} \phi^{\dagger} \, \gamma^0_{G}  \,
\tilde{ \gamma}^a\,( \hat{p}_a \phi)$  
$=  -i \frac{1}{4} \{ \phi^{\dagger} \, \gamma^0_{G}  \,
\tilde{ \gamma}^a\, \hat{p}_a \phi\, - \hat{p}_a  \phi^{\dagger}\, 
 \, \gamma^0_{G}  \,
\tilde{ \gamma}^a\, \phi\,\}$.}
\begin{eqnarray}
{\cal L}_{G}\,  &=&  \frac{1}{2} \phi^{\dagger} \, \gamma^0_{G} 
(\theta^a - \frac{\partial}{\partial \theta_a})\,( \hat{p}_a \phi) \nonumber\\
&=&  \frac{1}{4} \{ \phi^{\dagger} \, \gamma^0_{G}  \,
(\theta^a - \frac{\partial}{\partial \theta_a})\, \hat{p}_a \phi - \nonumber\\
&&(\hat{p}_a  \phi^{\dagger})  \gamma^0_{G} 
(\theta^a - \frac{\partial}{\partial \theta_a}) \phi\}\,,
\label{LDWeylGrass10}
\end{eqnarray}
leading to the equations of motion~\footnote{
Varying the action with respect to $\phi^{\dagger}$ and $\phi$ it follows:   
$\frac{\partial {\cal L}_{G}}{\partial \phi^{\dagger}} -  
 \hat{p}_{a} \,\frac{\partial {\cal L}_{G}}{\partial \hat{p}_a \phi^{\dagger}}  = 
0 =\frac{-i}{2} \gamma^0_{G}  \, \tilde{\gamma}^a\,\hat{p}_a\,\phi$, and  
$\frac{\partial {\cal L}_{G}}{\partial \phi} -  
 \hat{p}_{a} \,\frac{\partial {\cal L}_{G}}{\partial (\hat{p}_a \phi)}  = 0=
 \frac{i}{2}\hat{p}_a \,\phi^{\dagger} \gamma^0_{G}  \, \tilde{\gamma}^a$.}
\begin{eqnarray}
\label{Weylgrass}
\frac{1}{2}\,
\gamma^0_{G} \,(\theta^a - 
\frac{\partial}{\partial \theta_a})
\, {\hat p}_{a} \,|\phi>\,&= & 0\,,
\end{eqnarray}
as well as the ''Klein-Gordon''  equation,
 \[(\theta^a - \frac{\partial}{\partial \theta_a}) \,{\hat p}_{a} 
\,(\theta^b - \frac{\partial}{\partial \theta_b}) \, {\hat p}_{b} \,|\phi>=0 = 
{\hat p}_a {\hat p}^a \,|\phi>.\]

The eigenstates $\phi$ of the equations of motion for free massless "Grassmann 
fermions", Eq.~(\ref{Weylgrass}), can be found as the tensor product, 
Eq.(\ref{thetaptensor}), of the superposition 
 of $2^{d-1}$ Grassmann odd "basis vectors" $ {\hat b}^{\theta k \dagger}_{i}$ 
and   the momentum space, represented by plane waves, applied on the
 vacuum  state $|\,1> $. Let us remind that the ''basis vectors''  are the
 ''eigenstates''  of the Cartan subalgebra, Eq.~(\ref{cartangrasscliff}), fulfilling  
(on the algebraic level)  the anticommutation relations of Eq.~(\ref{ijthetaprod}).  
And since the oddness  of the Grassmann odd ''basis vectors'' guarantees the 
oddness of the tensor products  of the  internal part of ''Grassmann fermions'' 
and of the plane waves, we expect the equivalent 
 anticommutation relations also for the eigenstates of the Eq.~(\ref{Weylgrass}), 
 which define the single particle anticommuting states of ''Grassmann fermions''. 

The coefficients, determining the superposition, depend on momentum $p^a$, 
$a=(0,1,2,3,5,$ $\dots,d)$, $p^0 = |\vec{p}|$,  
of the plane wave solution $e^{-i p_a x^a}$. 

Let us therefore define the new creation operators and the corresponding single 
particle ''Grassmann fermion'' states as the tensor product of two spaces, the
Grassmann odd ''basis vectors'' and the momentum space basis 
\begin{eqnarray}
{\hat{\bf b}}^{\theta k \,s \dagger} (\vec{p})& \stackrel{\mathrm{def}}{=}& 
\sum_{i} c^{k s}{}_{ i}  (\vec{p})\, \hat{b}^{\theta k \dagger}_{i}\,*_{T}
\hat{b}^{\dagger}_{\vec{p}} \,,
 \;\;\; \qquad p^0 = |\vec{p}|\,,\nonumber\\
|\phi^{k s}_{tot} (\vec{p}, p^0)> 
&=& \hat{\bf b}^{\theta k s \dagger}(\vec{p}, p^0)\,*_{T}\,|0_{\vec{p}}>
|\, 1> \,, \;\; \qquad p^0 = |\vec{p}|\,,
\label{ptheta}
\end{eqnarray}
where $\hat{b}^{\dagger}_{\vec{p}}$ is defined in App.~\ref{continuous}, 
Eq.~(\ref{eigenvalue}),  $s$ 
represents different solutions of the equations of motion and $k$ different 
irreducible representations of the Lorentz group, $\vec{p}$ denotes the chosen 
vector ($p^0,\vec{p}$) in momentum space, and $|0_{\vec{p}}>$ is defined in 
Eqs.~(\ref{creatorp}, \ref{eigenvalue10}),and  also discussed in App.~\ref{continuous}. 


One has further  
\begin{eqnarray}
\label{phiksx}
|\phi^{k s} (x^0, \vec{x})> &=& \int_{- \infty}^{+ \infty} \,
\frac{d^{d-1}p}{(\sqrt{2 \pi})^{d-1}} \, 
\hat{\bf b}^{\theta k s \dagger} (\vec{p})|_{p^0 = |\vec{p}|} 
e^{-i p_a x^a}\,*_{T}\,|0_{\vec{p}}> |\,1>= \nonumber\\
&& \hat{\bf b}^{\theta k s \dagger} (\vec{x})\,*_{T}\,|0_{\vec{x}}>\,|\,1>\,.
\end{eqnarray}

The orthogonalized states $|\phi^{k s} (\vec{p})>$ fulfill  the relation 
\begin{eqnarray}
<\phi^{k s} (\vec{p})|\phi^{k' s'} (\vec{p'})>&=& \delta^{k k'}\, 
\delta_{s s'}\,\delta_{p p'}\,,
\;\;\quad  p^0 = |\vec{p}|\,,\nonumber\\
<\phi^{k' s'} (x^0, \vec{x'})|\phi^{k s} (x^0,\vec{x})>&=&
 \delta^{k k'}\, \delta_{s s'}\,
\delta_{\vec{x'}, \vec{x}}\,.
\label{ortpp'}
\end{eqnarray}
where we assumed the discretization of momenta $\vec{p}$ and coordinates 
$\vec{x}$.

%
%
In even dimensional spaces ($d=2(2n+1)$ and $4n$) there are $2^{d-1}$
Grassmann odd superposition of ''basis vectors'', which belong to different
irreducible representations, among them twice 
${\bf \frac{1}{2} \frac{d!}{\frac{d}{2}! \frac{d}{2}!}}$ of the kind presented in
Eqs.~(\ref{start(2n+1)2theta}, \ref{start4ntheta}) and presented 
in Table~\ref{Table grassdecuplet.} of
Sect.~\ref{propertiesGrass0}  for a particular case $d=(5+1)$, 
Ref.~\cite{n2019PIPII}. 
The illustration for the superposition ${\hat {\bf b}}^{\theta k\,s \dagger}
 (\vec{p})$
is presented, again for $d=(5+1)$, in Part I of Ref.~\cite{n2019PIPII}.

We introduced in Eq.~(\ref{ptheta}) the creation operators 
$ \hat {\bf b}^{\theta k\,s \dagger} (\vec{p}) $ as the tensor, $*_{T}$, 
product of the ''basis vectors'' of  Grassmann  algebra elements and the momentum 
basis. The Grassmann algebra elements transfer their oddness to the 
tensor products of these two basis. Correspondingly must 
$ \hat {\bf b}^{\theta k\,s \dagger} (\vec{p}) $
together with their Hermitian conjugated annihilation operators, 
$(\hat {\bf b}^{\theta k\,s \dagger} (\vec{p}) )^{\dagger}=$ 
$ \hat{\bf b}^{\theta k\,s } (\vec{p}) $, fulfill 
the  anticommutation relations equivalent to the anticommutation relations of 
Eq.~(\ref{ijthetaprod})  
\begin{eqnarray}
\{\hat{\bf b}^{\theta k\,s}\, (\vec{p}), 
\hat{\bf b}^{\theta k'\,s' \dagger}\, (\vec{p}{\,}') \}_{*_{T}+} 
\,|0_{\vec{p}}> |\,1> &=& \delta^{k k'}\; \delta_{s s'}
 \delta (\vec{p} -\vec{p}{\,}')\,|0_{\vec{p}}> |\,1>\,,\nonumber\\
\{\hat{\bf b}^{\theta k\,s}\, (\vec{p}), \hat{\bf b}^{\theta k'\,s'} (\vec{p}{\,}') 
\}_{*_{T}+} \,|0_{\vec{p}}> |\,1> &=& 0\;\cdot\,|0_{\vec{p}}> |\,1> \,,\nonumber\\
\{\hat {\bf b}^{\theta k\,s \dagger}\,  (\vec{p}) ,
\hat {\bf b}^{\theta k'\,s' \dagger}\, (\vec{p}{\,}') \}_{*_{T}+}  
\,|0_{\vec{p}}> |\,1>&=&0\;\cdot\,|0_{\vec{p}}> |\,1> \,,\nonumber\\
\hat {\bf b}^{\theta k\,s } \,(\vec{p})\,*_{T}\,|0_{\vec{p}}> |\,1>& =&
0\;\cdot\,|0_{\vec{p}}> |\,1> \,, \nonumber\\
|p^0|&=& |\vec{p}|\,.
\label{ijthetaprodgen}
\end{eqnarray}
$k$ labels different irreducible representations of Grassmann odd ``basis vectors'',
$s$ labels different --- orthogonal and normalized --- solutions of equations of motion 
and $\vec{p}$ represent different momenta fulfilling the relation $(p^0)^2 = (\vec{p})^2$.
Here we allow continuous momenta and take into account that 
\begin{eqnarray}
<\,1|\,<0_{\vec{p}}| \hat {\bf b}^{\theta k\,s } (\vec{p}) *_{T}
 \hat {\bf b}^{\theta k'\,s' \dagger} (\vec{p}{\,}') \,
|0_{\vec{p}}> |\,1>
 &=& \delta^{k k'} \delta^{s s'} \delta(\vec{p}- \vec{p}{\,}') \,,
\label{ortpp'con}
\end{eqnarray}
in the case of continuous values of $\vec{p}$ in even $d$-dimensional space.

For each momentum $\vec{p}$ there are $2^{d-1}$ members of the odd 
Grassmann character, belonging to different irreducible representations.
The plane wave solutions, belonging to different $\vec{p}$, are orthogonal,
defining correspondingly $\infty$ many degrees of freedom for each of 
$2^{d-1}$ ''fermion'' states, defined by 
$ \hat {\bf b}^{\theta k\,s\dagger } (\vec{p})$ for particular $\vec{p}$,
 when applying  on the vacuum state $|0_{\vec{p}}> |\,1>$, Eq.~(\ref{vactheta}). 

With the choice of the Grassmann odd ''basis vectors'' in the internal space of 
''Grassmann fermions'' and by extending these  ''basis states'' to momentum space 
to be able to solve the equations of motion, Eq.~(\ref{Weylgrass}),
we are able to define the creation operators 
$\hat {\bf b}^{\theta k\,s }(\vec{p})$ of the odd Grassmann character,
which together with their Hermitian conjugated partners annihilation operators, fulfill
the anticommutation relations of Eq.~(\ref{ijthetaprodgen}), manifesting the properties 
of the second quantized fermion fields. 
Anticommutation properties of creation and annihilation operators are due to the 
odd Grassmann character of the ''basis vectors''.

To define the Hilbert space of all possible ''Slater determinants'' of all possible 
occupied and empty fermion states, that is the tensor products $*_{T_H}$,
of any number of all possible single fermion states, and to discuss the 
application of 
$ \hat {\bf b}^{\theta k\,s } (\vec{p})$ and 
$ \hat {\bf b}^{k\,s \dagger} (\vec{p}) $ 
on the Hilbert space 
one can follow the procedure of Sect.~\ref{HilbertCliff0} or can
see Ref.~\cite{n2019PIPII}.
\begin{eqnarray}
\{ \hat {\bf b}^{\theta k\,s } (\vec{p})\,,  
\hat {\bf b}^{\theta k\,s \dagger} (\vec{p}{\,}') \}_{*_{T_H}+} 
{\cal H} &=& \delta^{k k'}\; \delta_{s s'} \delta (\vec{p} -\vec{p}{\,}')\;
{\cal H}\,,\nonumber\\
\{\hat {\bf b}^{\theta k\,s} (\vec{p}),  
\hat {\bf b}^{\theta k\,s \dagger} (\vec{p}{\,}') \}_{*_{T_H}+}\;  
{\cal H}&=& 0\;\cdot\,{\cal H} \,,\nonumber\\
\{\hat {\bf b}^{\theta k\,s \dagger} (\vec{p})\, ,
  \hat {\bf b}^{\theta k'\,s' \dagger}_ (\vec{p}{\,}') \}_{*_{T_H}+}\; {\cal H}&=
&  0\;\cdot\,{\cal H}\,.
\label{ijthetaprodgenHT}
\end{eqnarray}
%


Creation operators, $ \hat{\bf b}^{s f \dagger} (\vec{p})$, operating on a vacuum state, 
as well as on the whole Hilbert space, define the second quantized integer "fermion" 
states. 



%
%
 \subsection{Grassmann commuting "basis vectors" with integer spins}
 \label{evengrass}


\begin{small}

Grassmann even  "basis vectors" manifest the commutation relations, and not the 
anticommutation ones as it is the case for the Grassmann odd "basis vectors".
Let us use in the Grassmann even case, that is the case of superposition of an even 
number of $\theta^a$'s  
in $d=2(2n+1)$, the notation $\hat{a}^{\theta k \dagger}_j $, again chosen to be
eigenvectors of the Cartan subalgebra, Eq.~(\ref{cartangrasscliff}),  and let us start
with one  representative
\begin{eqnarray}
\hat{a}^{\theta 1 \dagger}_j {\bf :} &=&(\frac{1}{\sqrt{2}})^{\frac{d}{2}-1} \,
  (\theta^0 - \theta^3) (\theta^1 + i \theta^2) (\theta^5 + i \theta^6)\nonumber\\
   &&{}\cdots (\theta^{d-3} +
 i \theta^{d-2})  \theta^{d-1} \theta^d\,.
\label{start2(2n+1)thetaeven}
\end{eqnarray}
The rest of "basis vectors", belonging to the same Lorentz irreducible representation, follow by 
the application of ${\cal \bf{S}}^{ab}$. The Hermitian conjugated partner of 
$\hat{a}^{\theta 1 \dagger}_1$ is $\hat{a}^{\theta 1}_1 =
 (\hat{a}^{\theta 1 \dagger}_1)^{\dagger}$
\begin{eqnarray}
\hat{a}^{\theta 1}_{1} {\bf :} &=& (\frac{1}{\sqrt{2}})^{\frac{d}{2}-1}\,
\frac{\partial}{\;\partial \theta^{d}}
\frac{\partial}{\;\partial \theta^{d-1}} (\frac{\partial}{\;\partial \theta^{d-3}} -
                           i \frac{\partial}{\;\partial \theta^{d-2}})\nonumber\\
  &&{}\cdots (\frac{\partial}{\;\partial \theta^{0}}
+\frac{\partial}{\;\partial \theta^3})\,.
\label{start2(2n+1)thetaevenher}
\end{eqnarray}
%

If $\hat{a}^{\theta k \dagger}_{j}$ represents a Grassmann even creation operator, 
with index $k$ denoting different irreducible representations and index $j$ denoting a particular 
member of the  $k^{th}$ irreducible representation, while $\hat{a}^{\theta k }_{j}$ represents 
its Hermitian conjugated partner, one obtains by taking into account Sect.~\ref{propertiesGrass0},
the relations 
\begin{eqnarray}
\{ \hat{a}^{\theta k}_i, \hat{a}^{\theta k{'} \dagger}_{j} \}_{*_{A}-} 
|\,1> &=& \delta_{i j}\; \delta^{k k{'}}\;|\,1>\,,\nonumber\\
\{ \hat{a}^{\theta k}_i, \hat{a}^{\theta k{`}}_{j} \}_{*_{A}-}  |\,1>
&=& 0\;\cdot\, |\,1> \,,\nonumber\\
\{\hat{a}^{\theta k \dagger}_i,\hat{a}^{\theta k{'} \dagger}_{j}\}_{*_{A}-} \;|\,1>
&=&0\;\cdot\,|\,1> \,,\nonumber\\
\hat{a}^{\theta k}_{i} \, *_{A}\,|\,1>& =&0\;\cdot\,|\,1> \,,\nonumber\\
\hat{a}^{\theta k \dagger}_{i} \, *_{A}\,|\,1>& =&|\phi^{k}_{e\, i }>\,.
\label{ijthetaprodeven}
\end{eqnarray}
Equivalently to the case of Grassmann odd ''basis vectors''  here 
$\{ \hat{a}^{\theta k}_i, \hat{a}^{\theta l \dagger}_{j} \}_{*_{A}-}=$
$ \hat{a}^{\theta k}_i *_{A} \hat{a}^{\theta l \dagger}_{j} - 
\hat{a}^{\theta l}_j  *_{A}\hat{a}^{\theta k \dagger}_{i} $, with 
$\}_{*_{A}-}$ denoting commutation relations.

Also here, like in the Grassmann odd case, we can make the tensor
product of the internal space of even "basis vectors" and the basis of coordinate
or momentum space to form in this case the commuting creation and annihilation
operators.
\end{small}


\section{Trial to compare application of $\gamma^a$ matrices 
  in usual case and  in {\it spin-charge-family} theory}
\label{trial}
%
Sects.~\ref{internalspace} - \ref{fermionandgravitySCFT} show that
in the {\it spin-charge-family} theory  creation operators and their Hermitian 
conjugated annihilation operators anticommute, due to the odd character 
of the "basis vectors" describing the internal space of fermions.

The operators $S^{ab}$ and $\tilde{S}^{ab}$, connecting in the 
{\it spin-charge-family} theory the members of the same irreducible 
representation (the members of a particular family) 
and the members of different irreducible representations (the same family 
member of different families), respectively, having  an even Clifford character,
do not change the oddness of the "basis vectors". The application of the 
operators $\gamma^a$'s as well as $\tilde{\gamma^a}$'s, having 
 both an odd Clifford character,  do change 
the character of states from odd to even.

In the case of massive states the usual  way, the one of Dirac,  uses the 
$\gamma^a$'s to transform the left handed fermion states in $d=(3+1)$ into 
the right handed ones to generate basis for the massive states. Charges are 
treated separately --- by the choice of the charge groups.

We demonstrate in Sect.~\ref{TDN0} on the toy model of $d=(5+1)$ that 
breaking the staring symmetry, caused by the particular scalar fields appearing 
in the simple starting action, Eqs.~(\ref{wholeaction}, \ref{faction}), do lead to 
massless charged states in the manifold of $d=(3+1)$, while further break
of symmetry, when  scalar fields gain constant values, makes these massless
fermions and antifermions chargeless and massive with the properties of the
massive Majorana fermions, which are superposition of fermions and antifermions,
Eq.~(\ref{Majoranastates}).
 In this case if one forgets the spin part in the $5^{th}$ and $6^{th}$ dimensions,
manifesting as charges in $d=(3+1)$, which is now not conserved any longer,
there are only the spin parts manifesting in $d=(3+1)$.
 
Of course, the Dirac's states are not describing Majorana particles, but do also 
not pay attention on the oddness of the "basis states" describing the internal space 
of fermions as the {\it spin-charge-family} theory does. The vectors describing 
spins ($S^{03}$ and $S^{12}$, or equivalently, handedness and $S^{12}$) in 
the usual text books do not have the odd character, they are columns with
 numbers, which commute.

But if one looks at the solutions of  equations of motion for the massive 
states in the {\it spin-charge-family} theory, Eq.~(\ref{Weylmsolsimple}), as the
states without charges, the multiplication of this massless chargeless state by 
$\gamma^0$, if neglecting the charge part and correspondingly the oddness
of states, what the usual way of presenting  fermions does, then one can easier
understand the difference of the {\it spin-charge-family} theory way of the second 
quantization and the Dirac's way, presented in Sect.~\ref{creationtensorusual}.

We present in this Sect.~\ref{trial}, following Ref.~\cite{DMN}, the matrix 
representations of the operators $\gamma^a$'s,  $\tilde{\gamma^a}$'s, 
$S^{ab}$ and $\tilde{S}^{ab}$, manifesting how would matrices
look like if families, suggested by the use of the Clifford algebra in the 
{\it spin-charge-family} theory, would be taken into account.

 Since one can always embed the spin part in $d=(3+1)$ into $d\ge 5$, and if 
at the same time one requires  the Clifford oddness of "basis vectors" in the 
whole space $(d-1) +1$ so that spins in higher dimensions manifest as charges
in $d=(3+1)$, the corresponding
creation and annihilation operators fulfill the anticommutation postulates of Dirac. 
The single fermion states already anticommute.

We present  matrices for operators $\gamma^a$'s,  $\tilde{\gamma^a}$'s, 
$S^{ab}$ and $\tilde{S}^{ab}$ in Sect.~\ref{matrixCliffordDMN}.

 Sect.~\ref{evenclifford} demonstrates that  the application of the Clifford odd 
operators $\gamma^a$'s and  $\tilde{\gamma^a}$'s, Ref.~\cite{prd2018}, 
on the Clifford odd  anticommuting "basis vectors" transforms the Clifford 
odd "basis vectors" to the Clifford even "basis vector", which commute.


%

\subsection{Clifford even creation operators }
\label{evenclifford}
%



In Table~\ref{cliff basis5+1.} the Clifford odd "basis vectors" for $d=(5+1)$ case
are presented as suggested by the {\it spin-charge-family} theory, forming 
 four families, together with their Hermitian conjugated partners.  
Due to the oddness of the "basis vectors" the tensor products, $*_{T}$, of the 
"basis vectors" and the basis in the momentum/coordinate space fulfil together 
with their corresponding annihilation operators the anticommutation relations 
postulated by Dirac, explaining indeed the Dirac's anticommuting postulates.

The remaining $2^{d-1}$ "basis vectors" have an even Clifford character. They are 
presented in the lower half part of Table~\ref{Table Clifffourplet.}. The reader can 
immediately see that  the multiplication of any  odd "basis vector" by $\gamma^a$ 
 leads to 
an even "basis vector". For example, the application of $\gamma^0$ on the first 
family member of the first {\it odd I} family generates the first member of the
first {\it even I} family.

All the Clifford even "families" with "family" members of  
Table~\ref{Table Clifffourplet.} can be obtained also as algebraic products 
($*_{A}$) of the Clifford odd "basis vectors" from the upper half part
of the same table. 

All the even families have  one self adjoint "basis vector"; Being a product of 
projectors only this "basis vector" is its own Hermitian conjugated partner, 
enabling to generate and annihilate the state at the same time, when applying 
on the (any one) vacuum state.

Clifford even "basis vectors" commute.

The Clifford even operators $S^{ab}$ transform family members of one 
family among themselves, while $\tilde{S}^{ab}$ transform one family 
member of any family to the same family member of another family also in the
case of even "basis vectors". 

We shall comment these properties in Sect.~\ref{matrixCliffordDMN}.
 


\begin{table*}
\begin{center}
\begin{tiny}
\begin{minipage}[t]{16.5 cm}
\caption{$2^d=64$ "eigenvectors" of the Cartan 
subalgebra, Eq.~(\ref{cartangrasscliff}), of the Clifford  odd and even algebras in 
$d=(5+1)$ are presented, divided into four groups, each group with four 
"families", each ''family'' with four ''family'' members. Two of four groups are  
superposition of products of an odd number of $\gamma^a$'s.  The 
"basis vectors", ${\hat b}^{m \dagger}_f$,
Eq.~(\ref{start(2n+1)2cliffgammatilde4n}), in  {\it odd I} group, belong to four 
"families" ($f=1 (a),2(b),3(c),4(d)$) with four members ($m=1,2,3,4$), having 
their Hermitian conjugated partners, ${\hat b}^{m}_f$, among "basis vectors" of 
the {\it odd II} part, denoted with the corresponding "family" and "family" 
members ($a_m,b_m,c_m,d_m$) quantum numbers.
The "family" quantum  numbers, determined by 
$(\tilde{S}^{03}, \tilde{S}^{12},\tilde{S}^{56})$,
of ${\hat b}^{m \dagger}_f$ are written   above each "family".
The  two groups with the even number of $\gamma^a$'s in a product, {\it even I} 
and {\it even II}, have their Hermitian conjugated partners within their own group 
each. There are members in each of even groups, which are products of projectors only. 
Numbers --- 
$\stackrel{03}{\;\,}\;\;\,\stackrel{12}{\;\,}\;\;\,\stackrel{56}{\;\,}$ ---
denote the indexes of the corresponding Cartan subalgebra members 
($\tilde{S}^{03}, \tilde{S}^{12}, \tilde{S}^{56}$), Eq.~(\ref{cartangrasscliff}). 
In the columns $(7,8,9)$ the eigenvalues of the Cartan subalgebra members 
$(S^{03}, S^{12}, S^{56})$, Eq.~(\ref{cartangrasscliff}), are presented. The last
two columns  tell the handedness of "basis vectors" in $d=(5+1)$, 
$\Gamma^{(5+1)}$, and of $d=(3+1)$,  $\Gamma^{(3+1)}$, respectively, 
defined in Eq.(\ref{hand}).
}
\label{Table Clifffourplet.}
\end{minipage}
%
  \begin{tabular}{|c|c|c|c|c|c|r|r|r|r|r|}
\hline
$ odd \, I$&$m$&$ f=1 (a)$&$ f=2 (b)$&$ f=3 (c)$&
$ f=4 (d)$&$S^{03}$&$S^{12}$&$S^{56}$&$\Gamma^{(5+1)}$&
$\Gamma^{(3+1)}$\\
&&$(\frac{i}{2},\frac{1}{2},\frac{1}{2})$&$(-\frac{i}{2},-\frac{1}{2},
\frac{1}{2})$&
$(-\frac{i}{2},\frac{1}{2},-\frac{1}{2})$&$(\frac{i}{2},-\frac{1}{2},
-\frac{1}{2})$&
&&&&\\
&& 
$\stackrel{03}{\;\,}\;\;\,\stackrel{12}{\;\,}\;\;\,\stackrel{56}{\;\,}$&
$\stackrel{03}{\;\,}\;\;\,\stackrel{12}{\;\,}\;\;\,\stackrel{56}{\;\,}$&
$\stackrel{03}{\;\,}\;\;\,\stackrel{12}{\;\,}\;\;\,\stackrel{56}{\;\,}$&
$\stackrel{03}{\;\,}\;\;\,\stackrel{12}{\;\,}\;\;\,\stackrel{56}{\;\,}$&
&&&&\\
\hline
$$&$1$&$\stackrel{03}{(+i)}\stackrel{12}{(+)}\stackrel{56}{(+)}$ & 
$\stackrel{03}{[+i]}\stackrel{12}{[+]}\stackrel{56}{(+)}$ & 
$\stackrel{03}{[+i]}\stackrel{12}{(+)}\stackrel{56}{[+]}$ & 
$\stackrel{03}{(+i)}\stackrel{12}{[+]}\stackrel{56}{[+]}$&$\frac{i}{2}$&
$\frac{1}{2}$&$\frac{1}{2}$&$1$&$1$\\
$$&$2$&    $[-i][-](+) $ & $(-i)(-)(+) $ & $(-i)[-][+] $ & $[-i](-)[+] $ &
$-\frac{i}{2}$&$-\frac{1}{2}$&$\frac{1}{2}$&$1$&$1$\\
$$&$3$&    $[-i] (+)[-]$ & $(-i)[+][-] $ & $(-i)(+)(-) $ & $[-i][+](-) $&
$-\frac{i}{2}$&$\frac{1}{2}$&$-\frac{1}{2}$&$1$&$-1$ \\
$$&$4$&    $(+i)[-][-] $ & $[+i](-)[-] $ & $[+i][-](-) $ & $(+i)(-)(-) $&
$\frac{i}{2}$&$-\frac{1}{2}$&$-\frac{1}{2}$&$1$&$-1$ \\
\hline
$ odd\, II$&$$&$ $&$ $&$ $&
$ $&$S^{03}$&$S^{12}$&$S^{56}$&$\Gamma^{(5+1)}$&
$\Gamma^{(3+1)}$\\
&& 
$\stackrel{03}{\;\,}\;\;\,\stackrel{12}{\;\,}\;\;\,\stackrel{56}{\;\,}_{f_m}$&
$\stackrel{03}{\;\,}\;\;\,\stackrel{12}{\;\,}\;\;\,\stackrel{56}{\;\,}_{f_m}$&
$\stackrel{03}{\;\,}\;\;\,\stackrel{12}{\;\,}\;\;\,\stackrel{56}{\;\,}_{f_m}$&
$\stackrel{03}{\;\,}\;\;\,\stackrel{12}{\;\,}\;\;\,\stackrel{56}{\;\,}_{f_m}$&
&&&&\\
\hline
$$&$$&$(-i)(+)(+)_{d_4}$ & $[-i][+](+)_{d_3}$ & $[-i](+)[+]_{d_2}$ & 
$(-i)[+][+]_{d_1}$&$-\frac{i}{2}$&$\frac{1}{2}$&$\frac{1}{2}$&$-1$&$-1$ \\
$$&$$&$[+i][-](+)_{c_4}$ & $(+i)(-)(+)_{c_3}$ & $(+i)[-][+]_{c_2}$ & 
$[+i](-)[+]_{c_1}$&$\frac{i}{2}$&$-\frac{1}{2}$&$\frac{1}{2}$&$-1$&$-1$ \\
$$&$$&$[+i](+)[-]_{b_4}$ & $(+i)[+][-]_{b_3}$ & $(+i)(+)(-)_{b_2}$ & 
$[+i][+](-)_{b_1}$&$\frac{i}{2}$&$\frac{1}{2}$&$-\frac{1}{2}$&$-1$&$1$ \\
$$&$$&$(-i)[-][-]_{a_4}$ & $[-i](-)[-]_{a_3}$ & $[-i][-](-)_{a_2}$ & 
$(-i)(-)(-)_{a_1}$&$-\frac{i}{2}$&$-\frac{1}{2}$&$-\frac{1}{2}$&$-1$&$1$ \\
\hline\hline
$ even\, I$&$m$&$ $&$$&$ $&
$ $&$S^{03}$&$S^{12}$&$S^{56}$&$\Gamma^{(5+1)}$&
$\Gamma^{(3+1)}$\\
&&$(\frac{i}{2},\frac{1}{2},\frac{1}{2})$&
$(-\frac{i}{2},-\frac{1}{2},\frac{1}{2})$&
$(\frac{i}{2},-\frac{1}{2},-\frac{1}{2})$&
$(-\frac{i}{2},\frac{1}{2},-\frac{1}{2})$&
&&&&\\
&& 
$\stackrel{03}{\;\,}\;\;\,\stackrel{12}{\;\,}\;\;\,\stackrel{56}{\;\,}$&
$\stackrel{03}{\;\,}\;\;\,\stackrel{12}{\;\,}\;\;\,\stackrel{56}{\;\,}$&
$\stackrel{03}{\;\,}\;\;\,\stackrel{12}{\;\,}\;\;\,\stackrel{56}{\;\,}$&
$\stackrel{03}{\;\,}\;\;\,\stackrel{12}{\;\,}\;\;\,\stackrel{56}{\;\,}$&
&&&&\\
\hline
$$&$1$& $[-i](+)(+) $ & $(-i)[+](+) $ & $[-i][+][+] $ & $(-i)(+)[+] $ &
$-\frac{i}{2}$& $\frac{1}{2}$&$\frac{1}{2}$&$-1$&$-1$ \\ 
$$&$2$&    $(+i)[-](+) $ & $[+i](-)(+) $ & $(+i)(-)[+] $ & $[+i][-][+] $ &
$\frac{i}{2}$&$-\frac{1}{2}$&$\frac{1}{2}$&$-1$&$-1$ \\
$$&$3$&    $(+i)(+)[-] $ & $[+i][+][-] $ & $(+i)[+](-) $ & $[+i](+)(-) $&
$\frac{i}{2}$&$\frac{1}{2}$&$-\frac{1}{2}$&$-1$&$1$ \\
$$&$4$&    $[-i][-][-] $ & $(-i)(-)[-] $ & $[-i](-)(-) $ & $(-i)[-](-) $&
$-\frac{i}{2}$&
$-\frac{1}{2}$&$-\frac{1}{2}$&$-1$&$1$ \\
\hline
$ even\, II$&$m$&$ $&$ $&$ $&
$ $&$S^{03}$&$S^{12}$&$S^{56}$&$\Gamma^{(5+1)}$&
$\Gamma^{(3+1)}$\\ 
&&$(-\frac{i}{2},\frac{1}{2},\frac{1}{2})$&
$(\frac{i}{2},-\frac{1}{2},\frac{1}{2})$&
$(-\frac{i}{2},-\frac{1}{2},-\frac{1}{2})$&
$(\frac{i}{2},\frac{1}{2},-\frac{1}{2})$&
&&&&\\
&& 
$\stackrel{03}{\;\,}\;\;\,\stackrel{12}{\;\,}\;\;\,\stackrel{56}{\;\,}$&
$\stackrel{03}{\;\,}\;\;\,\stackrel{12}{\;\,}\;\;\,\stackrel{56}{\;\,}$&
$\stackrel{03}{\;\,}\;\;\,\stackrel{12}{\;\,}\;\;\,\stackrel{56}{\;\,}$&
$\stackrel{03}{\;\,}\;\;\,\stackrel{12}{\;\,}\;\;\,\stackrel{56}{\;\,}$&
&&&&\\
\hline
$$&$1$&$[+i](+)(+) $ & $(+i)[+](+) $ & $[+i][+][+] $ & $(+i)(+)[+] $ &
$\frac{i}{2}$&$\frac{1}{2}$&$\frac{1}{2}$&$1$&$1$ \\
$$&$2$&$(-i)[-](+) $ & $[-i](-)(+) $ & $(-i)(-)[+] $ & $[-i][-][+] $ &
$-\frac{i}{2}$&$-\frac{1}{2}$&$\frac{1}{2}$&$1$&$1$ \\
$$&$3$&$(-i)(+)[-] $ & $[-i][+][-] $ & $(-i)[+](-) $ & $[-i](+)(-) $&
$-\frac{i}{2}$&$\frac{1}{2}$&$-\frac{1}{2}$&$1$&$-1$ \\
$$&$4$&$[+i][-][-] $ & $(+i)(-)[-] $ & $[+i](-)(-) $ & $(+i)[-](-) $&
$\frac{i}{2}$&$-\frac{1}{2}$&$-\frac{1}{2}$&$1$&$-1$ \\ 
\hline
 \end{tabular}
\end{tiny}
\end{center}
\end{table*}

Let us conclude: The Grassmann algebra  offers the Grassmann odd 
creation operators  and their Hermitian conjugated partners annihilation operators, 
which anticommute~\footnote{Like in the Clifford case also in the Grassmann
case one can generate creation operators as superposition of tensor products,
$*_{T}$, of the Grassmann odd "basis vectors" and the ordinary, momentum 
or coordinate,  basis.},
as well as  Grassmann  even creation operators and their Hermitian conjugated 
partners annihilation operators, which commute. The Clifford algebra offers 
odd, anticommuting, creation and their Hermitian conjugated partners 
annihilation operators, Sect.~\ref{evengrass}, describing the second quantized 
fermions with spin, charges and families unified. The commuting Clifford even
creation operators (they have the Hermitian conjugated partners within 
the same group)  offer the description of the gauge vector fields to the
corresponding fermion fields.

\subsection{Relations between Clifford algebra and Dirac matrices in the
presence of families in $d=(3+1)$ and $d=(5+1)$}
\label{matrixCliffordDMN}
%

We learn in Sect.~\ref{internalspace}, in particular after reduction of the 
Clifford algebra space, Sect.~\ref{reduction}, and in Sects.~\ref{actionGrassCliff},~
\ref{HilbertCliff0}, that all the fermion states can be represented by the tensor
products, $*_T$, of the finite number of "basis vectors"  and the continuously 
infinite basis in the ordinary space, while the total Hilbert space consists of the 
tensor products, $*_{T_{H}}$, of all possible numbers of all possible anticommuting 
single particle creation operators applying on the vacuum state.
We also learn in Sects.~\ref{TDN1} and \ref{evenclifford} that "basis vectors" 
change oddness into evenness if we multiply them by Clifford odd operators 
$\gamma^a$'s and $\tilde{\gamma}^a$'s.

The Dirac's single particle states are not Clifford odd and vectors, solving the 
Dirac equations, also do not manifest oddness. The fermion states gain the 
anticommuting character with the second quantization postulates of Dirac.

In the {\it spin-charge-family} theory fermions already have the anticummuting 
nature due to the fact that the theory uses the products of an odd number of 
$\gamma^a$'s to describe the internal space of fermions. Correspondingly the 
"basis vectors" anticommute, fulfilling the anticommutation relations postulated 
by Dirac for the second quantized fermions, and the anticommuting  character 
of "basis vectors" transfers also to the states, which are superposition of 
tensor products of the "basis vectors" and basis in the momentum or coordinate
space, Eqs.~(\ref{Weylp0}, \ref{Weylp2}),  and are  solutions of equations
 of motion.
 Let us use the so far several time  discussed toy model in $d=(5+1)$ to 
illustrate the application of the operators  $\gamma^a$'s,
$\tilde{\gamma}^a$'s, $S^{ab}$ and $\tilde{S}^{ab}$ on the "basis vectors".

We use in Eq.~(\ref{weylgen0}) the Clifford odd part of "basis vectors" in  
$d=(5+1)$, Table~\ref{Table Clifffourplet.}, describing the anticommuting 
odd "basis vectors" in $d=(3+1)$ with definite handedness and spin and with
particular charge determined by the spin $S^{56}$, which in this case we neglect. 
These states, the solutions of equations of motions for free massless fermions 
in $d=(3+1)$, are correspondingly assumed to 
have no charge, and are the superposition of the first two members of the third 
family and the  first two members of the fourth families, respectively, with the 
charge part neglected.

The solutions of equations of motion for free massless particles in $d=(5+1)$ of
 Eq.~(\ref{weylgen05+1}) represent fermions and antifermions of the first 
family from Table~\ref{cliff basis5+1.} for $p^a=(p^0,  \vec{p},0)$.

The solutions of equations of motion in $d=(13+1)$ for free massless right 
handed $u$-quark  of the colour charge $(\frac{1}{2}, \frac{1}{2\sqrt{3}})$
and the "fermion charge" $\tau^4= \frac{1}{6}$ and its left handed antiquark, 
Eq.~(\ref{weylgen013+1}), manifest dynamics in $d=(3+1)$ only, if 
$p^a=(p^0,  \vec{p},0,0,0,0,\dots,0)$.

In Sect.~\ref{TDN1} the massless and massive solutions in $d=(5+1)$ are 
presented.

Here we look at the $d=(3+1)$ part of the "basis vectors" belonging to $d=(5+1)$ 
space, appearing correspondingly in four families. In spaces  with higher 
dimensions there would in general case be more families. 

The "basis vectors" are arranged as presented in Table~\ref{tableBasVecd4m}
\begin{table}
\begin{tiny}
\begin{center}
\begin{minipage}[t]{16.5 cm}    
\caption{In this table $(2^d =)16$ vectors, describing internal space 
of fermions in $d=(3+1)$, are presented. Each vector carries the family member quantum 
number  $m=(1,2,3,4)$ --- determined by $S^{03}$ and $S^{12}$, 
Eqs.~(\ref{cartangrasscliff}) 
 --- and the family quantum number $f=(I,II,III,IV)$ --- determined by 
$\tilde{S}^{03}$ and $\tilde{S}^{12}$, Eq.~(\ref{cartangrasscliff}). Vectors 
$\psi^m_f$ are obtained by applying $\hat{b}^{m \dagger}_f$ on the vacuum state,
Eq.~(\ref{vac1}, \ref{vac5+1}).
 Vectors, that is the family members of any family, split into even (they are sums of 
products of an even number of $\gamma^a$'s) and odd (they are sums of products of 
an odd number of $\gamma^a$'s). If these vectors are embedded into 
$d=(5+1)$ (by being multiplied by an appropriate nilpotent or projector so that they are
of an odd Clifford character), they "gain" charges as presented in Table~\ref{cliff basis5+1.}.} 
\label{tableBasVecd4m}
\end{minipage}
\begin{tabular}{|c| c |r r r r r r r r |r r r r r r |}
    \hline 
    & $\psi^f_m$ & $\gamma_0\,\psi^f_m$ & 
$\gamma_1\,\psi^f_m$ & $\gamma_2\,\psi^f_m$     &$\gamma_3\,\psi^f_m$ &
$\Tilde{\gamma}_0\,\psi^f_m$ &$\Tilde{\gamma}_1\,\psi^f_m$ & 
$\Tilde{\gamma}_2\,\psi^f_m$  &$\Tilde{\gamma}_3\,\psi^f_m$ & 
$S^{03}$ & $S^{12}$ &$\Tilde{S}^{03}$ &
$\Tilde{S}^{12}$ & $\Gamma^{(3+1)}$ & \vphantom{$b\biggm|b$} 
$\Tilde{\Gamma}^{(3+1)}$\\
    \hline
    $\psi^a_1$ & $(+i) (+)$ & $\psi^a_3$ & $\psi^a_4$ & $i\psi^a_4$ & $\psi^a_3$ &
      $-i\psi^b_1$ & $-i\psi^c_1$ & $\psi^c_1$ & $-i\psi^b_1$ 
                 & $\frac{i}{2}$ & $\frac{1}{2} $ & $\frac{i}{2} $ & $\frac{1}{2} $   
                 & $1 $ & $1 $ \\
    $\psi^a_2$ & $[-i] [-]$ & $\psi^a_4$ & $\psi^a_3$ & $-i\psi^a_3$ & $-\psi^a_4$ & 
                       $i\psi^b_2$ & $i\psi^c_2$ & $-\psi^c_2$ & $i\psi^b_2$
                  & $-\frac{i}{2}$ & $-\frac{1}{2} $ & $\frac{i}{2} $ & $\frac{1}{2} $   
                  & $1 $ & $1 $ \\
    $\psi^a_3$ & $[-i] (+)$ & $\psi^a_1$ & $-\psi^a_2$ & $-i\psi^a_2$ & $-\psi^a_1$ & 
      $i\psi^b_3$ & $i\psi^c_3$ & $-\psi^c_3$ & $i\psi^b_3$ 
                    & $-\frac{i}{2}$ & $\frac{1}{2} $ & $\frac{i}{2} $ & $\frac{1}{2} $   
                    & $-1 $ & $1 $ \\
    $\psi^a_4$ & $(+i) [-]$ & $\psi^a_2$ & $-\psi^a_1$ & $i\psi^a_1$ & $\psi^a_2$ &
     $-i\psi^b_4$ & $-i\psi^c_4$ & $\psi^c_4$ & $-i\psi^b_4$ 
                    & $\frac{i}{2}$ & $-\frac{1}{2} $ & $\frac{i}{2} $ & $\frac{1}{2} $   
                    & $-1 $ & $1 $ \\
    \hline
    $\psi^b_1$ & $[+i] (+)$ & $\psi^b_3$ & $-\psi^b_4$ & $-i\psi^b_4$ & $\psi^b_3$ &
     $i\psi^a_1$ & $i\psi^d_1$ & $-\psi^d_1$ & $-i\psi^a_1$ 
                   & $\frac{i}{2}$ & $\frac{1}{2} $ & $-\frac{i}{2} $ & $\frac{1}{2} $   
                   & $1 $ & $-1 $ \\
     $\psi^b_2$ & $(-i) [-]$ & $\psi^b_4$ & $-\psi^b_3$ & $i\psi^b_3$ & $-\psi^b_4$ &
      $-i\psi^a_2$ & $-i\psi^d_2$ & $\psi^d_2$ & $i\psi^a_2$ 
                  & $-\frac{i}{2}$ & $-\frac{1}{2} $ & $-\frac{i}{2} $ & $\frac{1}{2} $   
                  & $1 $ & $-1 $ \\
    $\psi^b_3$ & $(-i) (+)$ & $\psi^b_1$ & $\psi^b_2$ & $i\psi^b_2$ & $-\psi^b_1$ &
      $-i\psi^a_3$ & $-i\psi^d_3$ & $\psi^d_3$ & $i\psi^a_3$ 
                   & $-\frac{i}{2}$ & $\frac{1}{2} $ & $-\frac{i}{2} $ & $\frac{1}{2} $   
                   & $-1 $ & $-1 $ \\
    $\psi^b_4$ & $[+i] [-]$ & $\psi^b_2$ & $\psi^b_1$ & $-i\psi^b_1$ & $\psi^b_2$ &
      $i\psi^a_4$ & $i\psi^d_4$ & $-\psi^d_4$ & $-i\psi^a_4$ 
                   & $\frac{i}{2}$ & $-\frac{1}{2} $ & $-\frac{i}{2} $ & $\frac{1}{2} $   
                   & $-1 $ & $-1 $ \\
    \hline    
    $\psi^c_1$ & $(+i) [+]$ & $\psi^c_3$ & $-\psi^c_4$ & $-i\psi^c_4$ & $\psi^c_3$  &
      $i\psi^d_1$ & $-i\psi^a_1$ & $-\psi^a_1$ & $i\psi^d_1$ 
                  & $\frac{i}{2}$ & $\frac{1}{2} $ & $\frac{i}{2} $ & $-\frac{1}{2} $   
                  & $1 $ & $-1 $ \\
    $\psi^c_2$ & $[-i] (-)$ & $\psi^c_4$ & $-\psi^c_3$ & $i\psi^c_3$ & $-\psi^c_4$& 
      $-i\psi^d_2$ & $i\psi^a_2$ & $\psi^a_2$ & $-i\psi^d_2$ 
                  & $-\frac{i}{2}$ & $-\frac{1}{2} $ & $\frac{i}{2} $ & $-\frac{1}{2} $   
                  & $1 $ & $-1 $ \\
    $\psi^c_3$ & $[-i] [+]$ & $\psi^c_1$ & $\psi^c_2$ & $i\psi^c_2$ & $-\psi^c_1$ &
      $-i\psi^d_3$ & $i\psi^a_3$ & $\psi^a_3$ & $-i\psi^d_3$ 
                  & $-\frac{i}{2}$ & $\frac{1}{2} $ & $\frac{i}{2} $ & $-\frac{1}{2} $   
                  & $-1 $ & $-1 $ \\
    $\psi^c_4$ & $(+i) (-)$ & $\psi^c_{2}$ & $\psi^c_1$ & $-i\psi^c_1$ &  $\psi^c_{2}$ & 
     $i\psi^d_4$ & $-i\psi^a_4$ & $-\psi^a_4$ & $i\psi^d_4$  
                 & $\frac{i}{2}$ & $-\frac{1}{2} $ & $\frac{i}{2} $ & $-\frac{1}{2} $   
                 & $-1 $ & $-1 $ \\
    \hline
    $\psi^d_1$ & $[+i] [+]$ & $\psi^d_3$ & $\psi^d_4$ & $i\psi^d_4$ & $\psi^d_3$& 
       $-i\psi^c_1$ & $i\psi^b_1$ & $\psi^b_1$ & $i\psi^c_1$ 
                & $\frac{i}{2}$ & $\frac{1}{2} $ & $-\frac{i}{2} $ & $-\frac{1}{2} $   
                & $1 $ & $1 $ \\
    $\psi^d_2$ & $(-i) (-)$ & $\psi^d_4$ & $\psi^d_3$ & $-i\psi^d_3$ & $-\psi^d_4$& 
      $i\psi^c_2$ & $-i\psi^b_2$ & $-\psi^b_2$ & $-i\psi^c_2$ 
               & $-\frac{i}{2}$ & $-\frac{1}{2} $ & $-\frac{i}{2} $ & $-\frac{1}{2} $    
               & $1 $ & $1 $ \\
    $\psi^d_3$ & $(-i) [+]$ & $\psi^d_1$ & $-\psi^d_2$ & $-i\psi^d_2$ & $-\psi^d_1$& 
      $i\psi^c_3$ & $-i\psi^b_3$ & $-\psi^b_3$ & $-i\psi^c_3$ 
                & $-\frac{i}{2}$ & $\frac{1}{2} $ & $-\frac{i}{2} $ & $-\frac{1}{2} $    
                & $-1 $ & $1 $ \\
    $\psi^d_4$ & $[+i] (-)$ & $\psi^d_2$ & $-\psi^d_1$ & $i\psi^d_1$ & $\psi^d_2$&
     $-i\psi^c_4$ & $i\psi^b_4$ & $\psi^b_4$ & $i\psi^c_4$ 
                & $\frac{i}{2}$ & $-\frac{1}{2} $ & $-\frac{i}{2} $ & $-\frac{1}{2} $    
                & $-1 $ & $1 $ \\
    \hline 
  \end{tabular}
\end{center}
\end{tiny}
\end{table}
%

Using the Pauli matrices from Eq.~(\ref{pauli})  as  well as the unit 
$4 \times 4$ submatrix and the submatrix 
with all the matrix elements equal to zero 
\begin{equation}\label{pauli4x4}
     \mathbf{1}= 
   \begin{pmatrix}
       1 & 0 \\
       0 & 1 
    \end{pmatrix}  \, = \sigma^0\,,
\qquad\qquad
%
  %
    \mathbf{0}= 
   \begin{pmatrix}
       0 & 0 \\
       0 & 0 
    \end{pmatrix}\,,
  \end{equation}
and taking into account Table~\ref{tableBasVecd4m} the matrix representation for 
$\gamma^a$'s, $\tilde{\gamma}^a$'s, $S^{ab}$'s and $\tilde{S}^{ab}$'s follow. 
  %
 \begin{equation}\label{dn-mat-mgamma01}
 \gamma^{0} =
 \begin{pmatrix}
      \begin{smallmatrix}
          0 & \sigma^0\\
          \sigma^0 & 0 
        \end{smallmatrix} & \mathbf{0}  & \mathbf{0}  & \mathbf{0} \\
        \mathbf{0}  &
        \begin{smallmatrix}
          0 & \sigma^0\\
          \sigma^0 & 0 
        \end{smallmatrix} & \mathbf{0}  & \mathbf{0} \\
        \mathbf{0} & \mathbf{0}  &
        \begin{smallmatrix}
          0 & \sigma^0\\
          \sigma^0 & 0 
        \end{smallmatrix} & \mathbf{0} \\
        \mathbf{0} & \mathbf{0} & \mathbf{0} &
        \begin{smallmatrix}
          0 & \sigma^0\\
          \sigma^0 & 0 
        \end{smallmatrix}
  \end{pmatrix}\,, 
%
%
 \,\gamma^{1} =
 \begin{pmatrix}
      \begin{smallmatrix}
          0 & \sigma^1\\
          -\sigma^1 & 0
        \end{smallmatrix} & \mathbf{0} & \mathbf{0} & \mathbf{0}\\
        \mathbf{0} &
        \begin{smallmatrix}
          0 & -\sigma^1\\
          \sigma^1 & 0
        \end{smallmatrix} & \mathbf{0} & \mathbf{0}\\
        \mathbf{0} & \mathbf{0} &
        \begin{smallmatrix}
          0 & -\sigma^1\\
          \sigma^1 & 0
        \end{smallmatrix} & \mathbf{0} \\
        \mathbf{0} & \mathbf{0} & \mathbf{0} &
        \begin{smallmatrix}
          0 & \sigma^1\\
          -\sigma^1 & 0
        \end{smallmatrix}
  \end{pmatrix}\,, 
 \end{equation}
 \begin{equation}\label{dn-mat-mgamma23}
 \gamma^{2} =
 \begin{pmatrix}
      \begin{smallmatrix}
          0 & -\sigma^2\\
          \sigma^2 & 0 & 
        \end{smallmatrix} & 0 & 0 & 0\\
        \mathbf{0} &
        \begin{smallmatrix}
          0 & \sigma^2\\
          -\sigma^2 & 0 & 
        \end{smallmatrix} & \mathbf{0} & \mathbf{0}\\
        \mathbf{0} & \mathbf{0} &
        \begin{smallmatrix}
          0 & \sigma^2\\
          -\sigma^2 & 0 & 
        \end{smallmatrix} & \mathbf{0} \\
        \mathbf{0} & \mathbf{0} & \mathbf{0} &
        \begin{smallmatrix}
          0 & -\sigma^2\\
          \sigma^2 & 0 & 
        \end{smallmatrix}
  \end{pmatrix}\,,
%
 %
\,
 \gamma^{3} =
 \begin{pmatrix}
      \begin{smallmatrix}
          0 & \sigma^3\\
          -\sigma^3 & 0 & 
        \end{smallmatrix} & \mathbf{0} & \mathbf{0} & \mathbf{0}\\
        \mathbf{0} &
        \begin{smallmatrix}
          0 & \sigma^3\\
          -\sigma^3 & 0 & 
        \end{smallmatrix} & \mathbf{0} & \mathbf{0}\\
        \mathbf{0} & \mathbf{0} &
        \begin{smallmatrix}
          0 & \sigma^3\\
          -\sigma^3 & 0 & 
        \end{smallmatrix} & \mathbf{0} \\
        \mathbf{0} & \mathbf{0} & \mathbf{0} &
        \begin{smallmatrix}
          0 & \sigma^3\\
          -\sigma^3 & 0 & 
        \end{smallmatrix}
  \end{pmatrix}\,, 
 \end{equation}
manifesting  the $4 \times 4$ substructure (of Dirac matrices) along the diagonal of 
$16 \times 16$ matrices.

The representations of $\tilde{\gamma}^{a}$ do not appear in the Dirac case.
They manifest the off diagonal structure as follows
 \begin{equation}\label{dn-mat-mtgamma01}
 \Tilde{\gamma}^{0} =
 \begin{pmatrix}
      \mathbf{0} & 
      \begin{smallmatrix}
          -i\sigma^3 & 0 \\
          0 & i\sigma^3
        \end{smallmatrix} & \mathbf{0} & \mathbf{0} \\
        \begin{smallmatrix}
          i\sigma^3 & 0 \\
          0 & -i\sigma^3
        \end{smallmatrix} & \mathbf{0} & \mathbf{0} & \mathbf{0}\\
        \mathbf{0} & \mathbf{0} &  \mathbf{0} & 
        \begin{smallmatrix}
          i\sigma^3 & 0 \\
          0 & -i\sigma^3
        \end{smallmatrix}  \\
        \mathbf{0} & \mathbf{0}  &
        \begin{smallmatrix}
          -i\sigma^3 & 0 \\
          0 & i\sigma^3
        \end{smallmatrix} & \mathbf{0} 
  \end{pmatrix}
\,,
%
%
\, \Tilde{\gamma}^{1} =
 \begin{pmatrix}
        \mathbf{0} & \mathbf{0} &
        \begin{smallmatrix}
          -i\sigma^3 & 0 \\
          0 & i\sigma^3
        \end{smallmatrix} & \mathbf{0} \\
        \mathbf{0} & \mathbf{0} & \mathbf{0} & 
        \begin{smallmatrix}
          i\sigma^3 & 0 \\
          0 & -i\sigma^3
        \end{smallmatrix} \\
        \begin{smallmatrix}
          -i\sigma^3 & 0 \\
          0 & i\sigma^3
        \end{smallmatrix} & \mathbf{0}  & \mathbf{0} & \mathbf{0} \\
        \mathbf{0} & 
        \begin{smallmatrix}
          i\sigma^3 & 0 \\
          0 & -i\sigma^3
        \end{smallmatrix} & \mathbf{0} & \mathbf{0}
  \end{pmatrix}\,, 
 \end{equation}
 \begin{equation}\label{dn-mat-mtgamma23}
 \Tilde{\gamma}^{2} =
 \begin{pmatrix}
       \mathbf{0} & \mathbf{0} &
      \begin{smallmatrix}
           \sigma^3 & 0 \\
          0 & -\sigma^3
       \end{smallmatrix} & \mathbf{0} \\
        \mathbf{0} & \mathbf{0} & \mathbf{0} &
        \begin{smallmatrix}
          -\sigma^3 & 0 \\
          0 & \sigma^3
        \end{smallmatrix} \\
        \begin{smallmatrix}
          -\sigma^3 & 0 \\
          0 & \sigma^3
        \end{smallmatrix} & \mathbf{0}  & \mathbf{0} & \mathbf{0} \\
        \mathbf{0} & 
        \begin{smallmatrix}
          \sigma^3 & 0 \\
          0 & -\sigma^3
        \end{smallmatrix} & \mathbf{0} & \mathbf{0}
  \end{pmatrix}\,, 
%
 %
\, \Tilde{\gamma}^{3} =
 \begin{pmatrix}
        \mathbf{0} & 
       \begin{smallmatrix}
          -i\sigma^3 & 0 \\
          0 & i\sigma^3
        \end{smallmatrix} & \mathbf{0} & \mathbf{0} \\
        \begin{smallmatrix}
          -i\sigma^3 & 0 \\
          0 & i\sigma^3
        \end{smallmatrix} & \mathbf{0} & \mathbf{0} & \mathbf{0}\\
        \mathbf{0} & \mathbf{0} &  \mathbf{0} & 
        \begin{smallmatrix}
          -i\sigma^3 & 0 \\
          0 & i\sigma^3
        \end{smallmatrix}\\
        \mathbf{0} & \mathbf{0} & 
        \begin{smallmatrix}
          -i\sigma^3 & 0 \\
          0 & i\sigma^3
        \end{smallmatrix} & \mathbf{0}
  \end{pmatrix}\,.
 \end{equation}
%

Matrices $S^{ab}$ have again along the diagonal the $4 \times 4$ substructure,
 repeating, up to a phase, as expected, the corresponding Dirac matrices, 
since the Dirac $S^{ab}$ do not distinguish among families.
 \begin{equation}\label{dn-mat-mS0102}
 S^{01} = \frac{i}{2}
 \begin{pmatrix}
      \begin{smallmatrix}
          \sigma^1 & 0 \\
          0 & -\sigma^1
        \end{smallmatrix} & \mathbf{0} & \mathbf{0} & \mathbf{0}\\
        \mathbf{0} &
        \begin{smallmatrix}
          -\sigma^1 & 0 \\
          0 & \sigma^1
        \end{smallmatrix} & \mathbf{0} & \mathbf{0}\\
        \mathbf{0} & \mathbf{0} &
        \begin{smallmatrix}
          -\sigma^1 & 0 \\
          0 & \sigma^1
        \end{smallmatrix} & \mathbf{0} \\
        \mathbf{0} & \mathbf{0} & \mathbf{0} &
        \begin{smallmatrix}
          \sigma^1 & 0 \\
          0 & -\sigma^1
        \end{smallmatrix}
  \end{pmatrix}\,, 
%
%
\, S^{02} = \frac{i}{2}
 \begin{pmatrix}
      \begin{smallmatrix}
          -\sigma^2 & 0 \\
          0 & \sigma^2
        \end{smallmatrix} & \mathbf{0} & \mathbf{0} & \mathbf{0}\\
        \mathbf{0} &
        \begin{smallmatrix}
          \sigma^2 & 0 \\
          0 & -\sigma^2
        \end{smallmatrix} & \mathbf{0} & \mathbf{0}\\
        \mathbf{0} & \mathbf{0} &
        \begin{smallmatrix}
          \sigma^2 & 0 \\
          0 & -\sigma^2
        \end{smallmatrix} & \mathbf{0} \\
        \mathbf{0} & \mathbf{0} & \mathbf{0} &
        \begin{smallmatrix}
          -\sigma^2 & 0 \\
          0 & \sigma^2
        \end{smallmatrix}
  \end{pmatrix}\,,  
 \end{equation}
 \begin{equation}\label{dn-mat-mS0312}
 S^{03} = \frac{i}{2}
 \begin{pmatrix}
      \begin{smallmatrix}
          \sigma^3 & 0 \\
          0 & -\sigma^3
        \end{smallmatrix} & \mathbf{0} & \mathbf{0} & \mathbf{0}\\
        \mathbf{0} &
        \begin{smallmatrix}
          \sigma^3 & 0 \\
          0 & -\sigma^3
        \end{smallmatrix} & \mathbf{0} & \mathbf{0}\\
        \mathbf{0} & \mathbf{0} &
        \begin{smallmatrix}
          \sigma^3 & 0 \\
          0 & -\sigma^3
        \end{smallmatrix} & \mathbf{0} \\
        \mathbf{0} & \mathbf{0} & \mathbf{0} &
        \begin{smallmatrix}
          \sigma^3 & 0 \\
          0 & -\sigma^3
        \end{smallmatrix}
  \end{pmatrix}\,,  
%
%
\, S^{12} = \frac{1}{2}
 \begin{pmatrix}
      \begin{smallmatrix}
          \sigma^3 & 0 \\
          0 & \sigma^3
        \end{smallmatrix} & \mathbf{0} & \mathbf{0} & \mathbf{0}\\
        \mathbf{0} &
        \begin{smallmatrix}
          \sigma^3 & 0 \\
          0 & \sigma^3
        \end{smallmatrix} & \mathbf{0} & \mathbf{0}\\
        \mathbf{0} & \mathbf{0} &
        \begin{smallmatrix}
          \sigma^3 & 0 \\
          0 & \sigma^3
        \end{smallmatrix} & \mathbf{0} \\
        \mathbf{0} & \mathbf{0} & \mathbf{0} &
        \begin{smallmatrix}
          \sigma^3 & 0 \\
          0 & \sigma^3
        \end{smallmatrix}
  \end{pmatrix}\,, 
 \end{equation}
 \begin{equation}\label{dn-mat-mS1323}
 S^{13} = \frac{1}{2}
 \begin{pmatrix}
      \begin{smallmatrix}
          \sigma^2 & 0 \\
          0 & \sigma^2
        \end{smallmatrix} & \mathbf{0} & \mathbf{0} & \mathbf{0}\\
        \mathbf{0} &
        \begin{smallmatrix}
          -\sigma^2 & 0 \\
          0 & -\sigma^2
        \end{smallmatrix} & \mathbf{0} & \mathbf{0}\\
        \mathbf{0} & \mathbf{0} &
        \begin{smallmatrix}
          -\sigma^2 & 0 \\
          0 & -\sigma^2
        \end{smallmatrix} & \mathbf{0} \\
        \mathbf{0} & \mathbf{0} & \mathbf{0} &
        \begin{smallmatrix}
          \sigma^2 & 0 \\
          0 & \sigma^2
        \end{smallmatrix}
  \end{pmatrix}\,, 
%
%
 S^{23} = \frac{1}{2} 
 \begin{pmatrix}
      \begin{smallmatrix}
          \sigma^1 & 0 \\
          0 & \sigma^1
        \end{smallmatrix} & \mathbf{0} & \mathbf{0} & \mathbf{0}\\
        \mathbf{0} &
        \begin{smallmatrix}
          -\sigma^1 & 0 \\
          0 & -\sigma^1
        \end{smallmatrix} & \mathbf{0} & \mathbf{0}\\
        \mathbf{0} & \mathbf{0} &
        \begin{smallmatrix}
          -\sigma^1 & 0 \\
          0 & -\sigma^1
        \end{smallmatrix} & \mathbf{0} \\
        \mathbf{0} & \mathbf{0} & \mathbf{0} &
        \begin{smallmatrix}
          \sigma^1 & 0 \\
          0 & \sigma^1
        \end{smallmatrix}
      \end{pmatrix}\,.
\end{equation}
\begin{equation}\label{dn-mat-mGamma31}
  \Gamma^{(3+1)}=-4i S^{03}S^{12}=
 \begin{pmatrix}
      \begin{smallmatrix}
          1 & 0 \\
          0 & -1 
        \end{smallmatrix} & \mathbf{0} & \mathbf{0} & \mathbf{0}\\
        \mathbf{0} &
        \begin{smallmatrix}
          1 & 0 \\
          0 & -1 
        \end{smallmatrix} & \mathbf{0} & \mathbf{0}\\
        \mathbf{0} & \mathbf{0} &
        \begin{smallmatrix}
          1 & 0 \\
          0 & -1 
        \end{smallmatrix} & \mathbf{0} \\
        \mathbf{0} & \mathbf{0} & \mathbf{0} &
        \begin{smallmatrix}
          1 & 0 \\
          0 & -1 
        \end{smallmatrix}
  \end{pmatrix}\,.
\end{equation}
 
The matrices of the operators $\tilde{S}^{ab}$ have again off diagonal 
$4 \times 4$ substructure, except
$\tilde{S}^{03}$ and $\tilde{S}^{12}$, which~are~diagonal.
 \begin{equation}\label{dn-mat-mtS010203}
 \Tilde{S}^{01} = -\frac{i}{2}
 \begin{pmatrix}
      \mathbf{0} & \mathbf{0} & \mathbf{0} &
      \mathbf{1}\\
        \mathbf{0} & \mathbf{0} & 
        \mathbf{1} & \mathbf{0} \\
        \mathbf{0} & 
        \mathbf{1} & \mathbf{0}  & \mathbf{0} \\
        \mathbf{1} & \mathbf{0} & \mathbf{0} & \mathbf{0}
  \end{pmatrix}\,, 
%
 %
\, \Tilde{S}^{02} = \frac{1}{2} 
  \begin{pmatrix}
      \mathbf{0} & \mathbf{0} & \mathbf{0} &
       \mathbf{1}\\
        \mathbf{0} & \mathbf{0} & 
        \mathbf{1} & \mathbf{0} \\
        \mathbf{0} & 
        -\mathbf{1} & \mathbf{0}  & \mathbf{0} \\
        -\mathbf{1} & \mathbf{0} & \mathbf{0} & \mathbf{0}
  \end{pmatrix}\,,  
%
 %
\, \Tilde{S}^{03} = \frac{i}{2}
 \begin{pmatrix}
      \mathbf{1} & \mathbf{0} & \mathbf{0} & \mathbf{0}\\
        \mathbf{0} &
        -\mathbf{1} & \mathbf{0} & \mathbf{0}\\
        \mathbf{0} & \mathbf{0} &
        \mathbf{1} & \mathbf{0} \\
        \mathbf{0} & \mathbf{0} & \mathbf{0} &
        -\mathbf{1}
  \end{pmatrix}\,,  
\end{equation}
\begin{equation}\label{dn-mat-mtS121323}
\, \Tilde{S}^{12} = \frac{1}{2} 
 \begin{pmatrix}
       \mathbf{1} & \mathbf{0} & \mathbf{0} & \mathbf{0}\\
        \mathbf{0} &
        \mathbf{1}  & \mathbf{0} & \mathbf{0}\\
        \mathbf{0} & \mathbf{0} &
        -\mathbf{1}  & \mathbf{0} \\
        \mathbf{0} & \mathbf{0} & \mathbf{0} &
        -\mathbf{1} 
      \end{pmatrix}\,,  
%
 %
\, \Tilde{S}^{13} = \frac{i}{2}
   \begin{pmatrix}
      \mathbf{0} & \mathbf{0} & \mathbf{0} &
      -\mathbf{1}  \\
        \mathbf{0} & \mathbf{0} & 
        \mathbf{1} & \mathbf{0} \\
        \mathbf{0} & 
        -\mathbf{1} & \mathbf{0}  & \mathbf{0} \\
        \mathbf{1} & \mathbf{0} & \mathbf{0} & \mathbf{0}
  \end{pmatrix}\,, 
%
 %
\, \Tilde{S}^{23} = \frac{1}{2}
   \begin{pmatrix}
      \mathbf{0} & \mathbf{0} & \mathbf{0} &
      -\mathbf{1}  \\
        \mathbf{0} & \mathbf{0} & 
        \mathbf{1} & \mathbf{0} \\
        \mathbf{0} & 
        \mathbf{1} & \mathbf{0}  & \mathbf{0} \\
        -\mathbf{1} & \mathbf{0} & \mathbf{0} & \mathbf{0}
  \end{pmatrix}\,.
 \end{equation}

 \begin{equation}\label{dn-mat-mtGamma31}
   \Tilde{\Gamma}^{(3+1)}=-4i \Tilde{S}^{03}\Tilde{S}^{12}=
 \begin{pmatrix}
        \mathbf{1} & \mathbf{0} & \mathbf{0} & \mathbf{0}\\
        \mathbf{0} &
        -\mathbf{1} & \mathbf{0} & \mathbf{0}\\
        \mathbf{0} & \mathbf{0} &
        -\mathbf{1} & \mathbf{0} \\
        \mathbf{0} & \mathbf{0} & \mathbf{0} &
        \mathbf{1}
  \end{pmatrix}\,.
\end{equation}

Let us now conclude this appendix:
We start with the operators $\gamma^a$'s and $\tilde{\gamma}^a$'s and 
express the  "basis vectors" as  superposition of and an odd number of 
$\gamma^a$'s. 
Here  we present the matrix representation of the operators $\gamma^a$'s, 
$\tilde{\gamma}^a$'s, $S^{ab}$'s and $\tilde{S}^{ab}$'s, operating
on the "basis vectors", which are eigenvectors of the Cartan subalgebra of the
Lorentz algebra, $S^{ab}$'s and $\tilde{S}^{ab}$'s. This matrices can be 
compared with the Pauli matrices and 
$\gamma^m$'s and $S^{mn}$'s matrices of Dirac, after we neglect the charge
part in the $5^{th}$ and the $6^{th}$ dimension of the internal space of fermions.

Since there are several families the representations of the operators 
$\gamma^a$'s, $\tilde{\gamma}^a$'s, $S^{ab}$'s and $\tilde{S}^{ab}$'s
 manifest the existence of families. 

The states on which the Pauli matrices  and the Dirac's $\gamma^a$ matrices apply 
are just columns of numbers, with no anticommuting character, which "basis vectors"
 do have. 

The  Clifford algebra enables  to create creation operators which define, applying 
on the vacuum state, the single particle states with the odd Clifford character. 
Since the oddness of (the finite number of) "basis vectors" transfers in the 
tensor products, $*_{T}$,  with the continuously infinite basis in ordinary space
 to the creation and annihilation operators, making them Clifford odd, the Dirac 
postulates are not needed. 
The use of the Clifford algebra for the description of the internal space explains the
Dirac's second quantized postulates.

In Sect.~\ref{creationannihilationtensor} we compare the way of the second 
quantization of fermions in the {\i spin-charge-family} theory with the Dirac's one.



%
%
\section{ Understanding  triangle anomalies cancellation in standard model}
\label{appanomalies}
%


We clarify here the difference in explaining  the "miraculous" anomalies cancellation 
in the {\it standard model} in the $SO(10)$ unifying theories and the 
{\it spin-charge-family} theory. Since the $SO(10)$ group is subgroup of 
$SO(13,1)$ group, the difference occurs from the fact that in  the 
{\it spin-charge-family} theory the spins and charges are unified, while in the $SO(10)$
theories they are not. One sees in Table ~\ref{Table so13+1.} that handedness
and charges of quarks and leptons and of antiquarks and antileptons are uniquely 
determined --- fermions and anti-fermions belong to the same irreducible representation
of the Lorentz group, while in the unifying theories handedness and charges must be 
correlated "by hand". 
We follow here mainly Ref.~\cite{nh2017}.

In $d=(2n)$-dimensional space-time massless fermions contribute through the 
one-loop (n + 1)-angle diagram in general an anomalous (infinite) function, which 
causes the current non-conservation and contributes to the gauge non-invariance 
of the action~\cite{Bilal,AlvarezBondiaMartin}. 

To the  triangle anomalies the right-handed spinors (fermions) and the right handed 
anti-spinors (anti-fermions) contribute with the opposite sign than the left handed 
spinors and the left handed anti-spinors. Their common contribution to anomalies is 
proportional to~\cite{Bilal} 
\begin{eqnarray}
\label{anomaly0}
(\sum_{(A,i,B,j,C,k)_{L \,\bar{L}}} Tr [\tau^{Ai} \, \tau^{Bj} \,\tau^{Ck}] -
 \sum_{(A,i,B,j,C,k)_{R \,\bar{R}}} Tr [\tau^{Ai} \, \tau^{Bj} \,\tau^{Ck}]\,) \,,
\end{eqnarray}
where $\tau^{Ai}$ are in the {\it standard model} the generators of the infinitesimal 
transformation of the groups $SU(3), SU(2)$ and $U(1)$, while in the 
{\it spin-charge-family} theory $\tau^{Ai}$ are the infinitesimal generators of the 
irreducible subgroups of the starting orthogonal group $SO(2(2n+1)-1,1)$, $n=3$ 
(which include all the {\it standard model} groups, offering correspondingly the 
explanation for their origin), Eqs.~(\ref{so1+3}, \ref{so42}, \ref{so64}). 
The traces run over the representations of one massless family of the left handed 
fermions and anti-fermions, denoted by ${}_{L \,\bar{L}}$, and the right handed 
fermions and anti-fermions, denoted by ${}_{R \,\bar{R}}$.

Embedding the {\it standard model} groups into the orthogonal group $SO(13,1)$ 
explains elegantly the "miraculous" cancellation of the triangle anomalies in the 
{\it standard model}. Embedding the  {\it standard model} groups into the 
orthogonal group $SO(10)$ group explains the "miraculous" cancellation of the
 triangle anomalies if one correlates charges and handedness.

Table~\ref{Table so13+1.}  presents spinor handedness ($\Gamma^{(3,1)}$),
 their spin ($S^{12}$), weak charge $(\tau^{13})$, the second $SU(2)_{II}$ 
$(\tau^{23})$ charge (arising together with $SU(2)_{I}$ from $SO(4)$), their 
colour charge $(\tau^{33},\tau^{38})$  (arising together with $U(1)_{II}$ from 
$SO(6)$), and the "fermion charge" ($\tau^{4}$, the generator of $U(1)_{II}$). 
The hyper charge is $Y=(\tau^{23} +\tau^{4})$,  the electromagnetic charge is 
$Q=(\tau^{13}+ Y)$.


The triangle anomaly  of the {\it standard model}  occurs if the traces in 
Eq.(\ref{anomaly0}) are not zero for either the left handed quarks and leptons and 
anti-quarks and anti-leptons or the right handed quarks and leptons and anti-quarks 
and anti-leptons for the Feynman triangle diagrams in which the 
gauge vector fields of the type 
\begin{eqnarray}
\label{anomalyparticular}
&& U(1)\times U(1) \times U(1)\, ,\nonumber\\ 
&& SU(2)\times SU(2) \times U(1)\, ,\nonumber\\
&& SU(3)\times SU(3) \times SU(3)\, ,\nonumber\\ 
&& SU(3)\times SU(3) \times U(1)\, ,\nonumber\\  
&& U(1) \times {\rm gravitational}\,  
\end{eqnarray}
contribute to the triangle anomaly. 


Table~\ref{Table SMandSCFTspinors.} presents in the first seven columns (up to $||$) 
by the {\it standard model} assumed properties of the members of any massless 
family, running in the triangle. The last two columns, taken from 
Table~\ref{Table so13+1.}, describe additional properties which quarks and leptons and 
anti-quarks and anti-leptons would have, if the {\it standard model} groups 
$SO(3,1)$, $SU(2), SU(3)$ and $U(1)$ are embedded into the $SO(13,1)$ group.  
As already pointed out, the $SO(10)$ unifying theories do from the 
point of the charges  the same, but do  correlate handedness and charges in the 
same way as in the {\t standard model}. 
We  comment these last two columns in particular in Sect.~\ref{fermionsbosonslowE}.

In the {\it spin-charge-family} theory the family quantum numbers are determined by 
the second kind of the Clifford algebra objects $\tilde{S}^{ab}$, which commute 
with $S^{ab}$ describing spins and charges. Correspondingly the spins and charges 
are the same for all the families.
\begin{table}
\begin{center}
\begin{minipage}[t]{16.5 cm}
\caption{Properties  of the left handed quarks and leptons and 
of the left handed anti-quarks and anti-leptons (the first $16$ lines)  and of the right
handed quarks and leptons and the right handed anti-quarks and anti-leptons (the 
second $16$ lines), as assumed by the {\it standard model}, are presented in the 
first eight columns. In the last two columns the two quantum numbers are added, 
which fermions and anti-fermions would have if the {\it standard model} groups 
$SO(3,1)$, $SU(2)$, $SU(3)$ and $U(1)$ are embedded into the $SO(13,1)$ group. 
All the infinitesimal generators of the subgroups of the orthogonal group $SO(13,1)$, 
representing charges in $d=(3+1)$, are in the {\it spin-charge-family} theory the 
superposition of the generators $S^{st}$, 
Eqs.~(\ref{so1+3}, \ref{so42}, \ref{so64}). 
The handedness is defined in Eq.(\ref{hand}). 
The whole quark part appears, due to the colour charges, three times.  
These quantum numbers are the same for all the families.} 
\label{Table SMandSCFTspinors.}
\end{minipage}
\begin{tabular}{|r r | c r r r r r ||r r|}
\hline
           & & hand-   & weak    & hyper  & colour&charge & elm   & $SU(2)_{II}$&$U(1)_{II}$\\
            &&edness  & charge  & charge &        &          &charge&     charge     &  charge     \\
$i_{L}$&name    &$ \Gamma^{(3,1)}$&$ \tau^{13}$  & $ Y$ & $\tau^{33}$&$\tau^{38}$
&$Q$&$\tau^{23}$&$\tau^{4}$\\
\hline
$1_{L}$&$ u_{L} $&$ -1     $&$\frac{1}{2}$ &$ \frac{1}{6}$&$\frac{1}{2} $&$\frac{1}{2\sqrt{3}}$&
$ \frac{2}{3}$&0&$\frac{1}{6}$\\
$2_{L}$&$ d_{L} $&$ -1    $&$ -\frac{1}{2}$ &$ \frac{1}{6}$&$\frac{1}{2}$&$\frac{1}{2\sqrt{3}}$ &
$-\frac{1}{3}$&0&$\frac{1}{6}$\\
$3_{L}$&$ u_{L} $&$ -1     $&$\frac{1}{2}$ &$ \frac{1}{6}$&$-\frac{1}{2} $&$\frac{1}{2\sqrt{3}}$&
$ \frac{2}{3}$&0&$\frac{1}{6}$\\
$4_{L}$&$ d_{L} $&$ -1    $&$ -\frac{1}{2}$ &$ \frac{1}{6}$&$-\frac{1}{2}$&$\frac{1}{2\sqrt{3}}$ &
$-\frac{1}{3}$&0&$\frac{1}{6}$\\
$5_{L}$&$ u_{L} $&$ -1     $&$\frac{1}{2}$ &$ \frac{1}{6}$&$ 0 $&$-\frac{1}{\sqrt{3}}$&
$ \frac{2}{3}$&0&$\frac{1}{6}$\\
$6_{L}$&$ d_{L} $&$ -1    $&$ -\frac{1}{2}$ &$ \frac{1}{6}$&$ 0$&$-\frac{1}{\sqrt{3}}$ &
$-\frac{1}{3}$&0&$\frac{1}{6}$\\
\hline
$7_{L}$&$\nu_{L} $&$ -1  $&$ \frac{1}{2}$&$ -\frac{1}{2}$& 0&0&  0  &$0$&$ -\frac{1}{2}$  \\
$8_{L}$&$ e^{L}  $&$ -1  $&$-\frac{1}{2}$&$ -\frac{1}{2}$& 0&0&$-1$&$0$&$ -\frac{1}{2}$ \\
\hline \hline
$9_{L}$&$ \bar{u}{L} $&$ -1$&$  0 $ &$ -\frac{2}{3}$& $ - \frac{1}{2}$ &$-\frac{1}{2\sqrt{3}}$
&$- \frac{2}{3}$ &$ - \frac{1}{2}$ &$ - \frac{1}{6}$\\
$10_{L}$&$ \bar{d}{L} $&$ -1$&$  0 $ &$  \frac{1}{3}$& $ - \frac{1}{2}$ &$-\frac{1}{2\sqrt{3}}$
&$  \frac{1}{3}$ &$   \frac{1}{2}$ &$ - \frac{1}{6}$\\
$11_{L}$&$ \bar{u}{L} $&$ -1$&$  0 $ &$ -\frac{2}{3}$& $  \frac{1}{2}$ &$-\frac{1}{2\sqrt{3}}$
&$- \frac{2}{3}$ &$ - \frac{1}{2}$ &$ - \frac{1}{6}$\\
$12_{L}$&$ \bar{d}{L} $&$ -1$&$  0 $ &$  \frac{1}{3}$& $  \frac{1}{2}$ &$-\frac{1}{2\sqrt{3}}$
&$  \frac{1}{3}$ &$   \frac{1}{2}$ &$ - \frac{1}{6}$\\
$13_{L}$&$ \bar{u}{L} $&$ -1$&$  0 $ &$ -\frac{2}{3}$& $ 0 $&$ \frac{1}{\sqrt{3}}$
&$- \frac{2}{3}$ &$ - \frac{1}{2}$ &$ - \frac{1}{6}$\\
$14_{L}$&$ \bar{d}{L} $&$ -1$&$  0 $ &$  \frac{1}{3}$& $ 0 $&$\frac{1}{\sqrt{3}}$
&$  \frac{1}{3}$ &$   \frac{1}{2}$ &$ - \frac{1}{6}$\\
\hline
$15_{L}$&$\bar{\nu}_{L}$&$ -1$&$ 0 $&$ 0 $& 0&0&$0$&$- \frac{1}{2}$&$ \frac{1}{2}$  \\
$16_{L}$&$\bar{e}_{L}   $&$ -1$&$ 0 $&$ 1 $& 0&0&$1$&$  \frac{1}{2}$&$ \frac{1}{2}$  \\
\hline \hline\hline
$1_{R}$&$u_{R}   $&$ 1$& $0$ & $  \frac{2}{3}$&$ \frac{1}{2}$ &$\frac{1}{2\sqrt{3}}$
&$ \frac{ 2}{3}$&$ \frac{1}{2}$&$ \frac{1}{6}$ \\
$2_{R}$&$d_{R}   $&$ 1$& $0$ & $ -\frac{1}{3}$&$ \frac{1}{2}$ &$\frac{1}{2\sqrt{3}}$ 
&$-\frac{1}{3}$&$- \frac{1}{2}$&$ \frac{1}{6}$  \\
$3_{R}$&$u_{R}   $&$ 1$& $0$ & $  \frac{2}{3}$&$- \frac{1}{2}$ &$\frac{1}{2\sqrt{3}}$
&$ \frac{ 2}{3}$&$ \frac{1}{2}$&$ \frac{1}{6}$ \\
$4_{R}$&$d_{R}   $&$ 1$& $0$ & $ -\frac{1}{3}$&$- \frac{1}{2}$ &$\frac{1}{2\sqrt{3}}$ 
&$-\frac{1}{3}$&$- \frac{1}{2}$&$ \frac{1}{6}$  \\
$5_{R}$&$u_{R}   $&$ 1$& $0$ & $  \frac{2}{3}$&$ 0  $ &$-\frac{1}{\sqrt{3}}$
&$ \frac{ 2}{3}$&$ \frac{1}{2}$&$ \frac{1}{6}$ \\
$6_{R}$&$d_{R}   $&$ 1$& $0$ & $ -\frac{1}{3}$&$ 0  $ &$-\frac{1}{\sqrt{3}}$ 
&$-\frac{1}{3}$&$- \frac{1}{2}$&$ \frac{1}{6}$  \\
\hline
$7_{R}$&$\nu_{R}$&$ 1$& $0$  & $ 0$&0&0&   0 &$  \frac{1}{2}$&$- \frac{1}{2}$           \\
$8_{R}$&$ e_{R}  $&$ 1$& $0$  & $-1$&0&0&$-1$&$- \frac{1}{2}$&$- \frac{1}{2}$  \\
\hline\hline
$9_{R}$&$\bar{u}_{R}   $&$ 1$& $ -\frac{1}{2}$ & $ - \frac{1}{6}$&$ -\frac{1}{2}$ &$-\frac{1}{2\sqrt{3}}$
&$ - \frac{ 2}{3}$&0& $ - \frac{1}{6}$\\
$10_{R}$&$\bar{d}_{R}   $&$ 1$&$  \frac{1}{2}$ & $  - \frac{1}{6}$&$ -\frac{1}{2}$ &$-\frac{1}{2\sqrt{3}}$
&$    \frac{1}{3}$&0& $ - \frac{1}{6}$\\
$11_{R}$&$\bar{u}_{R}   $&$ 1$& $ -\frac{1}{2}$ & $ -  \frac{1}{6}$&$ \frac{1}{2}$ &$-\frac{1}{2\sqrt{3}}$
&$ - \frac{ 2}{3}$&0& $ - \frac{1}{6}$\\
$12_{R}$&$\bar{d}_{R}   $&$ 1$&$  \frac{1}{2}$ & $ -  \frac{1}{6}$&$  \frac{1}{2}$ &$-\frac{1}{2\sqrt{3}}$
&$    \frac{1}{3}$&0& $ - \frac{1}{6}$\\
$13_{R}$&$\bar{u}_{R}   $&$ 1$& $ -\frac{1}{2}$ & $ -  \frac{1}{6}$&$ 0 $ &$ \frac{1}{\sqrt{3}}$
&$ - \frac{ 2}{3}$&0& $ - \frac{1}{6}$\\
$14_{R}$&$\bar{d}_{R}   $&$ 1$&$  \frac{1}{2}$ & $  -  \frac{1}{6}$&$ 0 $ &$ \frac{1}{\sqrt{3}}$
&$    \frac{1}{3}$&0& $ - \frac{1}{6}$\\
\hline
$15_{R}$&$\bar{\nu}_{R}$&$ 1$& $ -\frac{1}{2}$ & $  \frac{1}{2}$&0&0&$0$&0&$\frac{1}{2}$        \\
$16_{R}$&$\bar{e}_{R}   $&$ 1$& $  \frac{1}{2}$ & $  \frac{1}{2}$&0&0&$1$&0&$\frac{1}{2}$ \\
\hline\hline
\end{tabular}
  \end{center}
%
\end{table}

To calculate the traces required in Eq.~(\ref{anomaly0}) for the triangle anomalies of 
Eq.~(\ref{anomalyparticular}) the quantum numbers of the left handed spinors and 
anti-spinors, as well as of the right handed spinors and anti-spinors, presented in 
Table~\ref{Table SMandSCFTspinors.},
are needed.

Let us calculate the traces, Eq.~(\ref{anomalyparticular}), for possible anomalous 
triangle diagrams in $d=(3+1)$. One must evaluate the trace of the product  of 
three generators and sum the trace over all the states of either the left handed 
members ---  $16$ states presented in the first part of 
Table~\ref{Table SMandSCFTspinors.} --- or the right handed members --- $16$ 
states presented in the second part of Table~\ref{Table SMandSCFTspinors.}.
Let us recognize again that in the case of embedding the {\it standard model} 
groups into $SO(13,1)$ we have $Y=(\tau^{4} + \tau^{23}) $. 

For the triangle Feynman diagram,  to which three hyper $U(1)$ boson fields 
contribute, we must evaluate  $\sum_{i} Tr (Y_{i})^3$, in which the sum runs 
over all the members ($i=(1,..,16)$) of the left handed spinors and anti-spinors, 
and of the right handed spinors and anti-spinors separately. 
In the case of embedding the {\it standard model} groups into $SO(13,1)$ we have
\begin{eqnarray}
\label{3Ytriangle}
\sum_{i_{L,R}}\, (Y_{i_{L,R}})^3&=&\sum_{i_{L,R}}\,
 (\tau^{4}_{i_{L,R}} + \tau^{23}_{i_{L,R}})^{3}  
\nonumber\\
 &=&\sum_{i_{L,R}}\, (\tau^{4}_{i_{L,R}})^{3} + \sum_{i_{L,R}}\, (\tau^{23}_{i_{L,R}})^{3}
\nonumber\\
&+&\sum_{i_{L,R}}\, 3 \cdot (\tau^{4}_{i_{L,R}})^{2} \cdot \tau^{23}_{i_{L,R}}
  +  \sum_{i_{L,R}}\, 3 \cdot \tau^{4}_{i_{L,R}} \cdot  (\tau^{23}_{i_{L,R}})^{2}\,,
\end{eqnarray}
for either the left, $i_{L}$, or the right, $i_{R}$, handed members. 
Table~\ref{Table SMandSCFTspinors.}  demonstrates  clearly that 
$ (Y_{i_{L,R}})^3=0$ without really making any algebraic evaluation. 
Namely, the last column of Table~\ref{Table SMandSCFTspinors.}  
manifests that $ \sum_{i_{L}}\,
 (\tau^{4}_{i_{L}})^{3} =0 $ [in details: $\sum_{i_{L}}\, (\tau^{4}_{i_{L}})^{3}  $ 
$ = 2 \cdot 3 \cdot (\frac{1}{6})^{3} + 2 \cdot 3 \cdot (-\frac{1}{6})^{3} +
2 \cdot (-\frac{1}{2})^{3} + 2 \cdot (\frac{1}{2})^{3}=0  $]. 
Table~\ref{Table SMandSCFTspinors.}  also demonstrates  (the last but one column)
 that $ \sum_{i_{L}}\, (\tau^{23}_{i_{L}})^{3} =0 $ 
$[=(3+1) \cdot ((-\frac{1}{2})^{3}  +  (\frac{1}{2})^{3}))]$, and that also
 $ \sum_{i_{R}}\,  (\tau^{23}_{i_{R}})^{3} =0 $ 
$[= (3+1) \cdot ((\frac{1}{2})^{3} +
 (- \frac{1}{2})^{3})]$. 

From Table~\ref{Table SMandSCFTspinors.}  one sees also (without calculating) that 
$\sum_{i_{L}}\, 3 \cdot (\tau^{4}_{i_{L}})^{2} \cdot \tau^{23}_{i_{L}} =0$, 
in particular 
$\sum_{i_{L}}\, 3 \cdot (\tau^{4}_{i_{L}})^{2} \cdot \tau^{23}_{i_{L}} $ 
$=3.\{ ((\frac{1}{2})^2 \cdot (-\frac{1}{2} +\frac{1}{2})+ 3 \cdot (-\frac{1}{6})^2  
\cdot (-\frac{1}{2} +\frac{1}{2})\}$, as well as  that $\sum_{i_{R}}\, 3 \cdot 
(\tau^{4}_{i_{R}})^{2} \cdot \tau^{23}_{i_{R}} =0$ [$=3\cdot\{ ((-\frac{1}{2})^2 
\cdot (\frac{1}{2} +
(-\frac{1}{2}))+ 3 \cdot (\frac{1}{6})^2  \cdot (\frac{1}{2} + (-\frac{1}{2}))\}$].

That the last term in Eq.~(\ref{3Ytriangle}) is zero for either the left or the 
right handed spinors can also easily be seen just by looking at 
Table~\ref{Table SMandSCFTspinors.} [or in details: 
$ \sum_{i_{L}}\, 3 \cdot \tau^{4}_{i_{L}} \cdot  (\tau^{23}_{i_{L}})^{2}=0$
$=3\cdot \{(\frac{1}{2} ((\frac{1}{2})^2 +(-\frac{1}{2})^2) + 3\cdot (-\frac{1}{6})
 ((\frac{1}{2})^2 +(-\frac{1}{2})^2))\} $, as well as that 
$ \sum_{i_{R}}\, 3 \cdot \tau^{4}_{i_{L}} \cdot  (\tau^{23}_{i_{R}})^{2}=0$
$=3\cdot \{(-\frac{1}{2} ((\frac{1}{2})^2 +(-\frac{1}{2})^2) + 3\cdot (\frac{1}{6})
 ((\frac{1}{2})^2 +(-\frac{1}{2})^2))\} $].

Since all the members belong to one spinor representation, it is straightforwardly 
that all the triangle traces are zero, if the {\it standard model}  groups are the 
subgroups of the orthogonal group  $SO(13,1)$. 

There is no need for a detailed calculations, since a look in 
Table~\ref{Table SMandSCFTspinors.} gives immediately the answer.

From only the {\it standard model} assumptions point of view the cancellation of 
the triangle anomalies does look  miraculous. For our $\sum_{i_{L,R}}\, (Y_{i_{L,R}})^3$ 
one obtains for the left handed members: 
[$3 \cdot2 \cdot (\frac{1}{6})^3 + 2 \cdot (-\frac{1}{2})^3 + 
3 \cdot  ((-\frac{2}{3})^3 + (\frac{1}{3})^3) + 1^3)$ ], and for the right handed 
members:  [$3 \cdot ((\frac{2}{3})^3 + (-\frac{1}{3})^3)+
(- 1)^3) +3 \cdot 2 \cdot (-\frac{1}{6})^3 + 2 \cdot (\frac{1}{2})^3 $ ].


%
\section{Norms in Grassmann space and Clifford space}
\label{normgrass}
%


Let us define the integral over the Grassmann  space~\cite{norma93} of two functions of the 
Grassmann coordinates $<{\cal {\bf B}}|\theta> <{\cal {\bf C}}|\theta>$,  $<{\cal {\bf B}}| \theta>= 
<\theta | {\cal {\bf B}}>^{\dagger}$,
\[<{\cal {\bf B}}|\theta>= \sum_{k=0}^{d} b_{a_1\dots a_k}
\theta^{a_1}\cdots \theta^{a_k},\]
 by requiring 
\begin{eqnarray}
\label{grassintegral}
&&\{ d\theta^a, \theta^b \}_{+} =0\,, \,\;\;  \int d\theta^a  =0\,,\,\;\; 
\int d\theta^a \theta^a =1\,,\;\; \nonumber\\
&&\int d^d \theta \,\,\theta^0 \theta^1 \cdots \theta^d =1\,,
\nonumber\\
&&d^d \theta =d \theta^d \dots d\theta^0\,,\,\;\; 
\omega = \prod^{d}_{k=0}(\frac{\partial}{\;\,\partial \theta_k} + \theta^{k})\,,
\end{eqnarray}
with $ \frac{\partial}{\;\,\partial \theta_a} \theta^c = \eta^{ac}$. We shall use the weight function~%
\cite{norma93} 
$\omega= \prod^{d}_{k=0}(\frac{\partial}{\;\,\partial \theta_k} + \theta^{k})$ to define the scalar
product  in Grassmann space $<{\cal {\bf B}}|{\cal {\bf C}} >$ 
\begin{eqnarray}
\label{grassnorm}
<{\cal {\bf B}}|{\cal {\bf C}} > &=&  \int 
 d^d \theta^a\, \,\omega 
 <{\cal {\bf B}}|\theta>\, <\theta|{\cal {\bf C}}> \nonumber\\
 &=& \sum^{d}_{k=0} \int 
\, b^{*}_{b_{1} \dots b_{k}} c_{b_1 \dots b_{k}}\,.%
\end{eqnarray}%
%

 To define norms in Clifford space Eq.~(\ref{grassintegral}) can be used as well.


%
\section{Expressions for scalar fields in term of $\omega_{s't s}$ and 
$\tilde{\omega}_{abs}$ }
\label{scalarsandvectorsapp}

As presented  in Sect.~\ref{scalar3+1} the {\it spin-charge-family} 
theory offers the explanation for the appearance of the scalar higss, with the
weak charge $\pm\frac{1}{2}$ and the hyper charge $\mp\frac{1}{2}$, and 
Yukawa couplings,  as well as for the matter/antimatter asymmetry in the 
universe. There are namely spin  connections of two kinds, $\omega_{abs}$
and $\tilde{\omega}_{abs}$, with the space index $s\ge 5$, which manifest
in $d=(3+1)$ as scalars with the weak charge $\pm\frac{1}{2}$ and the 
hyper charge $\mp\frac{1}{2}$ and with the space index $s=(7,8)$,  
and there are also scalars which manifest with respect to the space index
$t=(9,10,11,12,13,14)$ as colour triplets or antitriplets.

%
\subsection{Scalar fields with the space index $s=(7,8)$}
\label{scalar3+1higgsapp}
The scalar fields, responsible  for masses of the family members and 
of the heavy bosons~\cite{n2012scalars,JMP2013} after gaining 
nonzero vacuum expectation values and triggering the electroweak break, 
are presented in  the second line of Eq.~(\ref{faction}). These scalar fields
are included in the covariant derivatives of momenta as  
$-\frac{1}{2}$ $  S^{s' s"} \omega_{s' s" s} $ 
$- \frac{1}{2}  \tilde{S}^{a b}  \tilde{\omega}_{a b s}$, 
$s \in (7,8)$, $(a,b) $,  
$\in (0,\dots,3), (5,\dots,8)$.

One can express the scalar fields carrying the quantum numbers of the 
subgroups of the family groups, expressed in terms of 
$\tilde{\omega}_{a b s}$ (they contribute to mass matrices of 
quarks and leptons and to masses of the heavy bosons), if taking into 
account Eqs.~(%
\ref{so1+3}, \ref{so42}, \ref{YQY'Q'andtilde}), 
%
 \begin{eqnarray}
 \label{Atildeomegas}
\sum_{a,b} -\frac{1}{2}  \tilde{S}^{a b}\,\tilde{\omega}_{a b   s} &=&
 - (\vec{\tilde{\tau}}^{\tilde{1}}\, \vec{\tilde{A}}^{\tilde{1}}_{s} + 
  \vec{\tilde{N}}_{\tilde{L}}\, \vec{\tilde{A}}^{\tilde{N}_{\tilde{L}}}_{s} 
   + \vec{\tilde{\tau}}^{\tilde{2}}\, \vec{\tilde{A}}^{\tilde{2}}_{s} + 
  \vec{\tilde{N}}_{\tilde{R}}\, \vec{\tilde{A}}^{\tilde{N}_{\tilde{R}}}_{s})\,, 
  \nonumber\\
  \vec{\tilde{A}}^{\tilde{1}}_{s} &=& (\tilde{\omega}_{5 8 s}-
  \tilde{\omega}_{6 7 s},\, \tilde{\omega}_{5 7 s}+
  \tilde{\omega}_{6 8 s}, \,\tilde{\omega}_{5 6 s}-
  \tilde{\omega}_{7 8 s})\,, \nonumber\\
\vec{\tilde{A}}^{\tilde{N}_{\tilde{L}}}_{s} &=& 
(\tilde{\omega}_{2 3 s}+i\,  \tilde{\omega}_{0 1 s}, \,    \tilde{\omega}_{3 1 s}+
i\,  \tilde{\omega}_{0 2 s},  \,   \tilde{\omega}_{1 2 s}+i\,  \tilde{\omega}_{0 3 s})\,, \nonumber\\
 \vec{\tilde{A}}^{\tilde{2}}_{s} &=& (\tilde{\omega}_{5 8 s}+
  \tilde{\omega}_{6 7 s}, \,\tilde{\omega}_{5 7 s}-
  \tilde{\omega}_{6 8 s},\, \tilde{\omega}_{5 6 s}+
  \tilde{\omega}_{7 8 s})\,, \nonumber\\
\vec{\tilde{A}}^{\tilde{N}_{\tilde{R}}}_{s} &=& (\tilde{\omega}_{2 3 s}-
i\,  \tilde{\omega}_{0 1 s}, \,    \tilde{\omega}_{3 1 s}-i\,
  \tilde{\omega}_{0 2 s}, \,    \tilde{\omega}_{1 2  s}- 
  i\,  \tilde{\omega}_{0 3 s})\,,\nonumber\\
                                             & &(s\in (7,8))\,. 
\end{eqnarray}
Scalars, expressed in terms of $\omega_{abc}$ (contributing as well to
the mass matrices of quarks and leptons and to masses of the heavy bosons) follow, if using 
Eqs.~(\ref{so42}, \ref{so64}, 
\ref{YQY'Q'andtilde}) 
\begin{eqnarray}
\label{Aomegas}
\sum_{s', s''} \,-\frac{1}{2} S^{s's"}\, \omega_{s's"s}&=&
 - (g^{23}\, \tau^{23}\, A^{23}_{s} + 
                                             g^{13}\, \tau^{13}\, A^{13}_{s} + 
                                             g^{4 }\, \tau^{4} \, A^{4}_{s})\,,
\nonumber\\
g^{13}\,\tau^{13}\, A^{13}_{s} + g^{23}\,\tau^{23}\, A^{23}_{s} + g^{4}\,\tau^{4}\,
 A^{4}_{s} &=& 
g^{Q}\, Q A^{Q}_{s} + g^{Q'}\, Q' A^{Q'}_{s} + g^{Y'}\,Y'\, A^{Y'}_{s}\,, \nonumber\\
A^{4}_{s} &=& -(\omega_{9\,10\,s} + \omega_{11\,12\,s} + \omega_{13\,14\,s})\,,
\nonumber\\
A^{13}_{s}&=&(\omega_{56 s}- \omega_{78 s})\,, \quad A^{23}_{s}=(\omega_{56 s}+
\omega_{78 s})\,,\nonumber\\
A^{Q}_{s} &=& \sin \vartheta_{1} \,A^{13}_{s} + \cos \vartheta_{1} \,A^{Y}_{s}\,,\quad
A^{Q'}_{s}  = \cos \vartheta_{1} \,A^{13}_{s} - \sin \vartheta_{1} \,A^{Y}_{s}\,,\nonumber\\
A^{Y'}_{s}&=&\cos \vartheta_{2} \,A^{23}_{s} - \sin \vartheta_{2} \,A^{4}_{s}\,,\nonumber\\
          & &(s\in (7,8))\,. 
\end{eqnarray}
Scalar fields from Eq.~(\ref{Atildeomegas}) interact with quarks and leptons and 
antiquarks and antileptons through the family quantum numbers, while those 
from Eq.~(\ref{Aomegas})  interact through the family members quantum 
numbers. In Eq.~(\ref{Aomegas}) the coupling constants are explicitly written 
in order to see the analogy with the gauge fields of the {\it standard model}.
%


Let be shown that the scalar fields 
$A^{Ai}_{\scriptscriptstyle{\stackrel{78}{(\pm)}}}$  
are {\it triplets} as the gauge fields of the  family quantum numbers 
($\vec{\tilde{N}}_{R}, \,$ $\vec{\tilde{N}}_{L},\,$ $ \vec{\tilde{\tau}}^{2},\,$ 
$\vec{\tilde{\tau}}^{1}$;
 Eqs.~(\ref{so1+3}, \ref{so42}, \ref{bosonspin0})) or singlets as the gauge fields of 
$Q=\tau^{13}+Y, \,Q'= -\tan^{2}\vartheta_{1} Y$ $ + \tau^{13} $ and
 $Y' = -\tan^2 \vartheta_{2} \tau^{4} + \tau^{23}$, 
 for $\tilde{A}^{N_{L}i}_{\scriptscriptstyle{\stackrel{78}{(\pm)}}}$ 
and for $A^{Q}_{\scriptscriptstyle{\stackrel{78}{(\pm)}}}$, taking into account
Eq.~(\ref{so1+3}) (where we replace $S^{ab}$ by ${\cal S}^{ab}$) and 
Eq.~(\ref{bosonspin0}), while recognizing that 
$\tilde{A}^{N_{L}\spm}_{\scriptscriptstyle{\stackrel{78}{(\pm)}}}= $ 
$\tilde{A}^{N_{L}1}_{\scriptscriptstyle{\stackrel{78}{(\pm)}}}\, \smp\, i $ 
$\tilde{A}^{N_{L}2}_{\scriptscriptstyle{\stackrel{78}{(\pm)}}}$, and that 
\begin{eqnarray}
 \tilde{A}^{\tilde{N}_{L}\spm}_{\scriptscriptstyle{\stackrel{78}{(\pm)}}} &=& 
\{(\tilde{\omega}_{2 3 \scriptscriptstyle{\stackrel{78}{(\pm)}}} + i \,
   \tilde{\omega}_{01 \scriptscriptstyle{\stackrel{78}{(\pm)}}} ) \smp\,i \, 
  (\tilde{\omega}_{31 \scriptscriptstyle{\stackrel{78}{(\pm)}}}  + i\,
   \tilde{\omega}_{02 \scriptscriptstyle{\stackrel{78}{(\pm)}}})\}\,,\nonumber\\
\tilde{A}^{\tilde{N}_{L}3}_{\scriptscriptstyle{\stackrel{78}{(\pm)}}} &=& 
  (\tilde{\omega}_{12 \scriptscriptstyle{\stackrel{78}{(\pm)}}} +i \,
   \tilde{\omega}_{03 \scriptscriptstyle{\stackrel{78}{(\pm)}}})\,, \nonumber\\
A^{Q}_{\scriptscriptstyle{\stackrel{78}{(\pm)}}} &=&
 \omega_{56 \scriptscriptstyle{\stackrel{78}{(\pm)}}} -  
(\omega_{9\,10 \scriptscriptstyle{\stackrel{78}{(\pm)}}} + 
\omega_{11\,12 \scriptscriptstyle{\stackrel{78}{(\pm)}}} + 
\omega_{13\,14 \scriptscriptstyle{\stackrel{78}{(\pm)}}})\,, \nonumber
\end{eqnarray}
the scalar fields are chosen (with respect to the space index $s=(7,8)$) to be 
eigenstates of the weak charge, Eq.~(\ref{eigentau1tau2}).

One finds 
\begin{eqnarray}
\label{checktildeNL3Q1}
\tilde{N}_{L}^{3}\,
\tilde{A}^{\tilde{N}_{L}\spm}_{\scriptscriptstyle{\stackrel{78}{(\pm)}}} 
&=& \spm \tilde{A}^{\tilde{N}_{L}\spm}_{\scriptscriptstyle{\stackrel{78}{(\pm)}}}\,,
\quad \tilde{N}_{L}^{3}\,
\tilde{A}^{\tilde{N}_{L}3}_{\scriptscriptstyle{\stackrel{78}{(\pm)}}}=0
\,,\nonumber\\
Q \,A^{Q}_{\scriptscriptstyle{\stackrel{78}{(\pm)}}} &=&0\,,
\end{eqnarray}
%
%
taking into account $ Q={\cal S}^{56} + {\cal \tau}^{4}= {\cal S}^{56} -
\frac{1}{3}({\cal S}^{9\,10}+
{\cal S}^{11\,12} + {\cal S}^{13\,14})$, and with 
${\cal \tau}^{4}$ defined in Eq.~(\ref{so64}), if replacing 
$S^{ab}$ by ${\cal S}^{ab}$ from Eq.~(\ref{bosonspin0}). 
Similarly one finds properties with respect to the $Ai$ quantum 
numbers for all the scalar fields $A^{Ai}_{\scriptscriptstyle{\stackrel{78}{(\pm)}}}$. 
%
\subsection{Scalar fields with the space index $s=(9,\dots,14)$}
\label{scalar3+1matterantimatterapp}


Expressions for the vector gauge fields in terms of the spin connection fields 
and the vielbeins, which correspond to the colour charge ($\vec{A}^{3}_{m}$), 
the $SU(2)_{II}$ charge ($\vec{A}^{2}_{m}$), the weak $SU(2)_{I}$ charge 
($\vec{A}^{1}_{m}$) and 
the $U(1)$ charge originating in $SO(6)$ ($\vec{A}^{4}_{m}$), can be found 
by taking into account Eqs.~(\ref{so42}, \ref{so64}). Equivalently one finds 
the vector gauge fields in the "tilde" sector, or one just uses the expressions 
from Eqs.~(\ref{Aomegas}, \ref{Atildeomegas}), if replacing the 
scalar index $s$ with the vector index $m$. 


The expression for  $\sum_{t a b}\gamma^{t}\,$ 
$\frac{1}{2}\, \tilde{S}^{ab} \,\tilde{\omega}_{ab t}$, used in 
Eq.~(\ref{factionMaMpart10}) 
($\tilde{S}^{ab}$ are the infinitesimal generators of either $\widetilde{SO}(3,1)$ or 
$\widetilde{SO}(4)$, while $\tilde{\omega}_{ab t}$ belong to the corresponding 
gauge fields  with $t =(9,\dots,14)$), and
  obtained by using Eqs.~(\ref{so1+3} - \ref{plusminus}), are
\begin{eqnarray}
&&\sum_{a b t}\,\gamma^{t} \frac{1}{2}\, \tilde{S}^{ab} \,\tilde{\omega}_{ab t} = 
\sum_{+ - t t' a b}\,\stackrel{t t'}{(\cpm)} \, \frac{1}{2}\, \tilde{S}^{a b} \, 
\tilde{\omega}_{\scriptscriptstyle{ab \,\stackrel{t t'}{(\cpm)}}}= \nonumber\\
&&\sum_{+ - t t'}\stackrel{t t'}{(\cpm)} \,
\{\,\tilde{\tau}^{2+}\,\tilde{A}^{2+}_{\scriptscriptstyle{\stackrel{t t'}{(\cpm)}}} + 
\tilde{\tau}^{2-}\,\tilde{A}^{2-}_{\scriptscriptstyle{\stackrel{t t'}{(\cpm)}}} + 
\tilde{\tau}^{23}\,\tilde{A}^{23}_{\scriptscriptstyle{\stackrel{t t'}{(\cpm)}}} + \nonumber\\
& & \tilde{\tau}^{1+}\,\tilde{A}^{1+}_{\scriptscriptstyle{\stackrel{t t'}{(\cpm)}}} + 
\tilde{\tau}^{1-}\,\tilde{A}^{1-}_{\scriptscriptstyle{\stackrel{t t'}{(\cpm)}}} + 
\tilde{\tau}^{13}\,\tilde{A}^{13}_{\scriptscriptstyle{\stackrel{t t'}{(\cpm)}}} + \nonumber\\
& & \tilde{N}^{+}_{R}\,\tilde{A}^{N_{R}+}_{\scriptscriptstyle{\stackrel{t t'}{(\cpm)}}} + 
\tilde{N}^{-}_{R}\,\tilde{A}^{N_{R}-}_{\scriptscriptstyle{\stackrel{t t'}{(\cpm)}}} + 
\tilde{N}^{3}_{R}\,\tilde{A}^{N_{R}3}_{\scriptscriptstyle{\stackrel{t t'}{(\cpm)}}} + \nonumber\\
& &
\tilde{N}^{+}_{L}\,\tilde{A}^{N_{L}+}_{\scriptscriptstyle{\stackrel{t t'}{(\cpm)}}} +
\tilde{N}^{-}_{L}\,\tilde{A}^{N_{L}-}_{\scriptscriptstyle{\stackrel{t t'}{(\cpm)}}} + 
\tilde{N}^{3}_{L}\,\tilde{A}^{N_{L}3}_{\scriptscriptstyle{\stackrel{t t'}{(\cpm)}}}\,\}
\,,\nonumber\\ 
\tilde{A}^{N_{R}\spm}_{\scriptscriptstyle{\stackrel{t t'}{(\cpm)}}} &=& 
(\tilde{\omega}_{\scriptscriptstyle{23 \stackrel{t t'}{(\cpm)}}}- i\,
 \tilde{\omega}_{\scriptscriptstyle{01 \stackrel{t t'}{(\cpm)}}})\smp \, i 
(\tilde{\omega}_{\scriptscriptstyle{31 \stackrel{t t'}{(\cpm)}}}- i\,
 \tilde{\omega}_{\scriptscriptstyle{02 \stackrel{t t'}{(\cpm)}}})\,,\quad 
\tilde{A}^{N_{R}3}_{\scriptscriptstyle{\stackrel{t t'}{(\cpm)}}}= 
(\tilde{\omega}_{\scriptscriptstyle{12 \,\stackrel{t t'}{(\cpm)}}}- i\,
\tilde{\omega}_{\scriptscriptstyle{03 \stackrel{t t'}{(\cpm)}}})\,, \nonumber\\
\tilde{A}^{N_{L}\spm}_{\scriptscriptstyle{\stackrel{t t'}{(\cpm)}}} &=& 
(\tilde{\omega}_{\scriptscriptstyle{23 \stackrel{t t'}{(\cpm)}}}+ i\,
\tilde{\omega}_{\scriptscriptstyle{01 \stackrel{t t'}{(\cpm)}}})\smp \, i 
(\tilde{\omega}_{\scriptscriptstyle{31 \stackrel{t t'}{(\cpm)}}}+ i\,
 \tilde{\omega}_{\scriptscriptstyle{02 \stackrel{t t'}{(\cpm)}}})\,,\quad 
\tilde{A}^{N_{R}3}_{\scriptscriptstyle{\stackrel{t t'}{(\cpm)}}}= 
(\tilde{\omega}_{\scriptscriptstyle{12 \stackrel{t t'}{(\cpm)}}}+ i\,
 \tilde{\omega}_{\scriptscriptstyle{03 \stackrel{t t'}{(\cpm)}}})\,. 
\label{factionMaMpart20}
\end{eqnarray}
%

The term $\sum_{t t' t''}\gamma^t$ $\frac{1}{2}\,S^{t' t"}\,\omega_{t' t" t}$ in Eq.~(\ref{faction}) 
%
can be rewritten  with respect to the generators $S^{t' t"}$
and the corresponding gauge fields $\omega_{s' s" t}$  
as one colour octet scalar field and one $U(1)_{II}$ singlet scalar field (Eq.~\ref{so64})
\begin{eqnarray}
\sum_{t t'' t'''}\gamma^{t}\,\frac{1}{2}\, S^{t" t'"} \,\omega_{t" t'" t} &=&
\sum_{+,-}\,\sum_{(t\,t')}  \, \stackrel{t t'}{(\cpm)}\, 
\{\,\vec{\tau}^{3}\cdot \vec{A}^{3}_{\scriptscriptstyle{\stackrel{t t'}{(\cpm)}}}\, + \tau^{4}\cdot
A^{4}_{\scriptscriptstyle{\stackrel{t t'}{(\cpm)}}}\,\}\,, \nonumber\\
(t\,t') &\in& ((9\,10), 11\,12),13\,14))\,. 
\label{factionMaMpart21}
\end{eqnarray}
%


%

%
\section{Handedness in Grassmann and Clifford space}
\label{handednessGrassCliff}

The handedness $\Gamma^{(d)}$ is one of the invariants of the group $SO(d)$, 
with the infinitesimal generators of the Lorentz group $S^{ab}$,
defined as 
\begin{eqnarray}
\label{handedness}
\Gamma^{(d)}&=&\alpha\, \varepsilon_{a_1 a_2\dots a_{d-1} a_d}\, S^{a_1 a_2} 
\cdot S^{a_3 a_4} \cdots S^{a_{d-1} a_d}\,,
\end{eqnarray}
with $\alpha$, which is chosen so that $\Gamma^{(d)}=\pm 1$.

In the Grassmann case  $S^{ab}$  is defined in Eq.~(\ref{thetasab}), while in the Clifford case
Eq.~(\ref{handedness}) simplifies, if we take into account that $S^{ab}|_{a\ne b}= 
\frac{i}{2}\gamma^a \gamma^b$  and $\tilde{S}^{ab}|_{a\ne b}= 
\frac{i}{2}\tilde{\gamma}^a \tilde{\gamma}^b$, as follows
\begin{eqnarray}
\Gamma^{(d)} :&=&(i)^{d/2}\; \;\;\;\;\;\prod_a \quad (\sqrt{\eta^{aa}} \gamma^a), 
\quad {\rm if } \quad d = 2n\,. 
\nonumber\\
\label{hand}
\end{eqnarray}
%

%
\section{Discrete symmetries of vector and scalar gauge fields in $d=(3+1)$  from 
$d=((d-1) +1)$~\cite{nhds}}
\label{DSVS}


%
We follow here Ref.~\cite{nhds}.
In Sect.~\ref{CPT} our definition~\cite{nhds} of the discrete symmetry operators 
$\mathbb{C}_{{\cal N}}$, ${\cal P}^{(d-1)}_{{\cal N}} $ and $ \cal{T}_{{\cal N}}$,  
Eq.~(\ref{CPTNlowE}), is presented and their application on the second quantized states of 
fermions and antifermions, living in $d=((d-1)+1)$ spaces and manifesting in $d=(3+1)$ 
as the observed fermions and antifermions, is discussed. 

In this App.~\ref{DSVS} we apply  the discrete symmetry operators 
 $\mathbb{C}_{{\cal N}}$, ${\cal P}^{(d-1)}_{{\cal N}} $ and $ \cal{T}_{{\cal N}}$,  
Eq.~(\ref{CPTNlowE}), on vielbeins and spin connection fields, living in 
$d=((d-1)+1)$ spaces and manifesting  $d=(3+1)$ as the vector and scalar gauge 
fields (as it is in the Kaluza-Klein theories)~\cite{nd2017}. 
We again follow Ref.~\cite{nhds}.

Let us treat the simple starting Lagrange density for a spinor in $d=((d-1) +1)$ 
dimensional space, which carries, like in the Kaluza-Klein theories, the spins and no
 charges, Eqs.~(\ref{wholeaction}, \ref{faction}), Sect.~\ref{fermionandgravitySCFT} 
 %
%
\begin{eqnarray}
\label{lagrangespinsfamilies0}
{\cal L_f} &=&\frac{1}{2} \,E \, \Psi^{\dagger} \, \gamma^0 \, \gamma^a\, p_{0a}\,\Psi + \, h.c.\,,\nonumber\\
p_{0a}   &=& f^{\alpha}_{a} p_{\alpha} + \frac{1}{2E} \{p_{\alpha},f^{\alpha}_{a} E \}_{-}-
\frac{1}{2}\, S^{cd}\, f^{\alpha}_{a}\,\omega_{cd \alpha}\,. 
\end{eqnarray}
$f^{\alpha}_{a}$ are vielbein and $\omega_{cd \alpha}$ spin connection fields, the gauge fields of  
$p^{a}$ and $S^{ab}$, respectively. 
The families quantum numbers, determined by $\tilde{S}^{ab}$, 
commute, Eq.~(\ref{gammatildeantiher}), with in Sect.~\ref{CPT} defined 
discrete symmetries operators~\footnote{
The discrete symmetry operators for the "Grassmann fermions" are presented in Refs.~%
\cite{prd2018,n2019PIPII,2020PartIPartII}. }.
%
%
%
%
%

Let the vielbeins and spin connections be responsible for the break of symmetry of 
$M^{(d-1)+1}$ into $M^{3+1}\times$ $M^{d-4}$ so that the manifold $M^{d-4}$
 is (almost) compactified~\cite{NHD,ND012,familiesNDproc} and let the spinors (fermions)
manifest in $d=(3+1)$ the ordinary spin and the charges and the families
~\footnote{In Refs.~\cite{NHD,ND012,familiesNDproc} it is demonstrated on the
 toy model how such an almost compactification could occur.}.
Looking for the subgroups (denoted by $A$)  of the  $SO((d-1)+1)$ 
group and assuming no gravity in $d=(3+1)$, the Lagrange density of 
Eq.~(\ref{lagrangespinsfamilies0}) can be rewritten in a more familiar shape 
\begin{eqnarray}
\label{lagrangespinsfamilies1}
{\cal L} &=& \frac{1}{2} \,E \, \psi^{\dagger} \, \gamma^0 \, (\gamma^m \, 
p_{0m} + \gamma^s \, p_{0s})\,\,\psi \,+ h.c. \,,\nonumber\\
p_{0m}   &=&  p_{m} - \sum_{A}\,\vec{\tau}^{A}\,\vec{A}^{A}_{m}\,,\nonumber\\
p_{0s}   &=&  f^{\sigma}_{s} \,p_{\sigma} + \frac{1}{2E} \{p_{\sigma},
f^{\sigma}_{s} E \}_{-} - \sum_{B}\, \vec{\tau}^{B}\,\vec{A}^{B}_{s}\,,
\end{eqnarray}
with $m=(0,1,2,3)$,  $s=(5,6,\dots,d)$ and 
$\tau^{Ai}=        \sum_{st}\, c^{Ai}{}_{st}\, S^{st}\,$, 
$\sum_{A}\,\vec{\tau}^{A}\,\vec{A}^{A}_{m}= \frac{1}{2}\, \sum_{s t}\,S^{st}\, 
\omega_{st m}\,$,
$\sum_{A}\,\vec{\tau}^{A}\,\vec{A}^{A}_{s}= \frac{1}{2}\,\sum_{st}\, S^{st}\,
f^{\sigma}_{s} \omega_{st \sigma}\,$. 
%

Taking into account Eqs.~(\ref{so42}, \ref{so64}) (and recognizing that all $\tau^{Ai}
=\sum_{A,i} c^{Ai}{}_{st} \,S^{st}$ include only $S^{s t}= \frac{i}{2} \gamma^s 
\gamma^t $ with either $\gamma^s$ or $\gamma^t$ imaginary and the remaining 
one is real, Eq.~(\ref{complexgamatilde})) one easily finds  that
\begin{eqnarray}
\label{CNtransempt}
\mathbb{C}_{\cal N}\,\,  \tau^{Ai}\,\, \mathbb{C}_{\cal N}^{-1} &=& - \tau^{Ai}\,,
 \nonumber\\
\mathbb{C}_{\cal N}{\cal P}^{(d-1)}_{\cal N}\,\,  \tau^{Ai}\,\, (\mathbb{C}_{\cal N}
{\cal P}^{(d-1)}_{\cal N})^{-1} &=& - \tau^{Ai}\,, \nonumber\\
\mathbb{C}_{\cal N} {\cal P}^{(d-1)}_{\cal N} {\cal T}_{\cal N}\,\, \tau^{Bi}
\,(\mathbb{C}_{\cal N} {\cal P}^{(d-1)}_{\cal N} {\cal T}_{\cal N})^{-1}  
&=& -\tau^{Bi}\,.
\end{eqnarray}
Taking into account  how do the vector gauge  fields transform according to prescription
in Sect.~\ref{CPT}, Eq.~(\ref{CPTNlowE}), one further finds
\begin{eqnarray}
\label{CNtransempt1}
\mathbb{C}_{\cal N}\, \, A^{Ai}_{m}(x^0,\vec{x}_3)\,\, \mathbb{C}_{\cal N}^{-1} 
&=& -A^{Ai}_{m}(x^0,\vec{x}_3)\,, \nonumber\\
\mathbb{C}_{\cal N}{\cal P}^{(d-1)}_{\cal N}\,\,  A^{Ai}_{m}(x^0,\vec{x}_3) \,\,
(\mathbb{C}_{\cal N}{\cal P}^{(d-1)}_{\cal N})^{-1} &=& 
- A^{Ai\,m}(x^0,-\vec{x}_3)\,,\nonumber\\
\mathbb{C}_{\cal N} {\cal P}^{(d-1)}_{\cal N} {\cal T}_{\cal N}\,\, \tau^{Ai}\,
A^{Ai}_{m} (x)\,\,
(\mathbb{C}_{\cal N} {\cal P}^{(d-1)}_{\cal N} {\cal T}_{\cal N})^{-1}  
&=& (-\tau^{Ai})\,(-A^{Ai*}_{m}(-x))\,,
\end{eqnarray}
for $\tau^{Ai}$  from the Cartan subalgebra for each $A$. It  is always true that 
$\tau^{Ai}\,A^{Ai}_{m}$ 
 transforms either to $(-\tau^{Ai})\,(-A^{Ai}_{m})$  or to $\tau^{Ai}\,A^{Ai}_{m}$,  
for each $Ai$, 
all in agreement with the standard knowledge for the gauge vector fields and charges 
in $d=(3+1)$~\cite{Itzykson}.

One  can  check that
\begin{eqnarray}
\label{fieldsproperties}
\mathbb{C}_{\cal N} {\cal P}^{(d-1)}_{\cal N} {\cal T}_{\cal N}\, \gamma^a
(\mathbb{C}_{\cal N} {\cal P}^{(d-1)}_{\cal N} {\cal T}_{\cal N})^{-1}&=& 
\gamma^a \,,\nonumber\\ 
\mathbb{C}_{\cal N} {\cal P}^{(d-1)}_{\cal N} {\cal T}_{\cal N}  S^{ab}\,
(\mathbb{C}_{\cal N} {\cal P}^{(d-1)}_{\cal N} {\cal T}_{\cal N})^{-1}&=&- S^{ab}\,,
\nonumber\\
\mathbb{C}_{\cal N} {\cal P}^{(d-1)}_{\cal N} {\cal T}_{\cal N} \,
f^{\alpha}_{a}(x)\,p_{\alpha}\,
(\mathbb{C}_{\cal N} {\cal P}_{\cal N} {\cal T}_{\cal N})^{-1} &=& 
f^{\alpha\,*}_{a}(-x)\,p_{\alpha}\,,\nonumber\\
\mathbb{C}_{\cal N} {\cal P}^{(d-1)}_{\cal N} {\cal T}_{\cal N} \omega_{abc}(x)\,
(\mathbb{C}_{\cal N} {\cal P}^{(d-1)}_{\cal N} {\cal T}_{\cal N})^{-1}  
&=& - \omega^{*}_{abc}(-x) \,,\nonumber\\
\mathbb{C}_{\cal N} {\cal P}_{\cal N} {\cal T}_{\cal N}\, \tilde{\omega}_{abc} (x)\,
(\mathbb{C}_{\cal N} {\cal P}_{\cal N} {\cal T}_{\cal N})^{-1}&=&
 - \tilde{\omega}^{*}_{abc} (-x) \,.
\end{eqnarray}
%
%

From Eqs.~(\ref{CNtransempt}, \ref{CNtransempt1}, \ref{fieldsproperties}) it follows
 that 
$$\mathbb{C}_{\cal N} {\cal P}^{(d-1)}_{\cal N} {\cal T}_{\cal N}\,\tau^{Ai}\,
A^{Ai}_{s} (x) (\mathbb{C}_{\cal N} {\cal P}^{(d-1)}_{\cal N} {\cal T}_{\cal N})^{-1}  
= (-\tau^{Ai})\,(-A^{Ai*}_{s})(-x))\,,$$
%
concerning in $d=(3+1)$ the gauge scalar fields,  which determine in 
the {\it spin-charge-family} theory massless and massive solutions for fermions 
(quarks and leptons and antiquarks and antileptons) and, when some
of them gaining  constant values --- the one with the space
index $s=(7,8)$ --- contribute not only  to masses of 
spinors but also to those gauge fields, to which they couple.

The Hermiticity requirement for the Lagrange density (Eq.~(\ref{lagrangespinsfamilies1})),
${\cal L}^{\dagger}= {\cal L}$, leads to
\begin{eqnarray}
\label{hcreq}
\omega^{*}_{abc}(x)&=& (\mp)\,\omega_{abc}(x)\,;(-)\, {\rm if}\, a=c \,{\rm or}\, b=c\,, 
(+)\, {\rm otherwise}\,,
\end{eqnarray}
which is to be taken into account together with the $\mathbb{C}_{\cal N} {\cal P}_{\cal N} {\cal T}_{\cal N}$ 
invariance.

For additional arguments about the generality of our proposal for the discrete symmetry 
operators $\mathbb{C}_{\cal N}$, ${\cal P}^{(d-1)}_{\cal N}$ and ${\cal T}_{\cal N}$
the reader is invited to follow Ref.~\cite{nhds}.

\section{Symmetries of mass matrices on  tree level and beyond
 manifesting  $SU(2)\times SU(2) \times U(1)$ symmetry in  {\it 
spin-charge-family} theory~\cite{gn2013,gn2014}}
\label{M0SCFT}
%

Let states of massless quarks and leptons for two times four families --- 
presented in Table~\ref{Table III.} with creation operators, creating the 
internal part of the right handed  $u$-quark and right handed 
$\nu$-leptons if applying on the vacuum state $|\psi_{oc}>$ ---
be denoted by $\psi^{\alpha}_{f}$, $\alpha=(u,d,\nu, e)_{L,R}$ and 
$f$ is the family quantum number. 
Table~\ref{Table so13+1.} represents all the members of the first family 
of Table~\ref{Table III.}. All the members of any of two times four families 
follows by the application of $S^{ab}$.

The scalar gauge fields ($\vec{\tilde{\tau}}^{1}\,\cdot 
\vec{\tilde{A}}^{\tilde{1}}_{\scriptscriptstyle{\stackrel{78}{(\pm)}}}$,
$ \vec{\tilde{N}}_{L}\, \cdot 
\vec{\tilde{A}}^{\tilde{N}_{\tilde{L}}}_{\scriptscriptstyle{\stackrel{78}{(\pm)}}}$)
with the space index $s=(7,8)$ transform the right handed members into the 
left handed ones, causing as well transformations among families
of the first group. 

The scalar gauge fields ($\vec{\tilde{\tau}}^{2}\,\cdot 
\vec{\tilde{A}}^{\tilde{2}}_{\scriptscriptstyle{\stackrel{78}{(\pm)}}}$,
$ \vec{\tilde{N}}_{R}\,\cdot 
\vec{\tilde{A}}^{\tilde{N}_{\tilde{R}}}_{\scriptscriptstyle{\stackrel{78}{(\pm)}}}$)
behave equivalently on the second group of the four families.

The scalar gauge fields ($Q \cdot A^{Q}_{\scriptscriptstyle{\stackrel{78}{(\pm)}}},
Y \cdot A^{Q]Y}_{\scriptscriptstyle{\stackrel{78}{(\pm)}}},
\tau^4 \cdot A^{4}_{\scriptscriptstyle{\stackrel{78}{(\pm)}}}$) "see" only the 
family members quantum numbers, independently of the family quantum number.
They are diagonal and apply on both groups of four families equivalently.

Although the two groups of scalar fields might have different properties, different
different masses and different coupling constants, they bring to the two groups 
of four families the equivalent 
$SU(2)\times SU(2) \times U(1)$ symmetry~%
\footnote{The infinitesimal generators $\tilde{N}^{i}_{L}\,,\,i=(1,2,3)$ and 
$\tilde{N}^{i}_{R}\,,\,i=(1,2,3)$  determine the algebra 
of the two invariant subgroups of the $\widetilde{SO}(1,3)$ group, while 
$\tilde{\tau}^{1i}\,,\,i=(1,2,3)$ and $\tilde{\tau}^{2i}\,,\,i=(1,2,3)$ determine 
the two invariant subgroups of the $\widetilde{SO}(4)$ group. The first four 
families of Table~\ref{Table III.}, carrying the quantum numbers of  
$\tilde{N}^{i}_{L}$ and $ \tilde{\tau}^{1i}$, are influenced by the 
corresponding scalar gauge fields, the second  four families are influenced by 
the scalar fields of $\tilde{N}^{i}_{R}$ and $ \tilde{\tau}^{2i}$.}

Let us therefore treat only one group of four families, the first one, interacting with
($\vec{\tilde{\tau}}^{1}\,\cdot 
\vec{\tilde{A}}^{\tilde{1}}_{\scriptscriptstyle{\stackrel{78}{(\pm)}}}$,
$ \vec{\tilde{N}}_{L}\, \cdot 
\vec{\tilde{A}}^{\tilde{N}_{\tilde{L}}}_{\scriptscriptstyle{\stackrel{78}{(\pm)}}}$). 
The operators  
\begin{eqnarray}
\label{taunl}
\tilde{N}^{i}_{L}\,,\,i=(1,2,3)\,, && {\tau}^{1i}\,,\,i=(1,2,3)\,,
\nonumber\\ 
\{\tilde{N}^{i}_{L}, \tilde{N}^{j}_{L}\}_{-}= i\,\varepsilon^{ijk} \tilde{N}^{k}_{L}
\,,\;\;
\{\tilde{\tau}^{1i},\tilde{\tau}^{1j}\}_{-}&=& i\,\varepsilon^{ijk} \tilde{\tau}^{1k}
\,,\{\tilde{N}^{i}_{L}, \tilde{\tau}^{1j}\}_{-}=0\,,
\end{eqnarray}
where $\varepsilon^{ijk}$ is the totally antisymmetric tensor, transform the basic 
vectors $\psi^{\alpha}_{i}$, into one another  as follows
\begin{eqnarray}
\label{taunlonpsi}
&&\tilde{N}^{3}_{L}\, (\psi^{\alpha}_1, \psi^{\alpha}_2,\psi^{\alpha}_3,
\psi^{\alpha}_4)=
 \frac{1}{2}\, (-\psi^{\alpha}_1, \psi^{\alpha}_2,-\psi^{\alpha}_3,
\psi^{\alpha}_4)\,,\nonumber\\
&&\tilde{N}^{+}_{L}\, (\psi^{\alpha}_1, \psi^{\alpha}_2,\psi^{\alpha}_3,
\psi^{\alpha}_4)= 
 (\psi^{\alpha}_2, \;\;0,\psi^{\alpha}_4,\;\;0)\,,\nonumber\\
&&\tilde{N}^{-}_{L}\, (\psi^{\alpha}_1, \psi^{\alpha}_2,\psi^{\alpha}_3,
\psi^{\alpha}_4)= 
 (0\;\;, \psi^{\alpha}_1,\;\;0,\psi^{\alpha}_3)\,,\nonumber\\
&&\tilde{\tau}^{13}\, (\psi^{\alpha}_1, \psi^{\alpha}_2,\psi^{\alpha}_3,
\psi^{\alpha}_4)=
 \frac{1}{2}\, (-\psi^{\alpha}_1, -\psi^{\alpha}_2,\psi^{\alpha}_3,
\psi^{\alpha}_4)\,, \nonumber\\
&&\tilde{\tau}^{1+}\, (\psi^{\alpha}_1, \psi^{\alpha}_2,\psi^{\alpha}_3,
\psi^{\alpha}_4)= 
(\psi^{\alpha}_3, \psi^{\alpha}_4,\;\;0,\;\;0)\,,\nonumber\\ 
&&\tilde{\tau}^{1-}\, (\psi^{\alpha}_1, \psi^{\alpha}_2,\psi^{\alpha}_3,
\psi^{\alpha}_4)=
 (\;\;0, \;\;0,\psi^{\alpha}_1,\psi^{\alpha}_2) \,.
\end{eqnarray}
The three $U(1)$ operators ($Q,Y$ and $\tau^4$) (or any three superposition of
 them, like $Q, Q', Y'$) commute with the family operators 
$\vec{\tilde{N}}_{L}$ 
and $\vec{\tilde{\tau}}^{1}$, distinguishing only among family 
members $\alpha$.


%
\begin{eqnarray}
\label{QQ'Y'nltau}
&&\{ \tilde{N}^{i}_{L}\,, (Q,Q',Y') \}_{-} =(0,0,0)\,,\nonumber\\
&&\{ \tilde{\tau}^{1i}\,, (Q,Q',Y') \}_{-} =(0,0,0)\,,\nonumber\\
&&(Q\,, Y\,, \tau^4)\, (\psi^{\alpha}_1, \psi^{\alpha}_2,\psi^{\alpha}_3,
\psi^{\alpha}_4)= 
(Q^{\alpha}\,, Y^{\alpha}\,, \tau^{4 \alpha} )\,
 (\psi^{\alpha}_1, \psi^{\alpha}_2,\psi^{\alpha}_3,\psi^{\alpha}_4) \,,
\end{eqnarray}
giving the same eigenvalues for all the families.

The nonzero vacuum expectation values of the  scalar gauge fields of
 $\tilde{N}^{i}_{L}$  ($\tilde{A}^{\tilde{N}_{L}i}_{s}=$ $ 
\sum_{\tilde{m}\tilde{n}}\tilde{c}^{\tilde{N}_{L}i}{}_{\tilde{m}\tilde{n}}\, 
\tilde{\omega}^{\tilde{m}\tilde{n}}{}_{s}$, $(\tilde{m}, \,\tilde{n}) =(0,1,2,3)$), 
 of  $\tilde{\tau}^{1i}$  ($\tilde{A}^{\tilde{1}i}_{s}= \sum_{\tilde{t}\tilde{t'}}
\tilde{c}^{\tilde{1}i}{}_{\tilde{t} \tilde{t'}}\, \tilde{\omega}^{\tilde{t}\tilde{t'}}{}_{s}
$,  $(\tilde{t},\tilde{t'}) =(5,6,7,8)$) and of the three singlet scalar gauge fields of  
($Q,Y$ and $\tau^4$), which all are 
superposition of $\omega_{t,t',s}$ ($A^{Q}_{s} = \sum_{t\, t'} c^{Q}{}_{t t'}\,
\omega^{t t'}{}_{s}$\,,
 $A^{Y}_{s} =  \sum_{t\, t'} c^{Y}{}_{t t'}\,\omega^{t t'}{}_{s}$ and 
$A^{4}_{s} =  \sum_{t\, t'} c^{Y'}{}_{t t'}\,
\omega^{t t'}{}_{s}$, $s=(7,8)$, $(t,t') =(5,6,7,8)$), determine on the tree level, 
together with the 
corresponding coupling  constants,  the
 $SU(2) \times SU(2) \times U(1)$ symmetry and the strength of the mass matrix 
of each family member $\alpha$, Eq.~(\ref{M0}). 
In loop corrections  the scalar fields  
 $ \tilde{A}^{\tilde{N}_{L}i}_{s}$,  $\tilde{A}^{\tilde{1}i}_{s}$, 
$A^{Q}_{s}, A^{Y}_{s},A^{4}_{s}$  contribute to all the matrix elements, 
keeping the symmetry unchanged, Eq.~(\ref{M0}). The twice  two zeros on the 
tree level obtain in loop corrections the value $b$. 

One easily checks that a change of the phases of the left and the right
 handed members, there are $(2n-1)$ 
possibilities, causes  changes in phases of matrix elements in Eq.~(\ref{M0}). 
 
 All the scalars are doublets with respect to the weak charge,  
contributing to the weak and the hyper charge of the fermions so that they
 transform the right handed members into 
the left handed onces, Sect.~\ref{scalar3+1}, 
what is in the {\it standard model} just
required.


%
\subsection{Properties of non Hermitian mass matrices}
\label{nonhermitean}
%

This pedagogic presentation of well known properties of non Hermitian matrices can be 
found in many textbooks. 
We repeat this topic here only to make our discussions  transparent.

Let us take   a  non Hermitian mass matrix  $M^{\alpha}$ as it follows from the 
{\it spin-charge-family} theory, $\alpha$ denotes a family member. 

We always can diagonalize  a non Hermitian $M^{\alpha}$ with two unitary matrices, 
$S^{\alpha}$ ($S^{\alpha\, \dagger}\,S^{\alpha}=I$)
and $T^{\alpha}$ ($T^{\alpha\, \dagger}\,T^{\alpha}=I$)
\begin{eqnarray}
\label{diagnonher0}
S^{\alpha\, \dagger}\,M^{\alpha}\,T^{\alpha}&=& {\bf M}^{\alpha}_{d}\,=
(m^{\alpha}_{1}\, \dots m^{\alpha}_{i}\, \dots m^{\alpha}_{n}).
\end{eqnarray}
The proof is added below. 

Changing phases of the basic states, those of the left handed one and those 
of the right handed one, the new unitary matrices  $S'^{\alpha} = 
S^{\alpha} \,F_{\alpha S} $ and $T'^{\alpha} = T^{\alpha}\,F_{\alpha T}$
change the phase of the elements of diagonalized mass matrices
 ${\bf M}^{\alpha}_{d}$ 
\begin{eqnarray}
\label{diagnonher}
S'^{\alpha\, \dagger}\,M^{\alpha}\,T'^{\alpha}&=& F^{\dagger}_{\alpha S}\,{\bf M}^{\alpha}_{d}\,
F_{\alpha T}=\nonumber\\
& & diag(m^{\alpha}_{1} e^{i(\phi^{\alpha S}_{1}- \phi^{\alpha T}_{1})}\, \dots 
m^{\alpha}_{i}\, e^{i(\phi^{\alpha S}_{i}- \phi^{\alpha T}_{i})}\,, \dots m^{\alpha}_{n}\,
e^{i(\phi^{\alpha S}_{n}- \phi^{\alpha T}_{n})})\,,\nonumber\\
F_{\alpha S} &=& diag(e^{-i \phi^{\alpha S}_{1}},\,\dots\,,e^{-i \phi^{\alpha S}_{i}}\,,\dots\,,
e^{-i \phi^{\alpha S}_{n}})\,,\nonumber\\
F_{\alpha T} &=& diag(e^{-i \phi^{\alpha T}_{1}},\,\dots\,,
e^{-i \phi^{\alpha T}_{i}}\,,\dots\,, e^{-i \phi^{\alpha T}_{n}})\,.
\end{eqnarray}

In the case that the mass matrix is Hermitian $T^{\alpha}$ can be replaced by
 $ S^{\alpha}$, but only up to phases originating in the phases of the two basis,
 the left handed one and the right handed one, since they remain independent. 

 The non Hermitian mass matrices can be diagonalized 
in two ways, that is either one diagonalizes  
$M^{\alpha }M^{\alpha \,\dagger}$ or $M^{\alpha \dagger} M^{\alpha }$
\begin{eqnarray}
\label{diagMM}
(S^{\alpha \dagger} M^{\alpha} T^{\alpha})          (S^{\alpha \dagger} M^{\alpha } T^{\alpha})^{\dagger}&=&
S^{\alpha \dagger} M^{\alpha } M^{\alpha \,\dagger} S^{\alpha} = {\bf M}^{\alpha  2}_{d S}\,, \nonumber\\  
(S^{\alpha \dagger} M^{\alpha} T^{\alpha})^{\dagger}(S^{\alpha \dagger} M^{\alpha } T^{\alpha})&=&
T^{\alpha \dagger} M^{\alpha\, \dagger} M^{\alpha } T^{\alpha} = {\bf M}^{\alpha  2}_{d T}\,, \nonumber\\
{\bf M}^{\alpha\, \dagger }_{d S}&=& {\bf M}^{\alpha }_{d S}\, , 
\quad {\bf M}^{\alpha\, \dagger }_{d T}= {\bf M}^{\alpha}_{d T}\,.
\end{eqnarray}
The proof that ${\bf M}^{\alpha }_{d S}={\bf M}^{\alpha }_{d T}$ proceeds as 
follows.
Let us define two Hermitian ($H^{\alpha}_{S}\,$, 
$H^{\alpha}_{T}$) and two unitary matrices ($U^{\alpha}_{S}\,$, $H^{\alpha}_{T}$) 
\begin{eqnarray}
\label{proof1}
H^{\alpha}_{S} &=& S^{\alpha} {\bf M}^{\alpha }_{d S} S^{\alpha \, \dagger}\,, \quad \quad
H^{\alpha}_{T}  =  T^{\alpha} {\bf M}^{\alpha \dagger }_{d T} T^{\alpha \, \dagger}\,,\nonumber\\
U^{\alpha}_{S} &=& H^{\alpha -1}_{S} M^{\alpha } \,, \quad \quad U^{\alpha}_{T} = H^{\alpha -1}_{T} 
M^{\alpha \,\dagger } \,,
\end{eqnarray}
It is easy to show  that $H^{\alpha\, \dagger}_{S}= H^{\alpha}_{S}$, 
$H^{\alpha\, \dagger}_{T}= H^{\alpha}_{T}$, $U^{\alpha}_{S} \,U^{\alpha\, \dagger}_{S}=I$ and 
$U^{\alpha}_{T} \,U^{\alpha\, \dagger}_{T}=I$.
Then it follows  
\begin{eqnarray}
\label{proof2}
S^{\alpha \dagger}\, H^{\alpha}_{S} \,S^{\alpha} &=& {\bf M}^{\alpha }_{d S}= {\bf M}^{\alpha \,\dagger}_{d S}=
S^{\alpha \dagger}\,M^{\alpha }\,U^{\alpha \,-1}_{S} \,S^{\alpha}= S^{\alpha \dagger}\,M^{\alpha }\, T^{\alpha} \,,
\nonumber\\
T^{\alpha \dagger}\, H^{\alpha}_{T} \,T^{\alpha} &=& {\bf M}^{\alpha }_{d T}= {\bf M}^{\alpha \,\dagger}_{d T}=
T^{\alpha \dagger}\,M^{\alpha \, \dagger}\,U^{\alpha \,-1}_{T} \,T^{\alpha}= T^{\alpha \dagger}\,
M^{\alpha \dagger}\, S^{\alpha} \,,
\end{eqnarray}
where we recognized $U^{\alpha \,-1}_{S} \,S^{\alpha}= T^{\alpha}$ and $U^{\alpha \,-1}_{T} \,T^{\alpha}=S^{\alpha}$. 
Taking into account Eq.~(\ref{diagnonher}) the starting basis can be chosen so, that all diagonal 
masses are real and positive.

\section{Statements and proofs of statements~\cite{IARD2020}}
\label{proofs}
%

There are two kinds of the Clifford algebra objects, $\gamma^a$'s and 
$\tilde{\gamma}^a$'s, both expressible with the Grassmann algebra objects,
$\theta^a$'s and their derivatives $\frac{\partial}{\partial \theta_a}$'s.
In Grassmann $d$-dimensional space there are $d$ anticommuting operators $\theta^{a}$, 
$\{\theta^{a}, \theta^{b}\}_{+}=0$, $a=(0,1,2,3,5,..,d)$, and $d$ anticommuting derivatives 
with respect to $\theta^{a}$, $\frac{\partial}{\partial \theta_{a}}$, 
$\{\frac{\partial}{\partial \theta_{a}}, \frac{\partial}{\partial \theta_{b}}\}_{+} =0$, offering
together $2\cdot2^d$ operators, the half of which are superposition of products of  
$\theta^{a}$ and another half corresponding superposition of 
$\frac{\partial}{\partial \theta_{a}}$.
\begin{eqnarray}
\label{thetaderanti}
\{\theta^{a}, \theta^{b}\}_{+}=0\,, \, && \,
\{\frac{\partial}{\partial \theta_{a}}, \frac{\partial}{\partial \theta_{b}}\}_{+} =0\,,
\nonumber\\
\{\theta_{a},\frac{\partial}{\partial \theta_{b}}\}_{+} &=&\delta_{ab}\,, (a,b)=(0,1,2,3,5,\cdots,d)\,.
\end{eqnarray}
Defining~\cite{nh2018} 
\begin{eqnarray}
(\theta^{a})^{\dagger} &=& \eta^{a a} \frac{\partial}{\partial \theta_{a}}\,,\quad
{\rm it \, follows}\,,\quad
(\frac{\partial}{\partial \theta_{a}})^{\dagger}= \eta^{a a} \theta^{a}\,. 
\label{thetaderher}
\end{eqnarray}
The identity is the self adjoint member.
The signature $\eta^{a b}=diag\{1,-1,-1,\cdots,-1\}$ is assumed. 

One can define  
new operators, expressed with $\theta^{a}$'s  and 
$\frac{\partial}{\partial \theta_{a}}$'s 
\begin{eqnarray}
\label{cliffthetaapp}
\gamma^{a} &=& (\theta^{a} + \frac{\partial}{\partial \theta_{a}})\,, \quad 
\tilde{\gamma}^{a} =i \,(\theta^{a} - \frac{\partial}{\partial \theta_{a}})\,,\nonumber\\
\theta^{a} &=&\frac{1}{2} \,(\gamma^{a} - i \tilde{\gamma}^{a})\,, \quad 
\frac{\partial}{\partial \theta_{a}}= \frac{1}{2} \,(\gamma^{a} + i \tilde{\gamma}^{a})\,.
\end{eqnarray}
One  can make $2^d$  products of superpositions of $\gamma^a$'s and  $2^{d}$  products of superposition of $\tilde{\gamma}^{a}$'s, all together  $2\cdot 2^{d}$
objects.

{\bf  Statement 1.} $\gamma^a$'s  and $\tilde{\gamma}^{a}$'s define two independent Clfford algebras.

To prove this statement one only needs to take into account Eqs.~(\ref{thetaderher}, \ref{clifftheta}).
\begin{eqnarray}
\label{gammatildeantiherapp}
\{\gamma^{a}, \gamma^{b}\}_{+}&=&2 \eta^{a b}= \{\tilde{\gamma}^{a}, 
\tilde{\gamma}^{b}\}_{+}\,, \nonumber\\
\{\gamma^{a}, \tilde{\gamma}^{b}\}_{+}&=&0\,,\quad
 (a,b)=(0,1,2,3,5,\cdots,d)\,, \nonumber\\
(\gamma^{a})^{\dagger} &=& \eta^{aa}\, \gamma^{a}\, , \quad 
(\tilde{\gamma}^{a})^{\dagger} =  \eta^{a a}\, \tilde{\gamma}^{a}\,.
\end{eqnarray}
Eq.~(\ref{gammatildeantiherapp}) demonstrates that these two Clifford algebra 
objects obviously define two independent ''basis vectors''. Either $\gamma^a$'s or 
$\tilde{\gamma}^a$'s are, up to a sign, self adjoint operators.

The generators of the Lorentz transformations in the Grassmann algebra space are 
defined as follows
\begin{eqnarray}
{\cal {\bf S}}^{a b} &=& i \, (\theta^{a} \frac{\partial}{\partial \theta_{b}} - \theta^{b}
\frac{\partial}{\partial \theta_{a}})\,, \quad  ({\cal {\bf S}}^{a b})^{\dagger} = \eta^{a a}
\eta^{b b} {\cal {\bf S}}^{a b}\, . 
\label{thetasabapp}
\end{eqnarray} 

{\bf Statement 2:} 
The sum of the generators of the Lorentz transformations in 
each of the two Clifford algebra spaces, 
$S^{ab}=i \frac{1}{4} (\gamma^{a} \gamma^{b}- 
\gamma^{b}\gamma^{a})$ and 
$\tilde{S}^{ab}=i \frac{1}{4} 
(\tilde{\gamma}^{a}\tilde{\gamma}^{b}- \tilde{\gamma}^{b}\tilde{\gamma}^{a})$,
respectively, are equal to the generators of the Lorentz transformation in the 
Grassmann algebra space $\bf{\cal S}^{a b}=S^{ab} + \tilde{S}^{ab}$.

To prove this statement one only has to express in the sum $S^{ab} + \tilde{S}^{ab}$ first $S^{ab}$ with $\gamma^a$ and $\gamma^b$ and $\tilde{S}^{ab}$ with
$\tilde{\gamma}^a$ and $\tilde{\gamma}^b$, and then $\gamma^{a}$ and 
$\tilde{\gamma}^{a}$ with  $\theta^{a}$ and $\frac{\partial}{\partial \theta_{a}}$, 
using Eq.~(\ref{clifftheta}).

 One obtains  that 
 $S^{ab}=i \frac{1}{2}(\frac{\partial}{\partial \theta_{a}}\theta^b +
 \theta^a \frac{\partial}{\partial \theta_{b}})$ and $\tilde{S}^{ab}=i \frac{1}{2}(
 \theta^a \frac{\partial}{\partial \theta_{b}} +
 \frac{\partial}{\partial \theta_{a}}\theta^b)$, leading to $S^{ab} + \tilde{S}^{ab}=
{\cal{\bf S}}^{ab}$.
 
 We conclude
\begin{eqnarray}
 S^{ab} &=&\frac{i}{4}(\gamma^a \gamma^b - \gamma^b \gamma^a)\,,
\quad  
\tilde{S}^{ab} =\frac{i}{4}(\tilde{\gamma}^a \tilde{\gamma}^b - \tilde{\gamma}^b
 \tilde{\gamma}^a)\, , \nonumber\\
 {\cal {\bf S}}^{ab} &=&S^{ab} + \tilde{S}^{ab}\,, \quad
\{S^{ab}, \tilde{S}^{ab}\}_{-}=0\,,\nonumber\\
\{ S^{ab}, \gamma^c\}_{-}&=& i (\eta^{bc} \gamma^a - \eta^{ac} \gamma^b)\,, \nonumber\\
\{ \tilde{S}^{ab}, \tilde{\gamma}^c \}_{-}&=& i (\eta^{bc} \tilde{\gamma}^a - 
\eta^{ac }\tilde{\gamma}^b)\,,\nonumber\\
\{ S^{ab}, \tilde{\gamma}^c\}_{-}&=&0\,,\quad \{\tilde{S}^{ab}, \gamma^c\}_{-}=0\,.
\label{sabtildesabapp}
\end{eqnarray}

{\bf Statement 2a:} 
The eigenvectors of the operators ${\cal {\bf S}}^{ab}=
 i \, (\theta^{a} \frac{\partial}{\partial \theta_{b}} - \theta^{b}
\frac{\partial}{\partial \theta_{a}})\,$ can be  written as follows
\begin{eqnarray}
{\cal {\bf S}}^{ab} \,\frac{1}{\sqrt{2}}\, (\theta^a + \frac{\eta^{aa}}{i k} \theta^b) &=&
k\,\frac{1}{\sqrt{2}} (\theta^a + \frac{\eta^{aa}}{ik} \theta^b) \,, \nonumber\\
{\cal {\bf S}}^{ab} \,\frac{1}{\sqrt{2}}\, (1+ \frac{i}{k}  \theta^a \theta^b) &=&0\,,
\label{eigengrasscartanapp}
\end{eqnarray}
while the corresponding eigenvectors of $S^{ab}$ and  $\tilde{S}^{ab}$ in each 
of the two spaces are
\begin{eqnarray}
S^{ab} \frac{1}{2} (\gamma^a + \frac{\eta^{aa}}{ik} \gamma^b) &=& \frac{k}{2}  \,
\frac{1}{2} (\gamma^a + \frac{\eta^{aa}}{ik} \gamma^b)\,,\nonumber\\
S^{ab} \frac{1}{2} (1 +  \frac{i}{k}  \gamma^a \gamma^b) &=&  \frac{k}{2}  \,
 \frac{1}{2} (1 +  \frac{i}{k}  \gamma^a \gamma^b)\,,\nonumber\\
\tilde{S}^{ab} \frac{1}{2} (\tilde{\gamma}^a + \frac{\eta^{aa}}{ik} \tilde{\gamma}^b) &=& 
\frac{k}{2}  \,\frac{1}{2} (\tilde{\gamma}^a + \frac{\eta^{aa}}{ik} \tilde{\gamma}^b)\,,
\nonumber\\
\tilde{S}^{ab} \frac{1}{2} (1 +  \frac{i}{k}  \tilde{\gamma}^a \tilde{\gamma}^b) &=& 
 \frac{k}{2}  \, \frac{1}{2} (1 +  \frac{i}{k} \tilde{\gamma}^a \tilde{\gamma}^b)\,.
\label{eigencliffcartanapp}
\end{eqnarray}
with $k^2 = \eta^{aa} \eta^{bb}$.

The Eq.~(\ref{eigengrasscartan}) can be proven by applying ${\cal {\bf S}}^{ab} $
on $\frac{1}{\sqrt{2}}\, (\theta^a + \frac{\eta^{aa}}{i k} \theta^b)$, leading to

$i \frac{1}{\sqrt{2}} (- \theta^b \eta^{aa} +
\frac{\eta^{aa}\eta^{bb}}{ik} \theta^a)= \frac{ik^2}{ik \sqrt{2}} (\theta^a-
\frac{\eta^{aa} ik}{k^2} \theta^b)$. 

The application of ${\cal {\bf S}}^{ab} $ on either
a constant or on a constant$\cdot \theta^a \theta^b$ gives $0$. In all these cases
it is assumed that $a\ne b$.

The proof of Eq.(\ref{eigencliffcartanapp}) goes similarly, again $a\ne b$ is assumed:

$\frac{i}{2}\gamma^a \gamma^b \frac{1}{2}  (\gamma^a + 
\frac{\eta^{aa}}{ik} \gamma^b)=\frac{i}{2} \frac{1}{2}  (-\eta^{aa}\gamma^b + 
\frac{\eta^{aa}\eta^{bb}}{ik} \gamma^a) = \frac{k}{2}\frac{1}{2}(\gamma^a-
\eta^{aa}\frac{i}{k} \gamma^b)$.

$\frac{i}{2}\gamma^a \gamma^b \frac{1}{2}  ( 1+\frac{i}{k}\gamma^a \gamma^b)
=\frac{i}{2} \frac{1}{2}(\gamma^a \gamma^b -\frac{i}{k}\eta^{aa} \eta^{bb})=
\frac{k}{2} \frac{1}{2}(1 +  \frac{i}{k} \gamma^a \gamma^b)$.

Replacing $S^{ab}$ with $\tilde{S}^{ab}$, and $\gamma^a$'s with 
$\tilde{\gamma}^a$'s goes through the same steps.

{\bf Statement 2b:} 
The members of any irreducible representation of $S^{ab}$ follow
from the starting one by the application of $S^{cd}$, which do not belong
to the Cartan subalgebra of the Lorentz algebra. 

The proof follows if we apply $\gamma^a$ on nilpotents and projectors, since 
$S^{ab}=\frac{i}{2}\gamma^a \gamma^b$. 

\begin{eqnarray}
&&\gamma^a \stackrel{ab}{(k)}= \eta^{aa}\stackrel{ab}{[-k]},\; \quad
\gamma^b \stackrel{ab}{(k)}= -ik \stackrel{ab}{[-k]}, \; \quad
\gamma^a \stackrel{ab}{[k]}= \stackrel{ab}{(-k)},\;\quad 
\gamma^b \stackrel{ab}{[k]}= -ik \eta^{aa} \stackrel{ab}{(-k)}\,.
\label{snmb:gammatildegammaapp}
\end{eqnarray}
Correspondingly,  any transformation on  ''basis vectors'' of the kind, which
do not change sign as is required in Eq.~(\ref{snmb:gammatildegammaapp}),
lead to another irreducible representations.

{\bf Statement 3:} 
Postulating that  $\tilde{\gamma}^{a}$'s operate on 
$\gamma^a$'s as follows~%
\cite{nh03,norma93,JMP2013,normaJMP2015,nh2018}
\begin{eqnarray}
\tilde{\gamma}^a B &=&(-)^B\, i \, B \gamma^a\,,\nonumber
\end{eqnarray}
with $(-)^B = -1$, if $B$ is (a function of) an odd product of $\gamma^a$'s,
 otherwise $(-)^B = 1$~\cite{nh03}, the reduction of the Clifford space and 
 correspondingly also the reduction of the Grassmann space follows.
 
 Eq.~(\ref{tildegammareduced}) requires
\begin{eqnarray}
\label{partialthetazeroev}
&&[\tilde{\gamma}^a (a_0 + a_{bc}\gamma^b \gamma^c + a_{bcde}\gamma^b
 \gamma^c \gamma^d \gamma^e +\cdots) =  i  (a_0 + a_{bc}\gamma^b \gamma^c + a_{bcde}\gamma^b \gamma^c \gamma^d \gamma^e +\cdots) \gamma^a ]
 |\psi_{oc}>\,, \nonumber\\
&&[\tilde{\gamma}^a (a_{b}\gamma^b + a_{bcd}\gamma^b
\gamma^c \gamma^d +\cdots) = - i  (a_{b}\gamma^b + a_{bcd}\gamma^b
 \gamma^c \gamma^d + \cdots) \gamma^a ]
|\psi_{oc}>\,, \nonumber
\end{eqnarray}

 To prove {\it Statement 3.} let us evaluate what does Eq.~(\ref{tildegammareduced})
 require when we use Eq.(\ref{clifftheta})on $|\psi_{oc}>$: $ \gamma^{a} = (\theta^{a} + \frac{\partial}{\partial \theta_{a}})$ and $\tilde{\gamma}^{a} =i \,(\theta^{a} - \frac{\partial}{\partial \theta_{a}})$, with $|\psi_{oc}>$ expressed as well with $\theta^a$'s and 
$ \frac{\partial}{\partial \theta_{a}} $. 

Let us point out that $|\psi_{oc}>$, expressed in terms of $\theta^a$'s and
$ \frac{\partial}{\partial \theta_{a}}$'s is an even function of  $\theta^a$'s, 
$|\psi_{oc}>=(1 + \theta^a\theta^b +\cdots)$, while 
$ \frac{\partial}{\partial \theta_{a}} $, applying on identity, gives zero.

The proof is needed for any even and any odd summand of $B$, appearing in  Eq.~(\ref{tildegammareduced}) and for an arbitrary $|\psi_{oc}>$.
 

Let us start with $[\tilde{\gamma}^a a_0=i a_0 \gamma^a ] |\psi_{oc}>$, 
with $|\psi_{oc}>=(1 + \theta^a\theta^b +\cdots)$ and $a_{0}$ an arbitrary 
constant. This relation requires that
$[i(\theta^a- \frac{\partial}{\partial \theta_{a}})=
 i(\theta^a +\frac{\partial}{\partial \theta_{a}})] |\psi_{oc}>$, leading to $- 2i \frac{\partial}{\partial \theta_{a}}|\psi_{oc}>
\Rightarrow 0, \; \forall \frac{\partial}{\partial \theta_{a}}$.
This last relation can only be true if $\frac{\partial}{\partial \theta_{a}} |\psi_{oc}>
\Rightarrow 0,\;\forall \frac{\partial}{\partial \theta_{a}}$.  

Evaluating $[\tilde{\gamma}^a a_{bc}\gamma^b \gamma^c=
i a_{bc}\gamma^b \gamma^c \gamma^a] |\psi_{oc}>$ we end up again
with the requirement $\frac{\partial}{\partial \theta_{a}} |\psi_{oc}>
\Rightarrow 0,\;
\forall \frac{\partial}{\partial \theta_{a}}$.
Applying  $\tilde{\gamma}^a$ on any even products of $\gamma^a$'s  we end 
up with the same requirement $\frac{\partial}{\partial \theta_{a}} |\psi_{oc}>
\Rightarrow 0,\; \forall \frac{\partial}{\partial \theta_{a}}$.

Application of $\tilde{\gamma}^a$ on any odd products of $\gamma^a$'s,
while $|\psi_{oc}>=(1 + \theta^a\theta^b +\cdots)$,
%
%
%
\begin{eqnarray}
\label{partialthetazerood}
&&[\tilde{\gamma}^a (a_{b}\gamma^b + a_{bcd}\gamma^b
\gamma^c \gamma^d +\cdots) = - i  (a_{b}\gamma^b + a_{bcd}\gamma^b
 \gamma^c \gamma^d + \cdots) \gamma^a ]
|\psi_{oc}>\,, 
\end{eqnarray}
it follows again,  after expressing $\tilde{\gamma}^a=
i(\theta^a - \frac{\partial}{\partial \theta_{a}})$ and $\gamma^a=
(\theta^a + \frac{\partial}{\partial \theta_{a}})$ into Eq.~\ref{partialthetazerood}, 
that $\frac{\partial}{\partial \theta_{a}} |\psi_{oc}>\Rightarrow 0 $ is the only solution,
leading to . 
%
\begin{eqnarray}
\theta^a \Rightarrow \gamma^a\,, 
\nonumber
\label{tildegammareduced1}
\end{eqnarray}
which does not mean that $\theta^a$ is equal to $\gamma^a$ but rather that the whole 
Grassmann algebra reduces to only one of the two Clifford algebras, the one, in which
the ''basis vectors''  are superposition of products of (odd when describing fermions) 
number of $\gamma^a$'s. It also does not mean that $\theta^a$'s are equal to 
$i\tilde{\gamma}^a$'s, since the application of $\tilde{\gamma}^a$'s depend on 
properties of $B(\gamma^a)$, on which $\tilde{\gamma}^a$'s apply.

  {\bf Statement 3a:} 
  The relations of  Eq.~(\ref{gammatildeantiher}) remain valid also after the reduction of the   Clifford space.
  
 Let us check  Eq.~(\ref{gammatildeantiher}):
  $\{ \tilde{\gamma}^{a}, \tilde{\gamma}^{b}\}_{+}= 2\eta^{ab}=$
$\tilde{\gamma}^{a} \tilde{\gamma}^{b}+\tilde{\gamma}^{b}
\tilde{\gamma}^{a}=$ $ \tilde{\gamma}^{a} i\gamma^b +\tilde{\gamma}^{b} i \gamma^a=$
 $ i \gamma^b (-i)\gamma^a + i\gamma^a(-i)\gamma^b= 2\eta^{ab} $.
 $\{ \tilde{\gamma}^{a}, \gamma^b\}_{+}= 0=$
 $\tilde{\gamma}^{a} \gamma^b+\gamma^b \tilde{\gamma}^{a}=$
 $ \gamma^b (-i)\gamma^a+  \gamma^b i \gamma^a=0$.
For a particular case one has $\{ \tilde{\gamma}^{a}, \gamma^a\}_{+}= 0=$
$ \tilde{\gamma}^{a} \gamma^a + \gamma^a \tilde{\gamma}^a= $
 $\gamma^a (-i )\gamma^a +\gamma^a i \gamma^a=0$.
 
The application of $\tilde{\gamma}^a$  obviously depends on the space on which
 it applies, Eq.~(\ref{tildegammareduced}) namely requires: 
  $[\tilde{\gamma}^a (a_0 + a_b\gamma^b + 
  a_{bc} \gamma^b \gamma^c+ \cdots ) =
 ( i a_0 \gamma^a + (-i) a_b  \gamma^b \gamma^a + 
  i a_{bc} \gamma^b \gamma^c  \gamma^a +\cdots)] |\psi_{oc>}$. 
Statement 3b is proved.



{\bf Statement 3b:} 
Taking into account in Eq.~(\ref{tildegammareduced}) required application of $\tilde{\gamma}^a$'s on the Clifford space of $\gamma^a$'s (causing the reduction of the Clifford space and at the same time as well
the reduction of the Grassmann space), 
it follows that the eigenvalues of  $\tilde{S}^{ab}$ on the eigenvectors of $S^{ab}$
agree with the eigenvalues of $S^{ab}$ on nilpotents, while  the eigenvalues of
$\tilde{S}^{ab}$ and  $S^{ab}$ on projectors, which are eigenvectors of $S^{ab}$, have opposite sign.

Let us check Eq.(\ref{signature0}). 

$\tilde{S}^{ab}\frac{1}{2}(\gamma^a + \frac{\eta^{aa}}{ik}\gamma^b)= 
\frac{i}{2} \tilde{\gamma}^a \tilde{\gamma}^b \frac{1}{2}(\gamma^a +
 \frac{\eta^{aa}}{ik}\gamma^b) = \frac{i}{2} \frac{1}{2}(\gamma^a + 
\frac{\eta^{aa}}{ik}\gamma^b) \gamma^b \gamma^a= $
$\frac{i}{2} \frac{1}{2}(-\eta^{aa} \gamma^b +\frac{\eta^{aa} \eta^{bb}}{ik}
\gamma^a)$ $=\frac{k}{2} \frac{1}{2}(\gamma^a + 
\frac{\eta^{aa}}{ik}\gamma^b)$,

$\tilde{S}^{ab}\frac{1}{2}( 1+  \frac{i}{k} \gamma^a\gamma^b)= 
 \frac{i}{2} \frac{1}{2}(1 + \frac{i}{k}\gamma^a\gamma^b)
 \gamma^b \gamma^a= $ $\frac{i}{2} \frac{1}{2}(- \gamma^a\gamma^b + 
\frac{i}{k} \eta^{aa} \eta^{bb}) =- \frac{k}{2}\frac{1}{2} ( 1+  \frac{i}{k}
 \gamma^a\gamma^b)$,
 where it is taken into account that $k^2=\eta^{aa} \eta^{bb}$.\\
 This proves  {\it Statement 3b}.


{\bf Statement 4:} 
The algebraic product of 
$\hat{b}^m_{f}{}{*_{A}}\hat{b}^{m \dagger}_{f}$ is  the same for all
$m$ of a particular irreducible representation $f$.

To prove this we take into account that $\hat{b}^{m \dagger}_{f}$  follows 
from $\hat{b}^{m -1\dagger}_{f}$ by the application of a particular  $2i S^{eg}$. 
Then $\hat{b}^{m}_{f}{}{*_{A}}\hat{b}^{m \dagger}_{f}  =$
$\hat{b}^{m-1}_{f} (2\,S^{eg})^{\dagger}{}{*_{A}} (2\,S^{eg})
\hat{b}^{m-1 \dagger}_{f}=$
$\hat{b}^{m-1}_{f}{}{*_{A}}\hat{b}^{m-1 \dagger}_{f}$, due to the relation
$(-2i S^{ab})^{\dagger} (-2i S^{ab})=1$). Repeating this procedure for each $m$
proves the statement. 

{\bf Statement 5:}
 Each irreducible representation has its own algebraic product
$\hat{b}^m_{f}{}{*_{A}}\hat{b}^{m \dagger}_{f}$. 

We pay attention to the
Clifford odd representations, but the proof is as well valid for  the Clifford even
representations.

To prove this statement let us start with $d=2(2n+1)$ with the irreducible
representation $f$ and the member  $m$ equal to $\hat{b}^{m \dagger}_{f}=
\stackrel{03}{(+i)}\stackrel{12}{(+)}\cdots \stackrel{d-1\, d}{(+)}$. To obtain 
the $2^{\frac{d}{2}- 1}-1$  rest irreducible representations we must transform 
each of possible pairs  $\stackrel{ab}{(k)}\stackrel{cd}{(k)}$ into 
$\stackrel{ab}{[k]}\stackrel{cd}{[k]}$. Let us start with the first two. One obtains
$\hat{b}^{m \dagger}_{f'}=
\stackrel{03}{[+i]}\stackrel{12}{[+]}\cdots \stackrel{d-1\, d}{(+)}$.
The two algebraic products,  $\hat{b}^m_{f}{}{*_{A}}\hat{b}^{m \dagger}_{f}
=\stackrel{03}{[-i]}\stackrel{12}{[-]} \stackrel{56}{[-]}\cdots \stackrel{d-1\, d}{[-]}$ and $\hat{b}^m_{f'}{}{*_{A}}\hat{b}^{m \dagger}_{f'}
=\stackrel{03}{[+i]}\stackrel{12}{[+]} 
\stackrel{56}{[-]}\cdots \stackrel{d-1\, d}{[-]}$, 
distinguish in the first two projectors. When replacing a pair by a pair in
$\hat{b}^{m \dagger}_{f}$, we end up with $2^{\frac{d}{2}- 1}$ different 
$\hat{b}^m_{f}{}{*_{A}}\hat{b}^{m \dagger}_{f}$, differing in all possible 
pairs $\stackrel{ab}{[+]}\stackrel{a'b'}{[+]}$ replacing $\stackrel{ab}{[-]}\stackrel{a'b'}{[-]}$. 

For $d=4n$  and the Clifford odd representations we must start with 
$\hat{b}^{m \dagger}_{f}=
\stackrel{03}{(+i)}\stackrel{12}{(+)}\cdots \stackrel{d-1\, d}{[+]}$,
  and then repeat all steps. We shall again obtain $2^{\frac{d}{2}- 1}$  different
 $\hat{b}^m_{f}{}{*_{A}}\hat{b}^{m \dagger}_{f}$. 
 
{\bf Statement 5a:} 
There are  $2^{\frac{d}{2}- 1}$  different algebraic products
 $\hat{b}^m_{f}{}{*_{A}}\hat{b}^{m \dagger}_{f}$.
 
Since due to {\it Statement 4.}  all the members of a particular  irreducible 
representation have the same algebraic product 
$\hat{b}^m_{f}{}{*_{A}}\hat{b}^{m \dagger}_{f}$, we can conclude that there
are  $2^{\frac{d}{2}- 1}$  different algebraic products
 $\hat{b}^m_{f}{}{*_{A}}\hat{b}^{m \dagger}_{f}$. 
 
 {\bf Statement 5b:} 
 Each creation operator $\hat{b}^{m \dagger}_{f}$ gives 
 nonzero contribution when applied on $|\psi_{oc}>$.
  
  There is one summand in $|\psi_{oc}>$, namely, $\hat{b}^{m}_{f}
 \hat{b}^{m \dagger}_{f}=\hat{b}^{m'}_{f} \hat{b}^{m' \dagger}_{f}$, 
 $\forall m'$, on which $\hat{b}^{m \dagger}_{f}$ gives a nonzero contribution: 
 $\sum_{f'=1}^{2^{\frac{d}{2}-1}} \hat{b}^{m \dagger}_{f}\,$
 $\hat{b}^{m}_{f'}\, \hat{b}^{m \dagger}_{f'}=$ 
 $\sum_{f'=1}^{2^{\frac{d}{2}-1}} \,
 \delta^{f f'} \hat{b}^{m \dagger}_{f'} $.
  
 {\bf Statement 5c:} 
 In odd representations the algebraic product of any two
 annihilation operators  $\hat{b}^{m}_{f}{}_{*_A} \hat{b}^{m'}_{f'}$ gives 
zero, as also the algebraic product of any two creation operators
$\hat{b}^{m \dagger}_{f}{}_{*_A} \hat{b}^{m' \dagger}_{f'}$ gives 
zero. Correspondingly the application of the annihilation operators on 
$|\psi_{oc}>$ gives  zero contribution.   
 
All annihilation operators are ''orthogonal'', as also all the creation operators are:
$\hat{b}^{m}_{f}{}_{*_A} \hat{b}^{m'}_{f'}=0$, 
 $\hat{b}^{m \dagger}_{f}{}_{*_A} \hat{b}^{m' \dagger}_{f'}=0$.
Within the same irreducible representation at least one nilpotent of the two creation
operators or  of the two annihilation  operators are  the same.
Among different irreducible representations of each kind separately, either
one nilpotent is the same in both operators appearing in the product,  or 
$\stackrel{ab}{(k)}$ multiplies $\stackrel{ab}{[k ]}$ or 
$\stackrel{ab}{[-k]}$ multiplies $\stackrel{ab}{(k)}$ or 
$\stackrel{ab}{[k]}$ multiplies $\stackrel{ab}{[-k ]}$, since one irreducible representation differs from the other in a pair $\stackrel{ab}{(k)}\stackrel{ef}{(k')}$
going to   $\stackrel{ab}{[k]}\stackrel{ef}{[k']}$ or  
$\stackrel{ab}{(k)}\stackrel{ef}{[k']}$ going to  $\stackrel{ab}{[k]}
\stackrel{ef}{(k')}$ or $\stackrel{ab}{[k]}\stackrel{ef}{(k')}$ going 
to $\stackrel{ab}{(k)}\stackrel{ef}{[k']}$.

  
  
 {\bf Statement 6:} 
 The operator $\tilde{S}^{cd}$, which does not belong to the
 Cartan subalgebra of  Eq,~(\ref{cartangrasscliff}), generates after the reduction
 of the Clifford space  a new irreducible representation, carrying different 
 family quantum number.
 
 
 The proof of {\it Statement 5.} 
 contains also the proof for  {\it Statement 6.}.   All the members of one irreducible representation are reachable by  the application of $S^{ab}$'s. Let us start in 
 $d=2(2n+1)$ with the Clifford odd representation containing the member
 $\hat{b}^{m \dagger}_{f}=
\stackrel{03}{(+i)}\stackrel{12}{(+)}\cdots \stackrel{b d}{(+)} \cdots
\stackrel{b' d'}{(+)}\cdots\stackrel{d-1\, d}{(+)}$ (with the family quantum numbers
$(\frac{i}{2}, \frac{1}{2},\cdots, \frac{1}{2}, \cdots, \frac{1}{2}, \cdots, 
\frac{1}{2} )$, determined by $\tilde{S}^{ab}$'s from Eq.~(\ref{cartangrasscliff})).
Operator $\tilde{S}^{b b'}$  transforms   $\hat{b}^{m \dagger}_{f}$
 into $\stackrel{03}{(+i)}\stackrel{12}{(+)}\cdots \stackrel{b d}{[+]} \cdots
\stackrel{b' d'}{[+]}\cdots \stackrel{d-1\, d}{(+)}$. This  new creation operator 
belongs to new  irreducible representation (with the family quantum numbers
$(\frac{i}{2}, \frac{1}{2},\cdots,- \frac{1}{2}, \cdots, -\frac{1}{2}, \cdots, 
\frac{1}{2})$), since it is not reachable by $S^{ab}$
(which generate all the rest members of the same irreducible representation).
Transforming all pairs into the new ones, one obtains $2^{\frac{d}{2}-1}$
families.

For $d=4n$  we start for odd representations by $\hat{b}^{m \dagger}_{f}=
\stackrel{03}{(+i)}\stackrel{12}{(+)}\cdots \stackrel{b d}{(+)} \cdots
\stackrel{b' d'}{(+)}\cdots\stackrel{d-1\, d}{[+]}$ and repeat the above 
procedure (by taking into account that $\tilde{S}^{b b'}$  transforms 
$\stackrel{b d}{(+)} \cdots \stackrel{b' d'}{[+]}$ into $\stackrel{b d}{[+]} \cdots
\stackrel{b' d'}{(+)}$).

Similarly we can find all the families of the Clifford even representations, if
taking into account Eq.~(\ref{tildegammareduced})

 

 {\bf Statement 7:} 
 Creation operators $\hat{b}^{m \dagger}_{f}$ and their 
 Hermitian conjugated partners annihilation operators, appearing in
  Eq.~(\ref{almostDirac}), have the properties 
\begin{eqnarray}
\label{almostDirac1}
\hat{b}^{m}_{f} {}_{*_{A}}|\psi_{oc}>&=& 0\, \cdot|\psi_{oc}>\,,\nonumber\\
\hat{b}^{m \dagger}_{f}{}_{*_{A}}|\psi_{oc}>&=&  |\psi^m_{f}>\,,\nonumber\\
\{\hat{b}^{m}_{f}, \hat{b}^{m'}_{f'}\}_{*_{A}+}|\psi_{oc}>&=&
 0\,|\psi_{oc}>\,, \nonumber\\
\{\hat{b}^{m }_{f}, \hat{b}^{m' \dagger}_{f}\}_{*_{A}+}|\psi_{oc}>
&=&\delta^{mm'}|\psi_{oc}>\,,\nonumber\\
\{\hat{b}^{m  \dagger}_{f}, \hat{b}^{m' \dagger}_{f'}\}_{*_{A}+}|\psi_{oc}>
&=&0\cdot|\psi_{oc}>\,.
\end{eqnarray}
Let us prove this statement step by step:\\
a. The last line  of Eq.~(\ref{almostDirac1}) requires that $[\hat{b}^{m  \dagger}_{f}, \hat{b}^{m' \dagger}_{f'}=0] |\psi_{oc}>$. \\
a.i. The product of two equal creation operators is zero, since the product of two
nilpotents of the same kind, $(\stackrel{ab}{(k)})^2=0$, gives zero.\\
a.ii. All the creation operators can be obtained from the starting one by the
application $S^{cd}$.  Since at this application at least one nilpotent remains 
the same, it follows that all creation operators within the same irreducible 
representation are orthogonal 
($\hat{b}^{m \dagger}_{f}{}_{*_A}\hat{b}^{m' \dagger}_{f}=0$),
by themselves and then also on $|\psi_{oc}>$. \\
a.iii. The creation operators of two different irreducible representations $f\ne f'$ 
can be obtained from the starting one by replacing two nilpotents 
$\stackrel{ab}{(k)}{}_{*_A}\stackrel{ef}{(k')}$ by 
$\stackrel{ab}{[k]}{}_{*_A}\stackrel{ef}{[k']}$, 
a nilpotent $\stackrel{ab}{(k)}$ and a projector $\stackrel{ab}{[k]}$ 
by the projector $\stackrel{ab}{[k]}$ and the nilpotent $\stackrel{ab}{(k)}$,
or two projectors $\stackrel{ab}{[k]}$ $\stackrel{ab}{[k']}$ by two nilpotents.
Since at least one of the nilpotents remains the same, it follows that
all the same members of different irreducible representations are orthogonal,
($\hat{b}^{m \dagger}_{f}{}_{*_A}\hat{b}^{m \dagger}_{f'}=0$),
by themselves and then also on $|\psi_{oc}>$.\\
a.iv. The creation operators, belonging to two different irreducible representations
$(f, f')$ and to two different members $(m,m')$ have the property 
($\hat{b}^{m \dagger}_{f}{}_{*_A}\hat{b}^{m' \dagger}_{f'}=0$)
 (are  orthogonal), due to the way how they are created 
($\hat{b}^{m \dagger}_{f}{}_{*_A}\hat{b}^{m' \dagger}_{f'}
=$ $ S^{m m'}\hat{b}^{m' \dagger}_{f}{}_{*_A}\hat{b}^{m' \dagger}_{f'}=0$,
as it is proven under iii.); Either the two have the same nilpotent or 
there appear a product of two projectors of the same type with opposite $k$ 
($\stackrel{ab}{[k]}$, $\stackrel{ab}{[-k]}$), or it appears 
($\stackrel{ab}{[k]}{}_{*_A}\stackrel{ef}{(-k)}=0$) or 
($\stackrel{ab}{(k)}{}_{*_A}\stackrel{ef}{[k]}=0$), Eq.~(\ref{graficcliff}).\\
%
%
b. The first line requires $\hat{b}^m_{f}{}_{*_A}|\psi_{oc}>=0$, which is 
equivalent to requiring $\hat{b}^{m}_{f}{}_{*_A}\hat{b}^{m'}_{f'}{}_{*_A}
\hat{b}^{m' \dagger}_{f'}=0$, $\forall \,m' \, {\rm and}\, f'$, since the vacuum 
state is equal to $|\psi_{oc}>= \sum_{f=1}^{2^{\frac{d}{2}-1}} 
\hat{b}^m_{f}{}{*_{A}} \hat{b}^{m \dagger}_{f}$, independent of the 
choice of $m$. The proofs from a.i.-a.iv.
guarantee that $\hat{b}^m_{f}{}_{*_A}|\psi_{oc}>=0$, since 
$\hat{b}^m_{f}=(\hat{b}^{m \dagger}_{f})^{\dagger}$.
 \\
c. The second line requires that $\hat{b}^{m' \dagger}_{f'}{}_{*_A}|\psi_{oc}>
= |\psi^{m'}_{f'}>$.
Namely, $\hat{b}^{m' \dagger}_{f'}{}_{*_A}\hat{b}^{m}_{f'}{}_{*_A}
\hat{b}^{m\dagger}_{f'}$ $=\hat{b}^{m' \dagger}_{f'}$ $\forall m$,  what follows from the way how $|\psi_{oc}>$ is created, while 
$\hat{b}^{m' \dagger}_{f'}{}_{*_A}\hat{b}^{m}_{f}{}_{*_A} 
\hat{b}^{m \dagger}_{f}=0$, $\forall \, f\ne f'$, 
since the application of  $\hat{b}^{m' \dagger}_{f'}{}_{*_A}$ on 
$\hat{b}^{m}_{f}{}_{*_A}\hat{b}^{m\dagger}_{f'}$ gives zero due to the 
orthogonality of the members of different irreducible  representations.\\
d. The third line requires of Eq.~(\ref{almostDirac1}) requires that $[\hat{b}^{m}_{f}, \hat{b}^{m'}_{f'}=0] |\psi_{oc}>$.  Since $\hat{b}^{m'}_{f'}=
(\hat{b}^{m' \dagger}_{f'})^{\dagger}$, the proof is the same as in the case a..\\
e. The fourth line requires that $\{\hat{b}^{m }_{f}, \hat{b}^{m' \dagger}_{f}\}_{*_{A}+}|\psi_{oc}>=\delta^{mm'}|\psi^{m'}_{f}>$, which means that
$(\hat{b}^{m }_{f}{}_{*_A} \hat{b}^{m' \dagger}_{f}+ 
\hat{b}^{m' \dagger}_{f}{}_{*_A} \hat{b}^{m }_{f})_{*_{A}}|\psi_{oc}>=
\delta^{mm'}|\psi^{m'}_{f}>$. This proofs follows from a. and b..\\

Allowing however, that also $f\ne f'$, one finds that 
the term $\hat{b}^{m }_{f}{}_{*_A} \hat{b}^{m' \dagger}_{f'}{}_{*_A} 
|\psi_{oc}>$ gives a nonzero contribution --- what is an even Clifford object and not
 zero. But after reducing the Clifford algebra space, 
Eq.~(\ref{tildegammareduced}), the two irreducible representations $(f,f')$
carry two different family quantum numbers. The algebraic product has no 
meaning any longer: Two different families, reachable from each other 
by $\tilde{S}^{ab}$, are orthogonal in the sense of the tensor product.
 

(As an example let us demonstrate this on the case $d=(5+1)$, presented on 
Table~\ref{cliff basis5+1.}: $\hat{b}^{m =1}_{f=II}=
{\scriptstyle (-)} \stackrel{56}{(-)}|\stackrel{12}{[+]} \stackrel{03}{[+i]}$, 
$\hat{b}^{m=1 \dagger }_{f=I}= \stackrel{03}{(+i)}\,\stackrel{12}{(+)}| \stackrel{56}{(+)}$. One obtains $\hat{b}^{m =1}_{f=II}{}_{*_A}
\hat{b}^{m =1 \dagger}_{f=I}{}_{*_A} \,|\psi_{oc}>= \stackrel{03}{(+i)}\,\stackrel{12}{(+)}\stackrel{56}{[-]}$, what represents  a even Clifford object and not 
a nonzero contribution.)



%

 %
(c.i. Let us show as an example, that $\hat{b}^m_{f}{}_{*_A}\hat{b}^{m'}_{f}=0$. 
Each annihilation operator of the same irreducible representation $f$ and
different member $m'$ follows from $\hat{b}^m_{f}{}$,
$\hat{b}^{m'}_{f}=(S^{m' m} \hat{b}^{m\dagger}_{f})^{\dagger} =$
$\hat{b}^{m}_{f} (S^{m m'})^{\dagger}$. Correspondingly we have
$\hat{b}^m_{f}{}_{*_A}\hat{b}^{m'}_{f'}=
\hat{b}^m_{f}{}_{*_A}\hat{b}^{m}_{f} (S^{m m'})^{\dagger}=0$.)\\

{\bf Statement 8:} $
<\bf {\psi^{s' }}_{f'} (\vec{p'})| \bf {\psi^{s }}_{f} (\vec{p})>=$
$\delta^{s s'} \delta_{f f'}\,\delta(\vec{p}'-\vec{p})$.

 Let us prove this relation.

Since the ``basis vectors'' in internal space of fermions are orthogonal according to 
Eq.~(\ref{almostDirac}) 
\begin{eqnarray}
\label{ccorthogonal0app}
 \{\hat{b}^{ m}_{f}\,{}_{*_{A}}\,,\,
 \hat{b}^{ m' \dagger}_{f'}\,{}_{*_{A}}\}_{+}|\psi_{oc}>&=&
 \hat{b}^{ m}_{f}\;{}_{*_{A}}\, 
 \hat{b}^{ m' \dagger}_{f'}\,{}_{*_{A}}|\psi_{oc}> = \delta^{m m'} \delta_{f f'}\,
 |\psi_{oc}>\,,\nonumber
\end{eqnarray}
it follows by taking into account Eq.~(\ref{usefulcontinuous}) of 
App.~\ref{continuous}~\footnote{In Eq.~(\ref{usefulcontinuous}) one finds
the relation $<\vec{p}|f^{*}(\hat{\vec{p}})\, f(\hat{\vec{p\,'}})|\vec{p\,'}>=
 f^{*}(\hat{\vec{p}})\, f(\hat{\vec{p\,'}}) \delta(\vec{p}-\vec{p\,'})$.}, 
here we 
 leave out the tensor product between states of ordinary space and internal space
 of fermions.
\begin{eqnarray}
\label{ccorthogonal01app}
&&<0_{\vec{p}}|<\psi_{oc}|\hat{\bf b}^{s'}_{f'} (\vec{p}')
\hat{\bf b}^{s  \dagger}_{f} (\vec{p}) |\psi_{oc}>|0_{\vec{p}}>
\nonumber\\
&&= <0_{\vec{p}}|<\psi_{oc}|\sum_{m,m'} c^{s'm' *}{}_{f'} (\vec{p}') \,
\hat{b}_{\vec{p}'} \,\hat{b}^{m'}_{f'}\, c^{s m}{}_{f} (\vec{p}) \,
\hat{b}^{m \dagger }_{f}\,\hat{b}^{\dagger}_{\vec{p}}
\,|\psi_{oc}>|0_{\vec{p}}>\nonumber\\
&&=\sum_{m,m'} <\psi_{oc}|\hat{b}^{m'}_{f'} {}*_{A}
\hat{b}^{m  \dagger}_{f}  \,|\psi_{oc}> 
<0_{\vec{p}}| c^{s'm' *}{}_{f'} (\vec{p}') \, \hat{b}_{\vec{p}'} \,
\hat{b}^{\dagger}_{\vec{p}} \,c^{s m}{}_{f} (\vec{p}) |0_{\vec{p}}>=\nonumber\\
&&\sum_m\,  c^{ms *}{}_{f}  (\vec{p},|p^0|=|\vec{p}| )\;\; 
c^{m s' }{}_{f'} (\vec{p},|p^0|=|\vec{p}| )\, 
\delta_{f f'} \,\delta(\vec{p}'-\vec{p})\nonumber\\
&&= \delta^{s s'} \delta_{f f'}\,\delta(\vec{p}'-\vec{p})\,, 
 \end{eqnarray}
 since $<\psi_{oc}|\hat{b}^{m'}_{f'} {}*_{A}
\hat{b}^{m  \dagger}_{f}  \,|\psi_{oc}>=\delta^{m m'} \delta_{f f'}$
and
  $<0_{\vec{p}}|\hat{b}_{\vec{p}'} \hat{b}^{\dagger}_{\vec{p}} |0_{\vec{p}}>=
  \delta(\vec{p}'-\vec{p})$,  App.~\ref{continuous},  while
%
%
\begin{eqnarray}
\label{ccorthogonal1app}
&& \sum_{m} c^{ms *}{}_{f}  (\vec{p},|p^0|=|\vec{p}| )\;\; 
c^{m s' }{}_{f'} (\vec{p},|p^0|=|\vec{p}| ) =
 \delta^{s s'} \delta_{f f'}\, \,.
 \end{eqnarray}
 %



%
\section{Continuous spectra in momentum and coordinate space}
\label{continuous}




Creation  and annihilation operators for a single fermion state must include besides 
the internal space of fermions, described by the superposition of odd products 
of Clifford algebra operators $\gamma^a$'s, denoted by 
$\hat{b}^{m \dagger}_{f}$, also  the  momentum or coordinate part. We make 
the tensor product, $\,*_{T}\,$, of both parts, of the anticommuting internal one
and the commuting momentum or coordinate one, as suggested in 
Eq.~(\ref{wholespacegeneral}),
 \begin{eqnarray}
 \label{wholespacegeneralapp}
\{{\bf \hat{b}}^{m \dagger}_{f} (\vec{p}) \,&=& 
 \hat{b}^{m \dagger}_{f}\,\,*_{T}\, \hat{b}^{\dagger}_{\vec{p}}\}
\; |\psi_{oc}>\,*_{T}\, |0_{\vec{p}}> \,.                                                                                                 
 \end{eqnarray}
The quantum numbers $(m,f)$ represent  the internal part of anticommuting 
"basis states" $\hat{b}^{m \dagger}_{f}$ of the family member $m$ of the 
family $f$,   
the vacuum state written here as $|\psi_{oc}>\,*_{T}\, |0_{\vec{p}}>$ is  
the vacuum for the starting single particle state, from which one obtains  
other single particle states with the same internal part, by the operators,
$\hat{b}^{\dagger}_{\vec{p}}$, which pushes the momentum by an amount 
$\vec{p}$.

We study  free massless fermions in ($d=(d-1)+1$)-dimensional space, 
for any  $d$.
We treat  the continuous spectrum of  $\vec{p}$ assuming 
that solutions of the Euler-Lagrange equations relate momentum $\vec{p}$ 
and energy $p^0$, $(p^0)^2 =(\vec{p})^2$.

We follow up to Eq.~(\ref{eigenvalue}) more or less Ref.~\cite{Louisell}, while 
neither  Eq.~(\ref{eigenvalue}) nor the rest of equations are from 
Ref.~\cite{Louisell} or we have found in  the literature.

The commutation relations of the Hermitian operators $\hat{\vec{x}}$ 
and $\hat{\vec{p}}$, $\hat{\vec{x}}^{\dagger}=\hat{\vec{x}}$ 
and  $\hat{\vec{p}}^{\dagger}=\hat{\vec{p}}$,
in units in which $\hbar=1=c$,   are as follows
 \begin{eqnarray}
 \label{compx0}
\{\hat{x}^i\,, \hat{x}^j \}_- &=& 0\,, \quad 
\{\hat{x}^i\,, \hat{p}^j \}_-   =  \delta^{ij}\,,\quad
\{\hat{p}^i\,, \hat{p}^j \}_-   =  0\,.\quad
\end{eqnarray}
Let us  write down some useful relations, following from Eq.~(\ref{compx0}),
 \begin{eqnarray}
 \label{compx1}
\{\hat{\vec{x}}\,, f(\hat{\vec{x}}, \hat{\vec{p}})\}_- &=& 
i \frac{\partial f(\hat{\vec{x}}, \hat{\vec{p}})}{\partial \hat{\vec{p}}}\,, \quad 
\{\hat{\vec{p}}\,, f(\hat{\vec{x}}, \hat{\vec{p}})\}_-   =  
- i \frac{\partial f(\hat{\vec{x}}, \hat{\vec{p}})}{\partial \hat{\vec{x}}}\,,
\end{eqnarray}
Let us solve the eigenvalue equations for a state $| \vec{p}>$  and $| \vec{x}>$, 
both changing continuously with  $\vec{p}$ and $\vec{x}$, respectively,
 \begin{eqnarray}
 \label{eigenvalue0}
\hat{\vec{p}}\,| \vec{p}>&=& \vec{p}\,| \vec{p}>\,,\qquad \,
\hat{\vec{x}}\,| \vec{x}>  = \vec{x}\,| \vec{x}>\,,\nonumber\\
| \vec{p}>&=& e^{i \hat{\vec{x}}\cdot \vec{p}}\, |0_{\vec{p}}>\,,\quad\;
| \vec{x}>  =  e^{-i \hat{\vec{p}}\cdot \vec{x}}\, |0_{\vec{x}}>\,,\nonumber\\
< \vec{p}|&=& <0_{\vec{p}}|\,e^{-i \hat{\vec{x}}\cdot \vec{p}}\, ,\quad\;
< \vec{x}|  =  <0_{\vec{x}}|\,e^{i \hat{\vec{p}}\cdot \vec{x}}\, ,
\end{eqnarray}
the operator $ e^{i \hat{\vec{x}}\cdot \vec{p}}\, $ translates $|0_{\vec{p}}>$
to $|\vec{p}>$ and operator $e^{-i \hat{\vec{p}}\cdot \vec{x}}$ translates 
$|0_{\vec{x}}>$ to $|\vec{x}>$, where $|0_{\vec{p}}>$ and $|0_{\vec{x}}>$
are states with eigenvalues  $\vec{0}$ in both cases.

The continuous spectra are normalized as follows
 \begin{eqnarray}
 \label{eigenvalueorth}
<\vec{p}\,|\, \vec{p'}>&=& \delta(\vec{p}- \vec{p'})= 
\prod_{i=1, d-1} \delta(p^{i}-p^{'i})\,,\quad
<\vec{x}\,|\, \vec{x'}>  = \delta(\vec{x}- \vec{x'})=
\prod_{i=1, d-1} \delta(x^{i}-x^{'i})\,,
\end{eqnarray}
and fulfil the completeness relations 
 \begin{eqnarray}
 \label{eigenvalueorthcom}
\int |\vec{p}> d^{d-1} p <\vec{p}| &=& I\,, \quad
\int |\vec{x}> d^{d-1} x <\vec{x}|   = I\,.\\
\end{eqnarray}

Due to the fact that $\hat{\vec{p}}$ and $\hat{\vec{x}}$ do not commute, Eq.~(\ref{compx0}), we have to evaluate $<\vec{p}| \vec{x}>$, using Eq.~(\ref{eigenvalue0}) and the relations $ f(\hat{\vec{p}}) \,| \vec{p}>=
 f(\vec{p}) \,| \vec{p}>$ and $ f(\hat{\vec{x}}) \,| \vec{x}>=
 f(\vec{x}) \,| \vec{x}>$. It follows 
 \begin{eqnarray}
 \label{px0}
<\vec{p}\,|\, \vec{x}>&=& <\vec{p}\,| e^{-i \hat{\vec{p}}\cdot \vec{x}} 
|0_{\vec{x}}> = e^{-i \vec{p}\cdot \vec{x}} <\vec{p}\,|\, 0_{\vec{x}}>
\nonumber\\
&=& 
e^{-i \vec{p}\cdot \vec{x}} <0_{\vec{p}}\,|\, 
e^{-i \hat{\vec{x}}\cdot \vec{p}}|0_{\vec{x}}>=e^{-i \vec{p}\cdot \vec{x}} 
<0_{\vec{p}}\,|\, 0_{\vec{x}}>\,.
\end{eqnarray}
Taking into account $<\vec{p}\,|\, \vec{x}>^*=<\vec{x}\,|\, \vec{p}>$, and
the orthogonality and completeness relations, we 
evaluate the relation 
$|<0_{\vec{p}}\,|\, 0_{\vec{x}}>|=\frac{1}{(\sqrt{2\pi})^{d-1}}$,
 \begin{eqnarray}
 \label{px1}
\int <\vec{p}\,| \vec{x}> d\vec{x} <\vec{x}\,| \vec{p'}>&=& 
\delta(\vec{p}- \vec{p'}) = |<0_{\vec{p}}|0_{\vec{x}}>|^2
\int  e^{-i (\vec{p}-\vec{p}')\cdot \vec{x}} d\vec{x}\nonumber\\
& =& |<0_{\vec{p}}|0_{\vec{x}}>|^2
(2\pi)^{d-1}\,, 
\end{eqnarray}
from where we find 
 \begin{eqnarray}
 \label{px2}
<\vec{p}| \vec{x}>&=&e^{i \vec{p}\cdot \vec{x}} \frac{1}{\sqrt{(2 \pi)^{d-1}}}
=<\vec{x}| \vec{p}>^*
\end{eqnarray}

We started this section with the intention to define the creation operators
for fermions, which would include both spaces, the internal one and the
momentum (or coordinate) one, as presented in 
Eq.~(\ref{wholespacegeneralapp}). 

Therefore we rewrite Eqs.~(\ref{eigenvalue0}, \ref{eigenvalueorth}) as follows
 \begin{eqnarray}
 \label{eigenvalue}
| \vec{p}>&=& \hat{b}^{\dagger}_{\vec{p}}\,|0_{\vec{p}}>=
\hat{b}_{-\vec{p}}\,|0_{\vec{p}}>\,,\qquad \,
| \vec{x}>  = \hat{b}^{\dagger}_{\vec{x}}\,| 0_{\vec{x}}>=
\hat{b}_{-\vec{x}}\,| 0_{\vec{x}}>\,,\nonumber\\
< \vec{p}|&=& <0_{\vec{p}}|\,\hat{b}_{\vec{p}}\,=
<0_{\vec{p}}|\, \hat{b}^{\dagger}_{-\vec{p}}\,,\qquad \,
< \vec{x}|  = <0_{\vec{x}}| \, \hat{b}_{\vec{x}} = 
<0_{\vec{x}}| \, \hat{b}^{\dagger}_{-\vec{x}} \,,\nonumber\\
< \vec{p}| \vec{p'}>&=& <0_{\vec{p}}|\,\hat{b}_{\vec{p}}\,
\hat{b}^{\dagger}_{\vec{p'}}\,|0_{\vec{p}}>= \delta(\vec{p}-\vec{p'})\, , \;
< \vec{x}| \vec{x'}>= <0_{\vec{x}}|\,\hat{b}_{\vec{x}}\,
\hat{b}^{\dagger}_{\vec{x'}}\,|0_{\vec{x}}>= \delta(\vec{x}-\vec{x'})\, ,
\nonumber\\
\hat{b}_{\vec{p}}\,\hat{b}^{\dagger}_{\vec{p'}}&=&\delta(\vec{p}-\vec{p'})\,,
 \qquad
\hat{b}_{\vec{x}}\,\hat{b}^{\dagger}_{\vec{x'}}  =\delta(\vec{x}-\vec{x'})\,, 
\end{eqnarray}
with $<0_{\vec{p}}| 0_{\vec{p}}>=1 $ and $<0_{\vec{x}}| 0_{\vec{x}'}>=1 $
normalize to unity,
so that operating in the single particle space by $\hat{b}^{\dagger}_{\vec{p}}$ 
shifts the momentum by the amount $\vec{p}$, while the operator 
$\hat{b}_{\vec{p}}$ shifts the momentum by the amount $-\vec{p}$.
It then follows  
 \begin{eqnarray}
 \label{eigenvalue1}
 <\vec{p}| \vec{x}>&=&
<0_{\vec{p}}|\hat{b}_{\vec{p}}\,\hat{b}^{\dagger}_{\vec{x}}|0_{\vec{x}}>
=(<0_{\vec{x}}|\hat{b}_{\vec{x}}\,\hat{b}^{\dagger}_{\vec{p}} 
 |0_{\vec{p}}>)^{\dagger}\, \nonumber\\
 \{\hat{b}^{\dagger}_{\vec{p}},  \,\hat{b}^{\dagger}_{\vec{p'}}\}_{-}&=&
0\,,\qquad 
\{\hat{b}_{\vec{p}},  \,\hat{b}_{\vec{p'}}\}_{-}=0\,,\qquad
\{\hat{b}_{\vec{p}},  \,\hat{b}^{\dagger}_{\vec{p'}}\}_{-}=0\,,
\nonumber\\
\{\hat{b}^{\dagger}_{\vec{x}},  \,\hat{b}^{\dagger}_{\vec{x'}}\}_{-}&=&0\,,
\qquad 
\{\hat{b}_{\vec{x}},  \,\hat{b}_{\vec{x'}}\}_{-}=0\,,\qquad
\{\hat{b}_{\vec{x}},  \,\hat{b}^{\dagger}_{\vec{x'}}\}_{-}=0\,,
\nonumber\\
{\rm while}&&\nonumber\\
\{\hat{b}_{\vec{p}},  \,\hat{b}^{\dagger}_{\vec{x}}\}_{-}&=&
 e^{i \vec{p} \cdot \vec{x}} \frac{1}{\sqrt{(2 \pi)^{d-1}}}\,,\qquad,
\{\hat{b}_{\vec{x}},  \,\hat{b}^{\dagger}_{\vec{p}}\}_{-}=
 e^{-i \vec{p} \cdot \vec{x}} \frac{1}{\sqrt{(2 \pi)^{d-1}}}\,,
\end{eqnarray}
since $<\vec{p}| \vec{x}>= <0_{\vec{p}}|\hat{b}_{\vec{p}} \,
\hat{b}^{\dagger}_{\vec{x}}|0_{\vec{x}}>= 
e^{i \vec{p} \cdot \vec{x}} \frac{1}{\sqrt{(2 \pi)^{d-1}}}$
and $<0_{\vec{p}}|\hat{b}^{\dagger}_{\vec{x}} \,
\hat{b}_{\vec{p}}|0_{\vec{x}}>= 0$.
\\

Since the two spaces, the internal one and the ordinary one, are independent, 
it is proven 
that our assumption in Eq.~(\ref{wholespacegeneralapp})  is justified.

Let us present here some useful relations
\begin{eqnarray}
\label{usefulcontinuous}
f(\hat{\vec{p}})|\vec{p}>&=& f(\hat{\vec{p}})
 \hat{b}^{\dagger}_{\vec{p}}|0_{\vec{p}}>=f(\vec{p})
 \hat{b}^{\dagger}_{\vec{p}}|0_{\vec{p}}>\,, \quad
 f(\hat{\vec{x}})|\vec{x}>  =f(\hat{\vec{x}})
 \hat{b}^{\dagger}_{\vec{x}}|0_{\vec{x}}>=f(\vec{x})
 \hat{b}^{\dagger}_{\vec{x}}|0_{\vec{x}}>\,, \nonumber\\
 <\vec{p}|f^{*}(\hat{\vec{p}})\, f(\hat{\vec{p'}})|\vec{p'}>&=&
 f^{*}(\hat{\vec{p}})\, f(\hat{\vec{p'}}) \delta(\vec{p}-\vec{p'})\,, \quad
 <\vec{x}|f^{*}(\hat{\vec{x}})\, f(\hat{\vec{x'}})|\vec{x'}>=
 f^{*}(\hat{\vec{x}})\, f(\hat{\vec{x'}}) \delta(\vec{x}-\vec{x'})\,,\nonumber\\
 <\vec{p}|f(\hat{\vec{x}})|\vec{p'}>&=&
f(-i\frac{\partial}{\partial \vec{p}})\,\delta(\vec{p}-\vec{p'})\,, \quad
<\vec{x}|f(\hat{\vec{p}})|\vec{x'}>   = 
f(i\frac{\partial}{\partial \vec{x}})\,\delta(\vec{x}-\vec{x'})\,, \nonumber\\
<\vec{p}|\psi_{\vec{p}}>&=& \frac{1}{(\sqrt{2 \pi})^{d-1}}
\int  d^{d-1}x \, e^{-i \vec{p} \cdot \vec{x}}<\vec{x}|\psi_{\vec{x}}>\, ,
\nonumber\\ 
<\vec{x}|\psi_{\vec{x}}>&=& \frac{1}{(\sqrt{2 \pi})^{d-1}}
\int  d^{d-1}p \, e^{i \vec{p} \cdot \vec{x}}<\vec{p}|\psi_{\vec{p}}>\, ,
\end{eqnarray}
For
\begin{eqnarray} 
\label{usefulcontinuous1}
<\vec{p}|\psi_{\vec{p}}>&=& 
f(\hat{\vec{p}}) |\vec{p}>\, ,\quad
<\vec{x}|\psi_{\vec{x}}>  = 
g(\hat{\vec{x}}) |\vec{x}>\, ,\nonumber\\
&& {\rm it \;follows}\nonumber\\
<\vec{x} |\psi_{\vec{x}}>  &=& \frac{1}{(\sqrt{2 \pi})^{d-1}}
\int  d^{d-1}p \, e^{i \vec{p} \cdot \vec{x}} \, f(\vec{\hat{p}}) |\vec{p}>
=g(\hat{\vec{x}}) |\vec{x}> 
\,,\nonumber\\
<\vec{p} |\psi_{\vec{p}}>  &=& \frac{1}{(\sqrt{2 \pi})^{d-1}}
\int  d^{d-1}x \, e^{-i \vec{p} \cdot \vec{x}} \, g(\vec{\hat{x}}) |\vec{x}>
=f(\hat{\vec{p}}) |\vec{p}> 
\,.
\end{eqnarray}
%


Since the "basis vectors'' in internal space of fermions are orthogonal according to 
Eq.~(\ref{almostDirac}) 
\begin{eqnarray}
\label{ccorthogonal011}
 \{\hat{b}^{ m}_{f}\,{}_{*_{A}}\,,\,
 \hat{b}^{ m' \dagger}_{f'}\,{}_{*_{A}}\}_{+}|\psi_{oc}>&=&
 \hat{b}^{ m}_{f}\;{}_{*_{A}}\, 
 \hat{b}^{ m' \dagger}_{f'}\,{}_{*_{A}}|\psi_{oc}> = \delta^{m m'} \delta_{f f'}\,
 |\psi_{oc}>\,,
\end{eqnarray}
it follows by taking into account Eq.~(\ref{usefulcontinuous}) of 
App.~\ref{continuous}~~\footnote{In Eq.~(\ref{usefulcontinuous}) one finds
the useful relation $<\vec{p}|f^{*}(\hat{\vec{p}})\, f(\hat{\vec{p'}})|\vec{p'}>=
 f^{*}(\hat{\vec{p}})\, f(\hat{\vec{p'}}) \delta(\vec{p}-\vec{p'})$.}
\begin{eqnarray}
\label{ccorthogonal012}
&&<0_{\vec{p}}|{}*_{T}<\psi_{oc}|\hat{\bf b}^{s'}_{f'} (\vec{p'}){}*_{T}
\hat{\bf b}^{s  \dagger}_{f} (\vec{p}) \,{}*_{T}\,|\psi_{oc}>{}*_{T}|0_{\vec{p}}>
\nonumber\\
&&= <0_{\vec{p}}|{}*_{T}<\psi_{oc}|\sum_{m,m'} c^{s'm' *}{}_{f'} (\vec{p'}) \,
\hat{b}_{\vec{p'}} {}*_{T} \hat{b}^{m'}_{f'}\, c^{s m}{}_{f} (\vec{p}) \,
\hat{b}^{m \dagger }_{f}{}*_{T}\hat{b}_{\vec{p}}
\,{}*_{T}\,|\psi_{oc}>{}*_{T}|0_{\vec{p}}>\nonumber\\
&&=\sum_{m,m'} <\psi_{oc}|\hat{b}^{m'}_{f'} {}*_{A}
\hat{b}^{m  \dagger}_{f}  \,|\psi_{oc}> 
<0_{\vec{p}}| c^{s'm' *}{}_{f'} (\vec{p'}) \, \hat{b}_{\vec{p'}} \,
\hat{b}^{\dagger}_{\vec{p}} \,c^{s m}{}_{f} (\vec{p}) |0_{\vec{p}}>=\nonumber\\
&& \sum_{m} c^{ms *}{}_{f}  (\vec{p},|p^0|=|\vec{p}| )\;\; 
c^{m s' }{}_{f'} (\vec{p},|p^0|=|\vec{p}| ) \,\delta(\vec{p'}-\vec{p})\nonumber\\
&&= \delta^{s s'} \delta_{f f'}\,\delta(\vec{p'}-\vec{p})\,, 
 \end{eqnarray}
 %
due to
  $<0_{\vec{p}}|\hat{b}_{\vec{p}'} \hat{b}^{\dagger}_{\vec{p}} |0_{\vec{p}}>=
  \delta(\vec{p}'-\vec{p})$,  App.~\ref{continuous},  while
%
%
\begin{eqnarray}
\label{ccorthogonal1}
&& \sum_{m} c^{ms *}{}_{f}  (\vec{p},|p^0|=|\vec{p}| )\;\; 
c^{m s' }{}_{f'} (\vec{p},|p^0|=|\vec{p}| ) =
 \delta^{s s'} \delta_{f f'}\, \,.
 \end{eqnarray}
 %


\section{Acknowledgement}
 The author N.S.M.B. thanks Department of Physics, FMF, University of Ljubljana, 
Society of  Mathematicians, Physicists and Astronomers of Slovenia,  for supporting
 the research on the  {\it spin-charge-family} theory, the author H.B.N. thanks the 
Niels Bohr Institute for
 being allowed to staying as emeritus, both authors thank DMFA and 
 Matja\v z Breskvar of  Beyond Semiconductor for donations, in particular for 
sponsoring the annual workshops entitled  "What comes beyond the standard
 models" at Bled, in which the ideas and realizations, presented in this paper,
 were discussed.
Both authors thanks cordially the Editor Prof. A.Faessler and his Editorial board for
this opportunity. 


 

\end{document}